# High Efficiency Polymer based Direct Multi-jet Impingement Cooling Solution for High Power Devices

(MANUSCRIPT)

**Tiwei Wei**

Dissertation presented in partial fulfilment
of the requirements for the degree of
Doctor of Engineering Science (PhD):
Mechanical Engineering

In collaboration with:

imec

Examination Committee:
Prof. dr. A. Bultheel, chairman (KU Leuven)
Prof. dr. ir. M. Baelmans, promoter (KU Leuven)
Dr. ir. H. Oprins, co-promoter (imec)
Prof. dr. ir. R. Puers (KU Leuven)
Prof. dr. I. De Wolf (KU Leuven)
Prof. M. S.Bakir (Georgia Institute of Technology)
Dr. ir. E. Beyne (imec)

March 2020

# Abstract

To cope with the increasing cooling demands for future high-performance devices and 3D systems, conventional liquid cooling solutions such as (microchannel) cold plates are no longer sufficient. Drawbacks of these conventional cold plates are the presence of the thermal interface material (TIM), which represents a major thermal bottleneck, and the temperature gradient across the chip surface. Alternative advanced liquid cooling solutions have been proposed such as inter-tier and intra-tier cooling for 3D systems. These solutions are however not compatible with the fine pitch requirements for high bandwidth communication between different tiers of a 3D system.

Liquid jet impingement cooling is an efficient cooling technique where the liquid coolant is directly ejected from nozzles on the chip backside resulting in a high cooling efficiency due to the absence of the TIM and the lateral temperature gradient. In literature, several Si-fabrication based impingement coolers with nozzle diameters of a few tens of μm have been presented for common returns, distributed returns or combination of micro-channels and impingement nozzles. The drawback of this Si processing of the cooler is the high fabrication cost. Other fabrication methods for nozzle diameters of a few hundred μm have been presented for ceramic and metal. Low cost fabrication methods, including injection molding and 3D printing have been introduced for much larger nozzle diameters (mm range) with larger cooler dimensions. These dimensions and processes are however not compatible with the chip packaging process flow.

This PhD focuses on the modeling, design, fabrication and characterization of a micro-scale liquid impingement cooler using advanced, yet cost efficient, fabrication techniques. The main objectives are: (a) development of a modeling methodology to optimize the cooler geometry; (b) exploring low cost fabrication methods for the package level impingement jet cooler; (c) experimental thermal and hydraulic characterization and analysis of the fabricated coolers; (d) applying the direct impingement jet cooling solutions to different applications.

# Contents

# Abbreviations & Symbols

## List of abbreviations

| | |
|---|---|
| B.C | Boundary Conditions |
| BEOL | Back-end of Line |
| CAD | Computer-Aided Design |
| CFD | Computational Fluid Dynamics |
| CHT CFD | Conjugate Heat Transfer and Fluid Dynamics Simulations |
| COP | Coefficient of Performance |
| CPU | Central Processing Unit |
| DI | De-Ionized (water) |
| DI | De-Ionized (water) |
| DLP | Digital Light Processing |
| DMLS | Direct Metal Laser Sintering |
| DNS | Direct Numerical Simulation |
| DOE | Design of Experiments |
| DRIE | Si Deep reactive-ion Etching |
| FC-LPBGA | Flip-chip Low-Profile BGA |
| FDM | Fused Deposition Modeling |
| FEM | Finite element Model |
| FL | Flow Rate |
| FM | Full Cooler Level Model |
| GCI | Grid Convergence Index |
| GPU | Graphics Processing Unit |
| HDT | Heat Deflection Temperature |
| HS | Hotspots |
| IGBT | Insulated Gate Bipolar Transistor |
| ITR | Interfacial Thermal Resistance |
| LED | Light-Emitting Diode |
| LES | Large Eddy Simulations |
| LIGA | Lithography, Electroplating, and Molding |
| MEMS | Microelectromechanical System |
| MLC | Multilayer Ceramic Technology |
| MM | Micro-Machined |
| MMC | Manifold Microchannel (heat sink) |
| MOS | Metal Oxide Semiconductor |
| MOSFET | Metal–Oxide–Semiconductor Field-Effect Transistor |
| PCB | Printed Circuit Board |
| PCM | Phase-Change Materials |
| PTCQ | Packaging Test Chip Version Q |
| PVC | Polyvinyl Chloride |
| QUICK | Quadratic Upstream Interpolation for Convective Kinematics |
| RANS | Reynolds-averaged Navier–Stokes |
| RD | Output Reading |
| RTD | Resistance Temperature Detectors |
| SAM | Scanning Acoustic Microscopy |
| SiC | Silicon Carbide |
| SIMPLE | Semi Implicit Method for Pressure Linked Equations |
| SLA | Stereo-Lithography |
| SST | Shear Stress Transport |
| TCR | Temperature Coefficient of Resistance |
| TEC | Thermo-electric cooler |
| TIM | Thermal Interface Materials |
| TLC | Thermochromic Liquid Crystals |
| TMV | Through-Mold-Via |
| TPP | Two Photon Polymerization |
| TSP | Temperature Sensitive Paint |
| TSV | Through-Si Via |
| TTC | Thermal Test Chip |
| TTV | Thermal Test Vehicle |
| UC | Unit Cell Model |
| UV | Ultra-violet |
| VLSI | Very Large-Scale integration |

# Symbols

| Symbol | What | Unit |
|---|---|---|
| $\bar{T}_{chip}$ | average temperature of the heat source | ℃ |
| $T_{in}$ | coolant inlet temperature | ℃ |
| $\sigma$ | voltage versus temperature sensitivity of the diodes | |
| $\Delta T_{avg}$ | Chip temperature increases | ℃ |
| $R_{th}$ | thermal resistance | K/W |
| $Q_{heater}$ | heat generated in the heater cells | W |
| $T_{amb}$ | ambient temperature | ℃ |
| $Q_{loss}$ | Chip power loss | W |
| $\overline{T}_s$ | average cooling surface temperature | ℃ |
| $k_{si}$ | thermal conductivity of silicon | W/mK |
| $A_{heater}$ | rea of the heaters (8 mm × 8 mm × 75%) | $cm^2$ or $mm^2$ |
| $W_p$ | pumping power | W |
| ΔP | pressure difference between the inlet and outlet of the cooler | Pa |
| $\dot{V}$ | volumetric flow rate | min/L |
| $t_c$ | chip thickness | μm |
| $V_{heater}$ | voltage of all the connected heaters | V |
| $I_{heater}$ | current across the heaters | A |
| $R_{th}^*$ | normalized thermal resistance | $K.cm^2/W$ |
| $W_p^*$ | normalized pump power | $W/cm^2$ |
| A | chip area | $cm^2$ or $mm^2$ |
| $d_i$ | inlet diameter | μm |
| $d_o$ | outlet diameter | μm |
| H | standoff between the jet exit and the chip | μm |
| t | nozzle thickness | μm |
| L | unit cell length | μm |
| $\overline{V}_{in}$ | inlet velocity | m/s |
| $\overline{Nu_f}$ | Nusselt number based on average interface temperature of the chip | |
| $\overline{Nu_o}$ | Nusselt number based on stagnation temperature on the chip surface | |
| $\overline{Nu_j}$ | Nusselt number based on average junction temperature | |
| $Re_d$ | Reynolds number | |
| Pr | Prandtl number | |
| $k_{fl}$ | thermal conductivity of the fluid | W/mK |
| μ | dynamic viscosity | kg/(m s) |
| Cp | specific heat | J/(K kg) |
| f | friction loss coefficient | |
| k | pressure coefficient | |
| $Bi$ | Biot number | |

# Chapter 1

# 1. Introduction

### 1.1 Cooling solutions for high performance electronic applications

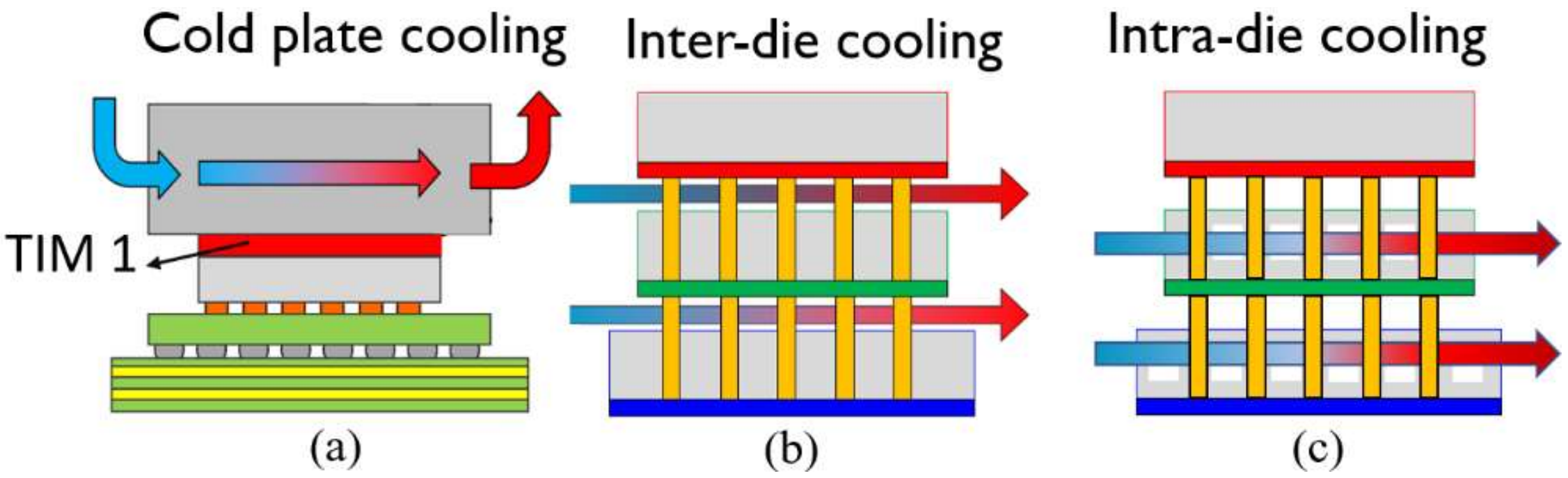

**Figure 1.1:** Schematic of the advanced liquid cooling solutions: (a) cold plate cooling; (b) inter-die cooling and (c) intra-die cooling.

Recent trends in high performance electronic applications show an increase in device power density due to the reduction in device size, combined with rising demands on control, energy efficiency and reliability [1]. For high power Insulated Gate Bipolar Transistors (IGBTs) modules, the heat flux in the transistor packages approaches 300 W/cm$^2$ [2], while for wide bandgap transistor electronics like GaN for radar and telecommunication applications, high performance applications such as CPU and GPU, the local power densities can be as high as 1 kW/cm$^2$ averaged over the chip [3]. To cope with the increasing heat flux challenge for future high power devices on the order of 600-1000 W/cm$^2$ [4], conventional liquid cooling solutions such as (microchannel) cold plates are no longer sufficient, and a transition towards direct liquid cooling, that achieves higher cooling rates, will be required. For the liquid cooling, the available solutions can be divided into two major parts: direct cooling or indirect cooling, which is shown in Figure 1.1 For the indirect cooling, the cooler is attached on the backside of the chip through thermal interface materials (TIM). While for the direct cooling, the coolant directly contacts the bare chip or component, even transfer through the interlayer of the 3D structure. This can be realized through microchannel etching on the chip backside or impingement jet cooling on the backside. For the immersion cooling, the whole package is immersed in the liquid. However, the main drawbacks of the available liquid cooling solutions are (1) the presence of the thermal interface material (TIM), which represents a significant thermal resistance contribution, and (2) the temperature gradient along the chip surface in the flow direction. Moreover, embedded cooling and interlayer cooling used in the 3D integration system is not compatible with the fine pitch requirement.

A comparison of the cooling performance between cold plates, microchannel cooling, and jet impingement cooling is presented in literature from an experimental [5] and numerical [6] perspective. The experimental study [5] shows that for a constant device junction temperature of 175°C the power dissipation capability of 60 W for a particular test case in the case of a cold plate can be increased to 99 W and 167 W respectively for a microchannel cooler and a jet impingement cooler. Moreover, CFD modeling studies [6] show that jet impingement cooling designs with interspersed fluid extraction powers yield lower average temperatures, improved temperature uniformity, and modest pressure drops compared to the microchannel, and jet arrays with edge fluid extraction. Liquid jet impingement cooling can be applied on the power module baseplate, on the substrate, or even on the bare die. Bare die liquid jet impingement cooling is the most efficient cooling option where the liquid coolant is directly ejected from nozzles on the heat source backside resulting in a high cooling efficiency due to the absence of the TIM and the lateral temperature gradient. Moreover, it can be used as hotspot targeted cooling by delivering high localized heat transfer rates at the location of the hot spot(s), which can improve the temperature uniformity. Recently, a cost-efficient cooling solution for a single MOSFET semiconductor based on a single jet direct impingement cooler was introduced [7]. This single chamber cooler with relatively larger nozzle diameter and simplified injection manifold can achieve average heat transfer coefficients of $1.2\times10^4$ $W/m^2K$ for a pumping power of 0.9 W for 8×8 mm chip size. Single jet cooling is however, limited to the efficient cooling of single hot spots since the obtained heat coefficient distribution is strongly non-uniform. The cooling efficiency quickly decays from the stagnation point towards the wall jet region. This can generate significant thermal gradients. Liquid multi-jet array impingement cooling, on the other hand, provides the scalability of the high heat coefficient area for areas from a few $mm^2$ to a few hundred $mm^2$, especially for multi-jets coolers with locally distributed outlets [8]. Furthermore, it is shown that cooling a hot spot array with a multi-jet cooler nozzle directed at hot spots is more efficient and achieves better temperature uniformity rather than cooling the complete surface area [9]. The main drawback of the multi-jet cooler is, however, the higher level of complexity to create the separate chambers for the fluid delivery and fluid extraction and the consequently added fabrication cost. The fabrication and thermal/hydraulic performance aspects of multi-jet coolers are reviewed in the next section.

## 1.2 State of the art liquid jet impingement coolers

In literature, a large variety of multi-jet impingement coolers fabricated with different materials and an extensive range of nozzle diameter values can be found. A selection of these multi-jet coolers has been summarized in Table 1, which lists the cooler material, nozzle array geometry, and the achieved thermal performance. Figure 1.2(a) shows the range of the nozzle diameters and the nozzle density for the considered impingement coolers, where the increase in nozzle density and the consequent reduction of the nozzle diameters result in an increasing complexity for the cooler fabrication. Multi-jet impingement coolers can achieve very high heat transfer coefficients. However, in the case of the small nozzle diameters, high pressure and consequently high pumping power is required. The relation between the total cooling power of the impingement coolers and the required pumping power, shown in Figure 1.2(b) for the available literature data, shows a clear correlation.

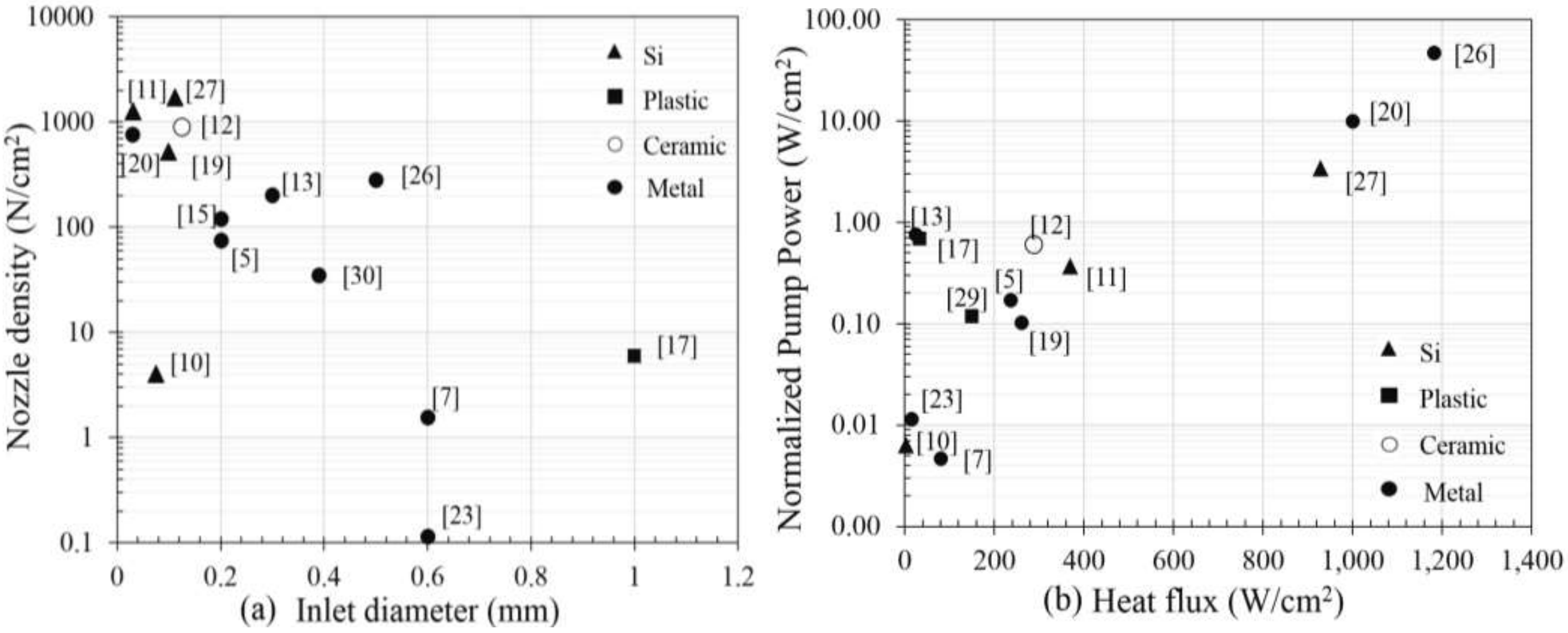

**Figure 1.2:** Graphical representation of the geometrical, thermal and hydraulic specifications of the literature survey from Table 1: (a) Trend of the nozzle density on the chip area as a function of the nozzle diameter. (b) The trend of the normalized required pump power in the cooler as a function of the dissipated heat flux in the chip. All quantities are normalized with respect to the chip area.

**Silicon** fabrication techniques, including etching, allow the very precise fabrication of nozzles with small diameters. Several Si-fabrication based impingement coolers with nozzle diameters of a few tens of µm have been presented for common returns, distributed returns or combination of micro-channels and impingement nozzles. Evelyn N. Wang [10] presented Si-based single jet and multi-jet impingement coolers with common returns with diameters ranging from 40 to 76 µm, achieving a heat transfer coefficient of $0.9\times 10^4$ $W/m^2K$ with a pump power of 6 mW for a chip size of 1 $cm^2$. In [11], Brunschwiler et al. demonstrated that Si processing could be used to fabricate performant and complex microjet array impingement coolers with branched hierarchical parallel fluid delivery and return architectures shown in Figure 1.3(a) with 50,000 inlet/outlet nozzles, allowing to increase the heat transfer coefficient up to $8.7\times10^4$ $W/m^2K$ with 1.43 W for a chip size of 4 $cm^2$. The main drawbacks of the Si-based coolers are the high pressure drop for the small diameter nozzles and the high fabrication cost. Other fabrication methods for nozzle diameters of a few hundred μm have been presented for ceramic and metal. G. Natarajan et al. [12] from IBM developed a microjet cooler with 1600 inlets and 1681 outlets using Multilayer **ceramic** technology (MLC) shown in Figure 1.3(b). Kyle Gould et al. [5] from Teledyne Scientific Company developed a compact jet impingement cooled **metal** heat exchanger with 48 200μm diameter jets for 600-V/50-A silicon carbide (SiC) power module used for bidirectional power conversion between a 28-V battery and a 300-V DC bus. Tolga Acikalin et al. [13] from Intel Labs paper developed a stainless steel direct liquid contact microchannel cold plate for bare die packages shown in Figure 1.3(d). Sheng Liu et al. [14] demonstrated a metal-based bottom-side microjet array cooling heat sink for the thermal management for both the active radar systems and high power density LEDs, in particular for LED array. International Mezzo Technologies [15] demonstrated a microjet cooler with a honeycomb structure, which contains microjets with 300 microns diameter, illustrated in Figure 1.3(c). A single jet metal cooler [7] with inlet diameter ranging from 0.6 mm to 1.6 mm was demonstrated on a bare MOSFET semiconductor device. Low-cost fabrication methods, including **injection molding** [16] and **3D printing** [17], have been introduced for much larger nozzle diameters (mm range) with larger cooler dimensions. B.P. Whelan et al. [17] developed a miniature 3D printed jet array water block using 49 individual 1 mm jets, as shown in Figure 1.4(a). Klaus Olesen et al. [16] demonstrated a module-level injection molded impingement cooler for the cooling of power electronic modules in hybrid electrical vehicle traction applications. These dimensions and processes are however, not compatible with the chip packaging process flow.

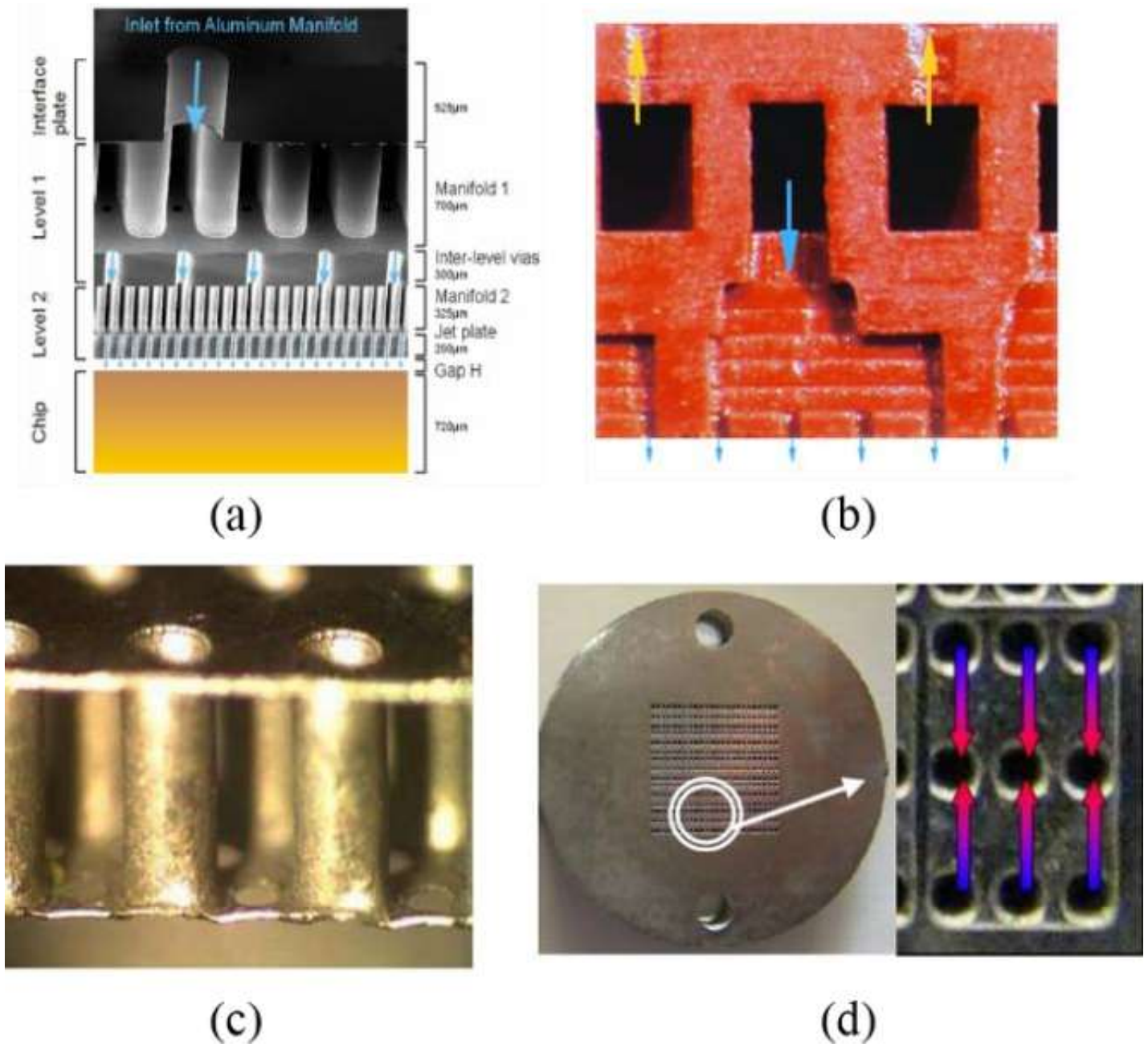

**Figure 1.3:** Impingement jet cooler with different fabrication techniques: (a) Silicon jet cooler [11]; (b) ceramic cooler [12]; (c) LIGA based cooler [15]; (d) metal-based cooler [13].

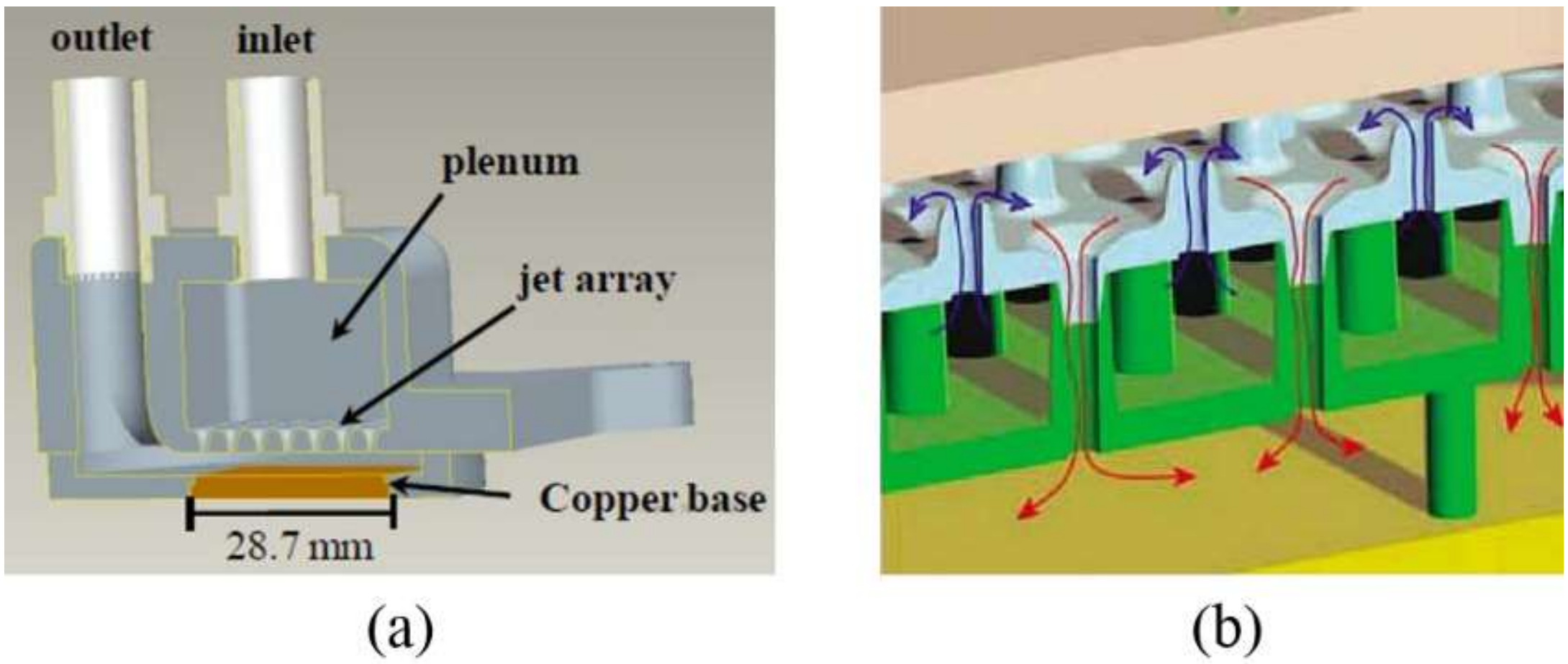

**Figure 1.4:** Impingement jet cooler with different fabrication techniques: (a) plastic cooler based on 3D printing techniques; (b) plastic cooler.

The thermal performance of the cooling solution can be further improved by modifying the contact surface of the semiconductor device on which the liquid coolant has impinged. Sidy Ndao et al. [18] experimentally investigated that the heat transfer can be as high as 3.03 or about a 200% increase by enhancing the target surface with a finned surface. This phenomenon can be exploited to develop "**hybrid**" micro-heatsinks, which contain impingement cooling channels as well as an array of fins created in the semiconductor device to achieve very high cooling rates, as shown in Figure 1.5. Yong Han et al. [19] from IME proposed a package-level hotspot cooling solution for GaN transistors using Si microjet/micro-channel hybrid heat sink, which can enable high spatially average heat transfer coefficient of 18.9 × $10^4$ W/$m^2$K with low pumping power of 0.17 W for a chip size of 0.49 $cm^2$. A.J. Robinson et al. [20] developed a micro heat sink designed with microchannels and an array of fins with integrated microjets using metallic additive manufacturing process, resulting in a heat transfer coefficient of 30×$10^4$ W/$m^2$K. These hybrid approaches are, however, a very disruptive cooling technology, requiring the etching of structures inside the device to be cooled. The overview of the literature study, summarized in Table 1, shows a large variety for the number of nozzles and the nozzle diameters for the liquid jet impingement coolers.

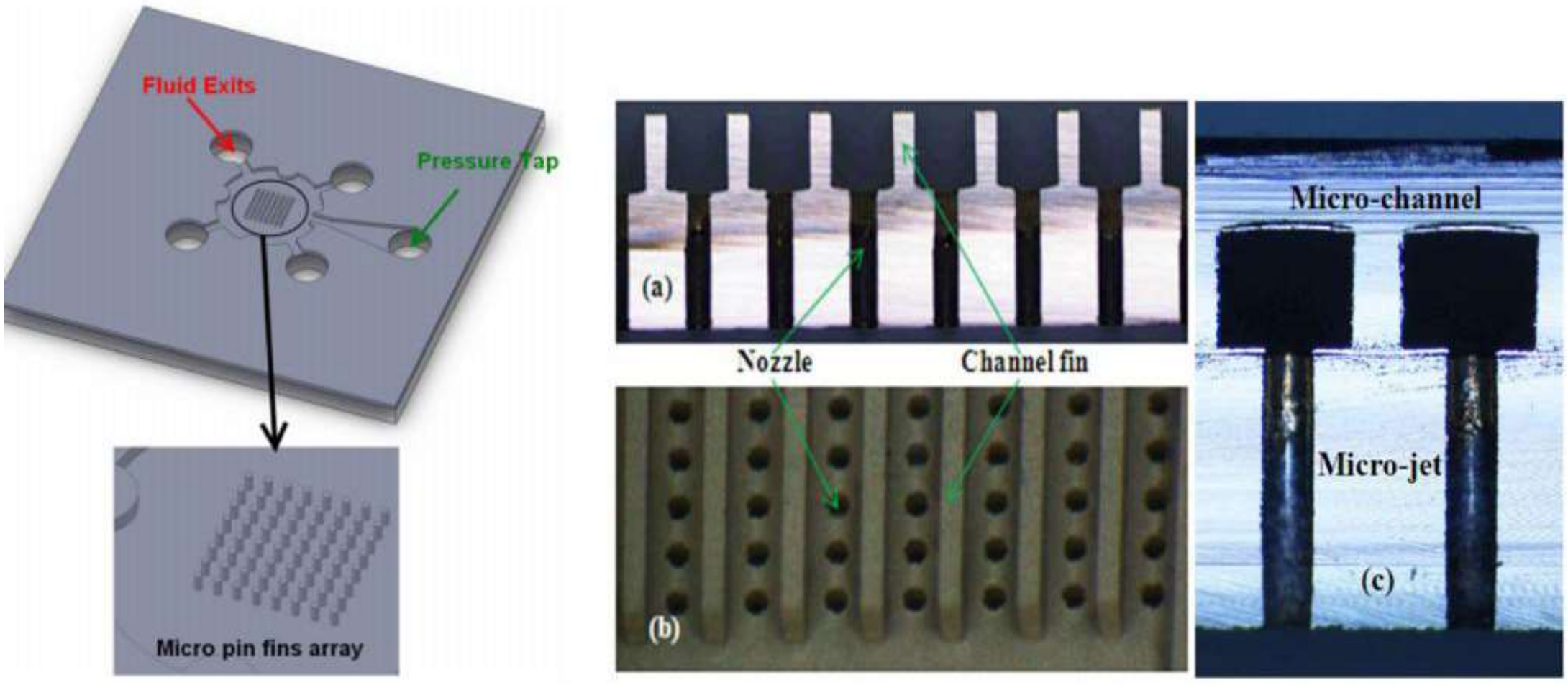

(a) (b)

**Figure 1.5:** "Hybrid" micro heatsinks which contain impingement cooling channels as well as an array of fins: (a) CAD structures; (b) fabricated micro heat sinks.

Table 1 Review of state of art cooling solutions part 1

| Material | Authors | Application | Coolant/ Phase | Number of jets | Nozzle diameter | Chip area | Power/Heat flux | Flowrate/Pressure drop/Pump power | Thermal Performance |
|---|---|---|---|---|---|---|---|---|---|
| Si | 2004,[10], E.N. Wang | TTV | Water/Tin=23° C | 4 | 76 μm | 1 $cm^2$ | 4W | FL=8 mL/min<br>$\Delta p$ = 47.57 KPa | $h_{max}$=4.4 W/($cm^2$K) |
| Si | 2010, [27] E.A.Browne | MEMS fabricated device- | Single phase Water Tin=23°C | 17 | 112μm | 1$mm^2$ | 930 W/$cm^2$ | $\Delta p$ = 348 KPa | $h_{avg}$=31.8 W/($cm^2$K) |
| Si | 2006, [11], T.Brunschwiler | CPU-Si die | Water/ single-phase | 50000 | 31-103 μm | 4 $cm^2$ | 370 W/$cm^2$ | FL=2.5 L/min<br>$\Delta p$ =35KPa<br>$Q_{Pump}$=1.46 W | $R_{avg}$= 0.17 K$cm^2$/W<br>$h_{av}$=8.7 W/($cm^2$K)<br>uniformity < 0.5°C |
| Si* Hybdrid | 2007, [28] E. G. Colgan | Bare die | Water@22°C | Channel pitch: 60-100 $\mu m$<br>450 $\mu m$ thick channel | | 4 $cm^2$ | 400 W/$cm^2$ | FL=1.25 L/min<br>$\Delta p$ =34.5 KPa<br>$Q_{Pump}$=0.72 W | $R_{avg}$=0.026 K/W |
| Si* Hybrid | 2015, [19] Yong Han | GaN device with diamond heat spreader | Single Phase Tin=25°C | 21×11 | 100μm | 0.49$cm^2$ | 110W;<br>260 W/$cm^2$ | $Q_{Pump}$=0.05W | $R_{avg}$= 0.09 K$cm^2$/W |
| Plastic | 2012, [17] B.P. Whelan | Heater block mimic CPU | Water Tin=15°C | 49 | 1mm | 8.24 $cm^2$ | 200W | FL=10 L/min<br>$\Delta p$ =37.5 KPa<br>$Q_{Pump}$=6.25 W | $R_{avg}$ =0.076 K/W |
| Plastic | 2015, [29], C.S. Sharma | Bare TTV die | Water Single phase Tin=20°C | Slot nozzle width | | 6.26 $cm^2$ | 285W;<br>150W/$cm^2$ | FL=1.2 L/min<br>$\Delta p$ =33 KPa<br>$Q_{Pump}$=0.66 W | $R_{avg}$= 0.2 K$cm^2$/W |
| Plastic | 2016, [2], A. S. Bahman | IGBT module | Glycol 50-water 50; Tin=20°C | 40 channels | 2.5-3mm wide Channel | --- | IGBT:50W<br>Diode: 25W | $\Delta p$ = 26 KPa<br>FL=5 L/min<br>$Q_{Pump}$=1.27 W | $R_{avg}$=0.2 K/W |
| Plastic | 2006,[9], Klaus Olesen | IGBT module | Ethylene-glycol/water 50%/50% | --- | ---- | 7.47$cm^2$ | --- | $\Delta p$ =0.19 bar<br>FL=12 L/min<br>$Q_{Pump}$=3.8 W | $R_{avg}$= 0.97 K$cm^2$/W |

Table 1 Review of state of art cooling solutions part 2

| Material | Authors | Application | Coolant/ Phase | Number of jets | Nozzle diameter | Chip area | Chip power /Heat flux | Flowrate /Pressure drop | Thermal performance |
|---|---|---|---|---|---|---|---|---|---|
| Ceramic | 2007, [12] G. Natarajan | CPU-Si die | liquid single-phase Tin=20°C | 1600 | 126 µm | 1.8$cm^2$ | 290W/$cm^2$ | FL=1.2 L/min<br>$\Delta p$ =53 kPa<br>$Q_{Pump}$=1.06 W | $h_{avg}$=5.2 W/($cm^2$K)<br>$R_{avg}$= 0.053 K/W |
| Metal | 2015, [5], Kyle Gould | SiC power module base plate | Water–ethylene glycol/ Tin=100 °C | 48 | 200 µm | 0.64 $cm^2$ (16 legs) | 151W | FL=195 ml/min<br>$\Delta p$ =34.47 kPa<br>$Q_{Pump}$=0.11 W | $T_j$ = 175 °C<br>$R_{avg}$=0.28 K$cm^2$/W |
| Metal | 2007, [15] Overholt MR | Electrical device | Water | 11×11 | 200-300µm | 1$cm^2$ | 1.5 kW/$cm^2$ | | $h_{avg}$=50 W/($cm^2$K)<br>$R_{avg}$=0.02 K/W |
| Metal | 2014, [13] Tolga Acikalin | VLSI Si device | Liquid/ Single phase Tin=22°C | 240 | 300µm | 1.18$cm^2$ Cooler | 40W | FL=1.18 L/min<br>$\Delta p$ =41.4kPa<br>$Q_{Pump}$=0.81 W | $R_{avg}$=0.24 K$cm^2$/W |
| Metal | 2008, [30], M. K. Sung | TTV | HFE 7100 - 40°C to 20°C | 5x14 | 390µm | 2 $cm^2$ | 16.1-304.9W/$cm^2$ | FL=6.82 to 45.5 mL/min | |
| Metal | 2012, [26], Skuriat, Robert | Si diode with AlN substrate | 40 °C water | 36 | 0.5mm | 12.7$mm^2$ | 150W<br>93 W/$cm^2$ | $Q_{Pump}$=6W | $R_{avg}$=0.2 K/W |
| Metal | 2017, [23], J. Jorg | IGBT-module | 22.5 °C water | 1 | 0.6 mm | 25mm×35mm | 125W | FL=300ml/min<br>$Q_{Pump}$=0.1W | $h_{avg}$=0.5 W/($cm^2$K) |
| Metal | 2018, [7], J. Jorg | MOSFET Device | 22.5 °C water | 1 | 0.6 mm | 0.64 $cm^2$ | 51W | FL=30 ml/min<br>$Q_{Pump}$=3mW | $h_{avg}$=1.2 W/($cm^2$K) |
| Metal* Hybrid | 2018, [20], A.J. Robinson | Heater block | 20°C water | 100 | 30$\mu m$ | 3.1×4.2 $mm^2$ | 1000 W/$cm^2$ | FL=0.5 L/min<br>$\Delta p$ =160 kPa<br>$Q_{Pump}$=1.3W | $h_{avg}$=30 W/($cm^2$K)<br>$R_{avg}$=0.03 K$cm^2$/W |

### 1.3 Objectives of the Dissertation

This Ph.D. work focuses on the modeling, design, fabrication, and characterization of the micro-scale liquid impingement cooler using advanced, yet cost-efficient, fabrication techniques. The above-presented literature study shows that direct liquid jet impingement cooling is an efficient cooling technique for high-performance electronic application that has been successfully applied with various materials including Si [10,11, 31], ceramic [12], metal [5, 7, 13, 14, 15] and plastic [16,17]. In addition, detailed reports cover experimental, theoretical and numerical analyses of different impingement jet configurations. These configurations range from single submerged jets [7, 23], to multiple submerged jets [33], and impinging jet cooling of electronic modules [5, 7, 26], configurations with a common return [10, 33] and with distributed returns [11]. However, a systematic study to determine the optimal nozzle array geometry in order to optimize the thermal, as well as the hydraulic performance is missing. The **first objective** is to investigate the thermal performance scaling trend of the required number of nozzles and the nozzle diameter for multi-jet impingement coolers by means of a numerical modeling approach, to derive a guideline for the cooler design and to predict the cooler's thermal and hydraulic performance for an arbitrary chip size.

The literature review shows that multi-jet impingement coolers can achieve very high cooling rates, but their major drawback is the complex and expensive fabrication. In this thesis, a cost-efficient high efficiency multi-jet impingement cooling solution is presented, fabricated using low-cost polymer fabrication techniques, targeted to directly cool the backside of high-power semiconductor devices. The schematic of the cooler for impingement jet cooling with distributed inlet and outlet channels for the delivery and removal of the coolant is shown in Figure 1.6. As shown in Figure 1.6(a), the inlet flow first goes through the inlet plenum and distributes the coolant to the individual inlet nozzles. After that, the fluid is ejected through the inlet nozzles and impinges on the heated chip surface shown in Figure 1.6(b). After striking on the chip surface, the fluid returns to the outlet plenum through the outlet nozzles. The **second objective** of this doctoral study is to demonstrate the feasibility of an integrated polymer impingement cooler and to benchmark its thermal performance with literature data for impingement coolers with various materials. In order to evaluate the thermal performance, the fabricated 3D-shaped polymer cooler is assembled to imec's 8×8 $mm^2$ thermal test chip [21] with integrated heaters and temperature sensors, which are used to mimic the power electronic chip or processor.

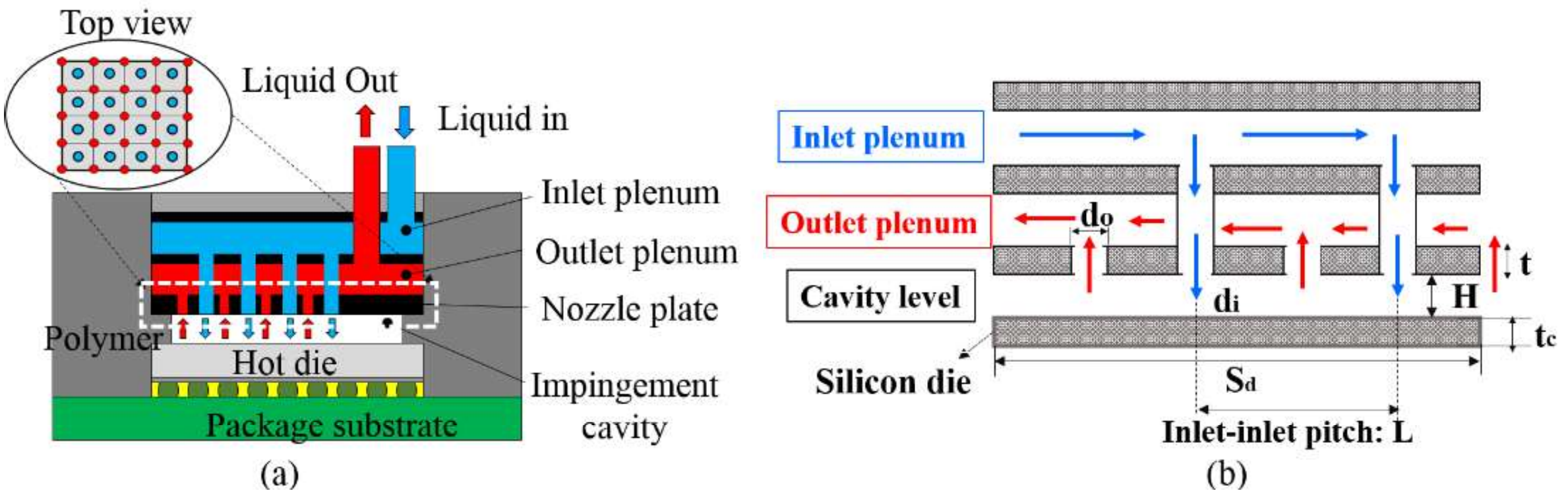

**Figure 1.6**: Chip level impingement jet cooler: (a) generic impingement cooler cross-section; (b) schematic with geometry parameters.

The **third objective** is to apply the impingement jet cooling solution to different applications, such as the hotspot target cooling, 2.5D interposer cooling, and large die cooling applications. Moreover, the reliability challenges will be discussed and addressed in this cooling design.

### 1.4 Original contributions of this work

The major contributions of this thesis are listed below:

- A multi-level modeling methodology based on unit cell model and full cooler level model has been introduced for the thermal and hydraulic assessment of impingement jet cooling. The novel unit cell model is firstly proposed to understand the scaling trend of jet number and jet diameter for multi-jet coolers. The full cooler level model including the manifold is used to extract the cooler level hydraulic performance such as the total pressure drop and flow uniformity, as well as the jet-jet interactions.
- A fast prediction method for the thermal and hydraulic performance of the cooler, based on dimensionless analysis, has been introduced. The *Nu-Re* and *f-Re* correlations have been extracted based on a large DOE of unit cell CFD simulations and have been validated by experiments using in-house demonstrators, and by literature data.
- The study of the full cooler level models indicates that polymer is a valuable alternative material for the fabrication of the impingement cooler instead of expensive Si-based fabrication methods. The modeling results show that the thermal conductivity of the cooler material has no impact on the thermal performance of the impingement cooler and that the heat transfer is dominated by the convection in the coolant, enabling the use of plastic materials with low thermal conductivity for the cooler.

- Polymer-based multi-jets impingement jet coolers including single jet cooler and multi-jet cooler have been designed and demonstrated based on mechanical machining and 3D printing. The benchmarking study with literature data for impingement coolers with a large range of inlet diameters shows a very good thermal performance of the fabricated polymer-based cooler for a low required pumping power.
- The hotspot targeted jet impingement cooling concept is successfully demonstrated with a chip-level jet impingement cooler with a 1 mm nozzle pitch and 300 μm nozzle diameter fabricated using additive manufacturing. The benchmarking study proves that the hotspot targeted cooling is much more energy-efficient than uniform array cooling, with lower temperature difference and lower pump power.
- Systematical investigation about the impingement jet cooling on the bare die and lidded package on the 2.5D interposer package has been performed. The study investigates the thermal coupling effects, TIM selection, as well as the manifold level design with lateral feeding scheme.
- The polymer-based impingement jet cooler using 3D printing, has been applied to a 23×23mm$^2$ large die with 1kW power dissipation. The measurement results show that the jet impingement cooling performance can be successfully described using a unit cell approach, allowing an easy scaling of the thermal performance for arbitrary die size applications. Long term thermal tests of 1000 h show a constant thermal performance and no degradation of the cooler material.
- The improvement of the manifold design has been assessed by means of a performance comparison of conceptual cooler manifold designs at the one hand and a topology optimization methodology at the other hand. For the conceptual designs, three different designs are proposed and compared, in terms of the average chip temperature, chip temperature gradient and pressure drop. A 2D topology optimization methodology has been introduced for the application of jet impingement flow, as a design tool to improve the inlet manifold geometry with respect to the coolant flow distribution over the nozzle array and the reduction of the pressure drop across the cooler.

## 1.5 Thesis organization

In **chapter 2**, the modeling approach and experimental tools used in this Ph.D. study are described. A multi-level modeling approach, including unit cell models and full cooler level models are used. Both models are based on conjugate heat transfer and fluid flow CFD modeling. The experimental tools, including the advanced thermal test chip and the thermo-fluidic measurement system, are introduced. Moreover, the cooler performance metrics with respect to the thermal and hydraulic performance are also defined in this chapter.

In **chapter 3**, the thermal and hydraulic behavior of the cooling unit cell is investigated by means of parametric analysis and dimensionless analysis. Firstly, the evaluation of the cooler performance is presented as a multi-objective optimization, illustrating the trade-off between thermal resistance and pumping power for all cooler designs. Secondly, using more than 1000 unit cell Computational Fluid Dynamics (CFD) simulations, the correlations for the dimensionless heat transfer (Nusselt number *Nu*) and the pressure coefficient in terms of the dimensionless flow velocity (Reynolds number *Re*) as a function of the normalized design parameters are developed.

In **chapter 4**, a single jet demonstrator is designed and fabricated as a proof of concept, for the fundamental understanding of the impingement jet cooling, from numerical modeling and experimental characterization point of view. With the successful experimental validation of the complex single jet model, the model assumptions, model simplifications, as well as the turbulence model comparisons are investigated.

In **chapter 5**, a multi-jet impingement cooler, fabricated using a mechanical micromachined process is demonstrated as a proof of concept. Firstly, the design considerations for the multi-jet coolers including the cavity height effects, inlet plenum thickness, cooler materials and liquid coolant are investigated systematically. In the second part, based on the thermal and hydraulic modeling results, a simplified board level polymer-based multi-jet cooler has been designed and. After that, the full cooler level CFD model is built to investigate the flow and thermal behavior of the multi-jet cooling. Moreover, the modeling results with uniform heating and quasi-uniform heating with hotspots are investigated and compared. Lastly, the thermal performance of the multi-jet demonstrator is compared with the performance of the single jet demonstrator, and benchmarked with state-of-the-art cooling solutions.

In **chapter 6**, an improved multi-jet cooler demonstrator based on 3D printing is introduced to exploit the unique capabilities of additive manufacturing. This leads to the manufacturability of customizable nozzle array, that can be matched to chip layout, as well as the possibility to create complex internal structures. Moreover, the defect measurement and manufacturing tolerance impact are analyzed and investigated. Finally, coolers with three different nozzle densities were fabricated and experimentally characterized.

In **chapter 7**, the hot spot targeted jet impingement cooling concept is successfully demonstrated with a chip-level jet impingement cooler fabricated using high-resolution 3D printing. The uniform array cooling and hot spot targeted cooling are compared using numerical modeling and experimental characterization. Moreover, detailed conjugate heat transfer CFD models have been used to assess the local flow distribution and temperature uniformity for the different coolers.

In **Chapter 8**, the package-level 3D printed direct liquid micro-jet array impingement cooling concept has been applied to a dual-chip module. The design, modeling, experimental characterization for the interposer cooler are discussed, including the comparison between the lidded cooling and lidless cooing, the thermal coupling effects and also the flow uniformity analysis. **Chapter 9** presents the design, fabrication, experimental characterization and reliability evaluation of a package level multi-jet cooler for large die sizes, fabricated using 3D printing. The first cooler version, referred to as the reference cooler, is the scaled-up design of the small chip size to much larger die size. The thermal and hydraulic performance of the reference large die cooler with and without lid is characterized and analyzed in section 9.3. The second version of the cooler, referred to as the improved cooler design, has an additional distribution layer to improve the coolant flow uniformity. In section 9.4, the thermo-hydraulic performance of the improved large die cooler is characterized and compared with the reference cooler. In the last section, a longer-term thermal measurement of 1000 hours for the reference large die cooler is performed and evaluated.

In **chapter 10**, the conceptual cooler designs' comparison and improved design based on a 2D topology optimization methodology are introduced. The objective is to improve the inlet manifold geometry with regard to the coolant flow distribution and pressure drop.

In **chapter 11**, an overview of the major findings and conclusions of this thesis are presented. Firstly, the material compatibility between cooler material, coolant, package materials and the reliability requirements of the application is discussed. Other aspects include the further continuation of the cooling design optimization, the experimental characterization and the cooling applications.

## References

[1]. H. Amano, Y Baines, et al., "The 2018 GaN power electronics roadmap," *J. Phys. D. Appl. Phys.*, vol. 51, no. 16, pp. 163001, March 2018.

[2]. A. S. Bahman and F. Blaabjerg, "Optimization tool for direct water cooling system of high power IGBT modules," in *Proc. 2016 European Conf. Power Electron. Appl.*, 2016, pp. 1–10.
[3]. S. V Garimella, T. Persoons, J. A. Weibel, and V. Gektin, "Electronics Thermal Management in Information and Communications Technologies: Challenges and Future Directions," *IEEE Trans. Components, Packag. Manuf. Technol.*, vol. PP, no. 99, pp. 1191–1205, September 2016.
[4]. S. Kandlikar, S. Garimella, D. Li, S. Colin, and M. R. King, *Heat transfer and fluid flow in minichannels and microchannels*, Oxford, Elsevier Ltd., 2014, Chap. 3.
[5]. K. Gould, S. Q. Cai, C. Neft, and S. Member, "Liquid Jet Impingement Cooling of a Silicon Carbide Power Conversion Module for Vehicle Applications," *IEEE Trans. Power Electronics*, vol. 30, no. 6, pp. 2975–2984, June 2015.
[6]. A. S. Rattner, "General Characterization of Jet Impingement Array Heat Sinks With Interspersed Fluid Extraction Ports for Uniform High-Flux Cooling," *J. Heat Transfer*, vol. 139, no. 8, p. 082201(1-11), August 2017.
[7]. J. Jorg, S. Taraborrelli, G. Sarriegui, R. W. De Doncker, R. Kneer, and W. Rohlfs, "Direct single impinging jet cooling of a mosfet power electronic module," *IEEE Trans. Power Electronics*, vol. 33, no. 5, pp. 4224–4237, May 2018.
[8]. A. Bhunia and C. L. Chen, "On the Scalability of Liquid Microjet Array Impingement Cooling for Large Area Systems," *J. Heat Transfer*, vol. 133, no. 6, 064501(1-7), January 2011.
[9]. K. Olesen, R. Bredtmann, and R. Eisele, "'ShowerPower' New Cooling Concept for Automotive Applications," in *Proc. Automot. Power Electron.,* no. June 2006, pp. 1–9.
[10]. E.N. Wang, L. Zhang, J.-M. Koo, J.G. Maveety, E.A. Sanchez, K.E. Goodson, and T.W. Kenny, "Micromachined Jets for Liquid Impingement Cooling for VLSI Chips," *J. Microeletromech. Sys*., vol. 13, no. 5, pp. 833-842, October 2004.
[11]. T. Brunschwiler et al., "Direct liquid jet-impingement cooling with micronsized nozzle array and distributed return architecture," in *Proc. IEEE Therm. Thermomechanical Phenom. Electron. Syst.*, 2006, pp. 196–203.
[12]. G. Natarajan and R. J. Bezama, "Microjet cooler with distributed returns," *Heat Transf. Eng.*, vol. 28, no. 8–9, pp. 779–787, July 2010.
[13]. T. Acikalin and C. Schroeder, "Direct liquid cooling of bare die packages using a microchannel cold plate," in *Proc. IEEE Therm. Thermomechanical Phenom. Electron. Syst.*, 2014, pp. 673–679.

[14]. S. Liu, T. Lin, X. Luo, M. Chen, and X. Jiang, “A Microjet Array Cooling System For Thermal Management of Active Radars and High-Brightness LEDs,” in *Proc. IEEE Electronic Components and Technology Conf.*, 2006, pp. 1634–1638.

[15]. Overholt MR, McCandless A, Kelly KW, Becnel CJ, Motakef S, “Micro-Jet Arrays for Cooling of Electronic Equipment,” in *Proc. ASME 3rd International Conference on Microchannels and Minichannels*, 2005, pp. 249-252.

[16]. [16] M. Baumann, J. Lutz, and W. Wondrak, “Liquid cooling methods for power electronics in an automotive environment,” in *Proc 2011 14th European Conference on Power Electronics and Applications*, 2011, pp. 1–8.

[17]. [17] B. P. Whelan, R. Kempers, and A. J. Robinson, “A liquid-based system for CPU cooling implementing a jet array impingement waterblock and a tube array remote heat exchanger,” *Appl. Therm. Eng.*, vol. 39, pp. 86–94, June 2012.

[18]. S. Ndao, H. J. Lee, Y. Peles, and M. K. Jensen, “Heat transfer enhancement from micro pin fins subjected to an impinging jet,” *Int. J. Heat Mass Transf.*, vol. 55, no. 1–3, pp. 413–421, January 2012.

[19]. Y. Han, B. L. Lau, G. Tang and X. Zhang, "Thermal Management of Hotspots Using Diamond Heat Spreader on Si Microcooler for GaN Devices," *IEEE Transactions on Components, Packaging and Manufacturing Technology*, vol. 5, no. 12, pp. 1740-1746, December 2015.

[20]. A. J. Robinson, W. Tan, R. Kempers, et al., “A new hybrid heat sink with impinging micro-jet arrays and microchannels fabricated using high volume additive manufacturing,” in *Proc. IEEE Annu. IEEE Semicond. Therm. Meas. Manag. Symp.*, 2017, pp. 179–186.

[21]. H. Oprins, V. Cherman, G. Van der Plas, J. De Vos, and E. Beyne, “Experimental characterization of the vertical and lateral heat transfer in three-dimensional stacked die packages,” *J. Electron. Packag.*, vol. 138, no. 1, pp. 10902, March 2016.

[22]. N. Zuckerman and N. Lior, “Jet Impingement Heat Transfer: Physics, Correlations, and Numerical Modeling,” *Advances In Heat Transfer*, Vol. 39, pp. 565-631, 2006.

[23]. J. Jorg, S. Taraborrelli, et al., “Hot spot removal in power electronics by means of direct liquid jet cooling,” in *Proc. IEEE Therm. Thermomechanical Phenom. Electron. Syst.*, 2017, pp. 471–481.

[24]. Skuriat , Robert, “Direct jet impingement cooling of power electronics,”. PhD thesis , University of Nottingham, 2012.

[25]. Y. Han, B. L. Lau, G. Tang, X. Zhang, and D. M. W. Rhee, “Si-Based Hybrid Microcooler with Multiple Drainage Microtrenches for High Heat Flux Cooling,”

*IEEE Trans. Components, Packag. Manuf. Technol.*, vol. 7, no. 1, pp. 50–57, 2017.
[26]. R. Skuriat and C. M. Johnson, “Thermal performance of baseplate and direct substrate cooled power modules,” in *Proc. 4th IET Int. Conf. Power Electron. Mach. Drives*, 2008, pp. 548–552.
[27]. E. A. Browne, G. J. Michna, M. K. Jensen, and Y. Peles, “Microjet array single-phase and flow boiling heat transfer with R134a,” *Int. J. Heat Mass Transf.*, vol. 53, no. 23–24, pp. 5027–5034, November 2010.
[28]. E. G. Colgan et al., “A practical implementation of silicon microchannel coolers for high power chips,” *IEEE Trans. Compon. Packag. Technol.*, vol. 30, no. 2, pp. 218–225, June 2007.
[29]. C. S. Sharma, G. Schlottig, T. Brunschwiler, M. K. Tiwari, B. Michel, and D. Poulikakos, “A novel method of energy efficient hotspot-targeted embedded liquid cooling for electronics: An experimental study,” *Int. J. Heat Mass Transf.*, vol. 88, pp. 684–694, September 2015.
[30]. M. K. Sung and I. Mudawar, “Single-phase hybrid micro-channel/micro-jet impingement cooling,” *Int. J. Heat Mass Transf.*, vol. 51, no. 17–18, pp. 4342–4352, August 2008.
[31]. Brunschwiler, T. et al. “Interlayer cooling potential in vertically integrated packages,” *Microsystem Technologies* 15, 57–74 (2009).
[32]. S. V. Garimella, R. a. Rice, “Confined and Submerged Liquid Jet Impingement Heat Transfer,” Journal of Heat Transfer. 117, 871 (1995).
[33]. P. S. Penumadu, A. G. Rao, “Numerical investigations of heat transfer and pressure drop characteristics in multiple jet impingement system,” *Appl Therm Eng*. 110, 1511–1524 (2017).

# Chapter 2

# 2. Modeling Approach and Experimental Tools

## 2.1 Multi Level Modeling

### 2.1.1 Introduction

In Figure 2.1(a), the schematic geometry of the multi-jet cooler is illustrated. Two different typical levels can be identified: local nozzle level and manifold level. As shown in Figure 2.1(b), the local level, where the heat and mass transfer occur, shows a scalable nozzle array with repeated unit cells and also includes the impingement cavity used to define the impingement jet region. The modeling of the local level is very important to understand the thermal and hydraulic behavior of the liquid jet as well as the jet-to-jet interactions in micro-jet cooling. This level determines the heat transfer coefficient applied on the chip surface.

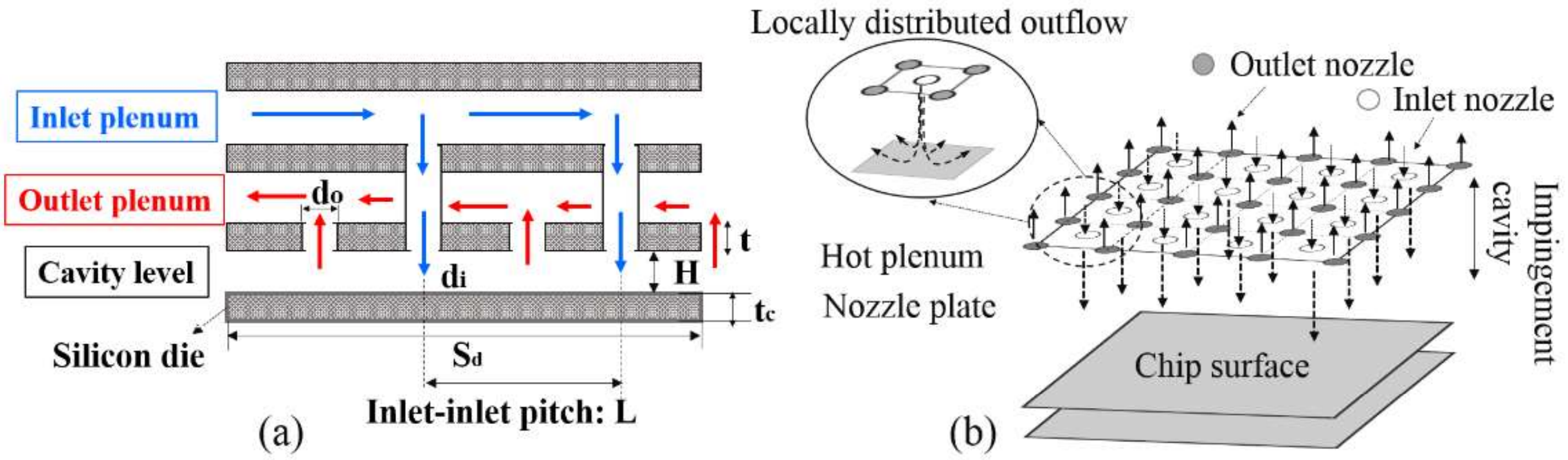

**Figure 2.1:** (a) Schematic of the impingement jet cooler and inside manifold fluid delivery system; (b) Typical flow regions of multi-jets impingement cooling.

As for the manifold level, there are two parts: one is the inlet manifold for delivering the incoming cold coolant flow; the other one is the outlet manifold for collecting the outcoming flow. The manifold level defines the hydraulic performance of the cooler, such as the pressure drop and flow uniformity. Moreover, full cooler level analysis is also necessary to study the system level optimization and cooling performance. From system level point of view, many factors need to be taken into account when designing a coolant distribution manifold, such as the placement of the coolant inlet and outlet, the designs to feed the coolant into the manifold inlet branches and the flow uniformity between different nozzles. Especially for the hot-spot target cooling, the 3D-full cooler level model should be considered to assess how to distribute the liquid coolant when multiple electronic modules with different power densities need to be cooled.

In our study, novel ***local level modeling with unit cell models*** is used to study the thermal performance of the core of the microjet cooling heat sink. As for the flow distribution through the manifold and the effect of flow distribution on the local heat transfer, a 3-D simulation of ***the full cooler level*** is necessary. In this thesis, a multi-level modeling approach with the combination of the jet nozzle level modeling and the manifold level modeling is implemented. In this section, the literature studies jet for impingement modeling and liquid manifold modeling will be reviewed and summarized. Moreover, a test case will be used to illustrate the multi-level modeling approach for a multi-jet cooler. The modeling methodology for the cooler and the interaction with the design, fabrication and experimental characterization activities is shown in Figure 2.2.

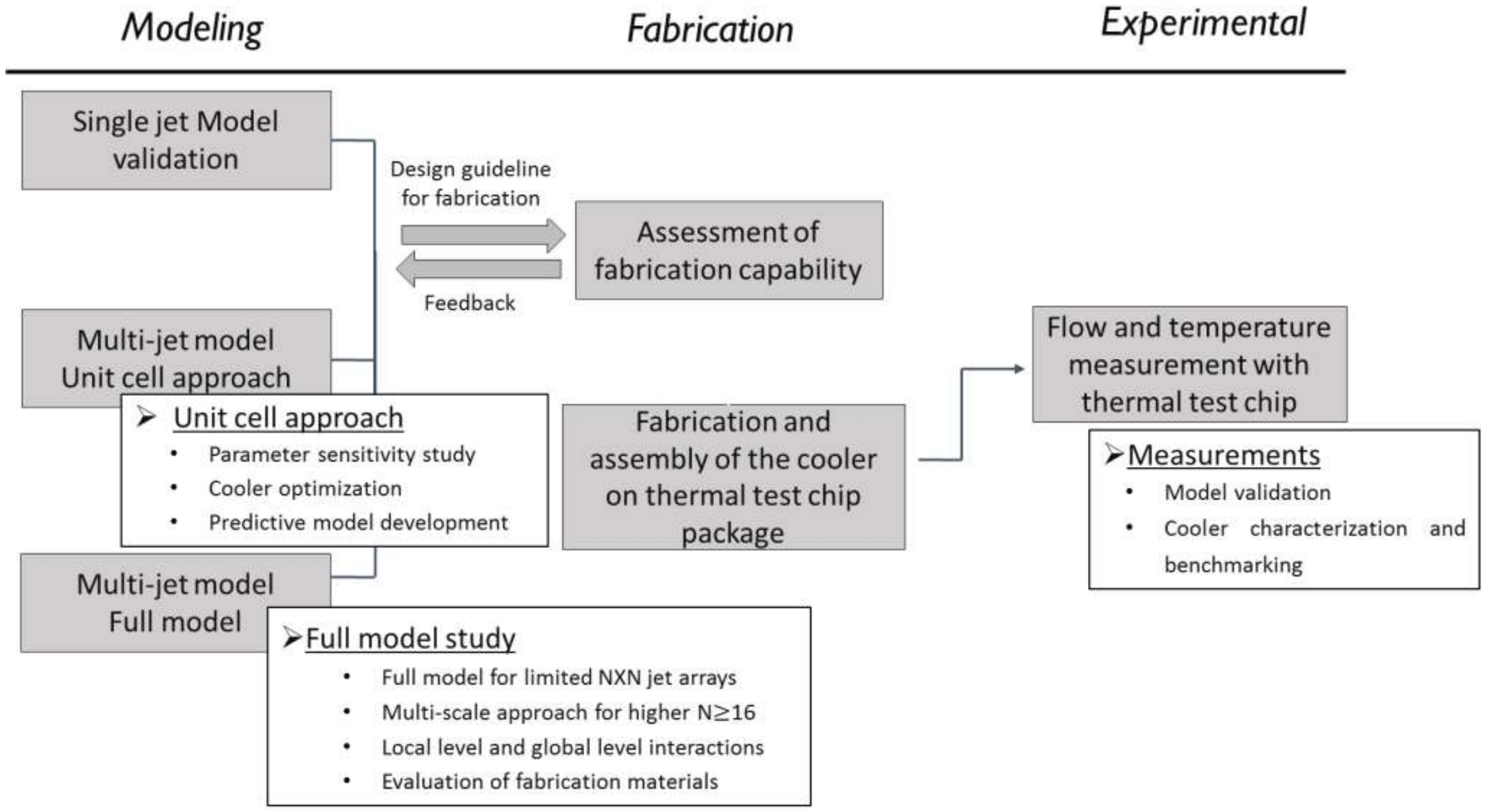

**Figure 2.2:** Modeling methodology for the cooler design, fabrication and experimental characterization.

### 2.1.2 State of art modeling overview

#### 2.1.2.1 Impingement jet modeling

The local level model is mostly based on the impingement jet cooling heat and mass transfer theory. For a typical impingement jet, five types can be differentiated: free surface jets, plunging jets, submerged jets, confined jets and wall jets (free surface) [1]. For the application for the cooling of electronic devices, jet impingement is mostly based on the confined jets shown in Figure 2.3, due to the limited available space. As illustrated in the schematic, there are four critical regions including the free jet region, the decaying jet region, the stagnation region and the wall jet region. The stagnation region is near the surface where the fluid changes direction, which produces the highest local heat transfer coefficient. After the fluid leaves the stagnation region, the heat transfer coefficient decreases with distance from the stagnation region, which is defined as the wall jet region.

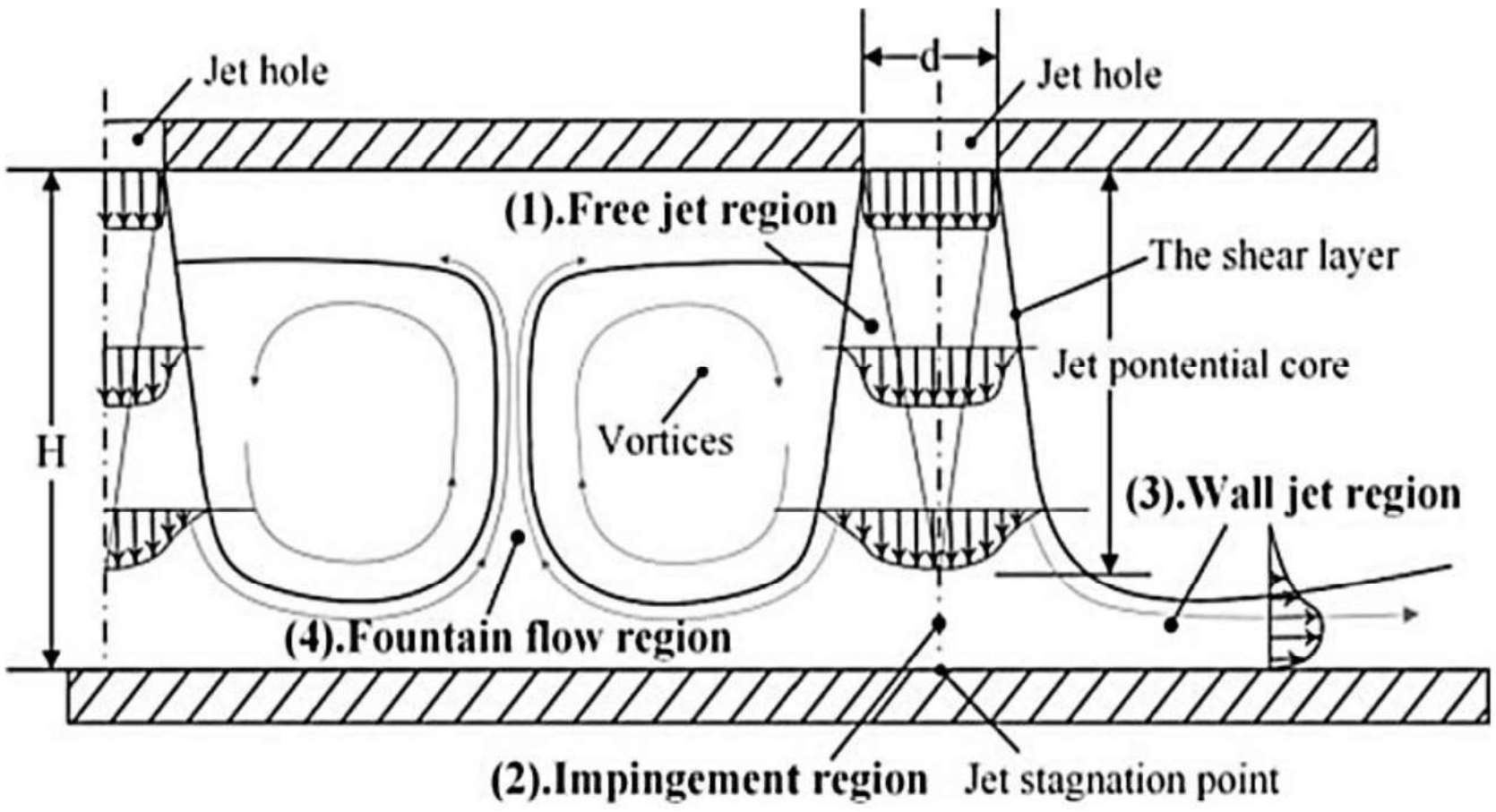

**Figure 2.3**: Free surface and confined-submerged jet array cooling configurations. [1]

For the modeling of impinging jets, there are mainly three types of methods: numerical modeling, analytical modeling and empirical modeling [2]. For the numerical modeling, the flow and thermal conjugate simulation can be implemented through CFD simulations. As for analytical modeling, the analysis is based on the mathematical models from the fluid and heat transfer equations. The empirical correlations based on Nusselt or Pressure drop correlations can be attained from available literature studies or through CFD simulation or experiments in real cases [2]. In this part, an overview of CFD simulation studies for jet impingement cooling is conducted and summarized.

In general, most of the studies are based on the two configurations with common return [3– 6] and locally distributed outlets [7–9]. While we focus on the Reynolds number, the underlying rationale for the increment in heat transfer arises due to the velocity fluctuations. This is best characterized in terms of the turbulence intensity parameter, usually considered to be proportional to $\mathrm{Re}_d < 1$ [10,11]. The jet Reynolds number $\mathrm{Re}_d$, is defined as $\mathrm{Re}_d = \frac{d_i \cdot V_{in}}{\nu}$, where $V_{in}$ is the mean inlet nozzle velocity, $d_i$ is the inlet nozzle diameter, and ν is the kinematic viscosity. For typical regions in impingement jet flow with a common outlet, the flow field exhibits laminar flow properties at $\mathrm{Re}_d \leq 1000$ based on the hydraulic nozzle diameter as representative length scale. At $\mathrm{Re}_d \geq 3000$ the flow has fully developed turbulent features. A transition region occurs with $1000 \leq \mathrm{Re}_d \leq 3000$ [12,13]. In literature, there are numerous articles concerning Computational Fluid Dynamics (CFD) numerical modeling of jet impingement cooling test cases using a large variety of turbulence models [14]. Narumanchi et al. [15] showed that the standard $k$-$\omega$ turbulence model can achieve less than 20% difference with experimental data for circular submerged jet configurations (Womac et al. [16]). For the submerged confined jet configuration (Garimella and Rice [17]), the difference can be as low as 10% over a wide range of Reynolds numbers. Isman et al. [18] showed that the overall performance of renormalization group (RNG) and standard k-ε models are better in comparison with other models by considering stagnation region and wall jet region. Esch and Menter [19] showed that the Menter's Shear Stress Transport (SST) model predicted heat transfer rates within 5% of those predicted by Durbin's $v^2$-f (V2F) model. In [20], John Maddox compared the transition SST and the V2F turbulence models, and finally selected the transition SST model for the 3 × 3 jet array with common outlet flow based on the computational cost considerations. Subrahmanyam et al. [21] used Large Eddy Simulations (LES) to study the unsteady flow and heat transfer characteristics of a single impinging jet at Reynolds number of 20,000 at four normalized nozzle-to-impinging plate distances ($0.5 \leq z/d \leq 2$). Sung [22] used the standard two-equation k-ε turbulence model to study the effects of the jet pattern on single-phase cooling performance of hybrid micro-channel/micro-circular-jet-impingement.

Several extensive review articles on CFD modeling are available. Polat et al. [23] reviewed the available numerical techniques to predict laminar and turbulent impingement heat transfer on a flat surface. Zuckerman and Lior [12] reviewed the suitability of different Reynolds-averaged Navier–Stokes (RANS) models in predicting average Nusselt number distribution and location and magnitude of the secondary peak in Nusselt number. The comparisons showed that direct numerical simulation (DNS)/LES time-variant models can accurately predict both Nu distribution and the secondary peaks. Moreover, the V2F and SST models also showed better predictions of fluid phenomenon in impinging jet flows while the standard $k$-$\varepsilon$ and $k$-$\omega$ models result in poor predictions. Behnia et al. [24] performed a critical review of important parameters in LES, DNS, and RANS-based techniques for computation of impinging flows. They concluded that the V2F model agrees very well with the experiments while $k$-$\varepsilon$ model highly overpredicts the rate of heat transfer and yield physically unrealistic behavior. Among all the models, LES model shows encouraging results and clarified the understanding for the unsteady flow and heat transfer characteristics of multiple impinging jets even though the high computing cost [25–31]. In [29] the objective and key findings of different LES studies dealing with impinging flows in recent times are reviewed. Cziesla et al. [30] demonstrated the ability of LES to predict local Nu under a slot jet within 10% of experimental measurements. Draksler et al. [31] carried out LES simulation to provide a detailed insight into unsteady flow mechanisms and the associated heat transfer process of multiple impinging jets. However, the computation cost of LES can be considerably reduced as compared to the DNS if sacrificing the accuracy with small-scale turbulence [31].

Based on the literature review, the transition SST model was selected as the most appropriate turbulence model for analysis of the multi-jet impingement cooling, in order to cover the laminar and transition regimes of the flow [2]. In appendix A, the comparison of different turbulence models is discussed using LES model as a benchmark. The comparison shows that the transition SST model and $k$-$\omega$ SST model both show excellent ability to predict the local or average Nu, as well as the local level pressure coefficient $f$ with less than 5% difference in the range of $30 < Re_d < 4000$, compared with the reference LES model.

**Table 2.1:** Comparison of CFD turbulence models used for jet impingement cooling [2]

| Turbulence model | Computational cost (time required) | Impingement jet transfer coefficient prediction | Ability to predict secondary peak |
|---|---|---|---|
| $k$-$\varepsilon$ | ★★★★<br>Low cost | ★<br>Poor: $Nu$ error of 15–60% | ★<br>Poor |
| $k$-$\omega$ | ★★★★<br>Low–moderate | ★★<br>Poor–fair: anticipate $Nu$ errors of at least 10–30% | ★★<br>Fair: may have incorrect location or magnitude |
| Realizable $k$-$\varepsilon$ and other $k$-$\varepsilon$ variations | ★★★★<br>Low | ★★<br>Poor–fair: expect $Nu$ errors of at least 15–30% | ★★<br>Poor–fair: may have incorrect location or magnitude |
| Algebraic stress model | ★★★★<br>Low | ★★<br>Poor–fair: anticipate $Nu$ errors of at least 10–30% | ★<br>Poor |
| Reynolds stress model (full SMC) | ★★<br>Moderate–high | ★<br>Poor: anticipate $Nu$ errors of 25–100% | ★★<br>Fair: may have incorrect location or magnitude |
| Shear stress transport (SST), hybrid method | ★★★<br>Low–moderate | ★★★<br>Good: typical $Nu$ errors of 20–40% | ★★<br>Fair |
| $v^2 f$ | ★★★<br>Moderate | ★★★★<br>Excellent: anticipate $Nu$ errors of 2–30% | ★★★★<br>Excellent |
| DNS/LES time-variant models | ★<br>Extremely high (DNS available for low $Re$ only) | ★★★★<br>Good–Excellent | ★★★★<br>Good–Excellent |

★: undesirable model characteristics
★★★★: excellent model characteristics

#### 2.1.1.2 Manifold level modeling

For the microchannel cooling heat sink, it is shown that the design of a manifold microchannel heat sink with alternating inlet and outlet channels has a big impact on the system pressure drop and thermal performance [32, 33]. With the manifold microchannel heat sink design, the flow length of the microchannel cooling will be reduced, while the flow length is determined by the length of the heat source. With shorter flow length, heat transfer coefficients can be enhanced by limiting the growth of the thermal boundary layers [34-36].

A lot of researchers conducted numerical simulations and experimental studies to optimize the geometry of the manifold microchannel heat sink. The experimental study shows that the thermal resistance of the manifold microchannel heat sink is approximately 35% lower than the traditional microchannel heat sink [32]. In literature [33], numerical simulation results indicate that an optimized manifold design can reduce the thermal resistance by 50% compared to a traditional microchannel heat sink. Moreover, a numerical study by Boteler et al. [37] indicates a more uniform flow distribution and lower pressure drop by as much as 97% for a 3D manifold design compared to a traditional microchannel design. Ryu et al. [38] reported that single phase manifold microchannel heat sinks can dissipate >50% higher heat fluxes than conventional microchannel heat sinks under the same pressure drop. Purdue university has published many studies [34, 35] on the design, fabrication, and experimental characterization of a compact hierarchical manifold for microchannel heat sink arrays with the two-phase cooling. They demonstrated that an intra-chip manifold microchannel heat sink can achieve extreme heat fluxes up to 910 W/cm$^2$ at pressure drops less than 162 kPa [34]. Moreover, for the hybrid cooling solution shown in Figure 1.5, it is very challenging to model each microjet accurately while considering the entire manifold with the microchannels. Therefore, porous-medium models are proposed for a system level flow and heat transfer analysis and optimization study of the manifold microchannel heat sink. In that methodology, the hydrodynamic performance of the heat sink is modeled via an equivalent permeability and porosity, without resolving the heat sink geometry down to scale of individual fins and channels [35].

The focus of this PhD is the multi-jet impingement cooling, the manifold level design is expected to be even more important than the microchannel cooler for the overall cooler performance, since it determines the flow uniformity, and system level pressure and thermal resistance, especially for large area die size applications.

### 2.1.3 Conjugate Heat Transfer and Fluid Flow Model

2.1.3.1 Microjet cooling test case

In order to illustrate the modeling methodology for the liquid jet impingement cooling solution, a multi-jet impingement cooler with a 4×4 inlet nozzle array is chosen as the test case for this study. Figure 2.4 shows the design of this cooler and the internal geometry revealing the inlet chamber, nozzle jets and outlet chamber. The liquid jets will impinge directly on the surface of the silicon die, resulting in a high convective cooling. The 14 mm × 14 mm × 8.7 mm cooler is mounted on the 8 mm× 8 mm thermal test chip [39]. The diameter of the inlet and outlet tube is 2 mm. The thickness of the inlet chamber is designed as 2.5 mm for uniform flow distribution. The cavity height is 0.7 mm. In the local level, the inlet and outlet nozzle diameters are designed as 0.6 mm, while the size of the nozzle array is designed to match the dimensions of the Si chip, in this case 8 × 8 $mm^2$. The chip thickness is 0.2 mm. In order to study the flow impact on the chip temperature distribution, the conjugate CFD model should include the fluidic part of the cooler, as well as the chip heat conduction part.

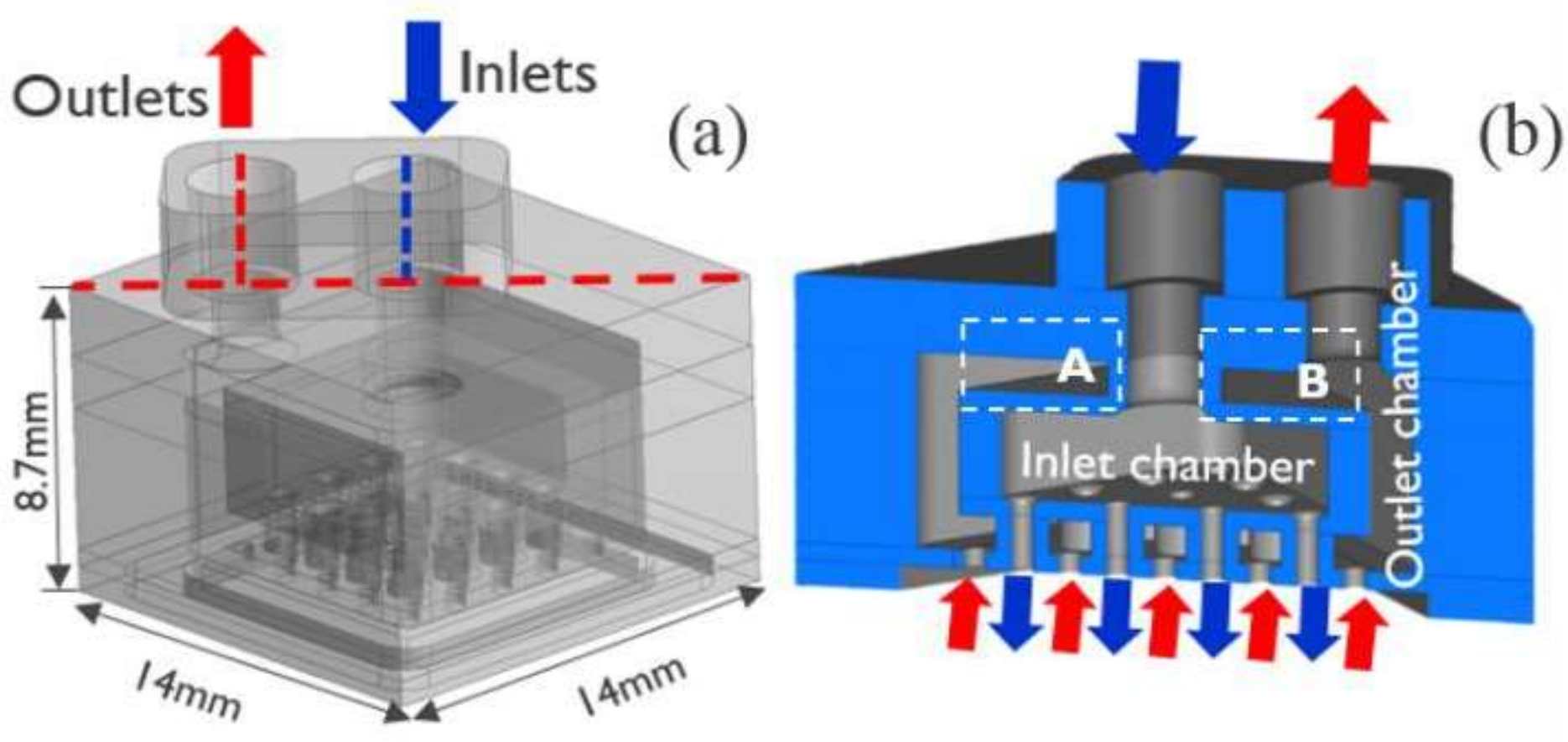

**Figure 2.4**: CAD structure of 3D printed N4 × 4 cooler: (a) transparent view; (b) cross section view with indications of the inlet chamber and outlet chamber.

2.1.3.2 CFD model introduction

Figure 2.5 shows the top view of the nozzle plate of the impingement cooler for an $N$×$N$ array where each inlet is surrounded by 4 outlets. This symmetrical nozzle array can be approximated by a unit cell of a single jet in Figure 2.5(b), the ignoring edge effects from the side walls of the device. Due to the symmetry of the structure, this unit cell can be further reduced to a 1/8 model allowing a drastic reduction of the computation time for the DOE, considering the following five design parameters for the unit cell: nozzle number $N$ for $N$×$N$ nozzle array, inlet diameter $d_i$, outlet diameter $d_o$, nozzle plate thickness $t$, chip thickness $t_c$ and cavity height $H$. For the unit cell model, the bottom package is not included in the model, which is represented by an equivalent heat transfer coefficient as a boundary condition.

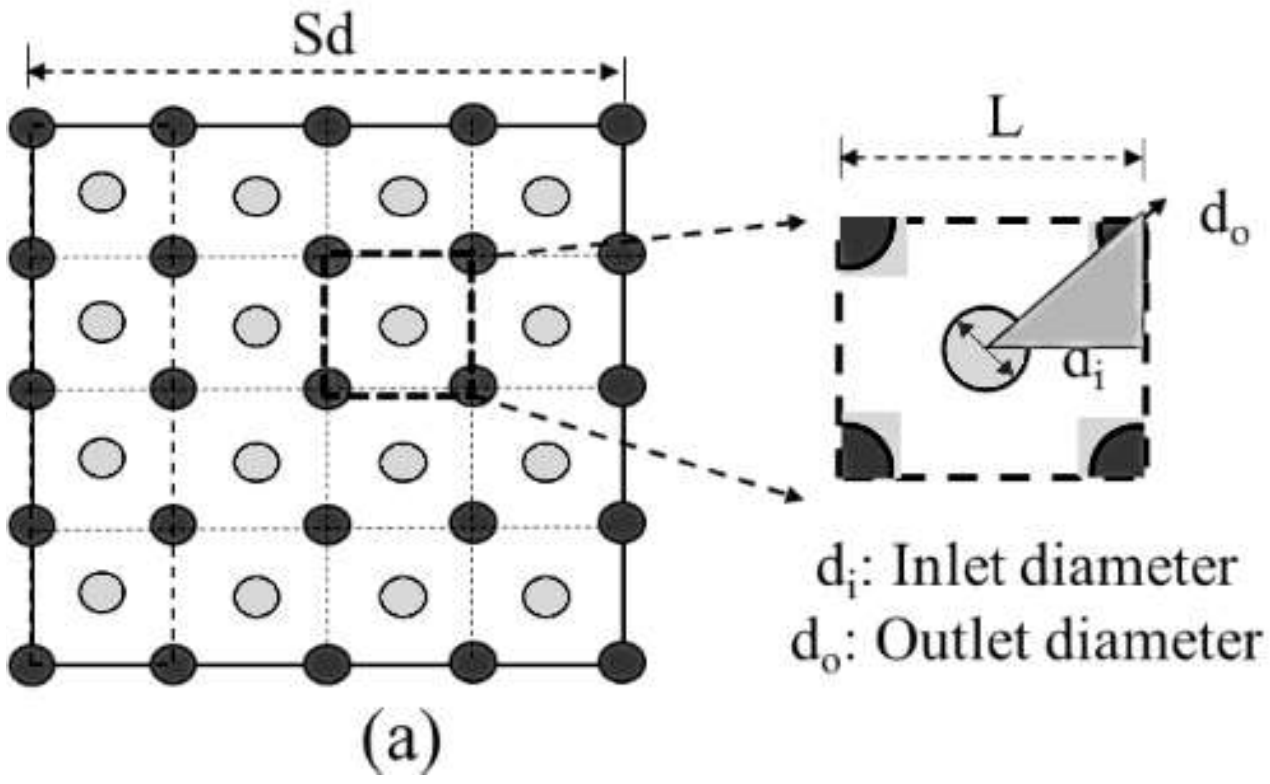

**Figure 2.5**: Unit cell modeling approach: (a) indication of cooler geometrical parameters and unit cell; (b) coupled flow and thermal simulation result from a 1/8 detailed model, simplified from the unit cell.

The unit cell model assumes an identical behavior for each cooling cell in the jet array, however in the cooler there are differences in the flow rate and chip temperature between central nozzles and corner nozzles. In order to study the system level behavior of the cooler, including the total pressure drop, flow uniformity, and full chip temperature distribution, the full cooler level CFD model is built. Since the cooler is fabricated with a polymer with very low thermal conductivity, the heat conduction through the cooler solid wall can be neglected, and only the fluidic parts of the cooler are included in the model, along with the solid domain for the Si chip.

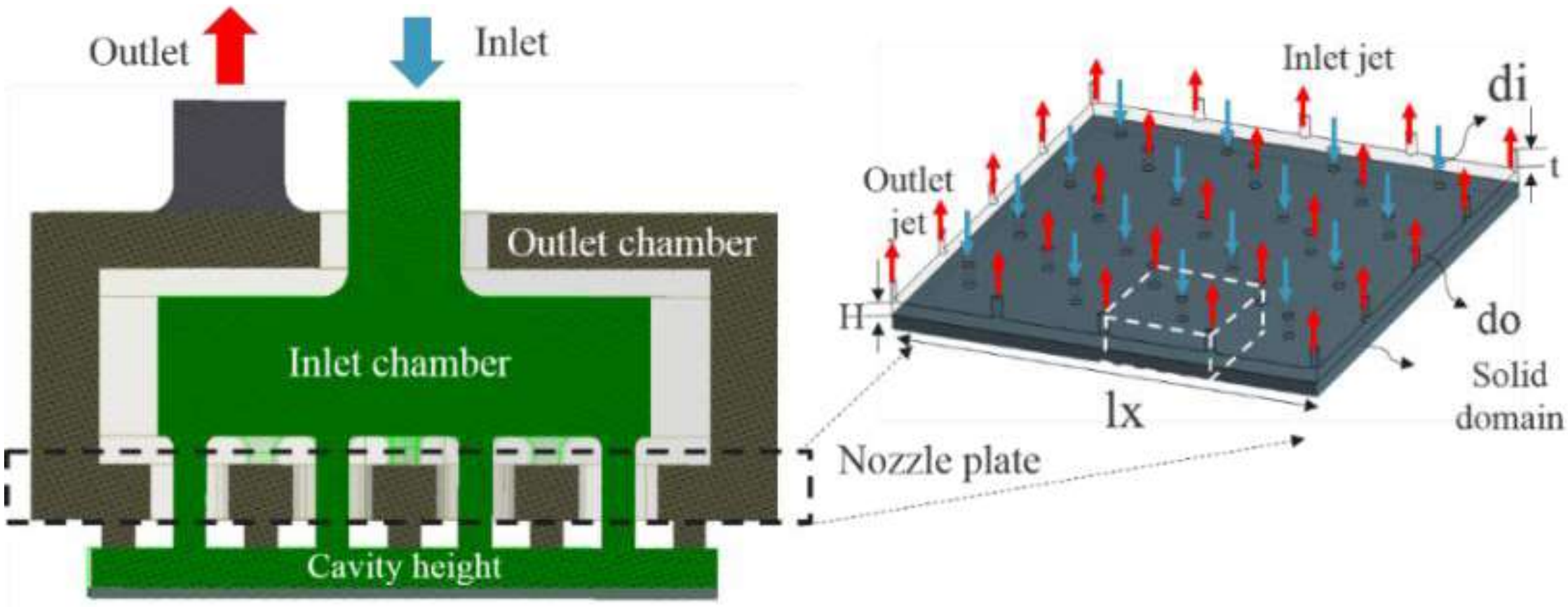

**Figure 2.6**: Full computational fluid dynamics (CFD) model with fluidic domain only of the impingement jet cooler with 4 × 4 nozzle array and inside manifold fluid delivery system.

As shown in Figure 2.6, the internal fluidic domain is extracted from the CAD model of the cooler, including the inlet chamber, outlet chamber, nozzle plate and the impingement cavity. Moreover, the nozzle array with 4 × 4 inlet nozzles and 5 × 5 outlet nozzles distribution is shown in the enlarged view of the nozzle plate in Figure 2.6.

2.1.3.3 Grid sensitivity study

For the meshing of the CFD models, hybrid meshing is chosen. The fluid domain mesh is chosen as tetrahedron mesh cells. Prism element cells are used for the meshing of the boundary layers with minimal meshing size of 0.002 mm. The latter is calculated from the $y^+<1$ constraint for the turbulence model near boundaries [40]. The number of boundary layer grid cells in the normal direction to solid walls is set to 15. For the solid domain mesh prism cells are used with a 20 μm mesh size. The grid convergence index (GCI) is used for the meshing sensitivity analysis. The GCI12 and asymptotic range of convergence are listed in Table 1 for both the unit cell and full cooler level model. Based on the GCI analysis, the final meshing details are shown in Table 2.2. For the unit cell model and the full cooler level model, the grid sensitivity analysis using the Richardson extrapolation [40] predicts a discretization error for the stagnation (minimum) temperature of 0.2% and 0.4% respectively. The details of the mesh for the full model and unit cell model are both shown in Figure 2.7.

The grid convergence order is defined as follows [41]:

$$p = \frac{\ln\left(\frac{f_3 - f_2}{f_2 - f_1}\right)}{\ln r} \tag{2.1}$$

where $p$ is the order of computational method. These solutions ($f_3$; $f_2$; $f_1$) are computed over three different grid levels ($\bar{h}_3$; $\bar{h}_2$; $\bar{h}_1$), which are subsequently refined according to a constant grid refinement ratio $r$, defined as $\bar{h}_1 = \frac{\bar{h}_2}{r} = \frac{\bar{h}_3}{r^2}$.

Once the order of convergence $p$ is known, the GCI can be calculated by using two subsequent results. In particular, if $f_3$ and $f_2$ are used and the final reported result is $f_3$, the one on the coarsest grid is defined as below:

$$\text{GCI} = \frac{F_s r^p}{r^p - 1}\left|\frac{f_3 - f_2}{f_2}\right| \tag{2.2}$$

where the $F_s$ is a safety factor. It is also important to be sure that the selected grid levels are in the asymptotic range of convergence for the computed solution. The check for asymptotic range is evaluated using the equation as below:

$$\frac{\mathrm{GCI}_{23}}{r^p \mathrm{GCI}_{12}} \approx 1 \quad (2.3)$$

where $\mathrm{GCI}_{23}$ and $\mathrm{GCI}_{12}$ are the values of GCI computed by considering, respectively, $f_2$; $f_3$ and $f_1$; $f_2$.

The $\mathrm{GCI}_{12}$ and asymptotic range of convergence are listed in Table 1 for both the unit cell and full cooler level model. Based on the GCI analysis, the final meshing details are shown in Table 2. The details of the mesh for the full model and unit cell model are both shown in Figure 4.

**Table 2.2:** Grid convergence index (GCI) meshing sensitivity analysis of full model

| Temperature | $\mathrm{GCI}_{12}$ | Asymptotic range of convergence |
|---|---|---|
| Stagnation Temp | 0.002 | 0.99 |
| Averaged chip Temp | 0.004 | 1.01 |

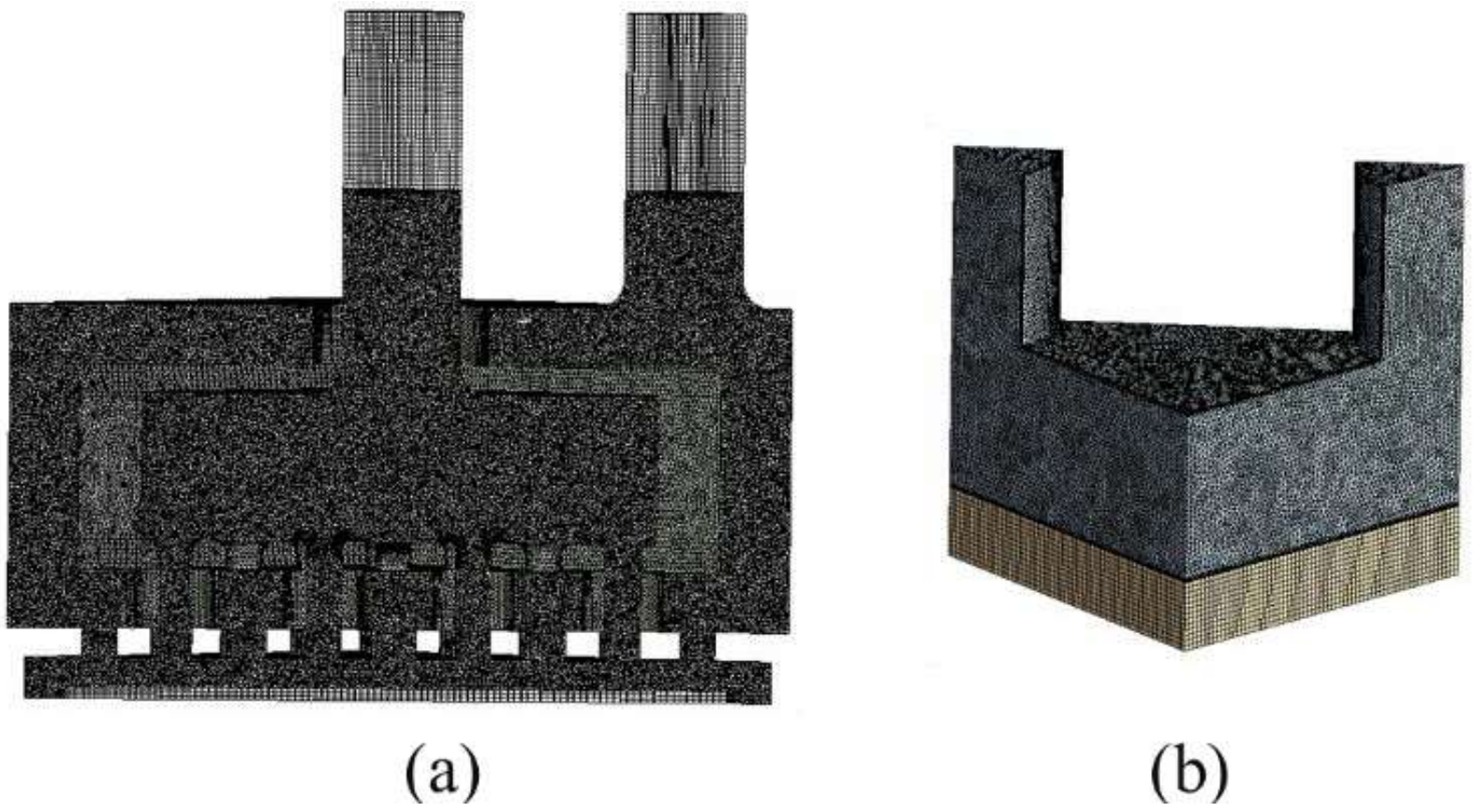

**Figure 2.7**: Meshing details of (a) full model combined manifold level and local nozzle array level; (b) 1/8 unit cell model with symmetry boundary conditions.

**Table 2.3:** Meshing comparison for the full model and unit cell model

| Model | Full model | Unit cell model （RANS） | Unit cell model (LES) |
|---|---|---|---|
| Elements | 8.5 M | 0.4 M | 3 M |
| Minimal Grid size | 80 µm | 20 µm | 1 µm |
| Computation time | 24 hours | 2 hours | 12 hours |

As shown in Table 2.3, the number of elements for the unit cell model is around 0.4 million, while the full model is around 8.5 million based on the meshing sensitivity study. In summary, the $GCI_{12}$ with meshing sensitivity analysis of the CFD model is smaller. The mesh size listed in Table 2.3 is sufficiently fine and will be used in the rest of this PhD.

2.1.3.4 Numerical Modeling Analysis

The conjugate heat transfer models consider conduction and convection in the liquid domain of the model and conduction in the solid domain. In this thesis, conjugate heat transfer and fluid dynamics simulations (CHT CFD) have been performed to assess the thermal and fluidic behavior of an impingement cooler with N×N nozzles array based on ANSYS FLUENT 18.0 [42]. The solid domain represents the silicon die only. A transition shear stress transport (SST) model is used for the CFD simulations, since this type of turbulence model offers a good compromise between accuracy and computational time for jet impingement modeling [2] and allows to cover the large range of Re numbers from laminar flow, over transitional flow to turbulent flow that is encountered in practical cooling design, as discussed in section 2.1.2.1. In this study, flow rates from 50 mL/min up to 1000 mL/min have been considered. This corresponds to a range from 10 to 3500 for the $Re_d$ number based on the nozzle diameter, while the reported laminar to turbulence transition range for liquid jet impingement is between 1000 and 3000 [2]. Based on this range of considered Re numbers from laminar to low Re turbulent flow, a RANS based transition SST model has been chosen, using the "Semi Implicit Method for Pressure Linked Equations (SIMPLE)" [43] algorithm as the solution method and the Quadratic Upstream Interpolation for Convective Kinematics (QUICK) scheme [2,44,45] for the numerical discretization. The power dissipation in the chip is represented as a heat flux boundary condition on the Si. The flow conditions are applied as a velocity condition at the inlet and a pressure outlet boundary condition for the outlet. For the model material properties, the density, viscosity and other material properties of the fluid/solid are assumed to be constant during the simulation. All cavities are assumed to be completely filled with the liquid coolant, without any presence of air (submerged jets). The physical property of the materials used in the numerical simulation are listed in Table 2.4.

**Table 2.4:** The physical property parameters of materials used in CFD simulation.

| **Material** | **Density** | **Specific heat** | **Thermal conductivity** | **Viscosity** | **Temperature** |
|---|---|---|---|---|---|
| Unit | $kg/m^3$ | J/(kg.k) | W/(m.K) | Kg/(m.s) | ºC |
| Silicon | 2329 | 556 | 149 | 0.1 | -- |
| Water-liquid | 999.7 | 4197 | 0.6 | 0.0013 | 10 |

A. Unit cell modeling analysis

For the boundary conditions of the unit cell modeling, a Dirichlet boundary condition is used. This means the velocity of the liquid at all fluid–solid boundaries is equal to zero (no slip condition). The boundary condition for the cooler inlet is set as a constant uniform inlet velocity while the static pressure for the outlet is set to 0 Pa, as a reference pressure. This means all pressure data obtained are specified relatively to the outlet pressure. As for the thermal boundary conditions, the coolant inlet temperature is set as a constant temperature. Moreover, constant heat flux is applied on the chip bottom to represent the power generation in the heating elements of the test chip. In addition, the bottom package of the chip is regarded as thermal insulation. This assumption will be explained in the single jet modeling study in chapter 4. The fluid and solid interface is set as a flow-thermal coupled boundary condition. Moreover, the residual which directly quantifies the error in the solution of the system of equations, is one of the most fundamental measures of an iterative solution's convergence. The convergence criteria for the unit cell modeling is set at $10^{-5}$ for continuity, $10^{-6}$ for energy and $10^{-6}$ for k, w and momentum (x, y and z velocities), respectively. Figure 2.8 shows the unit cell modeling results of 4×4 inlet jet arrays with flow streamline distribution for the flow rate of 37.5 mL/min per nozzle. The stagnation regions, wall jet region and recirculation region can be identified in the flow streamline distribution.

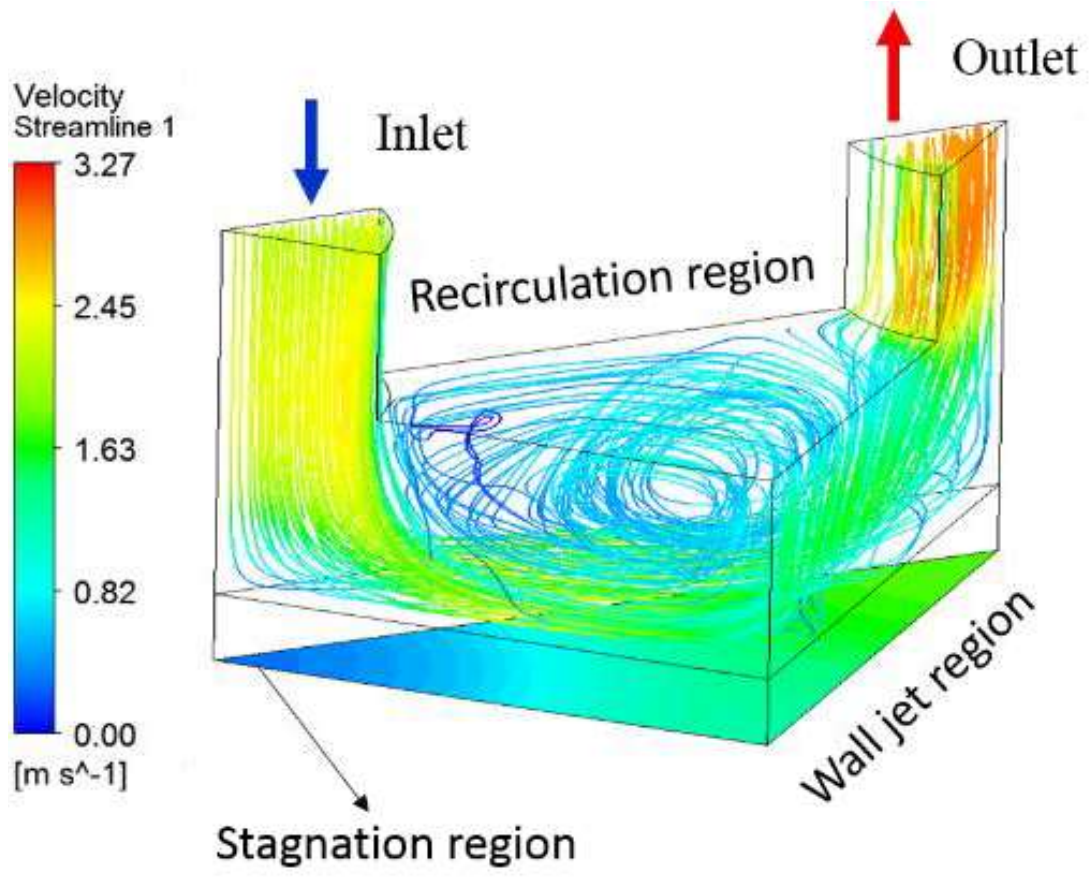

**Figure 2.8**: Flow regions for unit cell modeling results for the flow rate of 37.5 mL/min per nozzle.

Moreover, the temperature distributions of the unit cell model with different inlet nozzle diameter are shown in Figure 2.9. Also, the reconstructed temperature profile under different nozzle diameter are illustrated. Based on the unit cell approach, the parameter sensitivity study of the unit cell model for conditions raging from laminar flow to turbulent flow, including the grid sensitivity analysis and assessment of different turbulence models are studied systematically. The experimental validations of the unit cell model will be investigated in section 6.5 in chapter 6.

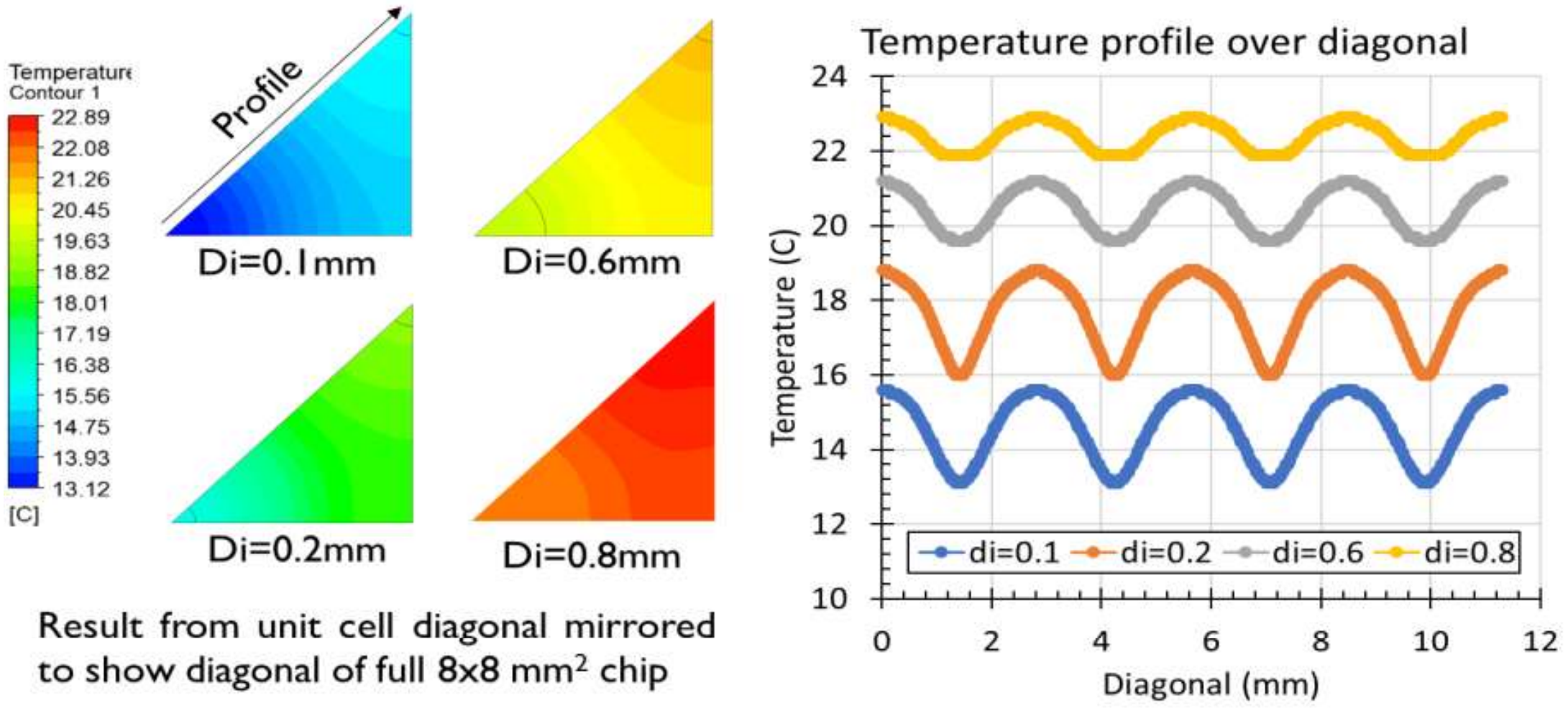

**Figure. 2.9:** Thermal modeling results for unit cell model: (a) temperature distribution for different nozzle diameter; (b) reconstructed temperature profile along x direction.

B. Full cooler level modeling analysis

With the modeling results of the full cooler level model, the flow and thermal behavior can be visualized and used to understand the physics behind. As shown in Figure 2.10, the flow streamline inside the manifold and temperature distribution across the chip are illustrated.

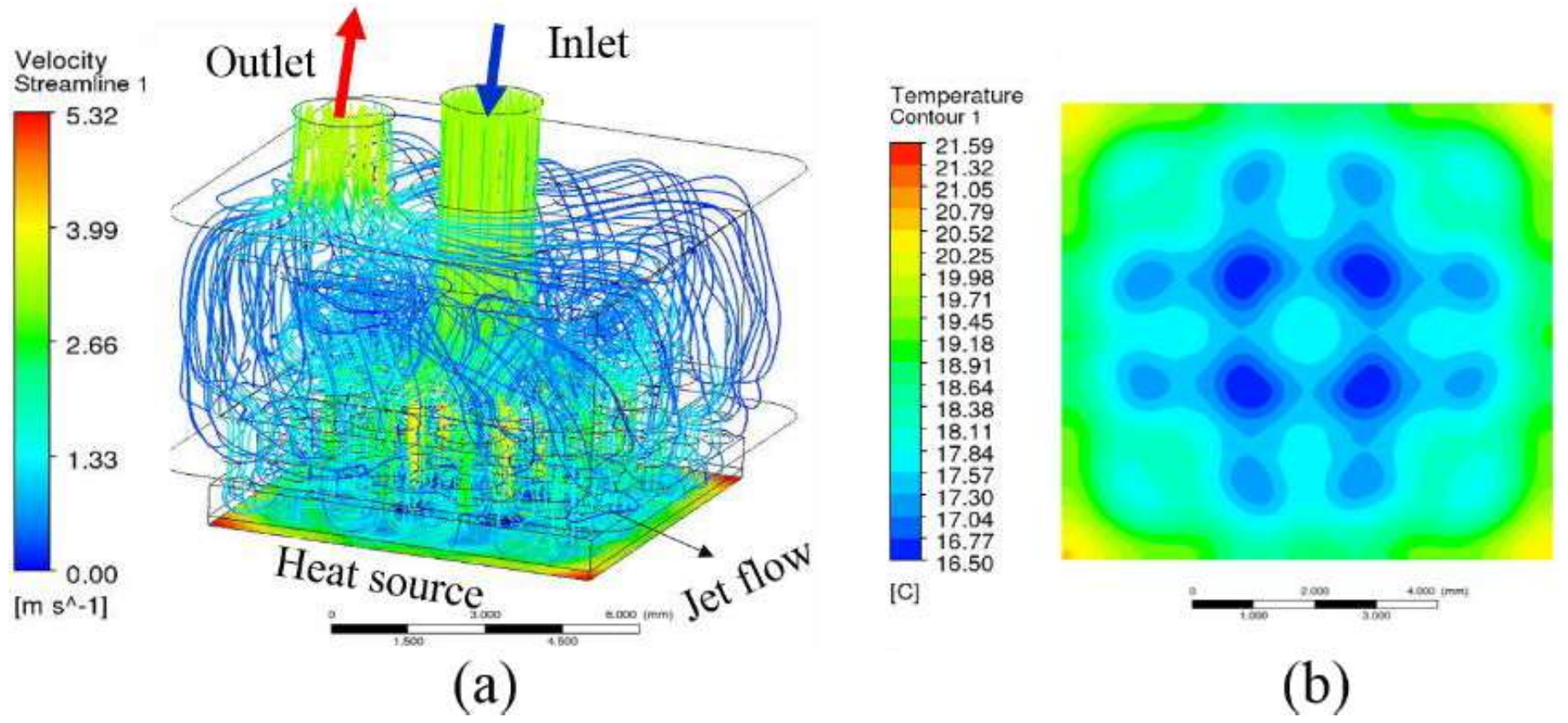

**Figure. 2.10:** Modeling results of full cooler level model: (a) flow streamline inside the cooler; (b) temperature distribution across the chip surface for a uniform chip power distribution.

The flow and temperature distributions are shown in Figure 2.11, for different flow rates ranging from 100 mL/min to 1000 mL/min. The velocity in Figure 2.11 is the tube velocity defined as the entrance velocity of the flow loop, while the $Re_d$ here is based on the inlet nozzle diameter and velocity. The corresponding $\mathrm{Re}_d$ number is between 130 and 1400. The chip temperature distribution map is linked to the velocity field inside the cooler. It can be seen that the temperature of the chip edge is higher for low $\mathrm{Re}_d$ number, due to the flow nonuniformity with higher velocity in the central nozzles and lower velocity at the edge nozzles. Moreover, the asymmetry of the temperature distribution is due to the asymmetric placement of the outlets, resulting in an asymmetric flow behavior.

| Flow rate | 100 mL/min | 300 mL/min | 650 mL/min | 1000 mL/min |
|---|---|---|---|---|
| Tube Velocity | 0.53 m/s | 1.59 m/s | 3.45 m/s | 5.3 m/s |
| $\mathrm{Re}_d$ | 135 | 406 | 879 | 1353 |

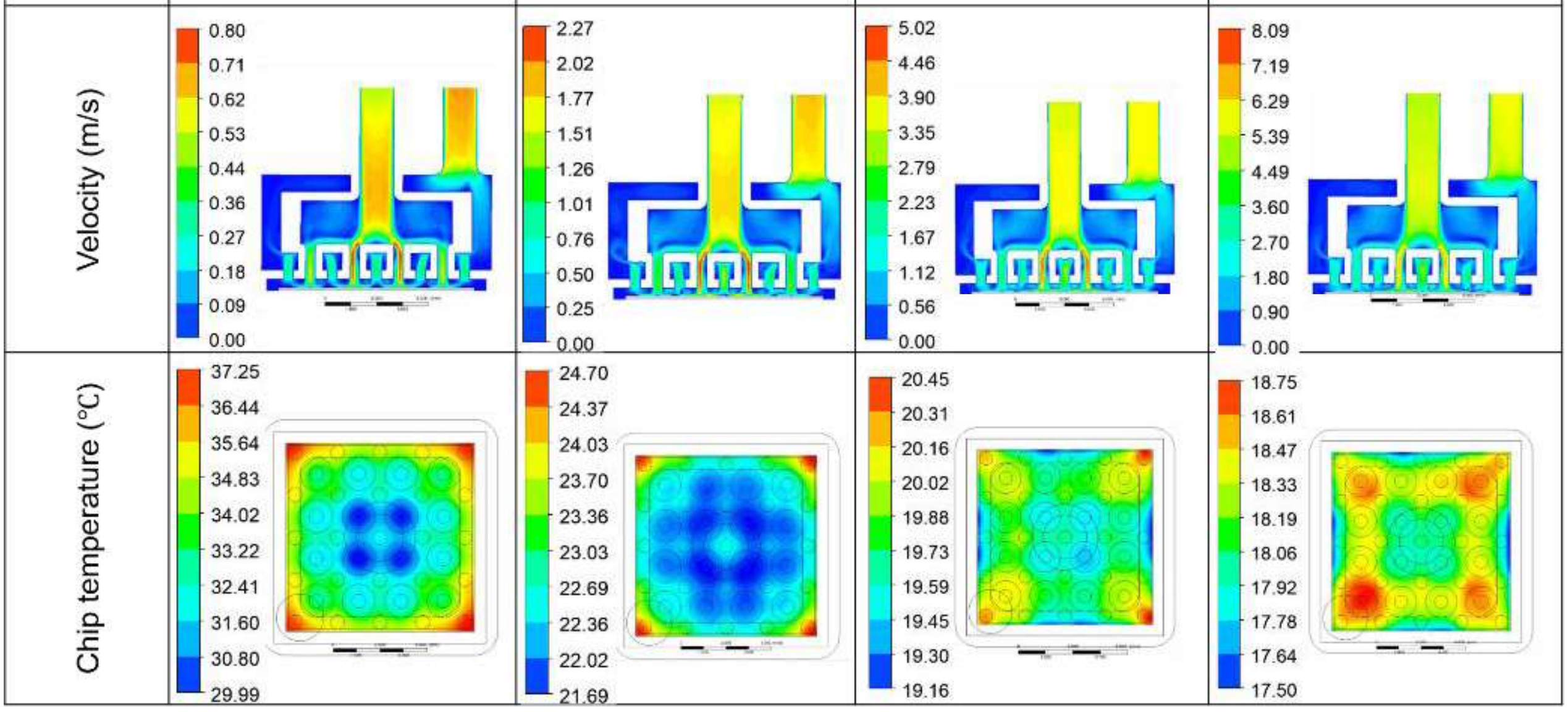

**Figure 2.11:** Conjugate flow and thermal modeling with full CFD model: top row-velocity streamline and bottom row-temperature distributions with different turbulent models (130≤ $\mathrm{Re}_d$ ≤ 1400): note that different velocity scale and temperature scale are sued for local distribution visualization.

The evolution of the total cooler pressure drop and the different contributions are shown in Figure 2.12, as a function of the total flow rate in the cooler. The cooler pressure drop and the pressure drop contributions all scale with the second power of the flow rate. It can be observed that the pressure drop in the outlet plenum is responsible for the major contribution to the total pressure, while the pressure drop contribution in the inlet plenum and the nozzle plate are considerably smaller. At the flow rate of 1000 mL/min, the pressure drop contribution of the outlet plenum amounts to 57%, while the inlet plenum and nozzle plate contributions to the pressure drop are 21% and 22% respectively. The reason for the dominance of the outlet plenum contribution is the large pressure drop associated with the flow through narrow gaps between the cylinder of the inlet/outlet divider, and the collection of the outlet flow in a single outlet connector.

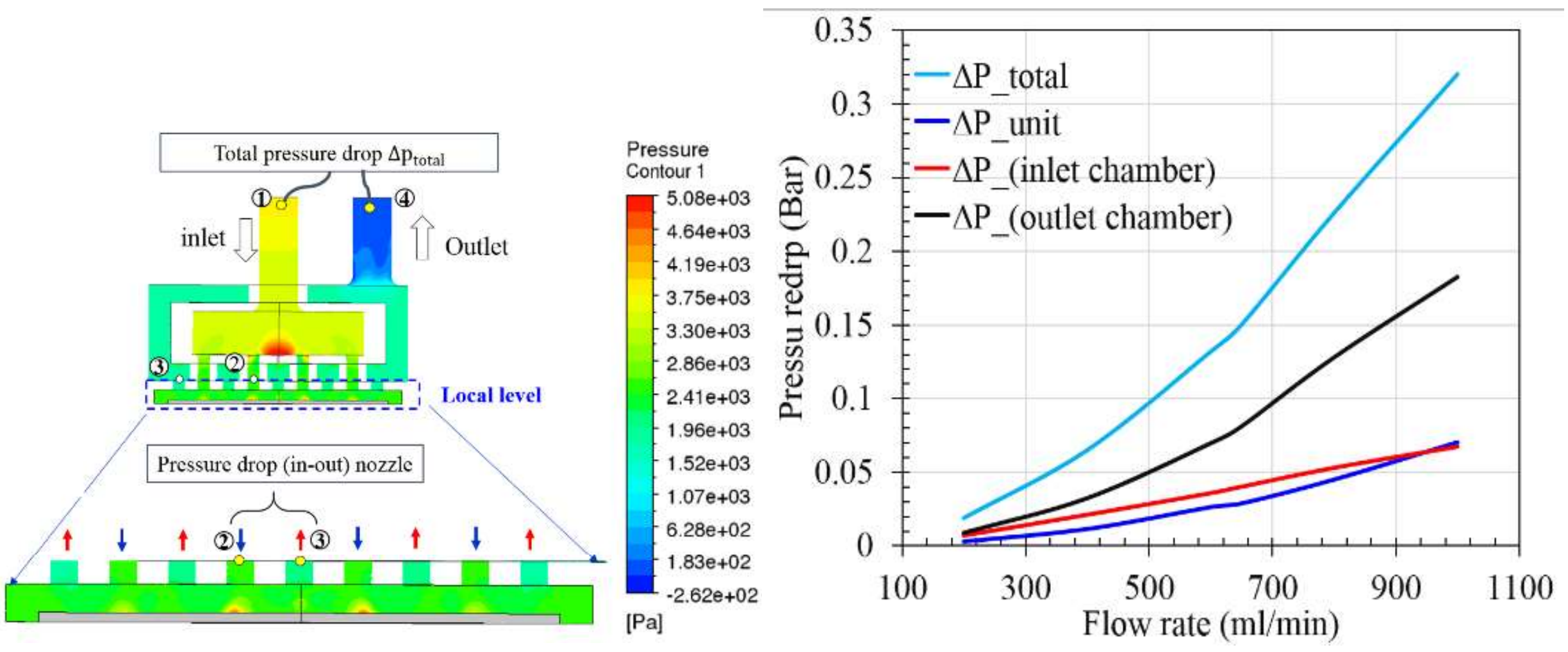

**Figure 2.12:** CFD simulation results for the pressure drop analysis of full cooler model and the indications of the pressure defined positions.

It is clear that, the flow distribution over the nozzles, the local temperature distribution and also the pressure drop inside the cooler can be investigated through the full cooler level model.

C. Unit cell validation versus full model

In order to evaluate the validity of the unit cell model, the temperature modeling data of unit cell are compared with the full cooler level model. In this comparison, the transition SST model is used for both the unit cell model and full cooler model since this model showed a good agreement with the reference LES model in the previous section and offers a good compromise between the model accuracy and computation cost, especially for the large simulation domain of the full cooler model.

Figure 2.13 shows the temperature distribution at the location of the heat sources in the Si chip, calculated by the full cooler level model for flow rate values from 100 mL/min to 1000 mL/min. The maximum, minimum and average chip temperature are extracted as a function of the different flow rates. In general, the accuracy of the unit cell model depends on the symmetry of the flow and temperature patterns. The comparison between the unit cell model and full cooler model results in Figure 2.13 shows a higher flow non-uniformity at low flow rate values with a higher local relative flow rate in the central nozzles, resulting in higher temperatures at the chip corners. For a moderate flow rate of 650 mL/min, the unit cell model shows a good agreement with the full model. The full profile comparison between the unit cell and full model are shown in Figure 2.14, which provides more information on the usability of the unit cell model. It also shows where the unit cell assumption is valid and that it can change with the flow rate.

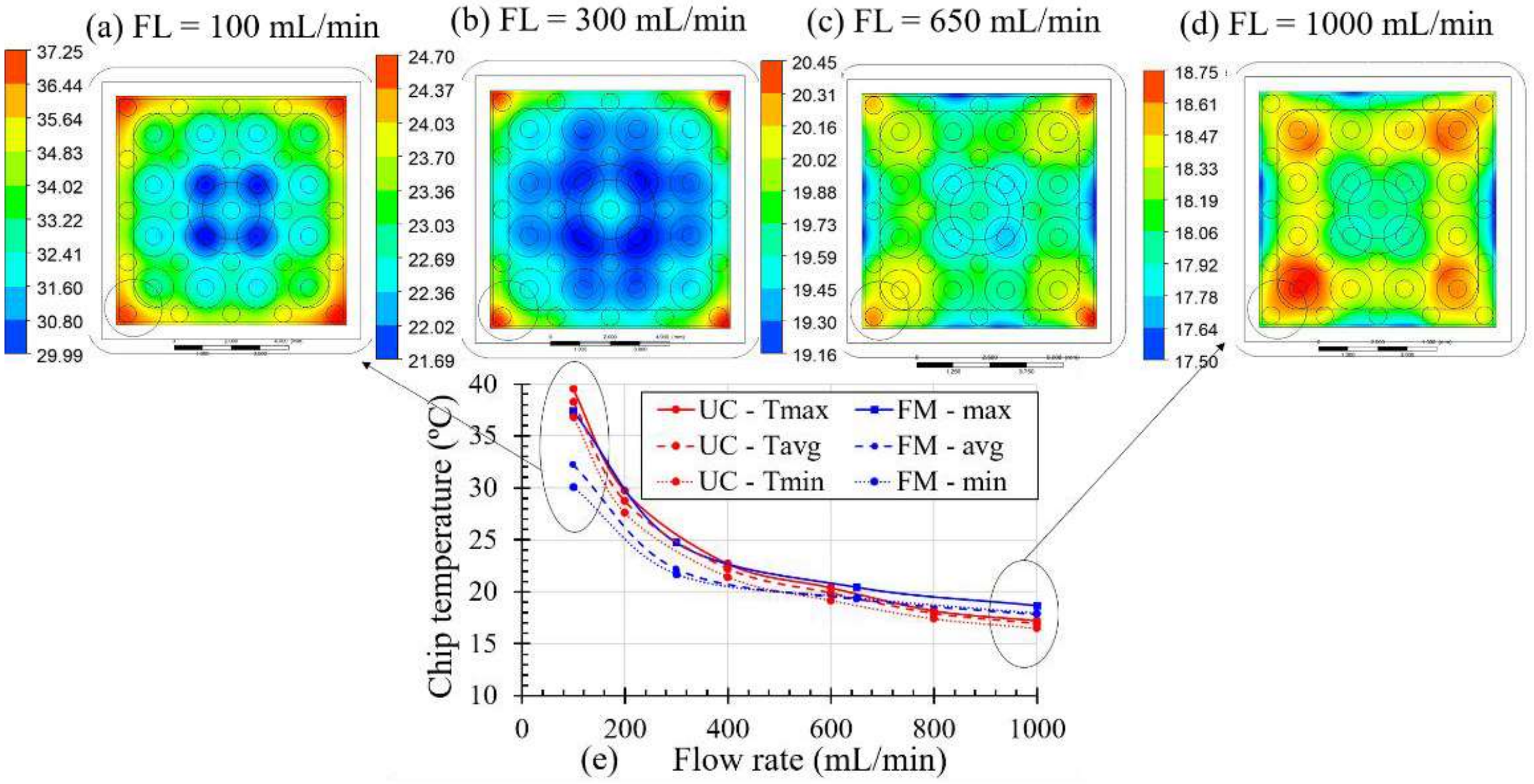

**Figure 2.13:** Temperature comparison between the unit cell model (UC) and full cooler level model (FM) with transition SST turbulence model for different flow rate values (FL): (a) FL = 100 mL/min; (b) FL = 300 mL/min; (c) FL = 650 mL/min; (d) FL = 1000 mL/min; (e) temperature comparison as a function of different flow rate. (note that the difference temperature scale is in use for local distribution visualization)

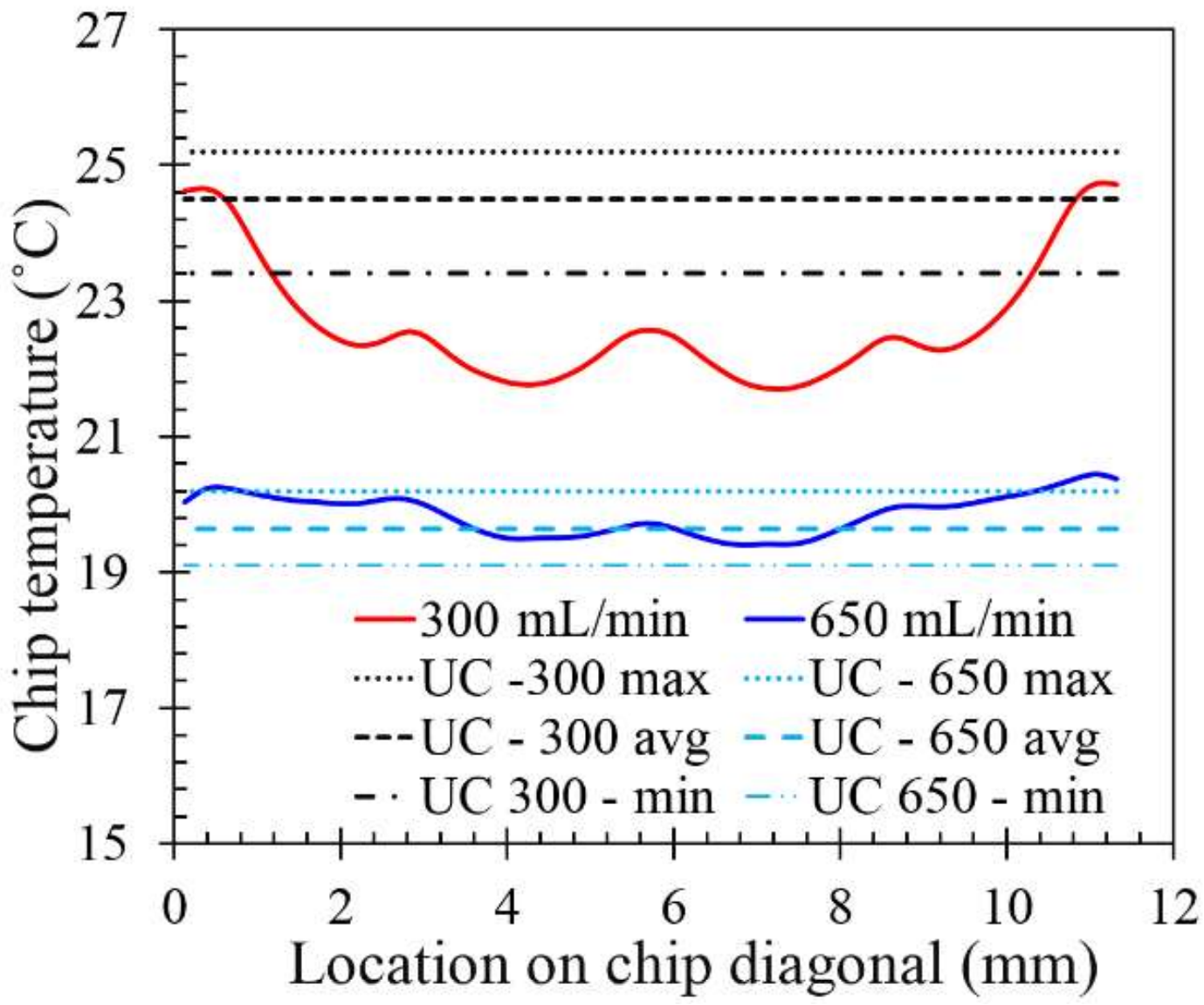

**Figure 2.14:** Temperature profiles along the chip diagonal: comparison for unit cell and full model with transition Shear Stress Transport (SST) model under the flow rate of 300 mL/min and 650 mL/min.

## 2.2 Experimental Tools

### 2.2.1 Introduction

Since impingement cooling can achieve very high heat transfer coefficients, accurate experimental studies with high spatial and temporal resolution are required to capture the local thermal impact of the cooling. The experimental characterization of jet impingement involves both the study of the flow behavior through visualization, as well as the heat transfer between the heated chip and the impinging coolant. The focus of this section is on the experimental characterization of the heat transfer. The experimental study of jet impingement as an electronic cooling solution typically involves two key elements: 1) the heat source to create a constant heat flux, and 2) the temperature measurement technique. In the next two paragraphs approaches to heat flux generation and temperature measurements techniques commonly used in literature will be reviewed.

For the characterization of the fundamental heat transfer phenomena, uniform heating in the surface is most appropriate since other effects such as thermal spreading in the silicon are minimized. In the real application however, the heated chip has non-uniform hot spots with peak heat fluxes up to 1000 W/cm$^2$ over very small areas (<0.25 mm$^2$) [46]. The measurements can either be performed in a mock-up apparatus of the integration of the cooling solution or using a more realistic test vehicle with integrated heaters and temperature sensors, where each approach has its advantages and limitations. Heating elements in the mock-up include films heaters [47], thin metal sheets [48], platinum serpentine heaters [49], Cu blocks [50] or coated heaters [51] in Incomel or stainless steel meshes on the heat transfer surface. The drawback of these additional heater materials is the introduction of additional thermal interfaces in the measurement structure, which can affect the temperature distribution, and the change in surface in case the heaters are deposited on the surface, which will impact both the flow behavior and the heat transfer [51].

The temperature measurement methods can typically be categorized in optical and electrical techniques. The optical techniques can produce the temperature map of the heat exchanging surface without making contact, and thus without disturbing the measurement. These techniques require however visual access to the surface which limits the integration options for the test structure. Examples of these optical techniques include thermochromic liquid crystals (TLC) [52], temperature sensitive paint (TSP) [53] and infrared thermography [54]. Electrical measurements techniques at the other hand require physical contact (resulting in an additional contact resistance and disturbance of the measurements [55]) to measure the temperature at the limited number of discrete locations of the sensors. Thermocouples are a commonly used method which are placed on or near the heated surface that is being cooled by impinging jets [56-59]. An example is the study by Maddox [60] where an array of twelve K-type thermocouples embedded in the measurement block with a pitch of 3 mm was used to capture the temperature and heat transfer coefficient peak. Other temperature sensors are resistance temperature detectors (RTDs), which can be deposited on the heater surface as RTD film or integrated separately with the heat source [61-63], and thermistors which are very sensitive (up to 100 times more than RTDs and 1000 times more than thermocouples) by measuring the change in resistance with temperature. However, thermistors have self-heating problems and have a slow response for transient thermal measurements.

Alternatively, thermal test chips or thermal test vehicles with integrated heaters and sensors can be used for steady-state and transient thermal measurements in real application conditions, including all realistic interfaces. The on-chip integrated temperature sensors can be metal resistors, RTDs or diodes, while the integrated heaters can be poly silicon heaters, transistors are metal resistors to create either a uniform power dissipation or a predefined hot spot pattern. The drawback of the test vehicles is the higher cost and the required processing or packaging to be used in the test set-up. These test vehicles can be fabricated using simplified processing of metal heaters and RTDs on Pyrex [64] or full CMOS Si processing. In literature only a small number of experimental studies using thermal test chips for liquid jet impingement cooling is available. Evelyn N. Wang et al. [65] used a 1 $cm^2$ Si thermal test chip with seven calibrated temperature sensors to study the performance of the microjet heat sinks, but the thermal test chip (TTC) used in the experimental investigations has a low spatial resolution. Bonner et al. [66] carried out thermal experiments for a flat spray cooling system with nozzles angled to the surface of a silicon chip using a Thermal Test Vehicle (TTV) with only four micro-heaters for delivering peak heat fluxes and 29 RTDs.

### 2.2.2 Advanced thermal test chip

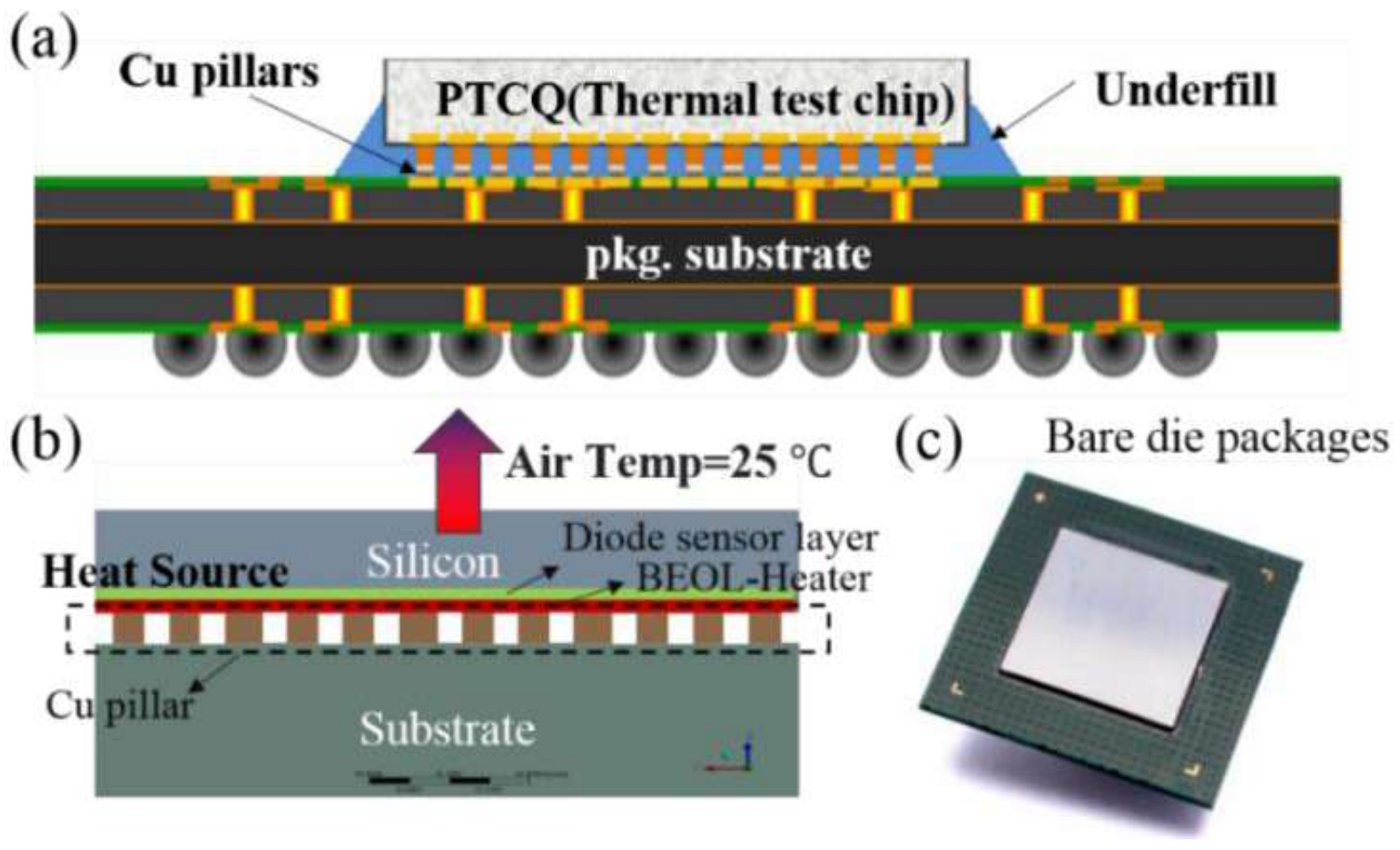

**Figure 2.15:** Details of the thermal test chip: (a) cross section view of the PTCQ package; (b) schematic of the heater and diode temperature sensor layer; (c) photograph of the bare die package.

In this thesis, a dedicated CMOS thermal test chip, named PTCQ (Packaging Test Chip Version Q) shown in Figure 2.15(c) is used to characterize the temperature response of liquid jet impingement cooling. This 8 × 8 mm$^2$ test chip includes integrated heaters to program a custom power map, and integrated sensors to measure the full temperature distribution map. As shown in Figure 2.15(a), the entire PTCQ package includes the thermal test chip, the Cu pillars and underfill material, the package substrate, the solder balls and the PCB. The dimensions and material properties are listed in Table 2.5. Moreover, the integrated diode temperature sensor layer and heater cell layer are illustrated in Figure 2.15(b). The size of the single diode temperature sensor is about 4.8 µm × 2.6 µm, which was fabricated using front-end of line (FEOL) semiconductor processing technology. Different from the temperature sensors, the heater cells were fabricated using back-end of line (BEOL) as resistors.

**Table 2.5:** Dimensions and material properties.

| **Layer (from top)** | **Dimensions (mm×mm×mm)** | **k or $k_x$, $k_y$, $k_z$ (W/mK)** |
|---|---|---|
| Silicon die | 8 × 8 × 0.2 | 150 |
| BEOL | 8 × 8 × 0.002 | 0.25 × 0.25 × 0.5 |
| Cu pillars and underfill | 8 × 8 × 0.1 | 0.4 × 0.4 × 8 |
| Substrate | 14 × 14 × 0.33 | 10 × 10 × 0.6 |
| PCB | 35 × 35 × 1.6 | 12 × 12 × 0.6 |

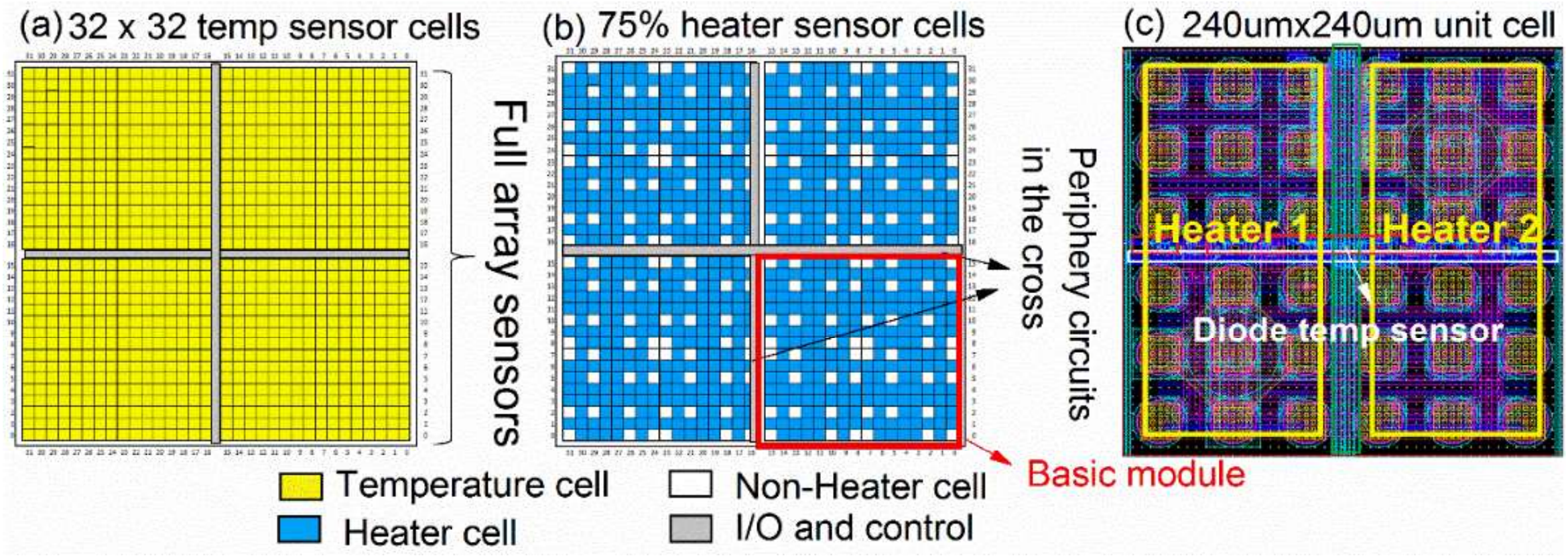

**Figure 2.16:** Floorplan of the 8 × 8 mm$^2$ PTCQ thermal test chip: (a) configurations of 32 × 32 array of temperature sensors; (b) configurations of 832 programmable heater cells; (c) details of the metal meander heaters within one cell (240 × 240 µm$^2$).

The test chip is divided into a 32 × 32 array of 240 × 240 µm$^2$ square cells with additional peripheral circuits with I/O and control cells in the central cross of the chip. The total number of the temperature sensor cells is 1024, marked with yellow color shown in Figure 2.16(a). All these cells contain a diode in the center of the cell as temperature sensor, resulting in a detailed temperature map measurement with 32×32 'thermal pixels' across the die surface. The voltage drop across the diode for a constant current is used as the temperature sensitive parameter of the sensor. The 95% confidence interval of the calibrated sensitivity is -1.55 ± 0.02 mV/ºC for a current of 5 µA in the temperature range between 10 and 75 ºC. This current level is sufficiently high to ensure stable operation of the diode as temperature sensor while it maintains the intrinsic power dissipation at a low level of 4 µW preventing it from self-heating.

As shown in Figure 2.16(b), the blue square elements represent the heater cells while white square elements stand for non-heater cells. Therefore, there are 832 cells indicated as 'heater cells' within the 32×32 array. The single heater cell is equipped with two 200×100 µm$^2$ metal meander heaters in the back-end of line (BEOL) shown in Figure 2.16(c). The maximal power dissipation of each cell is 100 mW for a voltage of 1 V. The calibrated resistance per heater cell is 10 Ohm. Including the periphery circuits with 192 grey square elements, the "heater cells" covers 75% of the chip area (832/1089=75%). Each of those cells is individually controlled by a local switch, resulting in a custom power map on the test chip ranging from quasi-uniform power dissipation with 75% coverage to localized hot spots. The other cells marked with white color in the test chip contain a variety of mechanical stress sensors. These stress sensors on the chip have been measured in our previous studies to evaluate the induced stress in the chip during the die stacking [67] and the chip packaging process [68]. Moreover, the stress caused by local hot spot power dissipation [69] has been also investigated by these stress sensors. By programming the heat cells in the PTCQ die, an example of power map distribution with hotspots is illustrated in Figure 2.17, which will be used for the hotspots target cooling studies in section 8.1.

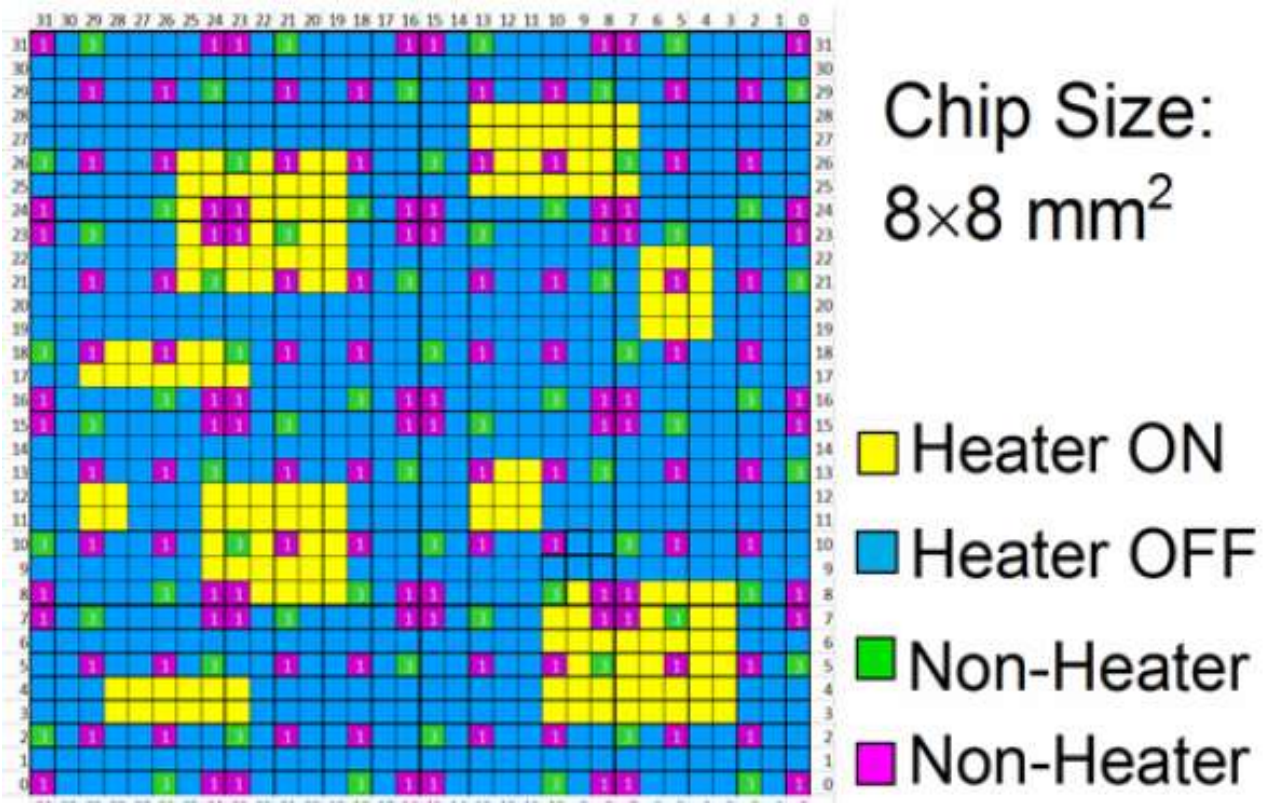

**Figure 2.17:** Example of power map distribution with hotspots in PTCQ test vehicle.

In order to supply the current to the test chip and to read out the data, the test chip needs to be packaged. To apply the test chip for the thermal evaluation of jet impingement cooling, the test chip is packaged face-down in a 14×14 $mm^2$ flip chip ball grid array package (FC-LPBGA). In the bare die package, the backside of the Si chip is exposed allowing direct contact of the liquid coolant to the heated chip.

### 2.2.3 Experimental thermo-fluidic test set-up

Figure 2.18 and Figure 2.19 show a schematic and a photograph of the dedicated experimental test set-up for the accurate flow and pressure measurements in the cooler and the temperature measurements in the test chip. All the sensors in the set-up are connected to and controlled by LabView, allowing operation of the flow loop either in a controlled mass flow rate mode or a controlled pressure mode. The flow loop contains a magnetically coupled gear pump with a maximum flow rate of 180 kg/h and a maximum pressure of 11.5 bar, a mini Cori-FLOW mass flow meter with a range of 0.1 to 3 kg/min and an accuracy of ± 0.2% RD, and a particle filter with a mesh size of 25 µm. A differential pressure gauge (EL-PRESS) is used to measure the pressure drop across the cooler with an accuracy of ± 0.5% FS in the range between 0.2 and 5 bar. Thermocouples with an accuracy of 2.2 °C are used to measure the coolant temperature before and after the cooler. A liquid-liquid heat exchanger is used to cool the coolant back to the set-point of 10 ºC.

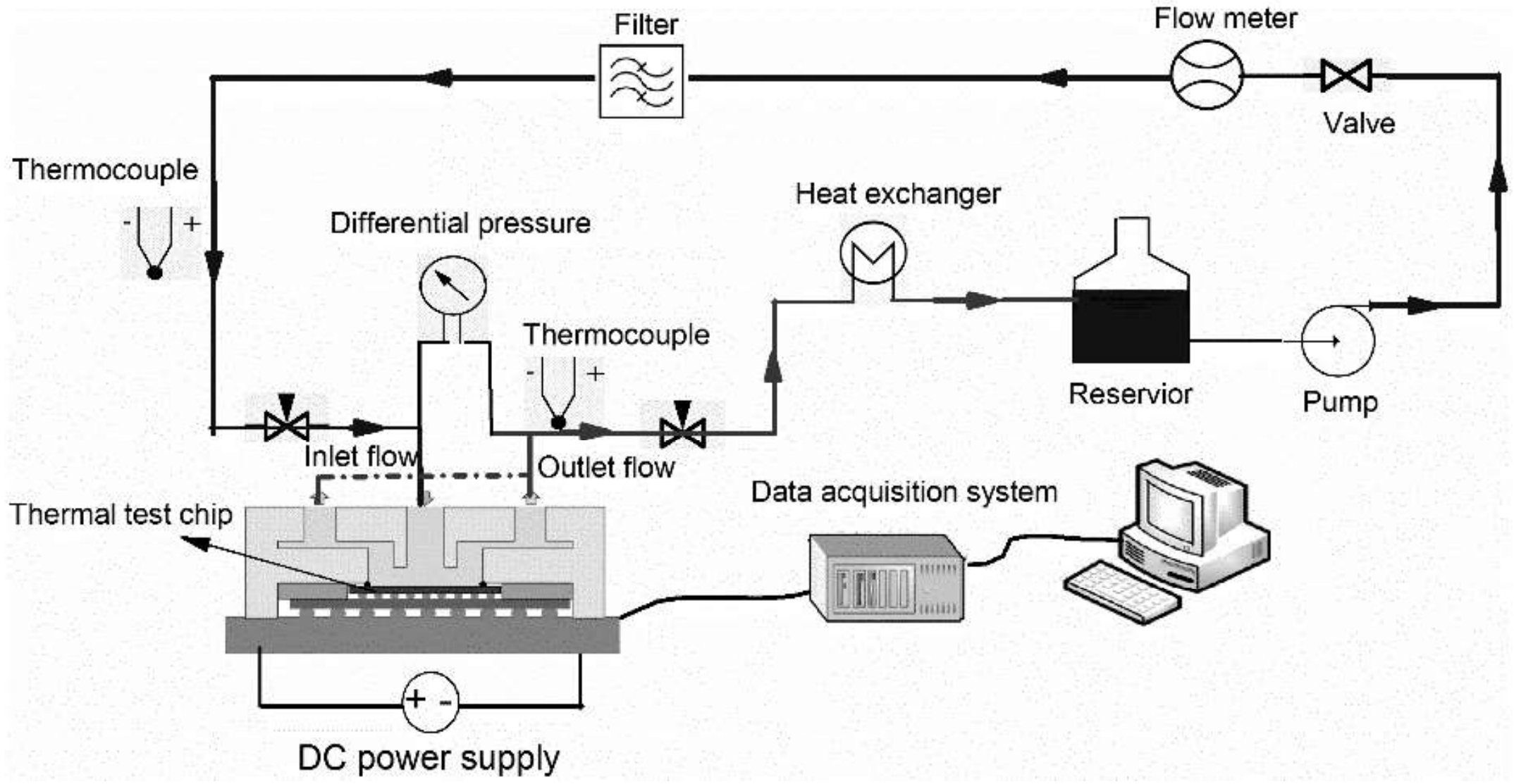

**Figure 2.18:** Schematic of the experimental flow loop: flow meter, valve, filter, heat exchanger, pressure drop transducer and pump.

As shown in Figure 2.18, the pressure drop of the inlet/outlet tube and connection is included in the measured pressure drop. The modeling results show that the pressure drop of the cooler is smaller than the tube connection part, therefore, a de-embedding technique can be used to measure the pressure of the cooler only, without the tube connection. Since the pressure drop over the tube is linearly proportional to the tube length, the pressure drop between the inlet and outlet connection of the cooler can be estimated by measuring the pressure drop for the different tube lengths and extrapolating to zero tube length.

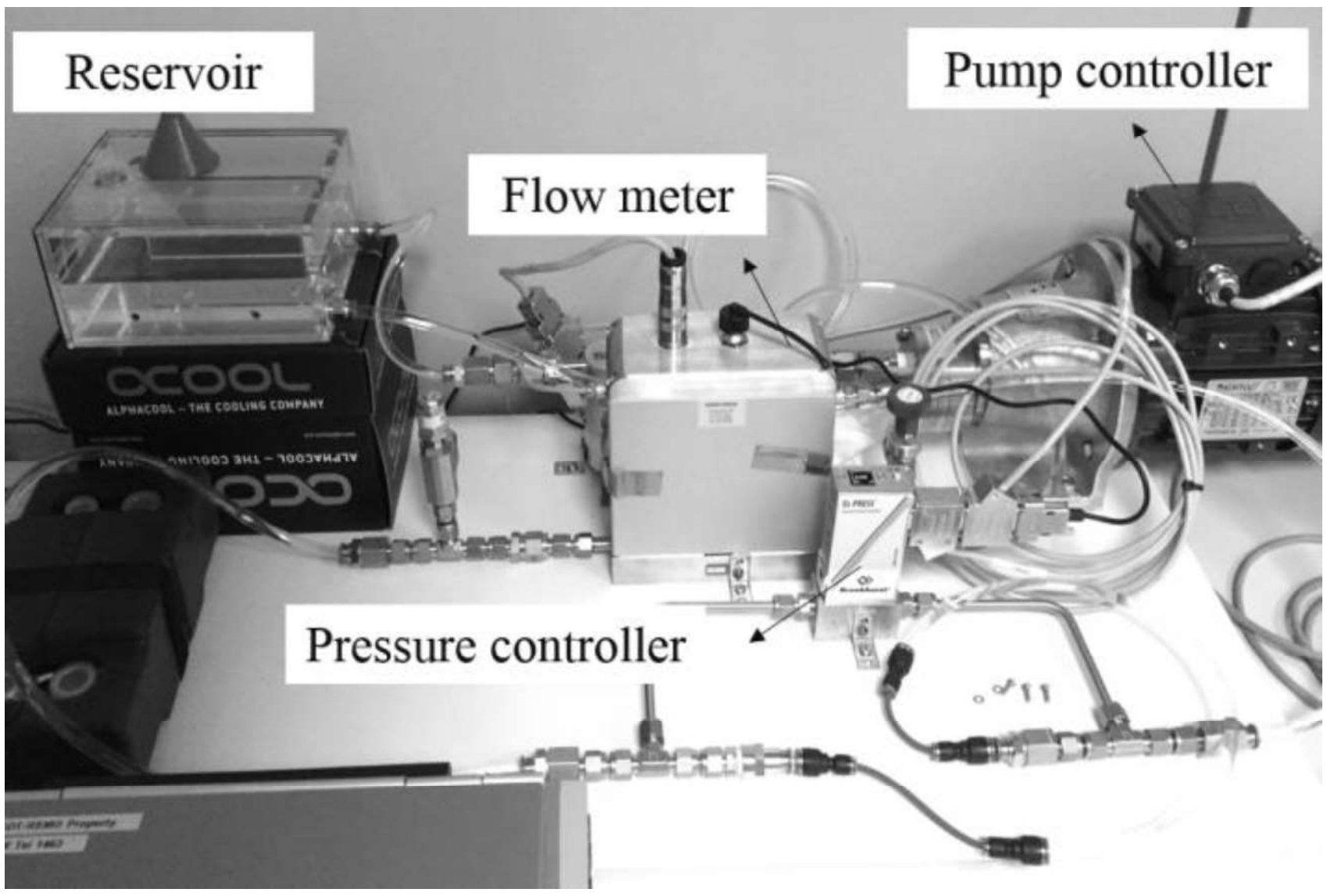

**Figure 2.19:** Experimental set-up for the hydraulic characterization of the microfluidic heatsink allowing accurate control and measurement of pressure drop and flow rate.

A liquid-liquid heat exchanger is used to cool the coolant back to the set-point of 10 ºC. In this work, DI-water is used as the coolant during the tests, with specified temperature at 10°C and ambient temperature is kept at 25 $\pm 1$ °C. During the measurement, the chilled water set with 10°C was applied to the cooling system without turning on the heater cells. After waiting 30 minutes, the steady-state chip surface temperature distribution was extracted by measuring the voltage across the 32 × 32 array of diode sensors. After that, the heaters with programmable pattern were turned on a waiting time of 30 minutes was used to achieve the steady-state regime. Finally, the temperature distribution map of the thermal test chip was measured.

## 2.3 Thermal performance characteristics

### 2.3.1 Thermal performance metrics

This section describes the design performance characteristics for the evaluation of the cooler performance.

The temperature measurements performed with the PTCQ test chip are used as relative temperature measurements, with respect to the case without power dissipation, rather than absolute temperature measurements. In the reference case without power generation in the test chip, the liquid cooling is already applied. Therefore, the initial temperature at zero power is assumed to be equal to the inlet temperature of the coolant. The measured chip temperature increases between the power-off state (diode voltage $V_{off}$) and the power-on state (diode voltage $V_{on}$) is defined as follows:

$$\Delta T_{avg} = \bar{T}_{chip} - T_{in} \quad (2.4)$$

$$\bar{T}_{chip} - T_{in} = \frac{V_{on} - V_{off}}{\sigma} \quad (2.5)$$

where $\bar{T}_{chip}$ is the measured average chip temperature, $T_{in}$ is the coolant inlet temperature and $\sigma$ is the voltage versus temperature sensitivity of the diodes. The value of the sensitivity is -1.55 mV/°C, which is defined as the calculated gauge factor of the temperature sensor, at the anode current fixed at 5μA. The 95% confidence interval for the diode temperature sensor sensitivity is ±1.8 %, with more than 500 diodes.

The average temperature increase $\Delta T_{avg}$ is defined as the average of the measured sensor temperature increase values of all the 1024 cells in the test chip. The calibration data in Figure.2.20 shows that the variation on the sensitivity, used for the relative temperature measurements is much smaller than the variation on the absolute voltage values for the different diodes.

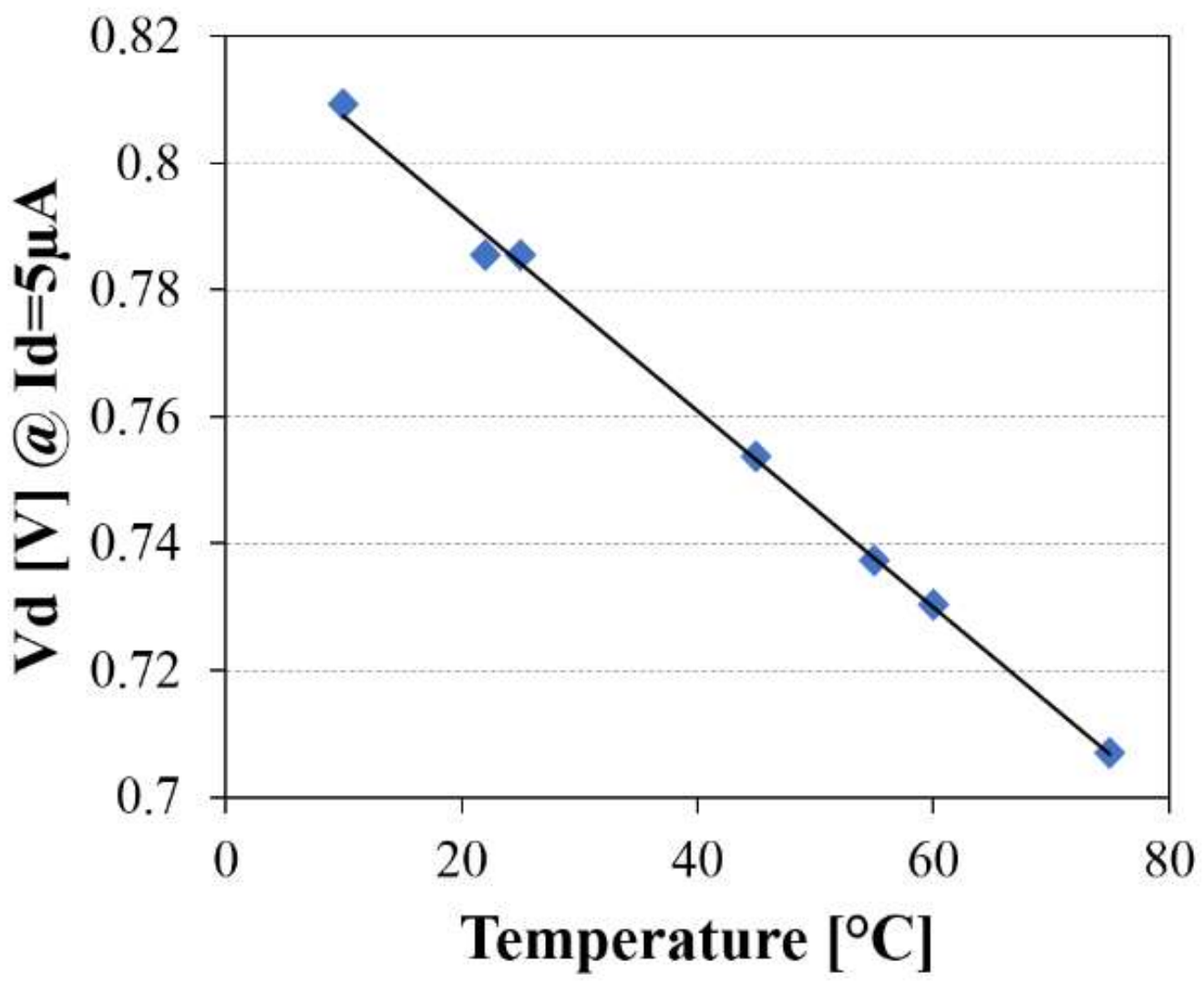

**Figure 2.20**: Voltage drop across diode as function of temperature at anode current of $I_d$=5µA.

The overall thermal performance of the cooler is expressed in terms of the thermal resistance defined as follows:

$$R_{th} = \frac{\Delta T_{avg}}{Q_{heater}} \tag{2.6}$$

where $\Delta T_{avg}$ is defined as the chip temperature increase and $Q_{heater}$ is the heat generated in the heater cells based on the measured electrical current and heater voltage. This thermal performance estimation of the assembled cooling solution also includes the heat losses through the cooler material into the ambient and the heat losses through the bottom side of the assembled test board.

In order to accurately estimate the heat transfer coefficient, the heat losses need to be characterized to identify the amount of heat absorbed by the coolant. Since the measurement of the coolant outlet temperature did not result in accurate results an alternative approach was used. The chip temperature profile was first measured for the assembled cooler without any coolant present in the cooler. This case, with an equivalent thermal resistance $R_{loss}$ of 16.8 K/W represents the heat removal from the generated heat through the considered heat losses only. For any liquid cooling measurement with the cooler, the heat losses can now be estimated as follows:

$$Q_{loss} = \frac{\bar{T}_{chip} - T_{amb}}{R_{loss}} \tag{2.7}$$

where $T_{amb}$ is the ambient temperature. Based on the net power ($Q_{heater}$ -$Q_{loss}$) and the assumed one-dimensional heat conduction across the chip thickness $t_c$, the average cooling surface temperature $\overline{T_s}$, defined as the interface temperature between the liquid coolant and chip surface, can be estimated as follows:

$$\overline{T_s} = \overline{T}_{chip} - \frac{(Q_{heater}-Q_{loss})*t_c}{A_{heater}*k_{si}} \tag{2.8}$$

where $\overline{T}_{chip}$ is defined as the average temperature of the heat source, $k_{si}$ is the thermal conductivity of silicon ($k_{si}$=149 W/mK), $A_{heater}$ is defined as the area of the heaters (8 mm × 8 mm × 75%).

Therefore, the area-averaged heat transfer coefficient $\overline{h}$ can be calculated as below:

$$\overline{h} = \frac{(Q_{heater}-Q_{loss})}{A_{heater}*\Delta T} \tag{2.9}$$

$$\Delta T = \overline{T_s} - T_{in} \tag{2.10}$$

where $\overline{T_s}$ is the average chip surface temperature, $T_{in}$ is the inlet temperature. The temperature difference ΔT is defined as the temperature increase of the cooling surface.

For the hydraulic performance of the cooler, the pumping power is defined as follows:

$$W_p = \dot{V} \,.\, \Delta P \tag{2.11}$$

where ΔP is defined as the pressure difference between the inlet and outlet of the cooler. $\dot{V}$ represents the volumetric flow rate.

### 2.3.2 Uncertainty analysis

For the uncertainty analysis of the reported quantities, the uncertainty of all the measurement devices has been considered, and the theory of measurement error propagation has been used. The considered measurement uncertainties and the measuring tools are shown in Table 2.6. The $\delta$ is defined as the measurement uncertainty of the different parameters.

The uncertainty of the chip power can be calculated as below:

$$\delta_{Q_{heater}}{}^2 = V_{heater}{}^2 + I_{heater}{}^2 \tag{2.12}$$

where the $V_{heater}$ is the voltage of all the connected heaters, and $I_{heater}$ is the current across the heaters. The uncertainty of the chip power is estimated to be ± 0.1%.

$$\delta_{\Delta T_{avg}}{}^2 = \delta_{V_{on}}{}^2 + \delta_{V_{off}}{}^2 + \delta_{\sigma}{}^2 \tag{2.13}$$

Based on the definition of $\Delta T_{avg}$ shown in equation 2.5, the uncertainty of $\Delta T_{avg}$ is estimated to be ± 1.5 %, which is dominated by the temperature sensitivity of the diode sensor (± 1.5 %).

The uncertainties in thermal resistance is listed as:

$$\delta_{R_{th}}{}^2 = \delta_{Q_{heater}}{}^2 + \delta_{\Delta T_{avg}}{}^2 \tag{2.14}$$

Based on the measurement uncertainty analysis, the analysis of the propagated measurement uncertainty results in a value of ± 1.5 % for the reported thermal resistance measurements.

**Table 2.6:** List of experimental tools in the test system and their accuracy information.

| Experimental tools | Affected parameters | Accuracy |
|---|---|---|
| Microscopy | Nozzle diameter $d_i$ | ± 3.5% |
| Thermocouple | Coolant inlet temperature $T_{in}$ | ± 1% |
| Thermal test chip | Temperature sensor V-T sensitivity $\sigma$ | ± 1.5 % |
| Voltage measurement | Diode voltage $V_{ON}$, $V_{OFF}$. | ± 0.07% |
| Pressure transducer | ΔP | ± 0.5% (0.2 -5 bar) |
| Current measurement | $I$ | ± 0.07% |
| Cori-FLOW mass flow meter | $\dot{V}$ | ± 0.2% |

**Table 2.7:** List of calculated uncertainty.

| Parameters | Symbols | Accuracy |
|---|---|---|
| Chip power | $Q_{heater}$ | ± 0.1 % |
| Chip power loss | $Q_{loss}$ | ± 2.13 % |
| Temperature increase | $\Delta T_{avg}$ | ± 1.5 % |
| Thermal resistance | $R_{th}$ | ± 1.51 % |
| Heat transfer coefficient | $h_{avg}$ | ± 2.61 % |

For the propagated measurement uncertainty for the reported heat transfer coefficient, the first item ΔT can be estimated as below:

$$\delta_{\Delta \mathrm{T}}{}^2 = \delta_{\overline{T_s}}{}^2 + \delta_{\mathrm{in}}{}^2 \tag{2.15}$$

$$\delta_{\overline{h}}{}^2 = \delta_{Q_{heater}}{}^2 + \delta_{Q_{loss}}{}^2 + \delta_{\Delta \mathrm{T}}{}^2 \tag{2.16}$$

Therefore, the propagated measurement uncertainty results in a value of ± 2.61 %. The pressure measurement uncertainties are based on the accuracy of the pressure transducer.

The flow rate measurement errors are based on the accuracy of the flow meter. The summary of the calculated uncertainties for different metrics is listed in Table 2.7.

### 2.3.3 Normalization

The normalized thermal performance represents the intrinsic thermal performance, independent of the chip and package area. This means that the thermal performance of the dual-chip package cooler in terms of absolute thermal resistance $R_{th}$ can be predicted based on the results from the single chip cooler if the same cooling cell dimensions are used (nozzle diameter and pitch, nozzle plate thickness). This normalization concept for the thermal resistance based on the chip area has been introduced in literature, to compare the thermal performance of different cold plates, with different sizes of heat sources at different flow rates [70-72].

The definitions of normalized thermal resistance $R_{th}^*$ and normalized pump power $W_p^*$ are respectively defined as:

$$R_{th}^* = \frac{(T_{\mathrm{avg}} - T_{\mathrm{in}}) \cdot A}{Q_{heater}} \quad (2.12)$$

$$V^* = \dot{V} / A \quad (2.13)$$

$$W_p^* = W_p / A \quad (2.14)$$

where A is the chip area. The validation of the normalization will be discussed in interposer package concept and large die application test cases.

### References

[1] Unnikrishnan K S1, Sabu Kurian2, Jijo Johnson, Study of Jet Impingement Heat Transfer, International Research Journal of Engineering and Technology (IRJET), Volume: 05 Issue: 04, April, 2018, pp. 1657-1664.

[2] N. Zuckerman, N. Lior, Jet impingement heat transfer: physics, correlations and numerical modeling, ASME. J. Heat Transfer. 2005, 127(5), pp.544-552.

[3] Bhunia, A.; Chen, C.L. On the scalability of liquid microjet array impingement cooling for large area systems. J. Heat Transf. 2011, 133, 064501.

[4] Bhunia, A.; Chandrasekaran, S.; Chen, C.L. Performance improvement of a power conversion module by liquid micro-jet impingement cooling. IEEE Trans. Compon. Packag. Technol. 2007, 30, 309–316.

[5] Robinson, A.J.; Schnitzler, E. An experimental investigation of free and submerged miniature liquid jet array impingement heat transfer. Exp. Therm. Fluid Sci. 2007, 32, 1–13.

[6] Molana, M.H.; Banooni, S. Investigation of heat transfer processes involved liquid impingement jets: A review. Braz. J. Chem. Eng. 2013, 30, 413–435.

[7] Han, Y.; Lau, B.L.; Zhang, H.; Zhang, X. Package-level Si-based micro-jet impingement cooling solution with multiple drainage micro-trenches. In Proceedings of the 2014 IEEE 16th Electronics Packaging Technology Conference (EPTC), Singapore, 3–5 December 2014; pp. 330–334.

[8] Brunschwiler, T.; Rothuizen, H.; Fabbri, M.; Kloter, U.; Michel, B.; Bezama, R.J.; Natarajan, G. Direct Liquid Jet-Imprringement Cooling with Micron-Sized Nozzle Array and Distributed Return Architecture. In Proceedings of the Thermal and Thermomechanical Proceedings 10th Intersociety Conference on Phenomena in Electronics Systems, San Diego, CA, USA, 30 May–2 June 2006; pp. 193–203.

[9] Natarajan, G.; Bezama, R.J. Microjet cooler with distributed returns. Heat Transf. Eng. 2007, 28, 779–787.

[10] Boldman, D.R.; Brinich, P.F. Mean Velocity, Turbulence Intensity, and Scale in a Subsonic Turbulent Jet Impinging Normal to a Large Flat Plate; NASA Lweis Center: Cleveland, OH, USA, 1977.

[11] S.B. Pope. Turbulent flows. Meas. Sci. Technol. 12(11) (2001).

[12] Zuckerman, N. Jet Impingement Heat Transfer: Physics, Correlations, and Numerical Modeling. Adv. Heat Transf. 2006, 39, 565–631.

[13] Viskanta, R. Heat transfer to impinging isothermal gas and flame jets. Exp. Therm. Fluid Sci. 1993, 6,111–134.

[14] Bernhard W. Multiple Jet Impingement—A Review. Heat Transf. Res. 2011, 42, 101–142, doi:10.1615/HeatTransRes.v42.i2.30.

[15] Narumanchi, S.V.J.; Hassani, V; Bharathan, D. Modeling Single-Phase and Boiling Liquid Jet Impingement Cooling in Power Electronics; National Renewable Energy Laboratory (NREL): Golden, CO, USA, 2005.

[16] Womac, D.J.; Ramadhyani, S.S.; Incropera, F.P. Correlating Equations for Impingement Cooling of Small Heat Sources With Single Circular Liquid Jets. ASME J. Heat Transf. 1993, 115, 106–115, doi:10.1115/1.2910635.

[17] Garimella, S.V.; Rice, R.A. Confined and submerged liquid jet impingement heat transfer. ASME J. Heat Transf. 1995, 117, 871–877.

[18] Isman, M.K.; Pulat, E.; Etemoglu, A.B.; Can, M. Numerical Investigation of Turbulent Impinging Jet Cooling of a Constant Heat Flux Surface. Numer. Heat Transf. Part A Appl. 2008, 53, 1109–1132, doi:10.1080/10407780701790078.

[19] Esch, T.; Menter, F. Heat transfer predictions based on two-equation turbulence models with advanced wall treatment. Turbul. Heat Mass Transf. 2003, 4, 633–640.

[20] Maddox, J.F. Liquid Jet Impingement with Spent Flow Management for Power Electronics Cooling. Ph.D. Thesis, Auburn University, Auburn, AL, USA, 2015.

[21] Prabhakar S., Arun K. Micro-Scale Nozzled Jet Heat Transfer Distributions and Flow Field Entrainment Effects Directly on Die. In Proceedings of the 18th IEEE ITHERM Conference, Las Vegas, NV, USA, 28–31 May 2019; pp.1082-1097.

[22] Sung, M.K.; Mudadar, I. Effects of jet pattern on single-phase cooling performance of hybrid micro-channel/micro-circular-jet-impingement thermal management scheme. Int. J. Heat Mass Transf. 2008, 51, 4614–4627.

[23] Polat, S.; Huang, B.; Majumdar, A.S.; Douglas, W.J.M. Numerical Flow and Heat Transfer under Impinging Jets: A Review. Annu. Rev. Heat Transf. 1989, 2, 157–197.

[24] Behnia, M.; Parneix, S.; Dur, P. Accurate modeling of impinging jet heat transfer. In Center for Turbulence Research, Annual Research Briefs 1997. Stanford University: Stanford, CA, USA, 1997, pp. 149–164.

[25] Gao, S.; Voke, P.R. Large-eddy simulation of turbulent heat transport in enclosed impinging jets. Int. J. Heat Fluid Flow 1995, 16, 349–356.

[26] Beaubert, F.; Viazzo, S. Large Eddy Simulation of a plane impinging jet. Mechanical Rendering, 2002, 330, 803–810.

[27] Thomas H¨allqvist. Large Eddy Simulation of Impinging Jets with Heat Transfer Licentiate Thesis. Ph.D. Thesis, Royal Institute of Technology, Stockholm, Sweden, 2006; ISSN 0348-467.

[28] Olsson, M.; Fuchs, L. Large eddy simulations of a forced semiconfined circular impinging jet. Phys. Fluids 1998, 10, 476–486.

[29] Anupam, D.; Rabijit D.; Balaji S. Recent Trends in Computation of Turbulent Jet Impingement Heat Transfer. Heat Transf. Eng. 2012, 33, 447–460.

[30] Cziesla, T.; Biswas, G.; Chattopadhyay, H.; Mitra, N.K. Large-eddy simulation of flow and heat transfer in an impinging slot jet. Int. J. Heat Fluid Flow 2001, 22, 500–508.

[31] Draksler, M.; Končar, B.; Cizelj, L.; Ničeno, B. Large Eddy Simulation of Multiple Impinging Jets in Hexagonal Configuration—Flow Dynamics and Heat Transfer Characteristics. Int. J. Heat Mass Transf. 2017, 109, 16–27.

[32] Drummond, K. P. et al. A hierarchical manifold microchannel heat sink array for high-heat-flux two-phase cooling of electronics. International Journal of Heat and Mass Transfer, February 2018, Vol.117, pp.319-330.

[33] Sharma, C. S. et al. A novel method of energy efficient hotspot-targeted embedded liquid cooling for electronics: An experimental study. International Journal of Heat and Mass Transfer, September 2015, Vol.88, pp.684-694.

[34] Drummond, K. P.; Back, D.; Sinanis, M. D.; Janes, D. B.; Peroulis, D.; Weibel, J. A.; and Garimella, S V., "A Hierarchical Manifold Microchannel Heat Sink Array for High-Heat-Flux Two-Phase Cooling of Electronics" (2018). CTRC Research Publications. Paper 320. http://dx.doi.org/10.1016/j.ijheatmasstransfer.2017.10.015.

[35] S. Sarangi, K. K. Bodla, S. V. Garimella, and J. Y. Murthy, “Manifold Microchannel Heat Sink Design Using Optimization Under Uncertainty,” International Journal of Heat and Mass Transfer, Vol. 69, pp. 92-105, 2014.

[36] K. P. Drummond, D. Back, M. D. Sinanis, D. B. Janes, and D. Peroulis, J. A. Weibel, and S. V. Garimella, “A Hierarchical Manifold Microchannel Heat Sink Array for High-Heat-Flux Two-Phase Cooling of Electronics,” International Journal of Heat and Mass Transfer, Vol. 117, pp. 319-330, 2018.

[37] De Oliveira PA, et al., Performance Assessment of Single and Multiple Jet Impingement Configurations in a Refrigeration-Based Compact Heat Sink for Electronics Cooling. Journal of Electronic Packaging, 09/01/2017, Vol.139(3), p.031005.

[38] J.H. Ryu, D.H. Choi, S.J. Kim, Three-dimensional numerical optimization of a manifold microchannel heat sink, International Journal of Heat and Mass Transfer, 46(9) (2003) 1553–1562.

[39] Oprins, H. et al., Experimental Characterization of the Vertical and Lateral Heat Transfer in 3D Stacked IC Packages, Journal of Electronic Packaging, 2016, Vol.138(1), pp.010902-1(10).

[40] Salim SM, Cheah SC. Wall y+ strategy for dealing with wall-bounded turbulent flows. In Proceedings of the International MultiConference on Engineerings and Computer Science, Hong Kong. 2009.

[41] Roache, P. J. Quantification of uncertainty in computational fluid dynamics. Annual Review of Fluid Mechanics 29, 1 (1997), 123–160.

[42] ANSYS Fluent Tutorial Guide. Release 18.0. ANSYS, Inc. January 2017. Southpointe. 2600 ANSYS Drive. Canonsburg, PA 15317.

[43] Penumadu, P. S. and Arvind G. Rao. Numerical investigations of heat transfer and pressure drop characteristics in multiple jet impingement system, (2017).

[44] A Heat Transfer Textbook 3rd Edition - By (John H. Lienhard IV & John H. Lienhard V).

[45] B.P.Leonard, A stable and accurate convective modelling procedure based on quadratic upstream interpolation, Computer Methods in Applied Mechanics and Engineering, Volume 19, Issue 1, June 1979, pp. 59-98.

[46] R. W. Bonner, et al., Local heat transfer coefficient measurements of flat angled sprays using thermal test vehicle. 24th Annual IEEE Semiconductor Thermal Measurement and Management Symposium, March 2008, pp.149-153.

[47] T. Brunschwiler et al. Direct liquid jet-impingement cooling with micron sized nozzle array and distributed return architecture. Thermal and Thermomechanical Proceedings 10th Intersociety Conference on Phenomena in Electronics Systems, 2006, pp.196-203.

[48] GA Kulkarni, et al., Jet impingement heat transfer of moving metal sheet. 13th International Conference on Heat Transfer, Fluid Mechanics and Thermodynamics (2017).

[49] KA Agbim, et al., Single-phase liquid cooling for thermal management of power electronic devices, thesis, Georgia Tech (2017).

[50] B. P. Whelan, R. Kempers, A. J. Robinson, A liquid-based system for CPU cooling implementing a jet array impingement waterblock and a tube array remote heat exchanger. Applied Thermal Engineering, June 2012, Vol.39, pp.86-94.

[51] 1.Carcasci, C., Cocchi, L.,et al., Impingement cooling experimental investigation using different heating elements. Energy Procedia, November 2016, Vol.101, pp.18-25.

[52] A. Terzis, S. Bontitsopoulos, et al., Improved Accuracy in Jet Impingement Heat Transfer Experiments Considering the Layer Thicknesses of a Triple Thermochromic Liquid Crystal Coating. Journal of Turbomachinery, 02/01/2016, Vol.138(2), p.021003.

[53] C. Y. Huang, et al., The application of temperature-sensitive paints for surface and fluid temperature measurements in both thermal developing and fully developed

regions of a microchannel. Journal of Micromechanics and Microengineering, 2013, Vol.23(3), p.037001 (7pp).

[54] H. D. Haustein, et al., Local heat transfer coefficient measurement through a visibly-transparent heater under jet-impingement cooling. International Journal of Heat and Mass Transfer, November 2012, Vol.55(23-24), pp.6410-6424.

[55] F. Xu, M. S. Gadala, Investigation of error sources in temperature measurement using thermocouples in water impingement cooling. Experimental Heat Transfer, 01 July 2005, Vol.18(3), p.153-177.

[56] B. P. Whelan, A. J. Robinson, Nozzle geometry effects in liquid jet array impingement. Applied Thermal Engineering, 2009, Vol.29(11), pp.2211-2221.

[57] T. B. Hoberg, et al., Heat transfer measurements for jet impingement arrays with local extraction. International Journal of Heat and Fluid Flow, 2010, Vol.31(3), pp.460-467.

[58] D. H. Rhee, et al., Local heat/mass transfer and flow characteristics of array impinging jets with effusion holes ejecting spent air. International Journal of Heat and Mass Transfer, 2003, Vol.46(6), pp.1049-1061.

[59] V. S. Patil, R. P. Vedula, Local heat transfer for jet impingement on a concave surface including injection nozzle length to diameter and curvature ratio effects. Experimental Thermal and Fluid Science, April 2018, Vol.92, pp.375-389.

[60] Maddox, J. F., Knight, et al., Liquid Jet Impingement with an Angled Confining Wall for Spent Flow Management for Power Electronics Cooling with Local Thermal Measurements. Journal of Electronic Packaging, 09/01/2015, Vol.137(3), p.031015.

[61] Drummond, K. P. et al. A hierarchical manifold microchannel heat sink array for high-heat-flux two-phase cooling of electronics. International Journal of Heat and Mass Transfer, February 2018, Vol.117, pp.319-330.

[62] Sharma, C. S. et al. A novel method of energy efficient hotspot-targeted embedded liquid cooling for electronics: An experimental study. International Journal of Heat and Mass Transfer, September 2015, Vol.88, pp.684-694.

[63] De Oliveira PA, et al., Performance Assessment of Single and Multiple Jet Impingement Configurations in a Refrigeration-Based Compact Heat Sink for Electronics Cooling. Journal of Electronic Packaging, 09/01/2017, Vol.139(3), p.031005.

[64] Vutha, A. K. et al. Spatial temperature resolution in single-phase micro slot jet impingement cooling. International Journal of Heat and Mass Transfer, March 2018, Vol.118, pp.720-733.

[65] E. N. Wang et al., Micromachined jets for liquid impingement cooling of VLSI chips. ournal of Microelectromechanical Systems, Oct. 2004, Vol.13(5), pp.833-842.

[66] Bonner, R. W., et al., Local heat transfer coefficient measurements of flat angled sprays using thermal test vehicle. 2008 Twenty-fourth Annual IEEE Semiconductor Thermal Measurement and Management Symposium, March 2008, pp.149-153.

[67] Cherman, V. et al. 3D stacking induced mechanical stress effects. IEEE 64th Electronic Components and Technology Conference, May 2014, pp.309-315.

[68] Cherman, V. et al. Effects of packaging on mechanical stress in 3D-ICs. 2015 IEEE 65th Electronic Components and Technology Conference, May 2015, pp.354-361.

[69] A. Salahouelhadj, M. Gonzalez, et al., Die thickness impact on thermo-mechanical stress in 3D packages. 2015 16th International Conference on Thermal, Mechanical and Multi-Physics Simulation and Experiments in Microelectronics and Microsystems, Budapest, 2015, pp. 1-6.

[70] Sauciuc, Ioan, et al., Thermal Performance and Key Challenges for Future CPU Cooling Technologies. IPACK2005 , Advances in Electronic Packaging, San Francisco, California, USA. July 17–22, 2005. pp. 353-364.

[71] Ralph Remsburg, Joe Hager, Direct Integration of IGBT Power Modules to Liquid Cooling Arrays, Electric Vehicle Symposium, 2007.

[72] Principles Elements of Power Electronics, barry W. Williams, Cooling of power switching semiconductor devices, 2006, pp 199-200.

# Chapter 3

# 3. Unit Cell Level Thermal & Hydraulic Analysis

### 3.1 Introduction

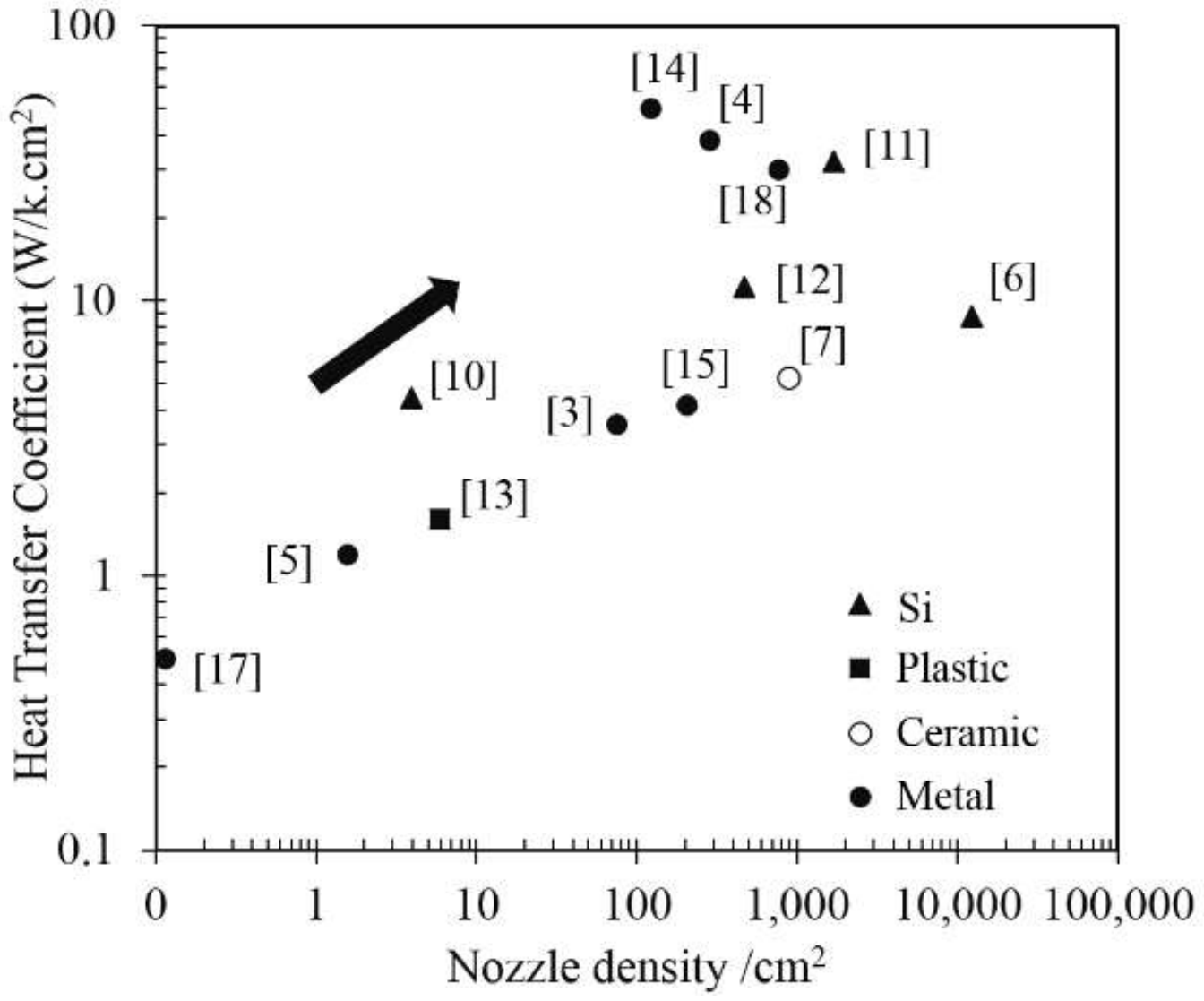

**Figure 3.1:** Trend of the dissipated heat flux in the chip as a function of the nozzle density of the cooler (adapted from [9]).

In chapter 1, a systematic review about microjet cooling techniques is presented and the thermal performance is benchmarked. Figure 3.1 summarizes the published performance data in terms of the reported heat transfer coefficient as a function of the jet nozzle density $\rho_N$=N$^2$/A, where the nozzle density is defined as the total nozzle number N$^2$ divided by the cooled chip area A. In general, the chart shows an increase of the heat transfer for high nozzle density values. However, the highest achieve heat transfer is not obtained from the finest nozzle array. Furthermore, high nozzle densities require more expensive fabrication techniques, such as the silicon and ceramic fabrication techniques, due to the complexity of the internal structure and the integration with the chip packaging. Therefore, a thorough investigation of the impact of the nozzle density is very important for the design of an efficient liquid impingement jet cooler and the selection of a cost-efficient fabrication technique.

As summarized in chapter 1, several studies have showed the demonstrations and measurements for different nozzle densities, based on jet nozzles with distributed returns. In [3], a metal based single-jet direct impingement cooler with nozzle density of 1.56 cm$^{-2}$ was demonstrated on a single MOSFET semiconductor showing a heat transfer coefficient of 1.2 × 104 W/m$^2$K for a pumping power of 0.9 W. However, the obtained heat coefficient distribution for single jet cooling is highly nonuniform, due to the different flow regions: stagnation region, wall jet region and decay region. In [6], Brunschwiler et al. demonstrated that Si based microjet array impingement coolers with 50,000 inlet/outlet nozzles, allow to increase the heat transfer coefficient to 8.7 × 10$^4$ W/m$^2$K with 1.43W pump power. The nozzle density of this cooler is 12500 cm$^{-2}$. Moreover, Natarajan and Bezama [7] developed a microjet cooler with 1600 inlets and 1681 outlets using multilayer ceramic technology, resulting a nozzle density of 888 cm$^{-2}$.

The thermal performance of the cooler is furthermore affected by the nozzle diameter d and cavity height H which are usually coupled with the nozzle density N$^2$/A. For a free-surface jet cooling, Womac et al. [20] observed that the H/d ratio has a negligible effect on the heat transfer based on the experimental study. For the single submerged and the confined jet cooling, Garimella [21,22] experimentally investigated the effect of the H/d, L/d, and flow rate on heat transfer. They found that H/d significantly affects the heat transfer performance of the system, especially for multiple confined jet impingement cooling. Aldabbagh and Sezai [23] also concluded that jet-to-plate spacing (H) significantly affects heat transfer performance. Afzal Husain [24] reported that at both low and high flow rates, the change in cavity height does not affect pressure drop significantly. However, the decrease in the spacing between the nozzle to the heated surface increases the heat transfer coefficient monotonously. Brunschwiler et al. defines four typical regimes based on the cavity height, which are the pinch-off regime ($H<H_{critical}$), the impingement regime, the transition regime, and the separation regime [6]. They observed that the both the heat transfer and the pressure drop increase rapidly for reducing cavity height in the pinch-off regime and that the heat transfer remains constant as a function of the cavity height in the impingement regime. In [25], the experimental study also shows that thermal performance was insensitive to the gap at large spacing, but below a specific gap it degraded with decreasing gap. However, the studies are only limited to a specific nozzle density.

From the discussion above, it is clear that there is an interaction between the impact of the nozzle density and the cavity height on the thermal and hydraulic performance of the impingement jet cooler. The focus of this chapter is to investigate the combined impact of the jet array design parameters on the thermo-hydraulic cooler performance. These parameters mainly include the nozzle density, the cavity height, and the nozzle diameter. The objective of this study is to provide guideline for predicting the thermal/hydraulic performance of the cooler based on user's constraint working conditions. For the structure of this chapter: in the first part in section 3.2, the parametric analysis based on the absolute parameters is performed. Moreover, the different optimization methodologies are compared and discussed for determining the optimal cooler. In the second part shown in section 3.3, the dimensionless analysis is introduced to understand the physics in the area of heat transfer and fluid mechanics for impingement jet cooling. Predictive models based on the thermal and hydraulic performance are developed for the fast prediction of the cooling performance.

## 3.2 Parametric analysis

### 3.2.1 Design of experiments

This section describes the parametric analysis of an impingement cooling geometry. Figure 3.2 shows a graphical representation of the geometrical parameters of the unit cell for an impinging system with an N×N array of inlet nozzles and distributed inlets: inlet diameter $d_i$, outlet diameter $d_o$, cavity height H, nozzle plate thickness t, chip thickness $t_c$ and unit cell size $L$. The unit cell size is defined as the ratio between the chip size $S_d$ and the nozzle row number $N$: $L = S_d / N$. In order to find the best combination of the design variables and study the scaling trends, an extensive parameter sensitivity study has been performed by varying the three variables: nozzle density: $N^2/A$, cavity height $H$ and nozzle diameter ratio $d_i/L$. The parameters ranges are listed in Figure 3.2. The chip area in this study is fixed to 8×8 cm$^2$, with the size of the thermal test chip introduced in chapter 2. Table 3.1 lists the nozzle density for different nozzle array with respect to the 8×8 cm$^2$.

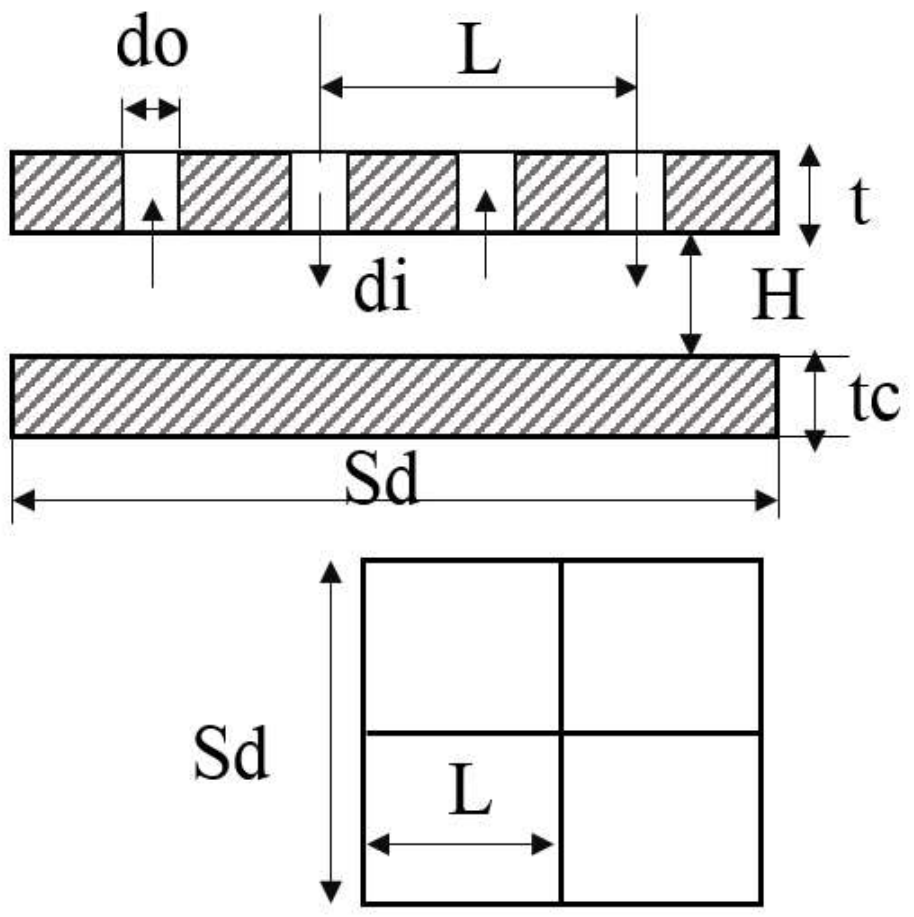

| Parameters | Range |
|---|---|
| N | 1-64 |
| $d_i/L$ | 0.025-0.4 |
| $d_o/L$ | 0.025-0.4 |
| H | 0.01-1 mm |
| t/L | 0.025-0.4 |
| $t_c$ | 0.2mm |
| $S_d$ | 8mm |

**Figure 3.2:** Cross-section of cooler configuration and parameter ranges for 8×8 cm$^2$ chip.

For each unit cell model in the DOE, the temperature distribution is calculated for the chosen input boundary conditions. From the temperature distribution in the active region of the Si chip, the thermal resistance of the impingement cooler is defined in chapter 2.

**Table 3.1:** Nozzle density with respect to the 8×8 $mm^2$ test chip

| N×N | Nozzle density | Cooling unit cell (mm) |
|---|---|---|
| 1×1 | 1.56 $cm^{-2}$ | 8 |
| 2×2 | 6.25 $cm^{-2}$ | 4 |
| 3×3 | 14.06 $cm^{-2}$ | 2.6 |
| 4×4 | 25 $cm^{-2}$ | 2 |
| 6×6 | 56.25 $cm^{-2}$ | 1.33 |
| 8×8 | 100 $cm^{-2}$ | 1 |
| 12×12 | 225 $cm^{-2}$ | 0.67 |
| 16×16 | 400 $cm^{-2}$ | 0.5 |
| 32×32 | 1600 $cm^{-2}$ | 0.25 |
| 64×64 | 6400 $cm^{-2}$ | 0.125 |

Based on the dimensionless inputs of the DOE, the inlet diameter ranges from 10 µm to 3.2 mm in absolute numbers. For the flow conditions and unit cell geometries used in the simulations, the Reynolds number $Re_d$ based on the nozzle diameter ranges from 10 to 5300. The turbulence transition region for liquid impingement flow is defined for $Re_d$ between 1000 and 3000 [26]. A benchmarking of laminar and turbulent models showed that accurate results were obtained with the transition SST model over the whole considered range of fully laminar, transitional and turbulent flow, as discussed in the modeling part of chapter 2. To compare the thermal and hydraulic performance of different cooler designs, a relevant basis for the comparison should be chosen. In literature, optimal thermal designs of the microchannel heat sink are performed and discussed under constraint conditions, such as constant pressure drops [27], constant coolant volumetric flow rates, and constant pumping power [28]. In this work, the impact of different constraint conditions on the impingement jet cooler design is also discussed. For the design of experiments, section 3.2.2 will investigate the combined effects of nozzle density and nozzle diameter; section 3.2.3 will discuss the combined effects between the nozzle density and cavity height.

### 3.2.2 Nozzle density versus nozzle diameter

#### 3.2.2.1 Single objective analysis

First, the modeling results of the DOE are analyzed for a constant pressure drop of 40 kPa (fixed pressure as limited by the strength of the assembly). The achieved thermal resistance and the required pumping power are shown in Figure 3.3 as a function of the nozzle row number $N$ and the absolute inlet diameter assuming a constant $H$ and $t_c$. From Figure 3.3(a), it can be observed that the thermal resistance reduces for an increasing inlet diameter. This happens however, at the expense of an increase in required pumping power due to the larger flow rate to maintain the constant pressure drop (Figure 3.3(b)). Alternatively, the cooling performance can be compared for a constant flow rate [28], which can be easily controlled in an experimental set-up. Figure 3.4 shows the modeling results of a subset of the DOE for a constant flow rate of 530 mL/min, assuming a constant $H$ and $t_c$. The analysis assuming a constant flow rate shows that the lowest thermal resistance values are obtained for the smallest nozzle diameters. This is caused by the much higher velocity that is reached in the smaller diameter channels, but at the expense of a much higher pumping pressure (Figure 3.4(b)).

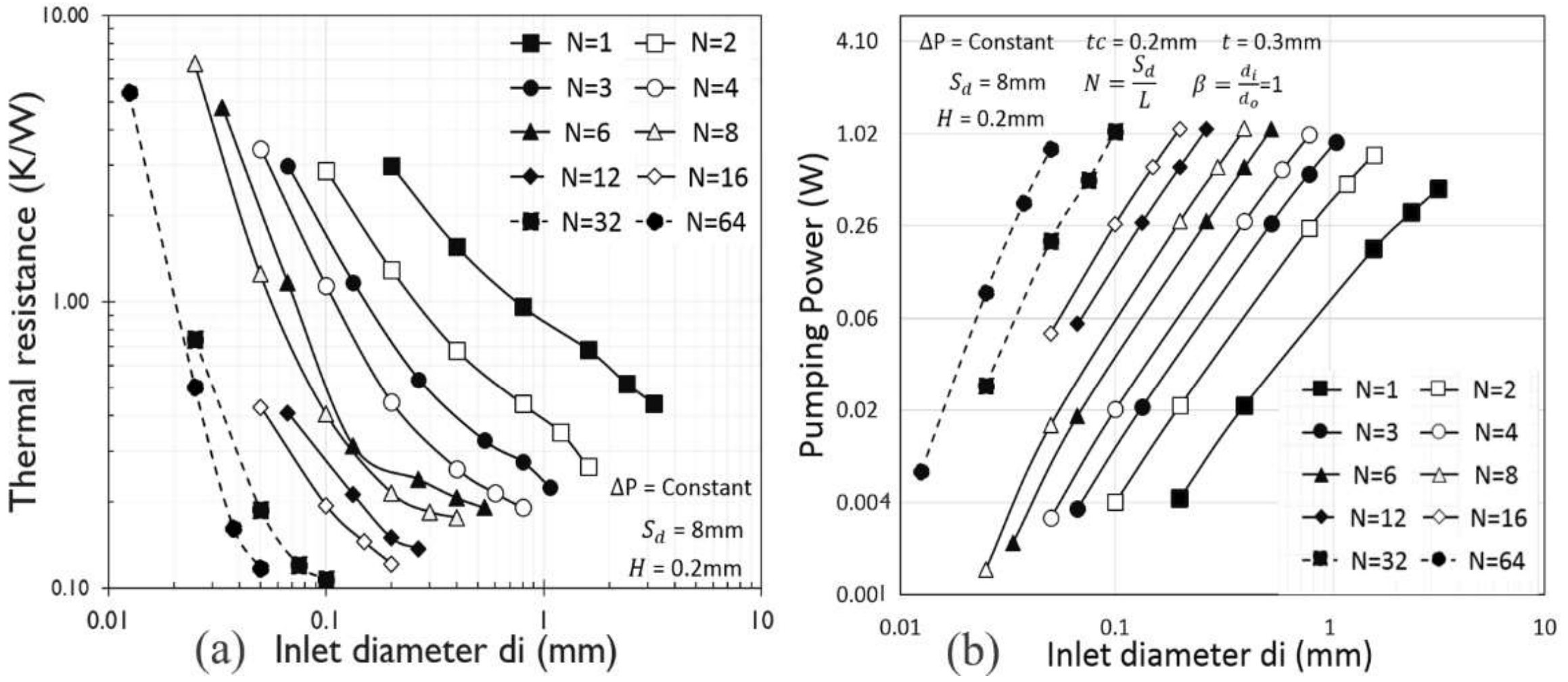

**Figure 3.3:** Thermal and hydraulic performance for a constant pressure drop $\Delta P$ = Constant: 40 kPa ($10 \leq Re \leq 3500$) for different nozzle arrays (from N=1 to N=64).

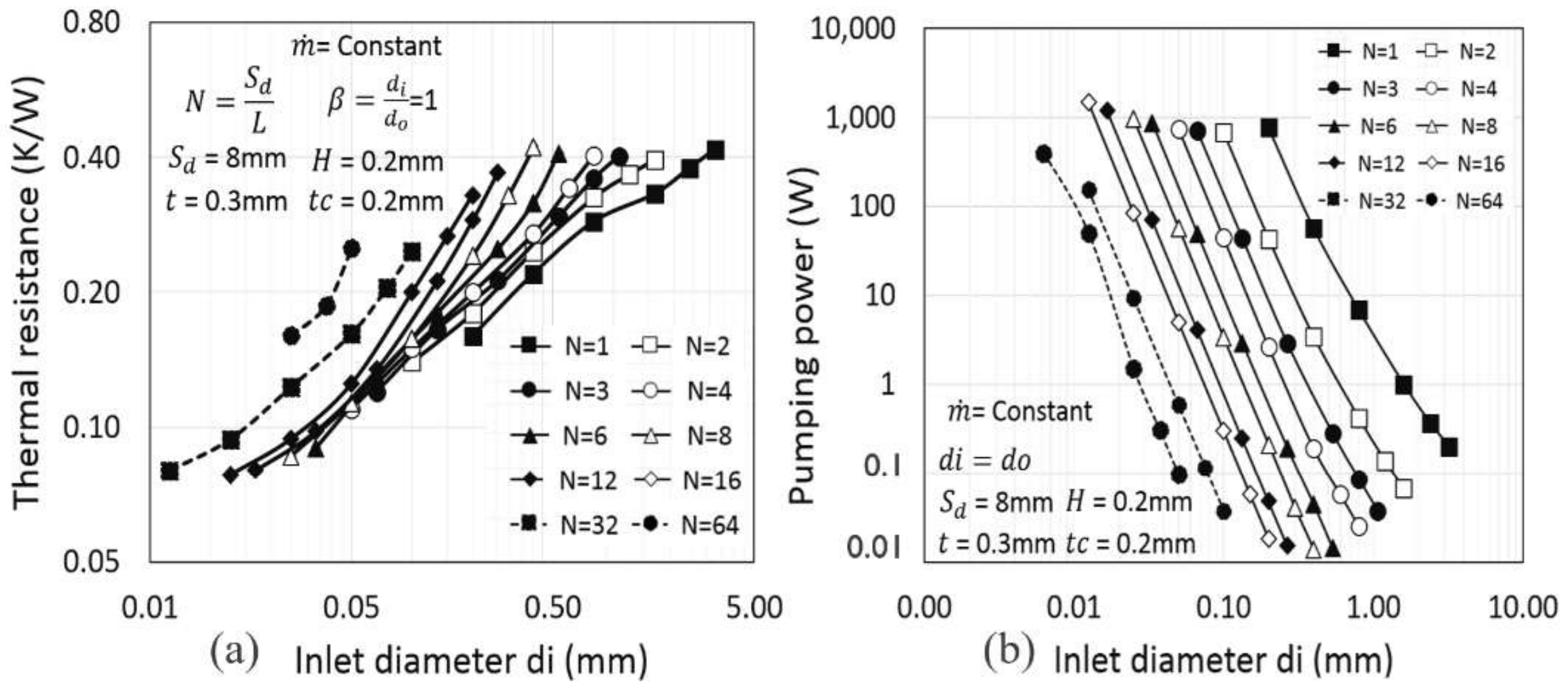

**Figure 3.4:** Thermal and hydraulic performance evaluation based on a constant flow rate of 530 mL/min ($10 \leq Re \leq 3500$).

These analyses show that the constraint of constant pressure drop will steer the optimal nozzle diameters towards larger values while the constant flow rate constraint favors small diameters, however both at the expense of the increased pumping power. Therefore, it is required to make the trade-off between the obtained thermal resistance value and the required pumping power in a single metric or chart.

3.2.2.2 Coefficient of performance

A metric that can be used to include both aspects is the coefficient of performance COP [29,30], defined as the ratio between the cooling power and the required pumping power. In the context of the impingement cooling, the COP can be defined as follows:

$$COP = \frac{Cooling\ power}{Required\ pumping\ power} = \frac{Max\ allowed\ temp\ uncrease/R_{th}}{Required\ pumping\ power} \tag{3.1}$$

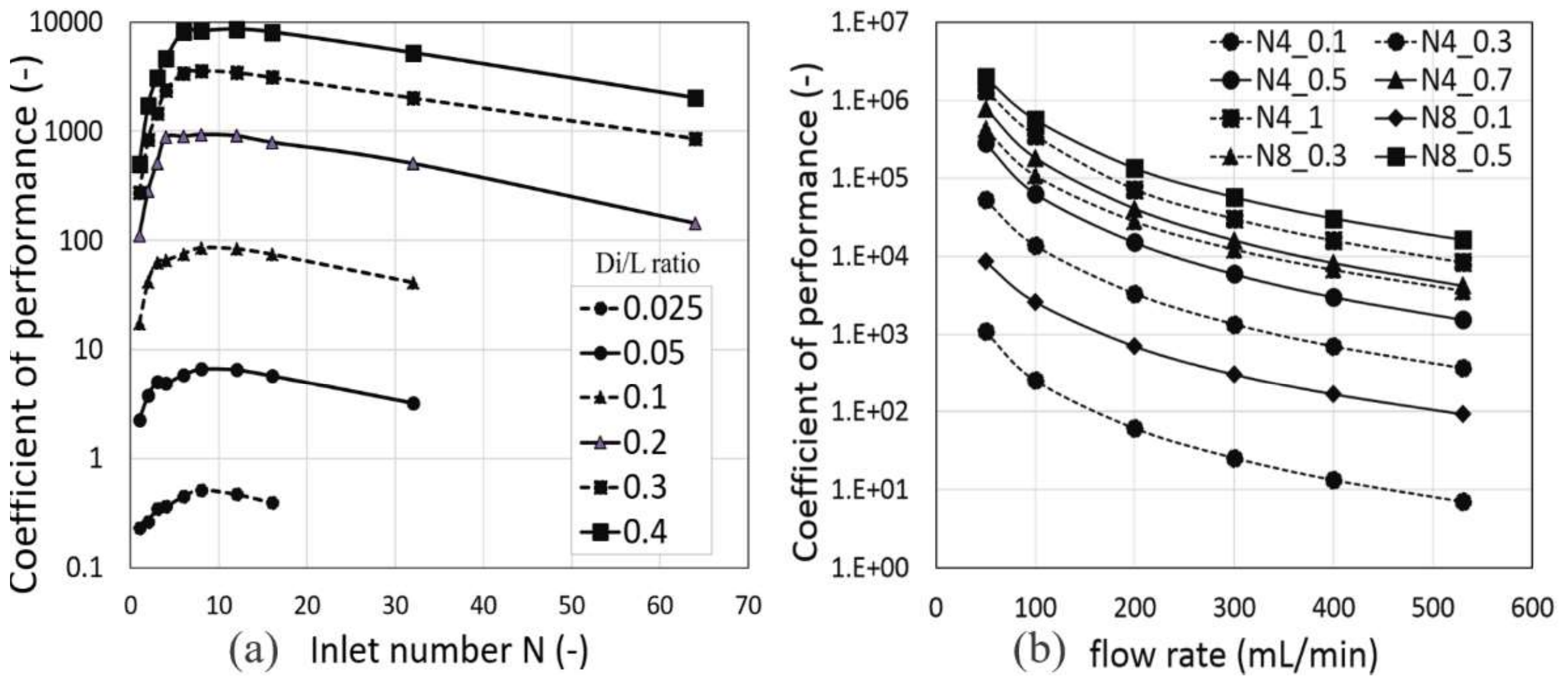

**Figure 3.5:** Thermal/hydraulic performance evaluation based on *COP* ($10 \leq Re \leq 3500$): (a) *COP* evaluations based on different inlet diameter for a fixed flow rate; (b) *COP* comparisons under different flow rates.

The results of the DOE are shown in Figure 3.5(a) expressed in terms of the *COP* as a function of the nozzle row number *N* and the inlet diameter ratio with a fixed flow rate. As the inlet row number increases, the *COP* firstly increases very fast until the range between 8 and 16 and then decreases. For the inlet diameter, we can also see that the *COP* is higher for large nozzle diameter values. Figure 3.5(b) shows however, that the *COP* increases rapidly for decreasing flow rate and that optimization for maximum *COP* would yield flow conditions with very small flow rates, which is not realistic for electronic cooling conditions. The question is whether the *COP* is an appropriate metric to compare the thermo-fluidic performance of impingement coolers under different flow conditions. It might turn out that this indeed is not the most desirable quantity to optimize.

3.2.2.3 Trade-off chart

Therefore, we analyzed the results in terms of thermal resistance and required pumping power independently. The thermal behavior of a specific cooler with certain dimensions for a large range of pressure drops and flow rates is represented by a curve in Figure 3.7 and Figure 3.8, respectively. In this chart, the thermal behavior of different cooler geometries can be compared for their full range of flow conditions; at each value of the pumping power, the obtained thermal resistance can be compared, or alternatively, the required pumping for different designs can be compared to obtain the desired thermal resistance value. The curve closest to origin has the best performance, shown as a Pareto front of the best possible thermal solutions.

The main trend from Figure 3.7 and Figure 3.8 is the saturation of the thermal resistance for increasing N under constant cavity height. Figure 3.7 shows the characteristic curves of the $R_{th}$-$W_p$ trade-off for different cooler arrays and different input flow rates. The flow rates range from 50 mL/min to 530 mL/min. The analysis is based on the constant $d_i$/L, resulting in a constant nozzle area when N is changed. Therefore, the velocity is constant for the constant flow rate. As for the constant flow rate, it shows that the thermal resistance decrease as the flow rate increases for a constant N, but the pumping power will increase on the other hand. Besides, the asymptotic behavior for higher N can be observed. It is also observed that the pumping power will first decrease and then increase for the increasing of N.

The opposite trend of the pumping power $W_p$ is observed in Figure 3.7 since there is an inverse relationship between flow rate and pressure drop. For the hydraulic performance, it can be seen that, along with constant pressure drop curves, for increasing nozzle number N, the pumping power $W_p$ will first increase and then decreases. As the N increases from N=1 to N=2, there is about 25% increment of the pumping power for constant pressure drop of 50 kPa. Along with the constant N curves, for increasing pressure drop, the pumping power will increase, and the thermal resistance will decrease. In the following part, a simplified pressure drop analysis is performed seen in Figure 3.6 to understand the trend.

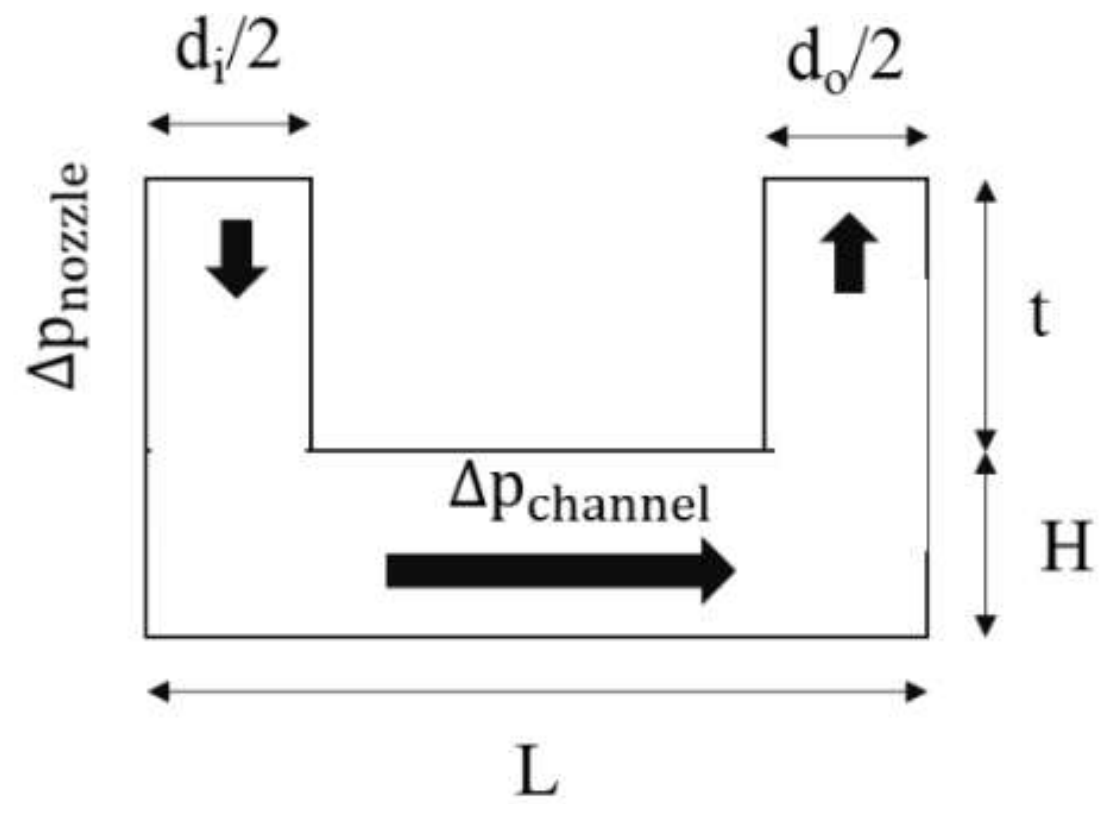

**Figure 3.6:** Schematic of the pressure drop contributions for the jet cooling within the unit cell model.

The pressure drop of the unit cell includes several parts: pressure drop along the inlet/outlet nozzle, $\Delta p_{in-nozzle}$ and $\Delta p_{out-nozzle}$ respectively, pressure drop along the cavity channel $\Delta p_{channel}$, the pressure drop generated by the jet cooling $\Delta p_{jet}$ due to flow expansion, contraction and bending. In order to understand the impact of the nozzle density on pressure drop, a simplified first-order analysis is performed, shown as below:

$$\Delta p_{tot} = \Delta p_{in-nozzle} + \Delta p_{out-nozzle} + \Delta p_{channel} + \Delta p_{jet} \quad (3.2)$$

For the pressure drop analysis inside the nozzles, the Hagen–Poiseuille equation for the pipe laminar flow is used. The equation is shown below:

$$\Delta p_{in-nozzle} = \frac{8\mu t \dot{V}_n}{\pi (d_i/2)^4}; \quad (3.3)$$

Where the nozzle length $t$ is defined as the length of the pipe flow, μ is the dynamic viscosity, $\dot{V}$ is the volumetric flow rate, and $d_i$ is the diameter of the nozzle. The $\dot{V}_n$ is the flow rate per nozzle, defined as $\frac{\dot{V}}{N^2}$.

For the pressure drop analysis inside the cavity channel,

$$\Delta p_{channel} \sim \frac{L\dot{V}_n}{H^4}; \quad (3.4)$$

The other pressure drops with the $\Delta p_{jet}$ are hard to estimate. Therefore, the pumping power of the whole nozzle array is shown below:

$$W_p = \dot{V} * \Delta p_{tot} = \dot{V} * (\Delta p_{in-nozzle} + \Delta p_{out-nozzle} + \Delta p_{channel} + \Delta p_{jet}) \quad (3.5)$$

In summary, the factors affecting the total pressure drop are the nozzle number, nozzle diameter, nozzle length, cavity height, and unit cell length L. As for the constant pressure drop, when the N is increasing, the parameters are changed as below:

- For the increasing of N, the unit cell length L will reduce, based on $L = \frac{8}{N}$;
- For the decreasing of L, the nozzle diameter $d_i$ decreases, based on $\frac{d_i}{L} = 0.1$;
- Since the $d_i$/L is kept as constant, the area of the total holes $A_{tot}$ is constant, based on the following equation:

$$A_{tot} = N^2 \frac{\pi {d_i}^2}{4} = \frac{64}{L^2} * \frac{\pi {d_i}^2}{4} = \frac{16\pi {d_i}^2}{L^2};$$

The changing of parameters for increasing N will result in the following trend:

- The decreasing of L will reduce the $\Delta p_{channel}$, as the H is kept constant;
- The decreasing of $d_i$ will increase the $\Delta p_{in-nozzle}$ and $\Delta p_{out-nozzle}$;

Therefore, the pumping power includes two main parts, with two opposite trends;

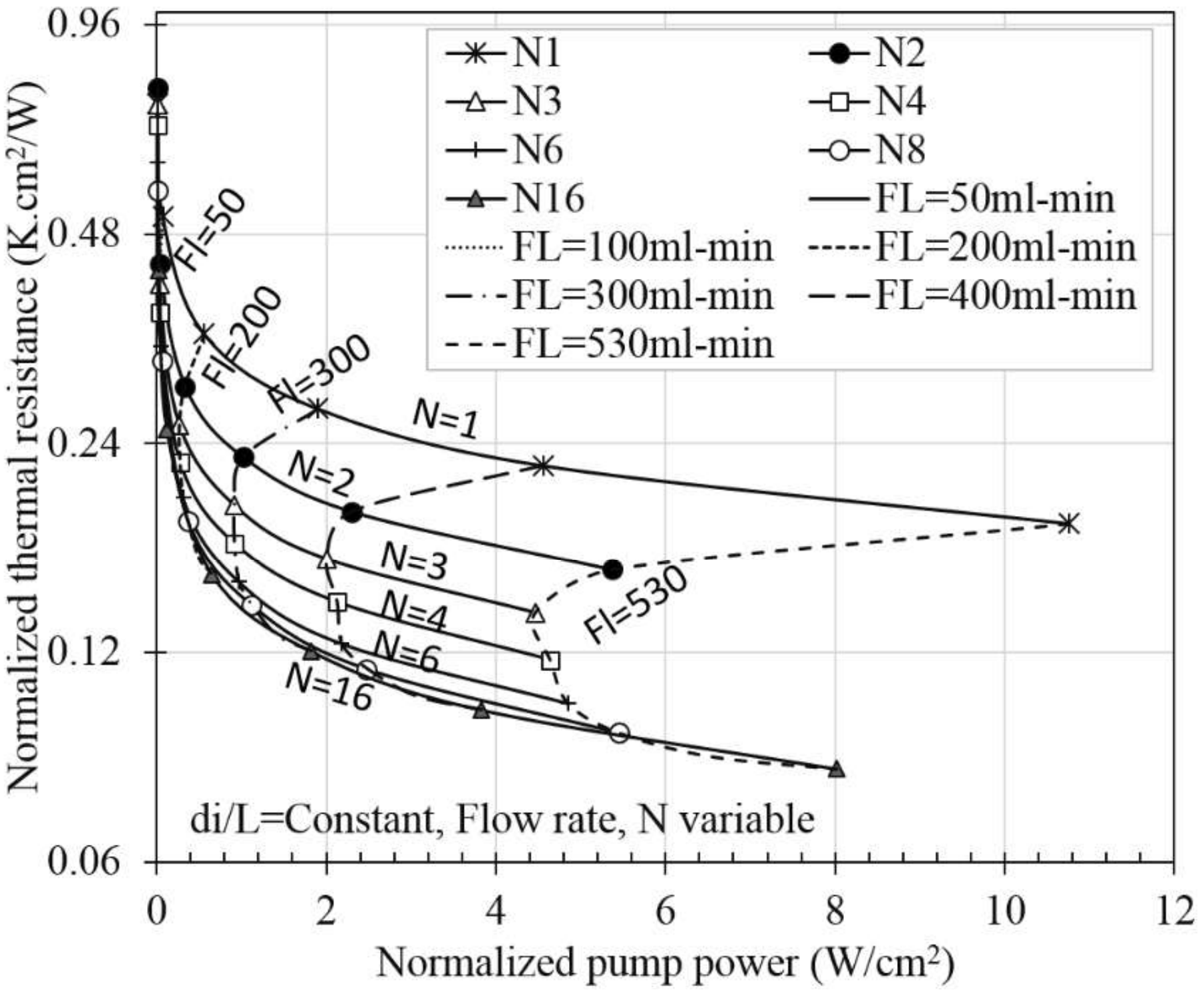

**Figure 3.7:** Characteristic curve of the cooler with different nozzle number and flow rate, with $d_i$/L kept constant ($d_i$/L=0.1, H=0.2 mm).

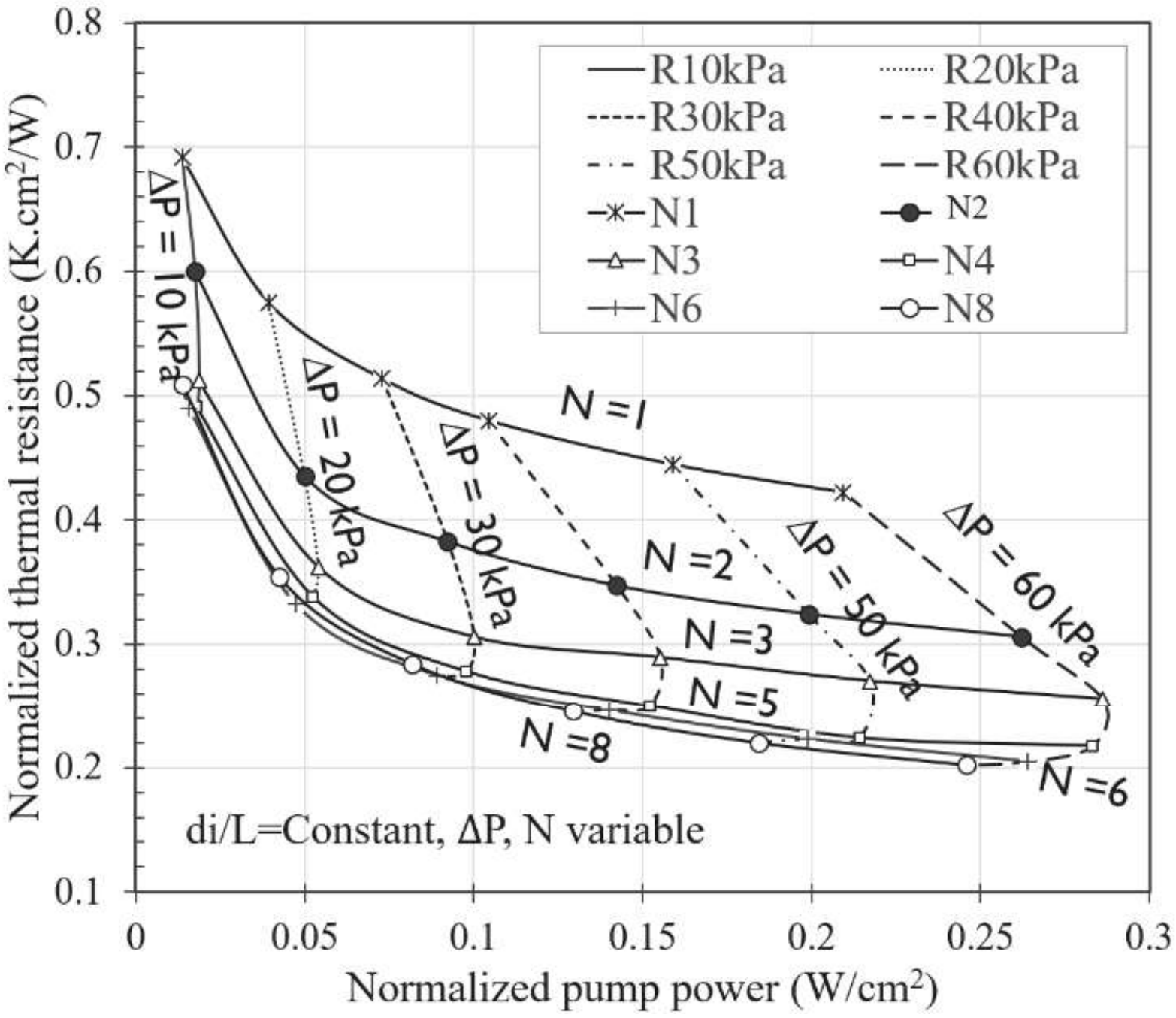

**Figure 3.8:** Characteristic curve of the cooler with different nozzle number and pressure drop, with $d_i$/L kept constant ($d_i$/L=0.1, H=0.2 mm).

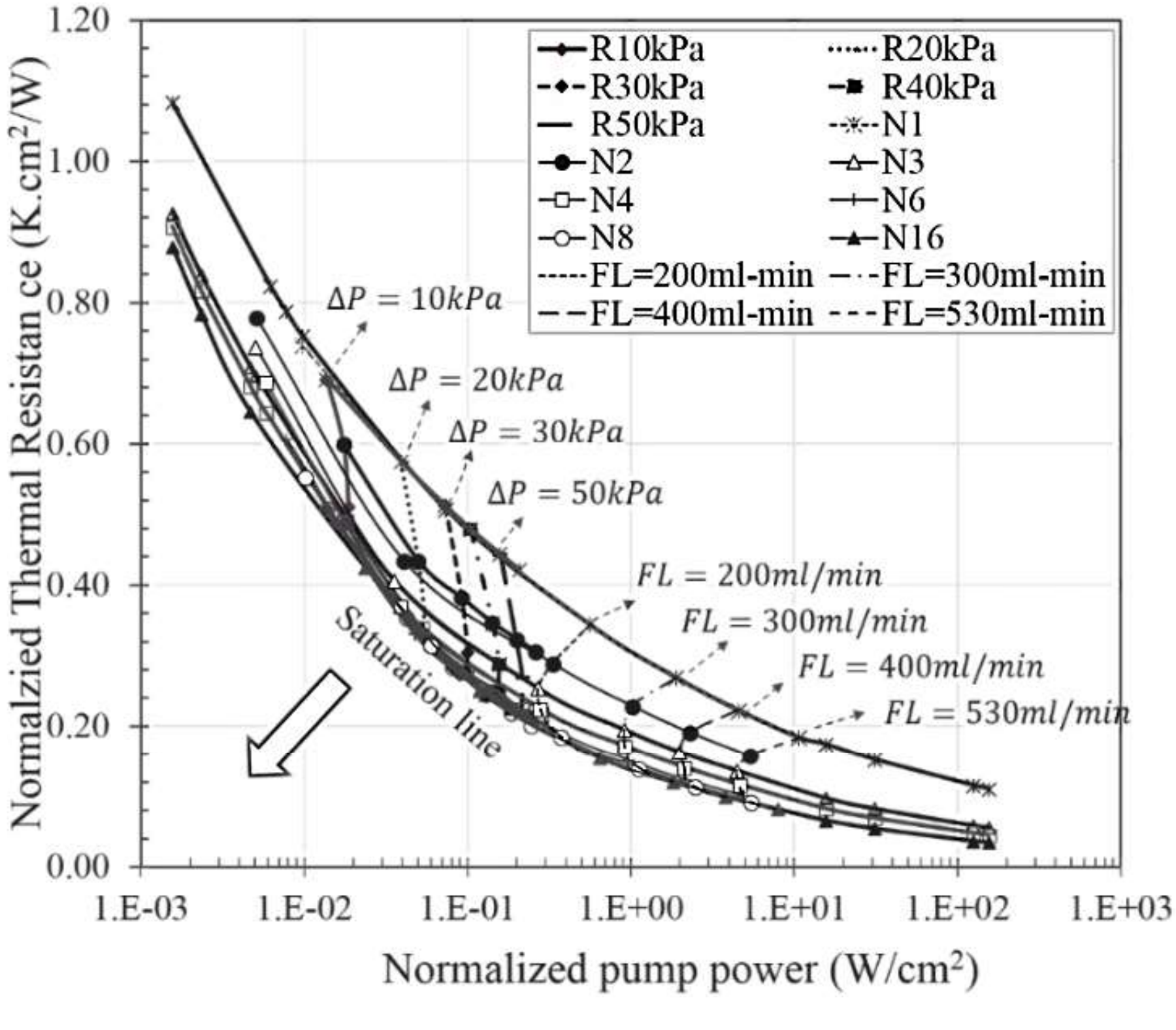

**Figure 3.9:** Characteristic curve of cooler under different boundary conditions: pressure drop=constant or flow rate=constant ($d_i$/L =0.1, H=0.2 mm).

In Figure 3.9, the analysis of Figure 3.7 and Figure 3.8 are combined to show the impact of the flow rate and pressure drop, where every line represents one designed cooler. In general, it can be seen that the higher N, the better performance for the cooler. However, the characteristic curve will saturate at a higher N number. Based on the characteristic curves, the designer can choose the optimal value based on the constrained flow rate or pressure drop.

In the next step, the impact of the nozzle diameter was also investigated as the trade-off chart for constant pressure drop and constant flow rate constrained. The nozzle diameter ratio $d_i$/L ranges from 0.025 to 0.4. The nozzle number N increases from N=1 to N=64. As shown in Figure 3.10, for constant nozzle number N and pressure drop $\Delta P$, the thermal resistance will reduce as the nozzle diameter becomes larger, and the pumping power will increase. This is due to the increase of inlet velocity as the nozzle diameter becomes larger. The inlet velocity has the inversed proportional relationship with nozzle diameter under constant pressure drop. For constant nozzle diameter, the thermal resistance decreases firstly and then increases as N increasing. On the other hand, the pumping power will decrease with N increasing. For small diameter $d_i$/L =0.025, the thermal resistance increases significantly from N=1 to N=8. This increase is mainly due to the decrease in nozzle diameter, which results in a velocity decrease when pressure drop is kept constant.

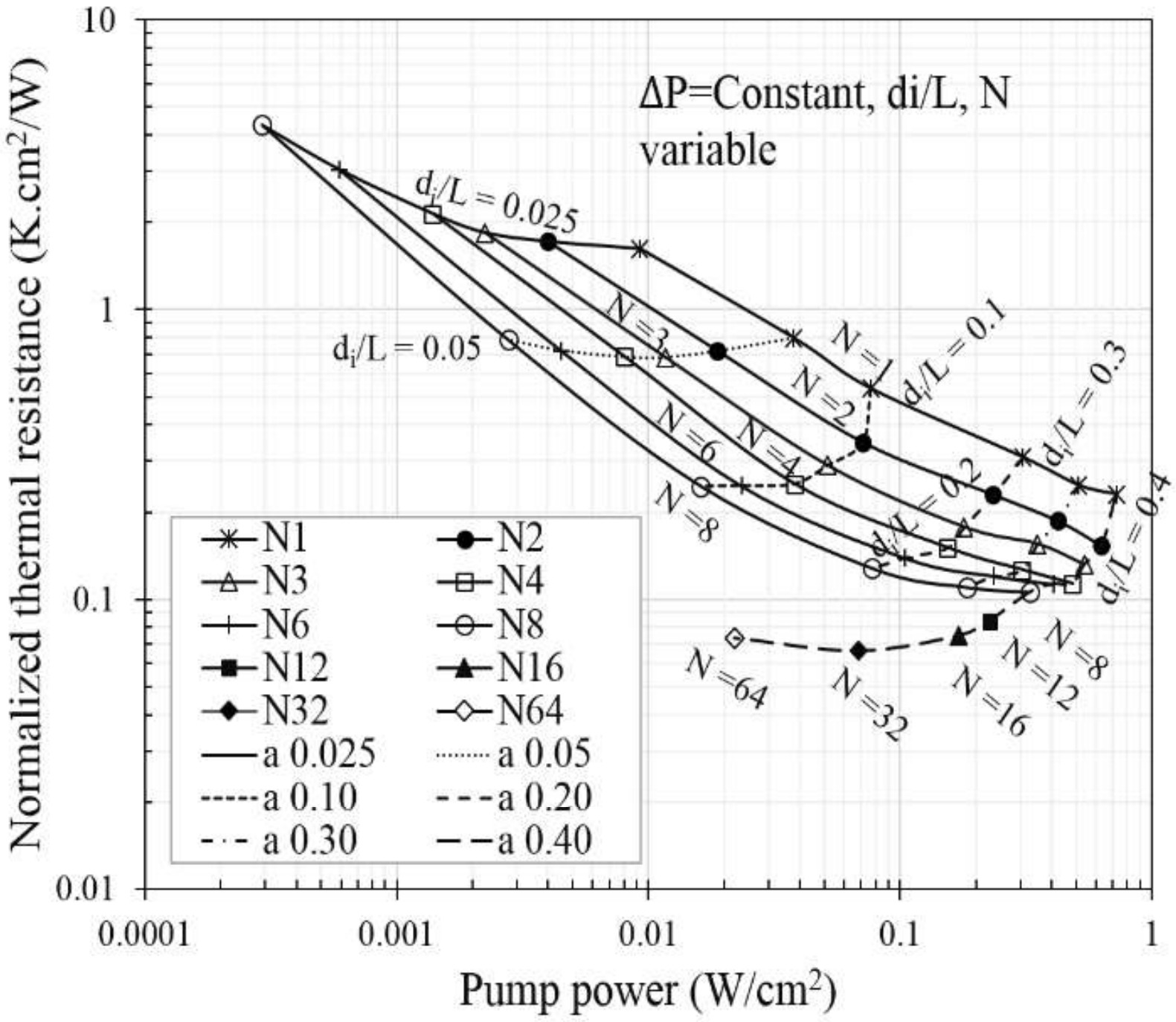

**Figure 3.10:** The characteristic curve with different nozzle diameter ratio and inlet number N (Pressure drop=40 kPa, H=0.2 mm). (Note: a in the chart represents $d_i$/L)

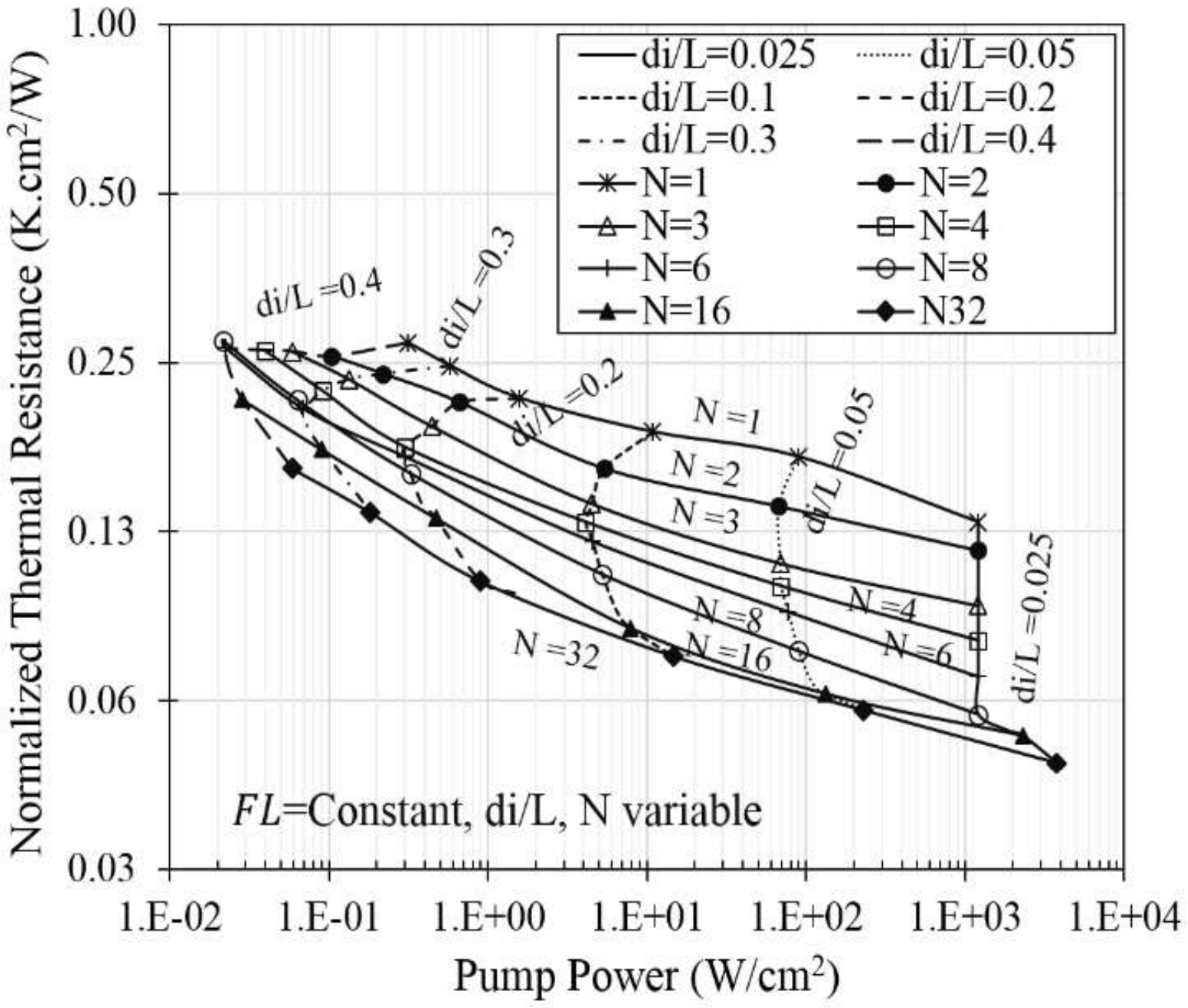

**Figure 3.11:** Characteristic curve with different nozzle diameter ratio and inlet number N (Flow rate=530 mL/min, H=0.2 mm).

Figure 3.11 shows a similar tradeoff chart for a constant flow rate. The trends are different compared with a constant pressure drop. For the constant N, the thermal resistance reduces as $d_i$/L decreases, while the pumping power increases. This is caused by the increase in inlet velocity. For constant $d_i$/L, $R_{th}$ decreases and $W_p$ decrease first and then increases with increasing N. Figure 3.12 shows the example of 4×4 and 8×8 with different inlet nozzle diameters. In summary, it can be seen that the 8×8 is much more energy efficient than 4×4. And also, the large nozzle diameter is more energy efficient for the cooler design.

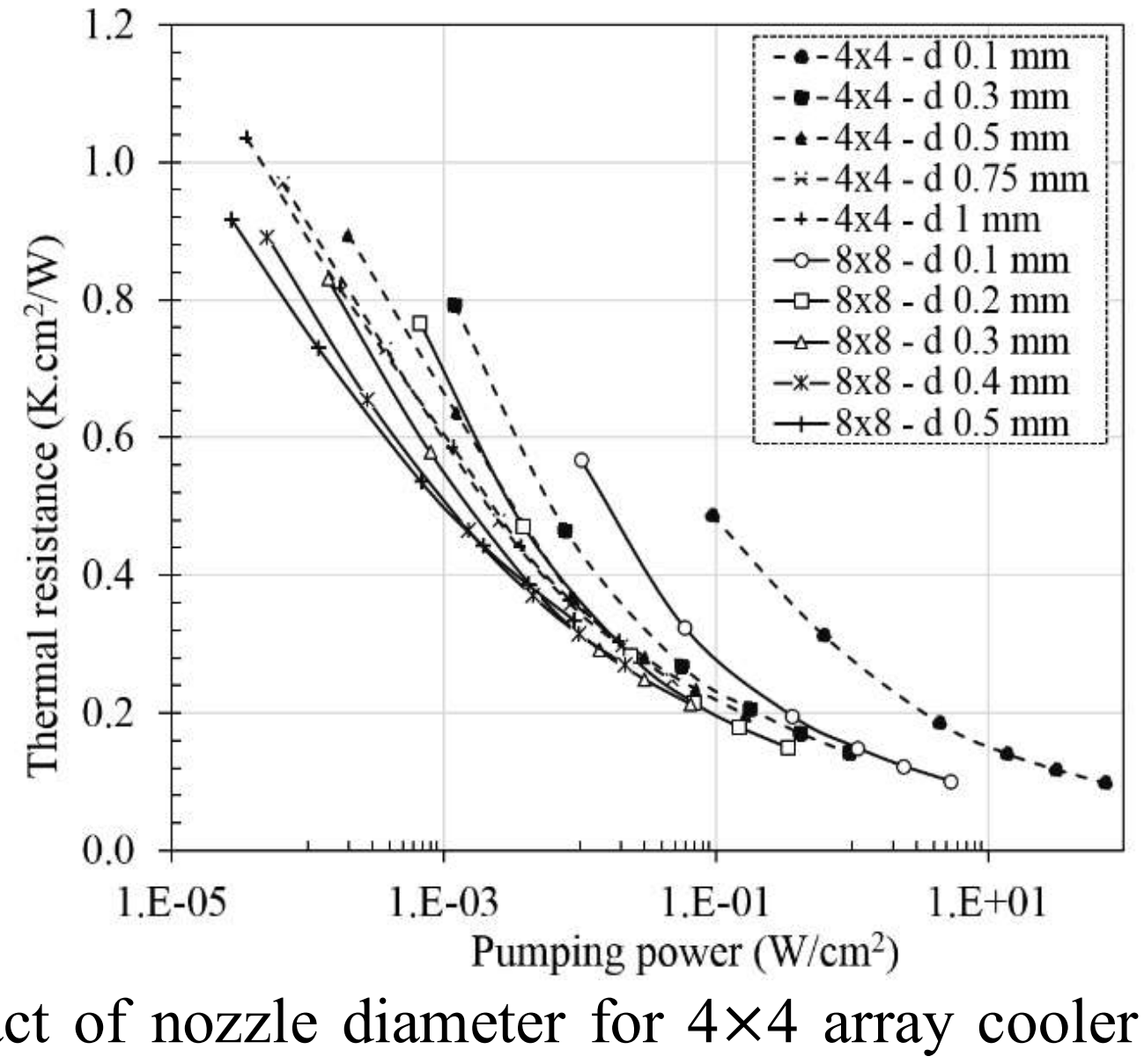

**Figure 3.12:** Impact of nozzle diameter for 4×4 array cooler and 8×8 array cooler (H=0.2mm).

### 3.2.3 Nozzle density versus Cavity height

3.2.3.1 Comparison 1: constant flow rate

In section 3.2.2, the investigation of the single variable shows that the inlet nozzle diameter is the dominant parameter, which should be optimized first. Next, the combined impact of nozzle density and cavity height are studied in this section.

In order to better understand the thermal and hydraulic behavior of the cooler, the temperature increase and pressure drop as a function of the different cavity heights are plotted in Figure 3.13, assuming a constant flow rate of 600 mL/min and a diameter ratio of $d_i/L$ = 0.3. It can be seen that the temperature increases rapidly as $H$ decreases when the cavity height is below 0.1 mm. This phenomenon can be explained by the "Pinch-off regime" observed by Brunschwiler et al. [6]. For the pinch-off regime with very thin cavity height, the pressure drop is very high since the flow is confined inside the thinner cavity channel. As the cavity height is higher than $H$=0.2 mm, the flow regime moves to the "impingement" regime, where the heat transfer performance and pressure drop both keep stable.

Besides, it is also observed that the average temperature for different nozzle densities are very close to each other in the "pinch-off" regime, shown in Figure 3.13(a). However, there is a large temperature difference in the "impingement" regime, showing that the higher nozzle density can achieve lower chip temperature for a constant flow rate.

As for the pressure drop in the "pinch-off" regime shown in Figure 3.13(b), the higher nozzle density (N=32) shows lower pressure drop. This is due to the short nozzle pitch L for higher nozzle density. The short channel length L along the wall jet region results in a pressure drop decrease in the thinner cavity channel, which is the dominating factor for the overall unit cell pressure. However, this trend changes inversely in the "impingement" regime, showing that the higher nozzle density can generate higher pressure drop. This is because the pressure in this regime is dominated by the nozzle channel pressure, where the nozzle diameter for higher N is very small.

In general, it can be seen that there are two different trends for the variation of the cavity height, from thermal and hydraulic point for view. In the next section, the interactions between the impact of the nozzle density and the cavity height on the thermal and hydraulic performance of the impingement jet cooler will be discussed.

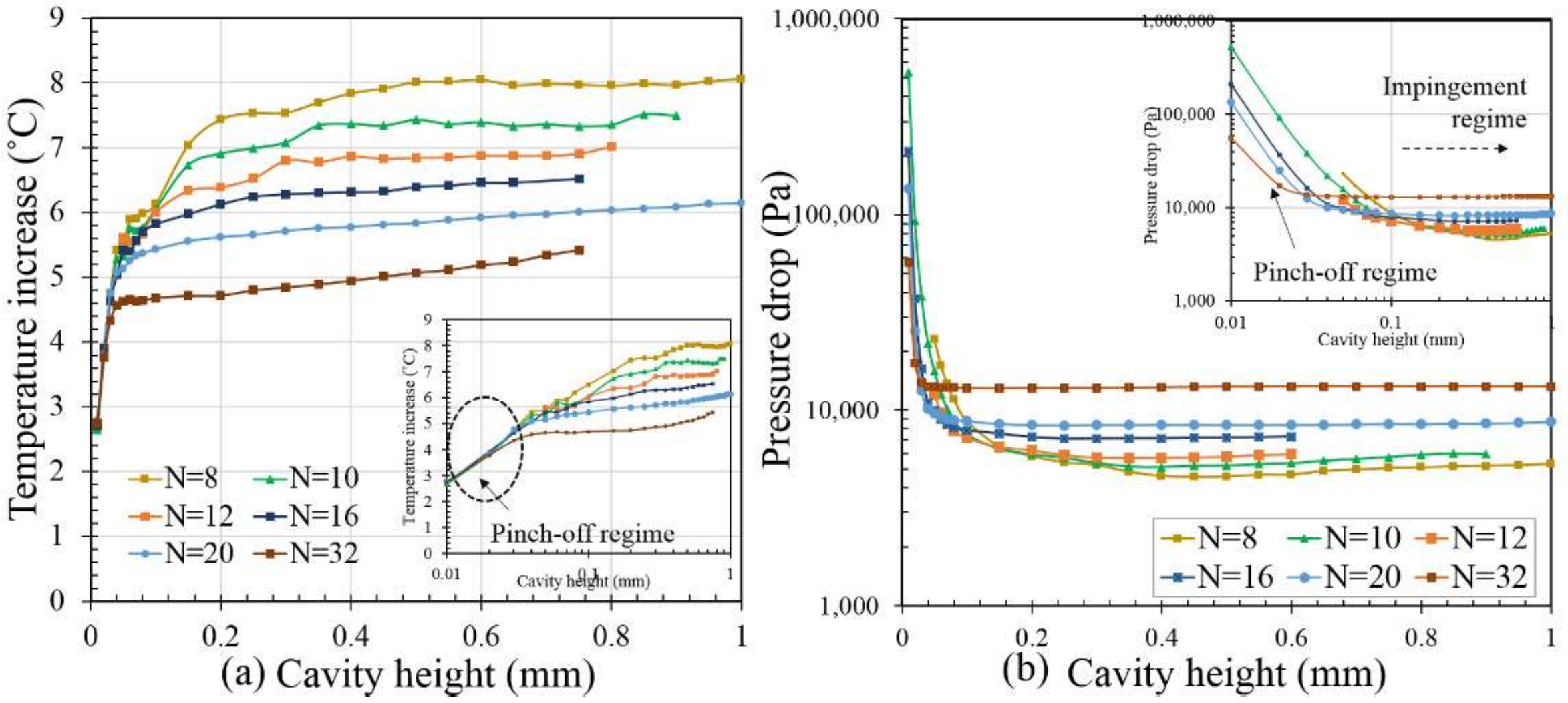

**Figure 3.13:** Unit cell modeling results: Pressure drop as function of cavity height between 0.01 mm and 1 mm ($d_i/L$ =0.3, FL=600 mL/min).

The final DOE results with the combined effects of the nozzle density and cavity height on the COP are summarized in Figure 3.14, for a flow rate of 0.3 L/min and 0.6 L/min. The opposing trends for the thermal and hydraulic performance result in a complex profile for the COP as a function of the nozzle density and cavity height, revealing a maximum for the COP in the middle range of the nozzle density. As shown in Figure 3.14(a) for the inlet diameter ratio of $d_i/L$ =0.3, the highest COP is found for the range between 30 cm$^{-2}$ and 300 cm$^{-2}$, and the cavity height effects are negligible between 0.15 mm and 0.6 mm. The region with high COP values becomes narrower as the flow rate increases to 0.6 L/min, as illustrated in Figure 3.14(b). The highest COP is now located between the range of 50 cm$^{-2}$ to 100 cm$^{-2}$.

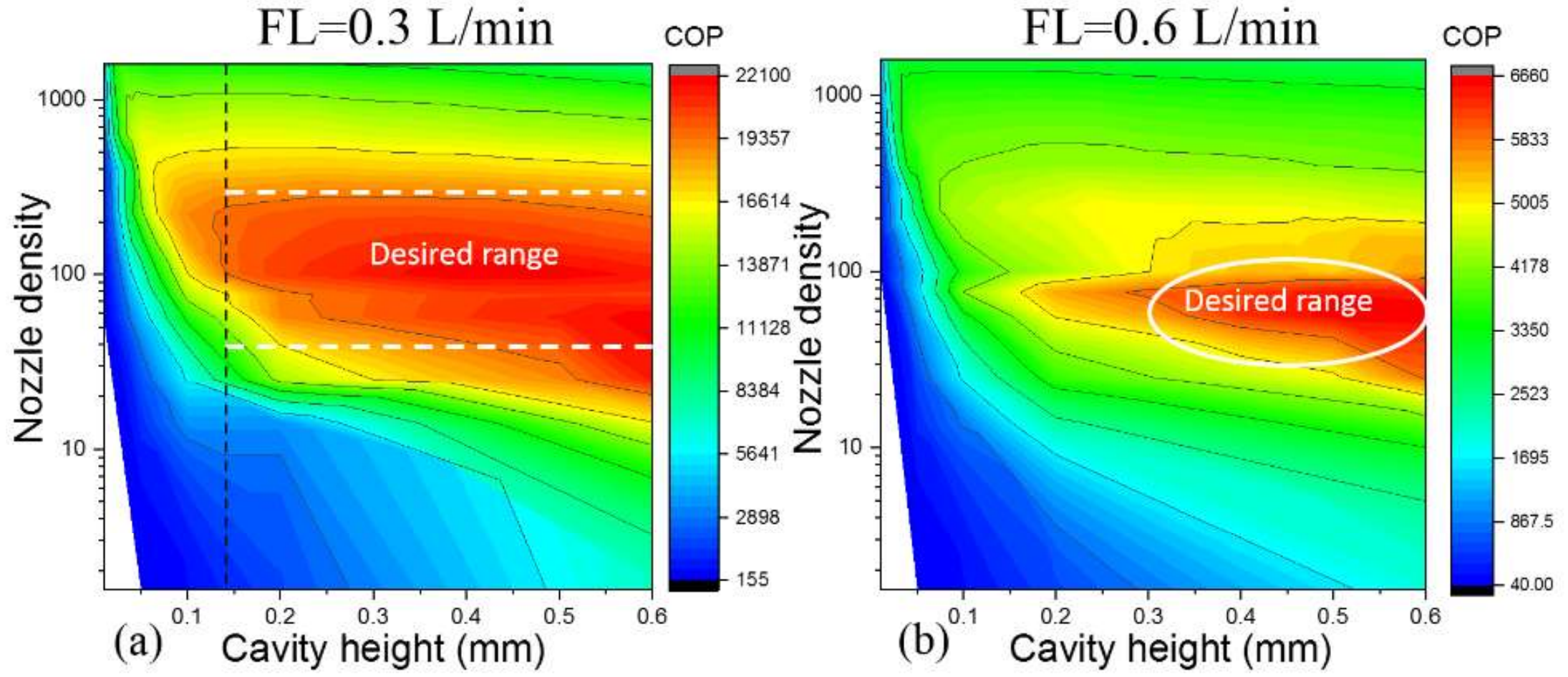

**Figure 3.14:** COP contour as function of the nozzle density and cavity height under different flow rate: FL=0.3 L/min and FL=0.6 L/min.

The profile of the COP surface can be explained based on the hydraulic analysis of the unit cell model, shown in Figure 3.6. For a constant flow rate, the pressure drop of the cooler $\Delta p_{tot}$ is very high at thinner cavity thickness H. The reason is that the dominated pressure drop is the pressure drop across the impingement cavity channel $\Delta p_{\text{channel}}$, which is inversely proportional to the cavity thickness H. On the other hand, for a fixed cavity height H, the pressure drop will be very high for low nozzle density, since the outlet drainage is far from the inlet. As the nozzle density is higher than 100 $cm^{-2}$, the pressure will also increase due to the scaling down of the nozzle diameter di, resulting in a higher pressure drop inside the nozzles $\Delta p_{\text{nozzle}}$.

3.2.3.2 Comparison 2: constant pump power

For a constant pumping power consideration, good thermal performance of the cooler is expressed by a low value of the normalized thermal resistance $R_{th}$. This is equivalent with a high value of the COP, as shown in equation 3.1. The nozzle diameter ratio $d_i/L$ is still kept as 0.3 in this study. The normalized thermal resistance contour is plotted as function of nozzle density and cavity height, for a constant pumping power of 0.1 $W/cm^2$ and 0.2 $W/cm^2$, shown in Figure 3.15. In general, for a constant cavity height, the thermal resistance decreases as the nozzle density increases. Moreover, it is observed that the lowest thermal resistance is found at the region with higher nozzle density and lower cavity height, which is located at the top left corner of the chart. To better understand the flow and thermal behaviors behind the phenomenon in Figure 3.15, the flow rate and pressure drop results are both extracted for the constant pump power of 0.2 $W/cm^2$, as illustrated in Figure 3.16.

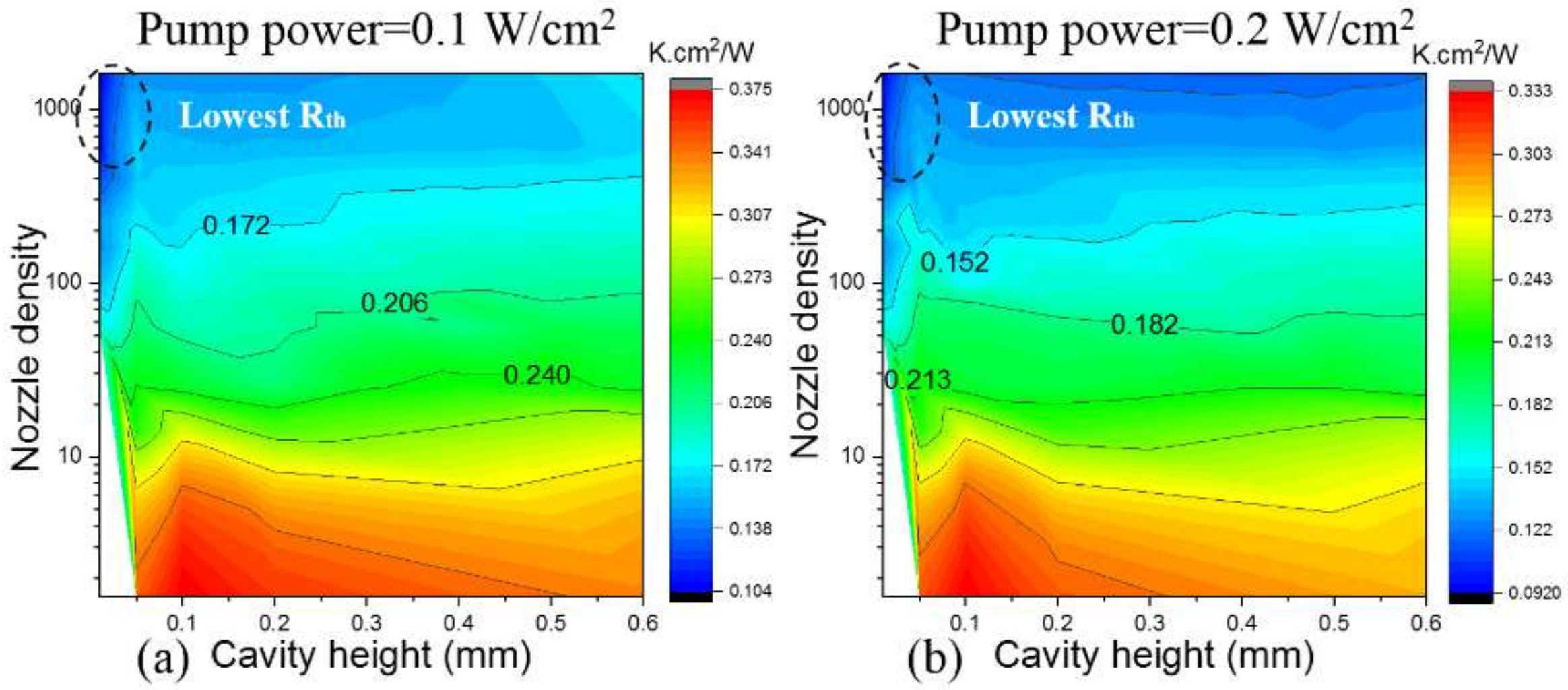

**Figure 3.15:** Characteristic contour of cooler under different required pump power for a constant $d_i/L$ =0.3: (a) normalized pump power of 0.1 $W/cm^2$; (b) normalized pump power of 0.2 $W/cm^2$.

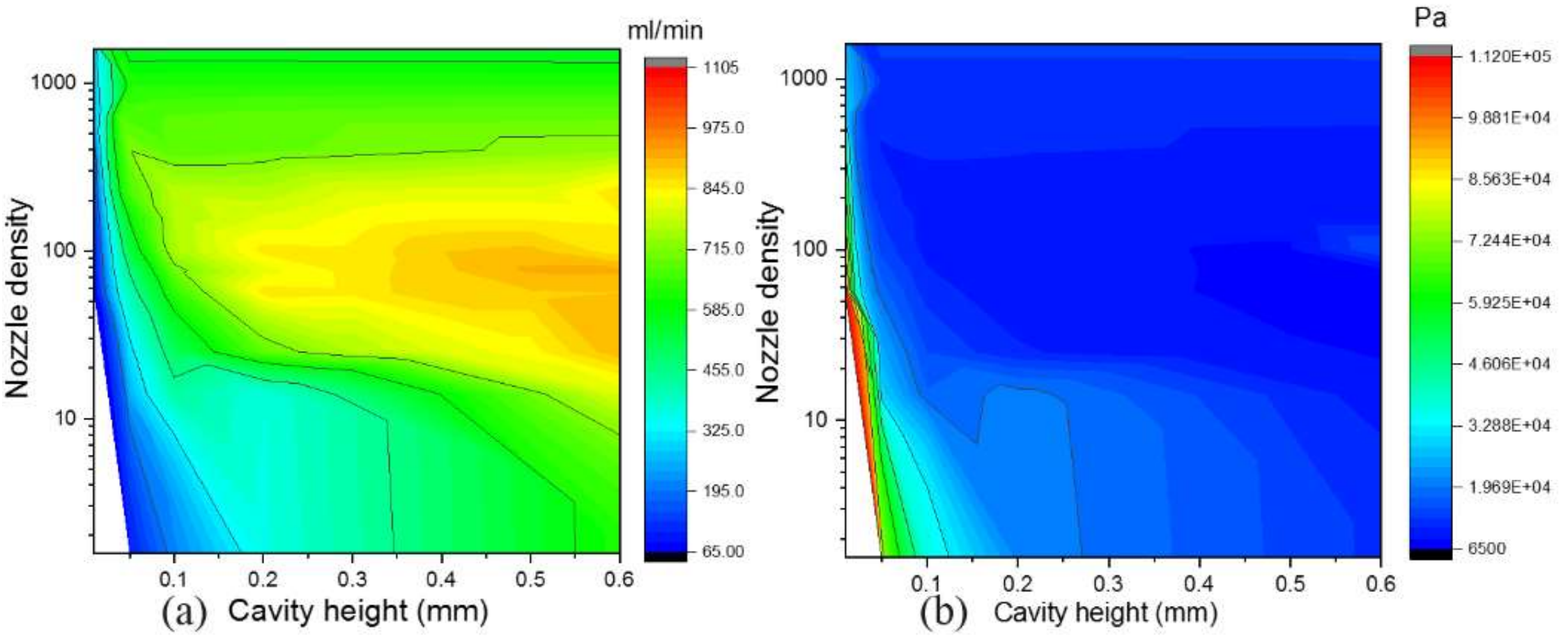

**Figure 3.16:** Characteristic contour of cooler under normalized pump power of 0.2 W/cm$^2$: (a) flow rate distribution; (b) pressure drop distribution.

Figure 3.17 shows the thermal resistance curves as function of the nozzle density and cavity height, for a constant pumping power of 0.2 W/cm$^2$. It shows that the lowest thermal resistance is in the thin cavity range ("pinch-off") around 10 µm, for different nozzle arrays. For higher cavity heights H, beyond 200 µm, defined as the impingement jet regime, the thermal resistance remains stable as a function of the cavity height, with small variations. However, the nozzle scaling trend is different for $d_i/L$ =0.1 and for $d_i/L$ =0.3, as illustrated in Figure 3.18. In general, the thermal resistance for the nozzle diameter ratio $d_i/L$ =0.3 is much lower than for $d_i/L$ =0.1, which means that a larger nozzle diameter ratio is better for the thermal cooling performance. Specifically, for constant cavity height at $d_i/L$ =0.1, it shows that the lowest thermal resistance is located in the middle range of the nozzle density around 100 cm$^{-2}$. For the larger inlet diameter ratio $d_i/L$ =0.3, the thermal performance can be further improved by increasing the nozzle density.

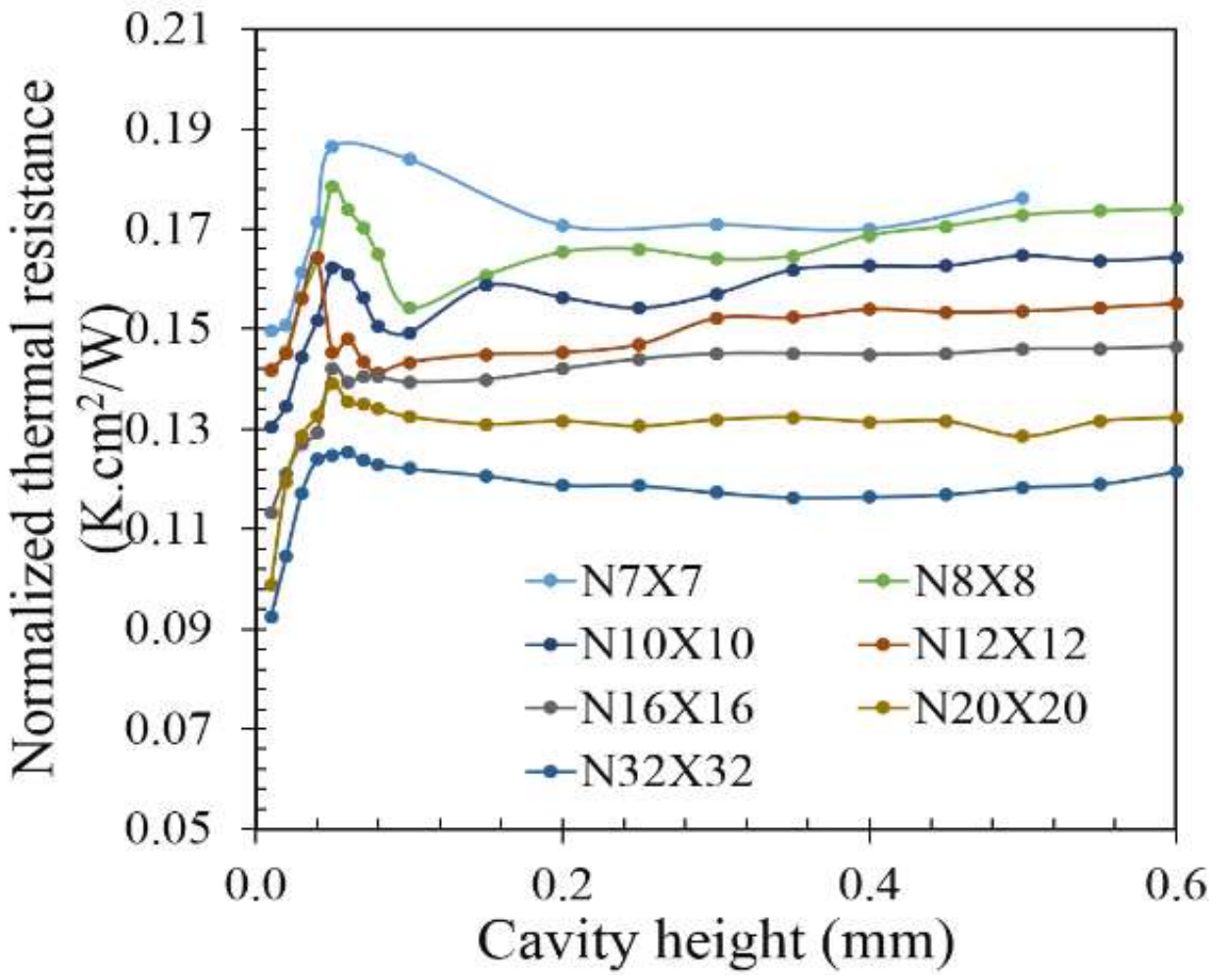

**Figure 3.17:** Thermal resistance curves as function of nozzle density with different inlet diameter ratio, for a constant pump power (Qpump=0.2 W/cm$^2$).

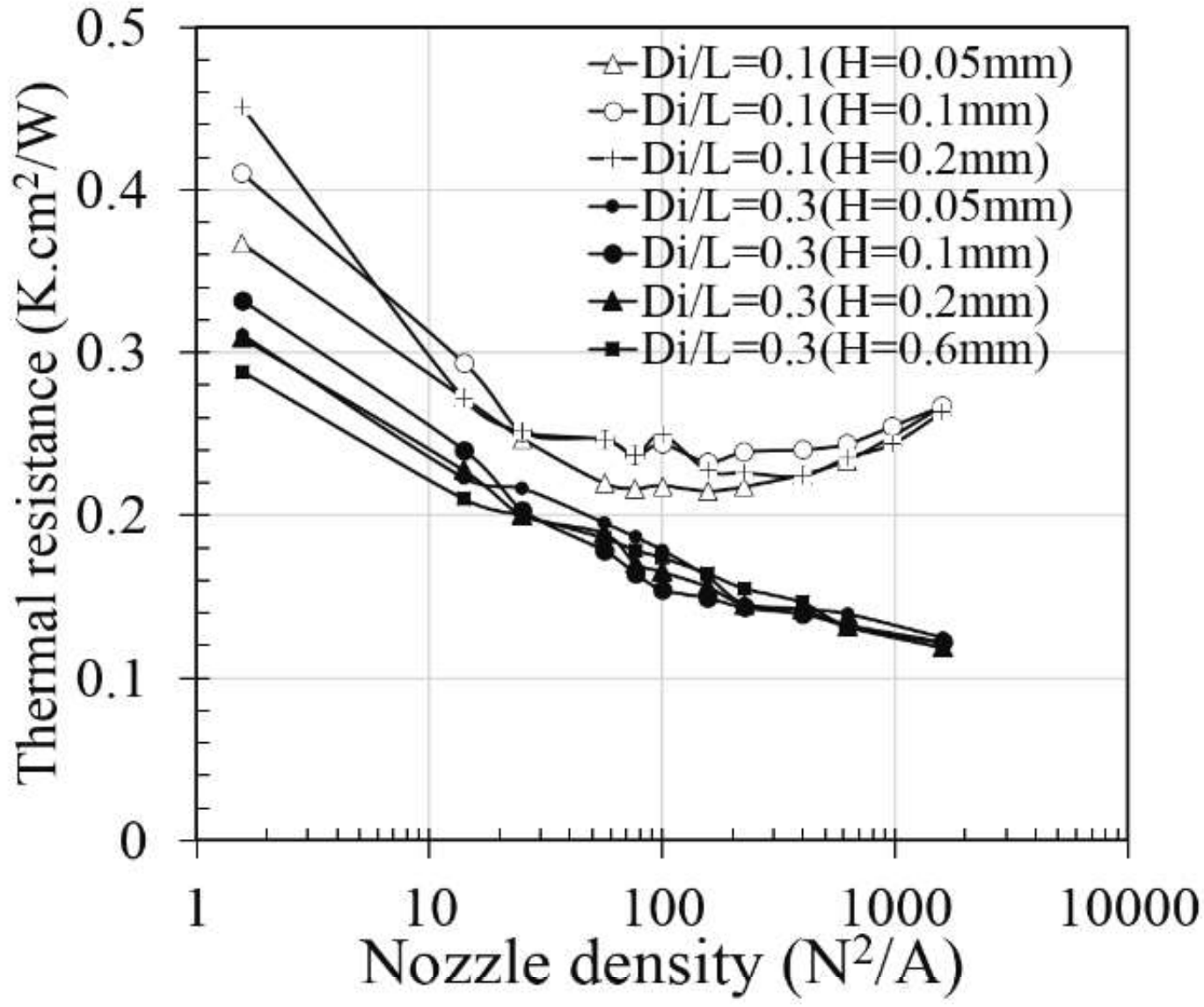

**Figure 3.18:** Thermal resistance curves as function of nozzle density with different inlet diameter ratio, for a constant pump power ($Q_{pump}$=0.2 W/cm$^2$).

In summary, it is found that a high as possible inlet diameter ratio $d_i/L=0.3$ is optimal for the cooler design, which is based on the design constraint. As for the impact of nozzle density for a constant pumping power, the optimal design for the nozzle density is around 1,000 cm$^{-2}$, with optimal cavity height range between 0.01and 0.05 mm, based on the cooler bonding techniques [31].

### 3.3 Dimensionless analysis

#### 3.3.1 Motivation and objective

The previous parametric analysis shows that there are a lot of parameters included in the cooler geometry. When the nozzle number N is scaling, the other parameters are changed too. Therefore, it is necessary to normalize the parameters to simplify the design. The dimensionless analysis is known as a very powerful tool to understand the physics in the area of heat transfer and fluid mechanics. It specifies that the normalized physical behavior of the impingement cooler is determined by the normalized proportions of the geometrical design parameters (the dimensionless parameters), and also the normalized flow conditions. This phenomenon can be exploited to generalize the obtained modeling results and to understand the fundamental behavior of the multi-jet impingement cooler.

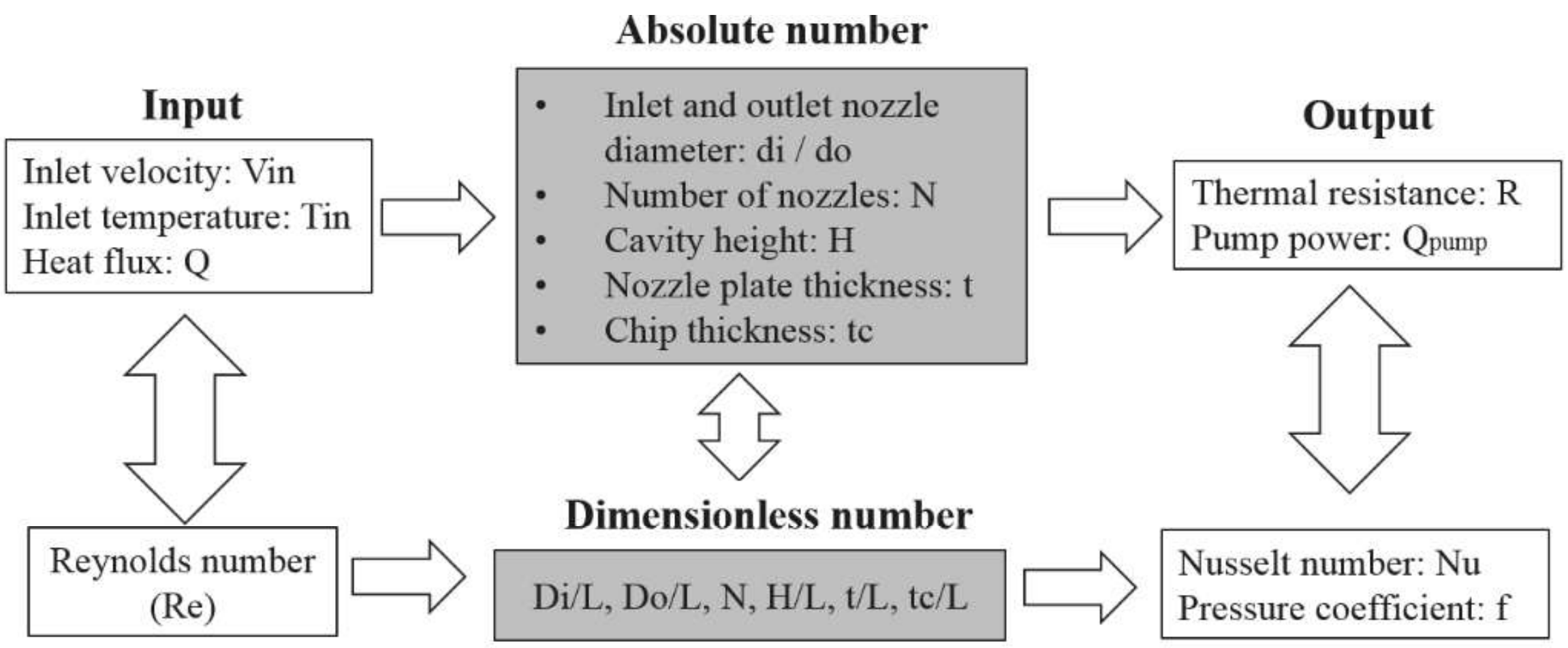

**Figure 3.19:** Relation between the absolute numbers and dimensionless numbers for the jet cooling analysis.

As discussed in chapter 1 in the impingement jet cooler with an N×N jet array, there are five design parameters needed to be considered for the cooler geometry design: $d_i$, $d_o$, H, t, $t_c$, L, where $d_i$ is the inlet diameter, $d_o$ is the outlet diameter, $t_c$ represents the chip thickness, H is the standoff between the jet exit and the heater, t is the nozzle thickness, and L is the unit cell length, which is defined as below:

$$L = \frac{A}{N \times N} \tag{3.6}$$

where A is the chip area. L also represents the spacing between the two neighboring inlet jets. Also, the flow and thermal parameters as the input conditions are listed as: $\overline{V}_{in}$ and $T_{in}$, where $T_{in}$ is the inlet temperature, and $\overline{V}_{in}$ is the inlet velocity. For the output parameters, the thermal resistance $R_{th}$ and pumping power $Q_{pump}$ are used for the cooler performance characterization. All the parameters are summarized in Figure 3.19.

Taking advantage of the Buckingham $\pi$ theorem, the abovementioned geometrical parameters and input/output parameters are transferred to the dimensionless form, shown in Figure 3.19. As a dimensionless number of the heat transfer using $d_i$ as the characteristic length scale, the Nusselt number in the unit cell is defined as below, with three definitions:

(1) Nusselt number based on average interface temperature of the chip:

$$\overline{Nu}_f = \frac{\overline{h}_f \, d_i}{k_{fl}} = \frac{\dot{Q}}{A \cdot \Delta T} \cdot \frac{d_i}{k_{fl}} = \frac{q \cdot d_i}{(\overline{T}_s - T_{in}) \cdot k_{fl}} \tag{3.7}$$

(2) Nusselt number based on stagnation temperature on the chip surface:

$$\overline{Nu}_o = \frac{\overline{h}_0 \, d_i}{k_{fl}} = \frac{\dot{Q}}{A \cdot \Delta T} \cdot \frac{d_i}{k_{fl}} = \frac{q \cdot d_i}{(T_0 - T_{in}) \cdot k_{fl}} \tag{3.8}$$

(3) Nusselt number based on average junction temperature:

$$\overline{\mathrm{Nu}}_{\mathrm{j}} = \frac{\dot{\mathrm{Q}}}{\mathrm{A}\cdot\Delta\mathrm{T}}\cdot\frac{d_i}{\mathrm{k_{fl}}} = \frac{\mathrm{q}\cdot d_i}{(\bar{T}_{chip}-\mathrm{T_{in}})\cdot\mathrm{k_{fl}}} \quad (3.9)$$

The Reynolds number and the Prandtl number are defined as following:

$$\text{Reynold number: } \mathrm{Re_d} = \frac{\rho d_i \bar{\mathrm{V}}_{\mathrm{in}}}{\mu}, \text{ Prandtl number: } \mathrm{Pr} = \frac{\mu\mathrm{Cp}}{\mathrm{k_{fl}}} \quad (3.10)$$

where $\mathrm{k_{fl}}$ is the thermal conductivity of the fluid, μ is the dynamic viscosity, and Cp is the specific heat. In addition, the $\bar{T}_s$ shown in $\overline{\mathrm{Nu}}_{\mathrm{f}}$ is the fluid and solid interface temperature, while $\bar{T}_{chip}$ is based on the junction temperature. Since the focus of this study is the geometrical aspect, the fluid properties are kept constant in this study. Therefore, the Prandtl number used in this thesis is fixed as 7.56, a representative value for DI water.

In order to generalize the parametric trend, we need to extract the relation between the geometrical flow parameters and normalized heat transfer in the following form:

$$\overline{\mathrm{Nu}}_{\mathrm{f}} = \mathrm{f}(\mathrm{Re_d}, \frac{d_i}{\mathrm{L}}, \frac{d_o}{\mathrm{L}}, \frac{\mathrm{H}}{\mathrm{L}}, \frac{\mathrm{t}}{\mathrm{L}})$$

$$\overline{\mathrm{Nu}}_{\mathrm{J}} = \mathrm{f}(\mathrm{Re_d}, \frac{d_i}{\mathrm{L}}, \frac{d_o}{\mathrm{L}}, \frac{\mathrm{H}}{\mathrm{L}}, \frac{\mathrm{t}}{\mathrm{L}}, \frac{\mathrm{t_c}}{\mathrm{L}}) \quad (3.11)$$

where $\overline{\mathrm{Nu}}_{\mathrm{f}}$ is the area averaged Nusselt number as function of the jet diameter $\frac{\mathrm{di}}{\mathrm{L}}$ and the other dimensionless variables. And also, the $\frac{\mathrm{t_c}}{\mathrm{L}}$ is not included in $\overline{\mathrm{Nu}}_{\mathrm{f}}$ function.

The dimensionless number for the friction factor f can be expressed as following:

$$\mathrm{f} = \mathrm{f}(\mathrm{Re_d}, \frac{\mathrm{d_i}}{\mathrm{L}}, \frac{\mathrm{d_o}}{\mathrm{L}}, \frac{\mathrm{H}}{\mathrm{L}}, \frac{\mathrm{t}}{\mathrm{L}}) \quad (3.12)$$

$$\mathrm{f} = \frac{\Delta\mathrm{P}}{(\frac{1}{2}\rho\cdot\bar{\mathrm{V}}_{\mathrm{in}}^{2})(\frac{\mathrm{t}}{\mathrm{d_i}})}; \quad (3.13)$$

$$\mathrm{k} = \frac{\Delta\mathrm{P}}{(\frac{1}{2}\rho\cdot\bar{\mathrm{V}}_{\mathrm{in}}^{2})}; \quad (3.14)$$

where f is the friction loss coefficient, k is the pressure coefficient, and t is the thickness of the nozzle plate. ΔP is defined as the pressure drop between the inlet and outlet nozzle at the unit cell level.

In Figure 3.20, the unit cell modeling results for the $R_{th}$-$\dot{V}$ curve in terms of absolute parameters and the $\overline{\mathrm{Nu}}_{\mathrm{f}}$-$\mathrm{Re_d}$ curve in terms of dimensionless parameters are compared. Figure 3.20(a) shows the different $R_{th}$-$\dot{V}$ curves for different nozzle numbers ranging from N=1 to N=64. However, the $\overline{\mathrm{Nu}}_{\mathrm{f}}$-$\mathrm{Re_d}$ curves for the same data based on the dimensionless numbers all collapse, as shown in Figure 3.20(b). This means that the intrinsic heat transfer and flow dynamics physics are the same in the same dimensionless parameter values.

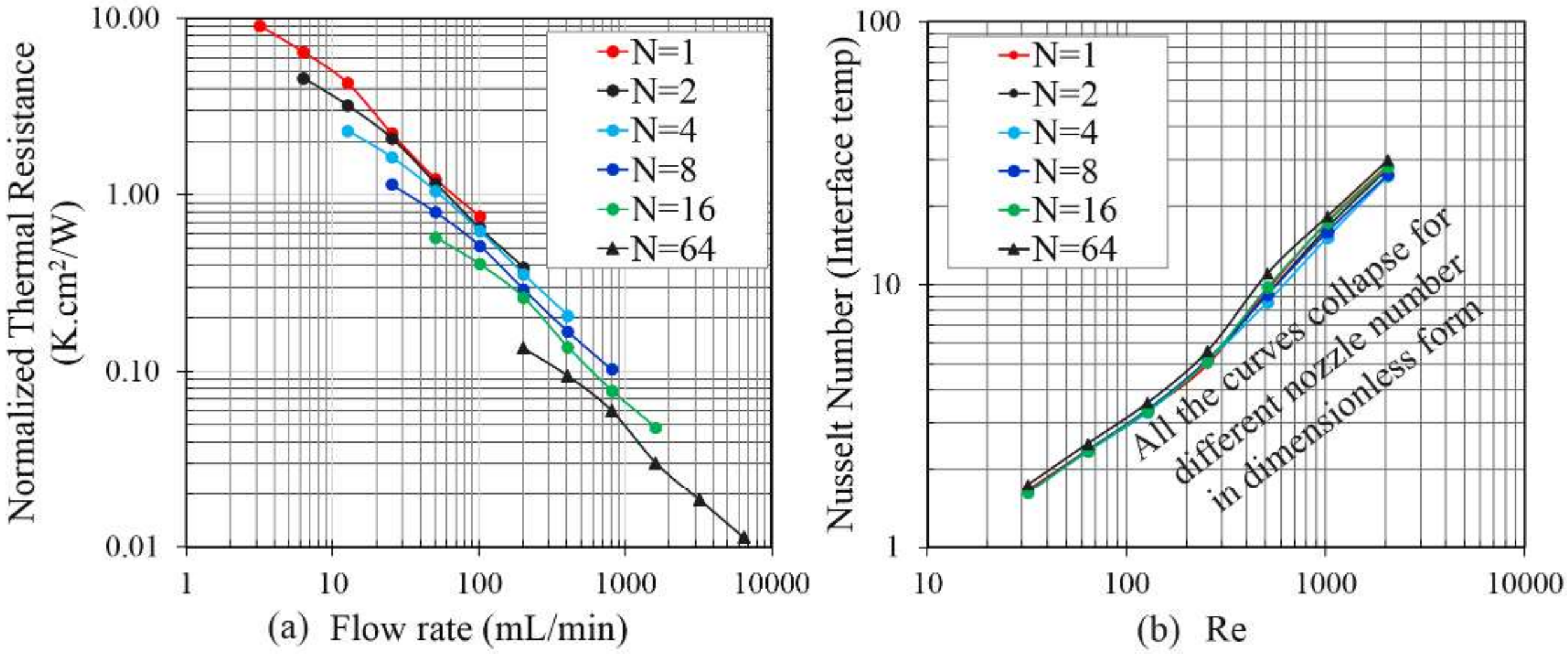

**Figure 3.20:** Unit cell modeling results based on $d_i$/L =0.1: (a) $R_{th}$-$\dot{V}$ curve for design variables with absolute number; (b) $\overline{\mathrm{Nu}}_\mathrm{f}$-$\mathrm{Re}_\mathrm{d}$ curve for dimensionless number ($\frac{di}{L} = \frac{d_o}{L} = \frac{H}{L} = \frac{t}{L} = 0.1$).

As discussed in this section, the $\overline{\mathrm{Nu}}_\mathrm{f}$-$\mathrm{Re}_\mathrm{d}$ curves are the same for different nozzle number, by using dimensionless number. Therefore, a single nozzle number N investigation can be used for the extraction of the correlations for $\overline{\mathrm{Nu}}_\mathrm{f}$-$\mathrm{Re}_\mathrm{d}$ and f-$\mathrm{Re}_\mathrm{d}$. The fitted predictive models can be used to extract the thermal and hydraulic performance for arbitrary nozzle numbers. In the next part, a literature study is performed for dimensionless heat transfer and pressure correlation for jet impingement.

### 3.3.2 Literature overview

As introduced in chapter 1, jet impingement cooling on the chip backside is very promising due to the high heat transfer rates and the absence of thermal interface material. With the literature study shown in section 2.1, the most commonly considered impingement jet cooling is based on the common outlets configuration, shown in Figure 3.21. In this configuration, the jet flow is injected through nozzle arrays and extracted through the outlets on the edges of the heat sink. However, the disadvantage of the jet cooling with a common return is that the heat transfer can be highly influenced by the "cross-flow effects" where the return flows interact with the jets flow [36,37,38]. The cooling performance of jet cooling can be significantly affected by a large number of jets, especially for large die area applications. Kercher and Tabakoff [39] and Florschuetz [40] experimentally examined the crossflow effects in reducing the heat transfer coefficient. JF Maddox [37] sought methods to manage the spent flow, such as angled confining wall and anti-crossflows (ACF) cooling structure or corrugated jet plane. Hollworth and Dagan [62] found that the convective coefficients can be improved with 20-30% by arranging the outlet nozzles through the impingement surface. However, this is not applicable for cooling on the electronic devices. Impingement jet cooling with alternating feeding and draining jets shown in Figure 3.21(b) is very promising for electronic cooling [6, 33], where the spent fluid can be extracted through the outlet nozzles very efficiently [6, 36, 37,38].

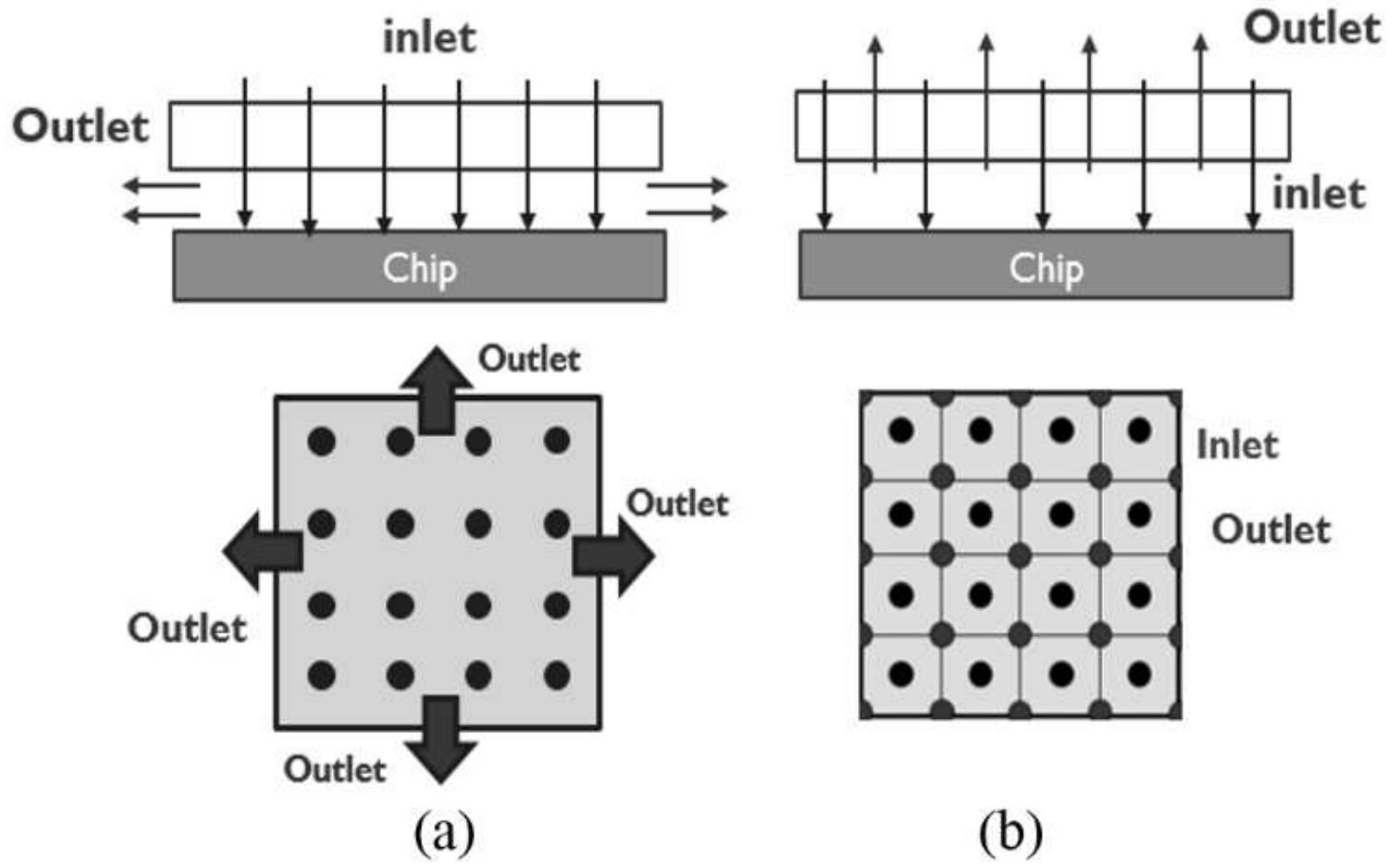

**Figure 3.21:** Impingement jet cooling configurations: (a) configuration A: common outlets and (b) configuration B: distributed outlets configurations.

3.3.2.1 Configuration A: common outlets

Empirical correlations for heat transfer and pressure coefficient are very important to understand the functional relations regarding different geometry parameters. Extensive literature studies about single impingement jet cooling correlations covering different nozzle geometries for both submerged and free-surface jet configurations are published in the last decades [41, 42 43, 60]. Garimella and Rice [21] developed $\overline{\mathbf{Nu_d}}$ correlation for a single confined circular submerged jet. Womac et al. [20] developed correlations for a single circular free-surface jet. The correlations with a single round nozzle, orifice, or pipe are developed by Martin [48].

Compared to single jet impingement, arrays of multiple jets can achieve a higher heat transfer rate and more uniform temperature distribution [61]. Weigand et al. [44] summarized and compared the existing empirical correlations of multiple impinging air jets for average and locally resolved heat transfer coefficients, respectively. Narumanchi [45] reported that there is a good match between CFD results and experimental data from Womac et al. [20] over a wide range of Reynolds numbers for confined and unconfined submerged jets. Whelan [55] reported that the confined submerged nozzles with contoured inlet or inlet/outlet are the suggested nozzle configurations. Florschuetz [40] developed the correlations for the inline and staggered nozzle patterns, and concluded that the staggered patterns resulted in smaller heat transfer coefficients than their inline counterparts. Besides, Royne and Dey [54] also investigated the effect of nozzle geometry on the heat transfer and pressure drop to confined-submerged jet arrays over a Reynolds number range of $1000 \leq Re_D \leq 7700$. It is reported that the sharp-edged and contoured nozzles can enhance the cooling performance in comparison to the conventional straight nozzle arrays for a given pumping power.

**Table 3.1:** State of the art Nusselt-Reynolds correlation for common outlets

| Source | Description | Provides | Methodology | Conditions | Reynolds exponent | Reynolds number, cavity height range |
|---|---|---|---|---|---|---|
| Martin, 1977 | Submerged, multiple, circular | $\overline{Nu}_d$ | Analytical | Tref=Tin | 0.67 | 2000<Re<100,000; 2≤H/D≤12 |
| Martin, 1977 | Single, circular, submerged | $\overline{Nu}_d$ | Analytical | Tref=Tin | 0.775 | 2000<Re<40,000; 2≤H/D≤12 |
| Womac, 1994 | Multiple jets, submerged, circular | $\overline{Nu}_{d+L}$ | Analytical | Tref=Tin | 0.5/0.8 | 5000<Re<200,000; 2≤ S/D≤4 |
| Womac, 1994 | Single, submerged, circular | $\overline{Nu}_{d+L}$ | Analytical | Tref=Tin | 0.5/0.8 | Re< 5000; 2≤ S/D≤4 |
| Elison, B. and Webb 1994 | Submerged, single jet, circular | $Nu_0$ | Experiment | Tref=Tin | 0.8 | 300 < Re < 7000; H/D<8 |
| Garimella and Rice 1995 | single, confined/submerged, circular | $Nu_0$ | Experiment | Tref=Tin | 0.585 | 4000<Re<23000; 1≤ S/D≤5 |
| Lee and Vafai, 1999 | Submerged, Multiple, circular | $\overline{Nu}_d$ | Analytical | Tref=Tin | 0.667 | 2000<Re<100000; 2≤H/D≤12 |
| Li and Garimella, 2001 | Single jet, circular, confined/submerged | $\overline{Nu}_{1.9d+d}/Nu_0$ | Experiment | Tref=Tin | 0.5555/ 0.515 | 4000 <Re< 23000; 1≤H/D≤5 |
| Robinson, Schnitzlaer 2007 | Submerged, jet array, circular | $Q_{pump}/\overline{Nu}_{d/2}$ | Experiment | Tref=Tin | 0.46 | 100 < Re < 10000; 2≤H/D≤ 20; 3 ≤ P/D ≤ 7 |
| Meola, 2009 | Air/water, confined, circular, jet arrays | $\overline{Nu}_d$ | Analytical | Tref=Tin | 0.68 | 200<Re<100,000; 1.6<*H/D*<20 |
| Peng Tie, 2011 | Submerged, jet arrays | $\overline{Nu}_d$ | Experiment | Tref=(Tin+ Tout)/2 | 0.51 | 1398.113≤Re ≤13440.4; 4.963<=Pr<= 9.311 |
| Yonehara 1982 | Free surface, liquid, multiple jets | $\overline{Nu}_d$ | Analytical | uniform temp | 0.67 | Re<48000 P/D>13.8 |
| Jiji and Dagan, 1988 | Free surface, liquid, single phase, multiple jets | $\overline{Nu}_L$ | Experiment | uniform heat flux | 0.5 | 0.5 mm and 1.0 mm 3 mm < z < 10 mm L=heater length |
| Fabbri and Dhir 2005 | Free surface, liquid, single phase, multiple jets | $\overline{Nu}_d/\Delta p$ | Experimental | water and FC40 | 0.78 | 73 < Re < 3813 65μm < dn < 250μm |
| D.T. Vader 1991 | Liquid, planar, confined, | $Nu_0$ | Numerical | water | 0.5 | 20000<Re<90000, 2.7<Pr<4.5 |
| X. Liu 1991 | Liquid jet, free surface, single phase | $Nu_0$ | Analytical and experimental | water | 0.5 | 0≤r/d<0.787 0.15≤Pr≤3 |
| Gregory J. Michna 2011 | Submerged/confined/microjet arrays/ single phase | $\overline{Nu}_d$ | Experiment | Tref=Tin | 0.55 | 50 < Red < 3500; D=54 and 112 μm |
| Tomasze Muszynski, 2016 | Confined, multiple | $\overline{Nu}_d$ | Experiment | LMTD | 0.65 | 500<Re<2500 |

Kaveh Azar et al. [46] and Molana [47] both present various average heat transfer coefficient for single-phase liquid correlations. Martin [48] developed correlations for multiple circular submerged jets. Lee and Vafai [49] proposed a criterion value $S_{NN}/d$ for negligible cross flow, and made a correction of Martin's correlation. The Womac et al. [50] correlation divided the entire heat transfer area into two separate regions: the ''impingement zone'' and the "wall-jet region" outside of the impingement zone. Experiments conducted for the confined-submerged liquid jet arrays found that the heat transfer coefficient was somewhat insensitive to jet to-target spacing within the range of $2 \leq H/D \leq 4$ due to the target surface being within the potential core of the issuing jets. Robinson and Schnitzler [51] conducted experiments investigating the impingement of water jet arrays under both free-surface and submerged conditions. For the submerged jets, it was found that heat transfer was insensitive to jet-to-target spacing changes in the range of $2 \leq H/D \leq 3$. A monotonic decrease in heat transfer was observed with increasing jet-to-target spacing in the range of $5 \leq H/D \leq 20$. It was also found that a stronger dependence on jet-to-jet spacing was encountered for smaller jet-to-target spacing. The effect of jet-to-jet spacing for jet arrays was more closely examined by Pan and Webb [52]. For the central jet module, the stagnation point heat transfer coefficient was found to be independent of jet-to-jet spacing. Conversely, a dependence on the jet-to-target spacing was discovered. The more recent work of Fabbri and Dhir [53] involved both heat transfer to the jet arrays and the associated pressure drop across the jet nozzle plate.

The Reynolds correlations for stagnation Nusselt number and average Nusselt number are summarized in Table 3.1, together with the methodology and the range for the Re and H/D. In general, there is an abundance of Nu-Re correlations for impinging jets cooling in the literature, and they generally show $Nu \sim a \cdot Re_d^b$, where the exponent b is typically in the range of 0.5-0.8. However, most of the correlations derived from the analytical predictions were based on the simplified assumption that each impinging jet formed an individual cell or module. The local and average heat transfer rates were determined for repeating modules surrounding each jet in the array. These correlations are valid when the jet-to-target distance and jet-to-jet spacing were larger, and the jet-to-jet interactions are negligible. Since the jets were well-drained, there was negligible crossflow between neighboring jets, and each jet established a cell that behaved thermally as a single isolated impinging jet.

3.3.2.2 Configuration B: distributed outlets

The correlation development with local extraction of the spent fluid to a plenum is very limited in the published research. The concept of a jet impingement array cooling with local effusion nozzles was first proposed by Huber and Viskanta [56]. They developed $\overline{Nu}_f$ correlations based on the experimental data for a confined $3 \times 3$ array with a center jet and spent air exit ports.

$$\overline{\mathrm{Nu}}_{\mathrm{f}} = 0.285{Re_D}^{0.710}Pr^{0.33}\left(\frac{H}{D}\right)^{-0.123}\left(\frac{X_n}{D}\right)^{-0.725} \quad (3.15)$$

The validated ranges of the parameters for the correlation are: 3400 < $Re_D$ < 20500, 4< $X_n/D$ < 8 and 0.25 < H/D < 6.0. The obtained experimental data can be applied to Martin's correlation [48] since both of them are based on the spent air exits and without considering the crossflow effect.

Rhee et al. [64] employed a naphthalene sublimation method to determine local heat/mass transfer coefficients on the target plate. They found that the heat/mass transfer for the smaller nozzle to target distance is improved significantly and the augmented values are 60% and 20% higher for H/D= 0.5 and 1.0, respectively than those without the effusion holes. However, the performance with the cooling performance with the effusion holes is similar to those without the effusion holes for large gap distances.

Onstad et al. [38, 57, 59, 65] showed that a geometry which incorporates local extraction with a large exhaust area ratio, $A_e/A_{jet}$, is preferred to maintain a high average heat transfer coefficient. Three different impingement arrays were studied, all of which had a jet-to-jet spacing of $Z_n/D$ = 2.34, jet-to-target spacing of H/D = 1.18, and extraction holes in the jet plane. The correlations are listed in Table 3.2.

$$\overline{\mathrm{Nu}}_{\mathrm{f}} = C_0 Pr^{1/3}{Re_D}^{b} \quad (3.16)$$

**Table 3.2:** Correlations for Onstad's empirical model.

| **Array** | **D(mm)** | **d(mm)** | **$A_e/A_{jet}$** | **$C_0$** | ***b*** |
|---|---|---|---|---|---|
| (1) | 8.46 | 7.29 | 2.23 | 0.376 | 0.586 |
| (2) | 8.46 | 5.08 | 1.08 | 0.436 | 0.579 |
| (3) | 2.82 | 1.69 | 1.08 | 0.602 | 0.531 |

Brunschwiler et al. [6] demonstrated and experimentally characterized the microscale liquid jet impingement array cooling will locally distributed outlets, where the number of inlet nozzles is up to 47,000. A simple heat transfer correlation was developed based on the experimental data with a ±9% confidence level.

$$\overline{\mathrm{Nu}}_{\mathrm{j}} = 0.78Re^{0.73} \quad (3.17)$$

The experimental data were measured at H/ D = 1.2, which is in the stable impingement regime. And also, the Reynolds number Re is below 800, which means the considered flow is laminar.

Hoberg et al. [58] evaluated a new nozzle array configuration with six small extraction ports centered around each injection nozzle. A Nu-Re correlation was proposed for laminar-to-turbulent flow, shown as below:

$$\overline{\mathrm{Nu}}_{\mathrm{f}} = 0.36Re^{0.59} \tag{3.18}$$

where the Reynolds number is in the range of 500–10,000. However, this correlation was only extracted at H/D=1.

Rattner et al. [63] developed new correlations for Nusselt number and pressure-drop k-factors based on 1000 randomized cases, shown as below:

$$\overline{\mathrm{Nu}}_{\mathrm{f}} = Pr^{0.29} \cdot 10^{\left\{\sum_{j=1}^{20} a_j Re_j^{\,b_j} \left(\frac{p}{D_j}\right)^{c_j} \left(\frac{th}{D_j}\right)^{d_j}\right\}} \tag{3.19a}$$

$$k = 10^{\left\{\sum_{j=1}^{20} a_j Re_j^{\,b_j} \left(\frac{p}{D_j}\right)^{c_j} \left(\frac{th}{D_j}\right)^{d_j}\right\}} \tag{3.19b}$$

The pressure-drop $k$ -factor is calculated based on the inlet and outlet boundary pressure difference, correcting for frictional losses in the injection and return channels. The new correlations for pressure drop ($k$ -factor) and heat transfer performance (Nusselt number, $\overline{\mathrm{Nu}}_{\mathrm{f}}$) are valid over a wide range of Reynolds number ($Re_j$= 20–500), fluid transport properties (Pr = 1–100), and component geometries ($p/D_j$=1.8–7.1 and $th/D_j$=0.1–4.0).

The objective of this study is to develop the predictive models for the Nusselt number and friction factor, as a function of Reynolds number and geometry parameters, for multiple impingement jet cooling wit locally distributed outlets. In section 6, a test vehicle with complex fluid routing is designed and fabricated to validate the predictive model.

### 3.3.3 Results and discussions

For the dimensionless analysis, the impact variation of the different dimensionless parameters is investigated in this section. Figure 3.20(b) illustrates that the $\overline{\mathrm{Nu}}_{\mathrm{f}}$-$\mathrm{Re}_{\mathrm{d}}$ curves collapse for different nozzle numbers if all the dimensionless geometric parameters are kept the same case with 0.1. In this section, an extensive design of experiments will be conducted by varying the dimensionless numbers. For the investigation of the individual dimensionless parameters ($d_i$/L, $d_O$/L, H/L, t/L, $t_c$/L), the nozzle number N=4 is chosen for the following investigations and correlation fittings. Moreover, the combined effects are also studied in this section with the chosen N=4. The study range for the dimensionless analysis parameter is listed in Table 3.3.

**Table 3.3:** List of dimensionless variables and range

| Parameter | Symbol | Range |
|---|---|---|
| $Re_d$ | $Re_d$ | 32, 64, 128, 216, 512, 1024, 2048 |
| $d_i/L$ | α | 0.01, 0.1, 0.2, 0.3, 0.4 |
| $d_o/L$ | β | 0.05-0.5 |
| t/L | γ | 0.1-1.2 |
| H/L | φ | 0.05-2 |

3.3.3.1 Impact of the chip thickness

In our test case, the thermal test chip is flip-chip bonded on the substrate, while the active heater region is at the bottom of the chip. Therefore, the junction temperature $T_j$ is higher than the interface temperature $T_s$ due to the heat conduction through silicon, resulting in an additional thermal resistance. In general applications, junction temperature $T_j$ can be measured. For the evaluation of the cooling performance, $T_s$ is needed. This section will discuss what is the relation between $\overline{Nu}$ and $\overline{Nu_f}$, and also how does it impact with the N scaling.

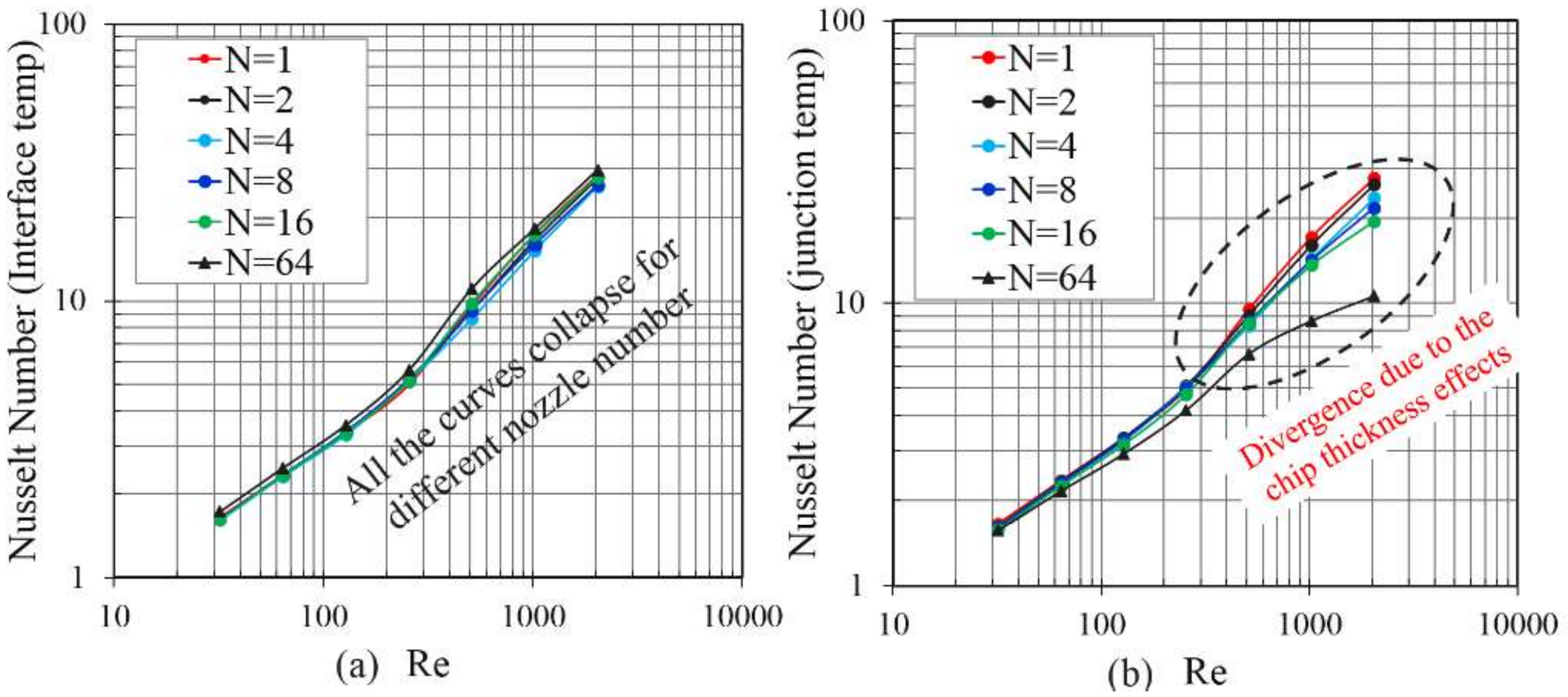

**Figure 3.22:** Impact of the chip thickness impact on $\overline{Nu_f}$-$Re_d$ relation curve: (a) Nusselt number based on solid-fluid interface temperature; (b) Nusselt number based on junction temperature.

As defined in section 3.3.3.1, $\overline{Nu_j}$ is based on the junction temperature while $\overline{Nu_f}$ is based on the fluid-solid interface temperature. This means that the chip thickness effect is not included in the $\overline{Nu_f}$. Figure 3.22 shows the $\overline{Nu_f}$-$Re_d$ curve and $\overline{Nu_j}$-$Re_d$ curve for different nozzle numbers. It can be seen that the chip size effects become larger for a higher Reynold number for the $\overline{Nu_f}$-$Re_d$ curves for a fixed chip thickness, shown in Figure 3.22(b). For the $\overline{Nu_j}$-$Re_d$ curve shown in Figure 3.22(a), the chip thickness effect is decoupled. The main reason for this divergence is the dominated factors between heat convection above the chip surface and heat conduction through the chip thickness. The contribution of the conduction increases as N and $\mathrm{Re_d}$ increase for a fixed chip thickness. There is a large difference between the $\overline{\mathrm{Nu_j}}$ and $\overline{\mathrm{Nu_f}}$ for smaller unit cell and higher cooling rates.

Through the thermal resistance network, the total thermal resistance between junction and coolant contains three parts: heat convection of the jet cooling, heat conduction and heat spreading effects through the silicon substrate. Thus, we can get the following equation:

$$R_{th} = \frac{1}{\bar{h}_f A} + \frac{t_c}{k_s A} + R_{spreading} \tag{3.20}$$

$$R_{th} = \frac{1}{\bar{h}_j A}$$

where $R_{th}$ is the thermal resistance based on the average junction temperature, defined in section 2.3 in chapter 2. A is the chip area. The equivalent heat transfer coefficient based on the average cooling interface temperature is defined as $\bar{h}_f$. And also, $\bar{h}_j$ is defined as the heat transfer coefficient based on the junction temperature. $R_{spreading}$ represents the heat spreading resistance from the cooling interface to the junction surface.

Therefore, the formula can be rewritten as:

$$\frac{A}{h_j A} = \frac{A}{\bar{h}_f A}\left(1 + \frac{\bar{h}_f A t_c}{k_s A} + f(t,\ \bar{h}_f, A, k_s)\right) \tag{3.21}$$

$$\frac{1}{h_j} = \frac{1}{\bar{h}_f}\left(1 + \frac{\bar{h}_f}{k_s} + f\left(t_c, \bar{h}_f,\ A,\ k_s\right)\right) \tag{3.22}$$

Since the Biot number ($Bi$) is defined as:

$$Bi = \frac{\bar{h}_f}{k_s} * t_c = \overline{\mathrm{Nu}}_\mathrm{f} * \frac{t_c}{di} * \frac{k_f}{k_s} \tag{3.23}$$

Where $k_s$ is the thermal conductivity of the silicon, and $k_f$ is the thermal conductivity of the fluid.

Substitute the $Bi$ into the above equation, we can get:

$$\frac{1}{h_j} = \frac{1}{\bar{h}_f}\left(1 + Bi + f\left(t_c, \bar{h}_f,\ A,\ k\right)\right) \tag{3.24}$$

Next, we use $g(Bi)$ to represent the terms in the parentheses:

$$g(Bi) = 1 + Bi + f(t_c, \bar{h}_f, A, k) \tag{3.25}$$

Thus, the $\overline{\mathrm{Nu}}_\mathrm{j}$-$\mathrm{Re_d}$ dimensionless formula can be improved as:

$$Nu_j = \frac{\overline{\mathrm{Nu}}_\mathrm{f}}{g(Bi)} = f(Re, \frac{d_i}{L}, \frac{d_o}{L}, \frac{H}{L}, \frac{t}{L}) \frac{1}{g(Bi)} \tag{3.26}$$

In order to get the relation between $Nu_j$ and $f(Bi)$, different values of chip thickness are studied. Figure 3.23 shows the impact of chip thickness on the heat source junction temperature. With the increase of the chip thickness $t_c/L$, the $\overline{\mathrm{Nu}}_\mathrm{j}$ decreases slightly. It can be seen that the heat conduction and spreading through the chip becomes more and more dominate than heat convection cooling. The final fitting curve for $g(Bi)$ is listed as below:

$$g(Bi) = 1 + Bi + (0.1Bi + 1.1Bi^2) \tag{3.27}$$

The relation $\overline{\mathrm{Nu}}_\mathrm{j}$ between $\overline{\mathrm{Nu}}_\mathrm{f}$ the can be expressed as below:

$$Nu_j = \frac{\overline{Nu}_f}{g(Bi)} = \frac{\overline{Nu}_f}{1+Bi+(0.1Bi+1.1Bi^2)} \tag{3.28}$$

Most of the effect can be explained by conduction in Si. Additional conduction for extra heat spreading is included in f($Bi$). According to the relation $Bi \sim \frac{t_c}{d_i}$, the $Bi$ is more pronounced for very thin Si and for small nozzle diameter, shown in Figure 3.23.

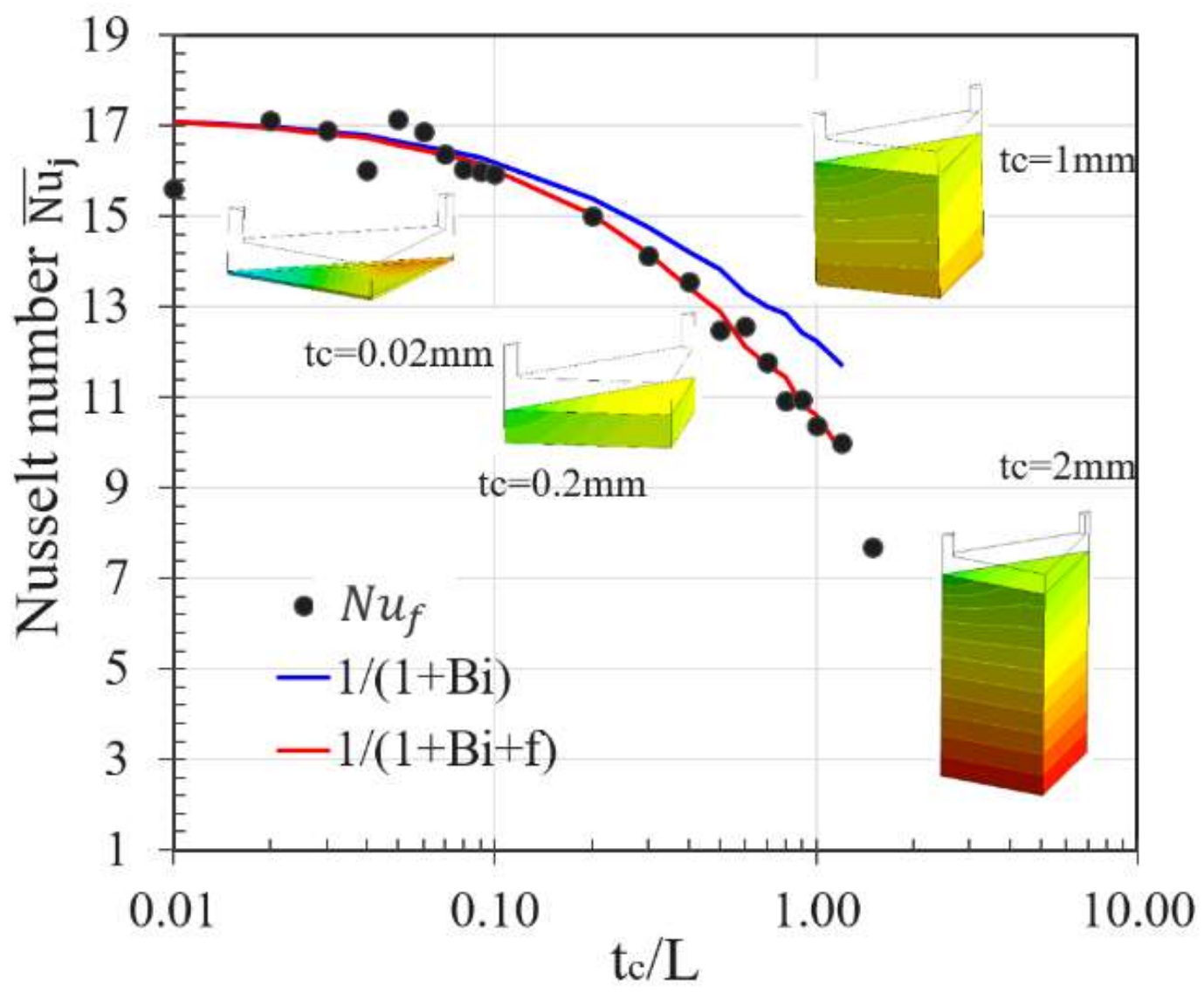

**Figure 3.23:** Effect of the chip thickness on $\overline{\mathrm{Nu}}_\mathrm{j}$ based on junction temperature: N=4, Re=1024.

Therefore, we can use the dimensionless analysis relation $\overline{\mathrm{Nu_f}}$-$\mathrm{Re_d}$ without considering the impact of the chip thickness. The final junction temperature can be calculated based on the equation $g(Bi)$. In this way, the design of the experiment for the cooler parameter analysis can be simplified.

3.3.3.2 Effects of nozzle scaling

For the investigation of the nozzle scaling, the same ratio is used for all the parameters. Figure 3.24 shows the impact of geometry parameters with different ratios, ranging from 0.01 to 0.4. As shown in Figure 3.24(a), the $\overline{\mathbf{Nu_f}}$-$\mathrm{Re_d}$ curves for all the nozzle number collapses for the same dimensionless number. According to the dimensionless theory, the physics should be the same if all the non-dimensional numbers related to geometry parameters and input parameters are kept constant. In this case, the dimensionless geometry parameters ($d_i$/L, $d_O$/L, t/L, H/L) and dimensionless velocity ($\mathrm{Re_d}$) are kept the same for different nozzle numbers, ranging from N=1 to N=64. However, the stagnation $\mathrm{Nu_o}$-$\mathrm{Re_d}$ correlation shown in Figure 3.24(b) scatters for smaller ratios 0.01 and 0.1, while the curve collapses for a higher dimensionless ratio beyond 0.2. For small $d_i$/L≤0.1, the heat spreading effect is more pronounced than larger $d_i$/L ≥0.2. This is due to the heat spreading effects in the Si substrate that is not included in this stagnation $\mathrm{Nu_o}$. Figure 3.24 also shows that the larger the inlet diameter ratio, the higher $\overline{\mathbf{Nu_d}}$ and $\mathrm{Nu_o}$. Therefore, the $\mathrm{Nu_o}$ represents the local effects which are impacted by the chip thickness $t_c$.

As for the $\overline{\mathbf{Nu_f}}$-$\mathrm{Re_d}$ curves, the relation function can be expressed as following:

$$\overline{\mathbf{Nu_f}} = f(x) * \mathrm{Re_d}^{0.48*\left(\frac{\mathrm{di}}{\mathrm{L}}\right)^{-0.16}} \tag{3.29}$$

where the exponent of $\mathrm{Re_d}$ is a function of $d_i$/L. The correlation $f$(x) is also a function of the other parameters: $d_i$/L, $d_O$/L t/L and H/L, and will be extracted later.

For the $\mathrm{f} - \mathrm{Re_d}$ correlation curve shown in Figure 3.25, the impact of the dimensionless variables is also studied when the inlet numbers scale from 1 to 64. It shows that all the $\mathrm{f} - \mathrm{Re_d}$ curves collapse for $\alpha$ is 0.4, and scatters for smaller values. In addition, it is also observed that the scattering is even pronounced for high Reynolds number. In order to study the impact of different variables, the inlet number N=4 is chosen for the DOE simulations.

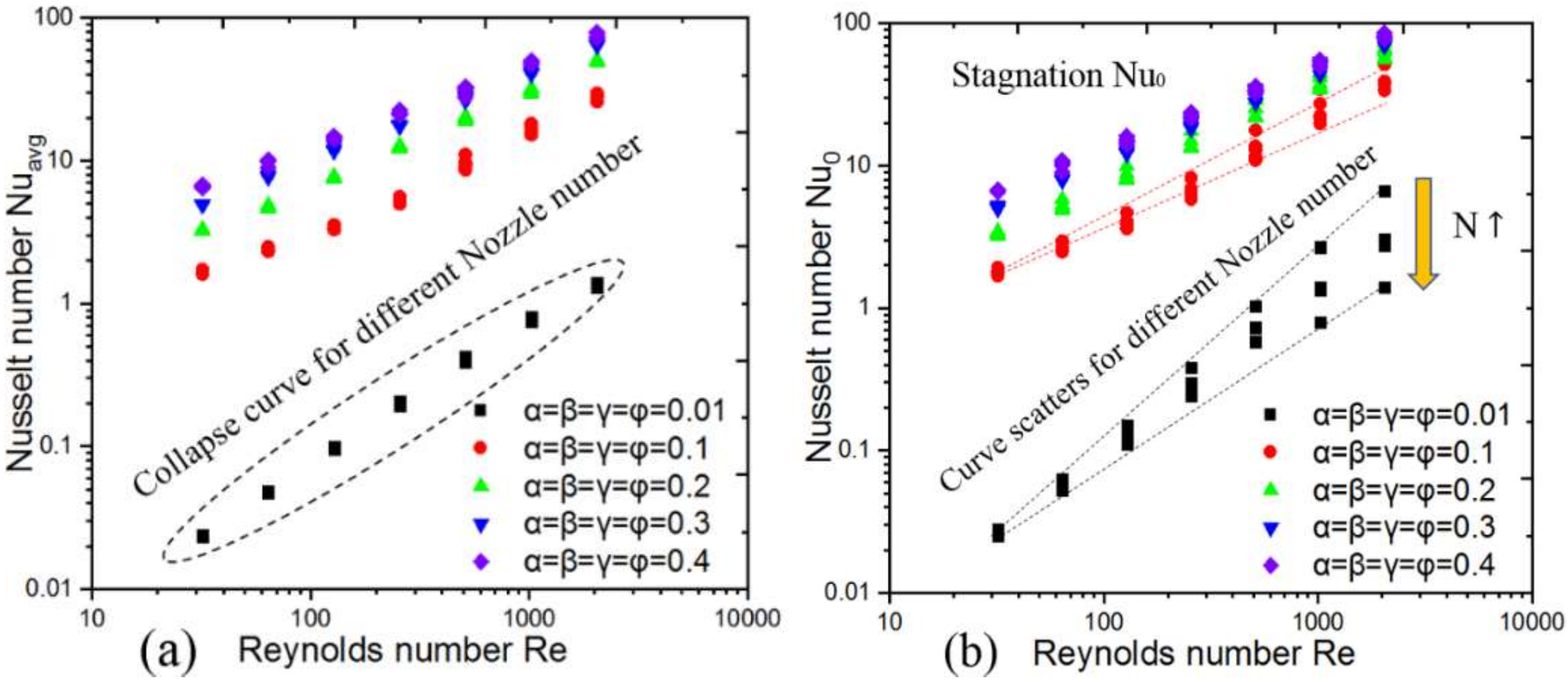

**Figure 3.24:** N scaling of the (a) average Nusselt and (b) stagnation Nusselt number as a function of Reynold number, under different inlet/outlet diameter ratios.

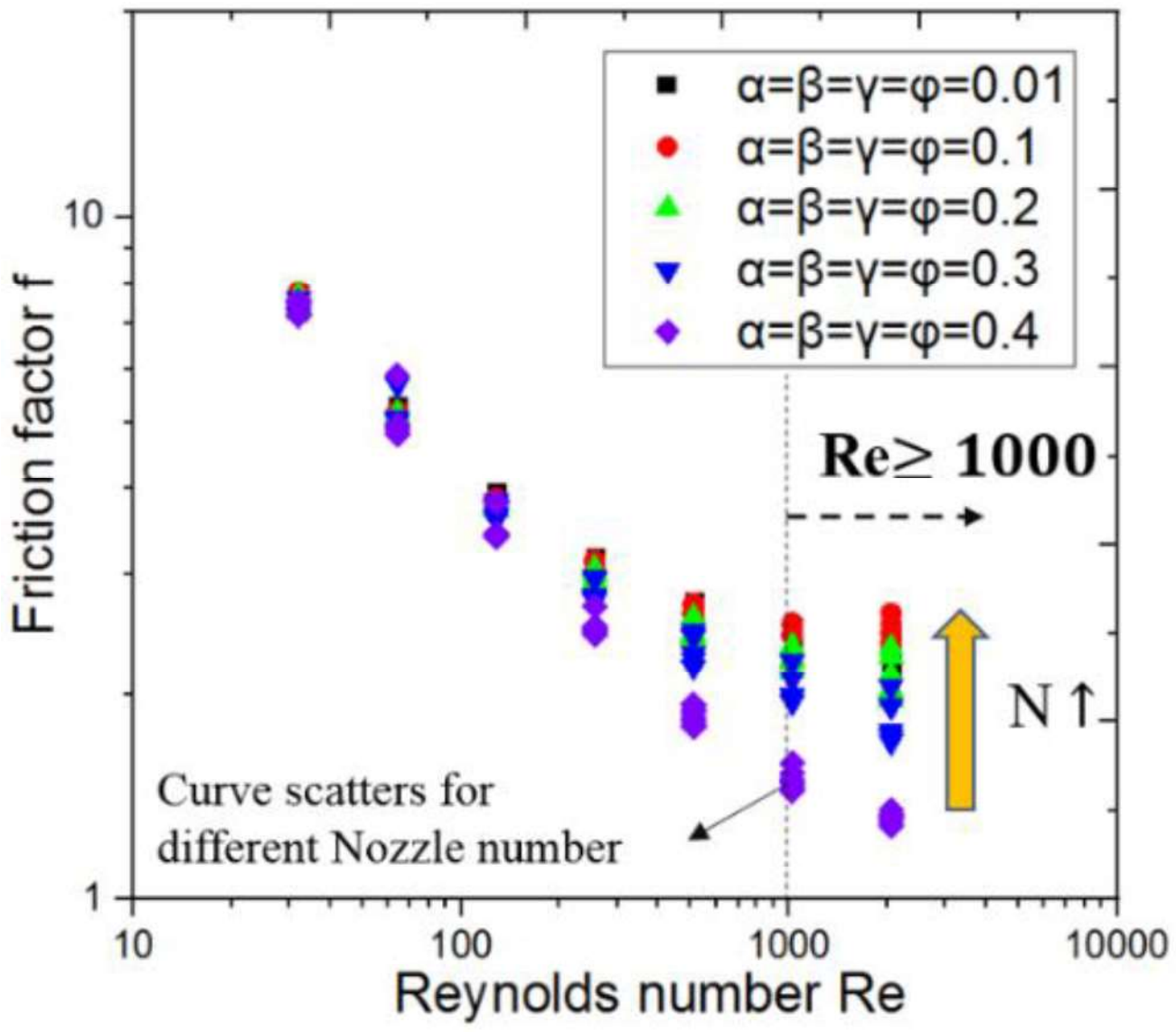

**Figure 3.25:** N scaling of the pressure coefficient with different inlet/outlet diameter ratios ($d_i/L \leq d_o/L$).

Conclusions can be summarized as below:

- The average Nusselt number $\overline{Nu}_d - Re_d$ curves for different nozzle numbers all collapse for the same dimensionless number ratio;
- The stagnation Nusselt number $Nu_0 - Re_d$ curves in the range ($d_i/L \geq 0.1$) collapse; however, at the small ratio with $d_i/L \leq 0.1$, the $Nu_0 - Re_d$ curves scatter due to the heat spreading effects in the chip, for which the chip thickness is not included in the dimensionless analysis;
- The friction factor $f - Re_d$ collapses below $Re_d \leq 1000$. At the Reynold number $Re_d$ higher than 1000, there are discrepancies, especially for smaller $\alpha$.

3.3.3.3 Effects of nozzle length

In this section, the impact of the nozzle length will be investigated. Literature [70] reported that the flow inside very short nozzle channels (t/L≤ 0.1) would not reach the developed flow regime. Therefore, in the present model, the dimensionless nozzle length is chosen beyond t/L>0.1. Figure 3.26 shows the impact of dimensionless nozzle length t/L on the $\overline{\mathrm{Nu_f}}$ and f. As illustrated in Figure 3.27(a), the $\overline{\mathrm{Nu_f}} - \frac{t}{L}$ curve shows like a linear relation, which can be expressed as below:

$$\overline{\mathrm{Nu_f}} = m\left(\frac{t}{L}\right) + n \tag{3.30}$$

where the range of m is in the range of 3-7. In general, the nozzle length has a small impact on the $\overline{\mathrm{Nu_f}}$.

This is due to that the increase of the $\frac{t}{L}$ can increase the pressure drop between the inlet and outlet, where the dominated pressure drop is inside the inlet/outlet nozzle channel. The raise of the overall pressure drop can, in turn, improve the cooling performance, resulting in a higher $\overline{\mathrm{Nu_f}}$. However, this increment can be negligible when the t/L is in the range of 0.1≤t/L≤ 1.2.

The $\mathrm{f} - \mathrm{Re_d}$ correlation is shown in Figure 3.26(b), which shows that the friction factor f decreases as the t/L becomes larger. The function between f and t/l for different $d_i$/L is shown as a power-law relation, as below:

$$\mathrm{f} = \mathrm{a}\left(\frac{t}{L}\right)^{b} \tag{3.31}$$

The correlation and exponent for different $d_i$/L are listed in Table 3.4.

**Table 3.4:** Correlations for different $d_i$/L ratio.

| $d_i$/L | a | b |
|---|---|---|
| 0.1 | 0.37 | -0.75 |
| 0.3 | 0.74 | -0.84 |
| 0.35 | 0.75 | -0.87 |
| 0.4 | 0.70 | -0.86 |
| 0.5 | 0.73 | -0.86 |

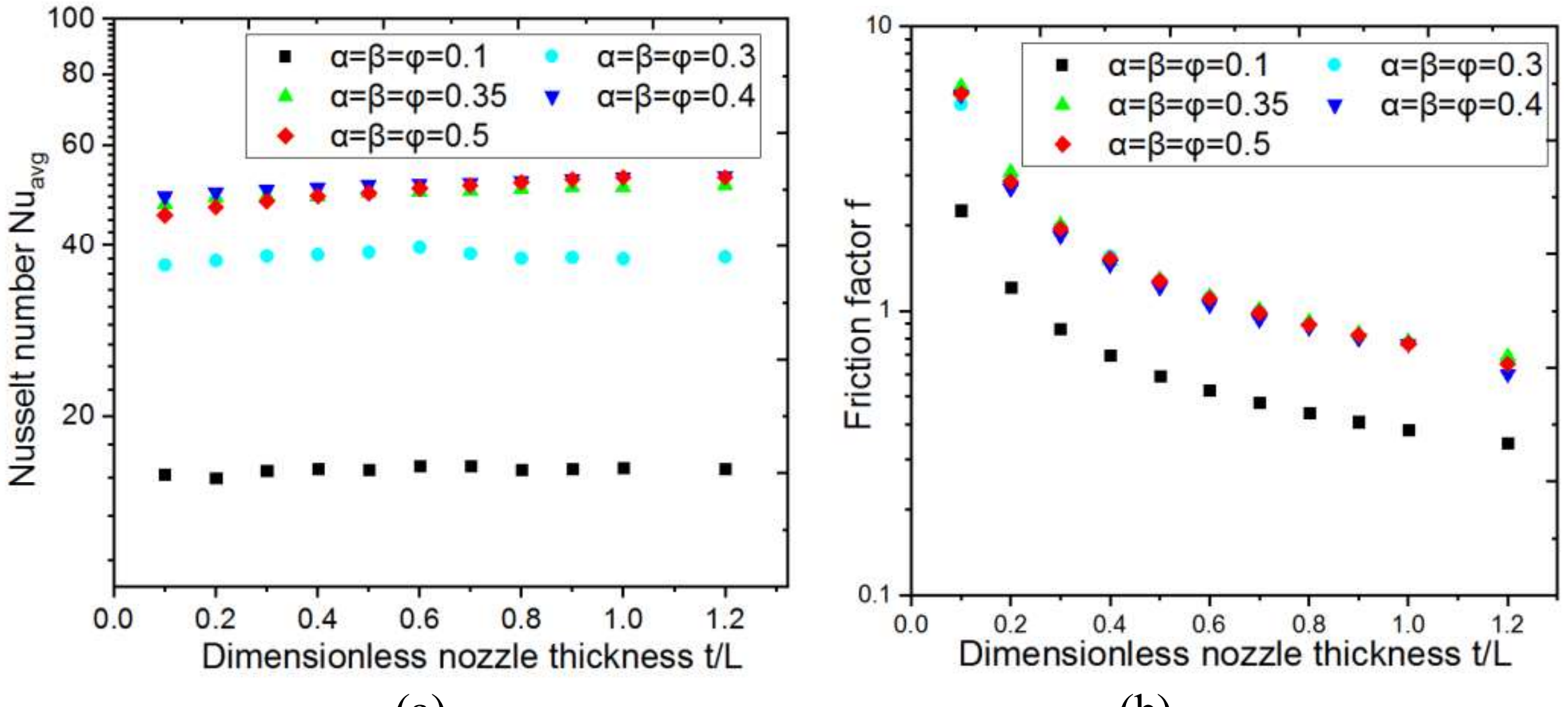

**Figure 3.26:** Impact of dimensionless t/L on the (a) Nusselt number $\overline{Nu_f}$ and (b) friction factor f under $Re_d$=1024.

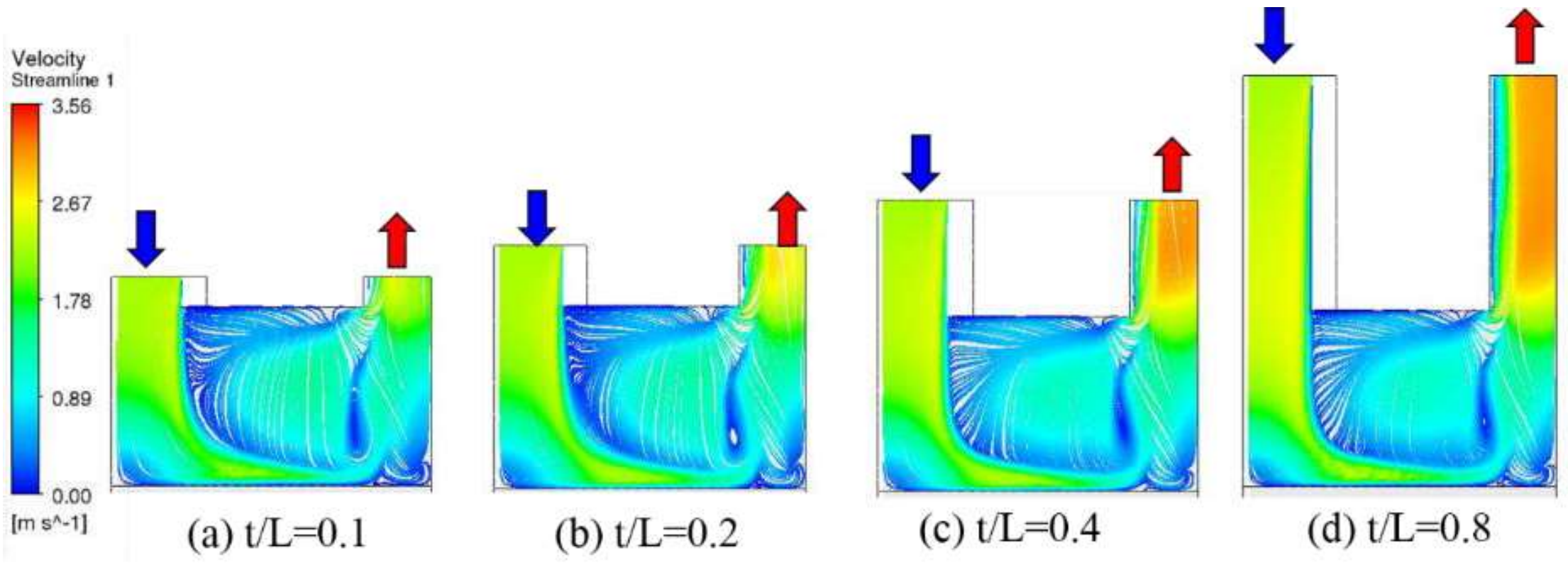

**Figure 3.27:** Flow distribution for different nozzle length: $Re_d$= 1024, $d_i/L$ =$d_o/L$ =0.3, H/L=0.3.

In order to understand the effect of nozzle plate thickness, the velocity distribution inside the jet cooling model is studied, as shown in Figure 3.27. The nozzle number is chosen as N=4, with an inlet/outlet diameter ratio of $d_i/L$ =0.3, under the $Re_d$=1024. It shows that the jet flow distribution at the stagnation and wall jet region does not change as the nozzle length increases, resulting in a stable cooling performance on the heating surface.

3.3.3.4 Effects of outlet diameter

The impact of dimensionless outlet diameter is investigated in this section, as shown in Figure 3.28. At the same time, the combined effects of $d_o$/L and H/L of the cavity height is also shown in Figure 3.28, indicated as the same color. All the results for different H/L plotted show only small scattering. In this study, the inlet diameter is kept smaller than the outlet diameter to reduce the system pressure drop, which is defined as $d_o/d_i \geq 1$. Therefore, a small $d_i$/L ratio of 0.05 is chosen as the reference value to guarantee that the $d_O$/L range can cover a larger range for $d_o$ in the analysis. As shown in Figure 3.28(a), the Nusselt number keeps stable when the $d_O$/L is increasing. It can also be seen that the changes of $\overline{Nu_f}$ are very small when the dimensionless cavity height H/L is varied from 0.08 to 0.6. In addition, $\overline{Nu_f}$ becomes larger when the Reynolds number $Re_d$ increases from 32 to 2048, as expected from the $\overline{Nu_f}$-$Re_d$ shown in Figure 3.24.

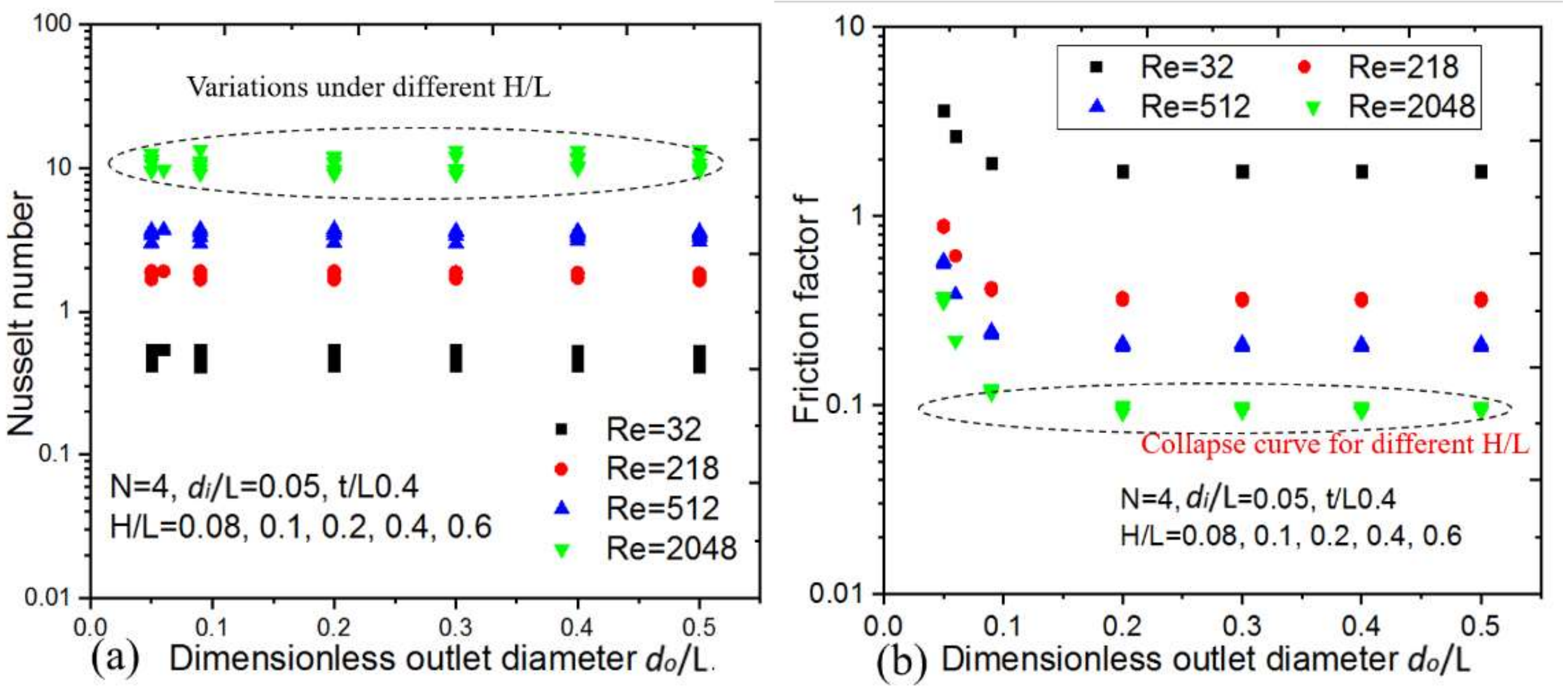

**Figure 3.28:** Impact of outlet diameter ratio on the $\overline{Nu_f}$ and friction factor: $d_i/L$ =0.05; $0.05 \leq d_o/L \leq 0.5$; t/L=0.4; $d_o/d_i \geq 1$.

The impact of dimensionless outlet diameter and cavity height on the friction factor is shown in Figure 3.28(b). It shows that the cavity height has negligible effects on pressure drop, where all the curves with the cavity height varying from 0.08 and 0.6 collapse. In general, the pressure drop decreases as the outlet diameter becomes larger. However, the influence of outlet diameter change becomes insignificant beyond $d_O$/L =0.1. The reason is that the pressure drop inside the outlet nozzles dominates when $d_O$/L is smaller than 0.1. When $d_O$/L is higher than 0.1, the pressure drop of the cooler is dominated by the pressure inside the inlet nozzles and impingement cavity.

For the Re=512, H/L∈ (0.08, 0.6), $d_i$/L =0.05, the function of f and $d_O$/L can be expressed as below:

$$f = e^{(-28.4*(\frac{d_O}{L}))} + \varphi(x) \tag{3.32}$$

where $\varphi(x)$ represents the effects of other parameters.The main conclusions can be summarized as below:

- $d_o/d_i \geq 1$; the variation of $d_o$ has no impact on $\overline{Nu_f}$-$Re_d$ relation;

- $d_o/d_i \geq 2$; the $d_o$ has no impact on f-$Re_d$; $d_o/d_i < 2$, the smaller $d_o/L$ has higher friction factor;

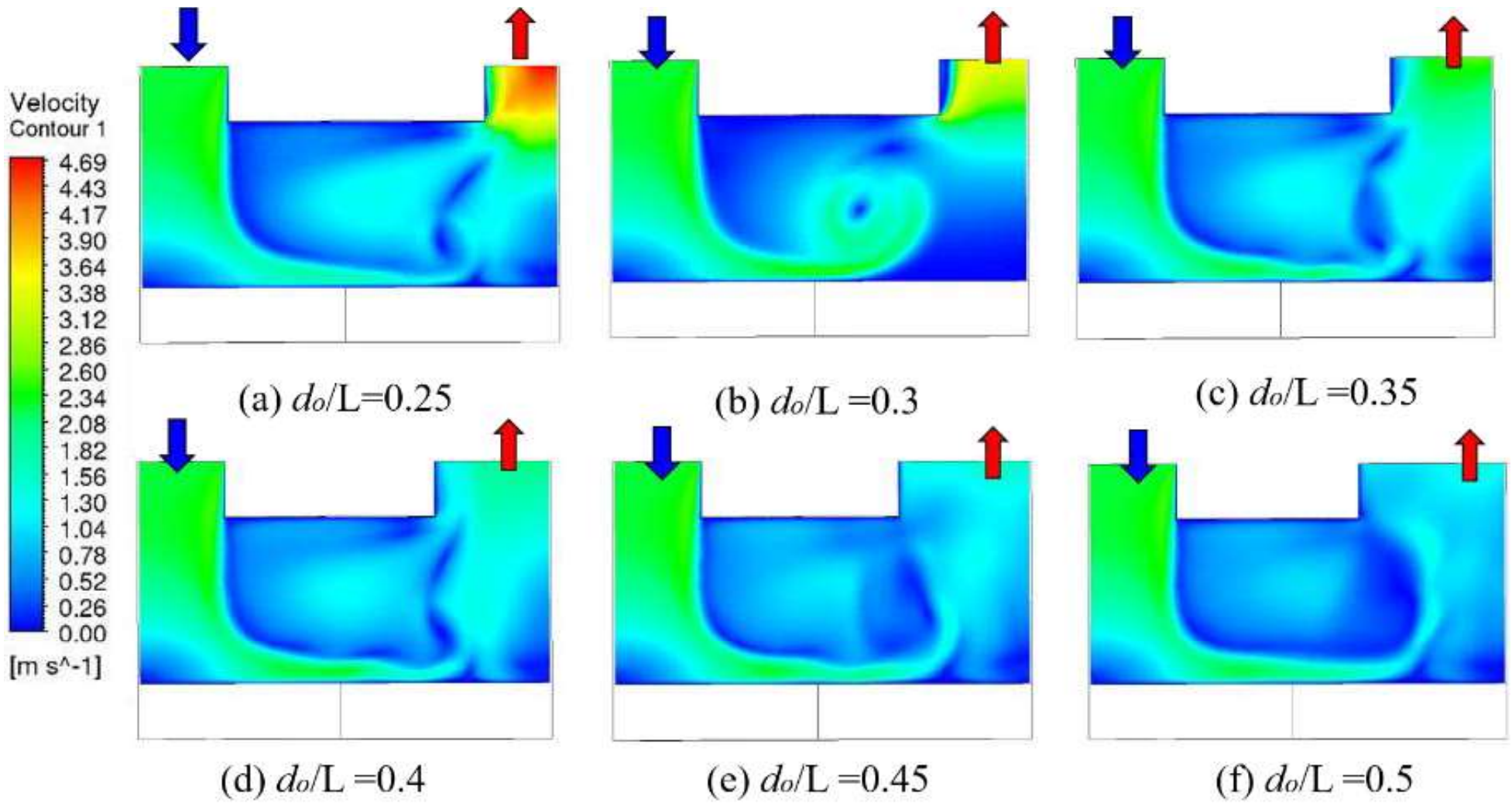

**Figure 3.29:** Impact of the outlet diameter on the flow distribution inside the unit cell model ($d_i/L$ =0.3, t/l=0.1, H/L=0.3, $Re_d$=1024).

The flow behaviors for different outlet diameters are analyzed from the modeling results, as shown in Figure 3.29. In this test case, the ratio $d_i/L$ is chosen as 0.3, where the ratio $d_o/L$ varies from 0.25 to 0.5. The confinement of the flow happens at the nozzle outlet, as the $d_o/L$ =0.25 is smaller than $d_i/L$ =0.3. As the $d_o$ is much smaller than $d_i$ ($d_o \ll d_i$), the pressure drop is higher. However, the pressure drop reduces as the $d_o/L$ becomes larger, shown in Figure 3.29(c). On the other hand, the flow regions with stagnation region and wall jet region still keep the same as the $d_o/L$ is increasing. In this study, the impact of the outlet diameter $d_o/L$ is not included in the predictive model within the range of 0.1 to 0.5 since the effects can be negligible under this range.

3.3.3.5 Effects of cavity height

Figure 3.30 shows the jet-to-target ratio H/L effects on heat transfer and pressure drop varying from 0.01 to 2. For the general trend, $\overline{Nu_f}$ and f are both higher for very small cavity heights. For the larger cavity height, the $\overline{Nu_f}$ and f keep constant. As for the heat transfer, the Nusselt number $\overline{Nu_f}$ increases with the raise of H/L as the cavity height ratio is below 0.2. This is due to the confining flow as the inlet diameter is much higher than the cavity height channel thickness, resulting in a higher pressure drop and higher flow velocity. There is a minimum value for $\overline{Nu_f}$ and f as the H/L is between 0.1 and 0.5.

For the friction factor f, there also exists a critical point $H_{critical}/L$ with a minimal f. The behavior is more significant for larger inlet diameter ratio $d_i/L$ while the f keeps constant after sharply decreases for smaller $d_i$/L. For both cases of the heat transfer and pressure drop, it can be seen that the critical point moves toward higher H/L with the increasing of $d_i$/L. In general, the impact of cavity height on average $\overline{Nu_f}$ and f are very small, for the ratio range: 0.3≤ H/L < 1, especially for the small nozzle diameter ratio. For the friction factor, the effects of H/L can be neglected when the H/L is higher than $H_{critical}$.

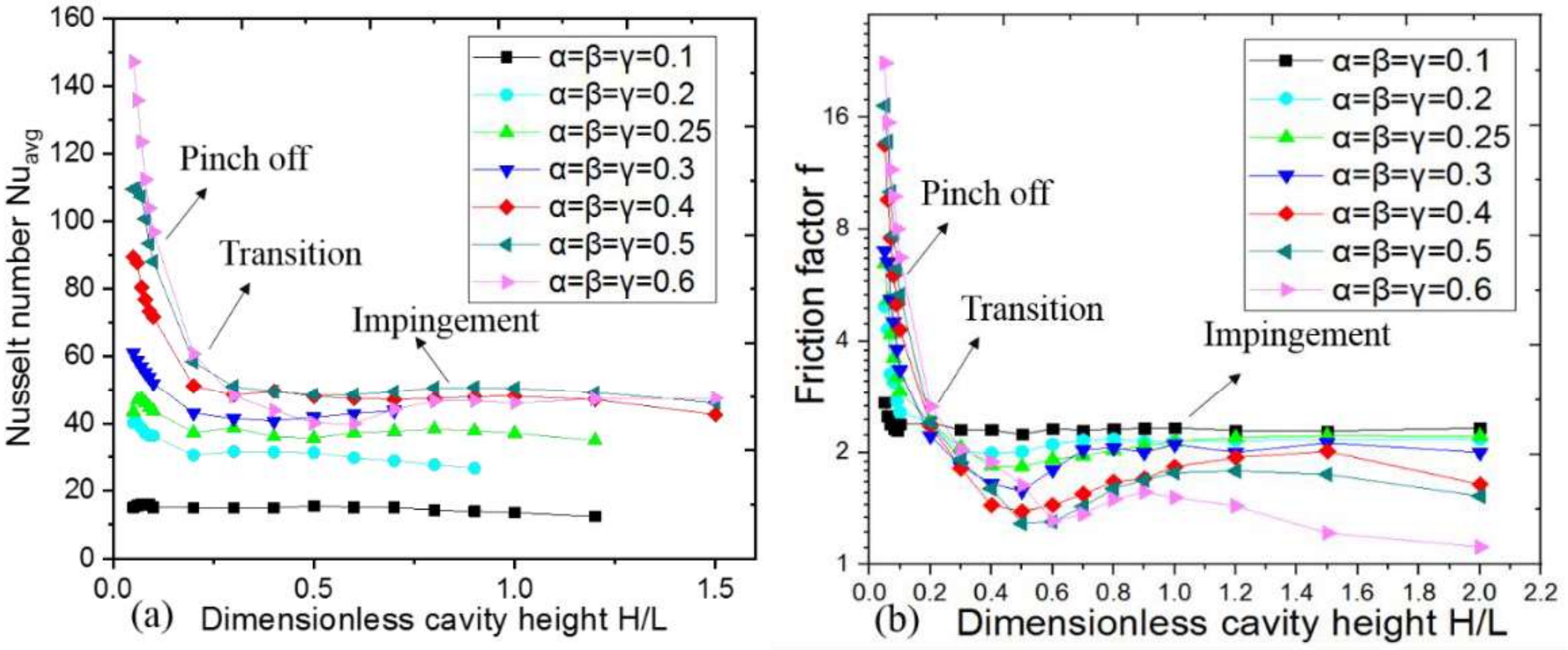

**Figure 3.30:** Effects of cavity height H/L on the $\overline{Nu_f}$ and friction factor f, for a fixed $Re_d$=1024 (N=4, $d_i$/L =0.3, Fl=600mL/min, $Re_d$=1024).

In order to provide insight into how the flow changes for different gap values, CFD simulation results with the unit cell model are shown in Figure 3.31. For H/L=0.01, the flow in the cavity height shows like the channel flow dominating most of the jet cooling pressure drop, which is defined as the “Pinch-off” regime in literature [6]. The heat transfer rate is higher inside the cavity height channel, since the boundary layer along the channel is thin. With the increasing of the H, the heat transfer decreases rapidly, as shown in Figure 3.31(a). For H/L=0.03, there is a hydraulic jump around the inlet nozzle region, which is defined as a “transition” regime [6]. The heat transfer will deteriorate due to the thickening of the flow boundary layer [6]. On the other hand, the pressure of the jet cooling decreases due to increasing channel thickness. As H is further increased, the hydraulic jump will move towards the outlet region, as shown in Figure 3.31(d). Since the boundary layer is thin before the hydraulic jump and becomes thicker afterward, the heat transfer rate along the chip surface is higher. In addition, there is also “recirculation” around the outlet region. For the cavity height H/L larger than 0.3, the negative effects of hydraulic jump will be reduced, and the heat transfer will keep constant. However, the recirculation flow along the wall jet region becomes more and more dominant, resulting in a higher pressure drop again, as shown in Figure 3.31(b).

The same physic phenomenon was also observed by Brunschwiler [6]. However, the flow physics used in their study is based on the laminar flow, with Reynolds number ranging from 11 to 402.6. And also, the dimensionless inlet diameter ranges from 0.1 to 0.3. This work extends the laminar flow to transition flow with Re between 32 and 2048. Moreover, this work also covers a wide range of the inlet nozzle diameter ratio $d_i$/L from 0.1 to 0.6. It is observed that the different flow regimes (pinch-off, transition, impingement) are different for smaller $d_i$/L and larger $d_i$/L.

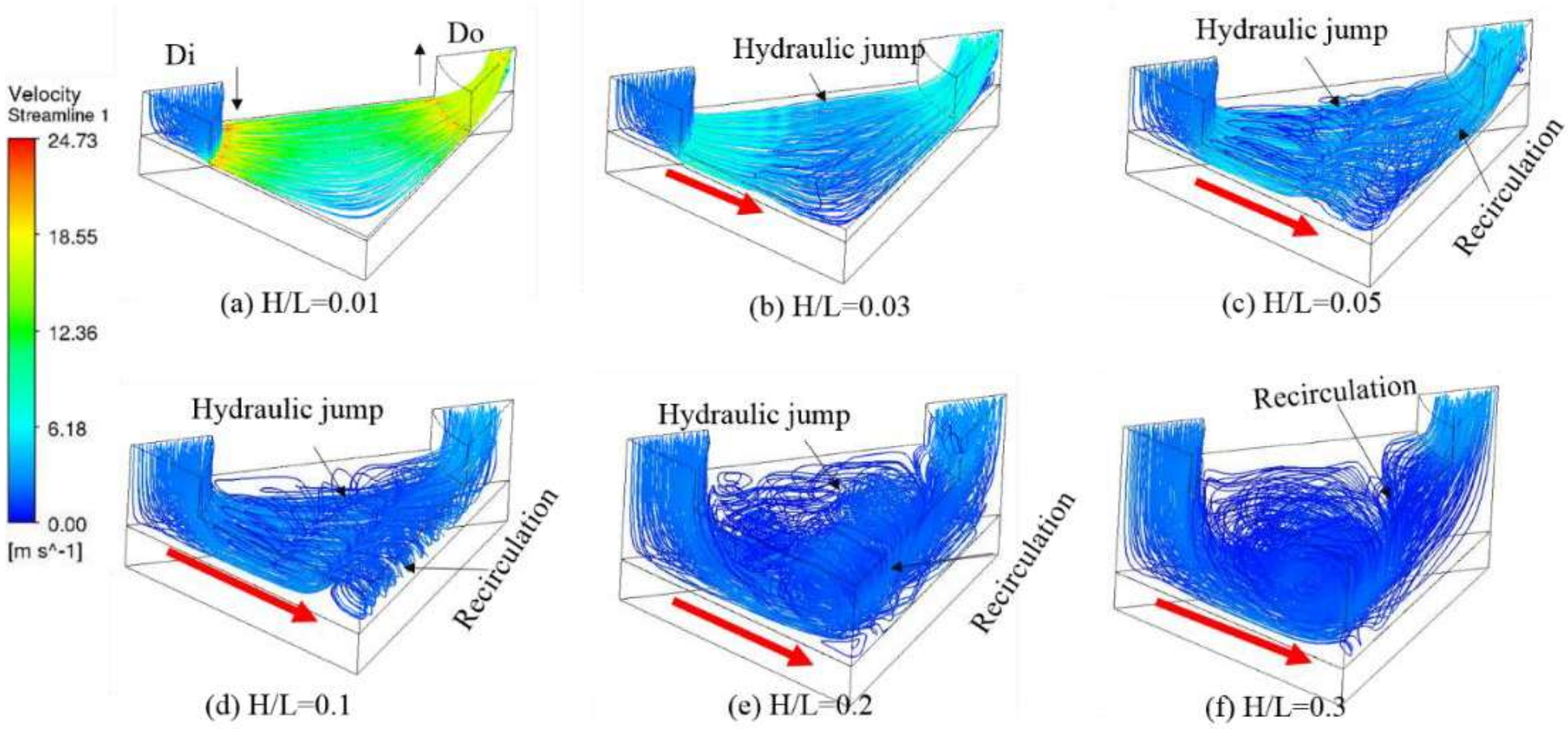

**Figure 3.31:** Impact of the jet-to-target at constant flow ratio (Reynolds number $\mathrm{Re_d}$=1024, $\mathrm{d_i/L}$ =$\mathrm{d_o/L}$ =0.3, t/l=0.1).

The effects of jet-to-target on heat transfer $\overline{\mathrm{Nu}}_\mathbf{f}$ and friction factor f with different Reynolds number ranging from 32 to 2048 are shown in Figure 3.32. The $\overline{\mathrm{Nu}}_\mathbf{f}$ keeps constant as the H/L is above 0.2, which is the impingement jet region. For H/L smaller than 0.2, the $\overline{\mathrm{Nu}}_\mathbf{f}$ increases as the H/L becomes smaller. This is due to the flow inside the cavity height level is dominated by the microchannel cooling. A similar trend is observed in Figure 3.33 with f-Re correlations. The friction factor becomes stable as 0.2 ≤H/L≥1. For the development of $\overline{\mathrm{Nu}}_\mathbf{f}$-$\mathrm{Re_d}$ and f-$\mathrm{Re_d}$, the effects of H/L will be included in the model.

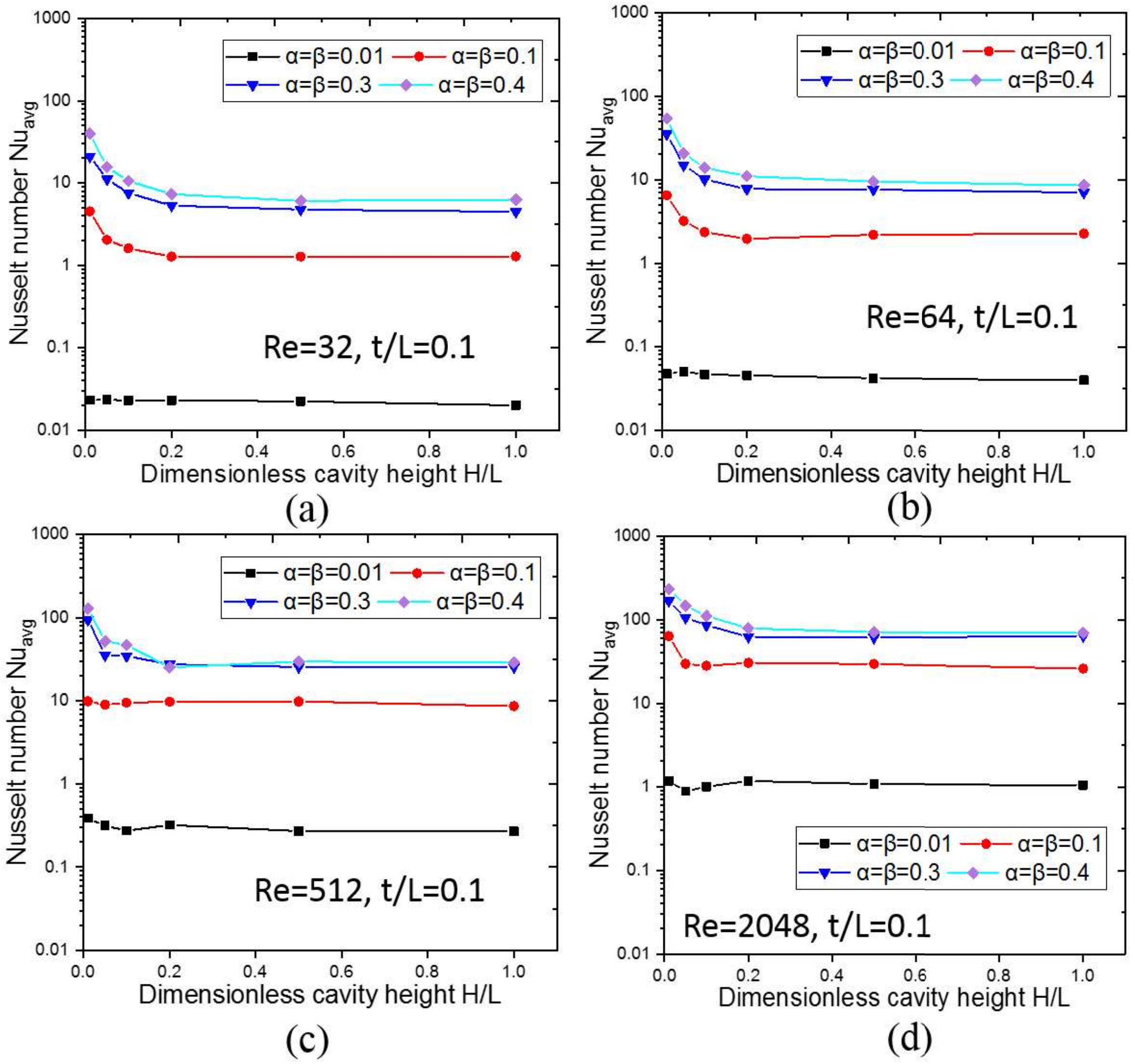

**Figure 3.32:** Impact of cavity height H/L on the average $\overline{Nu_f}$, udner different $Re_d$: (a) $Re_d$=32; (b) $Re_d$=64; (c) $Re_d$=512; (d) $Re_d$=2048; (t/L=0.1).

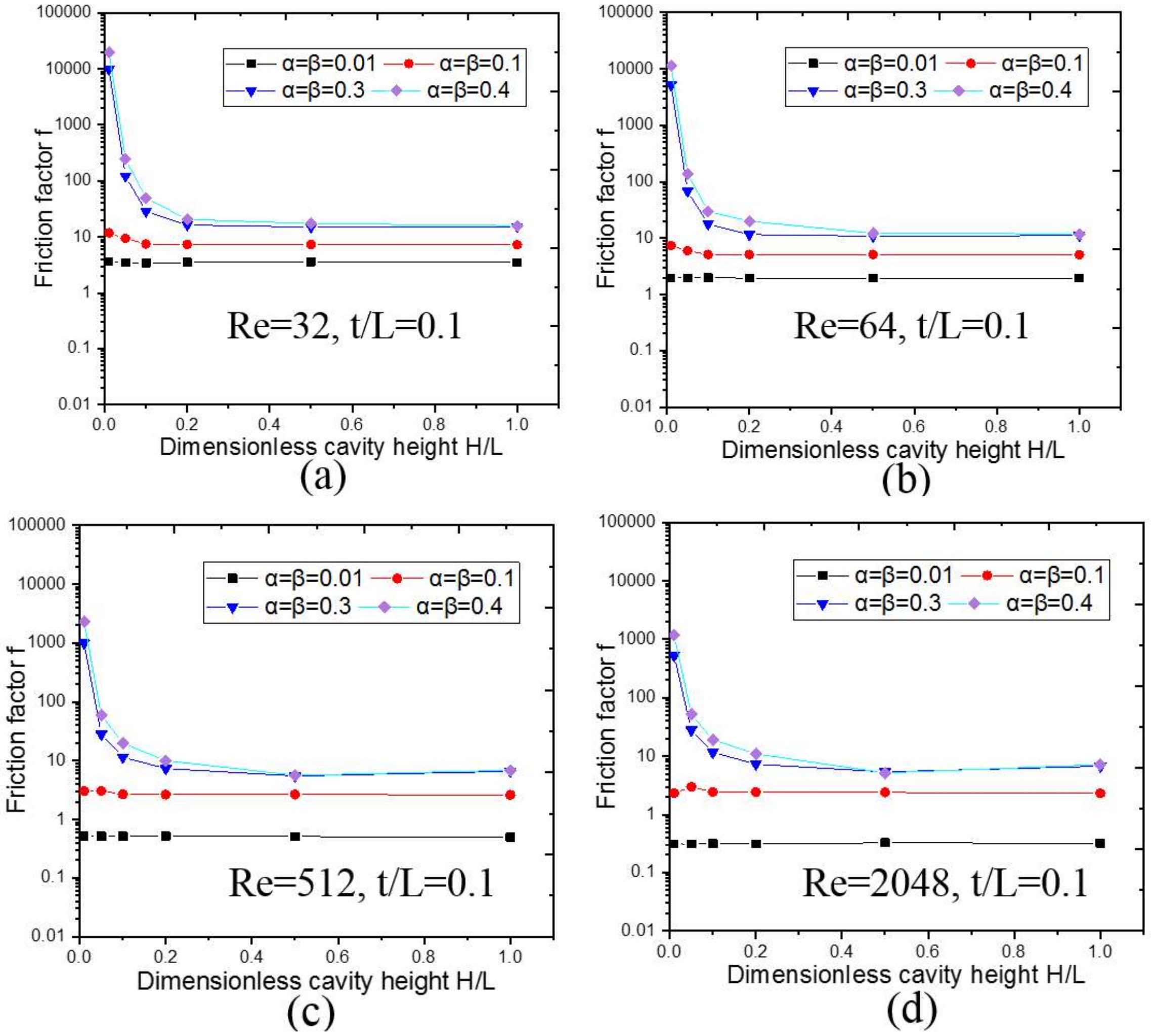

**Figure 3.33:** Impact of cavity height H/L on friction factor f, under different $Re_d$: (a) $Re_d$=32; (b) $Re_d$=64; (c) $Re_d$=512; (d) $Re_d$=2048; (t/L=0.1).

3.3.3.6 Effects of inlet diameter

As shown in Figure 3.34 and Figure 3.35, the impact of inlet nozzle diameter on the $\overline{Nu_f}$ and f is investigated, under smaller cavity height ratio H/L=0.5 and higher cavity height H/L=1. In general, the $\overline{\mathbf{Nu_f}}$ increases when the inlet diameter ratio $d_i$/L increases from 0.02 to 0.4. This is attributed to the stagnation region corresponding to the impingement surface becomes larger when the inlet nozzle diameter increases. It is found that the heat transfer coefficient decreases when the jet diameter becomes larger [42]. The impact of inlet diameter was also studied under different Reynolds number varying from 32 to 2048. In the general trend, the $\overline{\mathbf{Nu_f}}$-$d_i$/L presents a good linear function as below:

$$\overline{\mathbf{Nu_f}} = \alpha\left(\frac{d_i}{L}\right) + \beta \tag{3.33}$$

As for the friction factor, the friction factor shows power-law function with $d_i$/L, listed as below:

$$f \sim \left(\frac{d_i}{L}\right)^b \tag{3.34}$$

The modeling results shown Figure 3.34 and Figure 3.35 can also be aligned with the cavity height effects shown in Figure 3.30.

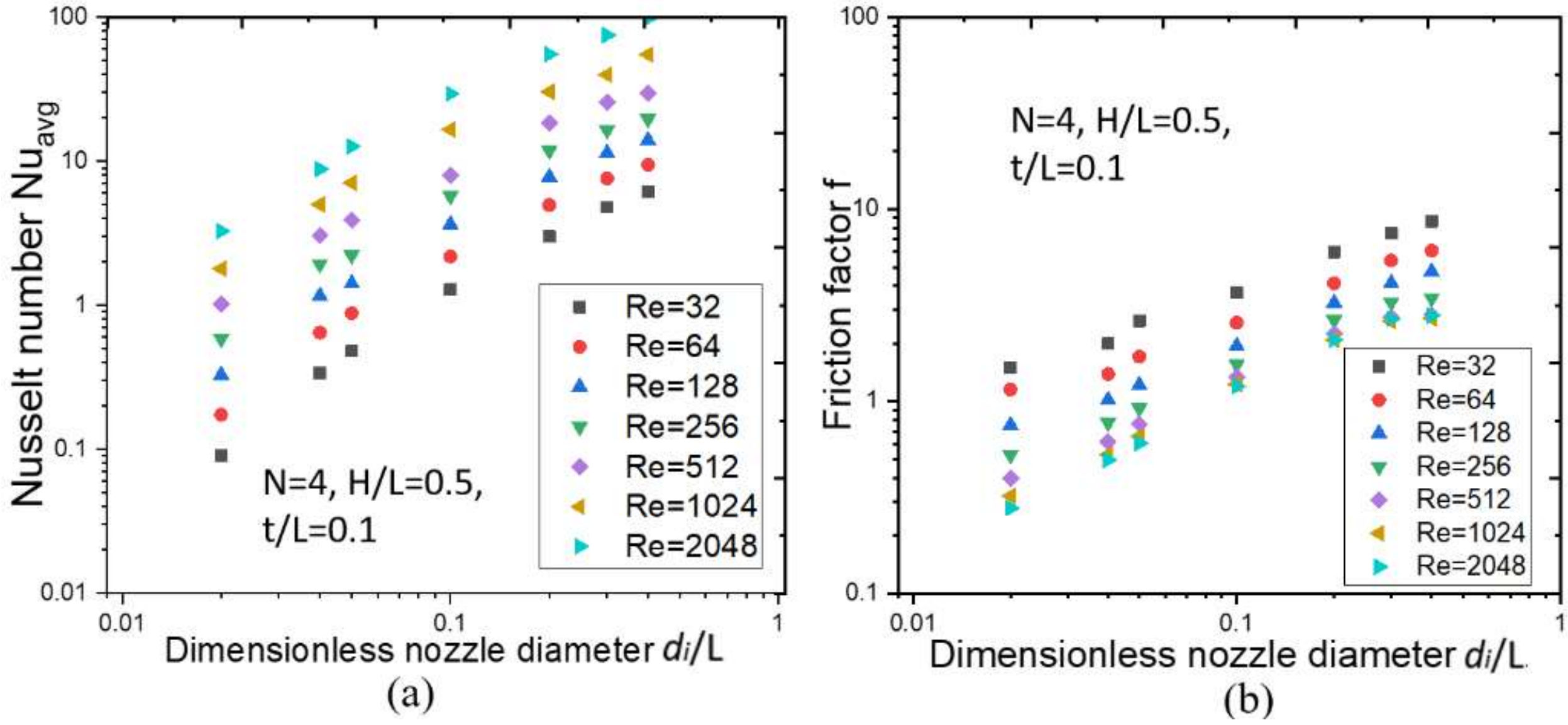

**Figure 3.34:** Nusselt number and friction factor as function of dimensionless inlet nozzle diameter: t/L=0.1 H/L=0.5 smaller cavity height ratio ($d_i$/L =$d_O$/L).

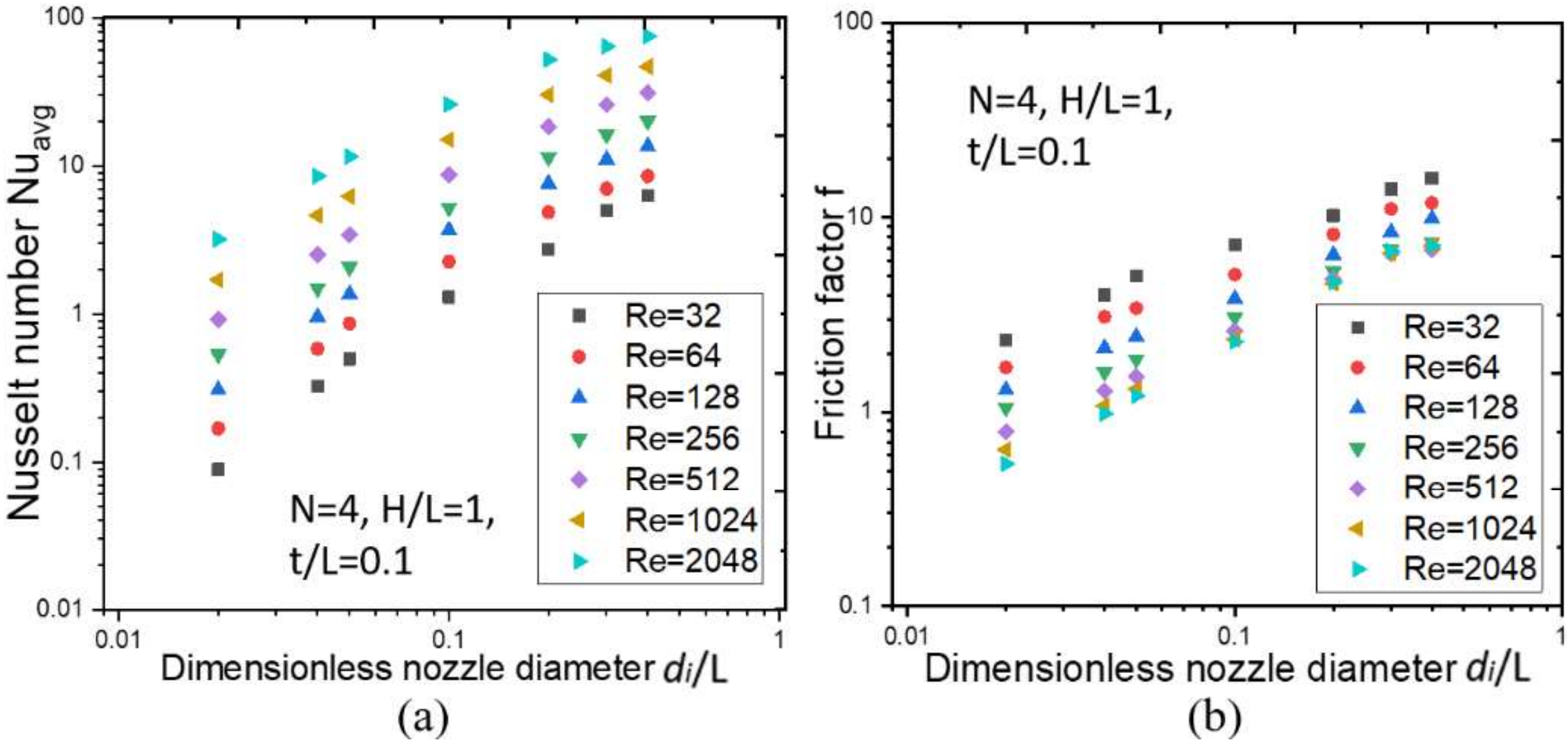

**Figure 3.35:** Nusselt number and friction factor as function of dimensionless inlet nozzle diameter: t/L=0.1 H/L=1, with larger cavity height ratio ($d_i$/L =$d_O$/L).

The details of the temperature distribution for different inlet diameter ratio is shown in Figure 3.36. It can be seen that the temperature reduces as the inlet diameter becomes larger. For small $d_i$/L, the stagnation region and wall jet region are very limited, resulting in a higher temperature at the outlet region. Moreover, the heat spreading through the silicon die dominates as the inlet diameter is very small. As the inlet diameter becomes larger, the jet cooling stagnation region becomes larger and the temperature drops.

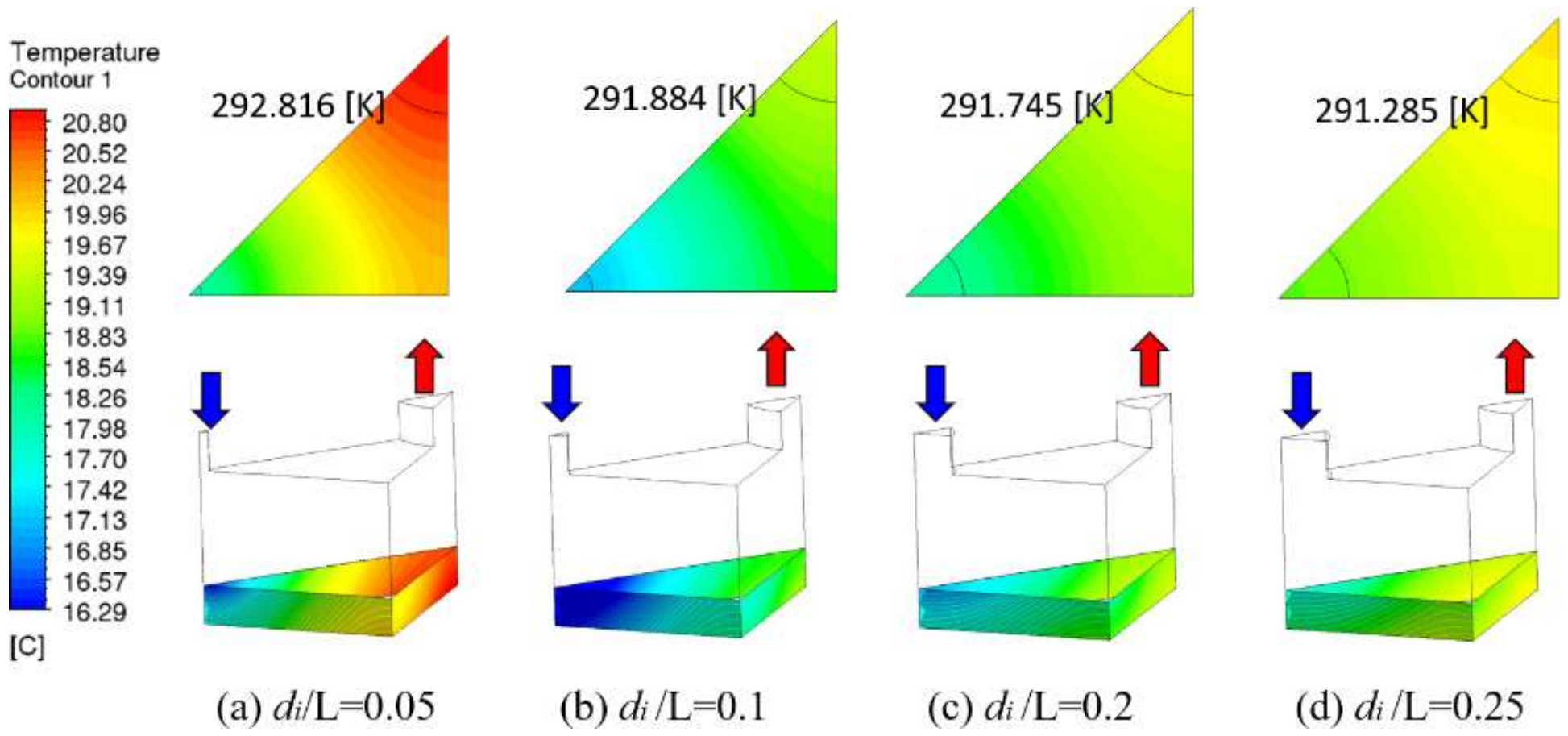

**Figure 3.36:** Temperature distribution for different inlet diameter $d_i$/L, $d_O$/L =0.3, t/L=0.1, and H/L=0.3, $Re_d$=1024: (a) $d_i$/L =0.05; (b) $d_i$/L =0.1; (c) $d_i$/L =0.2; (d) $d_i$/L =0.25.

### 3.3.4 Development of predictive models

As shown in Figure 3.24, the pressure drop is extremely high when the normalized inlet diameter ratio comes to 0.01. Therefore, the dimensional inlet diameter and outlet diameter ratio are chosen above 0.01 in order to keep the pressure drop in the reasonable region. Based on the multivariable regression analysis, the developed empirical models with $\overline{Nu_f}$-$Re_d$ and f-$Re_d$ are shown in Figure 3.37.

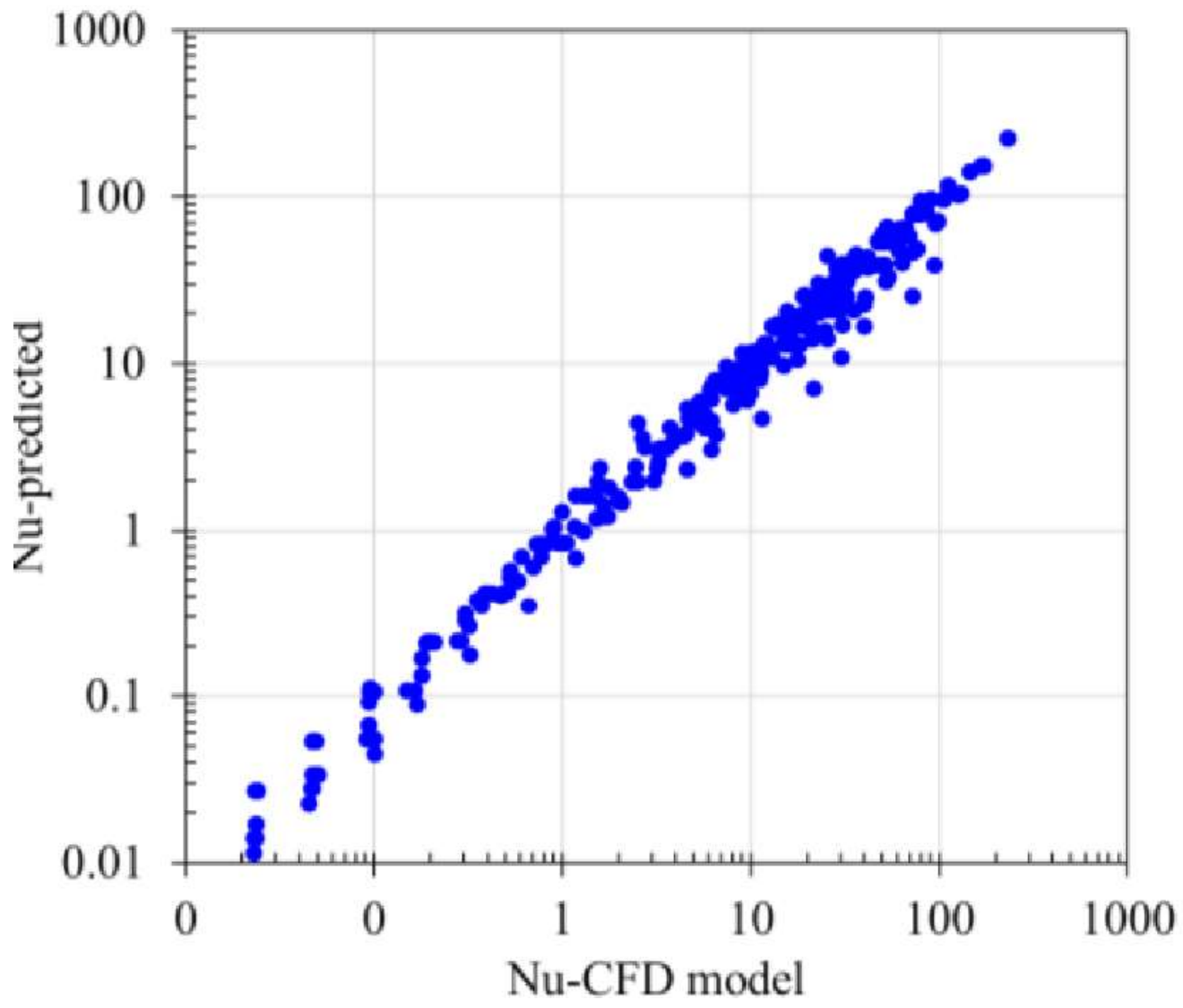

(a)

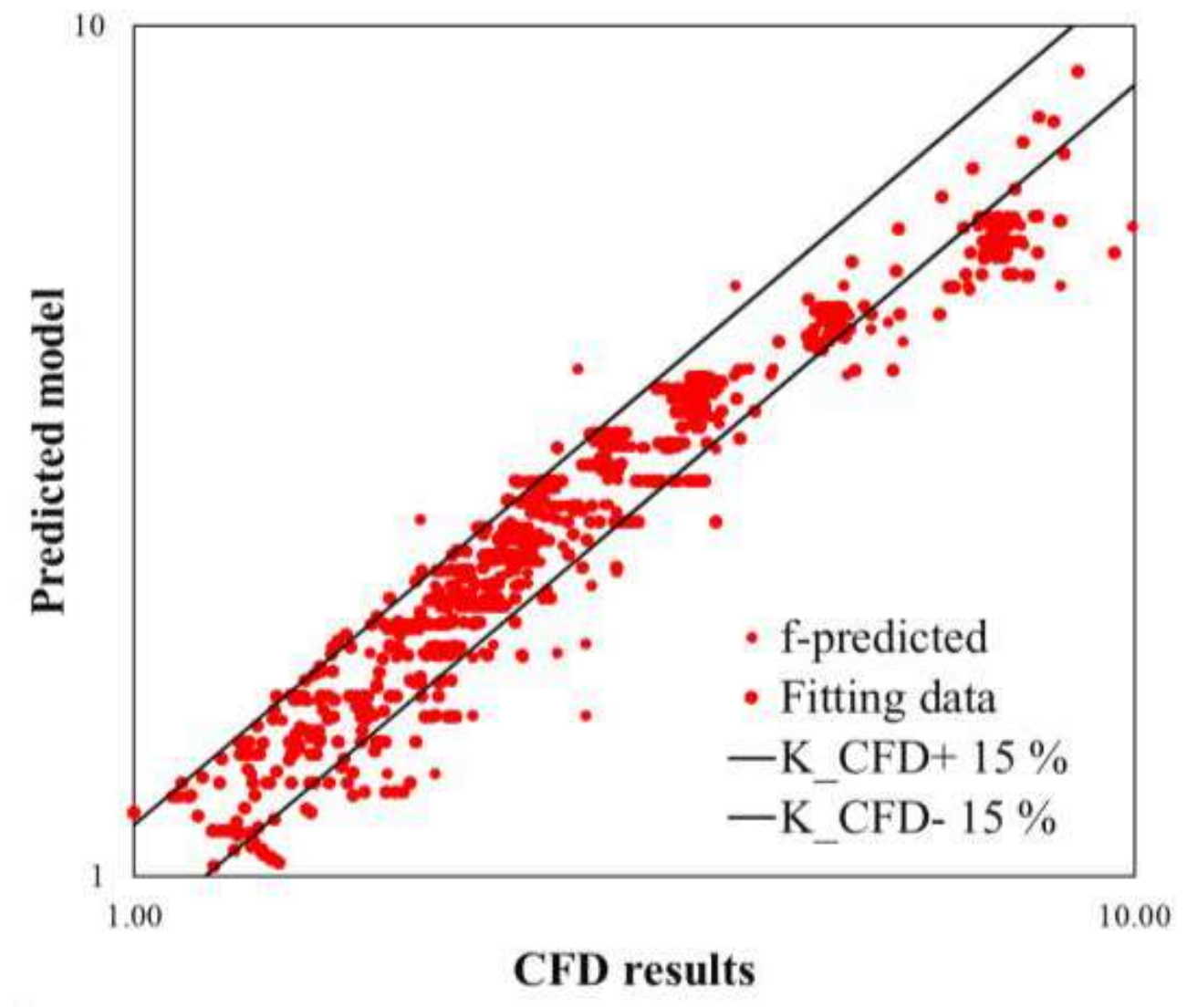

(b)

**Figure 3.37:** Correlations fitting for heat transfer and flow dynamics: (a) Nu-Re correlation comparison; (b) friction factor f-Re correlation comparison.

$$\overline{\mathrm{Nu}_\mathbf{f}} = \left(5.64\left(\frac{d_i}{\mathrm{L}}\right)^2 + 0.031\left(\frac{d_i}{\mathrm{L}}\right) - 0.000632\right)\left(\frac{\mathrm{H}}{\mathrm{L}}\right)^{-0.29} \mathrm{Re_d}^{0.48\left(\frac{d_i}{\mathrm{L}}\right)^{-0.16}} \tag{3.35}$$

$$\left(\frac{d_i}{\mathrm{L}} = \frac{d_o}{\mathrm{L}};\ 0.01 \leq \text{ a,b} \leq 0.4;\ 0.01 \leq \mathrm{H/L} \leq 0.4;\ 32 \leq \mathrm{Re_d} \leq 2048;\ 0.01 \leq \mathrm{t/L} \leq 0.4\right)$$

From the empirical model of $\overline{\mathrm{Nu}_\mathbf{f}} - \mathrm{Re_d}$, the variation is between ±25%. This model also first shows that the exponent of $\mathrm{Re_d}$ is as a function of $d_i$/L. Other parameters such as outlet diameter, cavity height and nozzle plate thickness are negligible when all the parameters are under the confined region

$$\mathrm{f} = \left(\left(21.2 * \left(\frac{d_i}{\mathrm{L}}\right)+14.5\right) \mathrm{Re_d}^{-0.73\left(\frac{d_i}{\mathrm{L}}\right)-0.26}\left(2.26\left(\frac{\mathrm{t}}{\mathrm{L}}\right)+0.89\right)\left(0.37\left(\frac{\mathrm{H}}{\mathrm{L}}\right)^{0.15}+0.55\right)+0.8\right)\left(\frac{t}{d_i}\right)^{-1} \tag{3.36}$$

$$\left(\frac{d_i}{\mathrm{L}} = \frac{d_o}{\mathrm{L}} = \mathrm{a};\ 0.05 \leq \text{ a} \leq 0.6;\ \mathrm{H}/d_i \geq 0.2;\ 32 \leq \mathrm{Re_d} \leq 2048;\ \mathrm{t/L} \geq 0.1\right)$$

where the confidence level is between ±15%. The friction factor f is defined as below:

$$f = \frac{\Delta P}{\left(\frac{1}{2}\rho\cdot V_{\mathrm{in}}{}^2\right)\left(\frac{t}{d_i}\right)} = \frac{k}{\left(\frac{t}{d_i}\right)} \tag{3.37}$$

For the $k$ -Re empirical model, it can be seen that $k$ has a linear relationship with the t/L. The effects of the cavity are also captured by the developed function.

## 3.4 Test case study using dimensionless analysis

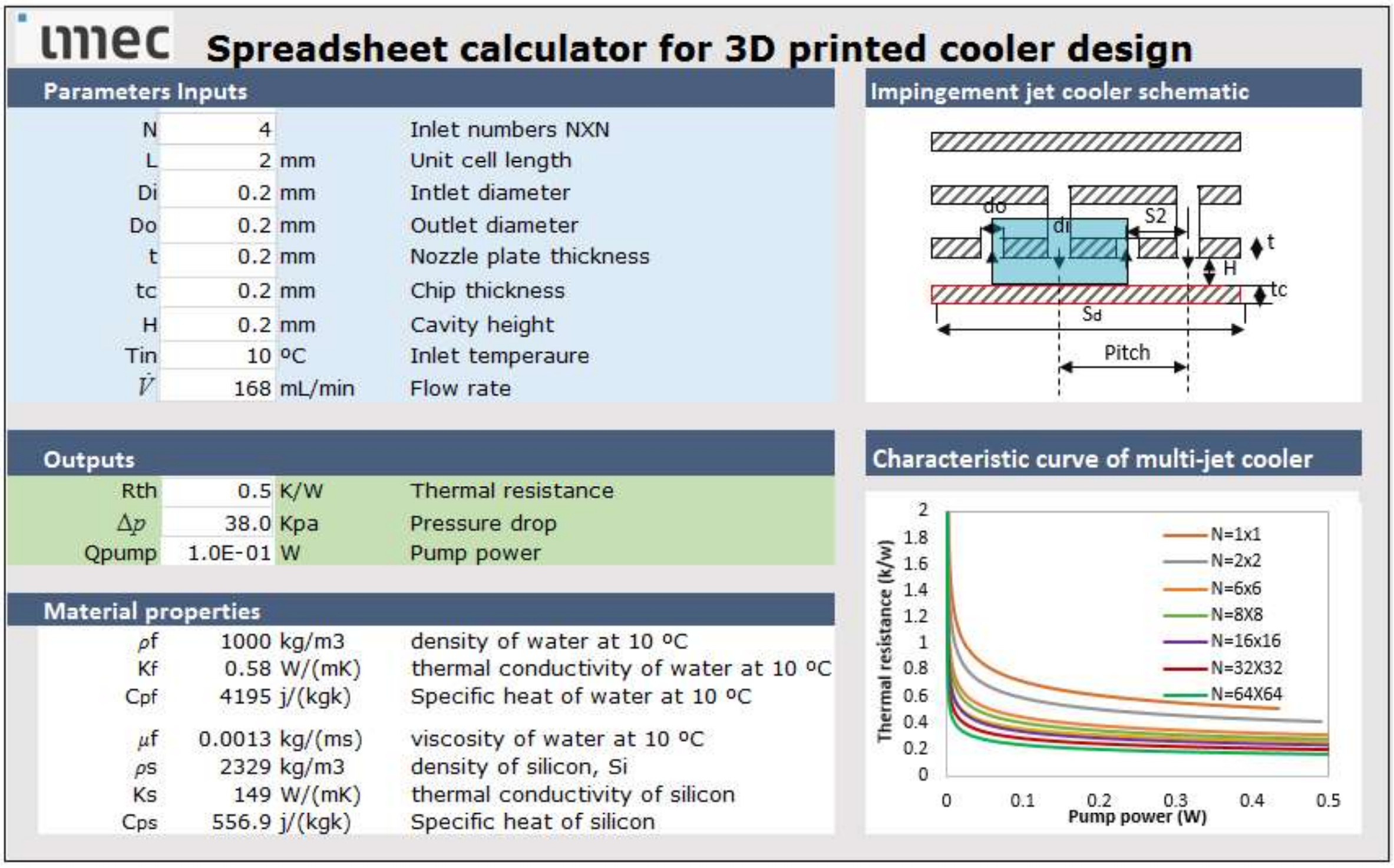

**Figure 3.38:** Spreadsheet tool implementation on the impingement jet cooling.

Based on the predictive model of the Nu and f calculation, the characteristic curve of the impingement jet cooler can be generated for the cooler design. As shown in Figure 3.38, a test case with a 4×4 nozzle array is implemented. The given parameters are listed in the parameter input region, including the nozzle number N, nozzle diameter di/do, nozzle length t, chip thickness $t_c$, and the flowrate. The material properties are also presented in the spreadsheet. The derived predictive models are embedded in the calculator. Therefore, the output parameters with the thermal resistance, pressure drop, and pumping power can be calculated based on the input parameters.

## 3.5 Conclusions

In this chapter, the parametric analysis of the absolute number and dimensionless analysis are both investigated. In the first part, the evaluation of the cooler performance is presented as a multi-objective optimization: the trade-off between thermal resistance and pumping power for the extensive design of experiments (DOE) of the unit cell CFD model is analyzed as a Pareto front in a thermal resistance versus pumping power chart for all cooler designs. For a fixed cavity height, a saturation of the thermal performance for scaling nozzles arrays is observed. Furthermore, the nozzle diameter in the chosen unit cell size should be as large as possible. The set of simulation results provides a guideline for the optimal thermal design of the impingement cooler in terms of a number of nozzles, nozzle diameter and pitch, and nozzle to chip distance. The trade-off chart also shows that multi-jets cooling is much more energy-efficient than single jet cooling.

For the second part, the predictive models with Nusselt number and friction factor are developed within 10% and 25% simulation confidence levels covering the laminar flow and transition flows. Moreover, the effects of the nozzle length, outlet diameter and cavity height are also studied in this work. The predictive model firstly includes the exponent of Re, which is the function of $d_i$/L, while some other works of literature do not include this part.

**References**

[1] Lee, H., Agonafer, D. D., Won, Y., Houshmand, F., Gorle, C., Asheghi, M., Goodson, K. E., 2016, "Thermal Modeling of Extreme Heat Flux Microchannel Coolers for GaN-on-SiC Semiconductor Devices," Journal of Electronic Packaging, Vol. 138, 010907.

[2] B.W.Webb, C.-F.Ma, "Single-Phase Liquid Jet Impingement Heat Transfer", Advances in Heat Transfer, Volume 26, 1995, Pages 105- 217.

[3] K. Gould, S. Q. Cai, C. Neft, and S. Member, "Liquid Jet Impingement Cooling of a Silicon Carbide Power Conversion Module for Vehicle Applications," IEEE Trans. Power Electronics, vol. 30, no. 6, pp. 2975–2984, June 2015.

[4] R. Skuriat and C. M. Johnson, "Thermal performance of baseplate and direct substrate cooled power modules," in Proc. 4th IET Int. Conf. Power Electron. Mach. Drives, 2008, pp. 548–552.

[5] J. Jorg, S. Taraborrelli, G. Sarriegui, R. W. De Doncker, R. Kneer, and W. Rohlfs, "Direct single impinging jet cooling of a mosfet power electronic module," IEEE Trans. Power Electronics, vol. 33, no. 5, pp. 4224–4237, May 2018.

[6] T. Brunschwiler et al., "Direct liquid jet-impingement cooling with micronsized nozzle array and distributed return architecture," in Proc. IEEE Therm. Thermomechanical Phenom. Electron. Syst., 2006, pp. 196–203.

[7] G. Natarajan and R. J. Bezama, "Microjet cooler with distributed returns," Heat Transf. Eng., vol. 28, no. 8–9, pp. 779–787, July 2010.

[8] N. Zuckerman and N. Lior, "Jet Impingement Heat Transfer: Physics, Correlations, and Numerical Modeling," Advances In Heat Transfer, Vol. 39, pp. 565-631, 2006.

[9] T. Wei et al., "High-Efficiency Polymer-Based Direct Multi-Jet Impingement Cooling Solution for High-Power Devices," in IEEE Transactions on Power Electronics, vol. 34, no. 7, pp. 6601-6612, July 2019.

[10] E.N. Wang, L. Zhang, J.-M. Koo, J.G. Maveety, E.A. Sanchez, K.E. Goodson, and T.W. Kenny, "Micromachined Jets for Liquid Impingement Cooling for VLSI Chips," J. Microeletromech. Sys., vol. 13, no. 5, pp. 833-842, October 2004.

[11] E. A. Browne, G. J. Michna, M. K. Jensen, and Y. Peles, "Microjet array single-phase and flow boiling heat transfer with R134a," Int. J. Heat Mass Transf., vol. 53, no. 23–24, pp. 5027–5034, November 2010.

[12] Y. Han, B. L. Lau, G. Tang and X. Zhang, "Thermal Management of Hotspots Using Diamond Heat Spreader on Si Microcooler for GaN Devices," IEEE Transactions on Components, Packaging and Manufacturing Technology, vol. 5, no. 12, pp. 1740-1746, December 2015.

[13] B. P. Whelan, R. Kempers, and A. J. Robinson, "A liquid-based system for CPU cooling implementing a jet array impingement waterblock and a tube array remote heat exchanger," Appl. Therm. Eng., vol. 39, pp. 86–94, June 2012.

[14] Overholt MR, McCandless A, Kelly KW, Becnel CJ, Motakef S, "Micro-Jet Arrays for Cooling of Electronic Equipment," in Proc. ASME 3rd International Conference on Microchannels and Minichannels, 2005, pp. 249-252.

[15] T. Acikalin and C. Schroeder, "Direct liquid cooling of bare die packages using a microchannel cold plate," in Proc. IEEE Therm. Thermomechanical Phenom. Electron. Syst., 2014, pp. 673–679.

[16] M. K. Sung and I. Mudawar, "Single-phase hybrid micro-channel/micro-jet impingement cooling," Int. J. Heat Mass Transf., vol. 51, no. 17–18, pp. 4342–4352, August 2008.

[17] J. Jorg, S. Taraborrelli, et al., "Hot spot removal in power electronics by means of direct liquid jet cooling," in Proc. IEEE Therm. Thermomechanical Phenom. Electron. Syst., 2017, pp. 471–481.

[18] A. J. Robinson, W. Tan, R. Kempers, et al., "A new hybrid heat sink with impinging micro-jet arrays and microchannels fabricated using high volume additive manufacturing," in Proc. IEEE Annu. IEEE Semicond. Therm. Meas. Manag. Symp., 2017, pp. 179–186.

[19] T. Wei et al., "Experimental Characterization of a Chip Level 3D Printed Microjet Liquid Impingement Cooler for High Performance Systems," in IEEE

Transactions on Components, Packaging and Manufacturing Technology. doi: 10.1109/TCPMT.2019.2905610.

[20] Womac, D. J., Ramadhyani, S., and Incropera, F. P., “Correlating Equations for Impingement Cooling of Small Heat Sources with Single Circular Liquid Jets,” Journal of Heat Transfer, Vol. 115, No. 1, 1993, pp. 106–115.

[21] Garimella, S. V., and Rice, R. A., “Confined and Submerged Liquid Jet Impingement Heat Transfer,” Journal of Heat Transfer, Vol. 117, No. 4, 1995, pp. 871–877. doi:10.1115/1.2836304

[22] Garimella, S. V., and Schroeder, V. P., “Local Heat Transfer Distributions in Confined Multiple Air Jet Impingement,” Journal of Electronic Packaging, Vol. 123, No. 3, 2001, pp. 165–172.

[23] Aldabbagh, L. B. Y., and Sezai, I., “Numerical Simulation of Three-Dimensional Laminar Multiple Impinging Square Jets,” International Journal of Heat and Fluid Flow, Vol. 23, No. 4, 2002, pp. 509–518. doi:10.1016/S0142-727X(02)00141-8

[24] Afzal Husain, et al., “Thermal Performance Analysis and Optimization of Microjet Cooling of High-Power Light-Emitting Diodes”, April 2013, Journal of Thermophysics and Heat Transfer 27(2):235-245, DOI: 10.2514/1.T3931.

[25] G. M. Chrysler, R. C. Chu and R. E. Simons, "Jet impingement boiling of a dielectric coolant in narrow gaps," Proceedings of 1994 4th Intersociety Conference on Thermal Phenomena in Electronic Systems (I-THERM), Washington, DC, USA, 1994, pp. 1-8.

[26] Lee, Jungho Lee, Sang-Joon. "Stagnation region heat transfer of a turbulent axisymmetric jet impingement." Experimental Heat Transfer 12.2 (1999): 137-156.

[27] Ndao, S., Peles, Y., & Jensen, M. K. “A Genetic Algorithm based Multi-objective Thermal Design Optimization of Liquid Cooled offset Strip Fin Heat Sinks,” ASME. Heat Transfer in Electronic Equipment, pp. 755-763, (2009).

[28] Lin, L., Chen, Y. Y., Zhang, et al., “Optimization of Geometry and Flow Rate Distribution for Double-layer Microchannel Heat Sink,” International Journal of Thermal Sciences, 78, pp. 158-168, (2014).

[29] Pablo A. de Oliveira, Jader R. Barbosa, “Novel two-phase jet impingement heat sink for active cooling of electronic devices”, Applied Thermal Engineering, Volume 112, 5 February 2017, pp. 952-964.

[30] S. K. Waye, S. Narumanchi, M, et al., "Advanced liquid cooling for a traction drive inverter using jet impingement and microfinned enhanced surfaces," Fourteenth Intersociety Conference on Thermal and Thermomechanical Phenomena in Electronic Systems (ITherm), Orlando, FL, 2014, pp. 1064-1073.

[31] Bae, Daniel, Raphael Mandel, and Michael Ohadi. "Effect of bonding structure and heater design on performance enhancement of feeds embedded Manifold-Microchannel cooling." InterPACK 2017-74158.

[32] Cristiano Santos, "Thermal performance of 3D ICs: Analysis and alternatives", 3D Systems Integration Conference (3DIC), (2014).

[33] T. Tiwei, H. Oprins, et al., High efficiency direct liquid jet impingement cooling of high power devices using a 3D-shaped polymer cooler, IEEE International Electron Devices Meeting (IEDM), (2017)

[34] D. S. Kercher, Jeong-Bong Lee, O. Brand, M. G. Allen and A. Glezer, "Microjet cooling devices for thermal management of electronics," in IEEE Transactions on Components and Packaging Technologies, vol. 26, no. 2, pp. 359-366, June 2003.

[35] Anuj K. Shukla, Anupam Dewan, Flow and thermal characteristics of jet impingement: comprehensive review, INTERNATIONAL JOURNAL OF HEAT AND TECHNOLOGY, Vol. 35, No. 1, March 2017, pp. 153-166 DOI: 10.18280/ijht.350121

[36] Michna G. J., et al., "The. Effect of Area Ration on Microjet Array Heat Transfer", International Journal of Heat and Mass Transfer, vol. 54, pp. 1782-1790, (2011).

-[37] JF Maddox, "Liquid Jet Impingement Cooling with Spent Flow Management for power electronics cooling", PhD thesis, (2015).

[38] Andrew J. Onstad, Terri B. Hoberg, Christopher J. Elkins, and John K. Eaton, Flow and Heat Transfer for Jet Impingement Arrays with Local Extraction, Sixth International Symposium on Turbulence and Shear Flow Phenomena Seoul, Korea, 22-24 June 2009, pp.315-320.

[39] Kercher DM, Tabakoff WW. Heat Transfer by a Square Array of Round Air Jets Impinging Perpendicular to a Flat Surface Including the Effect of Spent Air. ASME. J. Eng. Power. 1970;92(1):73-82. doi:10.1115/1.3445306.

[40] L.W. Florschuetz, D.E. Metzger, and C.R. Truman. Jet array impingement with crossflow - correlation of streamwise resolved flow and heat transfer distributions, NASA contractor report NASA CR-3373, 1981.

[41] Anuj K. Shukla, Anupam Dewan, Flow and thermal characteristics of jet impingement: comprehensive review, INTERNATIONAL JOURNAL OF HEAT AND TECHNOLOGY, Vol. 35, No. 1, March 2017, pp. 153-166 DOI: 10.18280/ijht.350121

[42] Button, B. L. and Jambunathan, K. 1989. Jet impingement heat transfer: a bibliography 1976-1985. Previews Heat Mass Transfer, 15, 149-178

[43] Jambunathan, et al. "A Review of Heat Transfer Data for Single Circular Jet Impingement." International Journal of Heat and Fluid Flow, vol. 13, no. 2, 1992, pp. 106–115.

[44] Bernhard Weigand, "Multiple Jet Impingement, A Review", DOI: 10.1615/HeatTransRes.v42.i2.30 , pages 101-142.

[45] S.V.J. Narumanchi, V. Hassani, and D. Bharathan, "Modeling Single-Phase and Boiling Liquid Jet Impingement Cooling in Power Electronics".

[46] Kaveh Azar et al., 2011 Jet impingement cooling, Part 2: Single phase

[47] Molana, M., and Salem Banooni. "Investigation of heat transfer processes involved liquid impingement jets: a review." Brazilian Journal of Chemical Engineering 30.3 (2013): 413-435.

[48] Martin, H., 1977, "Heat and mass transfer between impinging gas jets and solid surfaces", Advances in Heat Transfer, 13, pp. 1-60

[49] V.A. Patil, V. Narayanan, Spatially resolved heat transfer rates in an impinging circular microscale jet, Microscale Thermophys. Eng. 9 (2) (2005) 183–197.

[50] Womac DJ, Incropera FP, Ramadhyani SS. Correlating Equations for Impingement Cooling of Small Heat Sources With Multiple Circular Liquid Jets. ASME. J. Heat Transfer. 1994;116(2):482-486. doi:10.1115/1.2911423.

[51] A. J. Robinson and E. Schnitzler, An experimental investigation of free and submerged miniature liquid jet array impingement heat transfer, Experimental Thermal and Fluid Science 32 (2007) 1-13.

[52] Y. Pan and B.W. Webb, Heat transfer characteristics of arrays of free-surface liquid jets, Journal of Heat Transfer 117 (1993) 878-883.

[53] M. Fabbri and V.K. Dhir, Optimized heat transfer for high power electronic cooling using arrays of microjets, Journal of Heat Transfer 127 (2005) 760-769.

[54] A. Royne and C.J. Dey, Effect of nozzle geometry on pressure drop and heat transfer in submerged jet arrays, International Journal of Heat and Mass Transfer 49 (2006) 800-804.

[55] Brian P. Whelan, Anthony J. Robinson. Nozzle Geometry Effects in Liquid Jet Array Impingement. Applied Thermal Engineering, Elsevier, 2009, 29 (11-12), pp.2211. ff10.1016/j.applthermaleng.2008.11.003ff. ffhal-00511424f

[56] Huber, A. A. M., and Viskanta, R., 1994, "Effect of Jet–Jet Spacing on Convective Heat Transfer to Confined, Impinging Arrays of Axisymmetric Air Jets," Int. J. Heat Mass Transfer, 37(18), pp. 2859–2869.

[57] Andrew J. Onstad, Terri B. Hoberg, Christopher J. Elkins, and John K. Eaton, Flow and Heat Transfer for Jet Impingement Arrays with Local Extraction, Sixth International Symposium on Turbulence and Shear Flow Phenomena Seoul, Korea, 22-24 June 2009, pp.315-320.

[58] Hoberg et al, Heat transfer measurements for jet impingement arrays with local extraction, International Journal of Heat and Fluid Flow, Volume 31, Issue 3, June 2010, Pages 460-467

[59] Andrew J. Onstad, Terri B. Hoberg, Christopher J. Elkins, and John K. Eaton, Flow and Heat Transfer for Jet Impingement Arrays with Local Extraction, Sixth International Symposium on Turbulence and Shear Flow Phenomena Seoul, Korea, 22-24 June 2009, pp.315-320.

[60] Zuckerman N. and Lior N. "Jet Impingement Heat Transfer : Physics , Correlations , and Numerical Modeling", Advances in heat transfer 39,06, pp. 565–631, (2006).

[61] T.-W.Wei, et al., Experimental characterization and model validation of liquid jet impingement cooling using a high spatial resolution and programmable thermal test chip, Applied Thermal Engineering, Volume 152, April 2019, Pages 308-318.

[62] B. R. Hollworth and L. Dagan, Arrays of impinging jets with spent fluid removal through vent holes on the target surface part 1: average heat transfer, J. Engng Power 102,994999 (1980).

[63] Rattner, Alexander S. S. "General Characterization of Jet Impingement Array Heat Sinks with Interspersed Fluid Extraction Ports for Uniform High-Flux Cooling." Journal of Heat Transfer, vol. 139, no. 8, 2017, pp. 082201/1–082201/11.

[64] Rhee, D., Yoon, P., and Cho, H., 2003, "Local Heat/Mass Transfer and Flow Characteristics of Array Impinging Jets With Effusion Holes Ejecting Spent Air," Int. J. Heat Mass Transfer, 46(6), pp. 1049–1061.

[65] Onstad, A., Elkins, C., and Moffat, R., 2009, "Full-Field Flow Measurements and Heat Transfer of a Compact Jet Impingement Array With Local Extraction of Spent Fluid," ASME J. Heat Transfer, 131(8), p. 082201

[66] Husain, A., Al-Azri, N. A., and Al-Rawahi, N. Z. H., 2015, "Comparative Performance Analysis of Microjet Impingement Cooling Models With Different Spent-Flow Schemes," J. Thermophys. Heat Transfer, 30(2), pp. 466–472.

[67] Kashi, B. and Haustein, D. H., "Dependence of submerged jet heat transfer on nozzle length," Int. J. Heat Mass Transfer 121, 137–152 (2018).

[68] Levy, Yeshayahou, et al., "Pressure Losses for Jet Array Impingement With Crossflow." Proceedings of the ASME Turbo Expo 2012: Volume 4: Heat Transfer, Parts A and B. Copenhagen, Denmark. June 11–15, 2012. pp. 139-14.

[70] Kashi, B. and Haustein, D. H., "Dependence of submerged jet heat transfer on nozzle length," Int. J. Heat Mass Transfer 121, 137–152 (2018).

# Chapter 4

# 4. Single Jet Cooling: Fundamental Understanding

## 4.1 Single jet demonstrator

As discussed in chapter 3, the parametric and dimensionless analysis both show that multi-jet cooling is more energy efficient than single jet cooling. However, in order to understand the fundamental flow and thermal behavior for jet impingement cooling and to validate the modeling approach, the analysis of the thermal and fluidic behavior of a single central inlet jet and surrounding outlet jets in the corners is studied as the first step in the analysis. The demonstration of a relatively large single jet cooler (2 mm diameter nozzle) on the chip size of 8×8 $mm^2$ with 32×32 temperature sensors (240 μm spatial resolution of the sensors) allows to accurately capture the chip temperature distribution caused by the local cooling of the liquid impinging jet on the heated surface. These measurement profiles can be used for the validation of the modeling results. Therefore, in this chapter, a single jet demonstrator is designed and fabricated as a proof of concept, for the fundamental understanding of the impingement jet cooling, from numerical modeling and experimental characterization point of view.

### 4.1.1 Introduction

As illustrated in chapter 2, a generic jet impingement cooler with locally distributed outlets in between the inlet nozzles includes three functional levels: the inlet plenum level, the outlet plenum level, and the impingement jet cooling level. The inlet plenum is the flow distributor, which feeds the liquid coolant for all inlet nozzles. The outlet plenum is the collector that collects the liquid for drainage. The impingement cooling happens in the cavity region, defined by nozzle-to-chip distance. Figure 4.1 shows the design of the single jet cooling demonstrator that includes the same three functional layers. Figure 4.1(a) shows the cross-sectional view of the single jet cooler indicating the assembly of the different parts. The cooler is assembled on the organic package substrate. Figure 4.1(b) shows the details of the arrangement of the global central single inlet and six outlets.

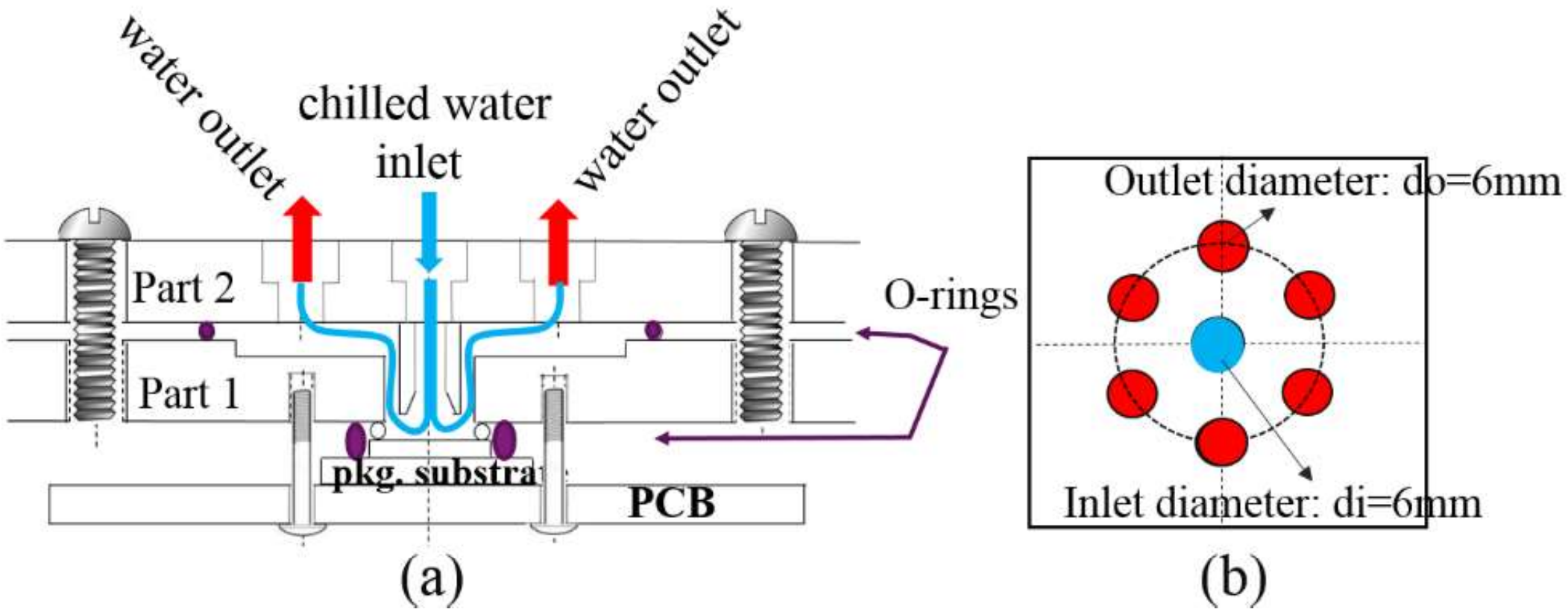

**Figure 4.1:** Chip level single impingement jet cooler: (a) schematic view of the cooler with different parts; (b) top view of a global single inlet and six outlets for the tube arrangement.

### 4.1.2 Fabricated demonstrator

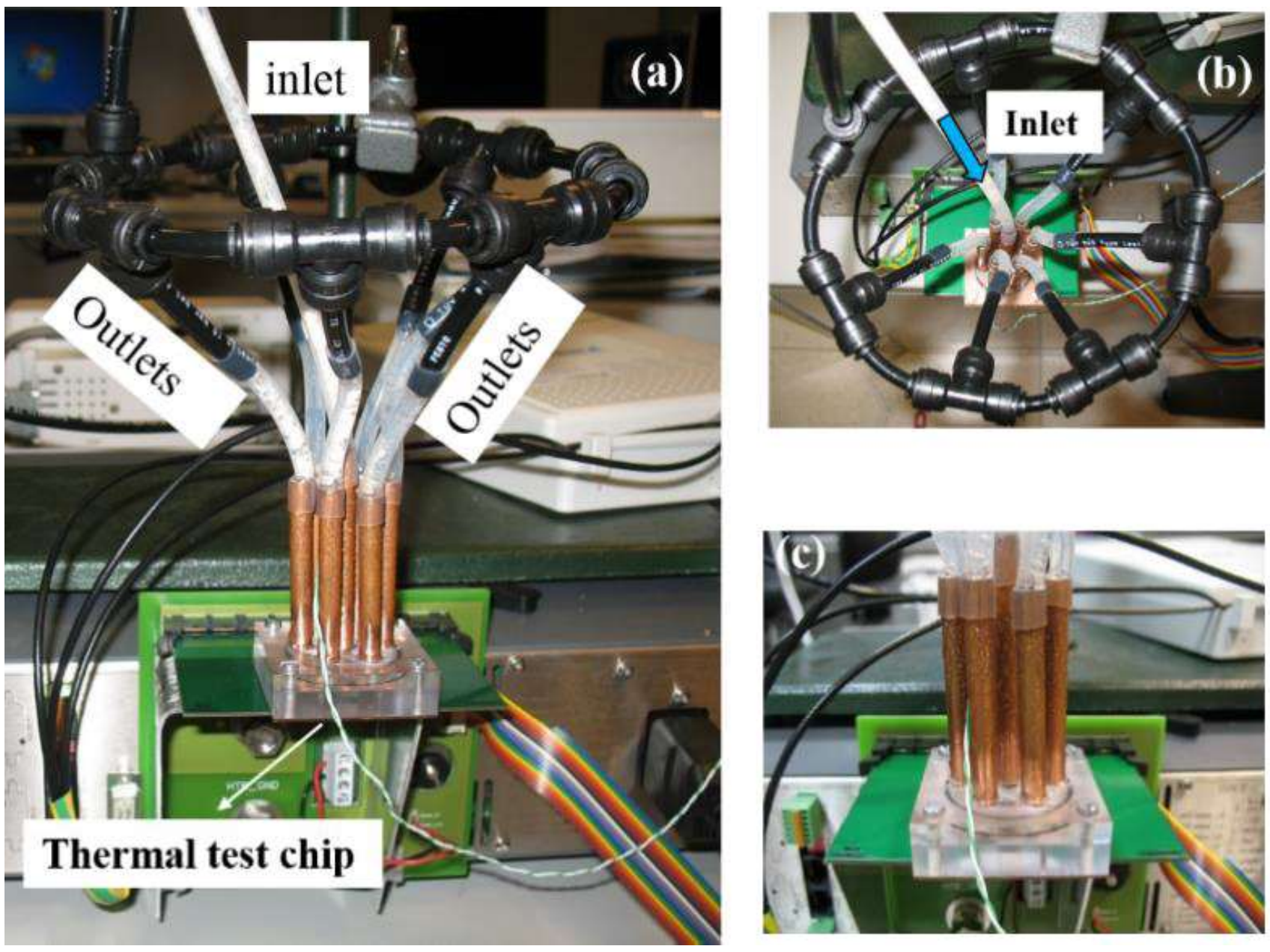

**Figure 4.2:** Demonstration of the chip level single impingement jet cooler: (a) experimental set-up photo of the single jet cooler; (b) top view of single inlet and six outlets; (c) side view of the cooler with different parts.

The single jet cooler demonstrator is fabricated in plexi-glass [1] with inlet and outlet tube diameters of 6 mm. The diameter of the inlet nozzle on top of the chip surface is 2 mm. The geometry parameters for the single jet cooler are listed in Table 4.1. The single jet cooler is mounted to the PTCQ thermal test chip assembly. The thermal test chip has programmable power dissipation and full chip temperature measurement, introduced in chapter 2. Based on this high-resolution thermal test chip, the demonstrator can capture the temperature distribution of the liquid impinging jet on the heated surface in detail. The final assembly of the single jet cooler on thermal test chip, as well as the outside tube connections, are shown in Figure 4.2. In this experimental set-up, copper tubes are used to connect the cooler to the flow loop system.

**Table 4.1:** Geometry parameters for single jet cooler

| Parameters | Single jet |
|---|---|
| N×N | 1 |
| $D_{i\text{-}tube}/D_{o\text{-}tube}$ | 6 mm / 6 mm |
| Nozzle diameter $d_i$ | 2 mm |
| $d_o$ | Common outlet |
| H | 2 mm |
| t | 7 mm |
| $t_c$ | 0.2 mm |
| L | 8 mm |

## 4.2 Modeling study of single jet cooler

### 4.2.1 Model description

In order to investigate the hydraulic and thermal phenomena in the cooler numerically, the full level CFD model for the single jet cooler is performed. Figure 4.3(a) shows the CFD model of the full cooler level model, indicating the inlet and six outlets. These simulations include the conduction and convection in the fluid domain for the coolant as well as the conduction in the solid domain. The solid domain includes the test chip, Cu pillars and underfills material, the package substrate, the solder balls, and the PCB, whereas the cooler material part is not included. The model dimensions of the test chip and package are summarized in Table 2.5, in chapter 2. For the boundary conditions of the single jet model shown in Figure 4.3(b), a constant velocity boundary condition is given at the top inlet feeding tube for flow rates between 200 and 600 mL/min. The boundary condition for the outlets is set as 'pressure out' with the shared outlet plenum. The inlet temperature is set to 10 °C. The heat flux boundary condition is applied to the locations that correspond to the activated heater cells in the test chip. All the boundary walls are set as an adiabatic wall since the cooler material is plexiglass with low thermal conductivity [1]. The detailed discussion about the cooler material impact and the justification for replacing the solid material of the cooler by an adiabatic boundary condition in the thermal model will be shown in chapter 5 for the test case of a 4×4 nozzle array multi-jet cooler.

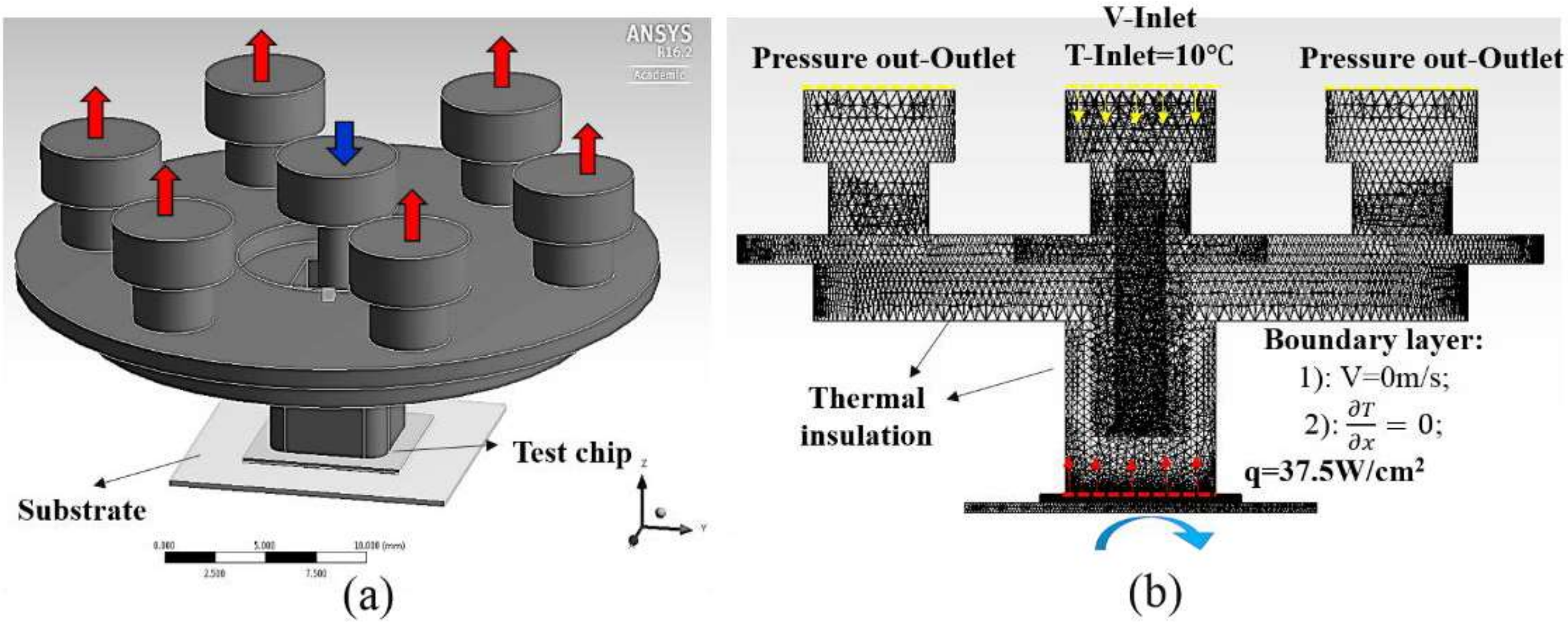

**Figure 4.3:** Full cooler level model: (a) CFD model of the single jet cooler and (b) the applied boundary conditions on the full single let cooler model.

**Table 4.2:** Grid convergence index analysis for the model of the single jet cooler

| Temperature | GCI12 | Asymptotic range of convergence |
|---|---|---|
| Stagnation Temp | 0.0019 | 0.9984 |
| Averaged Temp | 0.0043 | 1.0012 |

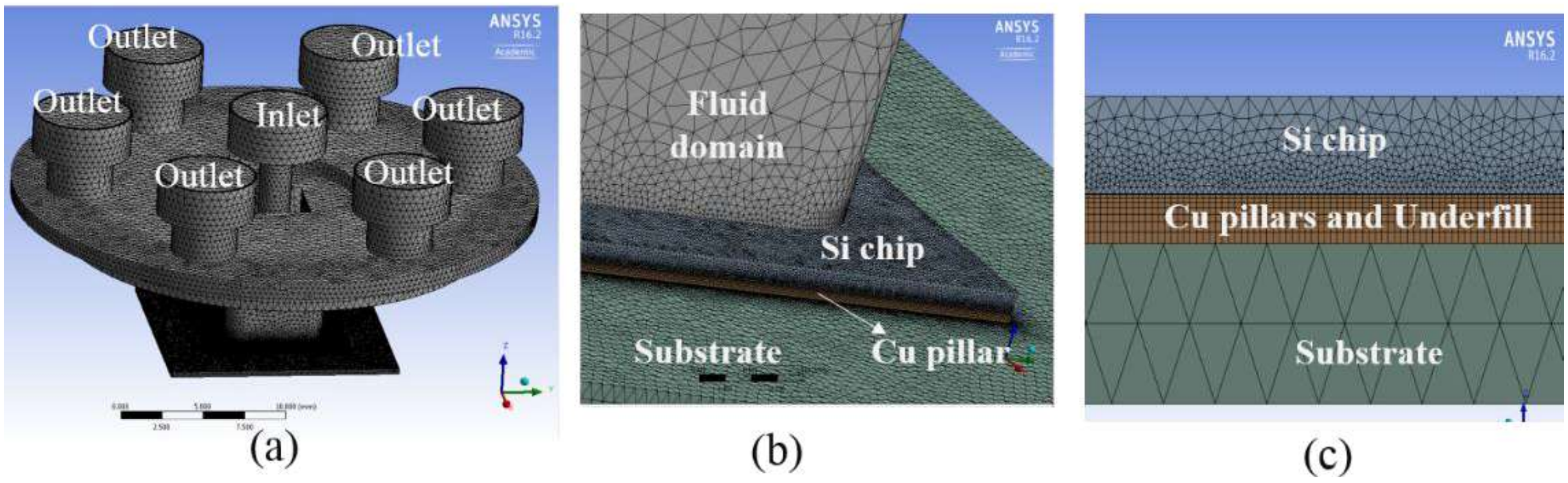

**Figure 4.4:** Meshing details of the full single jet cooler model: (a) fluid domain meshing with one inlet nozzle and 6 outlet nozzles; (b) and (c) details of the bottom package mesh including the Cu pillar and large size substrate.

The meshing and modeling methodologies for the full cooler level model are described in section 2.1.3 in chapter 2. For the single cooler model, a similar modeling methodology is used. As shown in Table 4.2, the grid convergence index analysis [2] for the model of the single jet cooler is listed. Based on the mesh sensitivity study, the number of elements for the full models is 2.5 million for the single jet model. Figure 4.4 shows the details of the single jet model, including the entire fluid/solid domain and bottom package.

### 4.2.2 Model simplifications

As illustrated in chapter 2, the thermal test chip package includes several parts, such as the Cu pillars and underfill material, the package substrate, the solder balls, and the large size PCB. The full cooler level model, including the bottom package, causes a high computation cost, especially for the large size PCB and heat sink. Therefore, it is necessary to investigate the thermal impact of the bottom package and simplify the model. In this part, the first step is to calibrate the bottom side boundary condition using the experimental data and to estimate the equivalent heat transfer coefficient applied to the substrate to represent the PCB and Al heat sink. The second step is to apply the equivalent convective boundary condition at the bottom of the Si substrate in the CFD model. The heat loss through the bottom package of the heat source will be investigated.

- **Step 1: Calibration of bottom side boundary condition using experimental data (No liquid cooling)**

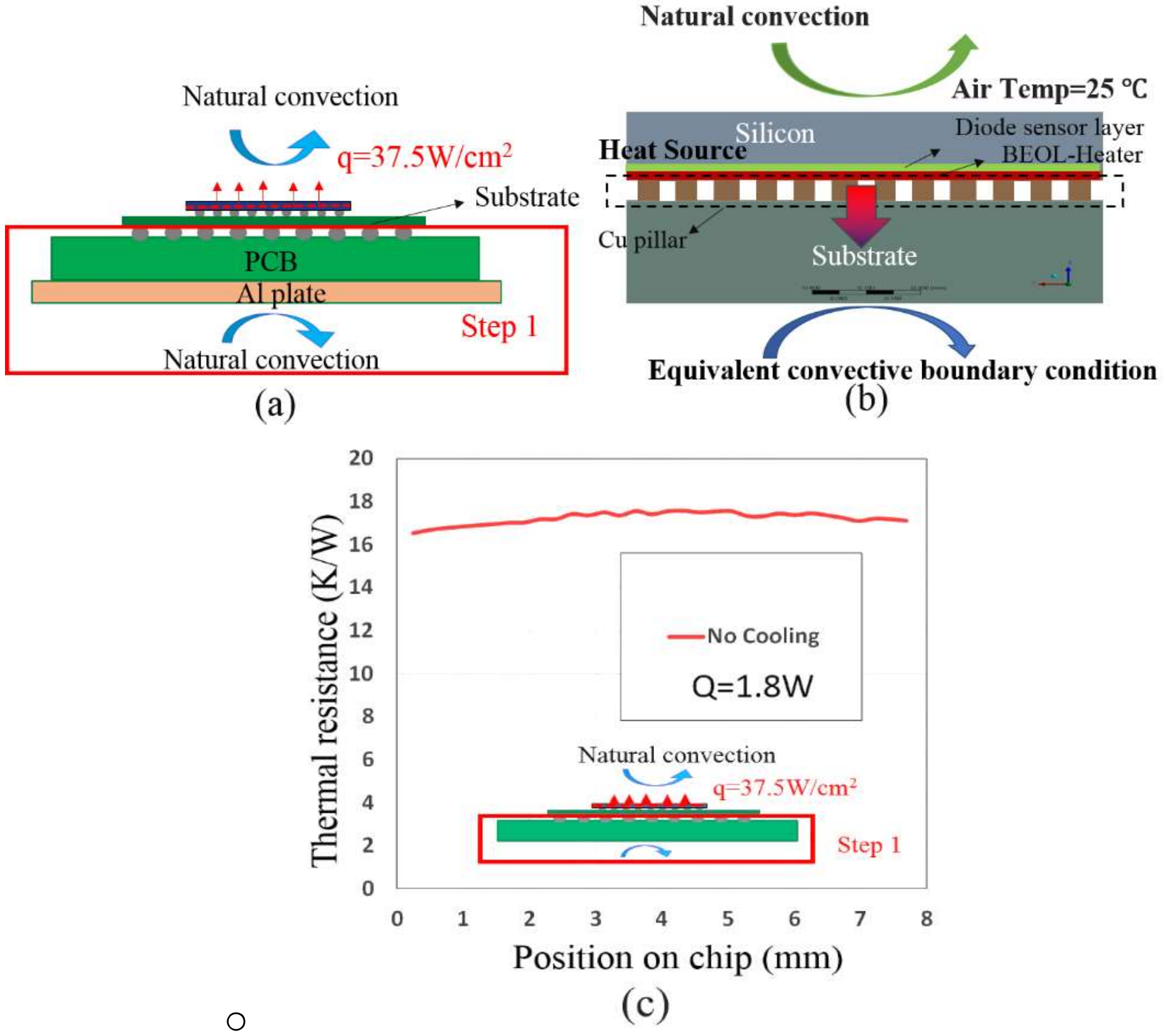

**Figure 4.5:** Boundary condition simplification: (a) thermal test chip package with PCB and heat sink; (b) equivalent convective boundary conditions applied at the bottom of the silicon substrate; (c) experimental measurement with the thermal test chip, at chip power of 1.8 W.

In the first step, the bottom side equivalent boundary condition will be extracted using an experimental analysis, where no liquid cooling is applied, as illustrated in Figure 4.5. The chip power is 1.8 W, and the heat is mainly removed through package substrate, PCB and Al heat sink. The measured data without liquid cooling is shown in Figure 4.5(c). The air temperature is measured at 25 °C . The natural convection heat transfer applied on top of the chip surface is estimated as 10 W/m$^2$K [3]. The equivalent heat transfer coefficient value for the PCB and Al heat sink, to match the measured temperature profile is 116.7 W/m$^2$K. It should be noted that this equivalent heat transfer coefficient is much smaller than the heat transfer coefficient for the impingement cooling on the chip (30,000 – 100,000 W/m$^2$K).

- **Step 2: Thermal insulation of chip back side**

For the model boundary conditions (B.C), the jet cooling is applied on the top of the chip surface, and the power is applied at the chip back side. Figure 4.6 shows two different assumptions for the model boundary condition simplification: (a) use equivalent B.C on the bottom of the package substrate to represent the large size PCB and Al heat sink; (b) use thermal insulation to replace the whole chip package including the silicon substrate, underfill, Cu pillar, large size PCB and Al heat sink.

Figure 4.7 shows the modeling results comparison of the chip temperature distribution with the bottom chip substrate and thermal insulation of the chip back side. The comparison shows that the bottom part of the package structure can be replaced with an insulation boundary condition to simplify the model, showing about a 3% difference.

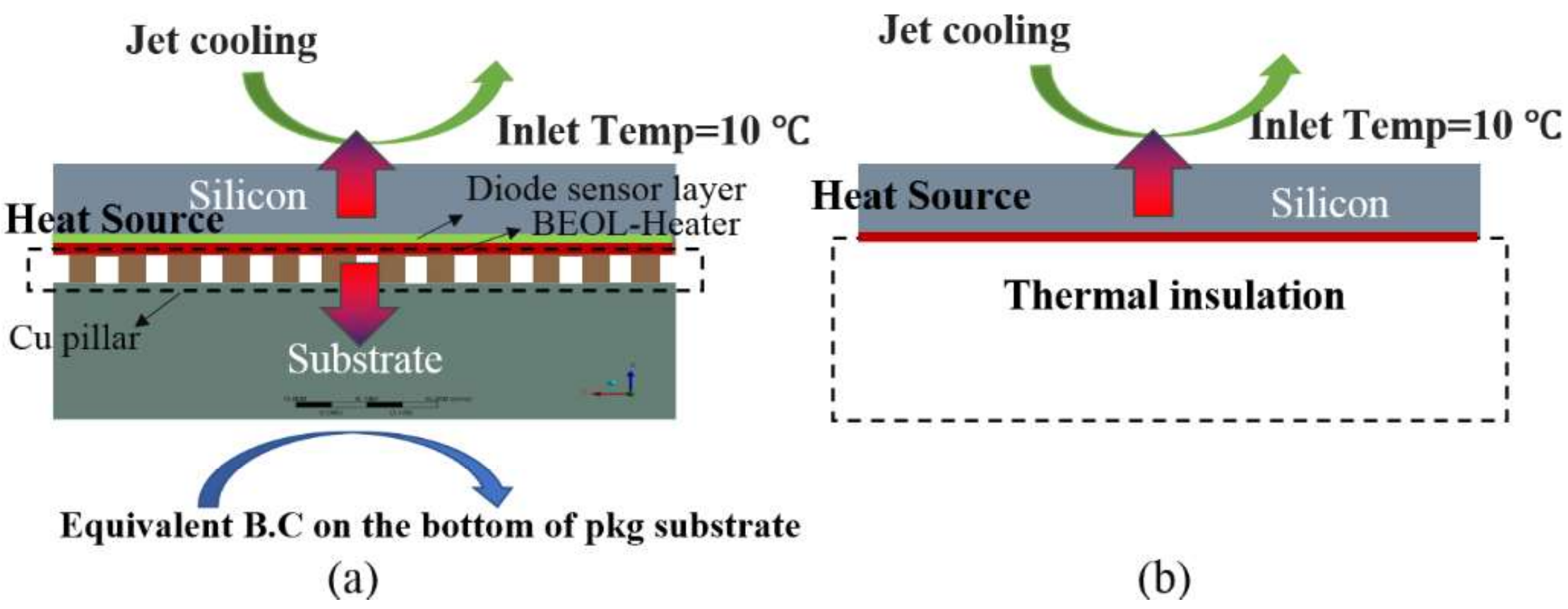

**Figure 4.6:** Boundary condition simplification: (a) silicon chip with the bottom packages; (b) thermal insulation on the silicon back side.

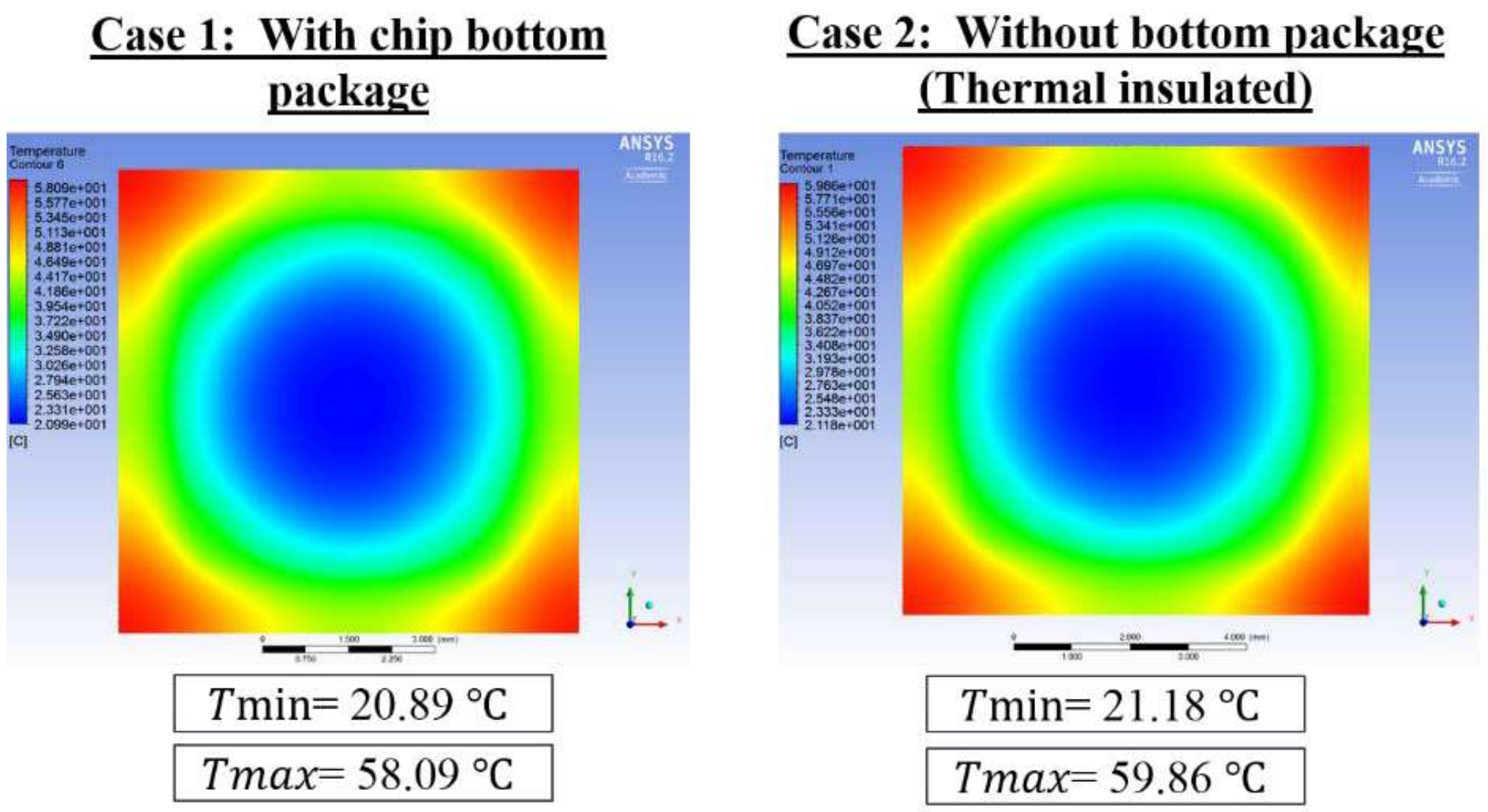

**Figure 4.7:** Comparison between the full cooler model with (a) bottom package or with (b) thermal insulation.

Moreover, the advanced thermal test chip, introduced in chapter 2, exhibits like a quasi-uniform power dissipation pattern, where the "heater cells" cover 75% of the chip area. However, including a large level of detail on the small heater cells in the CFD model will increase the number of elements, and therefore the computational cost. In section 4.3.2, the accuracy of the CFD model with quasi-uniform heating (75%), including the detailed location of all heater cells, and a model with uniform heating (100%) with the same total power, will be compared for different flow rates.

In summary, the critical simplification of the single jet CFD model covers the following two aspects:

- All the boundary walls are set as adiabatic wall since the cooler material is plastic with low thermal conductivity;
- The bottom part of the structure can be replaced with an insulation boundary condition to simplify the model;

### 4.2.3 Flow fields and thermal behavior

With the flow-thermal conjugate simulation for the single jet cooler model, the flow behavior in the cooler and temperature distribution in the chip can be studied. The first objective of this study is to identify the flow/thermal regions of the impingement jet cooling. In the considered experiment, the chip power is 24 W, and the coolant flow rate is 300 mL/min. Figure 4.8 shows the flow pattern in the fluid domain (arrows), including the stagnation region, the recirculation regions and the wall jet regions, and the temperature distribution in the Si chip [4]. As illustrated in Figure 4.8(b), the highest cooling efficiency is located at the stagnation point while the cooling performance decays along the wall jet region. Therefore, the higher temperature is discovered at the chip corner, resulting in a non-uniform chip temperature distribution.

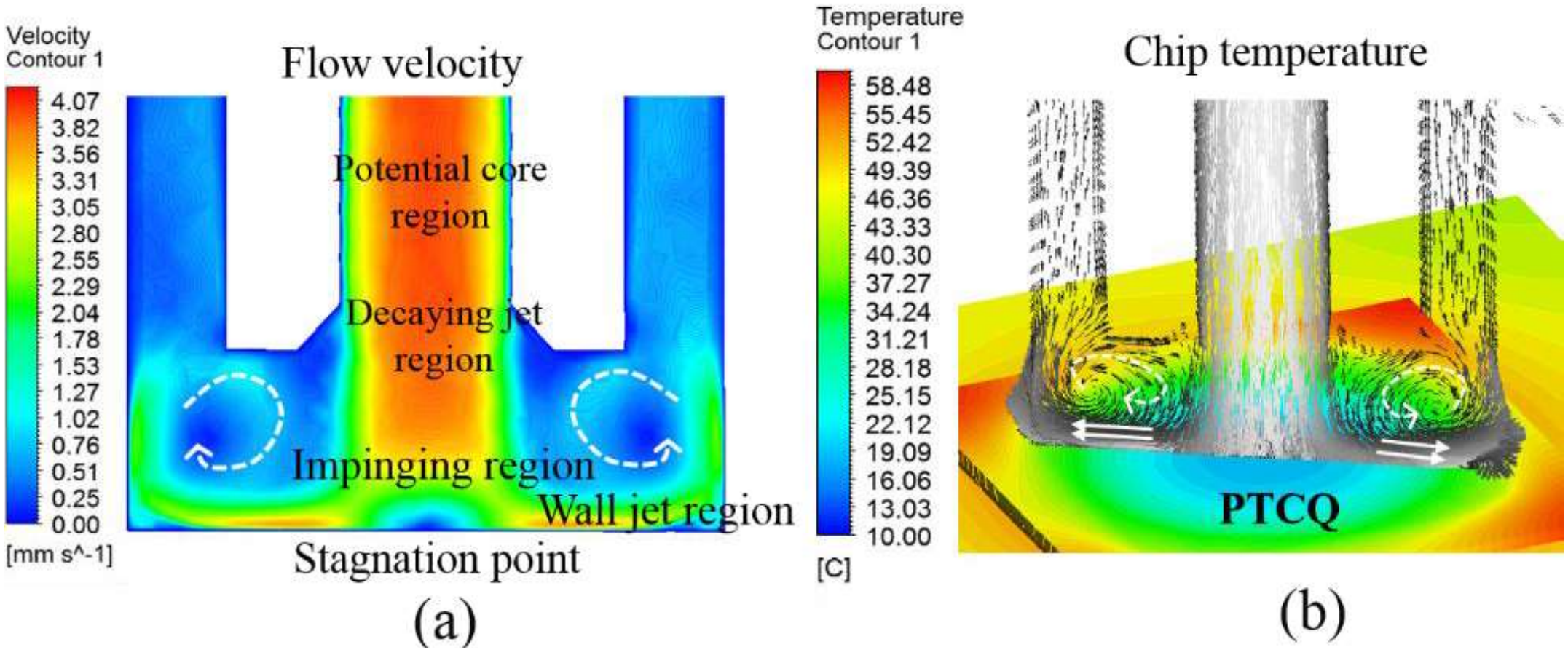

**Figure 4.8:** Single jet model validations: (a) Flow and thermal interactions (FL= 300 mL/min, chip power = 24 W); (b) Comparison of single jet modeling results and experiments data.

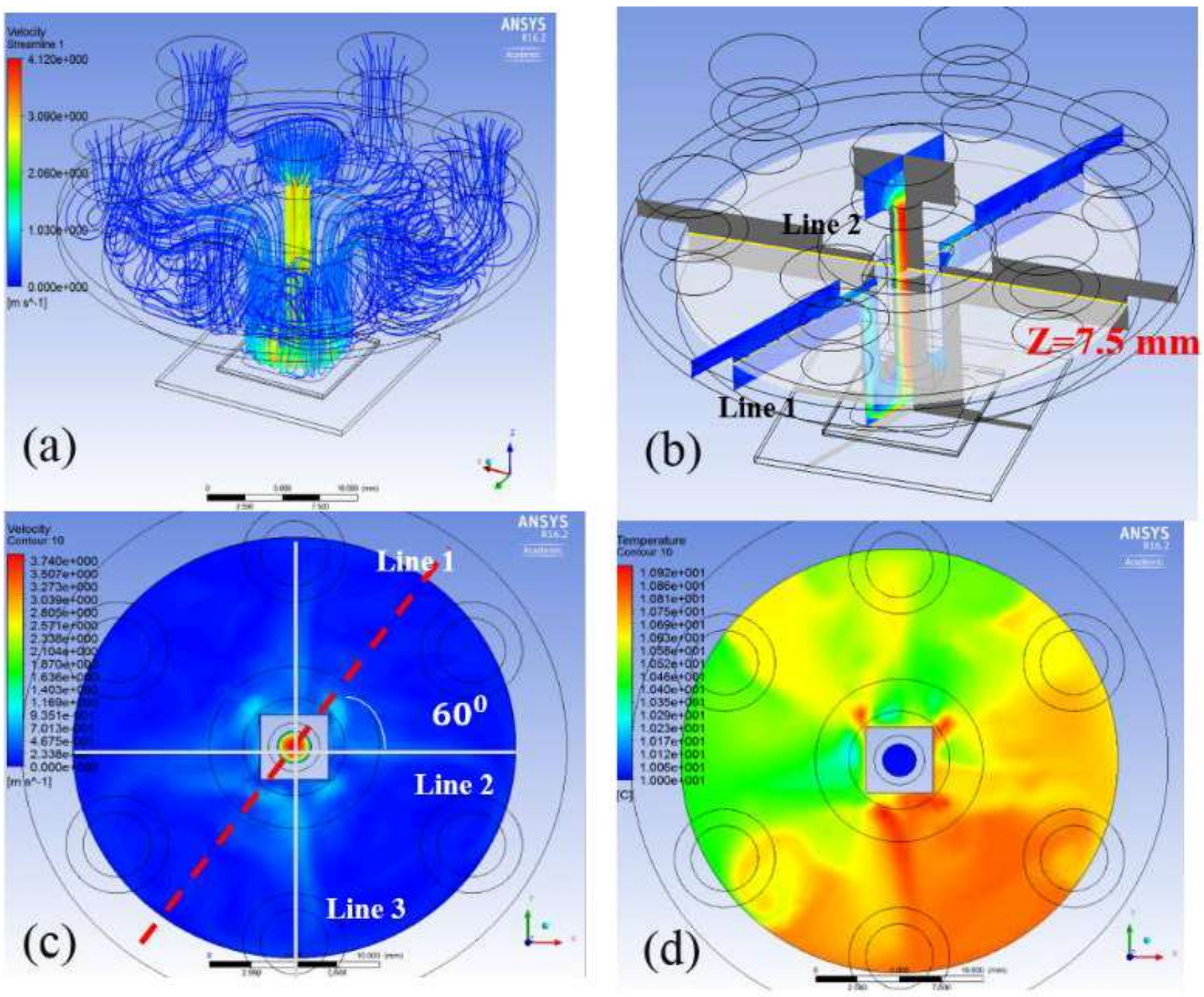

**Figure 4.9:** Investigation of the flow fields and thermal behavior: (a) flow streamlines inside the single jet cooler; (b) locations for the three studied profiles; (c) Flow and (d) temperature distributions in the outlet plenum at the location of Z=7.5 mm.

As shown in the top view of the single jet cooler, the inlet is located in the center of the cooler while the six outlets are distributed equally around the inlet. The second objective is to look at the symmetry or periodic behavior along with three studied profiles, as indicated in Figure 4.9(b). The profiles comparison of line 1 and line 2 between the nozzles are used to investigate symmetry or periodic behavior for a 1/6 model. The profiles comparison between line 2 and line 3 through the inlet/outlet nozzle are used to investigate the symmetry or periodic behaviors for the 1/4 model.

The velocity and temperature dirsubution inside the outlet plenum in the vertical direction of Z=7.5 mm is shown in Figure 4.9, where we can see the flow and temperature distributions at this level. The velocity and temperature profiles at Z=7.5 mm are plotted in Figure 4.10. At the Z direction with Z=7.5 mm, the velocity and temperature profile along the three studied lines shows good consistent, which means that the flow and temperature behaviors along the boundary of the studied lines are almost symmetric or periodic. It should be noted that the little asymmetric behavior is due to the mismatch of outlets and the square outlet chamber.

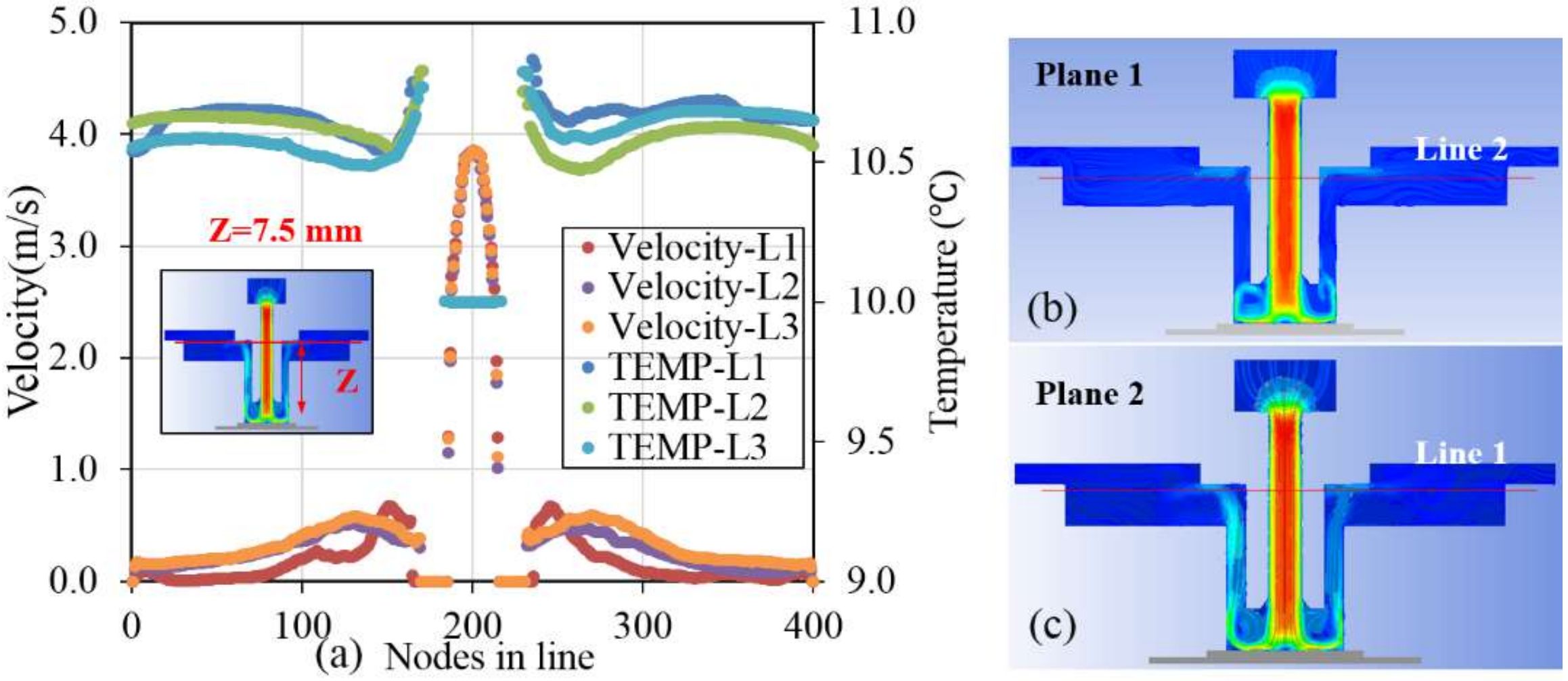

**Figure 4.10:** Investigation of the flow fields and thermal behavior: (a) locations for the plot; (b) Flow and temperature profile at the Z=7.5 mm.

Moreover, the temperature distribution profiles along the chip surface (Z=0 mm) with the three studied lines are shown in Figure 4.11. The flow shows similar behaviors for L2 and L3 in the wall jet region, and also the impinging region. Moreover, the temperature distribution along the wall jet region for every line shows symmetric behavior. It can be seen that the maximum chip temperature difference between L2 and L3 is 0.37 ℃, with no significant difference. This symmetry behavior shows that the 1/4 model and 1/6 model with symmetry boundary conditions can be used to simplify the full cooler model, even though there is an impact of the square chamber.

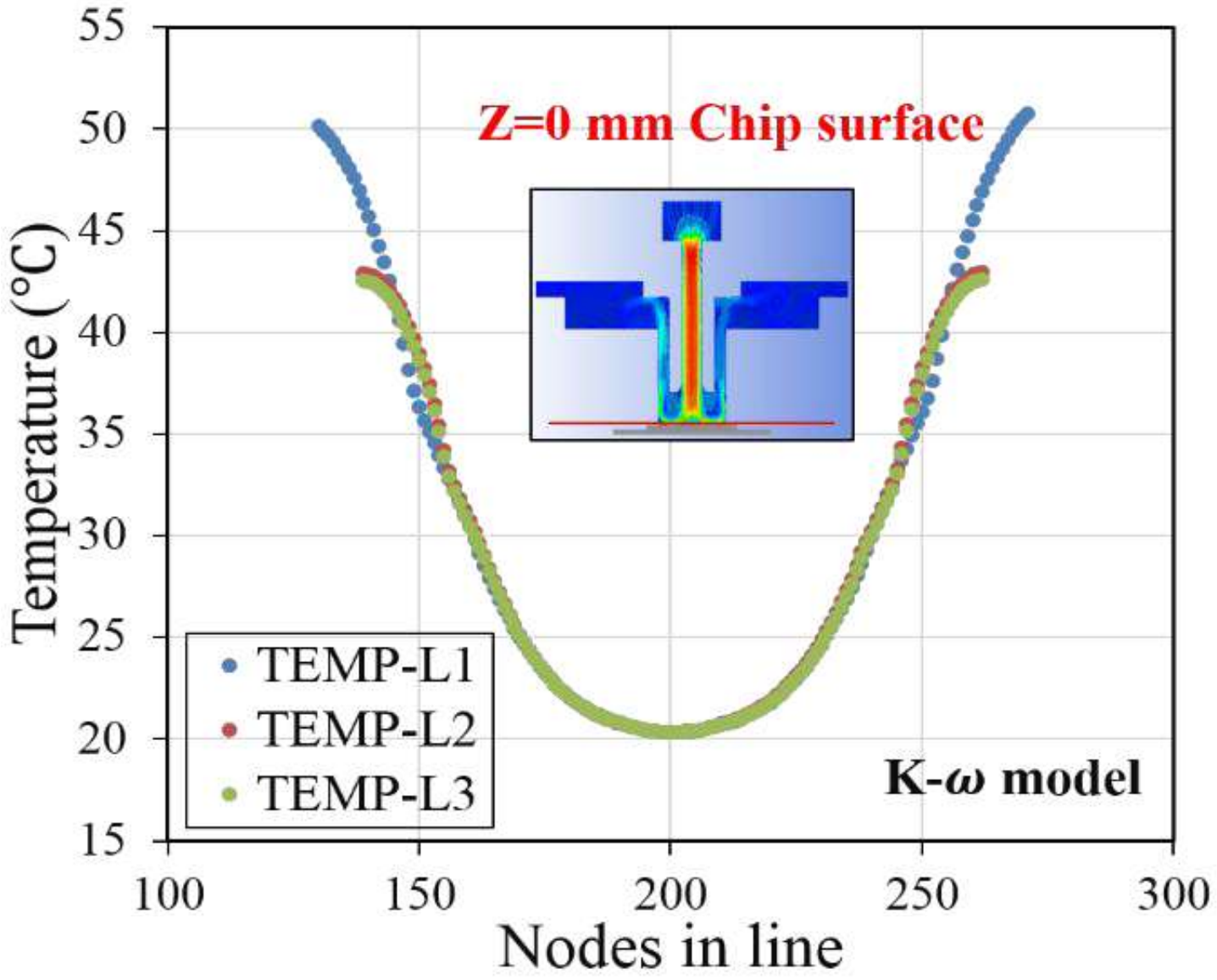

**Figure 4.11:** Temperature distribution profiles along the chip surface (Z=0 mm) with the three studied lines.

### 4.3 Experimental thermal validation

#### 4.3.1 Cooler thermal measurements

The dedicated experimental test set-up for the accurate flow and pressure measurements in the cooler and the temperature measurements in the test chip, introduced in chapter 2, is used for the thermal characterization of the single jet demonstrator.

A uniform power dissipation pattern is most suited to characterize the heat distribution map of a cooling solution since the impact of the thermal spreading in the Si chip is minimal. Figure 4.12(a) shows the temperature increase with regards to inlet temperature distribution map with single jet cooling for a flow rate of 530 mL/min. The detailed temperature map can be translated to heat transfer coefficient distribution based on the area of the heaters (8 mm × 8 mm ×75%), as shown in Figure 4.12(b) with a maximum heat transfer coefficient of 49000 W/($m^2$K) for a quasi-uniform heating of 24 W and a flow rate of 530 mL/min. The 240 µm resolution of the sensor array allows to accurately characterize the temperature profile below the liquid jet: the lowest temperature is observed in the stagnation region while the heat transfer decay along the wall jet region is also clearly visible. The measurement of the full-chip temperature distribution allows the evaluation of the maximum, the minimum, and the average temperature over the chip area, which exhibits a large temperature gradient in this case. Figure 4.12(c) presents the impact of the single jet cooler for different flow rates: the thermal resistance based on the average temperature improves by a factor of three when the flow is increased from 200 mL/min to 530 mL/min.

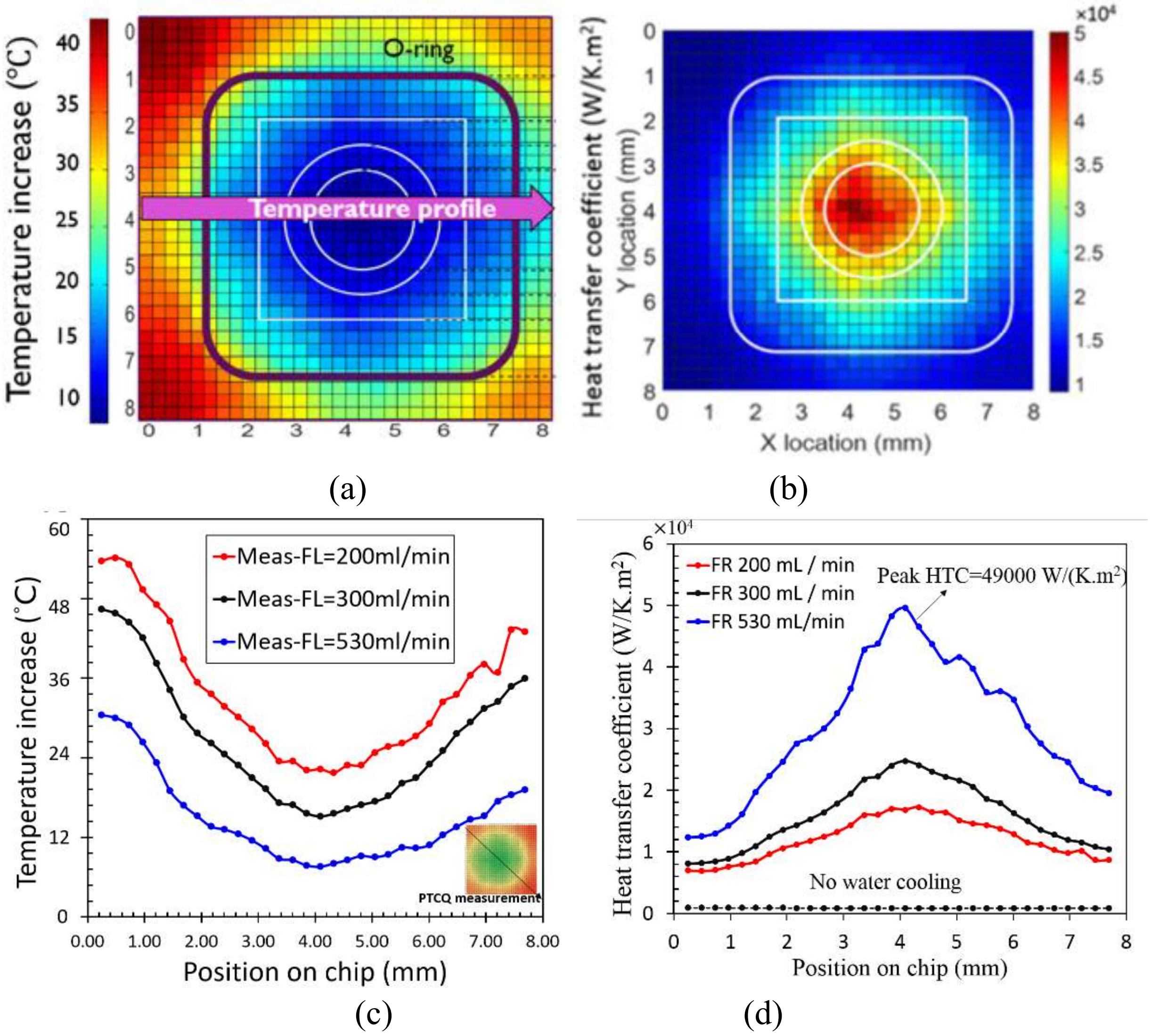

**Figure 4.12:** Single jet measurement with quasi-uniform heating: (a) Temperature increase with regard to inlet temperature map with single jet cooling (FL=530 mL/min) for 24W quasi-uniform heating; (b) Heat transfer coefficient map (FL=530 mL/min); (c) Thermal performance profile comparison for different flow rates; (d) Heat transfer coefficient profile comparison for different flow rates.

The thermal performance of the coolers can also be expressed in terms of the Nusselt number $\overline{\mathrm{Nu}}_{\mathrm{j}}$ and the Reynolds number $\mathrm{Re}_{\mathrm{d}}$, based on the nozzle diameter as characteristic length, shown in Figure 4.13. The following correlation can be extracted for the single jet cooler based on the experimental data:

$$\text{Single jet: } \overline{\mathrm{Nu}}_{\mathrm{j}} = 0.54 Re_d^{0.56} \text{ (experimental data)}$$

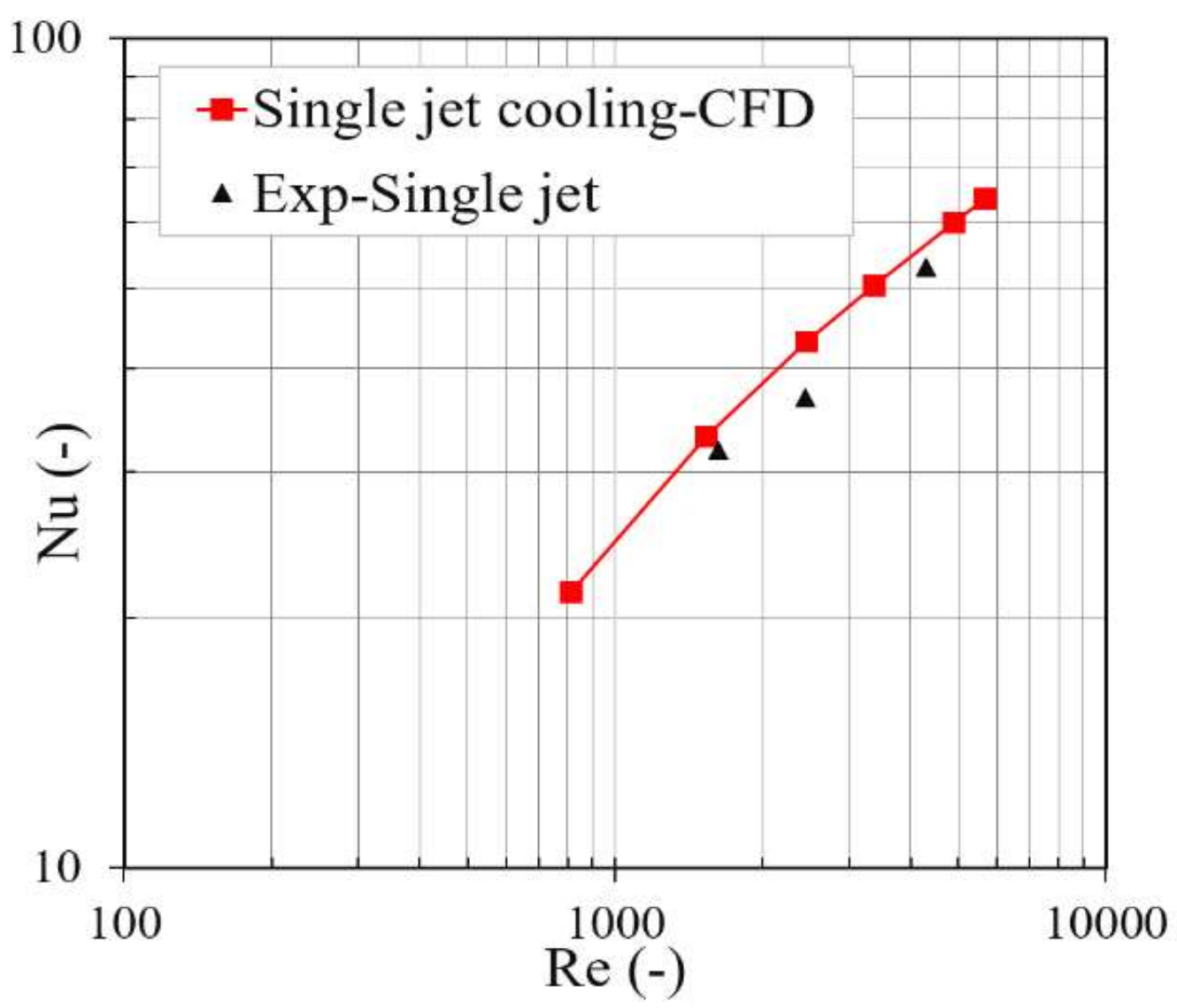

**Figure 4.13:** Correlation between Nusselt number and Reynolds number for the single jet with CFD modeling and experimental characterization.

### 4.3.2 Experimental model validation

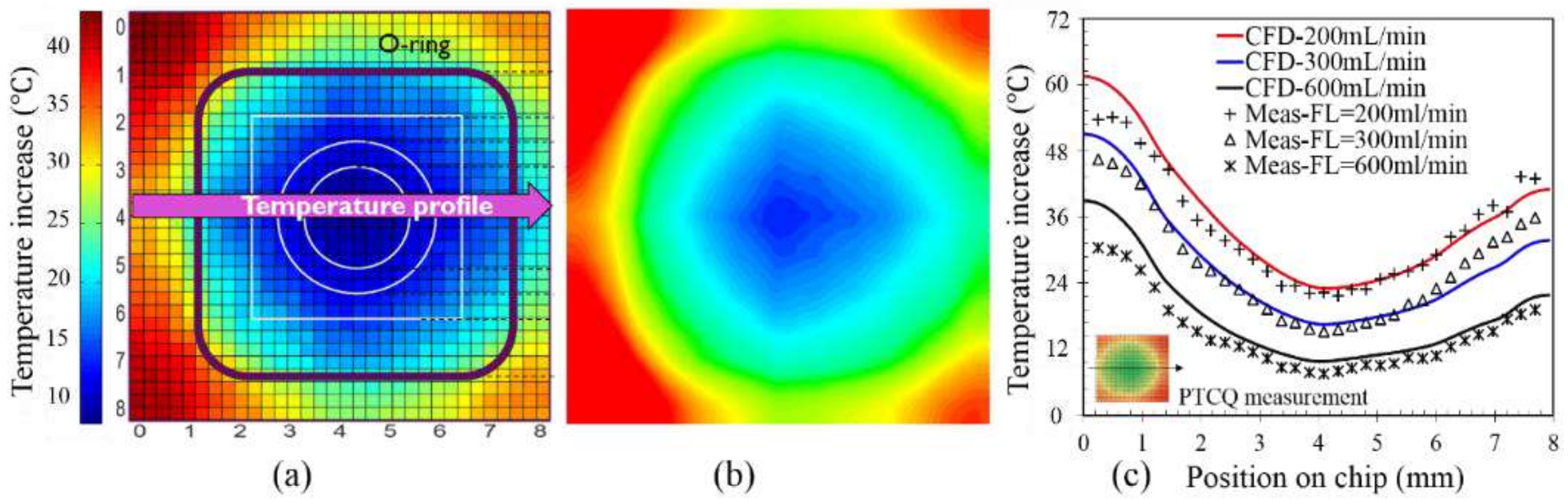

**Figure 4.14:** Single jet model validations (flow rate= 600 mL/min, chip power = 24 W): (a) temperature measurement results; (b) CFD modeling results; (c) comparison of single jet modeling results and experiments data.

In this section, the measurement results and the CFD modeling results will be compared and discussed. The measured full-chip temperature distribution is shown in Figure 4.14(a) with respect to the chip thickness of 0.2 mm. The measurement data show an asymmetrical temperature profile, which is caused by the misalignment of the cooler assembly. The same misalignment between the cooler and the test chip center of 0.24 mm has been included in the full cooler level CFD model of the single jet cooler. In general, the temperature distribution map of CFD modeling shows a good agreement with the experimental measurement. As illustrated in Figure 4.14(b), the CFD simulation results can accurately predict the stagnation temperature below the jet as well as the temperature increase along the wall jet region for the different flow rates ($Re_d$=4286 for 600 mL/min). In Figure 4.14(c), the temperature profiles from the test chip are compared for the CFD modeling results and the measurement data in the sensors for three different flow rates. It can be seen that the maximum errors for the stagnation temperature comparison between the single jet modeling results and experimental data are 3.9% (200 mL/min), 8.8% (300mL/min) and 10% (600mL/min), while the maximum errors are 13.4% (200 mL/min), 8.7% (300 mL/min) and 25.2% (600 mL/min) respectively at the chip edges. In summary, there is a good agreement between the experimental results and CFD modeling results, especially in the chip center, which is the most important region for the cooling.

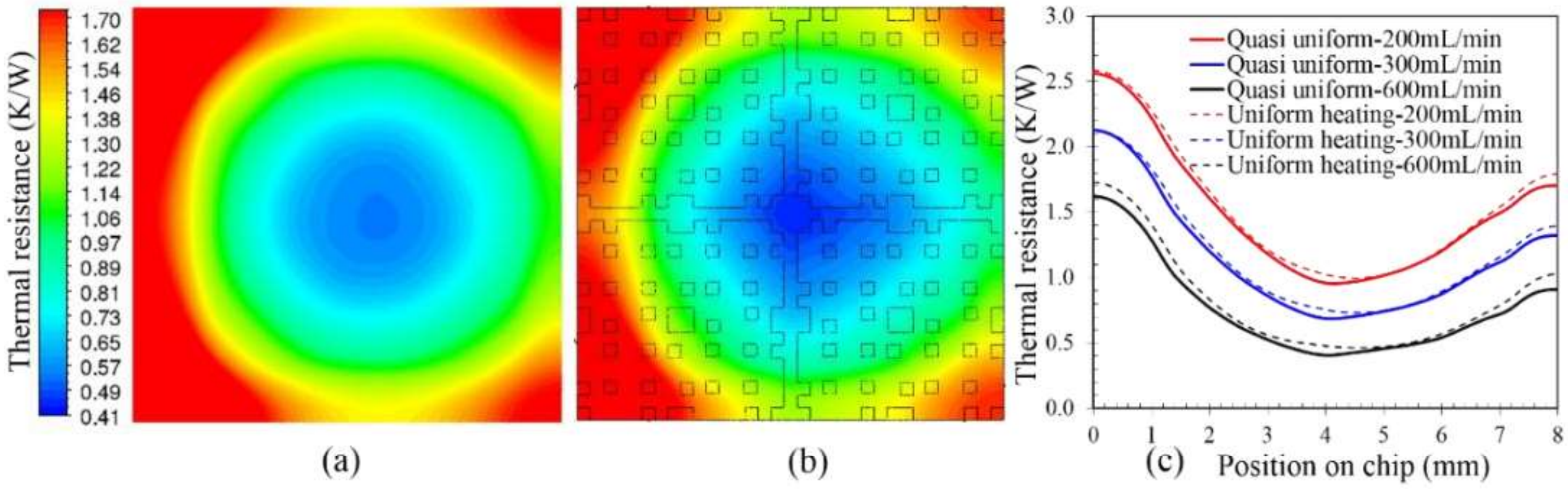

**Figure 4.15:** Comparison of the simulated thermal resistance distribution for single jet cooling with different heating configurations: (a) uniform heating modeling results for 600 mL/min; (b) quasi-uniform heating modeling results for 600 mL/min; (c) diagonal profile comparison between uniform heating and quasi-uniform heating (chip power=24 W).

Including a large level of detail on the small heater cells in the CFD model will increase the number of elements, and therefore the computational cost. The number of the meshing elements with quasi-uniform heaters is about 3,560,514, while the meshing number is 3510797 for a uniform heating model. In this section, the accuracy of the CFD model with quasi-uniform heating (75%), including the detailed location of all heater cells or uniform heating (100%) with the same total power, will be compared for different flow rates. In Figure 4.15, the comparison of thermal resistance distributions with uniform heating shown in Figure 4.15(a) and quasi-uniform heating in Figure 4.15(b) is shown for the single jet cooler case. For this moderate cooling condition, the introduction of the heater details in the model does not have a significant impact on the temperature distribution. The profiles for the uniform and quasi-uniform heating shown in Figure 4.15(c) are very similar, with only local differences of a 14.6 % and 6.7% at the locations where no heaters are present for the flow rates of 600 mL/min and 200 mL/min respectively. The comparison of the detailed temperature map measurements with the CFD modeling results indicates that for lower heat removal rates with the single jet cooler, a simpler model with uniform heating can be sufficient.

### 4.4 Conclusion

In this chapter, a single jet demonstrator is designed and fabricated as a proof of concept, for the fundamental understanding of the impingement jet cooling, from numerical modeling and experimental characterization aspects. The demonstration of the large single jet cooler on the chip size of 8×8 $mm^2$ with 32×32 temperature sensors allows to accurately capture the chip temperature distributions caused by the local cooling of the liquid impinging jet on the heated surface. For the modeling of the full cooler level single jet cooler, there is a good agreement between the experimental results and CFD modeling results. Moreover, the comparison of the detailed temperature map measurements with the CFD modeling results, indicates that for lower heat removal rates with the single jet cooler, a simpler model with uniform heating can be sufficient.

The results of this chapter are partially published in the following publication:

**Tiwei Wei**, Herman Oprins, et al., "Experimental characterization and model validation of liquid jet impingement cooling using a high spatial resolution and programmable thermal test chip [J]", Applied thermal engineering, 2019.

### References

[1] http://www.matweb.com/search/datasheet.aspx?bassnum=O1303&ckck=1

[2] Christopher J. Roy, Grid Convergence Error Analysis for Mixed-Order Numerical Schemes, AIAA Journal, 2003.

[3] Remsburg R. (1998) Convection Heat Transfer in Electronic Equipment. In: Advanced Thermal Design of Electronic Equipment. Springer, Boston, MA

[4] N. Zuckerman , N. Lior, "Jet Impingement Heat Transfer: Physics, Correlations, and Numerical Modeling", Advances in Heat Transfer, Volume 39, 2006, Pages 565-631.

# Chapter 5

# 5. Multi-jet Impingement Cooling: Proof of Concept

## 5.1 Design considerations for multi-jet coolers

In chapter 4, the single jet liquid impingement cooler is demonstrated and experimentally characterized to fundamentally understand the flow/thermal behavior of the jet impingement cooling. For the improvement of the jet impingement cooling design, it is observed that a nozzle array with higher numbers of nozzles can achieve lower thermal resistance and lower pumping power, as discussed in chapter 3 with the parametric and dimensionless analysis. However, for an N×N array, the performance will saturate beyond a certain number of N, under the assumption of a fixed cavity height. Figure 5.1 shows an example of the analysis for chapter 3 on the trend for the increasing of nozzle density, for a constant pumping power, where the inlet diameter ratio is kept as $d_i$/L=0.3. Therefore, multi-jet cooling is much more energy efficient than single jet cooling. In this chapter, the concept of multi-jet cooling will be demonstrated as a proof of concept to prove the improved energy efficiency of the multi-jet cooling compared to single jet cooling.

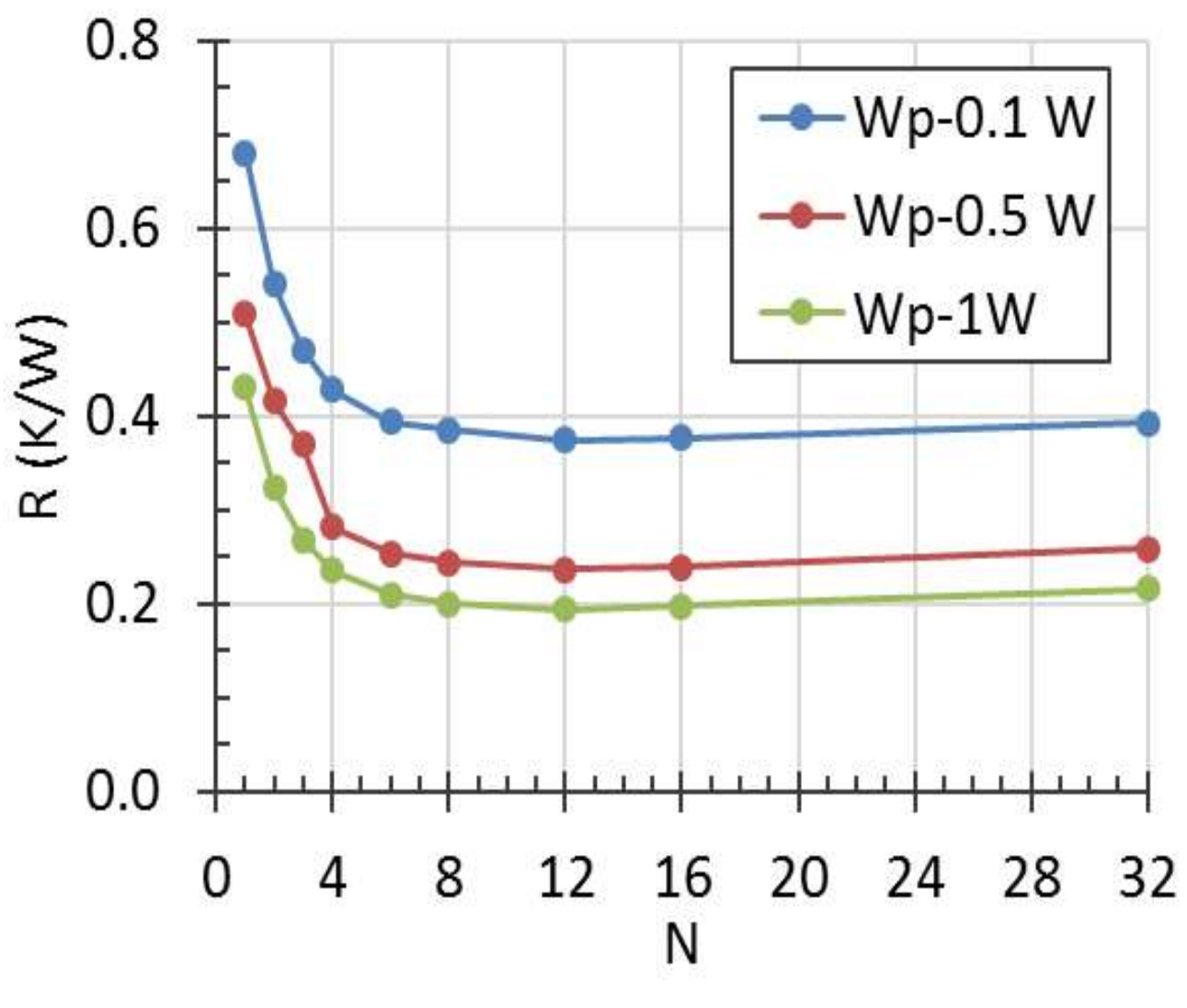

**Figure 5.1:** Saturation of the thermal resistance with the increasing of nozzle number for different constant values of the pumping power ($d_i$/L = 0.3).

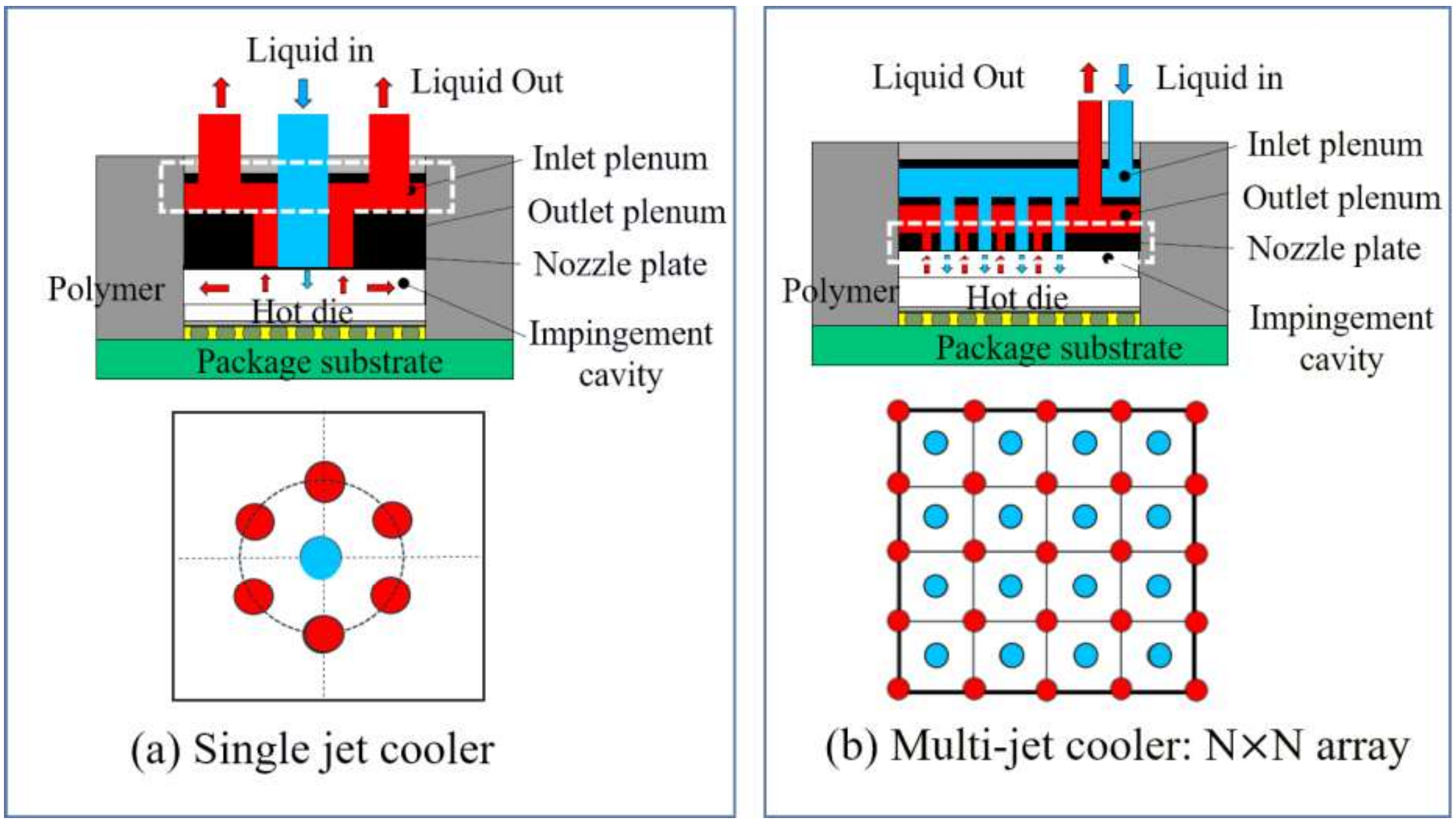

**Figure 5.2:** Schematic comparison of the (a) single jet cooling and (b) multi-jet cooling.

The comparison between the single jet cooling and the multi-jet cooling concept is illustrated in Figure 5.2. From the top view, the multi-jet cooler can be regarded as a scalable system with an N×N nozzle array, while the single jet cooler is a single unit. The functional layers for both the configurations are the same, including the cavity height, nozzle plate, inlet/outlet nozzles, and inlet/outlet plenum.

The choice of the number of nozzles and the nozzle diameter will have an impact on the required fabrication technology. Larger nozzle diameters will allow low-cost fabrication techniques, while finer nozzle diameters require more expensive processing options. Figure 5.3 shows the link between the inlet/outlet nozzle diameter and the nozzle number for the 8×8 $mm^2$ chip footprint and different inlet diameter ratios $d_i$/L. In the chart, the applicable range is indicated for three fabrication technologies: mechanical machining, 3D printing, and Si processing. Mechanical micro-machining, especially micro-milling, can be used to produce micro-features [1]. However, it is difficult to mill complex shape structures like cavities. Silicon processing has the advantage of fabricating small diameter holes below 10 μm with Deep Reactive Ion Etching (DRIE) technology. However, the cost of silicon processing is higher than the other fabrication methods. Furthermore, the modeling study showed that aggressive scaling of the nozzle diameter is not required due to the saturation of the thermal performance. The required optimal diameter for the considered structures is in the order of 100 μm to several hundred μm, which is discussed in chapter 3. Thanks to the advancements in recent years, 3D printing can be an interesting fabrication option to fabricate these structures with nozzle diameters ranging from 100 µm to 1 mm [3] and small cavities.

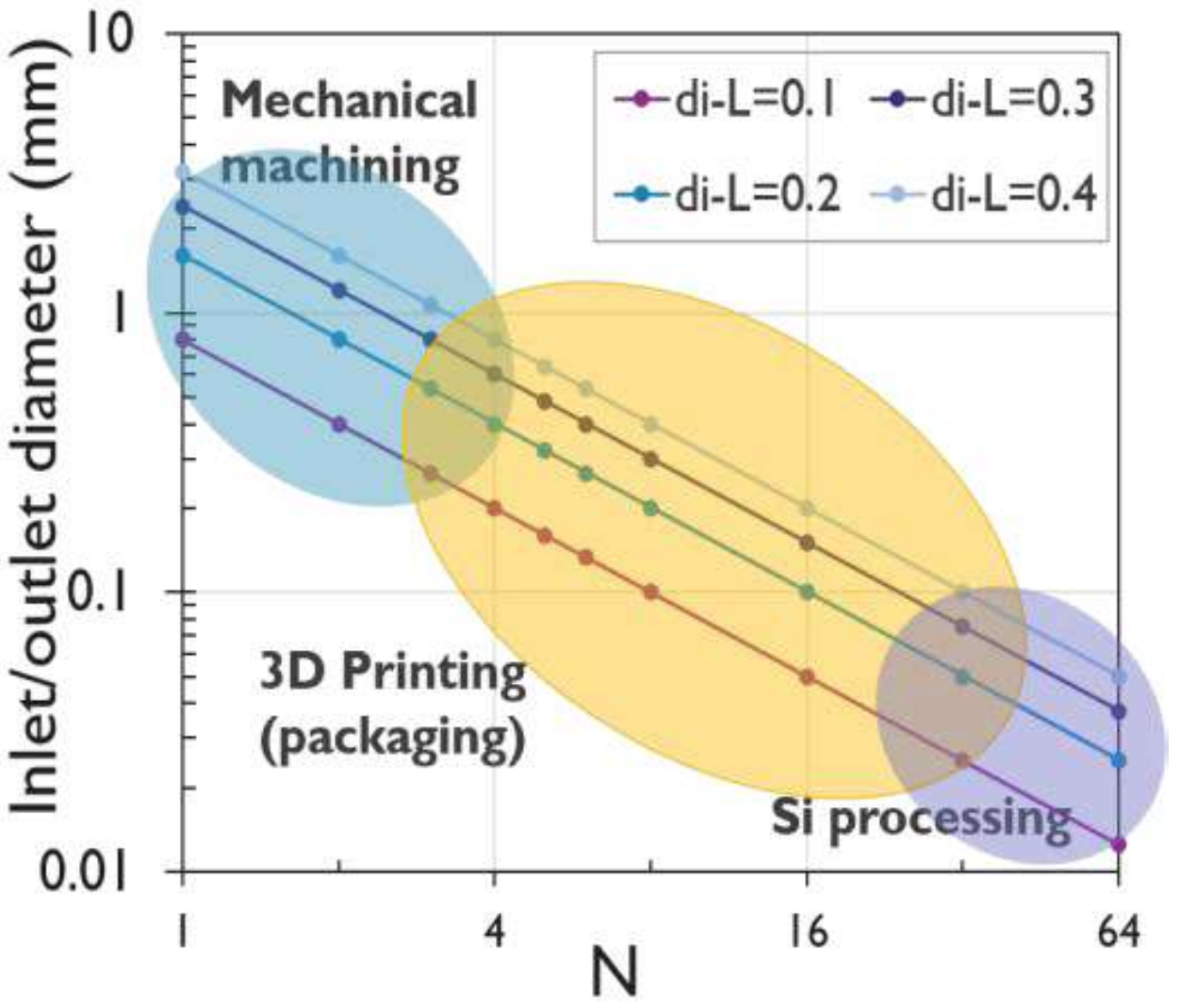

| N×N | Inlet/Outlet diameter | Manufacturing technology |
| --- | --- | --- |
| 1 | 2.4 mm | Mechanical machining process |
| 2×2 | 1.2 mm | |
| 4×4 | 600 µm | |
| 6×6 | 400 µm | Injection molding and 3D Printing and drilling **(low cost)** |
| 8×8 | 300 µm | |
| 16×16 | 150 µm | |
| 32× 32 | 75 µm | Silicon processing **(expensive)** |
| 64×64 | 38 µm | |

**Figure 5.3:** Link between the nozzle geometry and the fabrication technology options: mechanical machining process, 3D printing, and Silicon processing (N depends on the chip size).

In the first step, mechanical micromachining will be used to demonstrate the multi-jet cooler with a 4×4 nozzle array, as a proof of concept. In chapter 3, the geometrical parameter analysis and dimensionless analysis are both discussed using the unit cell model. In this chapter, the general considerations of the multi-jet cooling with a 4×4 nozzle array are discussed, covering two different aspects: geometrical aspects and material aspects. For the geometrical aspect, the impact of the cavity height and inlet plenum from the full cooler level model are discussed. For the material point of view, the thermal impact of the cooler material and liquid coolant will be investigated as the design guideline for the cooler fabrication. The investigations will be conducted based on the CFD model introduced in chapter 2.

### 5.1.1 Cavity height effects

For the assembly of the polymer-based cooler on the electronic devices, there are several bonding solutions such as thermal bonding, mechanical clamping, or adhesive bonding [4,5,6]. The choice of the cavity height can determine the assembly method of the cooler on the substrate. As discussed in chapter 3, the unit cell modeling results show that the heat transfer coefficient and the pressure drop are inversely proportional to the gap height in the pinch-off regime while the heat-transfer coefficient is constant in the impingement regime. In real applications, the cavity height is also limited by the whole cooler thickness and the package size. In this study, the cavity height range is chosen from 0.2 mm to 1.6 mm. Figure 5.4 shows the impact of the cavity height on the thermal resistance and pressure drop, based on the full cooler level CFD modeling. It is observed that the impact of the cavity height on the thermal resistance is negligible as the H is between 0.2 mm and 1.6 mm. The comparison between the full cooler model and unit cell model in Figure 5.4(a) shows good agreement. On the other hand, the impact on the pressure drop is also minimal, as the H is beyond 0.4 mm. In summary, the impact of the cavity height on the thermal and pressure drop can be negligible when the cavity height is higher than 0.5 mm. Therefore, mechanical clamping with a stand-off height of 0.6 mm is used in this study. The dimensions of the O-ring and groove can be determined to fit with this cavity height.

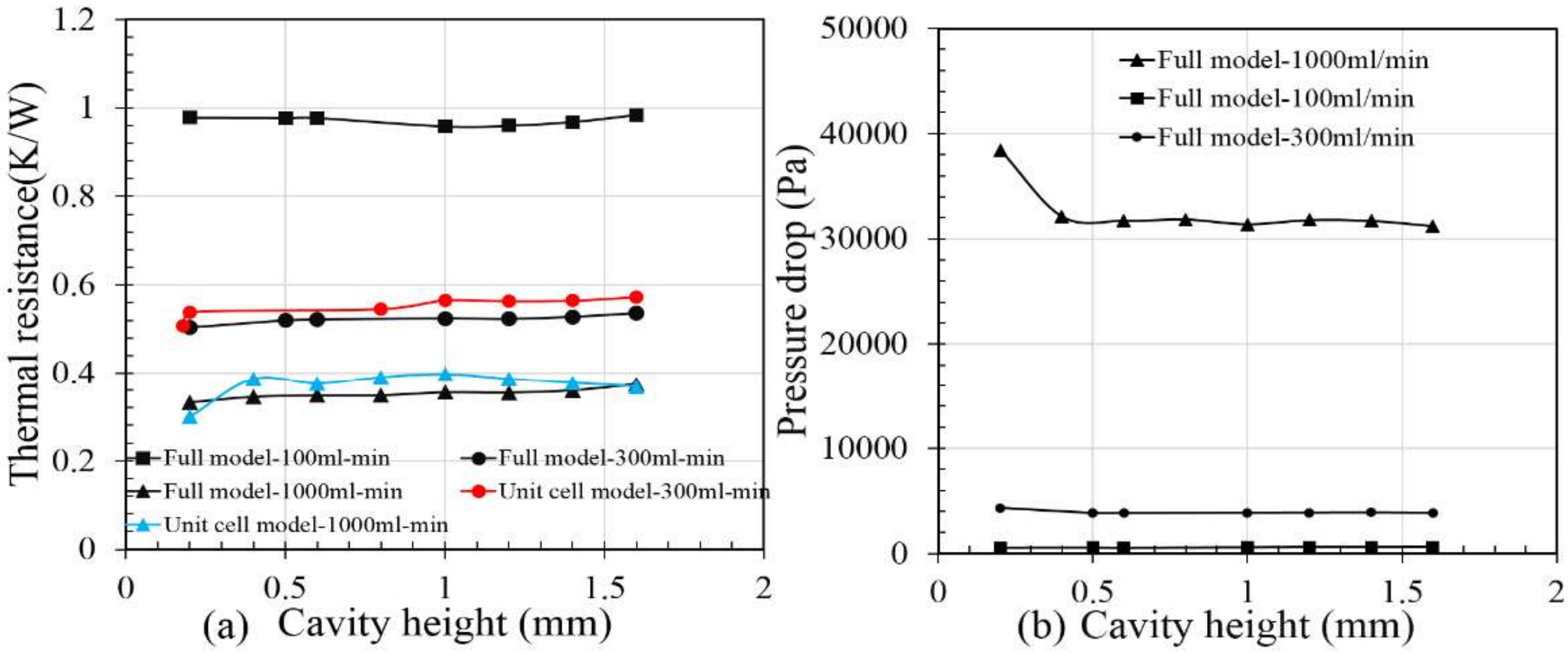

**Figure 5.4:** Impact of the cavity height on the (a) thermal resistance and (b) pressure drop (example for 4×4 nozzle array).

### 5.1.2 Impact of inlet plenum

As for the inlet plenum, the inlet flow comes directly from the top plenum, and the inlet tube is located above the center of the nozzle array. The flow distribution is determined by the pressure drop of the inlet plenum. When $\Delta p_{plenum}$ is smaller than the $\Delta p_{nozzle}$, the flow uniformity will be better. The impact of the inlet plenum thickness on the flow distribution in the nozzles of the 4×4 array cooler is shown in Figure 5.5 for three thickness values. A thin plenum with 1 mm height generates a significant flow maldistribution of more than 25% with higher velocity concentrating in the nozzles in the center of the cooler. This indicates that it is essential to balance the inlet diameter and plenum height when designing the impingement cooler. For the thicker inlet plenum with 5 mm thickness, the flow distribution is much more uniform. However, the use of a thicker plenum increases the total cooler thickness. From the thermal point of view, the temperature influence of the inlet plenum height is smaller when it is above 3 mm. A higher heat transfer coefficient can be observed in the middle part, as illustrated in Figure 5.6. Based on the modeling study and cooler size constraint, the inlet plenum height is chosen as 3 mm in the cooler design.

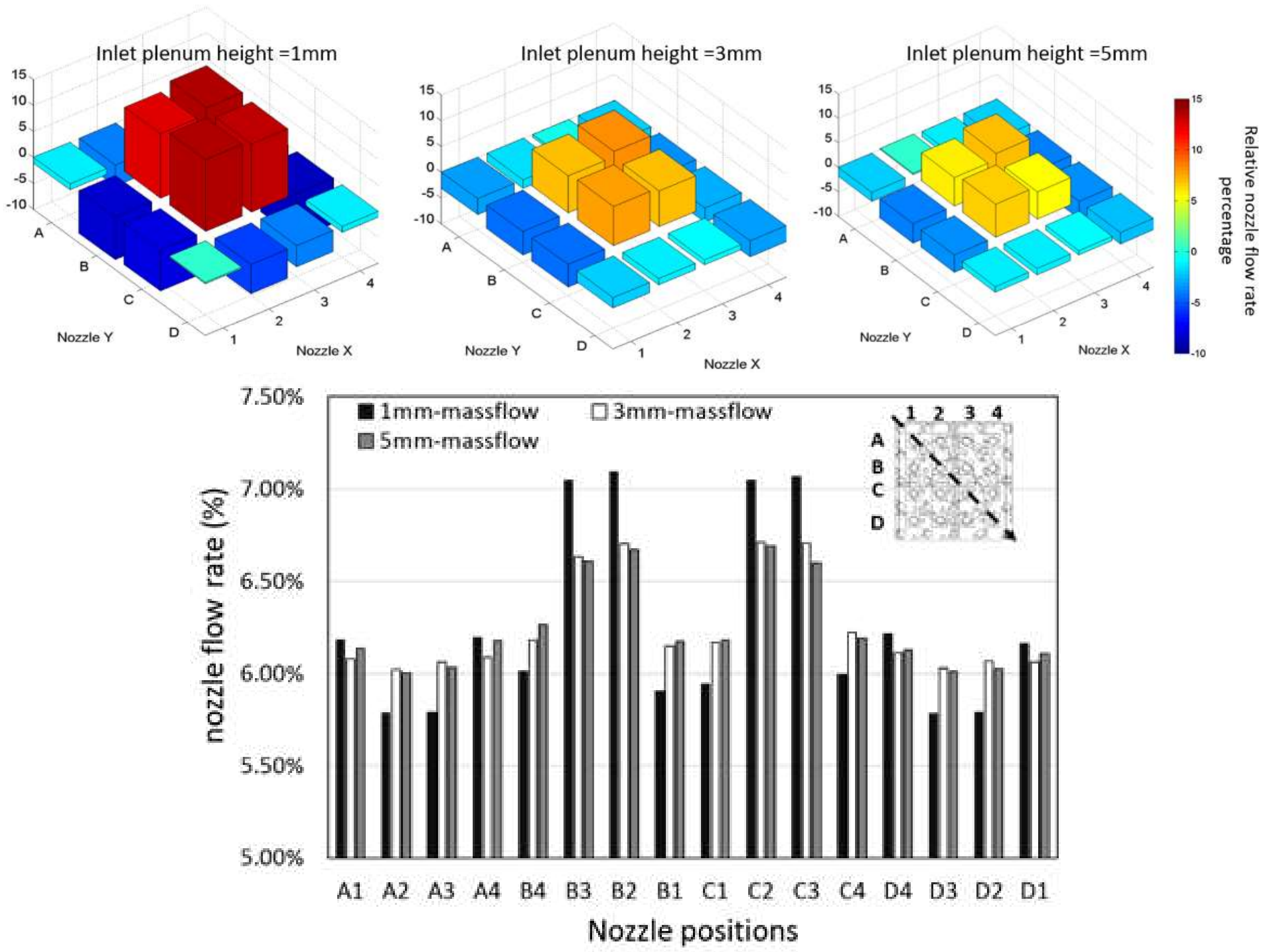

**Figure 5.5:** Impact of the plenum level thickness on the flow distribution in the 4×4 array of inlet nozzles: (top) relative nozzle flow rate percentage distributions; (bottom) plotted values are nozzle flow rate expressed as a percentile of total flow rate (100% for 16 outlets).

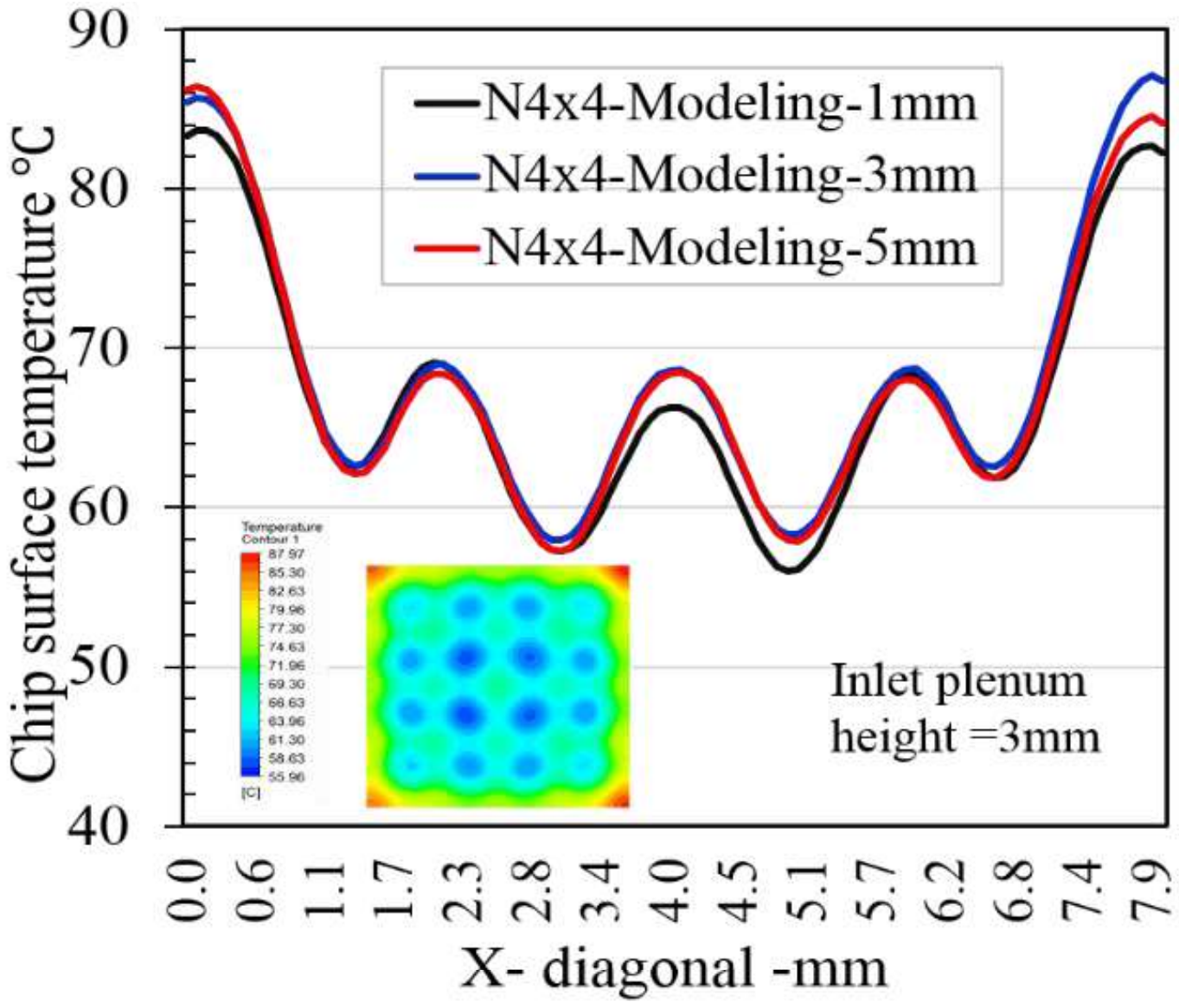

**Figure 5.6:** Impact of the plenum level thickness on the temperature profile for three different plenum thickness.

### 5.1.3 Impact of cooler material

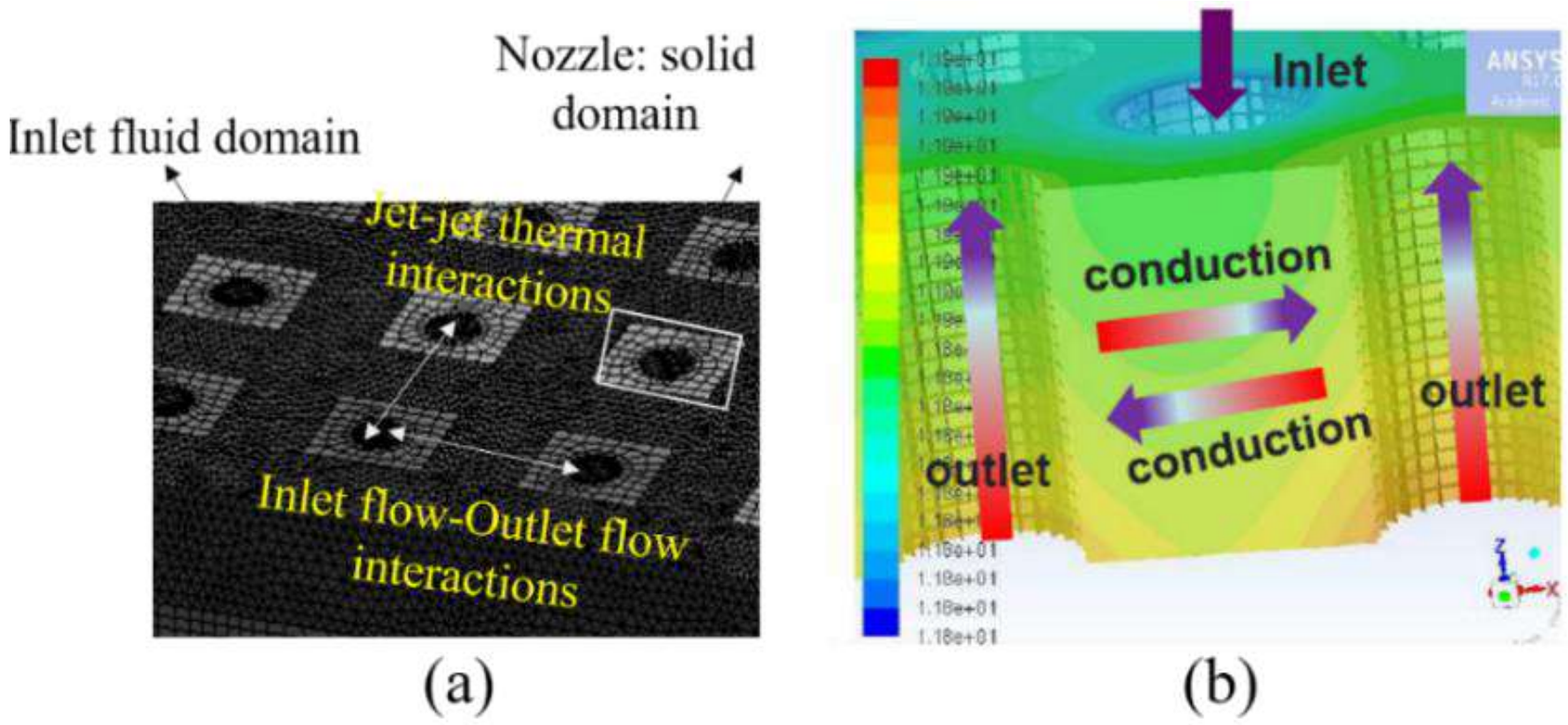

**Figure 5.7:** Full cooler CFD model with (a) meshing details of the fluid and solid domain, and (b) thermal interactions.

From a thermal point of view, the coolant in the inlet plenum can be heated up at a small flow rate due to the heat conduction between the hotter outlet flow and the cold inlet flow. The heat conduction through the cooler material is higher when the thermal conductivity of the cooler is larger. To study the impact on the chip temperature, the fluid and solid domain are both included in the CFD model for a large range of flow rates from 50 mL/min to 600 mL/min. The local mesh details of the jet-jet thermal interaction and inlet-outlet interactions are indicated in Figure 5.7. The considered materials are Cu (401 W/m-K), Si (1484 W/m-K) and polymer (0.2 W/m-K). The fluid domain contains the inlet water domain and outlet water domain. The temperature comparison shows that the impact of the cooler thermal conductivity on-chip temperature distribution can be neglected over a wide range of flow rates and chip power since the heat removal is dominated by the heat convection in the coolant. The trends, summarized in Figure 5.8, show that a polymer cooler has a similar performance as a Si or Cu cooler. The thermal resistance is defined based on the averaged chip temperature.

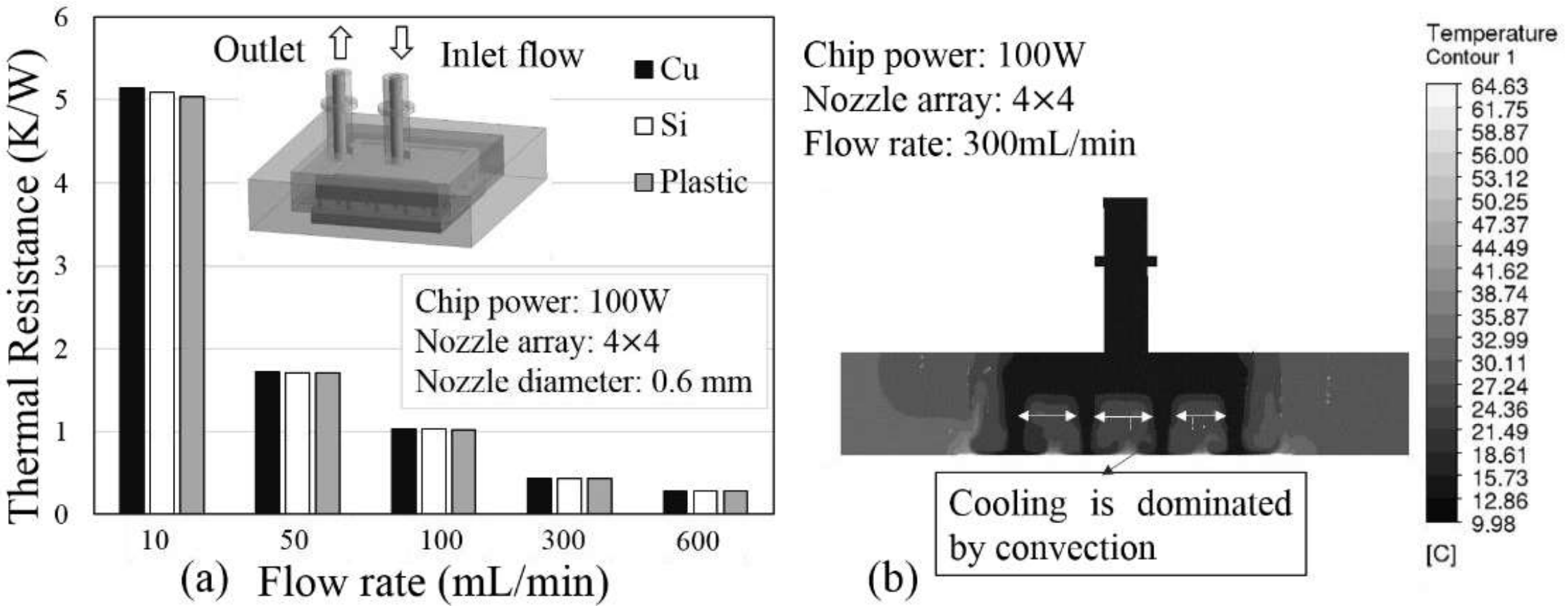

**Figure 5.8:** Contour plots of the full cooler CFD models showing the impact of the thermal conductivity of the cooler materials for different flow rates (chip power 100W).

Figure 5.9 shows the percentage change with regard to the Cu cooler. It can be seen that the impact of the cooler material is minimal. Even at very low flow rates, the difference between a Cu and a plastic cooler is only 2%, while the differences become much smaller at a higher flow rate. In summary, the cooler material impact is negligible at a higher flow rate and therefore offers opportunities for the use of polymer-based cost-efficient fabrication techniques.

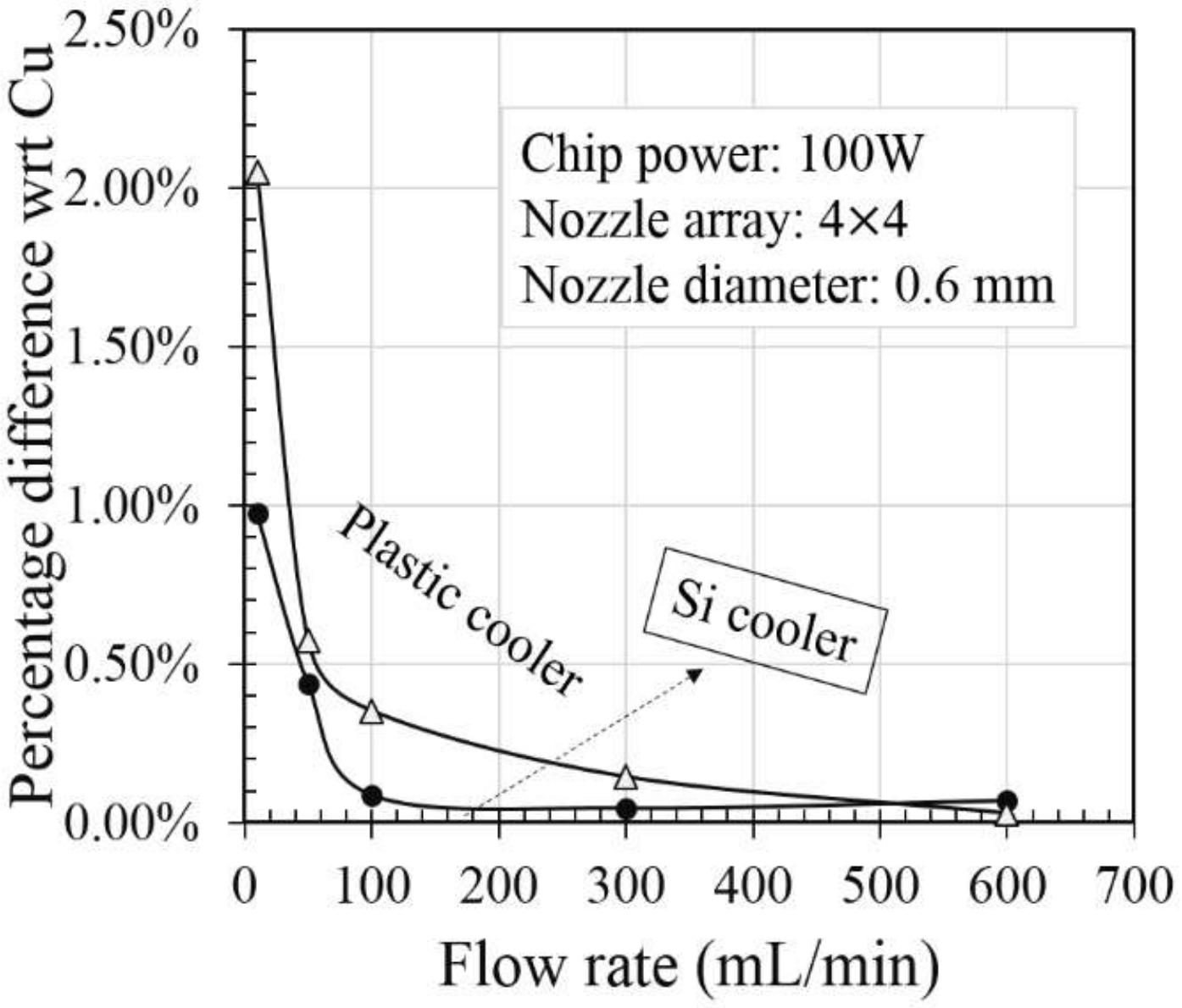

**Figure 5.9:** Thermal resistance percentage change with regard to the Cu based cooler.

### 5.1.4 Impact of liquid coolant

For the liquid coolant used in the application, DI water may not be an ideal cooling liquid due to potential freezing and the direct contact with the chip. In this section, modeling studies are performed to assess the thermal and hydraulic impact of coolant properties. In general, there is a very large variety of coolants and refrigerants. The following conditions should be considered during the selection of the liquid coolant in practical applications [7]:

- High ***boiling point*** desired for single-phase cooling
- Low ***freezing point*** desired for shipping and storage
- High ***reliability***, non-corrosive, inert, ...
- Ecological considerations, ...

Table 5.1 lists the material properties of the different commonly used liquid coolants, given in the literature [7]. The unit cell model introduced in chapter 2 is used for the evaluation of a set of popular single-phase coolants. The flow rate per nozzle is defined as the flow rate divided by the total number of inlet nozzles, rang from 1.6 mL/min to 16 mL/min. The extracted heat transfer coefficient is based on the averaged chip temperature. The heat transfer coefficient curves plotted as a function of different flow rate per nozzle, for the listed coolants are shown in Figure 5.10.

**Table 5.1:** Material properties of the different liquid coolant [7].

| Coolant | Density ($kg/m^3$) | Viscosity (kg/m.s) | Specific heat (J/kg.K) | Thermal cond. (W/m.K) |
|---|---|---|---|---|
| Water | 1000 | 8.90E-04 | 4217 | 0.68 |
| Coolanol 25R (Silicate-ester) | 900 | 0.009 | 1750 | 0.132 |
| Syltherm XLT (Silicone) | 850 | 0.0014 | 1600 | 0.11 |
| FC-77 (Fluorocarbon) | 1800 | 0.0011 | 1100 | 0.06 |
| Ethylene Glycol 50/50 (EG) | 1087 | 0.0038 | 3285 | 0.37 |
| Methanol/Water 40/60 | 935 | 0.002 | 3560 | 0.4 |
| Potassium Formate/ Acetate Solution 40/60 | 1250 | 0.0022 | 3200 | 0.53 |

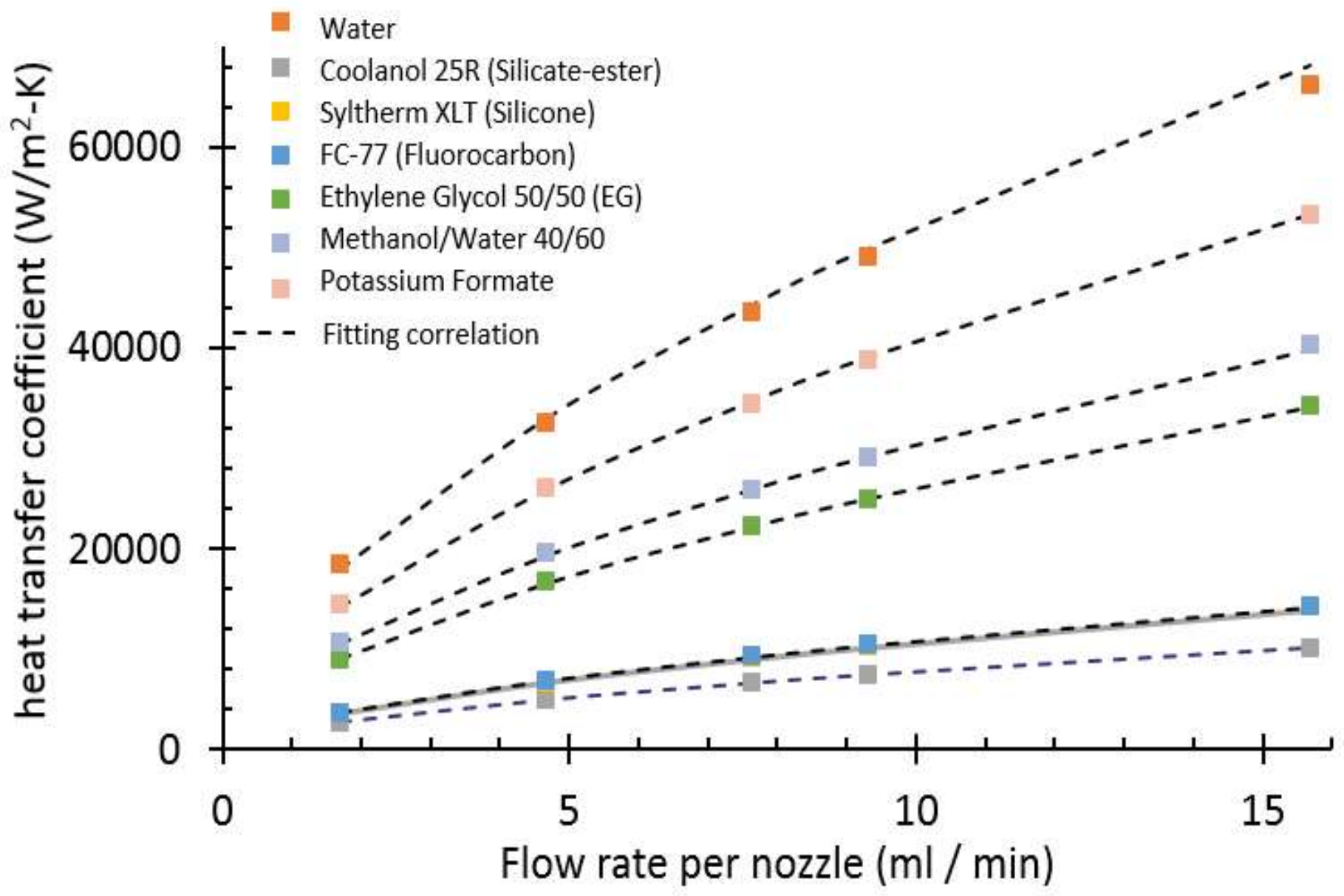

**Figure 5.10:** The heat transfer coefficient curves plotted as a function of different flow rate per nozzle, for the listed coolants.

Figure 5.11 shows the relative heat transfer rate comparison for the different coolants for constant flow rate and pumping power considerations. The relative heat transfer rate of the coolant is defined with respect to DI water. In general, the high heat transfer for water is due to its high conductivity and high specific heat. For other single-phase liquid coolant, the heat transfer typically drops by 40% or more for the constant flow rate comparison shown in Figure 5.11(a). Table 5.1 also shows that all the liquid coolants have a higher viscosity than DI water, resulting in a higher pressure drop than DI water. For a fixed pumping power, a lower flow rate is therefore required for the other coolants. The relative heat transfer rate will further decrease due to the lower flow rate. In summary, for both comparisons, DI water shows better cooling performance than other liquid coolants.

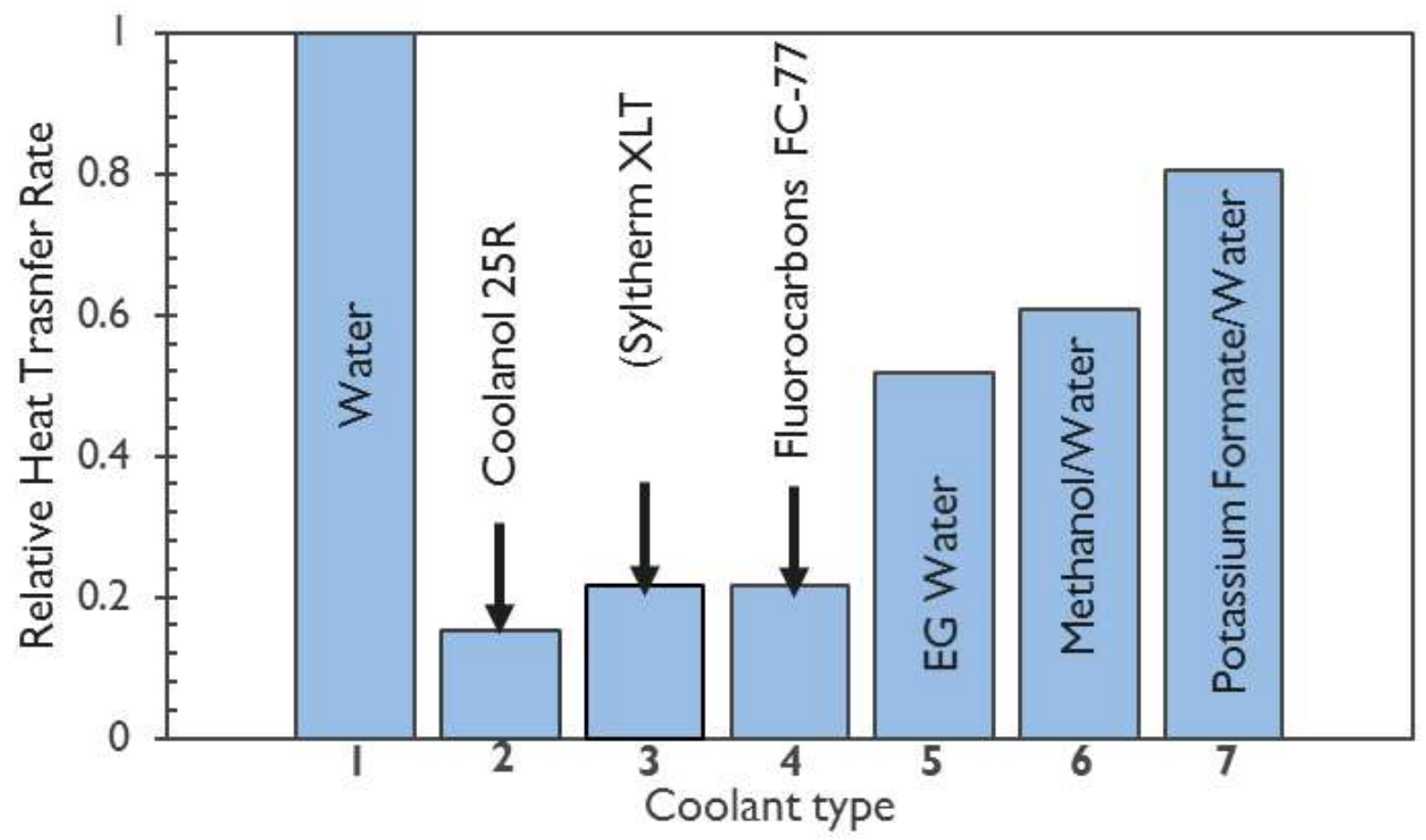

(a)

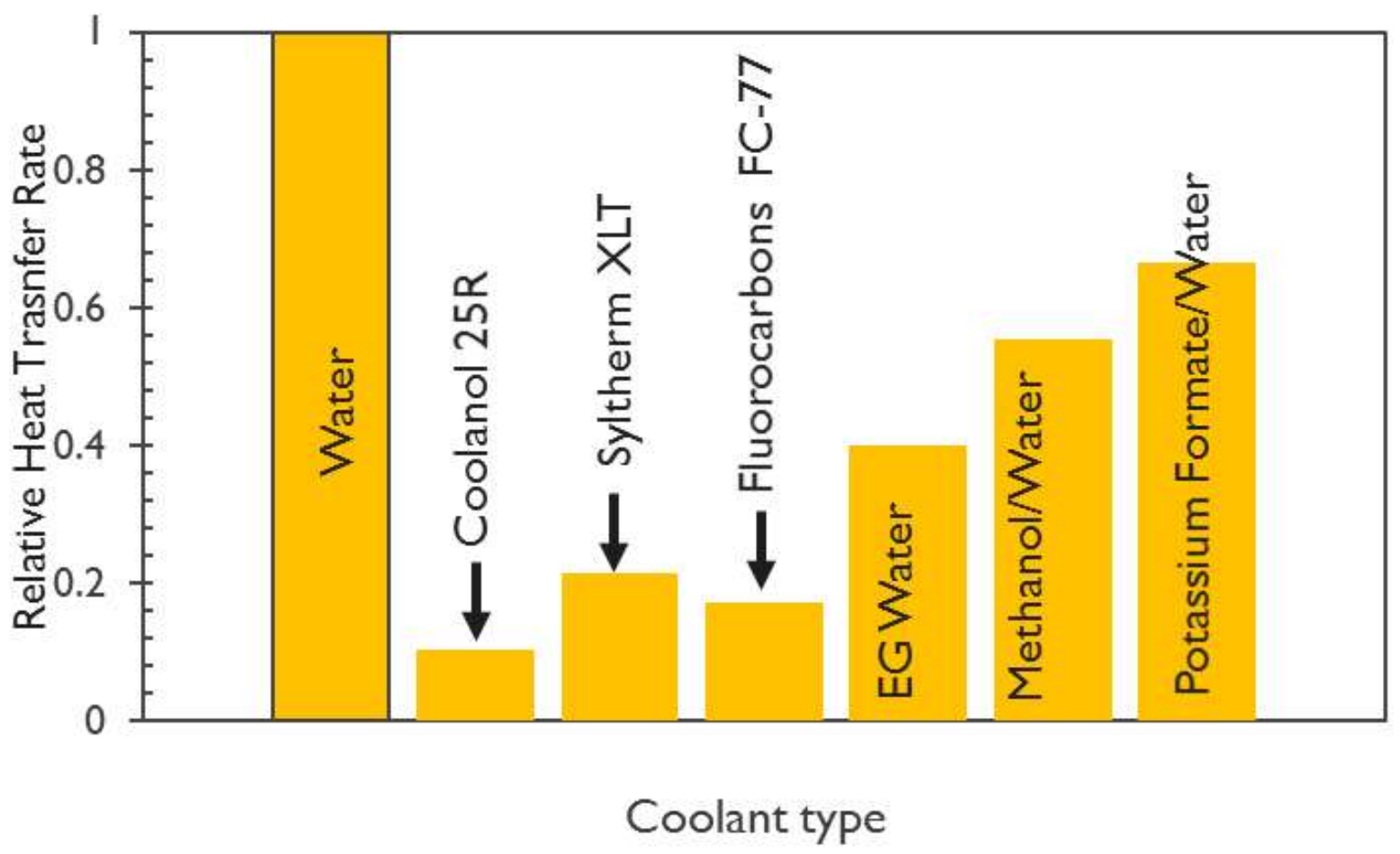

(b)

**Figure 5.11:** Comparison of the relative heat transfer rate between the different liquid coolant: (a) constant flow rate at 1 LPM; (b) constant pumping power at 0.003 W.

## 5.2 Proof of concept: mechanical micromachined demonstrator

Based on the thermal and hydraulic modeling results for the number of unit cells, inlet diameter (shown in chapter 3) and impact of material conductivity, and on the fabrication capabilities, a simplified board level polymer-based multi-jet cooler has been designed with a 4×4 inlet nozzle array. Figure 5.12 shows an exploded view of the design of the different parts of the cooler (cover layer, inlet/outlet plenum divider, nozzle plate, support structure, and copper spacer) that will be mounted on the test chip package and PCB. The inlet nozzle array is chosen as a 4×4 array while the outlets are organized in a 5×5 array in such a way that each inlet is surrounded by 4 outlets. The diameter of both inlets and outlets is set to 600 µm since larger diameters will result in a cooling performance saturation, as explained in chapter 3. In order to separate the inlet flow and outlet flow, an inlet/outlet plenum divider is needed, as indicated in Figure 5.12. The cavity height is chosen as 0.6 mm. The inlet chamber thickness is chosen as 3 mm. The designed diameters for both the inlet tube $D_{i\text{-}tube}$ and outlet tube $D_{o\text{-}tube}$ are both 6 mm. The designed geometry parameters of the single jet cooler and the multi-jet cooler are both summarized in Table 5.2.

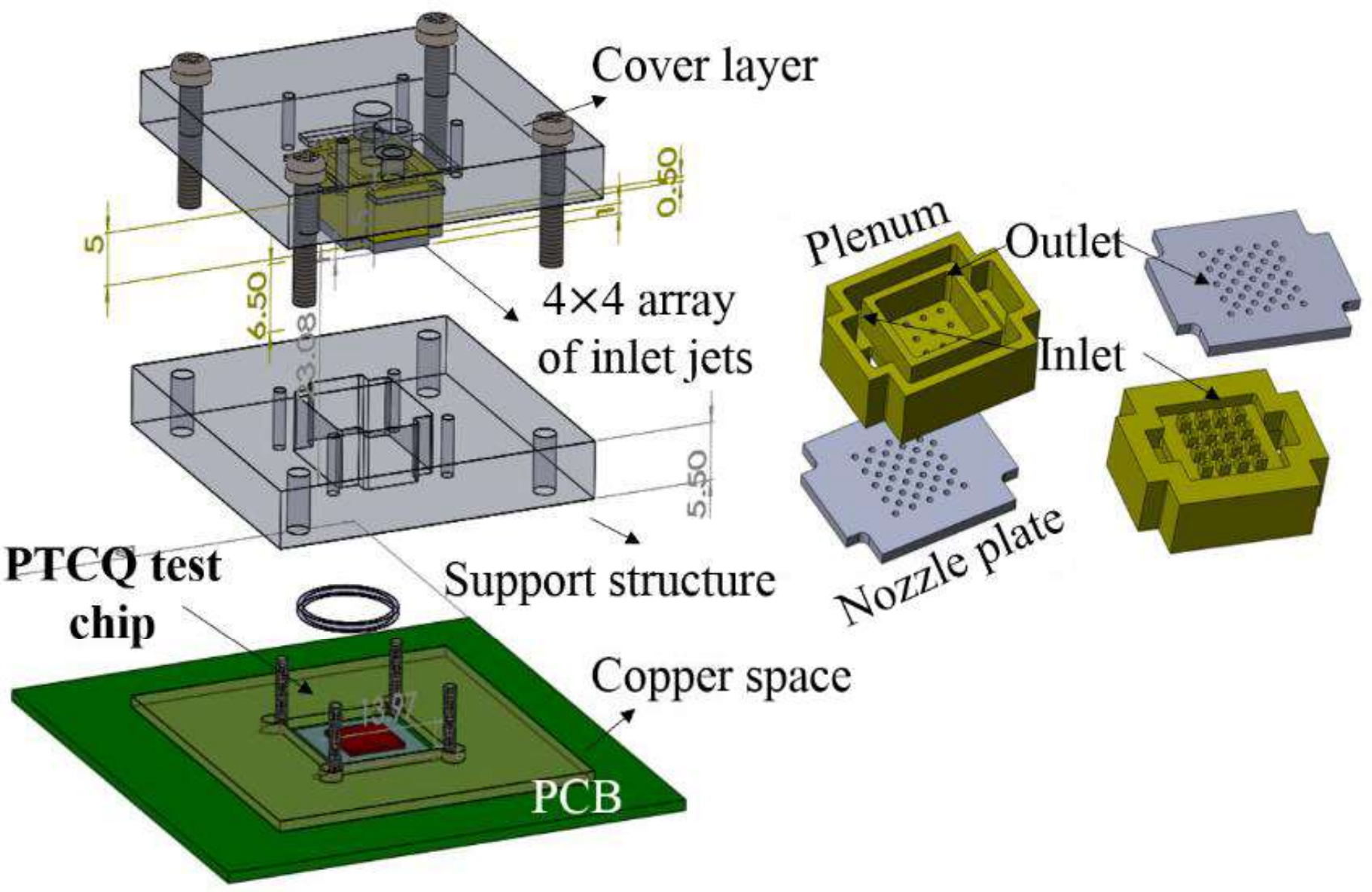

**Figure 5.12:** CAD design structure of the 4×4 array demonstrator of the impingement cooler and integration on the test chip and PCB.

**Table 5.2**: Geometry parameters comparison.

| **Parameters** | **Single jet** | **Multi-jet** |
|---|---|---|
| N×N | 1 | 4×4 |
| $D_{i\text{-tube}}/D_{o\text{-tube}}$ | 6 mm / 6 mm | 6 mm/ 6 mm |
| $d_i$ | 2 mm | 600 µm |
| $d_o$ | Common outlet | 600 µm |
| H | 2 mm | 600 µm |
| t | 7 mm | 1 mm |
| $t_c$ | 0.2 mm | 0.2 mm |
| L | 8 mm | 2 mm |

For the demonstration of the multi-jet cooler with a 4×4 nozzle array, mechanical micromachining [8] is used as the initial assessment. The multi-jet cooler is manufactured in Polyvinyl chloride (PVC) using micromachining and drilling. The mechanical drawing of the cooler with different cross-sections is shown in Figure 5.13. As indicated in Figure 5.13, there are several challenges for the fabrication of the micromachined cooler, which are listed below:

- The **aspect ratio** of milling tools, which can limit the thickness of the chamber thickness;
- The nozzle plate should be strong enough to withstand the **high pressure** at the interface with the nozzle plate;

- The pitch of the **inlet/outlet divider** is determined by the diameter of the micro-milling tool;
- **Alignment** between inlet holes with nozzle plate and inlet distributor is critical.

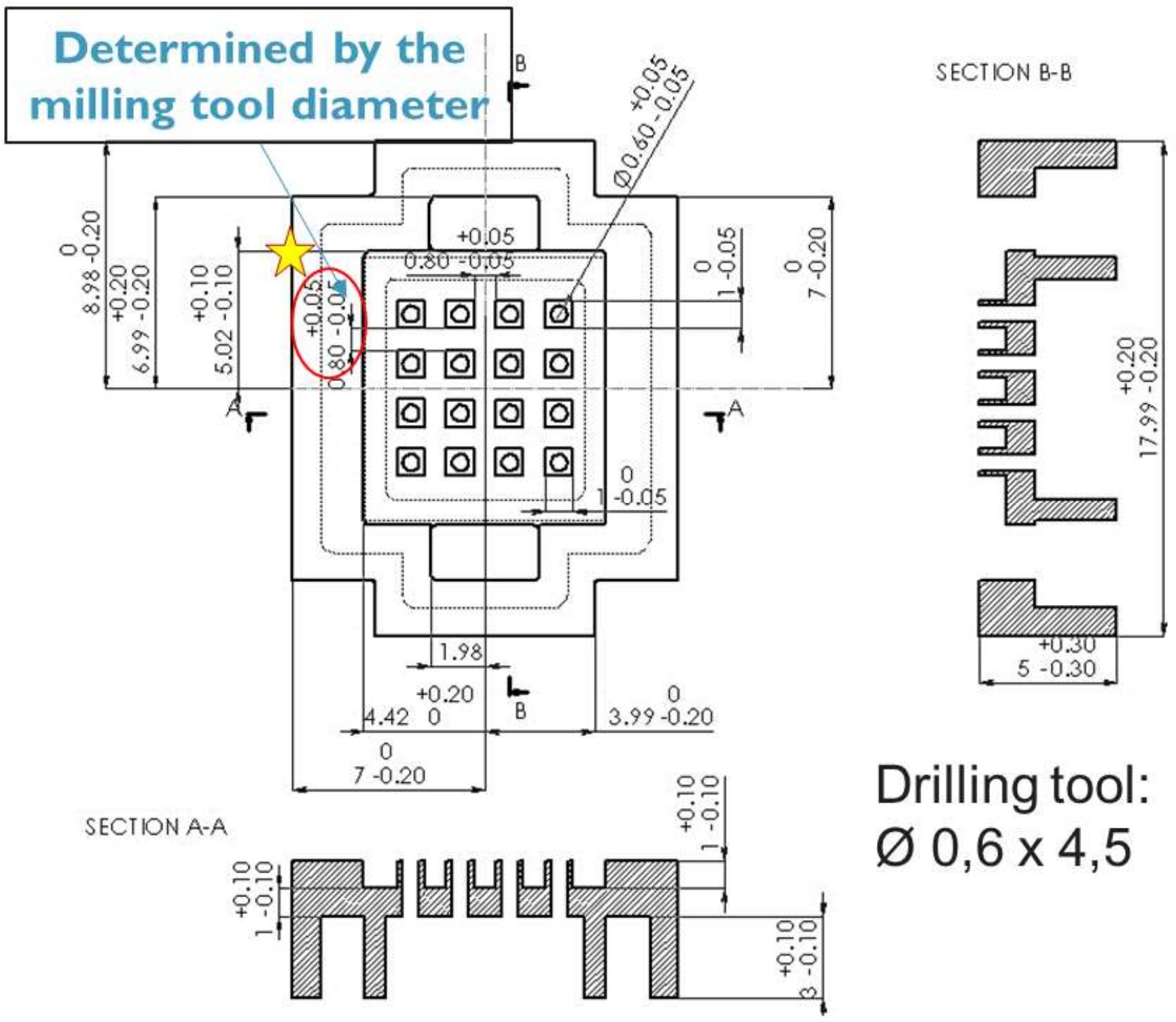

**Figure 5.13:** CAD structure with critical dimensions.

Figure 5.14 shows different fabricated parts of the 4×4 array demonstrator of the impingement cooler. The inlet/outlet nozzles have been fabricated using a 600 µm diameter tool for the micromachining and a 600 µm drill to create the inlets and outlets in the nozzle plate, creating a wall thickness of 200 µm for the inlets through the outlet plenum. For the fabrication micromachined cooler, the measured variation of the fabricated nozzle diameter is between 450 µm to 610 µm. It also shows the assembly of the cooler on the thermal test chip and PCB using an O-ring to prevent leakage of the coolant. The presence of the O-ring creates a stand-off of 600 µm between the electronic device and the nozzle plate, which is the cavity where the impingement takes place (cavity height). After assembly, the cooler has been successfully tested, and no leakage was observed. The assembled test board will be connected to the experimental measurement flow loop introduced in chapter 2.

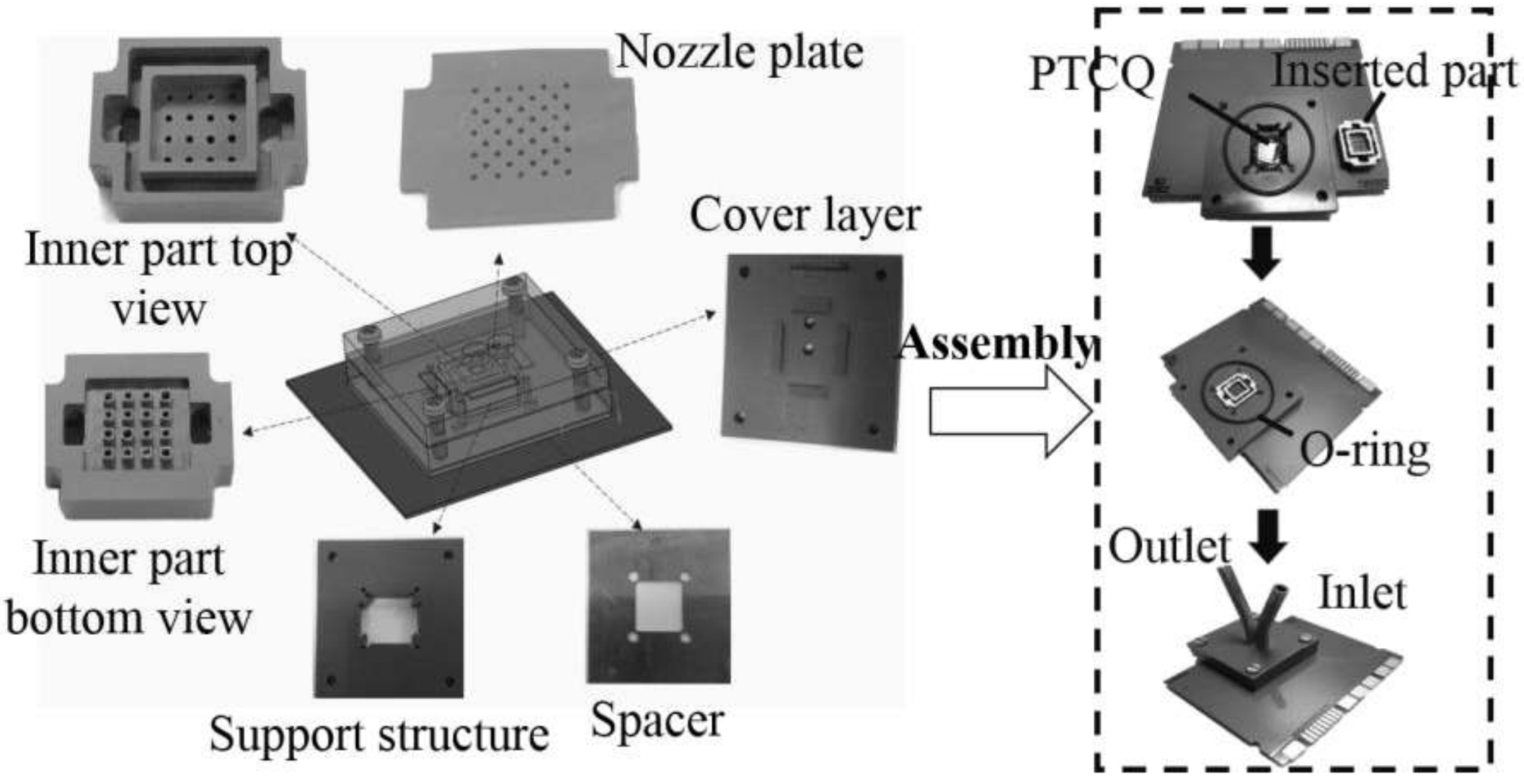

**Figure 5.14:** (a) Different fabricated parts of the 4×4 array demonstrator of the impingement cooler and (b) Final assembly of 4×4 array demonstrator of the impingement cooler and integration on the test chip and PCB.

## 5.3 Thermal characterizations and modeling validation

### 5.3.1 Micromachined cooler CFD model

Based on the fabricated micromachined (MM) cooler, the full cooler level CFD model is built to investigate the flow and thermal behavior of the multi-jet cooling. The CFD model is shown in Figure 5.15. Figure 5.15(a) shows the transparent view of the CAD structure, including the O-ring and the inlet/outlet plenum. After that, the fluid domain with the internal fluid delivery channel system is extracted from the CAD model shown in Figure 5.15(c) and Figure 5.15(d). Based on the meshing methodology introduced in chapter 2, the number of elements for the full models is 5.9 million, as shown in Figure 5.15(b).

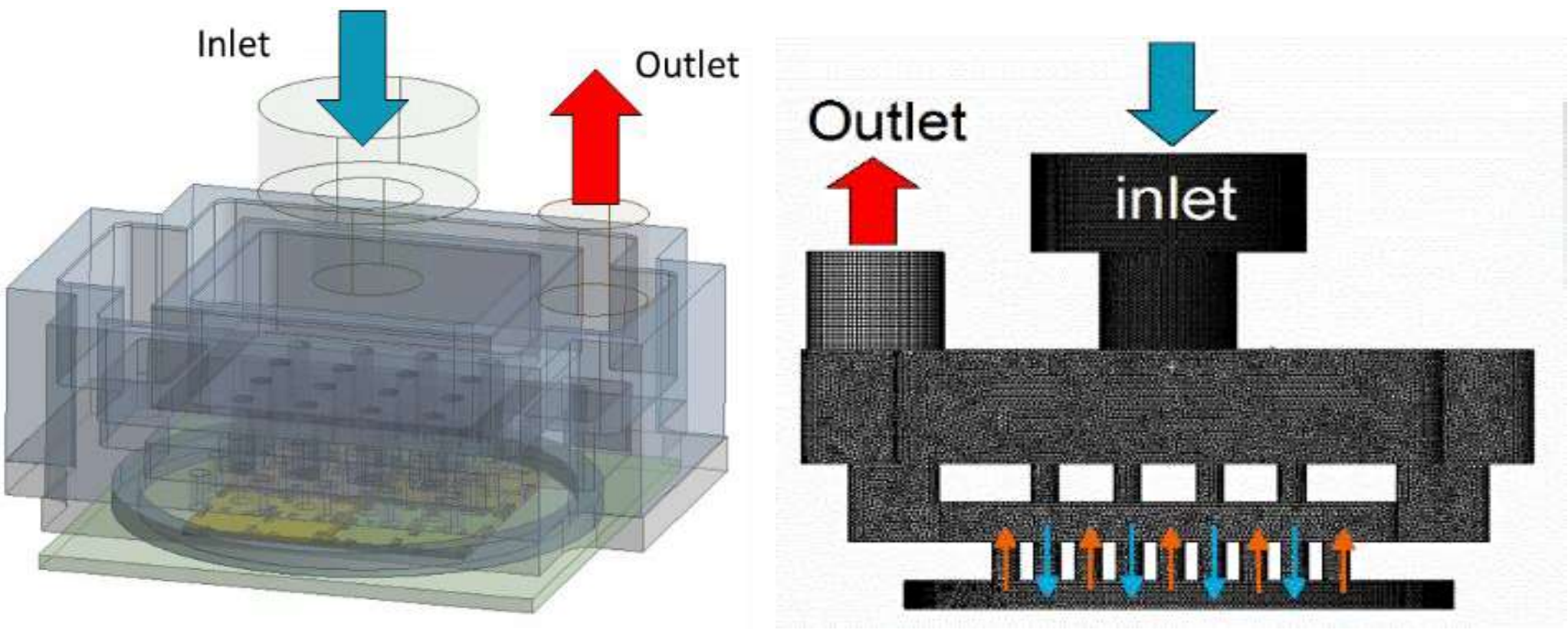

(a) (b)

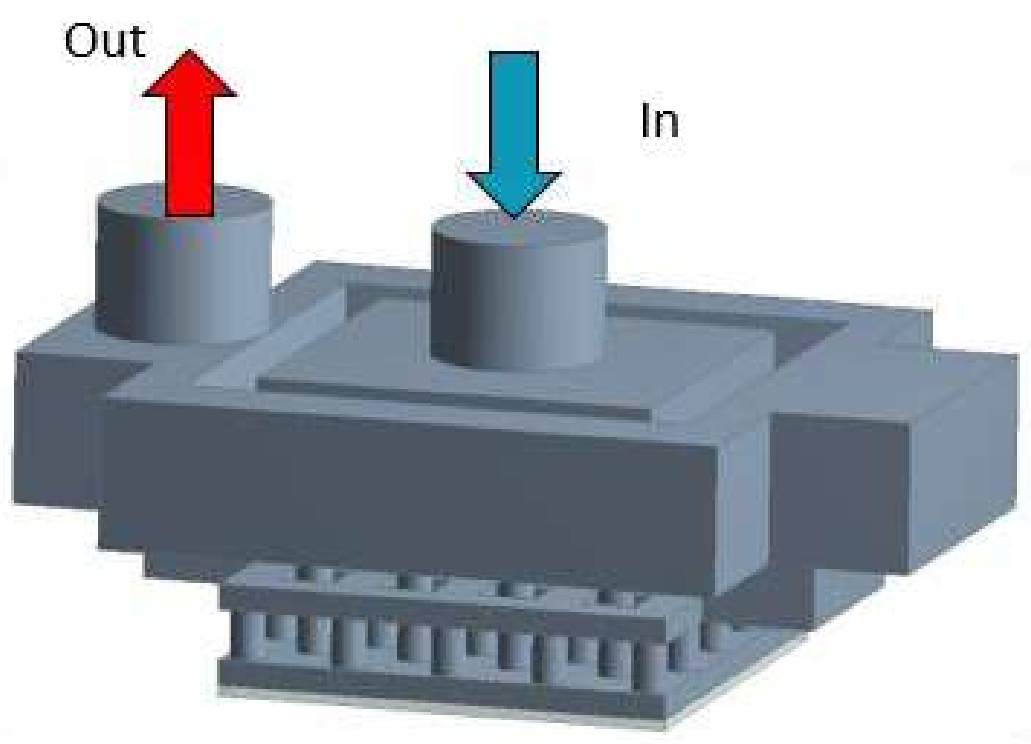

(c)

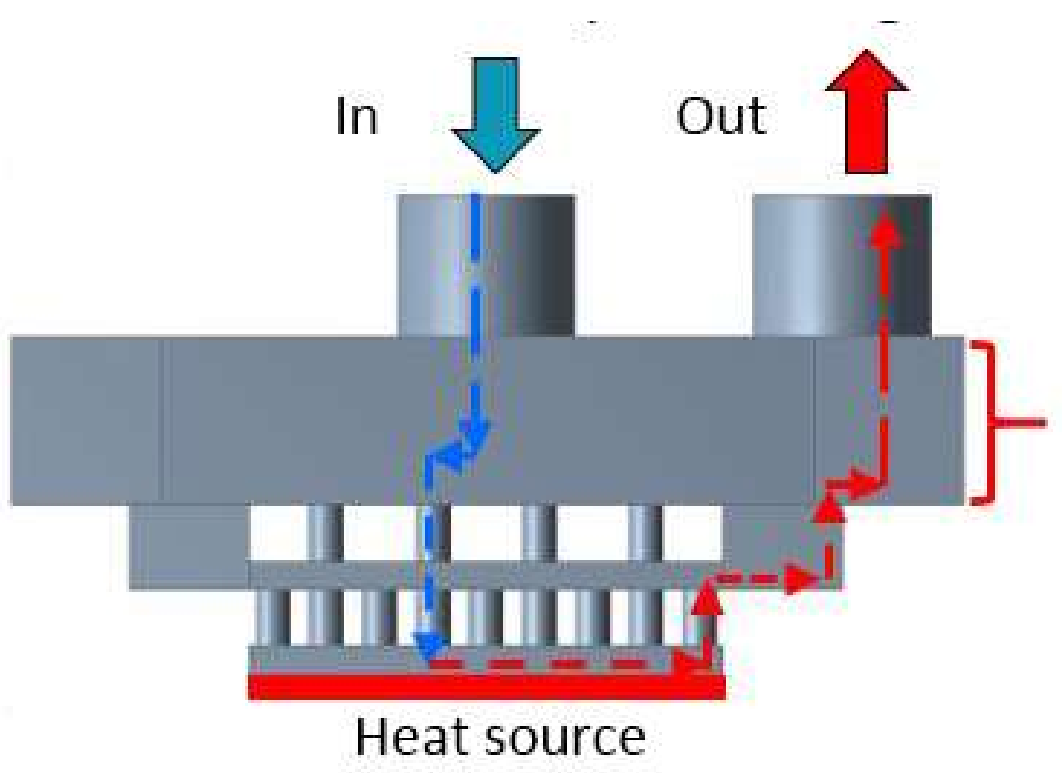

(d)

**Figure 5.15:** CFD model of the full impingement cooler: (a) CAD structure of the designed cooler; (b) Meshing of the full cooler level model; (c) side view and (d) cross-section view of the extracted fluid domain for the CFD model.

### 5.3.2 Quasi-uniform heating

The measured and simulated chip temperature increase maps for the $4 \times 4$ jet array cooler are shown in Figure 5.16(a) and Figure 5.16(b) for a power of 50 W and a flow rate of 600 mL/min. For the thermal performance characterization, the thermal resistance based on the average chip temperature is 0.25 K/W for a modeled pressure drop of 15 kPa. Figure 5.16(c) shows that the impact of the flow rate on the thermal resistance reduces by a factor of 1.7 by increasing the flow rate from 300 mL/min to 600 mL/min. As for the modeling validation, the comparison between the modeling and experimental results show a perfect agreement for the averaged chip temperature. The difference between the modeling and measurement results for the average chip temperature is only 4.86 % and 4.19 % for 300 mL/min and 600 mL/min flow rate, respectively.

Moreover, there is a local difference in the temperature distribution map in Figure 5.16(a) and Figure 5.16(b). The causes of the local difference will be investigated in this paragraph. First, the temperature asymmetry is mainly due to the asymmetrical flow since the outlet is located on one side of the cooler. The flow coming from the impingement zone has to be combined to the outside through the outlet tube. Moreover, the O-ring placed under the nozzle plate is fixed as a rectangle shape. Secondly, due to the higher heat transfer rate of the cooler compared to the single jet cooler, the location of the heated cells and non-heated cells is visible in the temperature map, revealing a minimum temperature in the central area of the chip where no heater cells are present and lower temperature around the chip periphery. This is caused by the presence of the coolant around the chip in the cavity defined on the chip package. The local minima and maxima of the temperature profile on the chip diagonal can be nicely matched to the location of the inlet and outlet nozzles in Figure 5.16(c). The cooling performance comparison between single jet cooler and multi-jet cooler will be discussed in section 5.5.1.

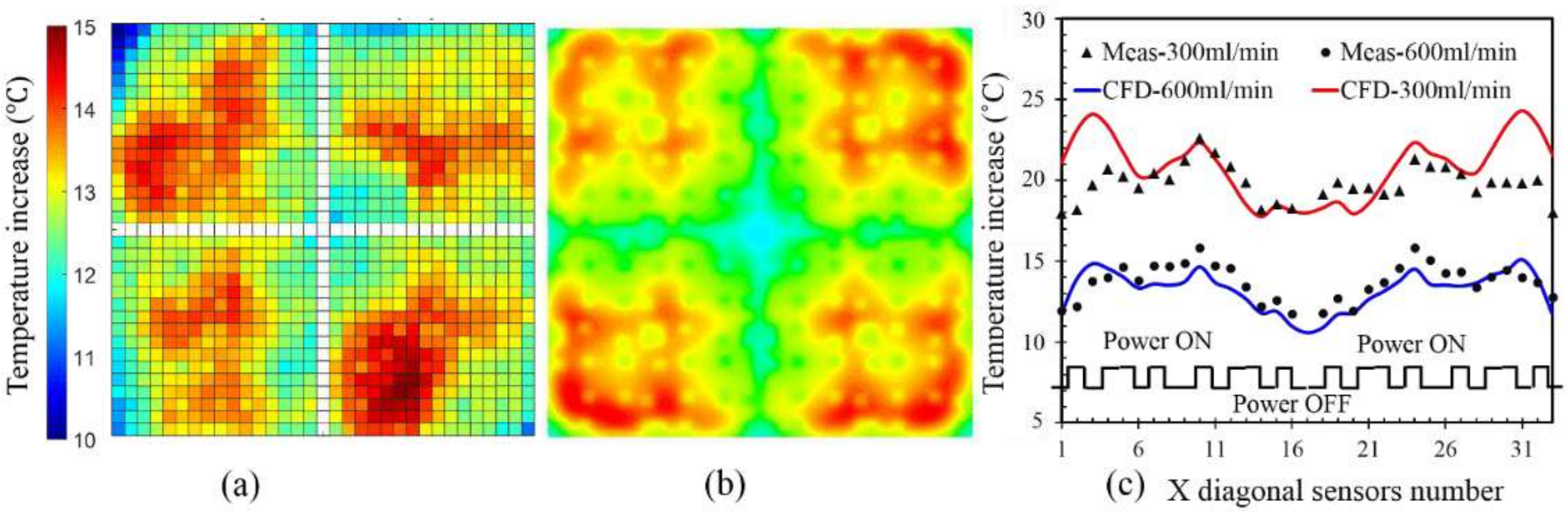

**Figure 5.16:** Modeling and experimental results of 4 × 4 multi-jet cooler ($Re_d$ = 1015 for 600 mL/min) (a) measurement, (b) CFD modeling, (c) temperature increase profile comparison between measurements and CFD modeling of multi-jet cooler (chip power = 50 W).

### 5.3.3 Uniform and quasi-uniform heating

Figure 5.17 shows the comparison between the uniform heating case and the quasi-uniform heating case for the modeling of the 4 × 4 multi-jet cooling. Although the difference for the average temperature is small, the temperature distribution maps look completely different. While in the case of uniform heating, shown in Figure 5.17(a), the nozzle pattern is visible, the pattern of heated and non-heated cells is visible in the case of quasi-uniform heating shown in Figure 5.17(b), due to the high cooling rate of the jet impingement on the surface of the Si chip. The comparison of the temperature profiles along with the chip diagonal with the modeling data in Figure 5.17(c), reveals that the uniform model is not capable of correctly predict the local temperature distribution. The quasi-uniform model with the complete details on the heater cells shows a much better agreement with the experimental data. This analysis for the single jet cooler and multi-jet cooler clearly shows the need to include sufficient details on the heater structures in the CFD model in order to accurately predict the local temperature distribution in case of high heat removal rates at the chip surface. Furthermore, the comparison highlights the importance of test structures with a high spatial resolution in order to detect these effects.

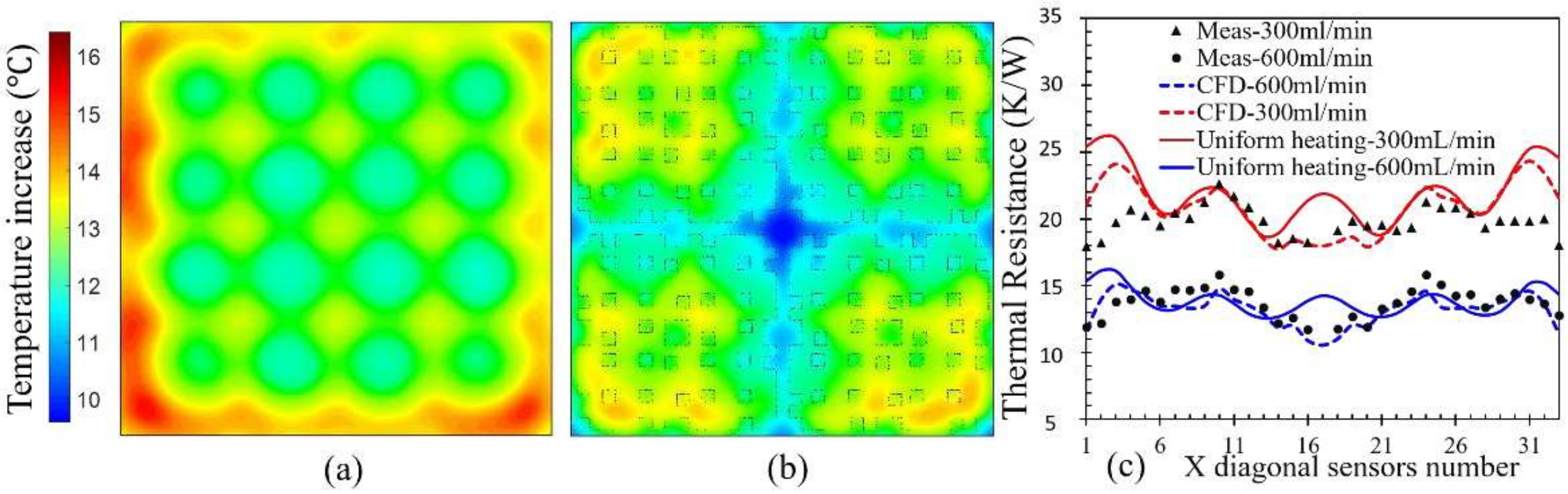

**Figure 5.17:** Comparison of simulated temperature increase comparing to inlet temperature distribution for multi-jet cooling with different heating configurations: (a) uniform heating modeling results for 600 mL/min; (b) quasi-uniform heating modeling results for 600 mL/min; (c) profile comparison between uniform heating and quasi-uniform heating (chip power=50 W).

## 5.4 Hot spots with uniform array cooling

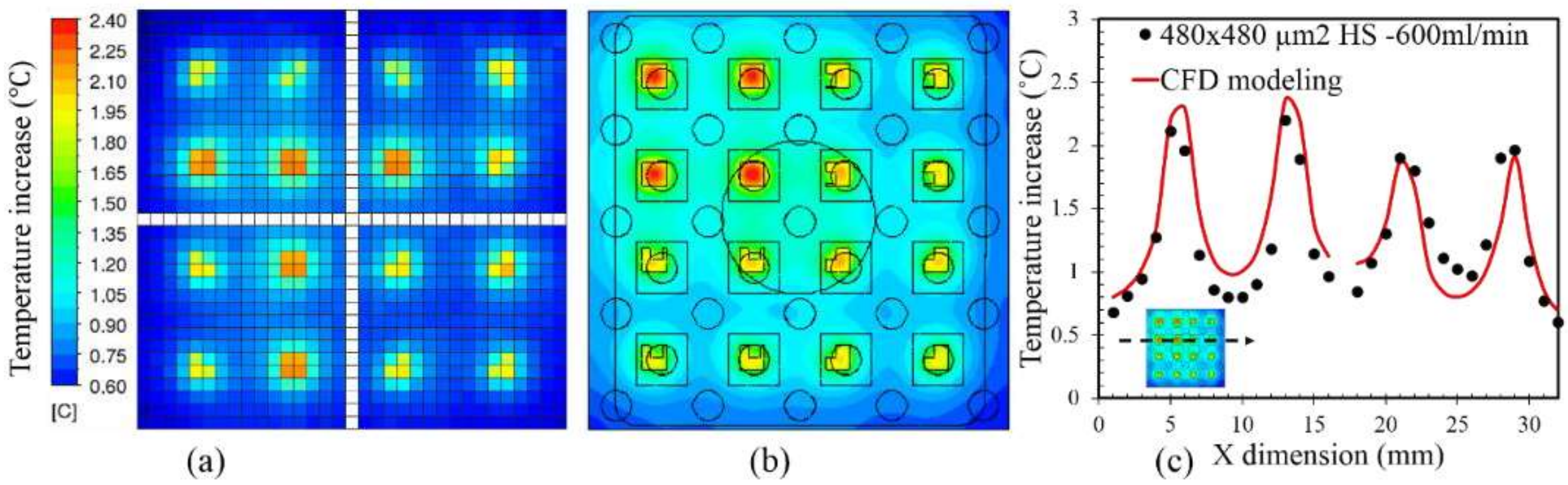

**Figure 5.18:** Experimental and modeling results of 480 × 480 µm2 hot spots under the chip power Q=3.3 W ($Re_d$=1015 for 600 mL/min): (a) experimental results of local temperature distribution; (b) CFD simulations of local temperature distribution; (c) experiments and modeling comparison under the same flow rate 600 mL/min, and same chip power: Q=3.3 W.

The heat sources in the test chip can also be programmed in a hot spot array pattern. Figure 5.18 shows the hot spot cooling results for a 4 × 4 array of hot spots aligned to the 4 × 4 inlet nozzle array: the hot spots consist of 2 × 2 heater cells (480 × 480 µm$^2$). The chip power is set as 3.3 W for a flow rate of 600 mL/min. Figure 5.18(a) and Figure 5.18(b) respectively show the measured and the modeled chip temperature increase distribution, while the comparison of the temperature profile across four hot spots is shown in Figure 5.18(c). The comparison shows that the test chip is capable of accurately capture the local temperature peak of the hot spots, as both the temperature peak values, as well as the valleys, are resolved. Overall, a good agreement between the modeling and measurement results is found, with a maximum difference of 10% at the peaks. Both the modeling and experimental results exhibit a similar asymmetrical pattern due to the presence of the outlet connector at only one side of the cooler.

The experimentally validated hot spot model can now be used to evaluate different cooler configurations:

1. Common outlets (indicated in Figure 3.21, chapter 3): in this case, there are no local outlets in between the inlet nozzles. The common outlets are located at the edges of the chip shown in Figure 5.19(a);
2. Locally distributed outlets in between the inlet nozzles that are aligned to the hot spots shown in Figure 5.19(b);
3. Locally distributed outlets between the inlet nozzles that are intentionally misaligned with the hot spots shown in Figure 5.19(c).

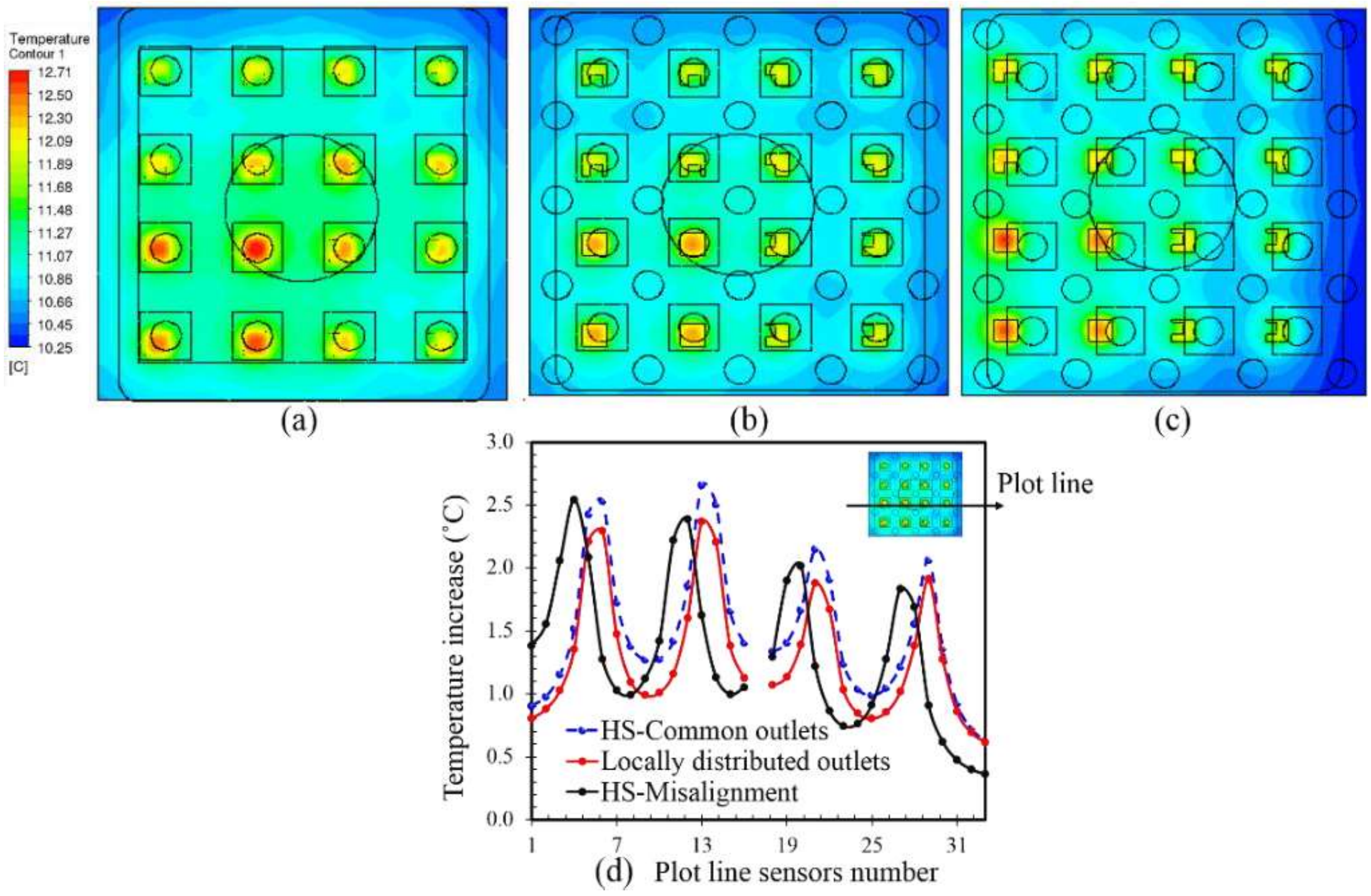

**Figure 5.19:** Hot spots cooling modeling for flow rate 600 mL/min: (a) hot spots cooling with common outlets, where the common outlets are located at the edges of the chip; (b) hot spots cooling with locally distributed outlets; (c) hot spots cooling with nozzle misalignment; (d) temperature profile comparison with different configurations.

The schematic of the common outlet configuration is explained in Figure 3.21(a) of chapter 3. Figure 5.19(d) shows that the hot spot cooling with locally distributed outlets can achieve a better cooling performance than common outlets. The main advantage of the locally distributed outlets is that the cross-flow effects can be reduced that are present in the common outlet flow. The simulation results for the aligned and misaligned hot spots with respect to the nozzle locations show that it is important to align the cooling nozzles with the hot spots, as shown in Figure 5.19(d). The temperature difference amounts to 10% between aligned and misaligned jet nozzles with the hot spot locations, illustrating the need for a matching design between the nozzle array and the chip floor plan.

## 5.5 Results and discussion

### 5.5.1 Single jet cooling versus multi-jet cooling

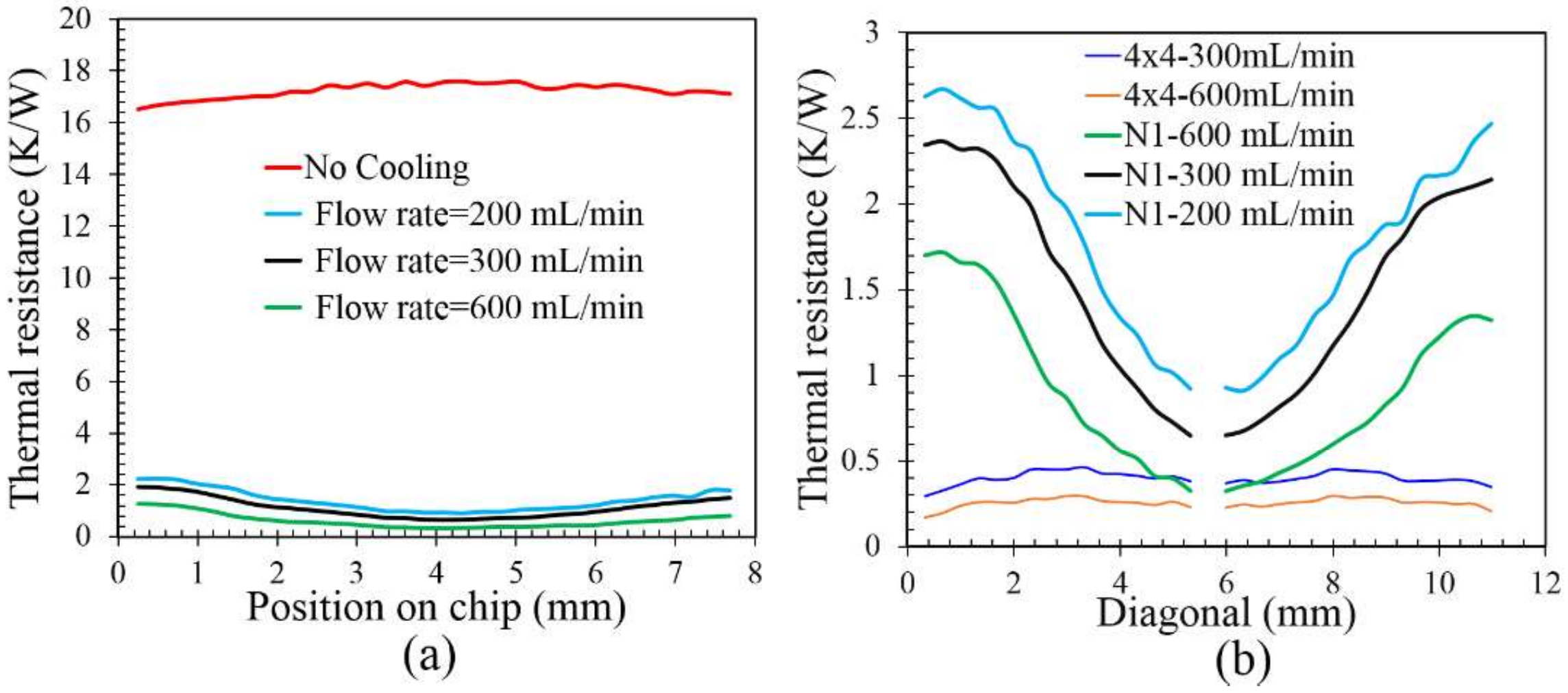

**Figure 5.20:** Normalized comparison of the measured temperature profiles for the cases of (a) single jet cooler with respect to the reference measurement without cooling; (b) comparison between the singe jet cooling and multi-jet 4×4 array cooling.

As shown in Figure 5.18(a), the thermal resistance without liquid cooling is taken as the reference case with regard to the single jet cooling for three different flow rates. The comparison with the single jet cooler on the same chip package in Figure 5.18(b), shows that the multi-jet impingement cooler results in lower thermal resistance and better temperature uniformity for the same flow rate. The thermal performance of the coolers can also be expressed in terms of the Nusselt number $\overline{\mathrm{Nu}}_{\mathrm{j}}$ and the Reynolds number $\mathrm{Re}_d$, based on the nozzle diameter as characteristic length, shown in Figure 5.19. The following correlations have been extracted for the three different considered coolers:

4×4 cooler: $\overline{\mathrm{Nu}}_{\mathrm{j}} = 1.63 Re_d^{0.57}$ (experimental data)

4×4 cooler (Common outlets): $\overline{\mathrm{Nu}}_{\mathrm{j}} = 1.34 Re_d^{0.59}$ (modeling data)

An extensive overview of heat transfer correlations for jet impingement cooling is available in chapter 3 in the form of:

$$Nu = c.\ Re_d^m$$

For all these correlations, the exponent m is within the range of 0.48 to 0.8. The obtained exponents of the correlations for the coolers studied in this work are within this range.

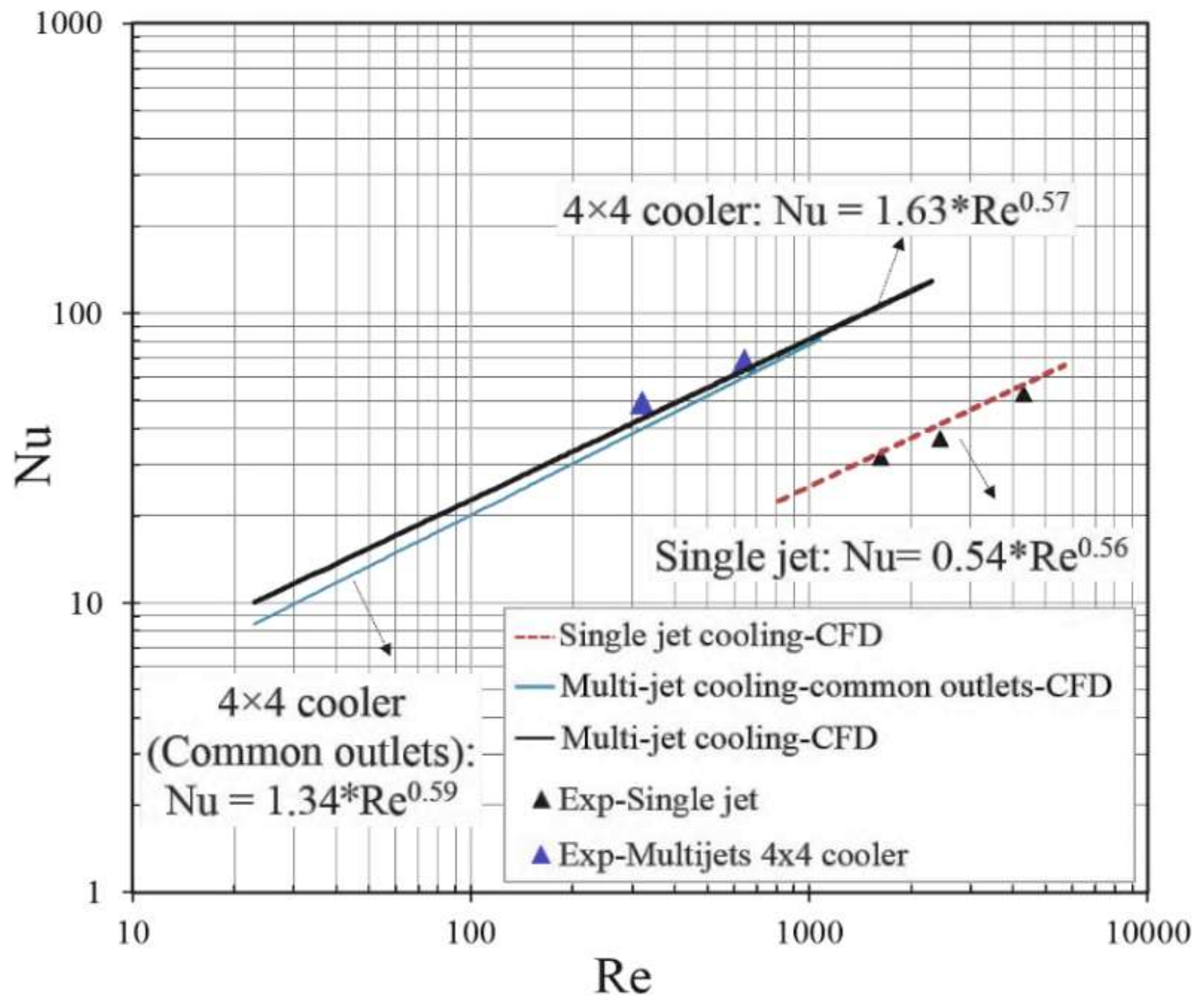

**Figure 5.21:** Correlations between Nusselt number and Reynolds number for single jet and multi-jet configurations.

### 5.5.2 Benchmarking with state-of-the-art cooling

The cooling performance in terms of thermal resistance and pumping power of the fabricated polymer cooler with a 4×4 array of inlets (0.25 K/W or 0.16 $cm^2$.K/W for a pumping power of 0.4 W) has been compared in Figure 5.22 with published data in literature for impingement coolers fabricated using various materials: Si [10,11], ceramic [12], metal [5, 7, 13, 14, 15] and plastic [16,17] presented in section 1.1 of chapter 1. To assess the trade-off between the cooling performance and the required pumping power for the liquid coolant, the results are compared in the thermal resistance versus pumping power chart, introduced in the parametric analysis of chapter 3. Since the literature measurement data of the cooling and hydraulic performance is reported for different chip sizes, the data needs to be normalized in order to compare the intrinsic cooling performance of the different coolers. The thermal resistance scales inverse proportionally with the chip size (resulting in lower thermal resistance values for large chips, while the pumping power scales proportionally with the area (resulting in high required pumping power for large chips). The proposed metrics for the benchmarking assessment in Figure 5.22 are, therefore, the normalized thermal resistance $R_{th}{}^{*} = R_{th}A$, and the normalized pumping power $W_p{}^{*} = W_p/A$. The two measurement points of this work (imec-polymer cooler) shown in this figure are based on the fabricated 3D-shaped polymer cooler with a flow rate of 280 mL/min and 600 mL/min. From the benchmarking chart in Figure 5.22, it can be observed that the presented 3D-shaped polymer cooler has a very good thermal performance, which is achieved with relatively low pumping power. The thermal performance of the polymer cooler is very similar to the Si-based cooler [11] with very fine pitch 25 µm diameter nozzles, however, that cooler requires expensive Si fabrication techniques. Comparison with the single jet cooler [7] applied on a single MOSFET semiconductor device shows that single jet impingement cooling can be operated at lower pumping power, especially for large nozzle diameters, but that the obtained thermal resistance of single jet cooling is much higher: the multi-jet cooler outperforms the single jet cooler by one order of magnitude. The comparison with the results of the cooler for IGBT devices shows the impact of the direct cooling on the semiconductor device: in [26], it is shown that impingement cooling on the base plate is more efficient than cooling on the substrate or using a cold plate. The use of direct liquid multi-jet impinging on the chip backside results in a further improvement of factor 4 compared to the cooling on the base plate. This benchmarking study clearly shows the potential of the presented multi-jet polymer-based cooler and proves that it is not necessary to scale down the nozzle diameters to a few tens of micron. Still, that very good thermal performance can be obtained with nozzle diameters in the range of several hundred micrometers, which is compatible with low-cost polymer fabrication.

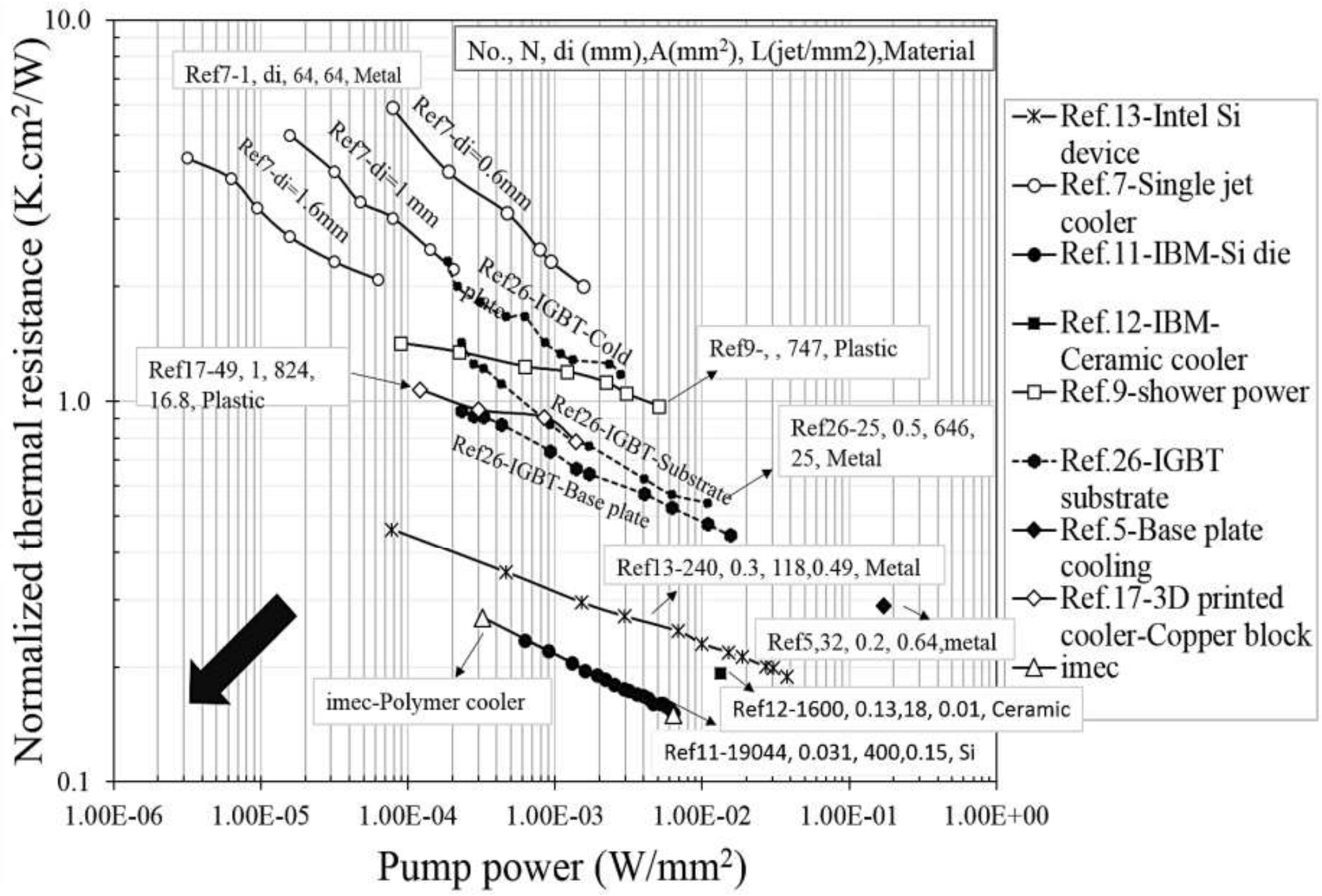

**Figure 5.22:** Thermal performance and pumping power comparison with state-of-the-art cooling solutions.

## 5.6 Conclusion

Liquid jet impingement cooling is known to be a very efficient cooling technology. State of the art highly efficient multi-jet cooling solutions rely on expensive Si or ceramic fabrication techniques, while cost-efficient cooling solutions have been proposed for less performant single jet impingement. In this chapter, we present the concept, modeling, design, fabrication, experimental characterization, and benchmarking with literature data of a novel multi-jet impingement based liquid cooling solution, fabricated using low-cost polymer fabrication techniques, targeted to directly cool the backside of high power devices. It is demonstrated that polymer is a valuable alternative material for the fabrication of the impingement cooler instead of expensive Si based fabrication methods. Unit cell thermal and hydraulic CFD models have been used to study the scaling trends for nozzles dimensions, while full cooler models have been applied to study the interactions between nozzles and the impact of the cooler material. The modeling results show that it is not necessary to scale up the number of unit cells and to shrink the nozzle diameter accordingly to improve the thermal performance for a fixed cavity height, making the required diameters compatible with polymer fabrication methods. Moreover, the simulations indicate that the thermal conductivity of the cooler material has no impact on the thermal performance of the impingement cooler. A 4×4 array jet impingement cooler with 600 µm nozzles has been fabricated using mechanical machining in PVC and has been assembled to a test chip package. The experimental characterization shows a very low thermal resistance of 0.25 K/W (0.16 $cm^2$.K/W) and good temperature uniformity across the chip surface. The experimental validation shows a good agreement between both the unit cell, the full cooler CFD models, and the experimental results. The benchmarking study with literature data for impingement coolers with a large range of inlet diameters shows a very good thermal performance of the fabricated polymer cooler for a low required pumping power. The benchmarking study confirms that multi-jet cooling is more efficient than single jet cooling and that direct cooling on the backside of the semiconductor device is more efficient than cooling the substrate or base plate.

Furthermore, the comparison of the detailed temperature map measurements with the CFD modeling results indicates the need to include sufficient details on the heater structures in the CFD model in order to accurately predict the local temperature distribution in case of high heat removal rates at the chip surface for the multi-jet cooler, while for lower heat removal rates with the single jet cooler, a simpler model with uniform heating can be sufficient.

The validated CFD model of the multi-jet coolers has been applied to evaluate different nozzle configurations for the hot spots test case. The analysis shows that the coolers with distributed outlets achieve better cooling performance than coolers with common outlets since the cross-flow effects can be reduced. Moreover, it is shown that the misalignment of the nozzles with the hot spot locations results in a temperature increase of 10%, indicating the need for a matching design between the nozzle array and the chip floor plan. Finally, the measurement results on the single jet and multi-jet cooler have been used to derive the Nusselt correlations after correction for the heat losses in the cooler assembly. The obtained correlations are $\overline{\mathrm{Nu}}_j = 1.63Re_d^{0.57}$ and $\overline{\mathrm{Nu}}_j = 0.54Re_d^{0.56}$ for the multi-jet and single cooler, respectively.

The results of this chapter are partially published in the following publication:

**Tiwei Wei**, Herman Oprins, Vladimir Cherman, Jun Qian, Ingrid De Wolf, Eric Beyne, Martine Baelmans, "High-Efficiency Polymer-Based Direct Multi-Jet Impingement Cooling Solution for High-Power Devices," in IEEE Transactions on Power Electronics, vol. 34, no. 7, pp. 6601-6612, July 2019.

**Tiwei Wei**, Herman Oprins, et al., “Experimental characterization and model validation of liquid jet impingement cooling using a high spatial resolution and programmable thermal test chip [J]”, Applied thermal engineering, 2019.

## References

[1] Brian Boswell, M. N. Islam, Ian J. Davies, “A review of micro-mechanical cutting”, The International Journal of Advanced Manufacturing Technology, January 2018, Volume 94, Issue 1–4, pp.789–806.

[2] Shashi Prakash, Subrata Kumar, “Fabrication of microchannels: A review”, Proceedings of the Institution of Mechanical Engineers, Part B: Journal of Engineering Manufacture, vol. 229, 8: pp. 1273-1288. , First Published June 17, 2014.

[3] Vaezi, M.; Seitz, H.; Yang, S. F., A review on 3D micro-additive manufacturing technologies. Int J Adv Manuf Tech 2013, 67, (5-8), 1721-1754.

[4] Temiz, Y.; Lovchik, R.D.; Kaigala, G.V.; Delamarche, E. Lab-on-a-chip devices: How to close and plug the lab? Microelectron. Eng. 2015, 132, 156–175.

[5] Yao Gong, Jang Min Park, and Jiseok Lim, “An Interference-Assisted Thermal Bonding Method for the Fabrication of Thermoplastic Microfluidic Devices”, Micromachines (Basel). 2016 Nov; 7(11): 211. doi: 10.3390/mi7110211.

[6] Tsao, C.W.; DeVoe, D.L. Bonding of thermoplastic polymer microfluidics. Microfluid. Nanofluid. 2009, 6, 1–16.

[7] Satish C. Mohapatra, “An Overview of Liquid Coolants for Electronics Cooling”, ElectronicsCooling, May 1, 2006.

[8] Zare Chavoshi, S., & Luo, X. (2015). Hybrid micro-machining processes: a review. Precision Engineering, 41, 1-23. https://doi.org/10.1016/j.precisioneng.2015.03.001

[5] K. Gould, S. Q. Cai, C. Neft, and S. Member, “Liquid Jet Impingement Cooling of a Silicon Carbide Power Conversion Module for Vehicle Applications,” *IEEE Trans. Power Electronics*, vol. 30, no. 6, pp. 2975–2984, June 2015.

[6] A. S. Rattner, “General Characterization of Jet Impingement Array Heat Sinks With Interspersed Fluid Extraction Ports for Uniform High-Flux Cooling,” *J. Heat Transfer*, vol. 139, no. 8, p. 082201(1-11), August 2017.

[7] J. Jorg, S. Taraborrelli, G. Sarriegui, R. W. De Doncker, R. Kneer, and W. Rohlfs, “Direct single impinging jet cooling of a mosfet power electronic module,” *IEEE Trans. Power Electronics*, vol. 33, no. 5, pp. 4224–4237, May 2018.

[8] A. Bhunia and C. L. Chen, “On the Scalability of Liquid Microjet Array Impingement Cooling for Large Area Systems,” *J. Heat Transfer*, vol. 133, no. 6, 064501(1-7), January 2011.

[9] K. Olesen, R. Bredtmann, and R. Eisele, "'ShowerPower' New Cooling Concept for Automotive Applications," in *Proc. Automot. Power Electron.,* no. June 2006, pp. 1–9.

[10] E.N. Wang, L. Zhang, J.-M. Koo, J.G. Maveety, E.A. Sanchez, K.E. Goodson, and T.W. Kenny, "Micromachined Jets for Liquid Impingement Cooling for VLSI Chips," *J. Microeletromech. Sys*., vol. 13, no. 5, pp. 833-842, October 2004.

[11] T. Brunschwiler et al., "Direct liquid jet-impingement cooling with micronsized nozzle array and distributed return architecture," in *Proc. IEEE Therm. Thermomechanical Phenom. Electron. Syst.*, 2006, pp. 196–203.

[12] G. Natarajan and R. J. Bezama, "Microjet cooler with distributed returns," *Heat Transf. Eng.*, vol. 28, no. 8–9, pp. 779–787, July 2010.

[13] T. Acikalin and C. Schroeder, "Direct liquid cooling of bare die packages using a microchannel cold plate," in *Proc. IEEE Therm. Thermomechanical Phenom. Electron. Syst.*, 2014, pp. 673–679.

[14] S. Liu, T. Lin, X. Luo, M. Chen, and X. Jiang, "A Microjet Array Cooling System For Thermal Management of Active Radars and High-Brightness LEDs," in *Proc. IEEE Electronic Components and Technology Conf.*, 2006, pp. 1634–1638.

[15] Overholt MR, McCandless A, Kelly KW, Becnel CJ, Motakef S, "Micro-Jet Arrays for Cooling of Electronic Equipment," in *Proc. ASME 3rd International Conference on Microchannels and Minichannels*, 2005, pp. 249-252.

[16] M. Baumann, J. Lutz, and W. Wondrak, "Liquid cooling methods for power electronics in an automotive environment," in *Proc 2011 14th European Conference on Power Electronics and Applications*, 2011, pp. 1–8.

[17] B. P. Whelan, R. Kempers, and A. J. Robinson, "A liquid-based system for CPU cooling implementing a jet array impingement waterblock and a tube array remote heat exchanger," *Appl. Therm. Eng*., vol. 39, pp. 86–94, June 2012.

[18] S. Ndao, H. J. Lee, Y. Peles, and M. K. Jensen, "Heat transfer enhancement from micro pin fins subjected to an impinging jet," *Int. J. Heat Mass Transf.*, vol. 55, no. 1–3, pp. 413–421, January 2012.

[19] Y. Han, B. L. Lau, G. Tang and X. Zhang, "Thermal Management of Hotspots Using Diamond Heat Spreader on Si Microcooler for GaN Devices," *IEEE Transactions on Components, Packaging and Manufacturing Technology*, vol. 5, no. 12, pp. 1740-1746, December 2015.

[20] A. J. Robinson, W. Tan, R. Kempers, et al., "A new hybrid heat sink with impinging micro-jet arrays and microchannels fabricated using high volume additive

manufacturing," in *Proc. IEEE Annu. IEEE Semicond. Therm. Meas. Manag. Symp.*, 2017, pp. 179–186.

[21] H. Oprins, V. Cherman, G. Van der Plas, J. De Vos, and E. Beyne, "Experimental characterization of the vertical and lateral heat transfer in three-dimensional stacked die packages," *J. Electron. Packag*., vol. 138, no. 1, pp. 10902, March 2016.

[22] N. Zuckerman and N. Lior, "Jet Impingement Heat Transfer: Physics, Correlations, and Numerical Modeling," *Advances In Heat Transfer*, Vol. 39, pp. 565-631, 2006.

[23] J. Jorg, S. Taraborrelli, et al., "Hot spot removal in power electronics by means of direct liquid jet cooling," in *Proc. IEEE Therm. Thermomechanical Phenom. Electron. Syst.*, 2017, pp. 471–481.

[24] Skuriat , Robert, "Direct jet impingement cooling of power electronics,". PhD thesis , University of Nottingham, 2012.

[25] Y. Han, B. L. Lau, G. Tang, X. Zhang, and D. M. W. Rhee, "Si-Based Hybrid Microcooler with Multiple Drainage Microtrenches for High Heat Flux Cooling," *IEEE Trans. Components, Packag. Manuf. Technol.*, vol. 7, no. 1, pp. 50–57, 2017.

[26] R. Skuriat and C. M. Johnson, "Thermal performance of baseplate and direct substrate cooled power modules," in *Proc. 4th IET Int. Conf. Power Electron. Mach. Drives*, 2008, pp. 548–552.

[27] E. A. Browne, G. J. Michna, M. K. Jensen, and Y. Peles, "Microjet array single-phase and flow boiling heat transfer with R134a," *Int. J. Heat Mass Transf.*, vol. 53, no. 23–24, pp. 5027–5034, November 2010.

[28] E. G. Colgan et al., "A practical implementation of silicon microchannel coolers for high power chips," *IEEE Trans. Compon. Packag. Technol.*, vol. 30, no. 2, pp. 218–225, June 2007.

[29] C. S. Sharma, G. Schlottig, T. Brunschwiler, M. K. Tiwari, B. Michel, and D. Poulikakos, "A novel method of energy efficient hotspot-targeted embedded liquid cooling for electronics: An experimental study," *Int. J. Heat Mass Transf.*, vol. 88, pp. 684–694, September 2015.

[30] M. K. Sung and I. Mudawar, "Single-phase hybrid micro-channel/micro-jet impingement cooling," *Int. J. Heat Mass Transf.*, vol. 51, no. 17–18, pp. 4342–4352, August 2008.

# Chapter 6

# 6. 3D Printed Multi-jet Cooling

## 6.1 Introduction

In chapter 5, we introduced a chip level 3D-shaped polymer cooler fabricated using mechanical micromachining for a 4 × 4 nozzle array with 600 μm diameter nozzles. The multi-jet cooler can achieve heat transfer coefficients up to $6.25 \times 10^4$ W/m$^2$K with a pump power as low as 0.3 W. The results show that cost-efficient polymer-based fabrication can be used to create a high-performance chip level cooler with sub-mm nozzle diameters. The benchmarking study confirms furthermore that multi-jet cooling is more efficient than single jet cooling (introduced in chapter 4) and that direct cooling on the backside of the semiconductor device is more efficient than cooling on the substrate or base plate. However, the mechanical micromachined cooler requires the different individual parts to be fabricated separately and then to be assembled together. Furthermore, it limits the design of the cooler to simple, straight geometries that can be fabricated using micromachining.

In recent years, 3D printing or additive manufacturing has become an emerging fabrication technique in electronic packaging [7,8,9,10], by providing the opportunities of the embedded electronic components in a single module, multiple materials printing, and 3D-Package geometries with circuitry and components printing [11]. Typical 3D printing methods include (1) Stereolithography, (2) Fused deposition modeling (FDM), (3) Selective laser sintering, and (4) Inkjet printing [12]. Stereolithography (SL or SLA) uses either galvo-scanners to guide the UV lasers or projectors to cure photopolymers layer by layer. Typically, the fabrication tolerance is limited to a few hundred μm in commercial systems and further development is required to fabricate more performant coolers. In research tools, smaller feature sizes are possible since the resolution depends on the size of the printing platform with a fixed number of pixels. Bijan Tehrani [13] demonstrated a series of "smart" microelectronic packages utilizing fully-additive inkjet and 3D printing fabrication technologies, which includes 3D square encapsulants, microfluidic channels, and through-mold-via (TMV) interconnects. B. Goubault [14] built encapsulation packages and lids onto the silicon substrate using stereolithography technology (SLA).

3D printing technology also has great potential for the application of electronic cooling solutions. The main advantages of additive manufacturing are that it can use low cost materials for the cooler fabrication and the whole geometry in one piece can be printed while creating complex internal geometries. It was first introduced for the fabrication of the complex shapes obtained from topology optimization of air-cooled heat sinks [15] and later for more advanced liquid cooling solutions such as microchannel heat sinks [16] and impingement coolers [17-18]. R Jenkins et al. [17] demonstrated an aluminum alloy (AlSi10Mg) microchannel heat sink with straight, parallel channels by using the Direct metal laser sintering (DMLS) process, where inlet nozzle diameter is 500 μm. The fabrication of impingement coolers has been demonstrated for coolers with a common return and with relatively large nozzle diameters of 1 mm for metal using laser sintered 3D printing technology [18] and UV curable acrylic plastic using SLA [19]. Robinson et al. [18] demonstrated a high efficient microjet array cooler using a micro metal additive manufacturing process with 30 μm diameter nozzles. However, metal-based bare die cooling solutions are expensive and have a high risk for device reliability due to the introduction of metal (Cu, Al) in VLSI devices.

Currently, the highest resolution by 3D printing can be achieved through the Two-Photon Polymerization (TPP) process [12], which is one of 3D micro/nanoscale manufacturing technologies for arbitrary 3D structures with sub-100 nm resolution. Most of the materials used for TPP are designed for conventional lithographic applications, including negative and positive photoresists. However, the TPP process is relatively slow and small for this application. Alternatively, the Stereolithography (SL or SLA) process, which uses similar materials as the TPP but use either galvo scanners to guild the UV lasers or projector (when a projector is used the process is called DLP, or Digital Light Processing) to cure photopolymers layer by layer. The resolution could result in micrometer range (for example, 1 micron in the Z direction (layer thickness) and a few to tens of microns in XY direction (pixel size). A comprehensive review of other micro Additive Manufacturing/3D printing technologies, such as SLM, Paste Extrusion, 3DP process, etc., could be found in the literature [12].

In this chapter, we present the design, fabrication, and experimental characterization of Multi-jet cooler with sub-mm diameter nozzles by using cost-effective 3D printing technologies. In order to evaluate the thermal performance, the fabricated 3D printed cooler is assembled to the 8 mm × 8 mm thermal test chip with integrated heaters and temperature sensors (introduced in chapter 2). In the first section of this chapter, the design of the impingement cooler with different nozzle arrays, and the design limitations related to the 3D printing fabrication tolerances are discussed. Next in Section 6.3, the demonstrations of the 3D printed cooler are investigated. Besides, an analysis of the tolerances of the fabricated coolers is presented, and a new defect detection metrology based on scanning acoustic microscope (SAM) measurements is introduced to detect defects inside the printed cooler. In Section 6.4, the experimental thermal and hydraulic characterization of the fabricated cooler are discussed. Moreover, the thermal performance of the printed cooler is benchmarked with the performance of a conventional air-cooling heat sink and the micromachined cooler. In the last section, the experimental results from the micromachined cooler and 3D printed cooler are used for the validation of the $\overline{\mathrm{Nu}_j}$-$\mathrm{Re_d}$ predictive model extracted in chapter 3.

## 6.2 Design of 3D printed cooler

### 6.2.1 Design constraints and critical parameters

The schematic of the 3D printed impingement jet cooler is the same as the micromachined cooler, including four critical layers: inlet plenum, outlet plenum, nozzle plate, and impingement cavity layer. The inlet plenum is used as the flow distributor to distribute the cold water inside the 4 × 4 inlet nozzles. After the fluid impinged onto the chip backside, the outlet plenum is used as a flow collector for the flow from the outlet nozzles. As shown in Figure 6.1(a), the geometrical design of the 3D printed microjet cooler should be taken into account the manufacturability of 3D printing technology, including printing resolutions, minimal feature size, and bridging of the cavity. The nozzle diameter $d_i$ and the nozzle plate thickness $t$ are the critical dimensions since they determine whether the excess liquid resin can be removed sufficiently through the tiny nozzle channels. This will form a constraint for designing nozzle diameters and plenum height in order to allow a successful draining of the excess resin. Therefore, the resin removal in the cooler design with small nozzle diameters and with limited plenum thickness is the major challenge for the use of additive manufacturing. Moreover, the nozzle wall with thickness $W$ used as the separation between the inlet nozzle and the outlet plenum should be strong enough to withstand high flow pressure and prevent leakage. As shown in Figure 6.1(b), the gap $S$ between the external walls of two adjacent inlet nozzles in the outlet plenum is also critical. In this outlet plenum layer, the walls of the inlet nozzles that go vertically through this layer act as a divider between the (vertical) inlet flow and the (horizontal) outlet flow. By decreasing the gap S (for a fixed nozzle pitch and internal diameter, thus increasing the wall thickness $W$), the available area for the coolant to flow in the horizontal outlet plenum reduces. Since the pitch and internal diameter do not change, there is no impact on the pressure drop for the vertical direction through the nozzles. However, the pressure drop in the outlet plenum where the coolant flows through the inlet/outlet dividers increases as the nozzle wall thickness $W$ becomes thicker for a fixed nozzle pitch.

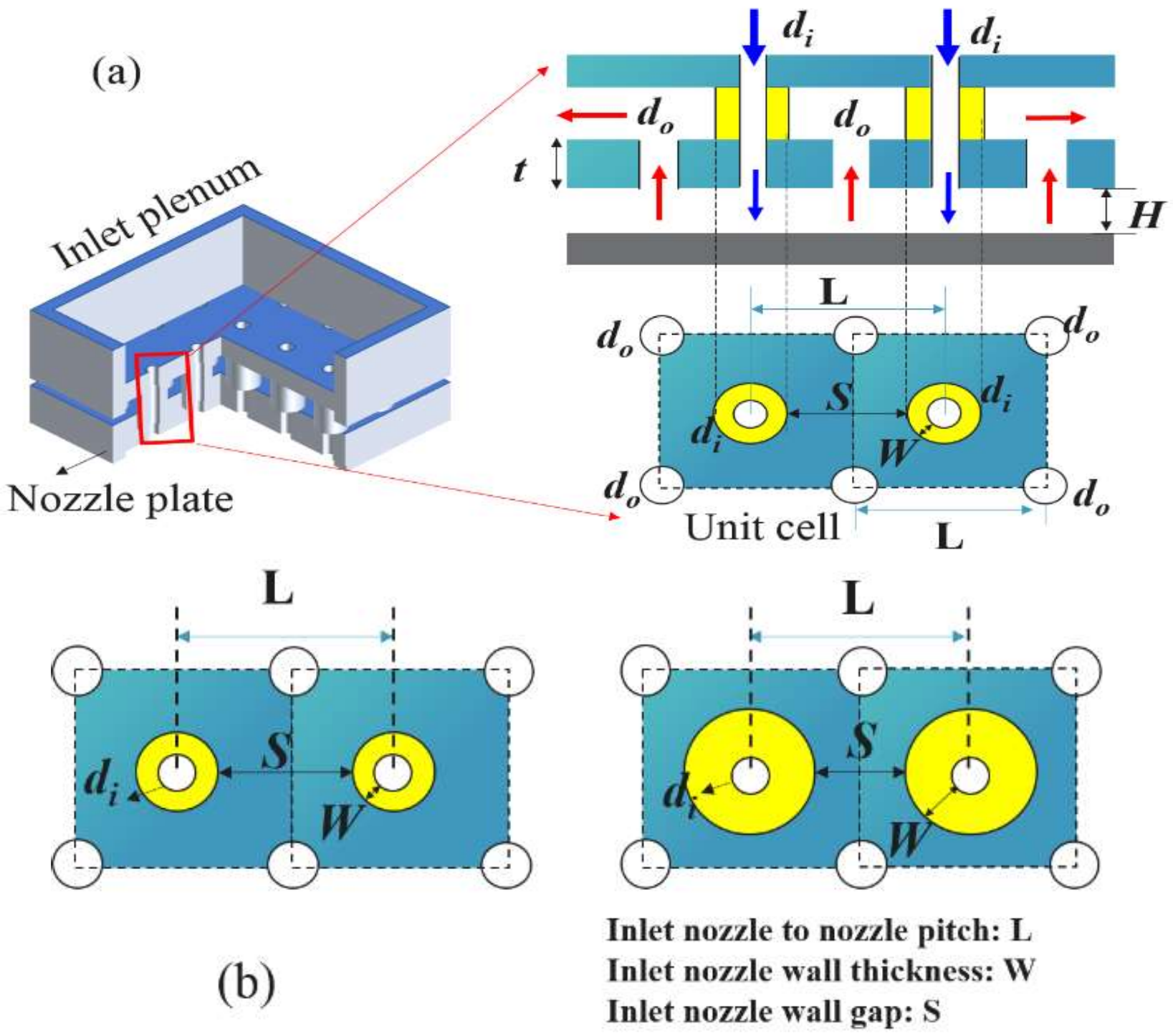

**Figure 6.1:** Design consideration: (a) critical design parameters limitations for the use of 3D printing for impingement coolers; (b) indication of the impact of wall thickness on the gap S in the outlet plenum between adjacent inlet nozzles.

Other design parameters such as tube connections, cavity height $H$, and the O-ring groove are also taken into account for a better design. The inlet connection is designed in the center of the cooler in order to improve the flow distribution over the different inlet nozzles. The rest of the available space in the cooler material can be used to improve the outlet chamber design to help the outlet flow evacuation. In chapter 5, the impact of the cavity height $H$ variation on the cooler thermal/hydraulic performance was investigated. The studies show that the cavity height above 0.5 mm has a negligible impact on cooling performance and pressure drop. From the material point of view, the material compatibility with the liquid coolant should also be taken into account, which determines the cooler reliability. Moreover, the cooler material should have a low water absorption ratio and high-temperature resistance.

Therefore, in order to design a 3D printed cooler with sub-mm dimensions, the critical design parameters can be listed as below:

- The nozzle inlet and outlet diameters $d_i$ and $d_o$ and the gap $S$ between two inlet nozzles can result in a resin removal issue due to the narrow gap;
- The nozzle sidewall, with thickness $W$, should be sufficiently strong to prevent the wall from breaking, which can result in "short-circuit" between inlet flow and outlet flow;
- The plenum cavity thickness should be sufficient to support the structure since a thin cavity wall can result in structural deformation;
- The cooler material should be compatible with the liquid coolant: low water absorption and high-temperature resistance, compatible mechanical properties to limit the reliability issues.

### 6.2.2 Cooler design for test chip

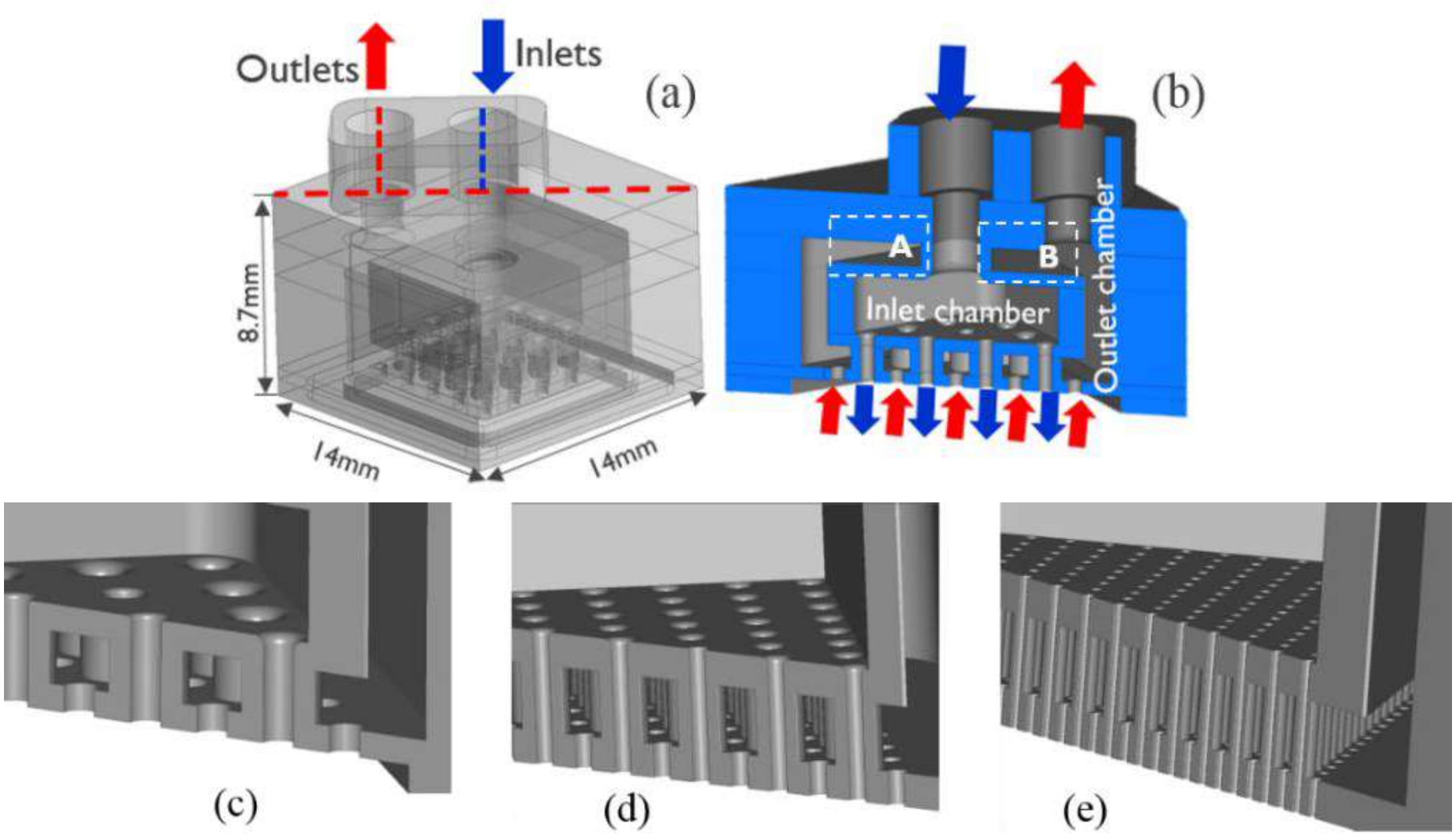

**Figure 6.2:** CAD structure of the designed 3D printed coolers: (a) CAD structure and (b) cross section view of the designed 3D printed 4 × 4 cooler; nozzle details of 3D printed cooler with (c) 3×3 nozzle array, (d) 4×4 nozzle array and (e) 8×8 nozzle array.

Based on the manufacturability of the 3D printing technology and the critical design parameters constraint listed in section 6.2.1, the 3D printed cooler with 4 × 4 inlet jet array is first demonstrated to compare with the micromachined cooler. For the comparison of the two different fabrication technologies, the critical design parameters for the 3D printed cooler is kept as the same with micromachined cooler. The proposed design of the 4 × 4 inlet jet array cooler with 600 µm diameter nozzles for the 8 mm × 8 mm thermal test chip is shown in Figure 6.2(a) and Figure 6.2(b). Moreover, two other different versions of impingement jet cooler with 3×3 and 8×8 nozzle arrays are also demonstrated to study the nozzle density impact. The details of the nozzle array with different nozzle densities are illustrated in Figure 6.2. All the nozzle arrays are designed on the 8×8 $mm^2$ PTCQ thermal test chip introduced in chapter 2. Therefore, the maximum nozzle density for the cooler demonstration is 100 /$cm^2$. For all the coolers, the same inlet diameter ratio $d_i$/L of 0.3 and the same ratio for the wall thickness are used, resulting in nozzle diameters ranging from 300 to 800 μm. This design takes into account the manufacturability aspects of the used 3D printer as well as the critical design parameters described above. For practical considerations, the cavity height chosen in this study is 0.6 mm in the impingement regime, where no significant impact on the heat transfer coefficient is observed.

Moreover, the wall thickness of the nozzles $W$ and the gap between 2 nozzles in the outlet plenum $S$ are both 400 μm. The impact of the inlet plenum dimensions has been investigated to improve the flow distribution uniformity shown in chapter 5. The results show that a lower plenum height can generate more flow maldistribution, with higher velocity concentrating in the nozzles in the center of the cooler. The inlet diameter and plenum height both should be considered when designing the impingement cooler. Therefore, the inlet chamber thickness is chosen as 2.5 mm, based on the tradeoff between the cooler size and flow distributions. Moreover, both in the inlet chamber and the top part of the outlet chamber, support pillars have been added to ensure the structural integrity. The geometry dimensions comparison between the micromachined cooler and 3D printed cooler are summarized in Table 6.1.

**Table 6.1:** Geometry comparison between MM cooler and 3D printed cooler.

| Geometry | | MM cooler | 3D printed Cooler |
|---|---|---|---|
| Nozzle array | $N$ | 4×4 | 4×4 |
| Inlet chamber height | | 3 mm | 2.5 mm |
| Inlet diameter | $d_i$ | 0.6 mm | 0.6 mm |
| Outlet diameter | $d_o$ | 0.6 mm | 0.6 mm |
| Cavity height | $H$ | 0.6 mm | 0.6 mm |
| Nozzle plate thickness | $t$ | 1 mm | 0.55 mm |
| Cooler size | $x,y,z$ | 46×46×13 ($mm^3$) | 14×14×8.7 ($mm^3$) |

Figure 6.3 shows the internal structure of the micromachined cooler and the 3D printed cooler side by side. In general, the process flow of the 3D printed cooler is simple since all the parts can be printed as a whole as a single part while creating the complex 3D geometries, including cavities and O-ring grooves. For the micromachined cooler, all the individual parts have to be fabricated separately and glued together, which might cause a higher risk of water leakage. Furthermore, the jet nozzles must be drilled by micro-drilling, where the nozzle diameter is limited by the mill tool diameter. As indicated in Figure 6.3(b), inlet and outlet divider can be printed as hollow cylinders to separate the inlet flow and outlet flow, which can significantly reduce the pressure drop compared to square shape dividers shown in Figure 6.3(a). In summary, compared to micromachining prototype, the 3D printed cooler has a lower fabrication cost, more flexible and customizable design, and finer resolution of the internal structures.

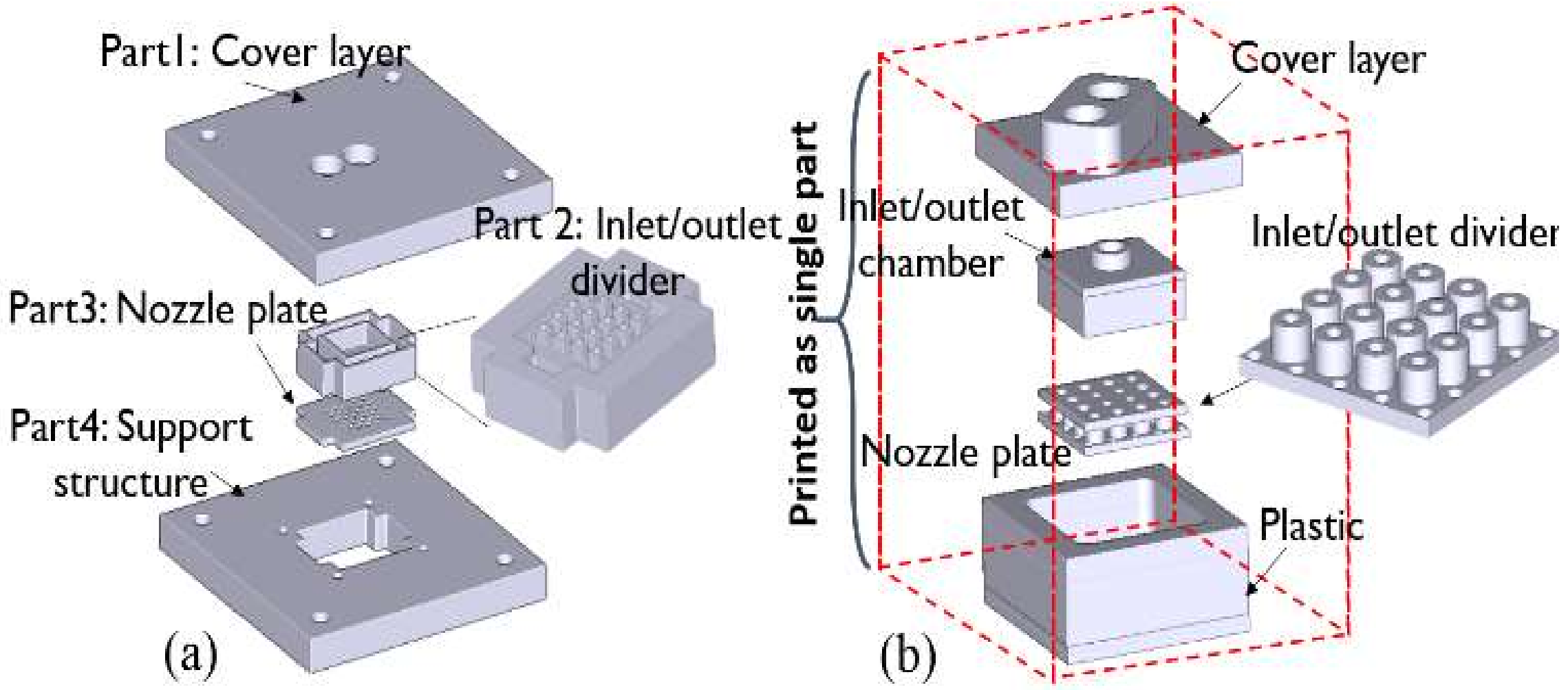

**Figure 6.3:** Internal CAD structures comparison between (a) mechanical micromachined cooler [6] and (b) 3D printed cooler.

Figure 6.4 shows a cross-section of the printed cooler to visualize the flow inside the internal structure. It can be seen that, by the use of 3D printing, more outlet chamber space can be designed in order to improve the evacuation of the outlet flow at a reduced pressure drop.

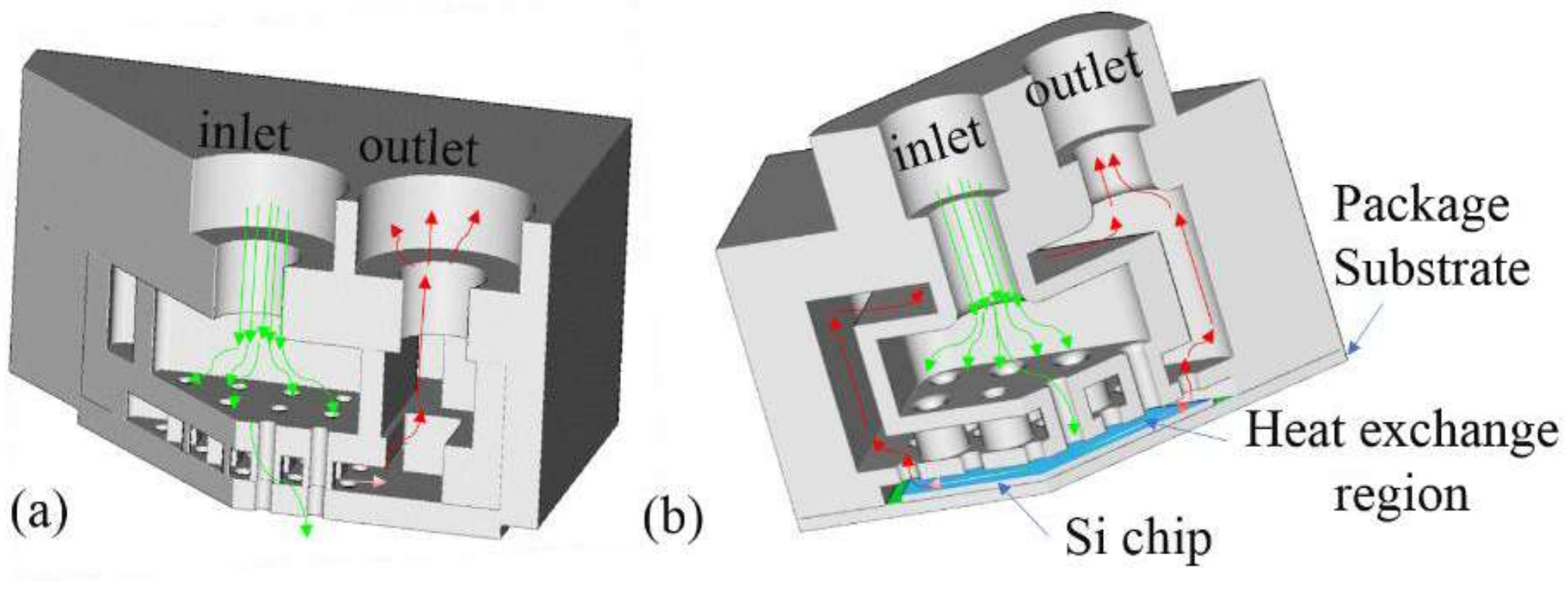

**Figure 6.4:** Cooler structure with pressure reduction through improved internal geometry: (a) micromachined cooler; (b) 3D printed cooler.

### 6.2.3 Modeling study: micromachined and 3D printed cooler comparison

In order to evaluate and compare the thermal and hydraulic performance of the two coolers with different fabrication techniques, computational fluid dynamics (CFD) modeling is used during the initial design stage. The meshing details of CFD models based on MM cooler and 3D printed cooler are both shown in Figure 6.5. It should be noted that Figure 6.5(a) and Figure 6.5(b) are shown on the same scale, highlighting the compact design of the printed cooler. An adiabatic wall is assumed for the CFD cooler model. We assume that there is no heat loss through the plastic material of the cooler due to the low thermal conductivity. Previous simulations discussed in chapter 4 have shown that there is no significant difference for the chip temperature between the model with a fully modeled cooler with low conductivity material and the model where only the fluidic channels without cooler were considered.

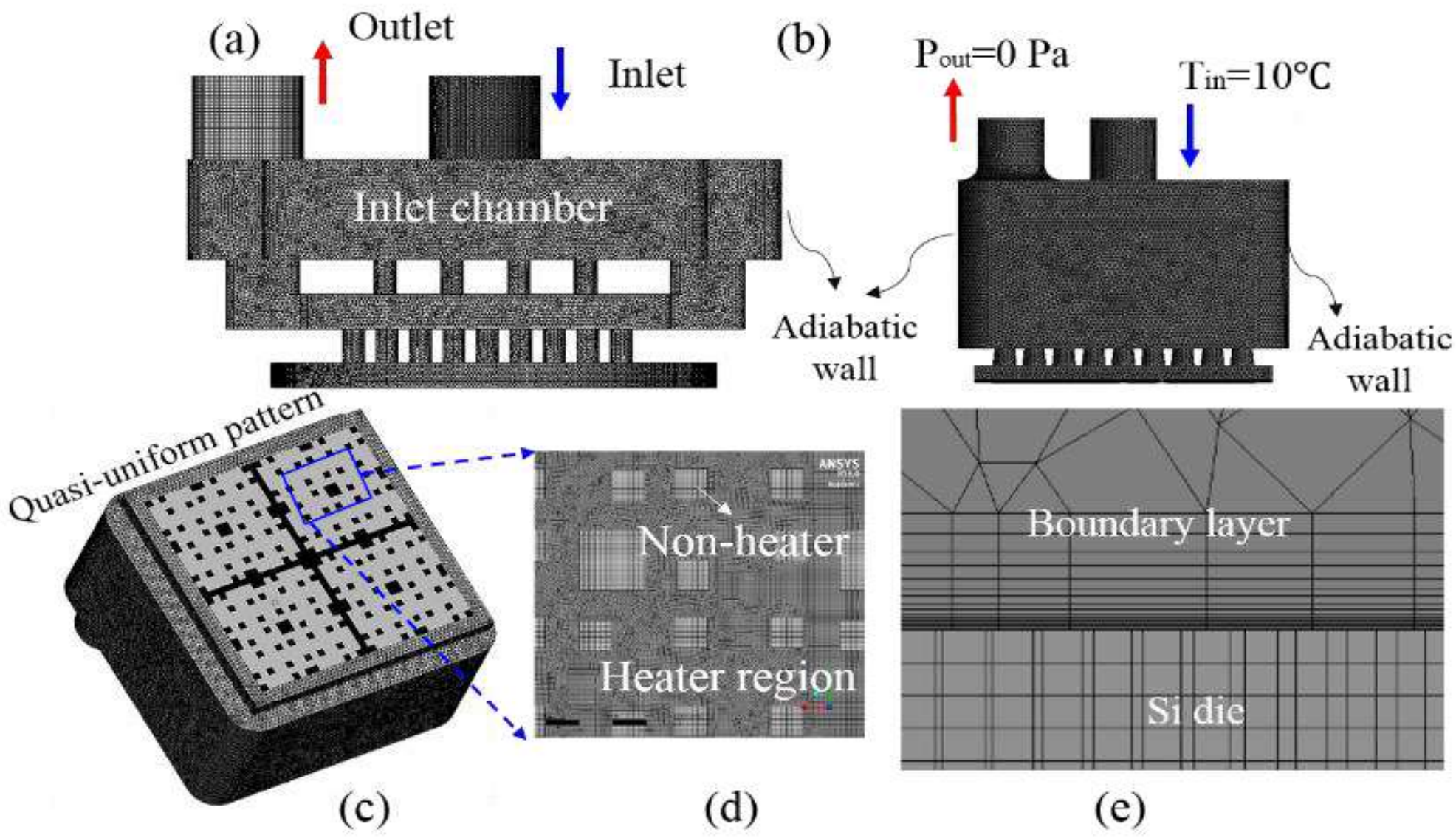

**Figure 6.5:** CFD models with mesh for the comparison of the micromachined (MM) and 3D printed coolers: (a) MM cooler model; (b) 3D printed cooler; (c) bottom view of 3D printed cooler model mesh with quasi-uniform heater cells in the test chip; detailed view of heater cell meshing (d) and boundary layer meshing (e).

**Table 6.2:** GCI meshing sensitivity analysis of full model.

| Temperature | $GCI_{12}$ | Asymptotic range of convergence |
|---|---|---|
| Stagnation Temp | 0.002 | 0.99 |
| Averaged chip Temp | 0.004 | 1.01 |

Based on the mesh sensitivity study, the number of elements for the micromachined cooler demonstrator and the 3D printed cooler is 6 million and 4 million, respectively. The *Re* numbers based on the nozzle diameter considered in this study are in the range from 10 to 3500. At the maximum considered flow rate, the flow inside the cooler is, therefore, slightly turbulent or in the transitional regime. Therefore, the transition shear stress transport (SST) model is still used as the turbulence model in this chapter. The liquid used in this study is deionized (DI) water, and the inlet temperature is kept at 10°C. For both cases, the chip power is 50W, with the chip heated area of 8 × 8 × 75% $mm^2$. The details about the chip and its heating elements are already introduced in chapter 2.

Figure 6.6 shows the pressure distribution results from the full cooler level simulations for both coolers. The comparison shows a significantly lower pressure drop for the printed cooler. Exploiting the capabilities of 3D printing to design the internal cooler geometry results in a reduction of the pressure drop by 24% compared to the geometry of micromachined cooler.

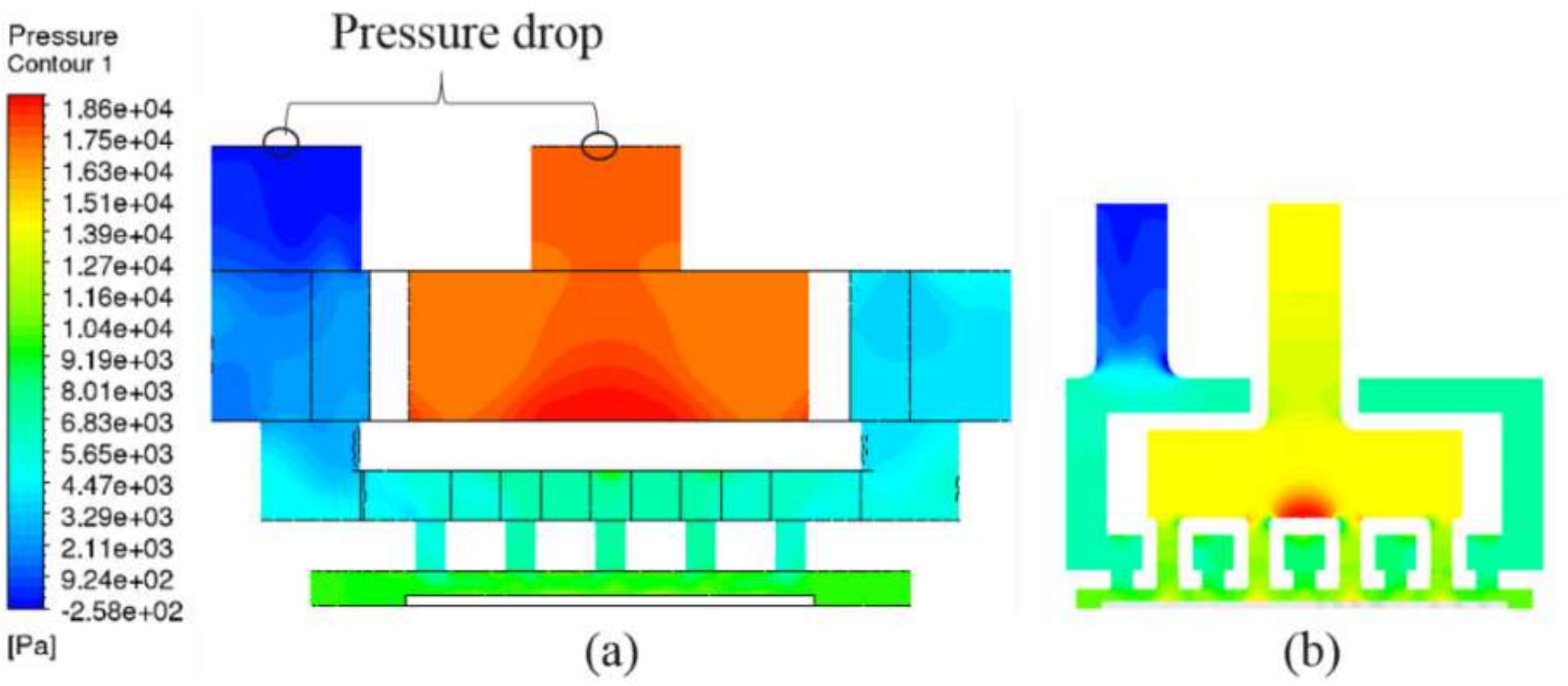

**Figure 6.6:** Pressure comparison between micromachined and printed cooler for a flow rate of 650 mL/min drawn in the same scale: (a) micromachined cooler with pressure drop $\Delta P_{total}$= 0.18 bar; (b) 3D printed cooler with pressure drop $\Delta P_{total}$ = 0.15 bar.

It should be noted that the main reason for the pressure reduction is the difference in internal geometry. As shown in Table 6.1, the critical nozzle parameters are the same for both coolers (at least the nominal design values): inlet/outlet nozzle diameters, cavity height, nozzle array. However, other parameters are different (nozzle plate thickness, outlet chamber, and inlet chamber). These values are linked to the fabrication capabilities, and therefore, they are inherently part of the comparison. Other aspects that contribute to the pressure reduction are the cylindrical shape of the inlet/outlet (less pressure drop) and rounded inlet/outlet nozzle transitions from/to the plenum. 3D printing allows making those features with less pressure drop compared with the micromachined demonstrator.

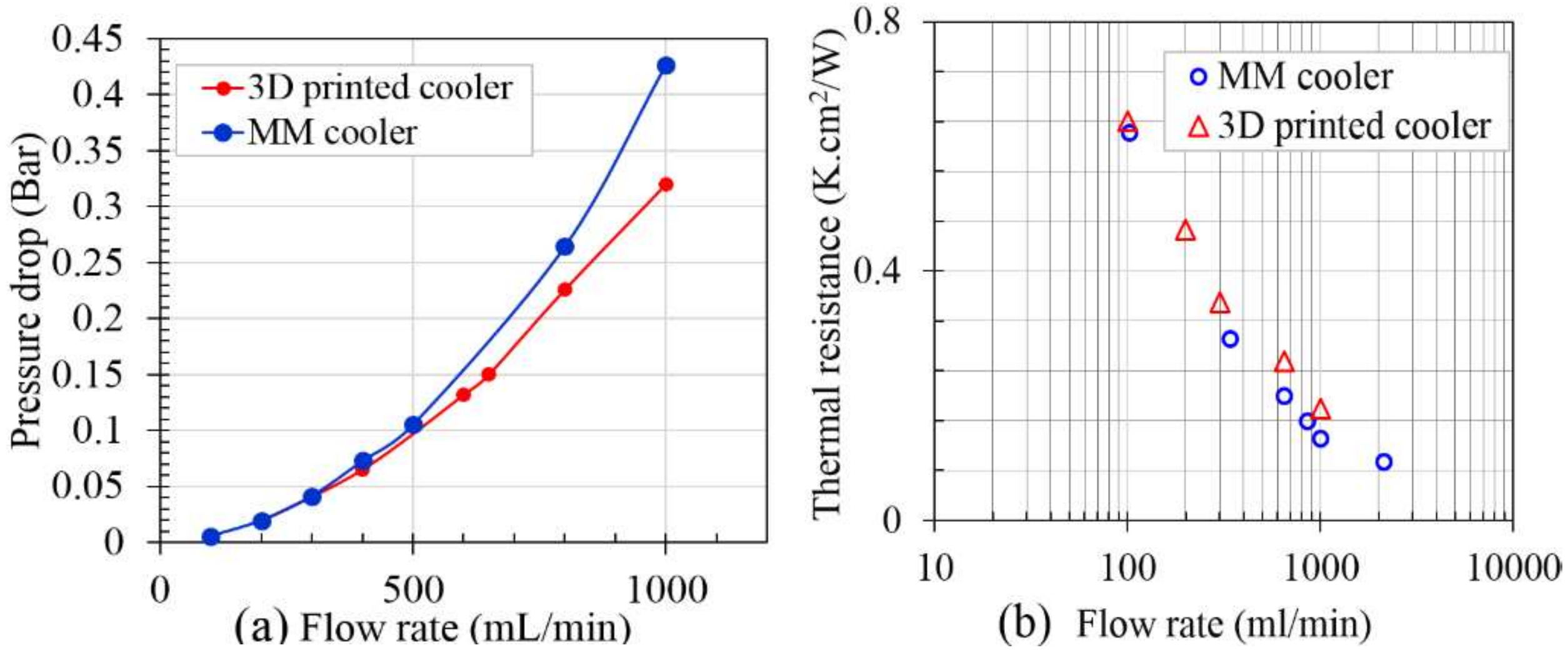

**Figure 6.7:** Comparison between MM cooler and 3D printed cooler based on modeling study: (a) pressure drop comparison as a function of flow rate; (b) thermal resistance comparison.

In Figure 6.7, the thermal resistance and pressure drop as a function of flow rate are compared for the micromachined cooler, and 3D printed cooler. The general trends show that the pressure drop can be reduced by using 3D printing, while the thermal resistance is similar due to the same design of the nozzle plate.

## 6.3 3D printed cooler demonstration

### 6.3.1 Cooler fabrications

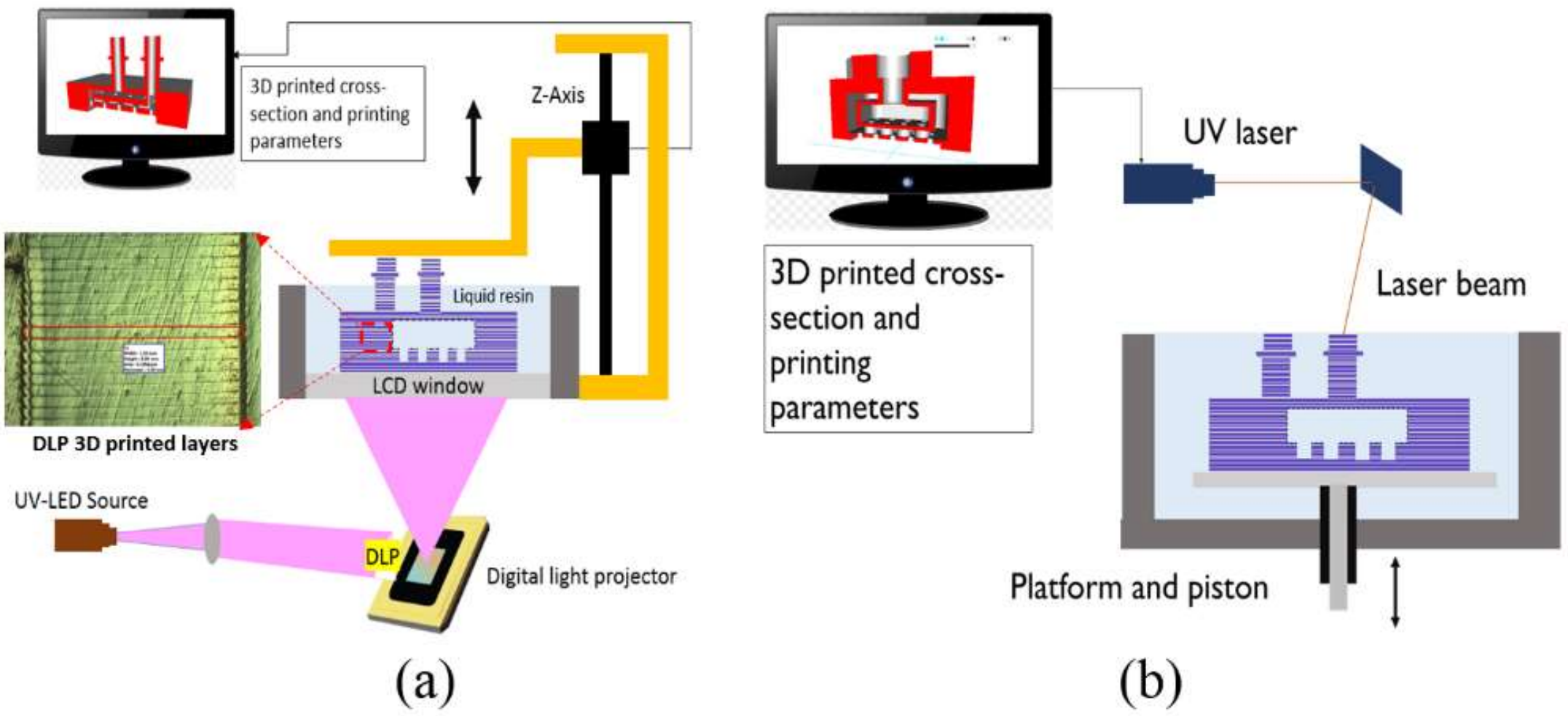

**Figure 6.8:** Principle of the 3D printing technology: (a) 3D printed cooler with DLP process; (b) 3D printed cooler by SLA. (figure adapted from Texas Instruments [31])

For the manufacturability investigation of 3D printing technology, two different methods are implemented for the demonstration: Stereolithography (SLA) and digital light processing (DLP), which are the two most common processes for polymer-based 3D printing. The SLA process uses either galvo scanners to guild the UV lasers while DLP used a projector to cure photopolymers layer by layer. The resolution could result in the micrometer range and a few to tens of microns. In this section, the cooler demonstrations based on these two technologies, including the manufacturability, and material compatibility are investigated.

6.3.1.1 First demonstrator: DLP with ABS

In the first attempt, DLP (digital light projector) was used with a standard ABS (acrylic-based plastic) material. In the DLP process, is used to curing photo-reactive polymer (liquid resin). The minimal feature size is 200 μm in optimal conditions. In order to investigate the manufacturability of the DLP based 3D printed cooler, two different cooler versions are designed, as illustrated in Figure 6.9: a cooler with a 4×4 inlet nozzle array with feature sizes beyond the claimed capabilities of the 3D printing tool and a more conservative cooler design with a 3×3 inlet nozzle array with more relaxed dimensions. The designed nozzle diameter for 4×4 nozzle array cooler is 0.6 mm and 1 mm for 3×3 nozzle array cooler. For the cooler material, acrylic-based plastic is selected based on the "bridging" capabilities and limitations of DLP capabilities.

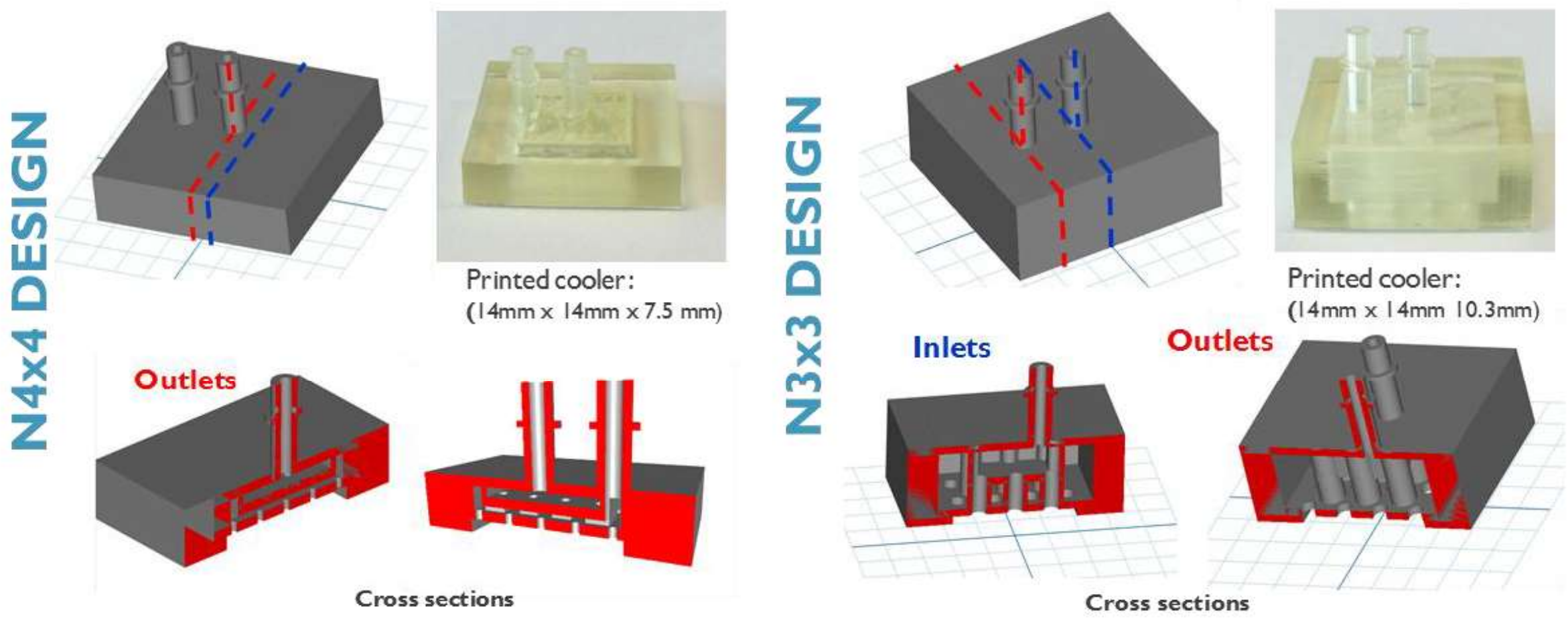

**Figure 6.9:** 3D printed cooler designs and demonstrations with Digital light processing (DLP): (a) 4×4 nozzle array; (b) 3×3 nozzle array.

The fabrication results are also shown in Figure 6.9. In general, the liquid resin could not be sufficiently removed from 4×4 printed cooler through the small dimension nozzles, while for the 3×3 design with 1 mm nozzles, the uncured resin removal from the inside of the cooler was successful. The microscope image with a cross-section of the fabricated 3×3 using DLP is shown in Figure 6.10. The inlet nozzle channels, inlet delivery channel, outlet chamber, and the cavity height are all indicated in Figure 6.10. The photographs clearly show the individual layers of 100 μm of the 3D printing process, the successful fabrication of the internal geometry with the different plenums.

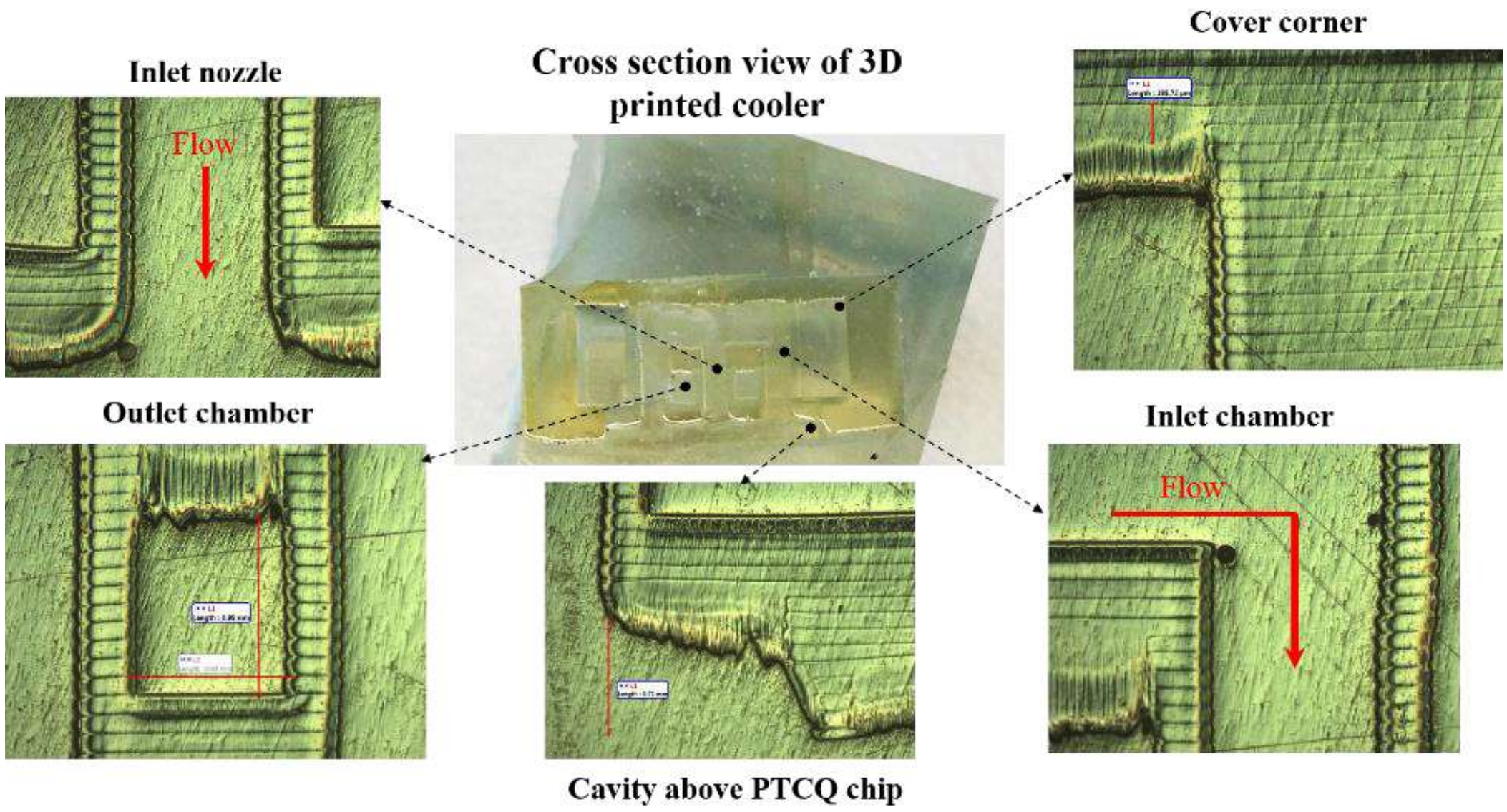

**Figure 6.10:** Cross-section view of 3D printed cooler indicating different locations inside the cooler.

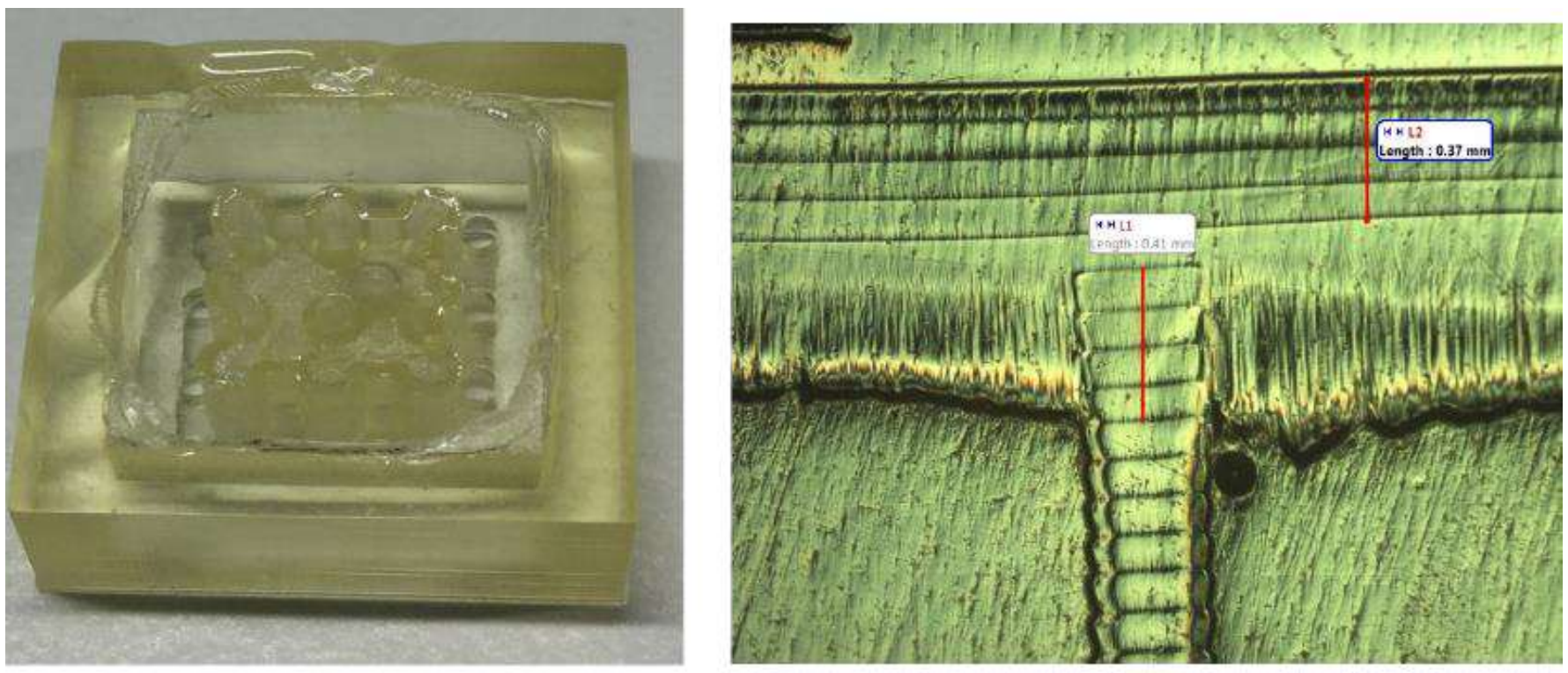

(a) (b)

**Figure 6.11:** Issues with 3D printed 3×3 cooler: (a) water absorption test results; (b) defects at the cover layer.

After that, the issues with the 3D printed 3×3 array cooler using acrylic based plastic are investigated. The 3D printed cooler is immersed inside DI water for 24 hours. The water absorption test results are shown in Figure 6.11. The first issue we observed is that the plastic material absorbs water during the test. The material becomes soft in the top part, which has potential defects, resulting in leaks. Another issue we identified is the defects at the cover layer of the cooler, which is not well structured. The conclusion from the first experiment shows that polymer-based 3D printing can create the complex internal geometries for package-level impingement coolers, but that the material aspects (defect-free fabrication and water resistance) are very important. Therefore, other materials with better coolant compatibility should be investigated to improve the cooler reliability.

6.3.1.2 SLA: watershed material

In the next step, other cooler materials are studied. The selected polymer material, Somos WaterShed XC [32], is a water-resistant material, which shows ABS-like properties and excellent temperature resistance. The material properties of this material are listed in Table 6.3. The heat deflection temperature (HDT) of the printed material is around 60°C. Therefore, the temperature of the coolant should be below 60°C to remain in the safe temperature range for the cooler material. In this experiment, the 3D printed cooler was printed using stereolithography (SLA) with a reported minimal feature size of 130 μm and with 50 μm layer thickness. The printed demonstrator is shown in Figure 6.12. The bottom view of the nozzle plate with inlet/outlet nozzles shows that all the holes are printed uniformly across the cooler. The cross-section analysis of the printed 14×14 $mm^2$ cooler confirmed the successful printing of the cooler and its internal structure and revealed no left-over resin residuals.

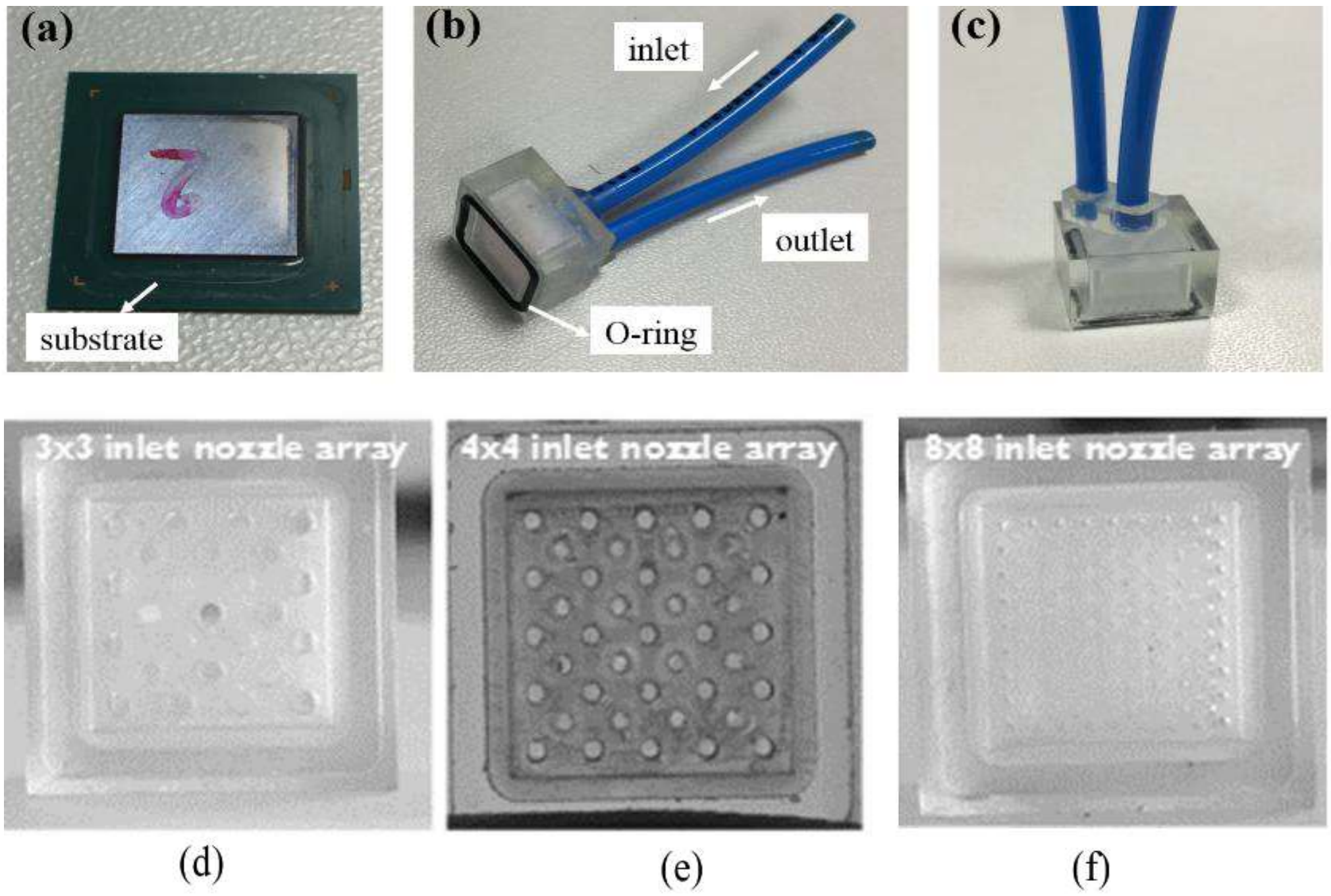

**Figure 6.12:** Demonstration of 3D printed cooler: (a) (b) and (c) photograph of the 3D printed cooler; (d)-(f) bottom view of the nozzle plate with inlet/outlet nozzles for different nozzle arrays.

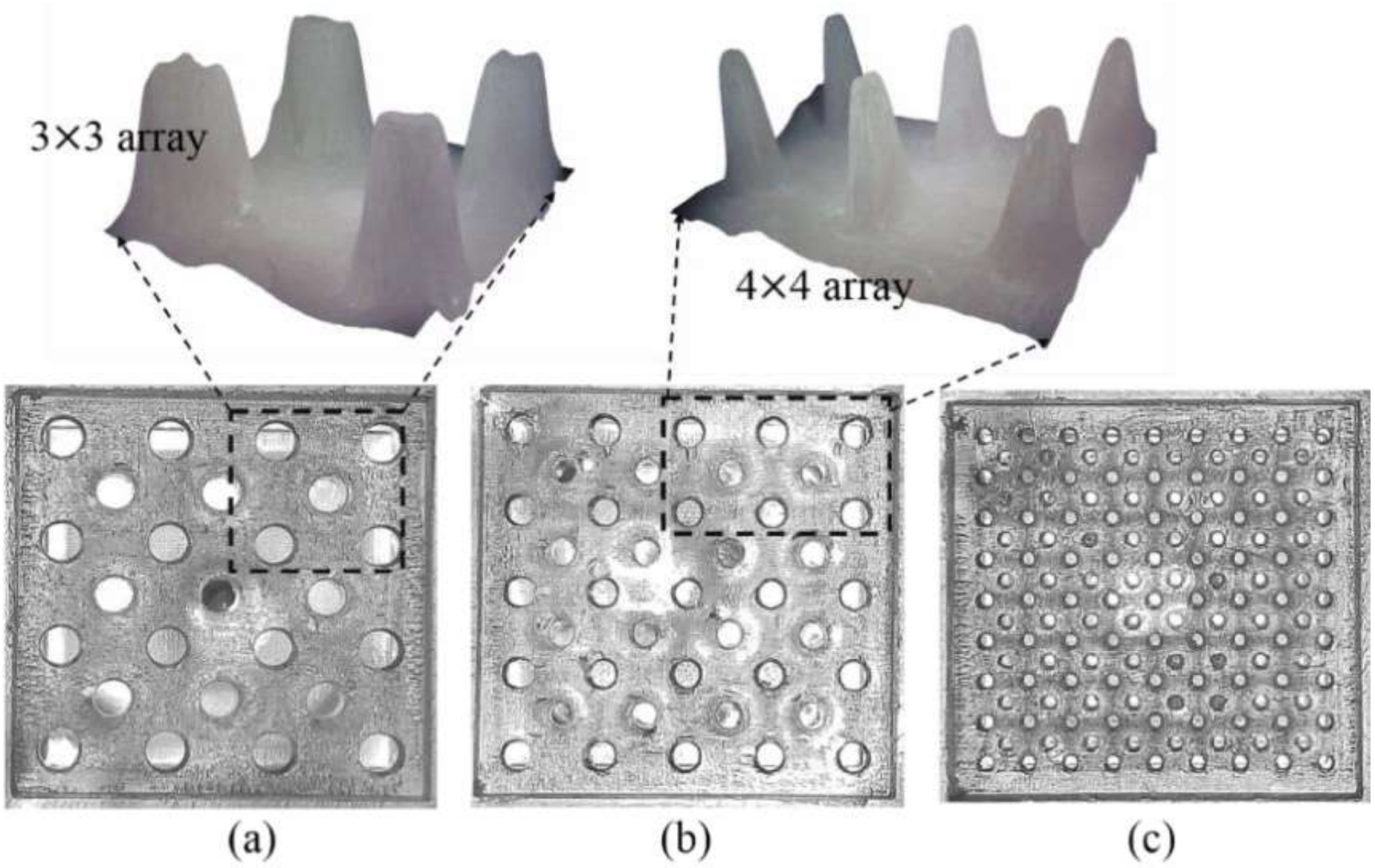

**Figure 6.13:** Demonstration and bottom view of the nozzle plate of the fabricated 3D printed coolers: (a) 3×3 (b) 4×4 and (c) 8×8.

**Table 6.3:** Material properties of 3D printed cooler.

| **Material properties** | **Water absorption** | **Tg (ºC)** | **HDT@0.48MPa HDT@ 1.81MPa** | **CTE ppm/K** | **Dielectric constant** |
|---|---|---|---|---|---|
| Watershed | 0.35% | 39 – 46 | 45.9 - 54.5ºC 49.0 - 49.7ºC | 90 - 96 | 3.9 - 4.1 (60HZ) |

With the high-resolution SLA, the coolers with 3×3, 4×4 and 8×8 inlet nozzle arrays are successfully printed using one single process without assembly of the individual parts. The bottom view of different fabricated coolers is illustrated in Figure 6.13. The 3D reconstructed microscope images are also shown in Figure 6.13. It can be seen that all nozzles are functional with no blockages or trapped resin observed even for the smallest 8×8 design. Figure 6.14 shows the distribution of the measured nozzle diameters for the different nozzle arrays. The measured average nozzle diameter is 950 µm for 3×3 (nominal design of 800 µm), 575 µm for 4×4 (nominal design of 600 µm), and 380 µm for 8×8 (nominal design of 300 µm). In general, the actual nozzles diameters are within an 18% difference than the nominal design values for 4×4 and 8×8. However, the difference with 26% is higher for 3×3 with the nominal nozzle diameter of 300 µm. This is due to the higher measurement uncertainty for the small nozzle diameter with relatively larger roughness. The fabricated larger nozzle diameter is expected to have a higher thermal resistance under the same flow rate but exhibits a lower pressure drop (larger diameters) and therefore, will require a lower pumping power. The fabrication tolerance and its impact on the thermal and hydraulic performance will be investigated in Section 6.3.3.

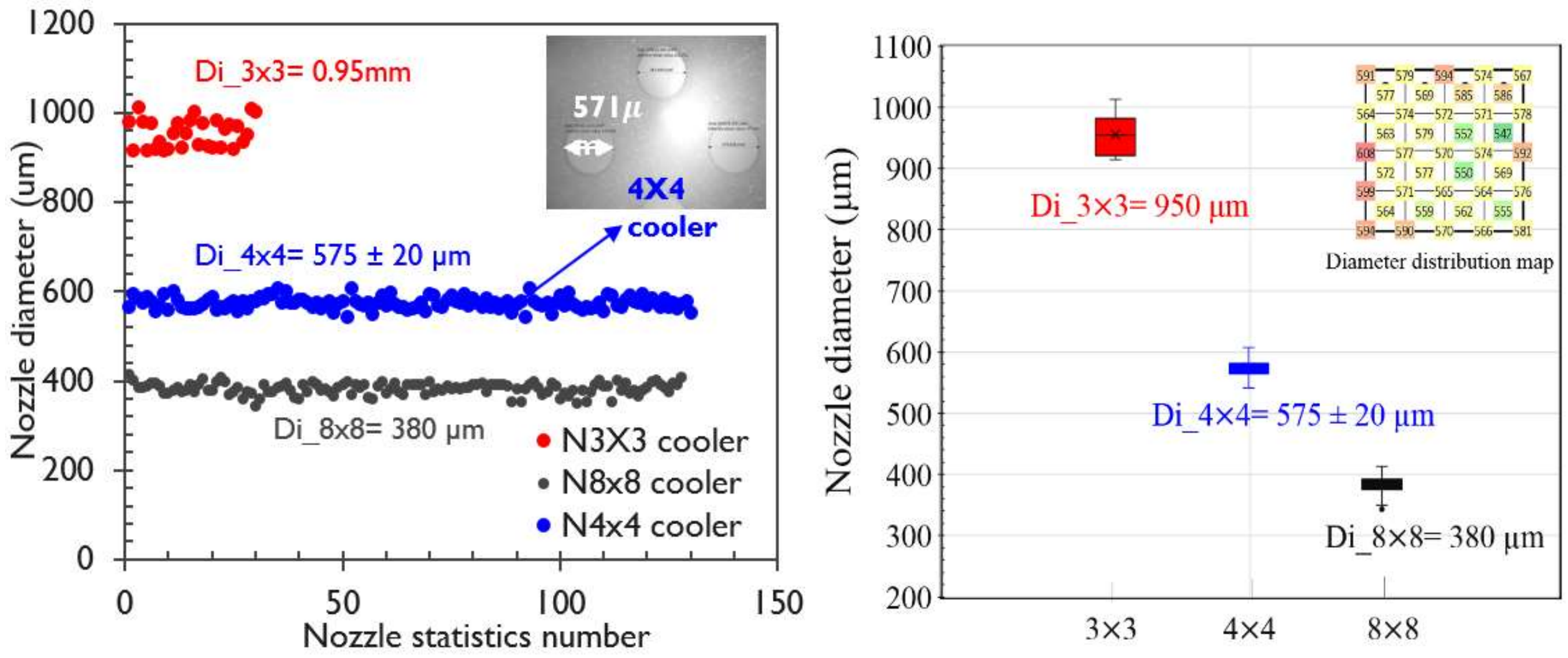

**Figure 6.14:** Nozzle diameter statistics with different nozzle array.

### 6.3.2 Defect measurements and analysis

Microscopy measurement and cross-section analysis is useful methods for the evaluation of the cooler, as shown in Figure 6.15. For the microscopy measurement, the bottom view of the nozzle plate is measured, as shown in Figure 6.15(b). However, it can only check the open nozzles in the nozzle plate. As for the cross-section analysis in Figure 6.15(c), it can help to check the internal channel, but it is a destructive measured technique.

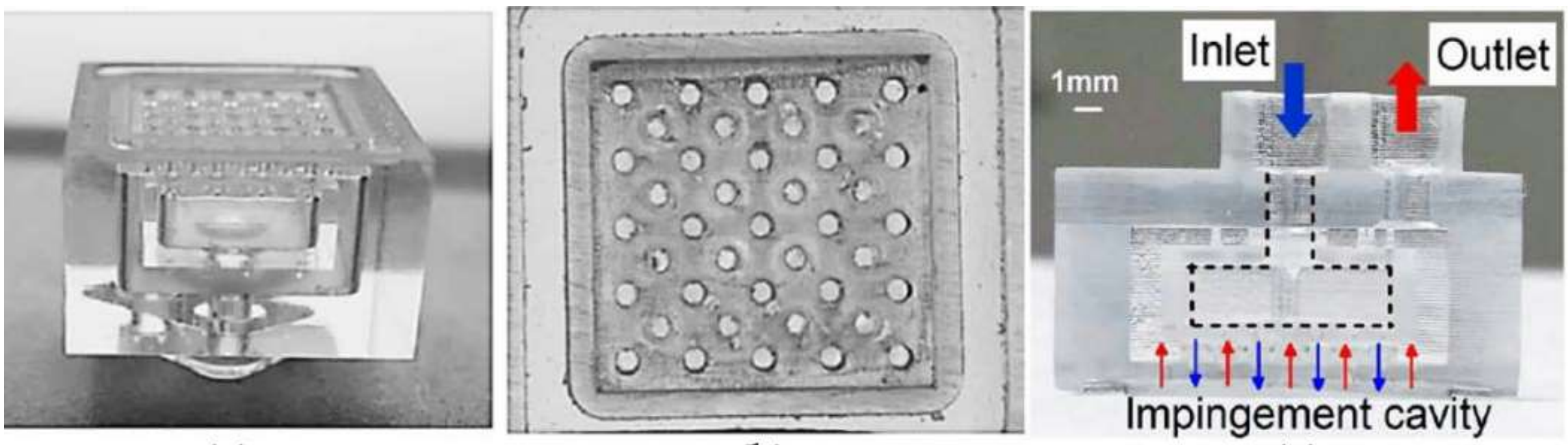

(a) (b) (c)

**Figure 6.15:** Demonstration of 3D printed cooler: (a) photograph of the 3D printed cooler; (b) bottom view of the nozzle plate with inlet/outlet nozzles; (c) detail of the channels in cross-section of the cooler.

In the first step of the fabrication assessment, specific geometry details are compared between the fabricated cooler and the designed geometry. As indicated in Figure 6.16, four different regions are marked, including inlet nozzle opening, nozzle plate, inlet/outlet nozzles, and cavity height. After the post-cure process, the designed straight corners are rounded due to polymer shrinkage. The angled nozzle walls and the rounded corners of the nozzles will have an impact on the flow distribution, the chip temperature profile, and the pressure drop. The details of this impact will be shown in the modeling section. From the cross-section can furthermore be observed that the nozzle plate is 0.55 mm thick compared to the design value of 0.5 mm (printed layers are 50 µm thick). For the cavity height, the actual height is approximately 0.65 mm compared to the nominal design value of 0.6 mm. In summary, the cross-section analysis shows a good quality of the fabrication.

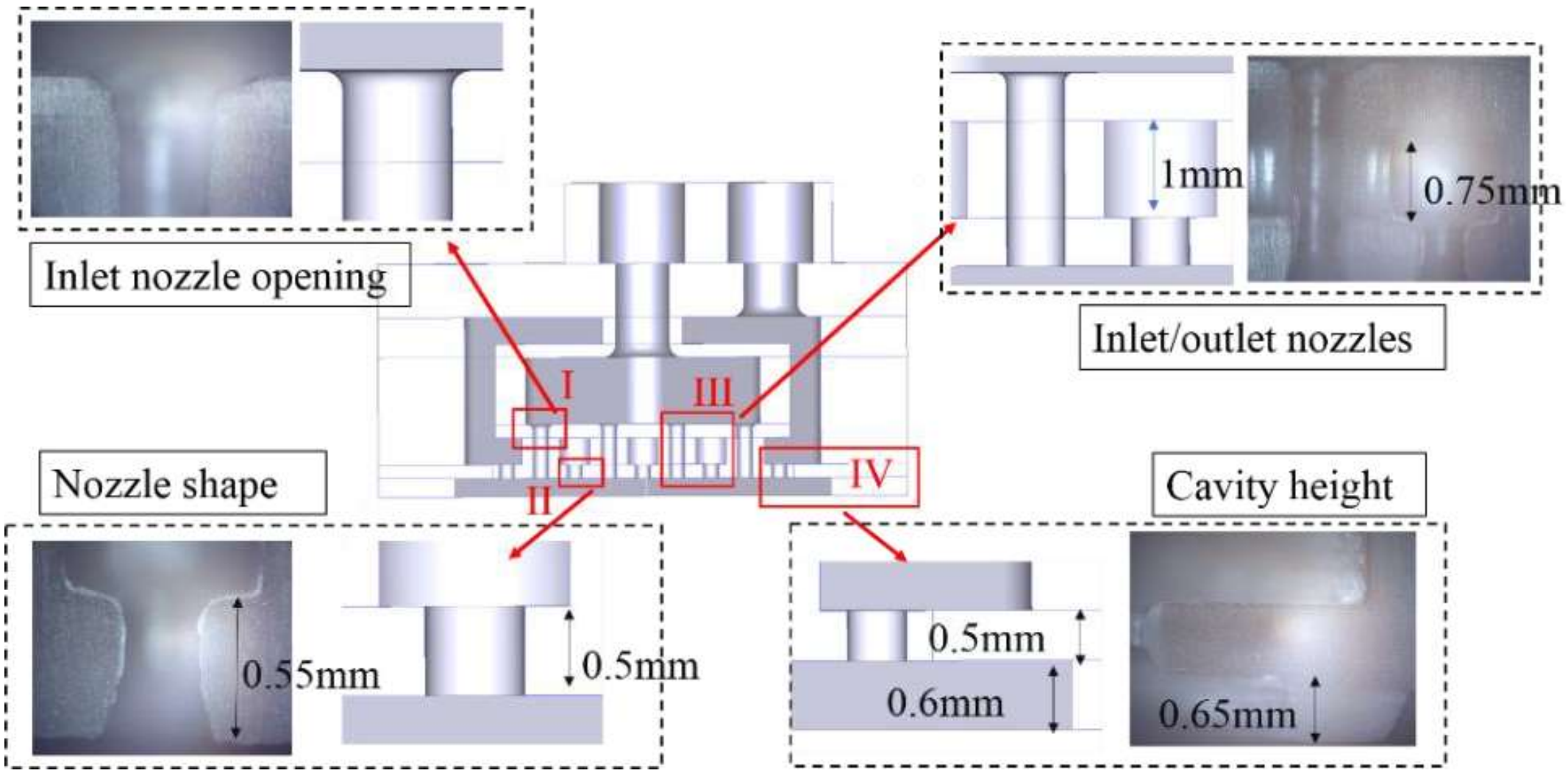

**Figure 6.16:** Comparison between the nominal design and the fabricated 4×4 jet array cooler at four different locations.

Since the 3D printed cooler is printed as a single part, it is difficult to check for potential internal blockages with residual uncured resin. For this application, we demonstrated that the Scanning Acoustic Microscopy technique (SAM) could be used to evaluate the cooler quality. SAM is a non-destructive technique used for micro-inspection [23]. By adapting the focus depth, the potentially blocked resin inside the nozzles can be detected at different layers in the structure. Figure 6.17 shows two examples of the SAM analysis of a printed cooler with or without defects: from the SAM images, it is possible to differentiate between the open nozzles without resin residues (A), open nozzles with tapered edges (B) and the presence of blocked nozzles (C). Figure 6.17(a) shows the 3 × 3 cooler with blocked nozzles and tapered nozzles (Section 6.3.1.1), while Figure 6.17(b) shows the 4 × 4 cooler without defects (Section 6.3.1.2). A 'Time of flight' scan on the cooler from the bottom side of the cooler can be used to assess the depth of the nozzles inside the cooler from the cooler plot. As shown in Figure 6.18, the blue color is set as the reference depth as 0 µm. The green color indicates a depth level of 2.2 mm, while the red color represents the height level as 5.1 mm, corresponding to the top surface of the inlet and outlet chamber, respectively. The 3 blue spots that appear in the SAM images at the corner locations of the O-ring groove correspond to the location of the support pins in the SAM tool to submerge the buoyant cooler in the water.

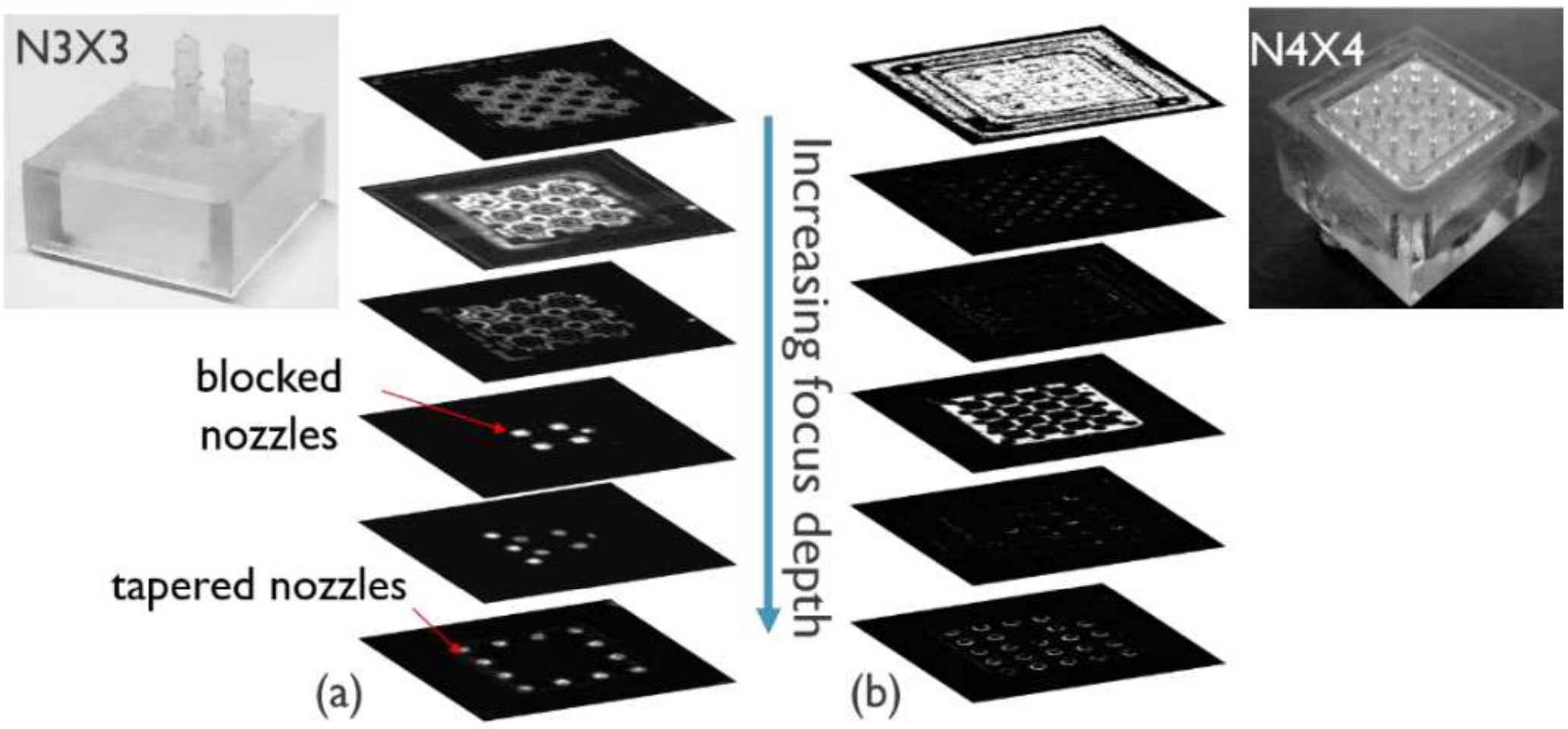

**Figure 6.17:** Defect measurements of the 3D printed coolers using SAM inspection. (a): printed cooler with blocked nozzles and tapered nozzles; (b) printed cooler without blocked nozzles.

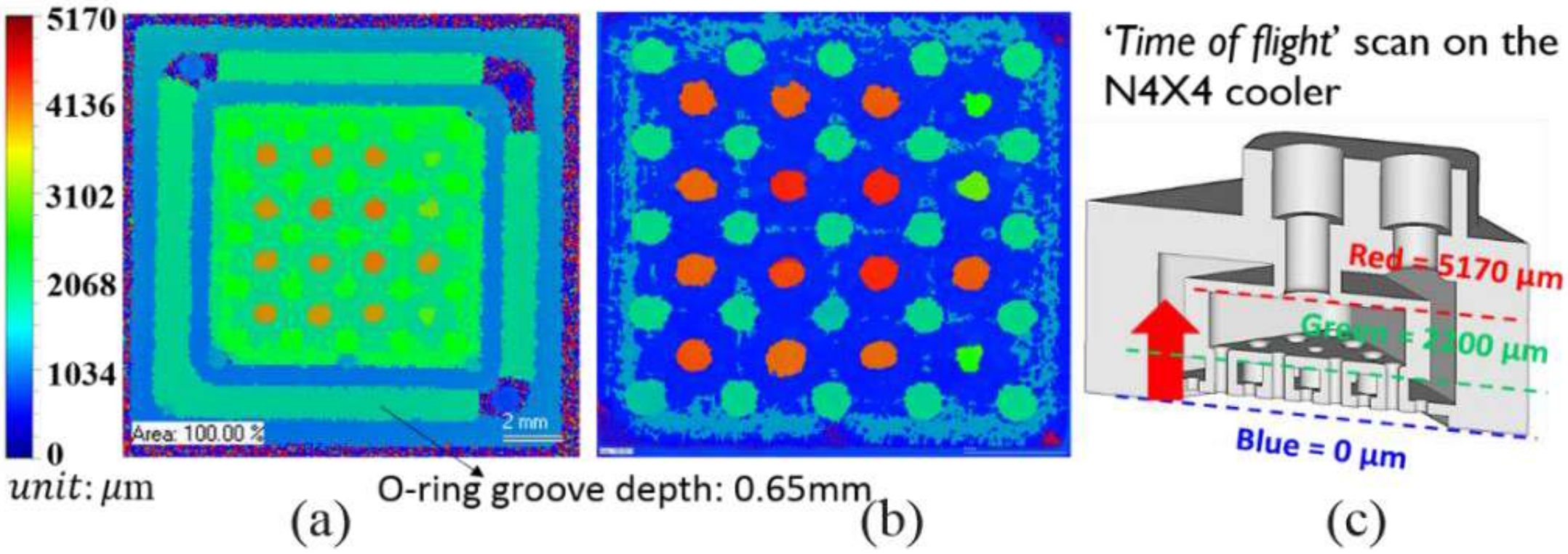

**Figure 6.18:** SAM measurement for assessing the depth of the printed nozzles: (a) SAM image with focused depth on the cooler bottom layer and (b) nozzle plate layer; (c) Indication of the depth of different layers.

Finally, the surface roughness of the groove surface is a crucial factor for the sealing ability with the sealing ring. For the early additive manufacturing technologies, the surface roughness is quite high due to the low resolution of the 3D printing machine. For the current 3D printing tools with high resolution, the surface roughness is drastically reduced. The 3D printing technology used in this study is SLA with 50 µm layer thickness. The roughness is expected to be smaller than a layer thickness. The designed groove has a depth of 600 µm. The SAM depth measurement from the cooler's bottom side in Figure 6.7 shows a smooth groove surface as an indication of limited roughness.

### 6.3.3 Manufacturing tolerance impact analysis

6.3.3.1 Impact of nozzle diameter deviation

As discussed in the Section 6.3.1, the deviation between the measured printed nozzle (575 µm ± 20 µm) and the nominal design value of 600 µm is only 5% for 4×4 array cooler. In order to understand the impact of the 3D printing fabrication tolerance on the cooler thermal/hydraulic performance, the impact for a nozzle geometry of a 4×4 cooler with a cooling unit cell area of 2×2 $mm^2$, and a nominal nozzle diameter of 600 µm is investigated numerically. The unit cell modeling approach introduced in chapter 2 is used to assess the impact of the geometry deviation on the temperature and pressure drop based on the 4×4 array cooler.

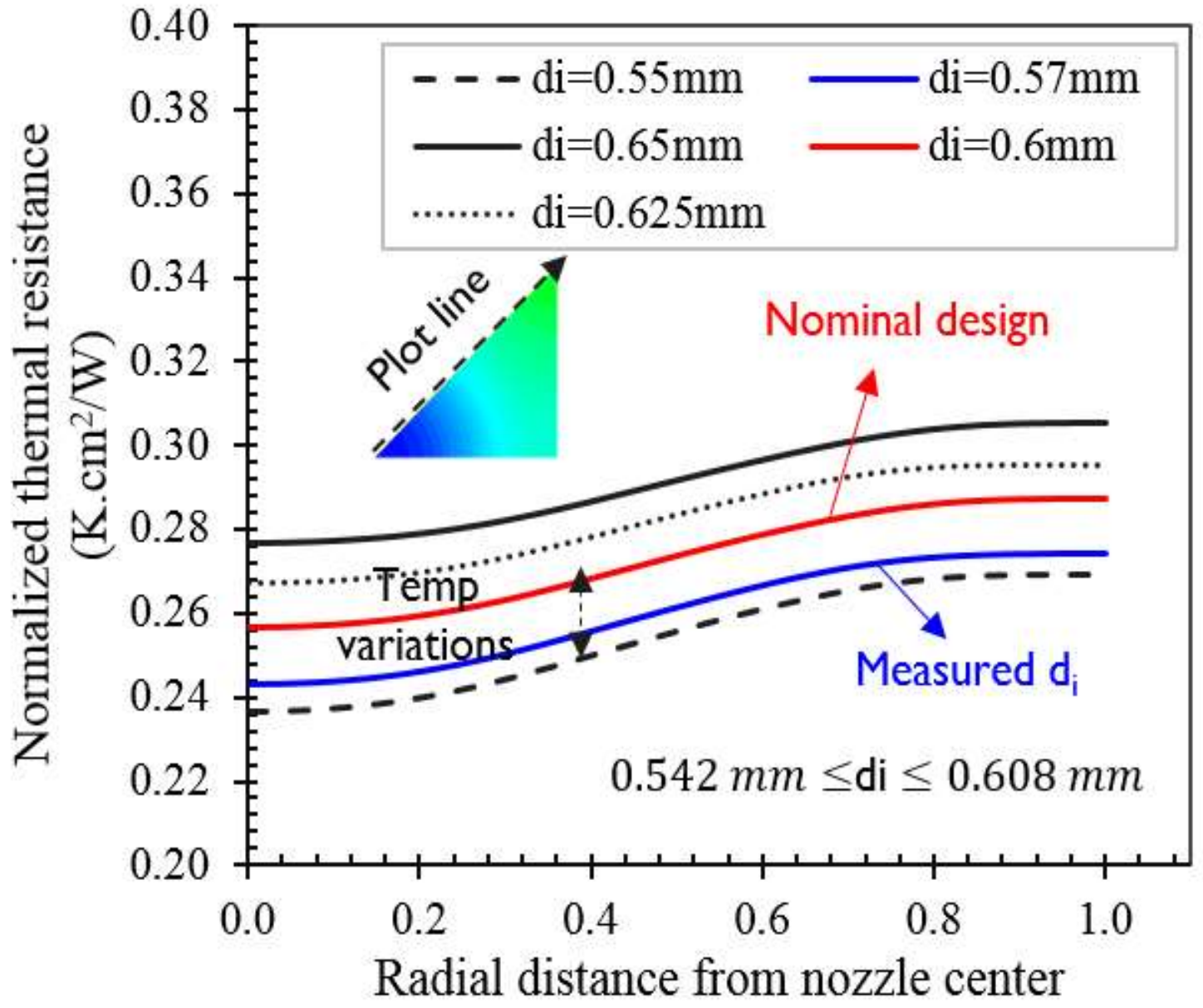

**Figure 6.19:** Impact of the nozzle diameter deviation on the temperature distributions for 2×2 $mm^2$ cooling unit cell area with a nominal design nozzle diameter of 600 µm. (flow rate = 600 mL/min, Q = 50 W).

Figure 6.19 shows that the normalized thermal resistance will drop down for a decrease of the nozzle diameter at a constant flow rate. The reason is that the inlet nozzle velocity will increase due to the reduction of the nozzle diameter for the fixed flow rate. For the impingement jet cooling, the chip temperature is dominated by the stagnation point where the inlet jet nozzles are targeted. The stagnation temperature in the temperature profile shows about a 7.7% variation for the nozzle diameter changing from 0.55 mm to 0.6 mm for the 4×4 cooler. The reduction of the nozzle diameter can reduce the chip temperature, however, at the expense of a higher pressure drop. The thermal and hydraulic comparison between the nominal design and actual measured values are illustrated in Figure 6.20. The modeling study shows that the nozzle diameter deviation of 5% at a flow rate of 600 mL/min results only in a 4.7% reduction for the averaged chip temperature and a 23% increase of the pressure drop.

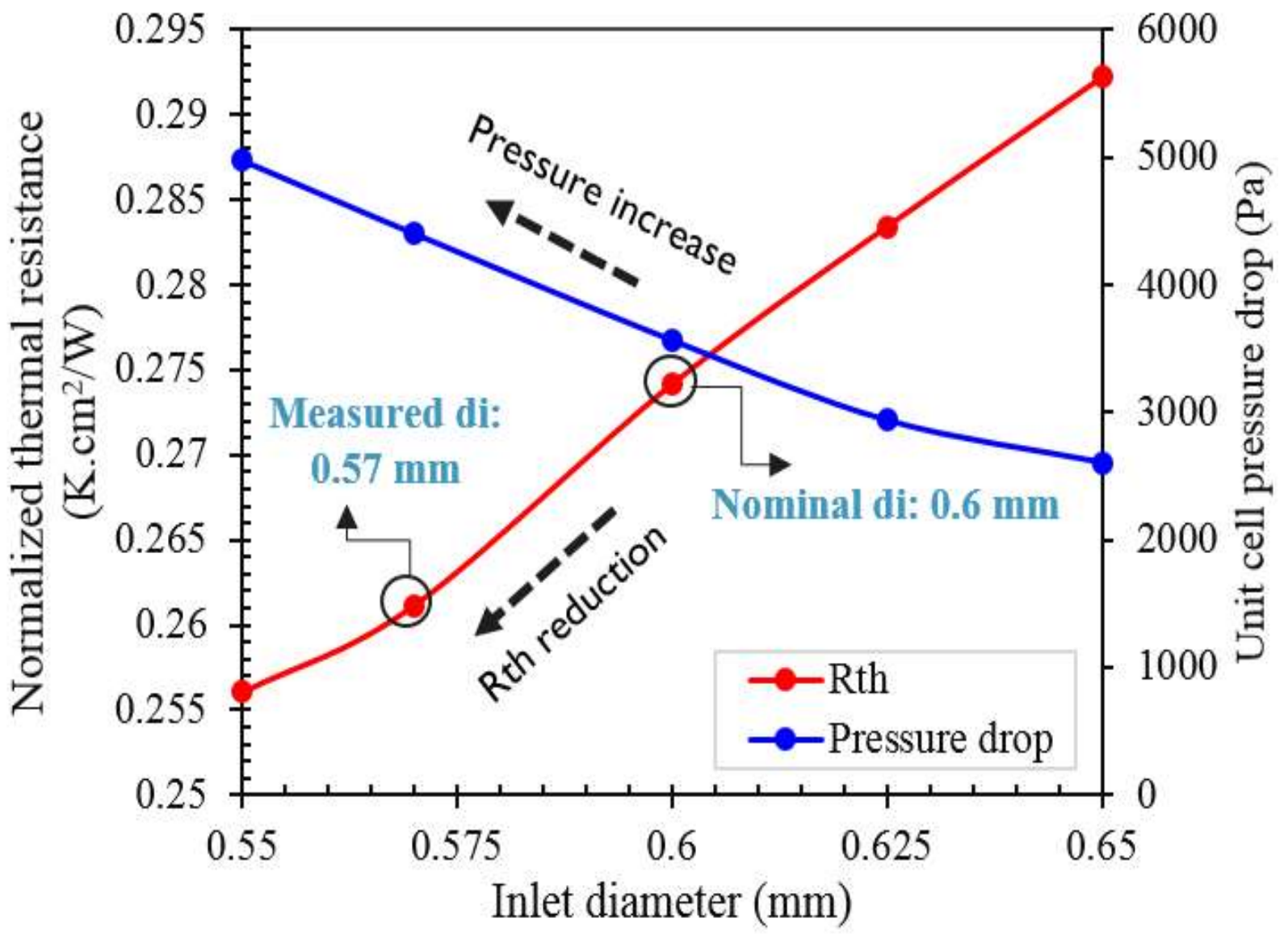

**Figure 6.20:** Impact of the inlet/outlet nozzle diameter on the averaged chip temperature and pressure drop for 2×2 $mm^2$ cooling unit cell area with a nominal design nozzle diameter of 600 µm. (flow rate =600 mL/min, Q = 50 W).

6.3.3.2 Impact of nozzle angle deviation

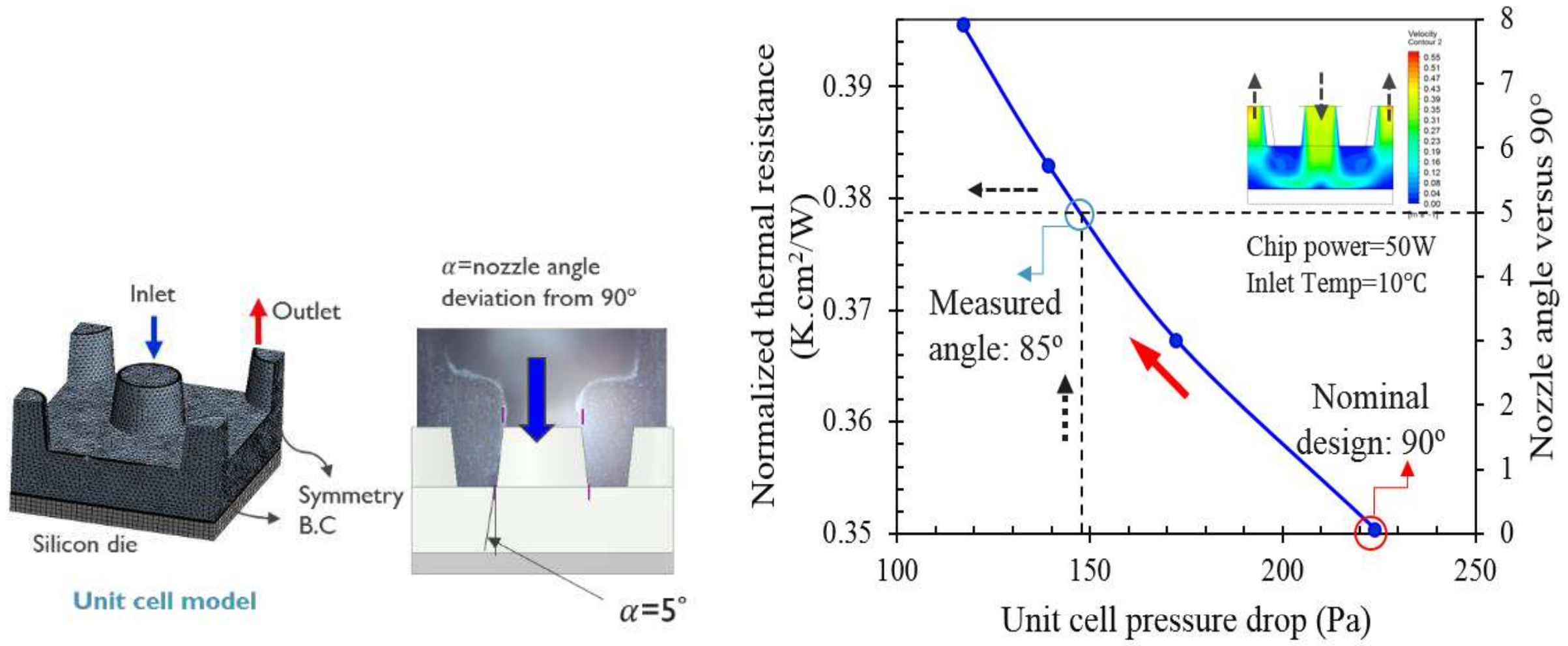

**Figure 6.21:** Unit cell modeling study on the impact of nozzle angle on the thermal and hydraulic performance for 2×2 $mm^2$ cooling unit cell area with a nominal design nozzle diameter of 600 µm. (flow rate =300mL/min, Q=50W)

The cross-section pictures of the printed cooler show that the nozzle shapes are slightly tapered instead of straight. The tapered nozzle can reduce the cooling performance due to the less concentrated flow targeted at the stagnation point, resulting in higher local chip temperature. On the other hand, the tapered inlet nozzle and outlet nozzle shape can help to reduce the pressure drop. As shown in Figure 6.21, the modeling study shows that a nozzle diameter deviation of 5º (85º instead of 90º) only results in an 8% difference for the averaged chip temperature but caused a 34.2% reduction for the local pressure drop on the unit cell level.

6.3.3.3 Impact of nozzle-to-chip distance deviation

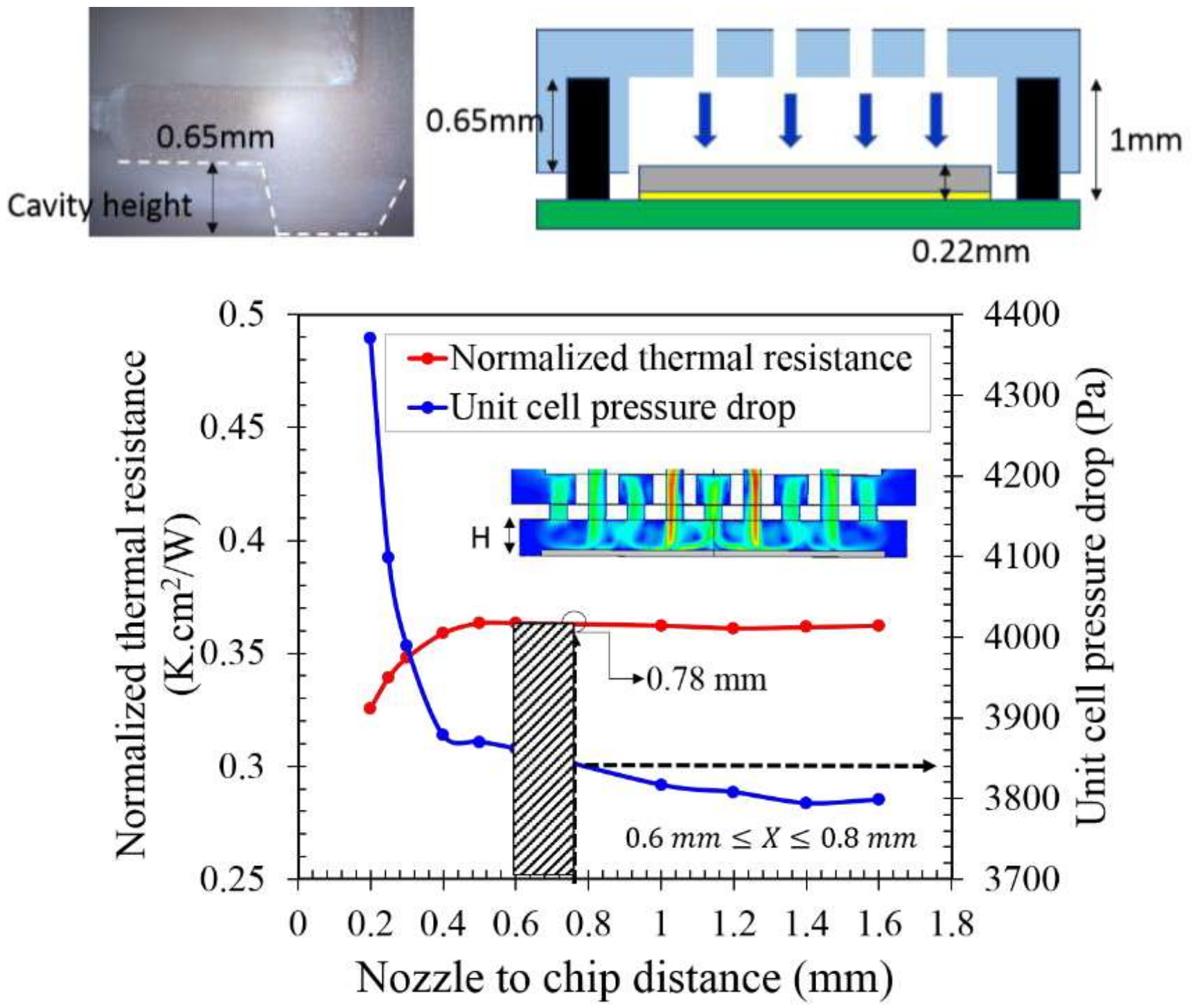

**Figure 6.22:** Impact of the nozzle-to-chip distance variations for the 2×2 mm$^2$ cooling unit cell area with a nominal design nozzle diameter of 600 µm. (flow rate =300mL/min, Q=50W).

The deviation of the nozzle-to-chip distance depends on the assembly pressure, O-ring and groove design. In order to define the deviation of the nozzle-to-chip distance *H*, the groove depth, the thickness of the O-ring, and the fabrication tolerance of the cavity height should be considered. For the cavity height and groove, the actual depth is about 0.65 mm compared to the nominal design value of 0.6 mm. The thickness of the O-ring is 1 mm, which will be placed on the organic substrate. The chip thickness is 0.2 mm. The thickness of the micro-bump used to connect the thermal test chip, and the organic substrate is 0.02 mm. Taking account of the O-ring thickness without compression, the distance between the nozzles and chip cooling surface is 0.78 mm. Therefore, the nozzle-to-chip backside distance variation is expected to between 0.6 mm to 0.8 mm.

The impact of the nozzle-to-chip distance above the chip cooling surface is shown in Figure 6.22. The modeling study shows that the impact on the thermal resistance is negligible beyond 0.6 mm, while the impact on the cooler pressure drop will result in a difference of 1.1 % between the range of 0.6 mm and 0.8 mm. Therefore, the nozzle-to-chip distance variation does not have a significant impact on the chip averaged temperature when the nozzle-to-chip distance ratio is above H/L > 0.25.

## 6.4 Experimental characterization and model validations

### 6.4.1 Thermal experimental characterization

The coolers and the package are placed in a measurement socket on the PCB and held together by screws. An O-ring is placed in the foreseen groove at the bottom of the 3D printed cooler to create a sealing between the cooler and the package, as shown in Figure 6.23. The thermal experiments are performed using DI water as a coolant and for 50 W as quasi-uniform power dissipation in the 8 mm × 8 mm chip area, according to the heater map. The reliable water-resistant polymer material allows performing 48 hours measurement without observing cooler deformations or leaks.

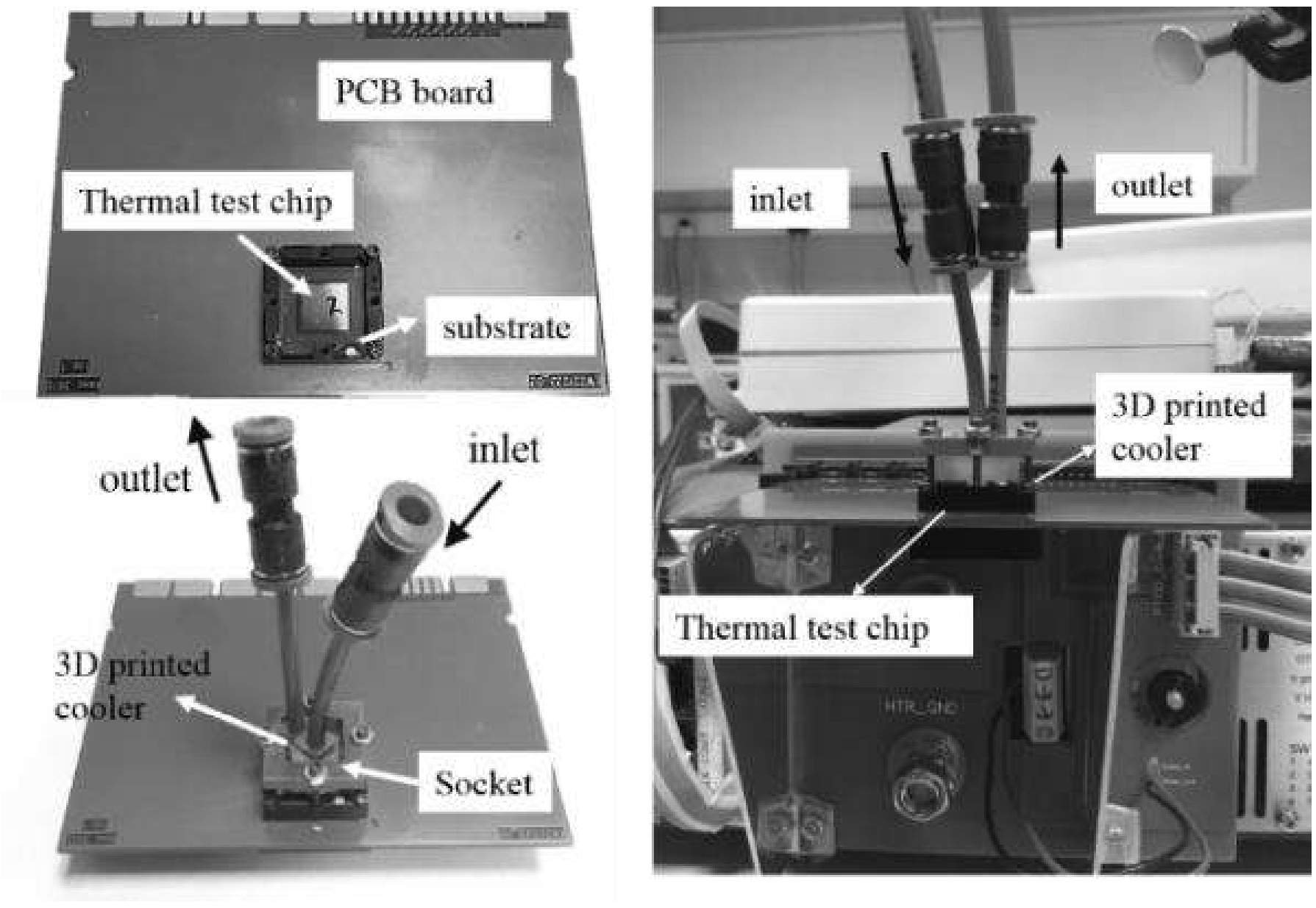

**Figure 6.23:** Cooler assembly and experiment set up for the temperature measurement of the microjet cooling.

In the first parts, the thermal and hydraulic performance of the 4×4 printed cooler is characterized and analyzed for different flow rates. As shown in Figure 6.24(a), the normalized temperature profile along the chip diagonal is plotted for different flow rates ranging from 100 mL/min to 1000 mL/min. The measurements show that the cooling performance, as well as the temperature uniformity across the chip surface, improve for increasing flow rate. Due to the high cooling rate of the multi-jet cooler, the measured temperature profile reveals the pattern of the heated and non-heated cells in the thermal test chip. The relation between the thermal resistance based on the average chip temperature and the total inlet flow rate is also shown in Figure 6.24(b). It can be seen that the thermal resistance $R_{th}^*$scales with flow rate $\dot{V}$ according to the following power-law behavior $R_{th}^* \propto \dot{V}^{-0.54}$. The exponent of 0.54 is in line with the exponents for published heat transfer correlations for multi-jet impingement cooling, which typically show a range between 0.5 and 0.8 [21]. The achieved minimal thermal resistance of the 3D printed 4 × 4 cooler is 0.16 $cm^2$.K/W for a flow rate of 1000 mL/min.

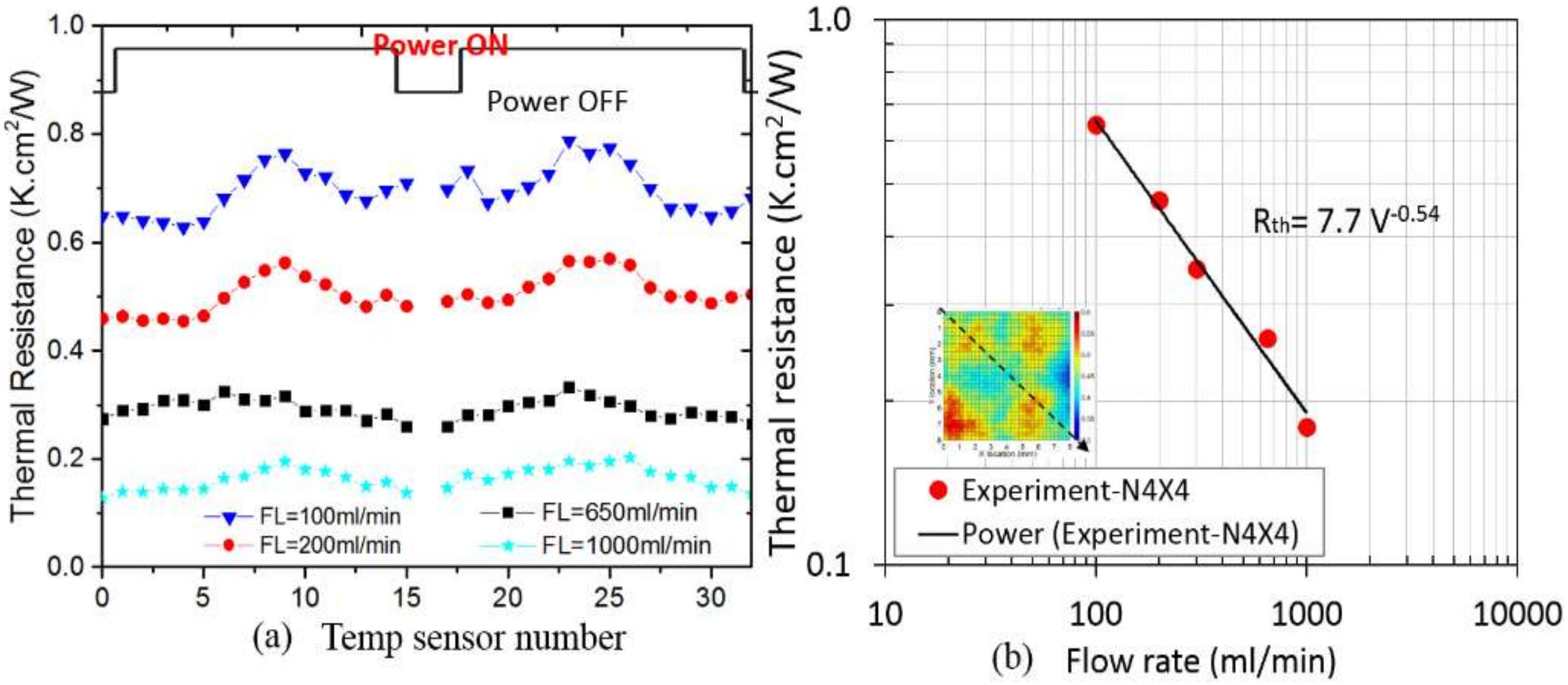

**Figure 6.24:** Thermal characterization of the 3D printed cooler: (a) measured temperature profile under different flow rates; (b) validation of predicted model based on corrected nozzle diameter.

The thermal performance measured by the thermal test chip with the coolers with 3 different nozzle arrays is shown in Figure 6.25 for a constant flow rate of 1000 mL/min. It can be seen that the temperature reduces with increasing nozzle density. The observed trend with increasing performance can be concluded as $R_{3\times3} < R_{4\times4} < R_{8\times8}$ . The temperature profile along the chip diagonal is also compared in Figure 6.26. The comparison shows that an excellent thermal performance $Rth^*$ for the 8×8 cooler with 1×1 $mm^2$ cooling cells can be achieved as 0.13$cm^2$-K/W, based on the average chip temperature.

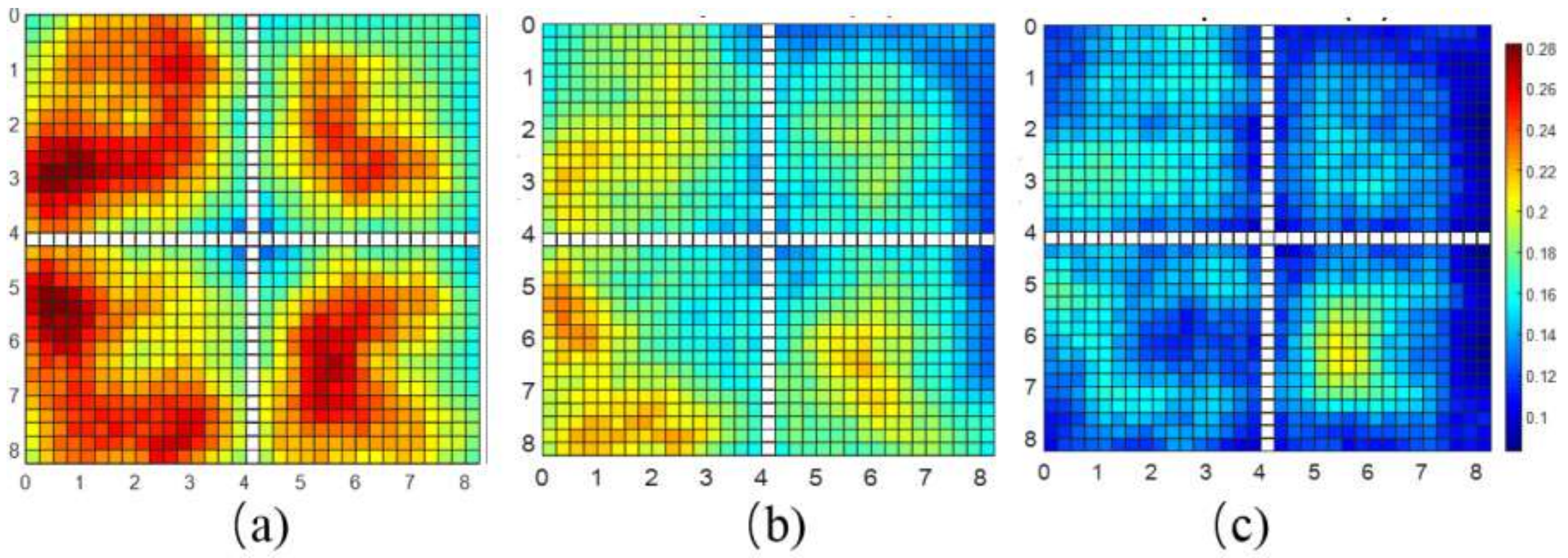

**Figure 6.25:** Comparison of the temperature distribution for (a) 3×3 (b) 4×4 and (c) 8×8 nozzle array.

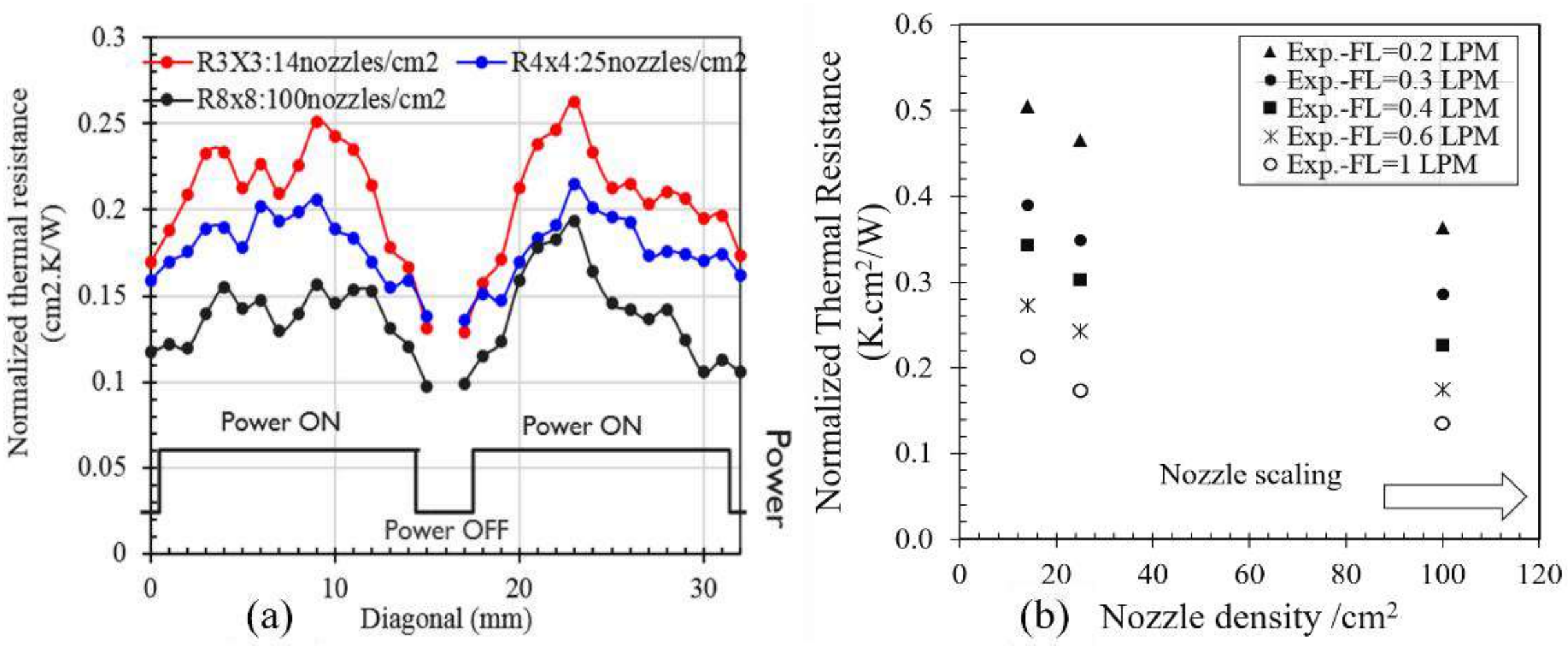

**Figure 6.26:** Temperature measurements for 3×3, 4×4 and 8×8 nozzle array as a function of flow rate: (a) temperature profiles along the chip diagonal; (b) thermal resistance as function of different flow rates (H=0.6 mm, Q=40W, FL=1000 mL/min).

### 6.4.2 Experimental model validation

In this section, the full cooler level CFD simulations are updated with the measured nozzle diameter of 575 μm and the nozzle plate thickness of 550 μm and the modeling results are validated by experimental data for different flow rates. Figure 6.27(a) shows the experimental and modeling results of the chip temperature along with the chip diagonal for different flow rates. To better capture the local level temperature difference, the detailed heaters are included in the full CFD model, showing quasi-uniform heating. This part has been discussed in chapter 5 with the discussion about the uniform heating and quasi-uniform heating in the full CFD model. In Figure 6.27(b), the average thermal resistance of the chip is compared for the modeling results and the measurement data as a function of the flow rate. It can be seen that the maximum error between modeling results, and experiments is 12% at 100 mL/min, while the difference reduces to 4% at the high flow rate of 1000 mL/min. In conclusion, the CFD model used in this study is validated and shows good agreement with the experimental results.

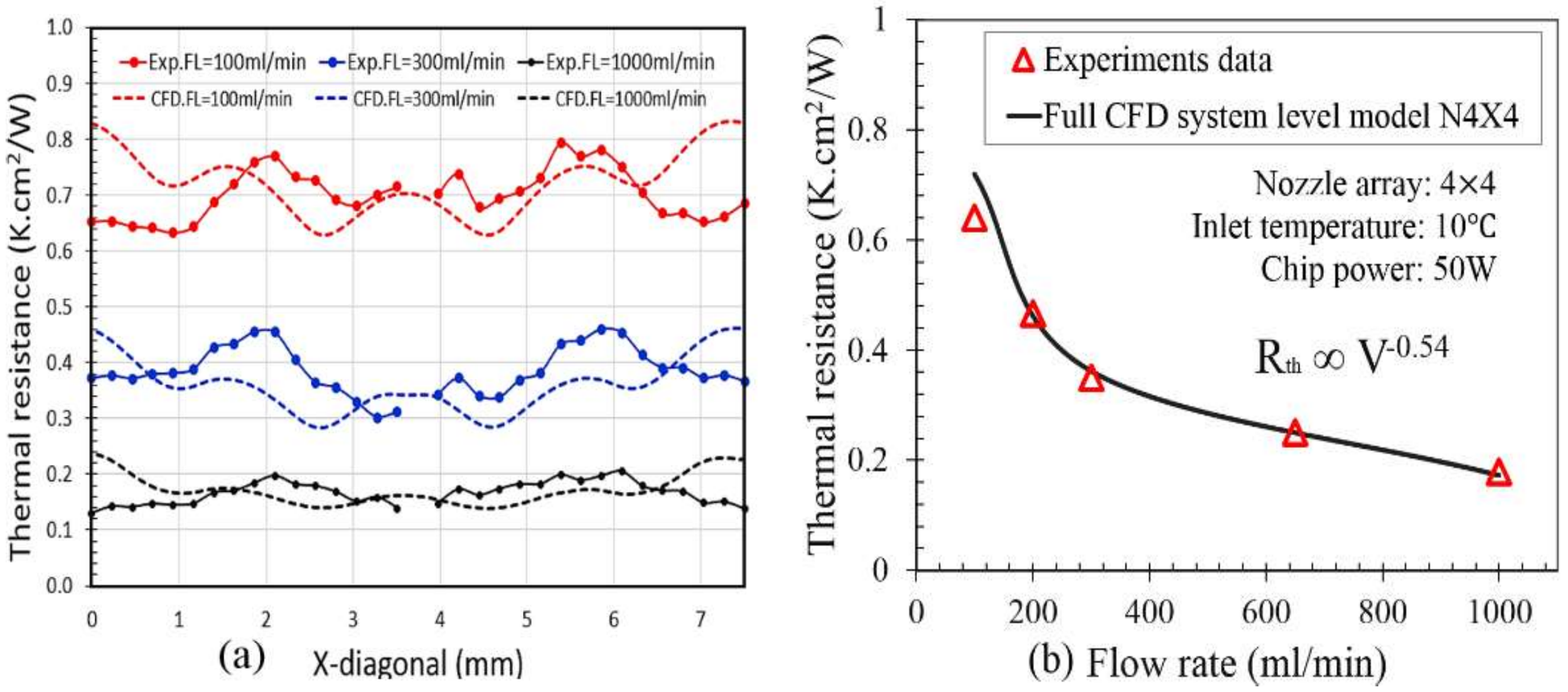

**Figure 6.27:** System-level CFD modeling results of the 3D printed 4 × 4 cooler: (a) temperature profile comparison between the experimental and modeling results; (b) full CFD model validation with experimental results (quasi-uniform heating with detailed heaters is used in the CFD model).

### 6.4.3 Benchmarking study

To put the thermal performance and size of the 3D printed cooler in perspective, the measured chip temperature map is compared for a conventional heat sink – fan combination, the micromachined polymer liquid jet cooler [6] (600 mL/min) and the 3D printed liquid jet cooler (600 mL/min) for a custom power dissipation map. The photography of the coolers is shown in Figure 6.28. As mentioned before, the cooler is part of a closed flow loop system, and this comparison only considers the size of the cooler on the chip, since the intention of the proposed cooler is to replace a bolt-on liquid cold plate in the existing infrastructure with the pump, tubing and heat exchanger. It should be noted that the size of the additional parts in the closed-loop is not considered here.

The chosen power map with different hot spot sizes that is applied to the thermal test chip is shown in Figure 6.28(b). The measured chip temperature increase map is shown in Figure 6.28(c), for the case of the 3D printed cooler. Figure 6.28(d) shows the comparison of the temperature profiles along with the chip diagonal for the three different coolers. The comparison shows that both liquid jet coolers achieve a 2.7× lower peak temperature difference and a 3.5× lower average temperature difference with respect to the inlet temperature compared to the conventional air-cooling heat sink.

Moreover, the temperature profile of the 3D printed cooler is a bit lower than the micromachined cooler. This difference is due to the smaller inlet nozzle diameter of 575 $\mu$m fabricated by 3D printing than the micromachined cooler with around 600 $\mu$m. As discussed in Section 6.3.3.1, the smaller inlet nozzle diameter can result in a lower average chip temperature for a constant flow rate. The size of the coolers is respectively 8 cm × 8 cm, 46 mm × 46 mm and 18 mm × 18 mm for the heat sink and fan combination, the micromachined cooler, and the 3D printed cooler respectively. This clearly shows that the 3D cooler offers a considerable reduction in the cooler size, matching the footprint of the chip package.

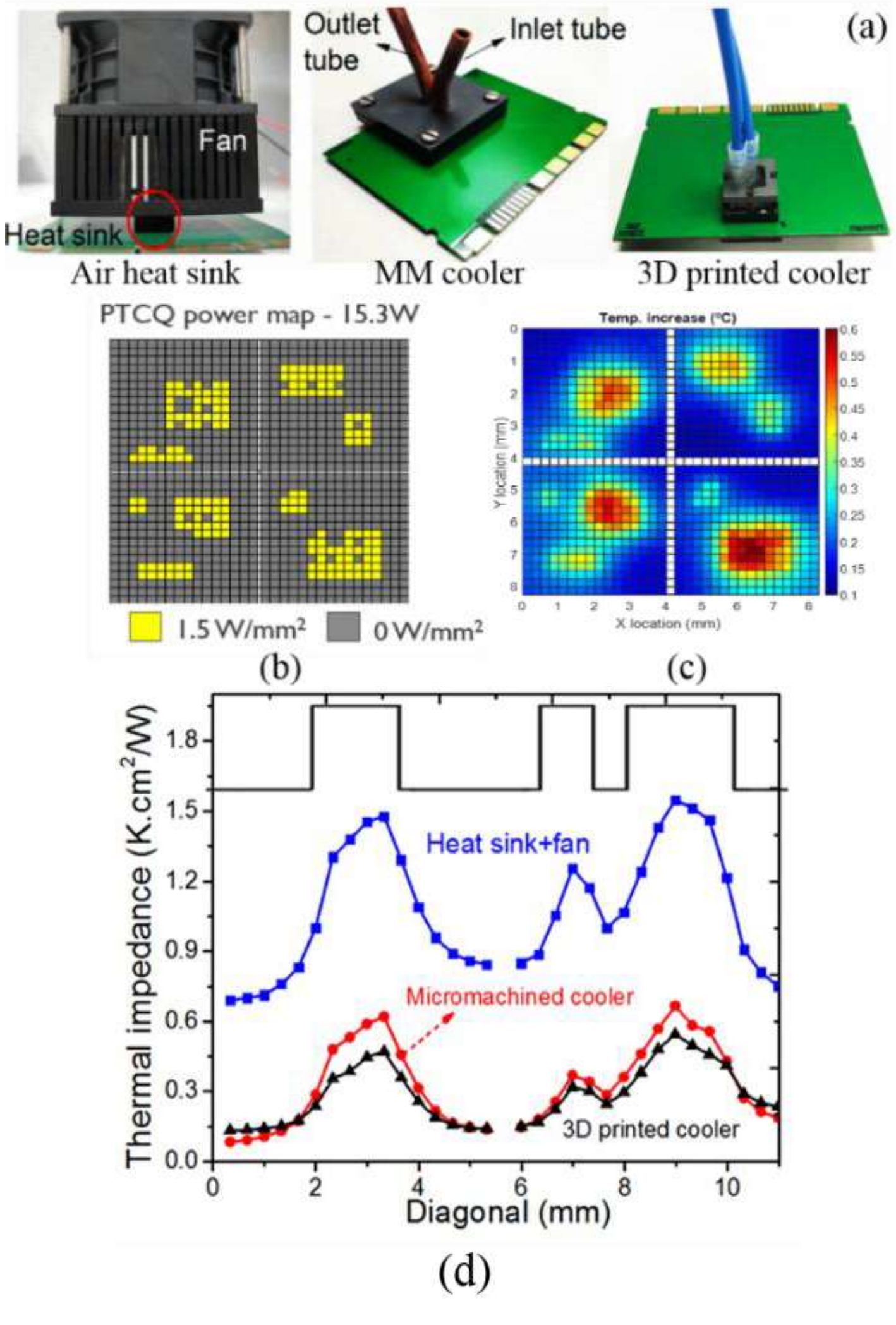

**Figure 6.28:** Comparison of the measured chip temperature map for the heat sink, micro-machined cooler and the 3D printed cooler: (a) three demonstrators' comparison; (b) the defined PTCQ hot spots power map; (c) measured temperature map with 3D printed cooling; (d) measured temperature profile comparison (600 mL/min for liquid coolers).

## 6.5 Model Validations: unit cell

The measured cooling performance of the 3D printed coolers with 3×3, 4×4, and 8×8 inlet nozzle arrays are characterized in this chapter. Table 6.4 lists the comparison between the designed value and measured value of the nozzle diameters for the three coolers. Also, the dimensionless number $d_i$/L with the measured values are also listed. Since the cavity height designed for all the three coolers are measured as the same value with 650 $\mu$m, the dimensionless number H/L used in the predictive model is H/L=0.33. In general, it can be seen that the measured $d_i$/L is a bit lager than the designed value of $d_i$/L =0.3. For the predictive model validation, the measured thermal resistance and flow rate are all transformed to the $\overline{Nu_j}$-$Re_d$ correlations. The plotted $\overline{Nu_j}$-$Re_d$ relations for the coolers with three different inlet nozzle arrays are used to validate the predictive model, developed in chapter 3, based on the dimensionless analysis. It can be seen that the predictive model based on the measured $d_i$/L value listed in Table 6.29 shows a good agreement with the experimental results.

**Table 6.4:** Comparison between the designed and fabricated parameters (unit: mm).

| NXN array | Unit cell | Di-(Design) | Di-(Meas.) | $d_i$/L (Meas.) | H/L (Meas.) |
|---|---|---|---|---|---|
| 3×3 | 2.67 | 0.8 | 0.95 | 0.36 | 0.33 |
| 4×4 | 2 | 0.6 | 0.75 | 0.375 | 0.33 |
| 8×8 | 1 | 0.3 | 0.38 | 0.38 | 0.33 |

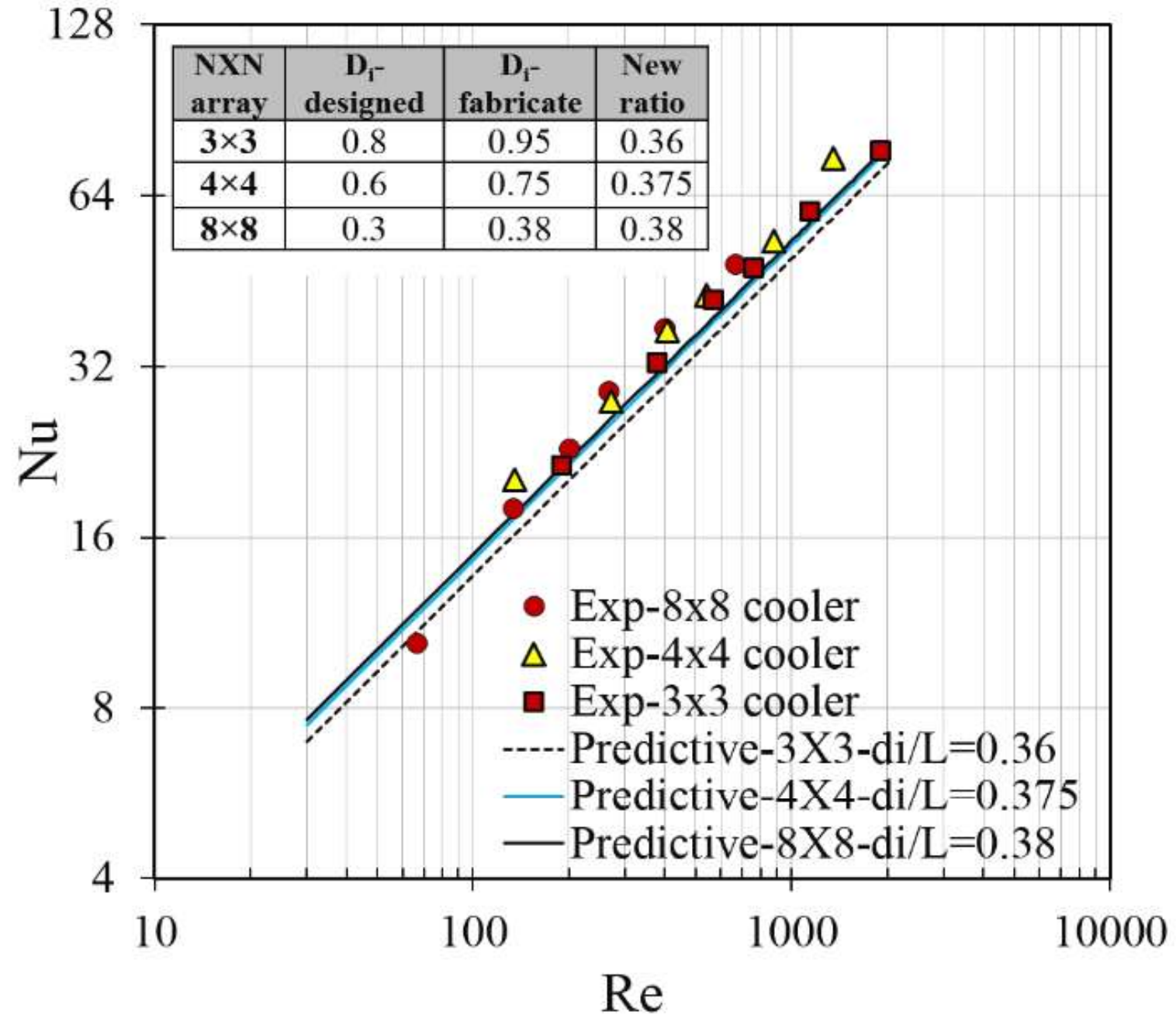

**Figure 6.29:** Comparison of the measurements data and developed predictive model $\overline{Nu_j}$-$Re_d$ for the 3D printed coolers with 3×3, 4×4 and 8×8 inlet nozzle arrays.

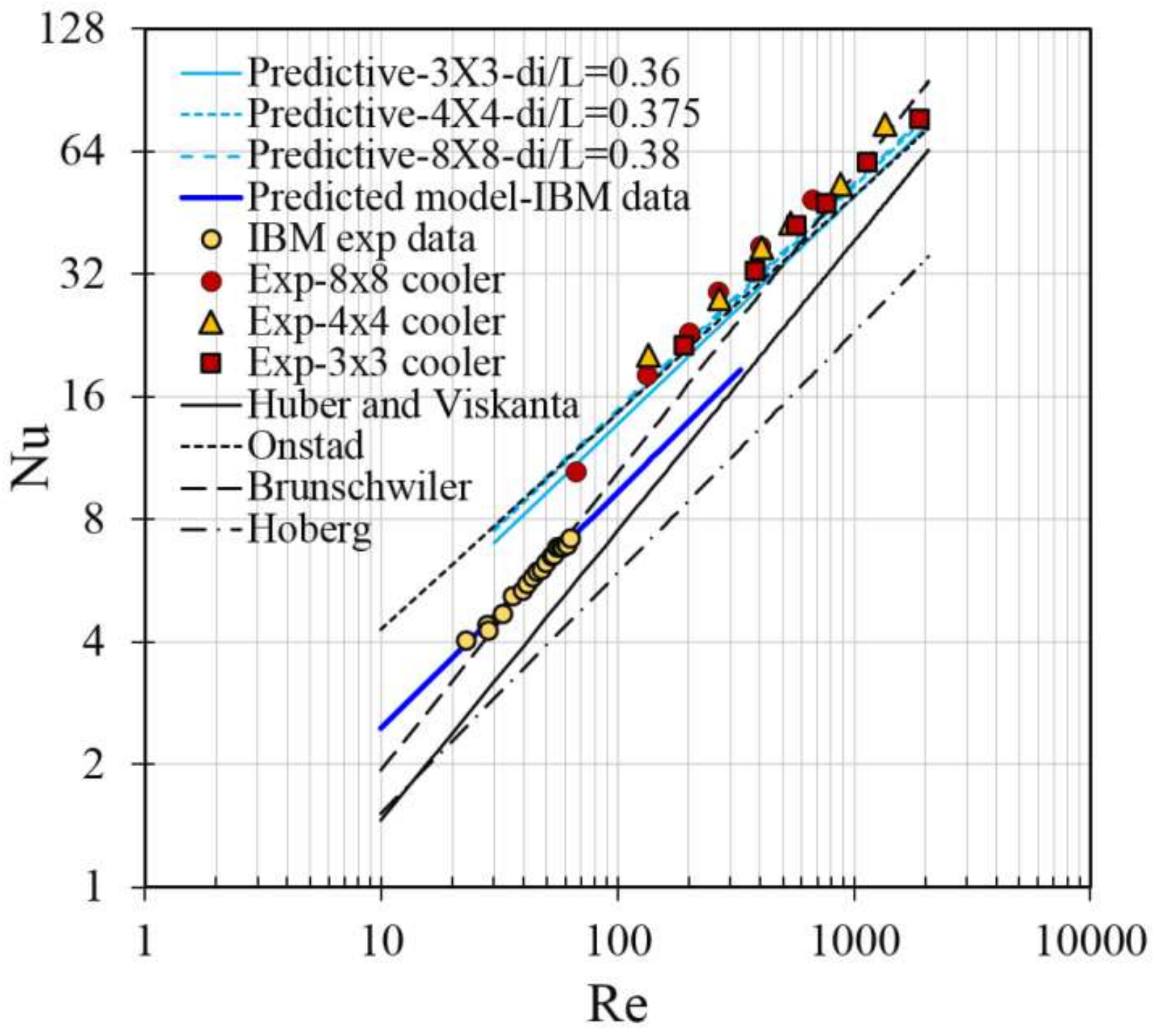

**Figure 6.30:** Comparison of the developed predictive model with literature experimental data and $\overline{Nu}_j$-$Re_d$ correlations.

Moreover, the measurement data from IBM are also used to validate our developed predictive model, as shown in Figure 6.30. The cooler parameters from IBM [11] are imported into the predictive model, where the nozzle diameter is 43 $\mu$m, unit cell length is 150 $\mu$m, and the nozzle number is 19044. It can be seen that the predicted $\overline{Nu}_j$-$Re_d$ curve based on the geometry parameters shows good agreement with the IBM experimental data in the literature [11], and this agreement is better than their $\overline{Nu}_j$-$Re_d$ correlation. This is because the impacts of the $d_i$/L and H/L are included in our predictive model. Furthermore, the extracted $\overline{Nu}_j$-$Re_d$correlation is also compared with the state of art $\overline{Nu}_j$-$Re_d$ correlations in the literature based on the 3D printed cooler configurations, as reviewed in chapter 3. It can be seen that the most matched developed model is Onstad's model, while the difference is larger for other models. However, the empirical constants for the power-law relationship of $\overline{Nu}_j$ versus $Re_d$ for Onstad's model are only limited to three different nozzle diameter values. It should be noted that our developed model is based on the jet cooling with locally distributed outlets, which includes the dimensionless term of $d_i$/L and H/L. As illustrated in chapter 3, the developed $\overline{Nu}_j$-$Re_d$ model applies to different $d_i$/L and H/L ratios, under the range (0.01 $\leqslant$ $d_i$/L $\leqslant$0.4; 0.01$\leqslant$H/L$\leqslant$ 0.4). In summary, the predictive model including the effects of $d_i$/L and H/L matches well with our in-house developed experimental results and also shows good agreement with the available experimental IBM data.

## 6.6 Conclusion

This chapter focuses on the design, demonstration and experimental characterization of the chip level 3D printed microjet cooler using high-resolution additive manufacturing technology. Firstly, the design constraints and critical parameters for the 3D printing are analyzed. Based on the analysis, three 3D printed coolers with 3×3, 4×4 and 8×8 inlet nozzle arrays are designed. Moreover, the modeling studies are conducted to compare the micromachined cooler and 3D printed cooler, illustrating the advantages of the 3D printed coolers. In the second part, the cooler materials based on two different 3D printing techniques: DLP and SLA are investigated in detail, showing that the material aspects (defect-free fabrication and water resistance) are very important for the cooler fabrication. Also, the defect measurement techniques and the manufacturing tolerance impact are discussed in this chapter. Thirdly, the experimental studies show that a very good thermal performance for 8×8 cooler with $1\times1 mm^2$ cooling cells can be achieved as 0.13 $cm^2$-K/W under the same flow rate at 1000 mL/min. The observed trend with increasing performance is $R_{3\times3} < R_{4\times4} < R_{8\times8}$. The comparison between experiments and CFD modeling results shows good agreement with the maximum difference below 15%.

Finally, the measurement data based on the different 3D printed coolers, and also the experimental results from the literature are compared with the predicted results extracted from the developed $\overline{\mathrm{Nu}}_\mathrm{j}$-$\mathrm{Re}_\mathrm{d}$ correlations, showing a good agreement. In summary, the comparisons show that the validated $\overline{\mathrm{Nu}}_\mathrm{j}$-$\mathrm{Re}_\mathrm{d}$ correlation including the impact of the $d_i$/L and H/L can be used as a fast-predictive model for the cooler design.

The results of this chapter are partially published in the following publication:

**Tiwei Wei**, Herman Oprins, Vladimir Cherman, Ingrid De Wolf, Eric Beyne, Martine Baelmans, " Experimental Characterization of a Chip Level 3D Printed Microjet Liquid Impingement Cooler for High Performance Systems," in IEEE Transactions on Components, Packaging and Manufacturing Technology, 2019.

## References

[1] H. Amano, Y Baines, et al., “The 2018 GaN power electronics roadmap,” J. Phys. D. Appl. Phys., vol. 51, no. 16, pp. 163001, March 2018.

[2] A. S. Bahman and F. Blaabjerg, “Optimization tool for direct water cooling system of high power IGBT modules,” in Proc. 2016 European Conf. Power Electron. Appl., 2016, pp. 1–10.

[3] S. V Garimella, T. Persoons, J. A. Weibel, and V. Gektin, “Electronics Thermal Management in Information and Communications Technologies: Challenges and Future Directions,” IEEE Trans. Components, Packag. Manuf. Technol., vol. PP, no. 99, pp. 1191–1205, September 2016.

[4] S. Kandlikar, S. Garimella, D. Li, S. Colin, and M. R. King, Heat transfer and fluid flow in minichannels and microchannels, Oxford, Elsevier Ltd., 2014, Chap. 3.

[5] K. Gould, S. Q. Cai, C. Neft, and S. Member, “Liquid Jet Impingement Cooling of a Silicon Carbide Power Conversion Module for Vehicle Applications,” IEEE Trans. Power Electronics, vol. 30, no. 6, pp. 2975–2984, June 2015.

[6] A. S. Rattner, “General Characterization of Jet Impingement Array Heat Sinks With Interspersed Fluid Extraction Ports for Uniform High-Flux Cooling,” J. Heat Transfer, vol. 139, no. 8, p. 082201(1-11), August 2017.

[7] J. Jorg, S. Taraborrelli, G. Sarriegui, R. W. De Doncker, R. Kneer, and W. Rohlfs, “Direct single impinging jet cooling of a mosfet power electronic module,” IEEE Trans. Power Electronics, vol. 33, no. 5, pp. 4224–4237, May 2018.

[8] A. Bhunia and C. L. Chen, “On the Scalability of Liquid Microjet Array Impingement Cooling for Large Area Systems,” J. Heat Transfer, vol. 133, no. 6, 064501(1-7), January 2011.

[9] K. Olesen, R. Bredtmann, and R. Eisele, “‘ShowerPower’ New Cooling Concept for Automotive Applications,” in Proc. Automot. Power Electron., no. June 2006, pp. 1–9.

[10] E.N. Wang, L. Zhang, J.-M. Koo, J.G. Maveety, E.A. Sanchez, K.E. Goodson, and T.W. Kenny, “Micromachined Jets for Liquid Impingement Cooling for VLSI Chips,” J. Microeletromech. Sys., vol. 13, no. 5, pp. 833-842, October 2004.

[11] T. Brunschwiler et al., “Direct liquid jet-impingement cooling with micronsized nozzle array and distributed return architecture,” in Proc. IEEE Therm. Thermomechanical Phenom. Electron. Syst., 2006, pp. 196–203.

[12] G. Natarajan and R. J. Bezama, “Microjet cooler with distributed returns,” Heat Transf. Eng., vol. 28, no. 8–9, pp. 779–787, July 2010.

[13] T. Acikalin and C. Schroeder, “Direct liquid cooling of bare die packages using a microchannel cold plate,” in Proc. IEEE Therm. Thermomechanical Phenom. Electron. Syst., 2014, pp. 673–679.

[14] S. Liu, T. Lin, X. Luo, M. Chen, and X. Jiang, “A Microjet Array Cooling System For Thermal Management of Active Radars and High-Brightness LEDs,” in Proc. IEEE Electronic Components and Technology Conf., 2006, pp. 1634–1638.

[15] Overholt MR, McCandless A, Kelly KW, Becnel CJ, Motakef S, “Micro-Jet Arrays for Cooling of Electronic Equipment,” in Proc. ASME 3rd International Conference on Microchannels and Minichannels, 2005, pp. 249-252.

[16] M. Baumann, J. Lutz, and W. Wondrak, “Liquid cooling methods for power electronics in an automotive environment,” in Proc 2011 14th European Conference on Power Electronics and Applications, 2011, pp. 1–8.

[17] B. P. Whelan, R. Kempers, and A. J. Robinson, “A liquid-based system for CPU cooling implementing a jet array impingement waterblock and a tube array remote heat exchanger,” Appl. Therm. Eng., vol. 39, pp. 86–94, June 2012.

[18] S. Ndao, H. J. Lee, Y. Peles, and M. K. Jensen, “Heat transfer enhancement from micro pin fins subjected to an impinging jet,” Int. J. Heat Mass Transf., vol. 55, no. 1–3, pp. 413–421, January 2012.

[19] Y. Han, B. L. Lau, G. Tang and X. Zhang, "Thermal Management of Hotspots Using Diamond Heat Spreader on Si Microcooler for GaN Devices," IEEE Transactions on Components, Packaging and Manufacturing Technology, vol. 5, no. 12, pp. 1740-1746, December 2015.

[20] A. J. Robinson, W. Tan, R. Kempers, et al., “A new hybrid heat sink with impinging micro-jet arrays and microchannels fabricated using high volume additive manufacturing,” in Proc. IEEE Annu. IEEE Semicond. Therm. Meas. Manag. Symp., 2017, pp. 179–186.

[21] H. Oprins, V. Cherman, G. Van der Plas, J. De Vos, and E. Beyne, “Experimental characterization of the vertical and lateral heat transfer in three-dimensional stacked die packages,” J. Electron. Packag., vol. 138, no. 1, pp. 10902, March 2016.

[22] N. Zuckerman and N. Lior, “Jet Impingement Heat Transfer: Physics, Correlations, and Numerical Modeling,” Advances In Heat Transfer, Vol. 39, pp. 565-631, 2006.

[23] J. Jorg, S. Taraborrelli, et al., “Hot spot removal in power electronics by means of direct liquid jet cooling,” in Proc. IEEE Therm. Thermomechanical Phenom. Electron. Syst., 2017, pp. 471–481.

[24] Skuriat , Robert, “Direct jet impingement cooling of power electronics,”. PhD thesis , University of Nottingham, 2012.

[25] Y. Han, B. L. Lau, G. Tang, X. Zhang, and D. M. W. Rhee, “Si-Based Hybrid Microcooler with Multiple Drainage Microtrenches for High Heat Flux Cooling,” IEEE Trans. Components, Packag. Manuf. Technol., vol. 7, no. 1, pp. 50–57, 2017.

[26] R. Skuriat and C. M. Johnson, “Thermal performance of baseplate and direct substrate cooled power modules,” in Proc. 4th IET Int. Conf. Power Electron. Mach. Drives, 2008, pp. 548–552.

[27] E. A. Browne, G. J. Michna, M. K. Jensen, and Y. Peles, “Microjet array single-phase and flow boiling heat transfer with R134a,” Int. J. Heat Mass Transf., vol. 53, no. 23–24, pp. 5027–5034, November 2010.

[28] E. G. Colgan et al., “A practical implementation of silicon microchannel coolers for high power chips,” IEEE Trans. Compon. Packag. Technol., vol. 30, no. 2, pp. 218–225, June 2007.

[29] C. S. Sharma, G. Schlottig, T. Brunschwiler, M. K. Tiwari, B. Michel, and D. Poulikakos, “A novel method of energy efficient hotspot-targeted embedded liquid cooling for electronics: An experimental study,” Int. J. Heat Mass Transf., vol. 88, pp. 684–694, September 2015.

[30] M. K. Sung and I. Mudawar, “Single-phase hybrid micro-channel/micro-jet impingement cooling,” Int. J. Heat Mass Transf., vol. 51, no. 17–18, pp. 4342–4352, August 2008.

[31] What is a DLP 3D Printer? https://www.makepartsfast.com/what-is-dlp-3d-printer/, APRIL 13, 2015.

[32] Somos® WaterShed XC 11122 – Protolabs, https://www.protolabs.com/media/1010884/somos-watershed-xc-11122.pdf.

# Chapter 7

# 7. Hotspot Target Cooling

In the previous chapter, the multi-jet cooling with different nozzle densities is demonstrated using 3D printing. It shows that the high-resolution stereolithography can realize a 1 mm nozzle pitch and 300 μm nozzle diameters. This small cooling cell of 1×1 mm$^2$, can enable to focus of the cooling solution at the location where it is needed. This chapter will present the design, fabrication, experimental characterization and modeling analysis of a chip-level hotspot targeted liquid impingement jet cooling. The hotspot targeted jet impingement cooling concept is successfully demonstrated with a chip-level jet impingement cooler with a 1 mm nozzle pitch and 300 μm nozzle diameter fabricated using high-resolution stereolithography (additive manufacturing). The CFD modeling and experimental analysis show that the improved hotspot targeted cooler design with fully open outlets can reduce the on-chip temperature difference by 70% compared with the full array cooler at the same pumping power of 0.03 W. The local heat transfer coefficient can achieve 15 × 10$^4$ W/m$^2$ K with a local flow rate per nozzle of 40 mL/min, requiring a pump power of 0.6 W. The benchmarking study proves that the hotspot targeted cooling is much more energy-efficient than uniform array cooling, with lower temperature difference and lower pump power.

### 7.1 Introduction

Thermal management is becoming a primary design concern for high power devices with the continuous scaling of the transistor size and increasing power density [1]. The localized heat flux can achieve values above 1 kW/cm$^2$ for sub-millimeter areas. These concentrated high heat flux values can cause localized hotspots (HS) with very high peak temperature [2], which can adversely impact the device performance and reliability [3,4]. In works of literature, many cooling solutions are investigated to minimize the maximum chip temperature [5], such as liquid cooling based microchannel [6] and microjet heatsinks [7], that can be further enhanced by increasing the contact area with fin arrays [8] or porous media [9]. In addition, some compact two-phase cooling systems such as micro heat pipes [10] are studied. However, these uniform cooling solutions for the entire chip surface or base plate area can result in excessive cooling to keep the maximum junction temperature below the specified maximum value in concentrated heat flux cases. Therefore, more energy-efficient cooling techniques should be developed by providing the targeted cooling on the local hotspots directly.

In the literature, several approaches have been proposed to eliminate the hotspots with high heat fluxes. To dissipate the high concentrated heat flux on the hotspots, diamond [11] or graphene [12] heat spreaders are applied to enhance the effective heat spreading capability. However, the cooling capacity is limited for high power devices. Embedded thermoelectric cooling (TEC) with small size, high reliability and low noise has great potential to provide reliable and localized cooling at hotspots [13] as they can be integrated into the heat spreader [14] or lid [15], embedded in the 2.5D/3D stacked chip package [16,17], or placed directly on the backside of the device [18]. Droplet-based cooling of electronic hotspots without external pumps has been demonstrated with the control of electrostatically actuated droplets, referred to as digital microfluidics using planar [19] or vertical integration [20, 21] schemes. However, the drawback of the TEC cooling and droplet cooling is the overall low cooling efficiency, the high energy consumption [14] and the complex integration in the chip package.

In addition, liquid-based cooling solutions have been investigated to deal with the hotspots, including manifold microchannel (MMC) heat sinks with embedded microchannels [22]. The hotspot targeted cooling is achieved by optimizing the microchannel array: narrow channels are designed over the hotspot locations, whereas coarse channels are present at the locations with lower background power dissipation, used as flow throttling zones to regulate flow in the different regions. The optimized cooler of [22] can reduce the maximum chip temperature nonuniformity by 61% to 3.7 °C for an average steady-state heat flux of 150 W/cm$^2$ in core areas (hotspots) and 20 W/cm$^2$ over remaining chip area (background). Microchannel cooler designs can be further optimized by varying the fin length and fin pitch in the heat sink according to the local hotspot heat flux [23, 24]. The thermal performance of the microchannel coolers can be further improved by combining the microchannels with an impinging microjet array. This type of hybrid Si heat sink has been introduced as a package-level hotspot cooling solution [11, 25], for GaN-on-Si Device in combination with a diamond heat spreader, achieving a high spatially average heat transfer coefficient of $18.9 \times 10^4$ W/m$^2$ K with low pumping power of 0.17 W. However, these microchannel based cooling technologies require expensive Si-based fabrication techniques such as etching and lithography.

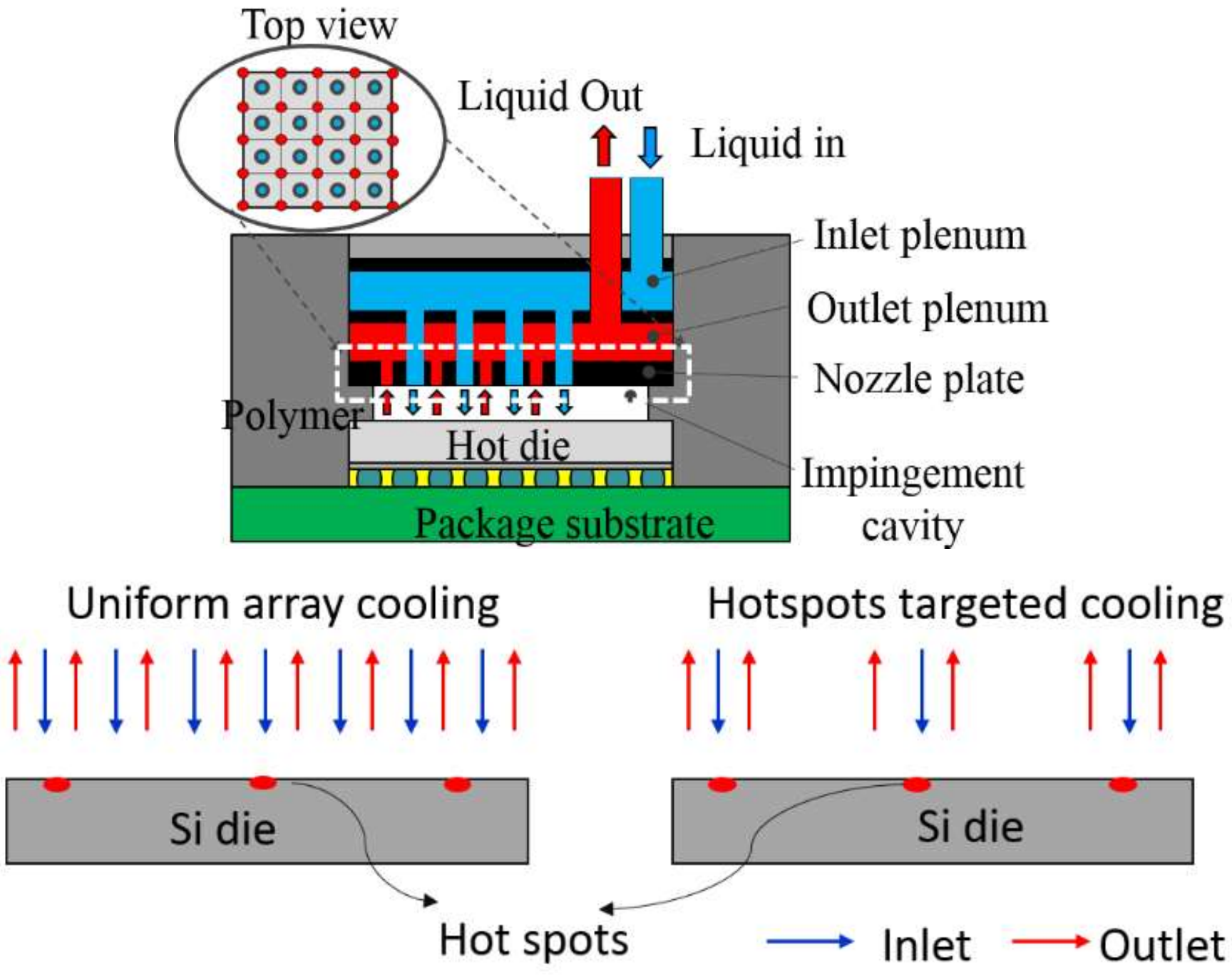

**Figure 7.1:** Concept of bare die jet impingement cooling with uniform array cooling [27].

In this chapter, jet impingement cooling technology is applied to concentrate the cooling on the hotspots, as illustrated in Figure 7.1. In section 7.2, the novel hotspot targeted cooling concept is introduced in detail and compared with the reference uniform array cooling. In section 7.3, the hotspot targeted coolers are demonstrated and experimentally characterized as a proof of concept, benchmarking the improved performance with respect to the reference cooler. Next, in section 7.4, full cooler level CFD models are introduced for the detailed analysis of the flow and temperature distribution inside the cooler in order to investigate the internal thermal and flow behavior in detail. Next, the CFD modeling results are experimentally validated by thermal and hydraulic measurements. Finally, in section 7.5, the flow and heat transfer characteristics are analyzed based on the validated CFD models, and further improvements of the cooler geometry are discussed to increase energy efficiency.

## 7.2 On-chip hotspot targeted cooling

### 7.2.1. Reference cooler: Uniform array cooling

In chapter 5, bare die jet impingement cooling has been demonstrated with mechanical micromachining, showing high cooling efficiency with a low pressure drop and low thermal resistance. The benchmarking study proves that multi-jet array cooling is more energy-efficient than other states of art liquid cooling solutions. It is also shown that the thermal conductivity of the cooler material has no big impact on the thermal performance of the impingement cooler, allowing to use a low-cost polymer-based cooler. 3D printing technology shows great advantages to fabricate low-cost polymer microjet coolers with complex internal 3D geometries comparing with mechanical micromachining techniques, which has been introduced in chapter 6. With the high-resolution 3D printing, microjet coolers with 8 × 8 nozzle array and 0.3 mm nozzle diameter have been demonstrated in chapter 6. In this section, the new measurement results for different flow rates are reported by using our 8 mm × 8 mm thermal test chip (introduced in chapter 2) for quasi-uniform heating with integrated heaters and temperature sensors. Figure 7.2(a) shows the $\overline{\mathrm{Nu}_\mathrm{j}}$-$\mathrm{Re}_d$ correlation for the 8 × 8 nozzle array cooling. The extracted conclusion is:

$$\overline{\mathrm{Nu}_\mathrm{j}} = 1.24Re_d^{0.67} \tag{7.1}$$

where $\overline{\mathrm{Nu}_\mathrm{j}}$ is based on the measured averaged chip temperature. The hydraulic characteristic lengths of the dimensionless number $\overline{\mathrm{Nu}_\mathrm{j}}$ and $\mathrm{Re}_d$ are both based on the inlet nozzle diameter $d_i$.

With the $\overline{\mathrm{Nu}_\mathrm{j}}$ - $\mathrm{Re}_d$ correlation shown in equation (7.1), the local heat transfer coefficient *htc* can be expressed as a function of the local flow rate per nozzle $\dot{V}$ with a power-law trend with an exponent of 0.67, where the local flow rate is calculated based on the total flow rate, listed as below:

$$\dot{V} = \frac{\dot{V}_{tot}}{N \times N} \tag{7.2}$$

where the analysis is based on the assumption of the unit cell behaviors of the multi-jet cooling. Also, the $\dot{V}_{tot}$ is the total flow rate for the cooler, and $\dot{V}$ is the local flow rate per nozzle. Also, N × N is the inlet nozzle array. The parameter N is a fixed number equal to 8 throughout the analysis since the uniform nozzle array design is an 8×8 nozzle array in the considered case. As shown in Figure 7.2(b), the measured maximum heat transfer coefficient can be achieved as $7.39 \times 10^4$ $\mathrm{W/m^2\,K}$ with a local flow rate per nozzle of 15.63 mL/min, resulting in a total flow rate of 1000 mL/min. The pressure drop measurement result for the full nozzle array cooling with an 8×8 nozzle array is shown in Figure 7.10(b) in section 7.3. In the next part, a hotspot targeted cooling concept will be introduced for non-uniform power, which can be made even more energy efficient.

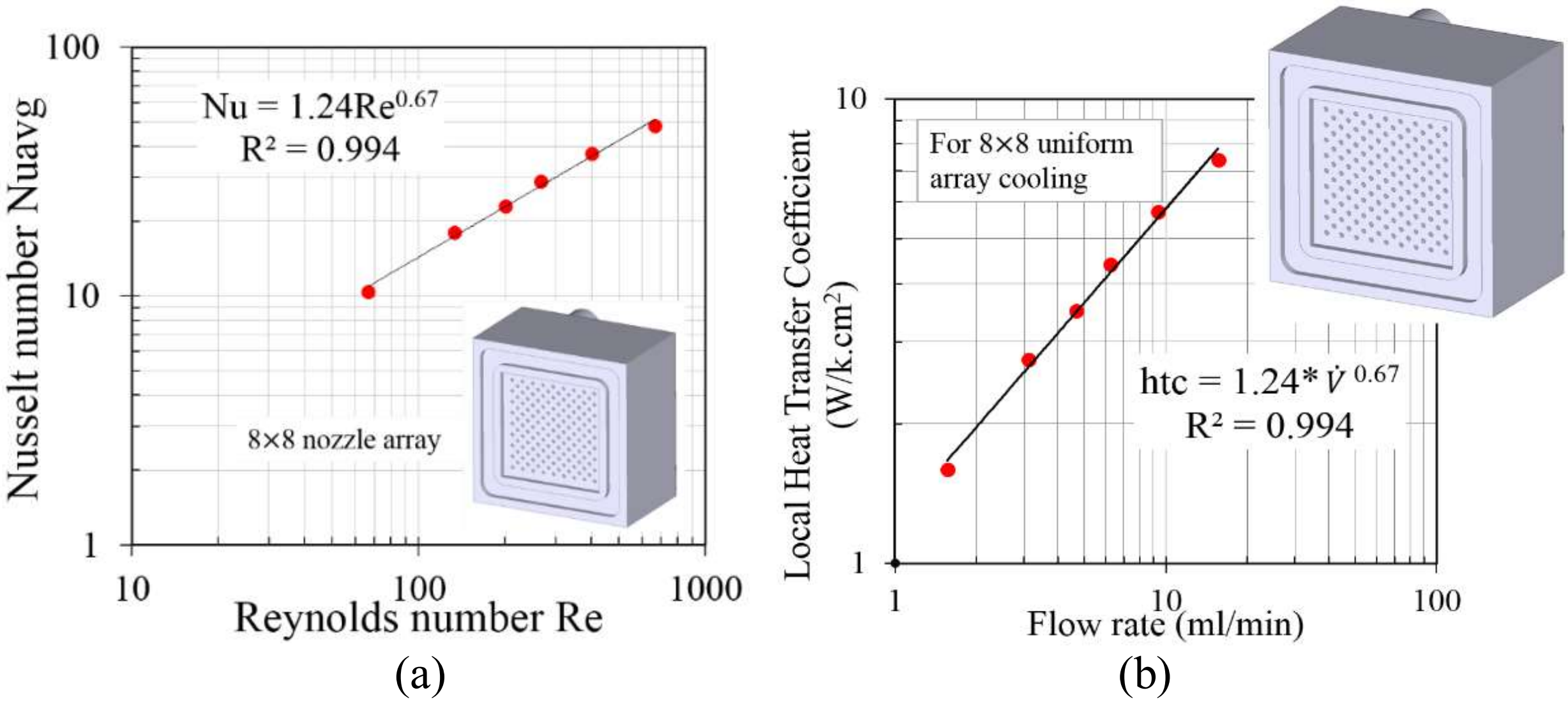

**Figure 7.2:** Measurement results with uniform array cooling for 8×8 nozzle array cooler under different local nozzle flow rate: (a) Nusselt number as a function of Reynolds number; (b) local heat transfer coefficient as a function of flow rate per nozzle.

### 7.2.2. Hotspot targeted cooling concept

Additive manufacturing enables the customization of the cooler design to match the power dissipation pattern of the chip in order to increase cooling efficiency. In the case of hotspot power dissipation patterns, the location of the impinging jet nozzles that eject the coolant onto the chip can be aligned to the location of the hotspots. The main idea of hotspot targeted cooling is to focus the cooling solution, at the location where it is needed. In the areas outside the hotspots, a lower nozzle density is designed to cover the area with lower heat flux values for the background power dissipation. In the extreme case where no background power is present, the nozzles outside the hotspot area can be omitted since no power generation is present. In this way, a higher local cooling flow rate will be provided to the chip locations with higher power densities, resulting in a selective cooling of the chip area, rather than a uniform cooling across the whole chip surface. Since the constriction of the coolant to these selected regions will result in higher heat transfer rates as well as higher required pressure drop values, a detailed experimental and numerical analysis is presented in the next sections. The concept of the hotspot targeted liquid impingement jet cooling is schematically shown in Figure 7.3 for the case with several hotspots and no background power.

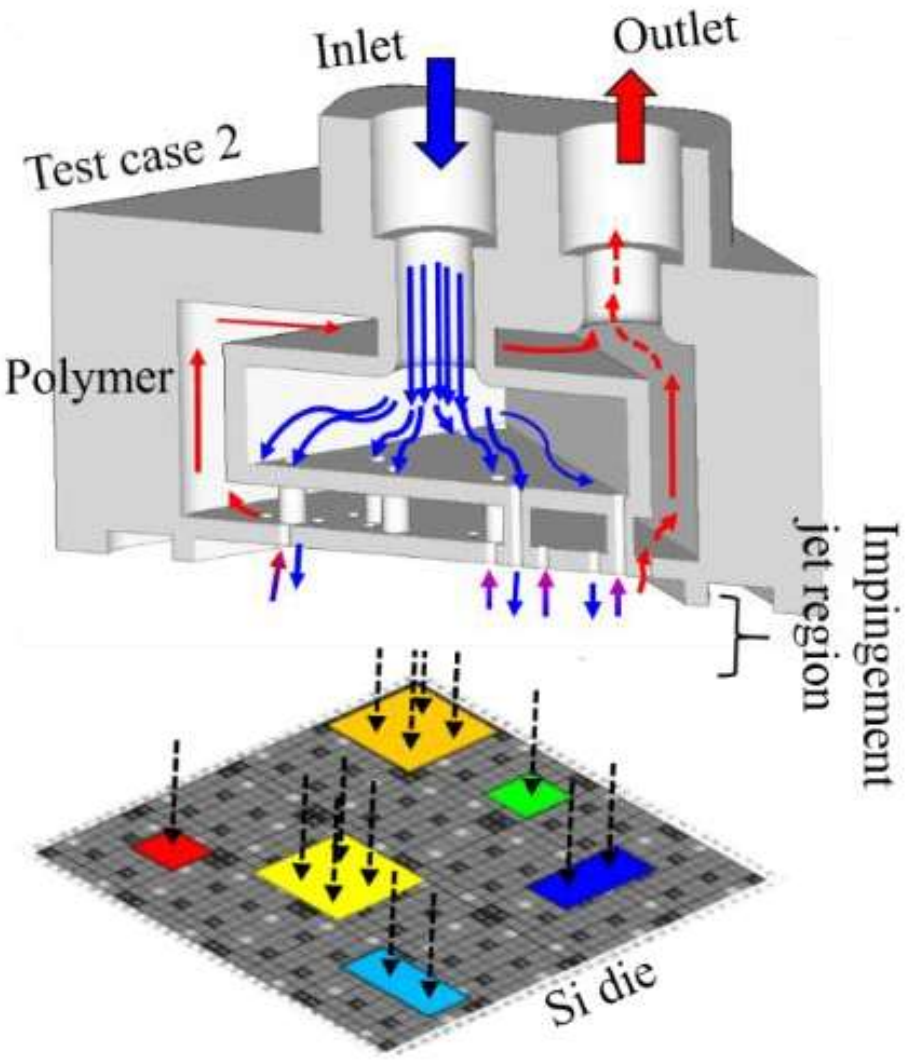

**Figure 7.3:** Concept of hotspot targeted liquid impingement cooling with several hotspots and no background power.

### 7.2.3. Hotspot pattern

For the assessment of the hotspot targeted cooling, two hotspot case studies have been defined, based on the heat generation capabilities of the test chip:

- test case 1 with a regular hotspot pattern, mimicking the design of a multi-core processor (Figure 7.4(b));
- test case 2 with various hotspot sizes, mimicking a power electronics die (Figure 7.4(c));

For test case 1 with the regular hotspot pattern, there are 72 heater cells activated with a total heater area of 4.15 $mm^2$ for the 24 heat sources. For test case 2 with various hotspot sizes, the total number of the activated heater cells is 127, with a heater area of 7.32 $mm^2$. The power density scale for the three different power maps (for 1V)is shown in Figure 7.4. The power and power density are different for all three cases. The test chip is powered by applying a voltage, and by choosing which heater cells are activated. In the experiments, a constant voltage of 1V is applied at the package. The actual power in the heater cells (and the local voltage drop) depends on the series connection of parasitic resistance and heater resistance array, which acts as a voltage divider, and on the connections in the package substrate depending on the metal line connections between the heater cell and contact pad. For the example of 1V of applied voltage on the package, the actual measured power dissipation in the heater cells is:

1) Hot spot test case 1: Total power is 4.1 W, power density is 98 $W/cm^2$;

2) Hot spot test case 2: Total power is 5.5 W, power density is 75 W/cm$^2$;
3) Uniform reference case: Total power is 30 W, power density is 62.5 W/cm$^2$;

The power values/densities in this study are limited to a small value (<10W), which is not representative of the modern high-power CPUs/GPU in 100W-200W. However, the extracted thermal resistance or heat transfer coefficient can be scaled to higher power values.

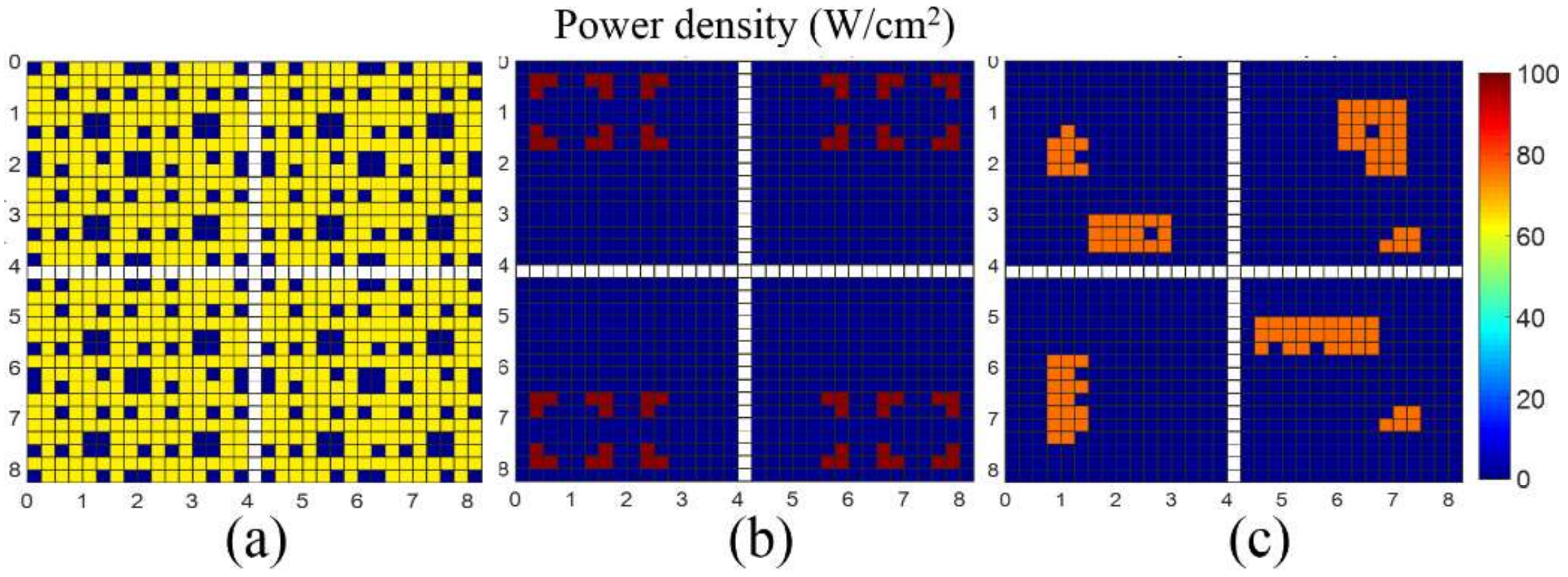

**Figure 7.4:** Test cases for the hotspot cooling with different power density maps: (a) reference case with the quasi-uniform heating pattern; (b) test case 1 with the regular pattern; (c) test case 2 with various hotspot sizes.

## 7.3 Proof of concept: Hotspot target cooling

### 7.3.1. Demonstration of 3D printed cooler

The hotspot targeted cooler demonstrator is fabricated using the polymer-based high-resolution stereolithography (SLA) as for coolers discussed in chapter 6. This results in the successful fabrication of the cooler with nozzles diameter of 300 µm and a pitch of 1 mm using the water-resistant Somos WaterShed XC material [31], which shows ABS-like properties. As discussed in chapter 6, the temperature of the coolant should be below 60°C to remain in the safe temperature range for the cooler material. For a cooler size of 14 mm × 14 mm × 8 mm, the total time required to produce the part is about 8 hours. Schematics of the designed hotspot targeted cooler versions for the two test cases are shown in Figure 7.5, revealing the internal cooler geometry. The cavity height is designed as 0.6 mm. The number of the inlet nozzles for test case 1 is 24, while it is only 15 for test case 2, compared to 64 for the full array cooler. The location of the nozzles has been aligned to the location of the hotspots of Figure 7.4(b) and Figure 7.4(c). The top row of Figure 7.6 shows a bottom view of the designs of the nozzle plate, while the bottom row of the figure shows a comparison with the photographs of the actually fabricated demonstrators. The uniformity of the printed nozzle diameter can be measured from the bottom view of the cooler, showing only a 5% difference.

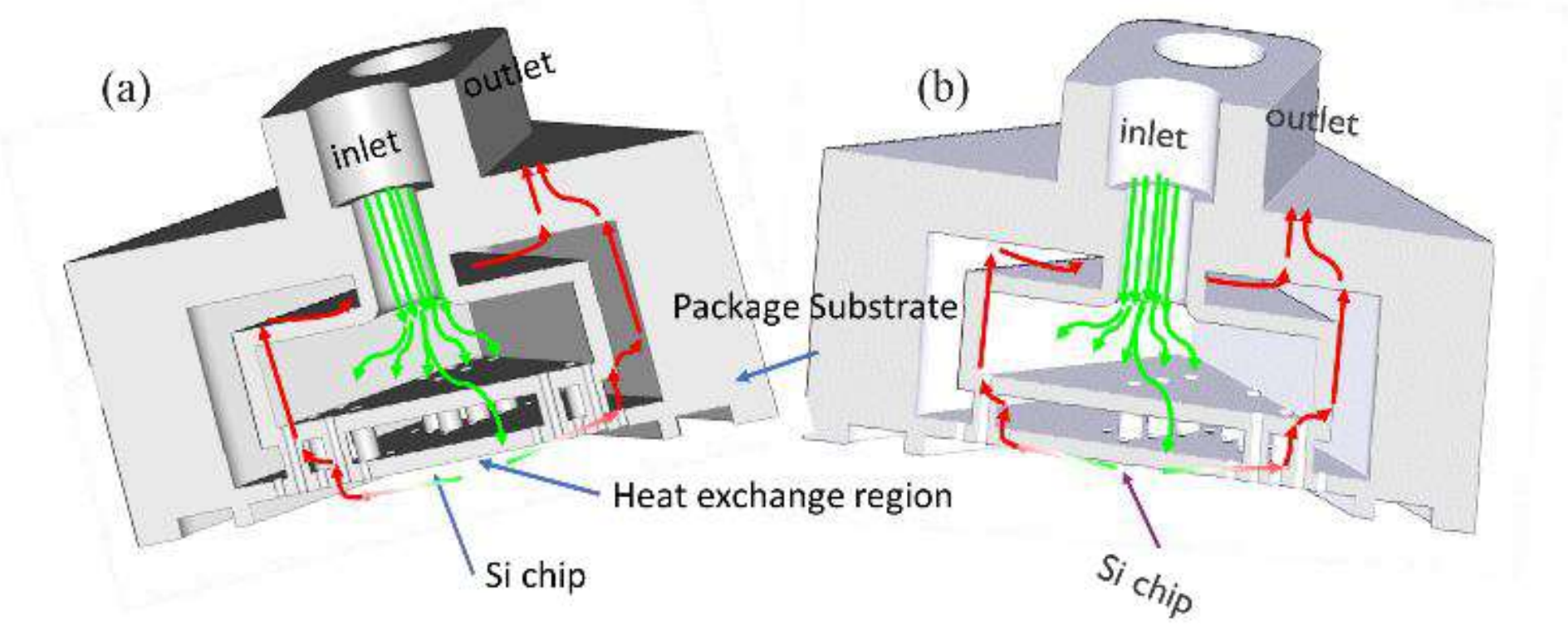

**Figure 7.5:** Cross-section of the CAD designs of the two test cases: (a) test case 1; (b) test case 2.

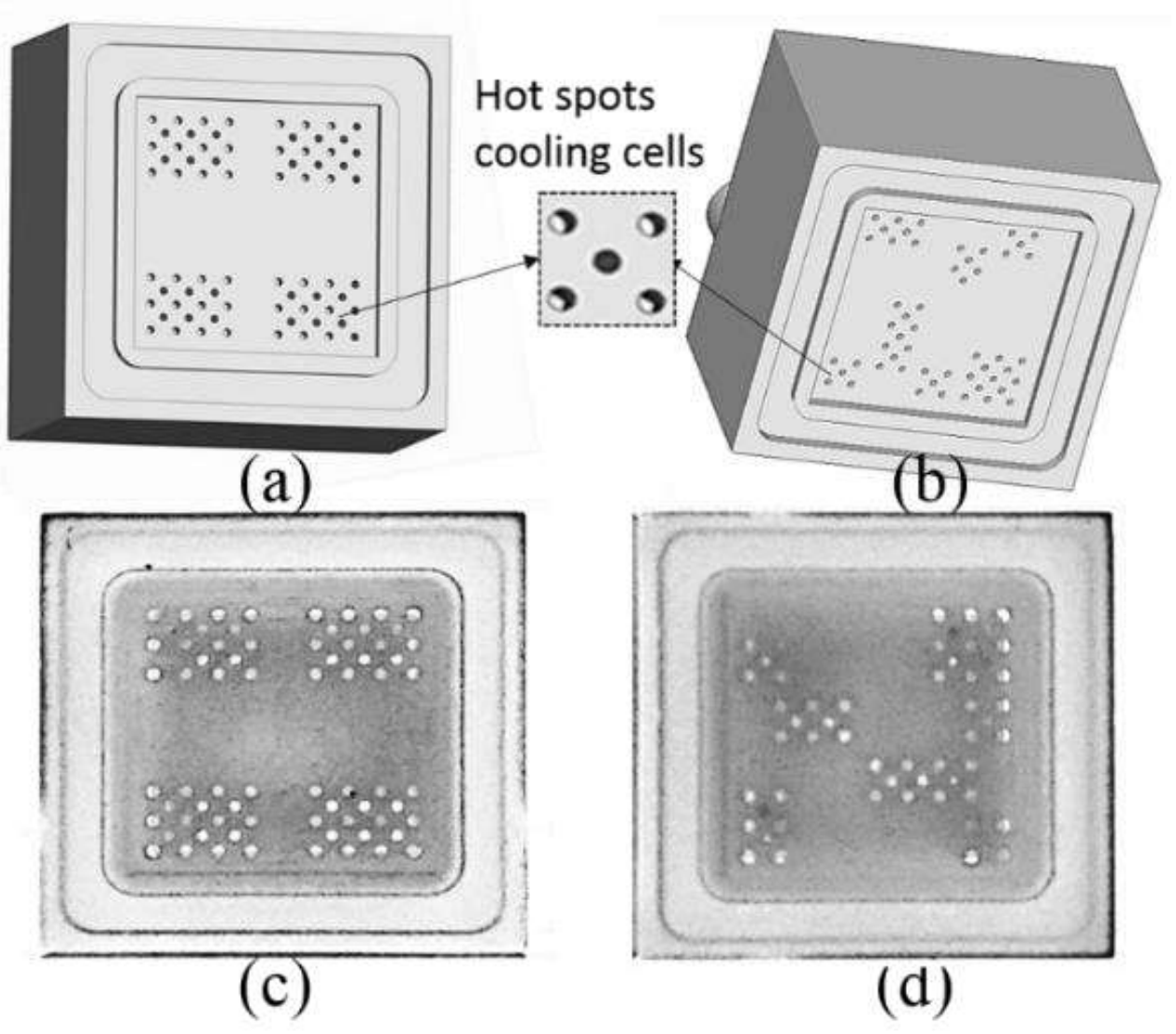

**Figure 7.6:** Top row: bottom view of the hotspot targeted coolers revealing the nozzle array for test case 1 (a) and 2 (b). Bottom row: photographs of the nozzle plate of the fabricated coolers for test case 1 (c) and test case 2 (d).

### 7.3.2. Thermal characterization

The dedicated hotspot coolers are mechanically assembled on the package substrate with the advanced thermal test chip by using a plastic socket, illustrated in Figure 7.7. The assembled cooler is finally connected into the closed-loop test set-up enabled with accurate flow rate and pressure drop measurement systems, as introduced in chapter 2.

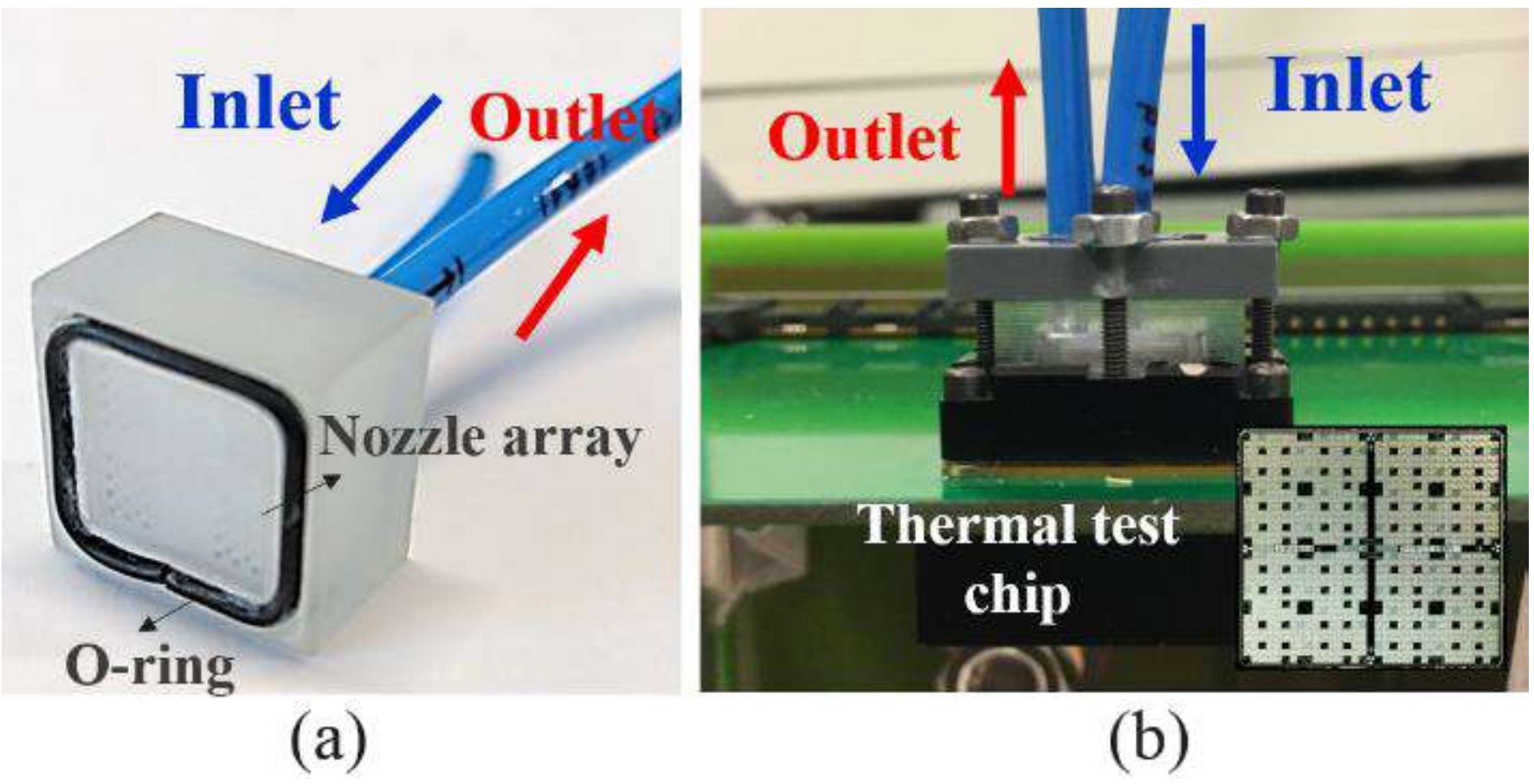

**Figure 7.7:** Cooler assembly: (a) hotspot cooler for the regular pattern with O-ring placement; (b) assembly of the cooler on the thermal test chip and PCB test board.

In the first set of experiments, the heat dissipation patterns on the chip, shown in Figure 7.5(b) or Figure 7.5(c), are activated while the full-chip temperature map is measured for a specific flow rate once the steady-state condition has been reached. The chip temperature is extracted at the chip FEOL, which is the same location as where the diodes are, and the junction in the application.

The measured total chip power for test case 1 with a regular hotspot pattern is 4.1 W. For test case 2 with various sizes of the hotspot, the measured full chip power is 5.5 W. The power for the reference case with uniform heating is set as 30W. For both test cases, the chip temperature profile is compared between the reference full array cooler (Section 7.2.1) and the respective hotspot targeted cooler (Section 7.3.1) for the same coolant flow rate.

For the coolant heat removal percentages, we performed the thermal measurements of the packages without cooling applied. In the experiments, the results show that the percentage of heat loss through the package is limited to only 2 – 5 % and the majority of the heat is removed through the top side of the chip, and since the power values for test case 1 and test case 2 are only 4.1 and 5.5 W respectively, the heat loss through the package and convection can be considered very similar. At these temperature values (15 °C average chip temperature), radiation can be neglected.

In Figure 7.8, the measured temperature maps are compared for a total flow rate of 600 mL/min. For a more detailed comparison, the temperature profile is plotted across the test chip diagonal, shown in Figure 7.9. The temperature measurements show a peak temperature reduction of 16% and 42% at a flow rate of 600 mL/min compared to the full array cooler for the targeted hotspot coolers of test case 1 and test case 2 respectively, indicating that concentrating the liquid coolant on the locations where it is needed, can result in a significant reduction of the chip peak temperature due to the locally increased coolant flow rate. This is, however, achieved at the cost of an increase in the required pressure drop. The pressure drop will be characterized experimentally in the next section, while the flow distribution impact will be discussed in Section 7.5.1.

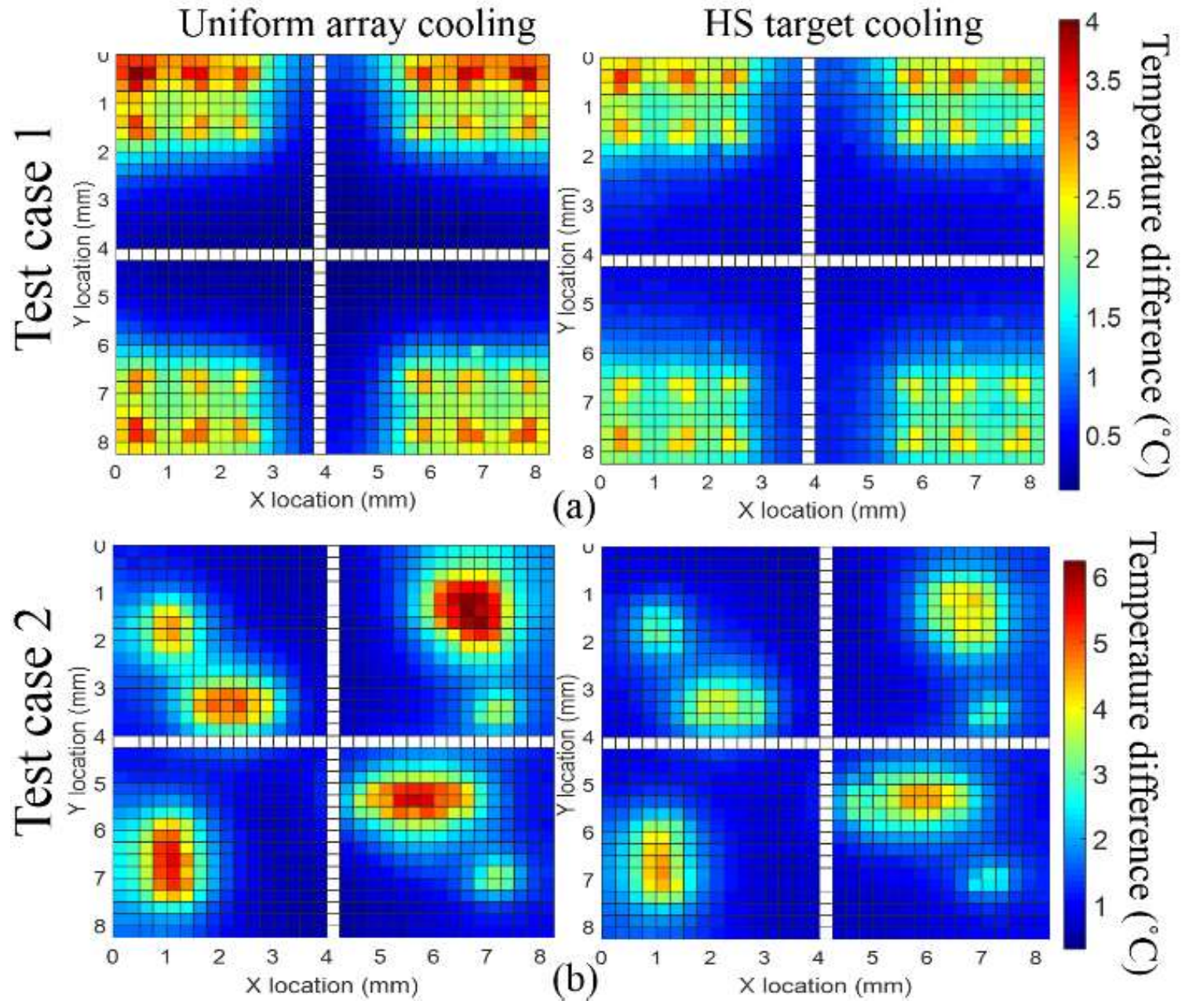

**Figure 7.8:** Measured temperature distribution for uniform array cooling and HS cooling with (a) test case 1 regular hotspot pattern and (b) test case 2 with various hotspot sizes at a flow rate of 600 mL/min.

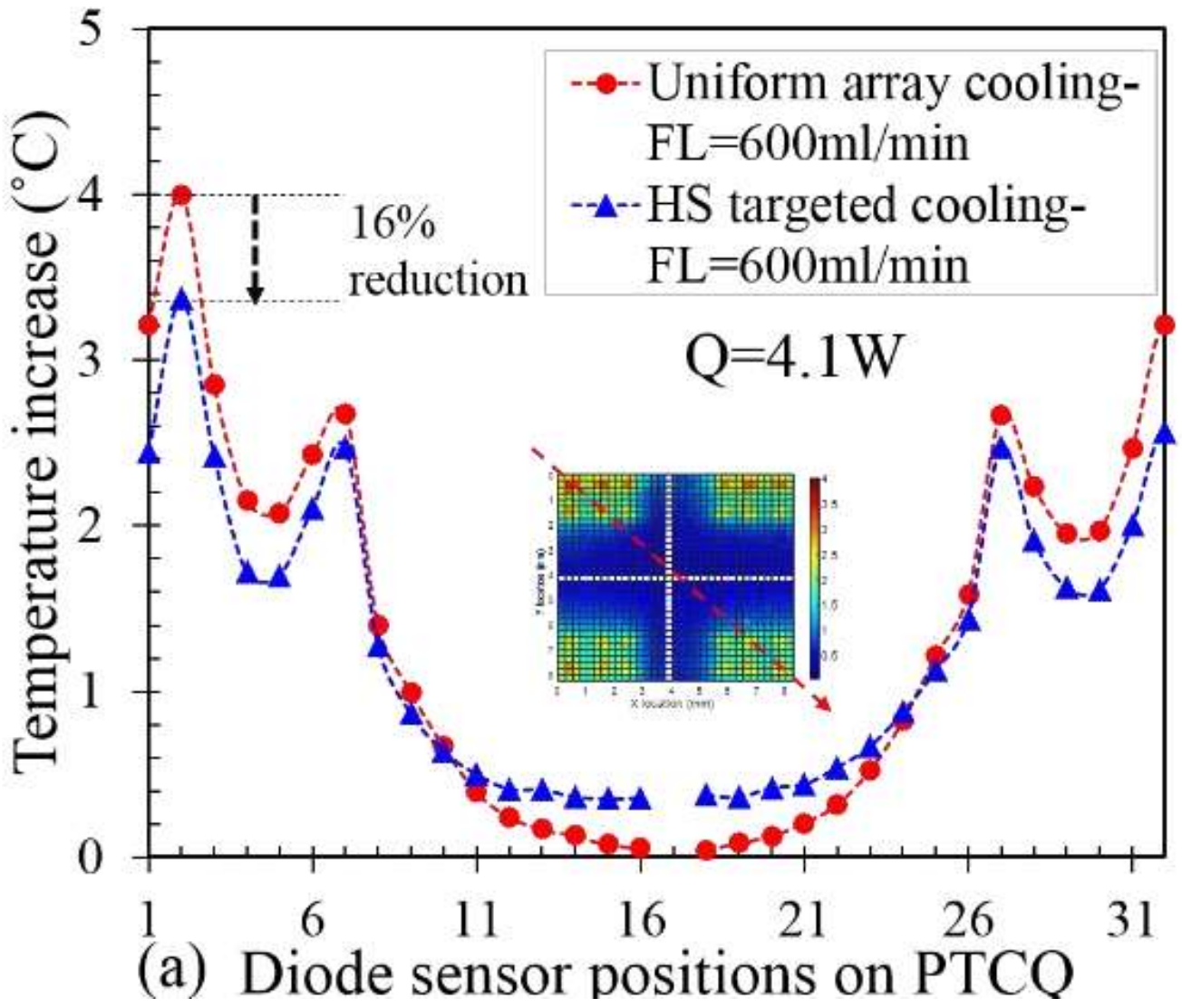

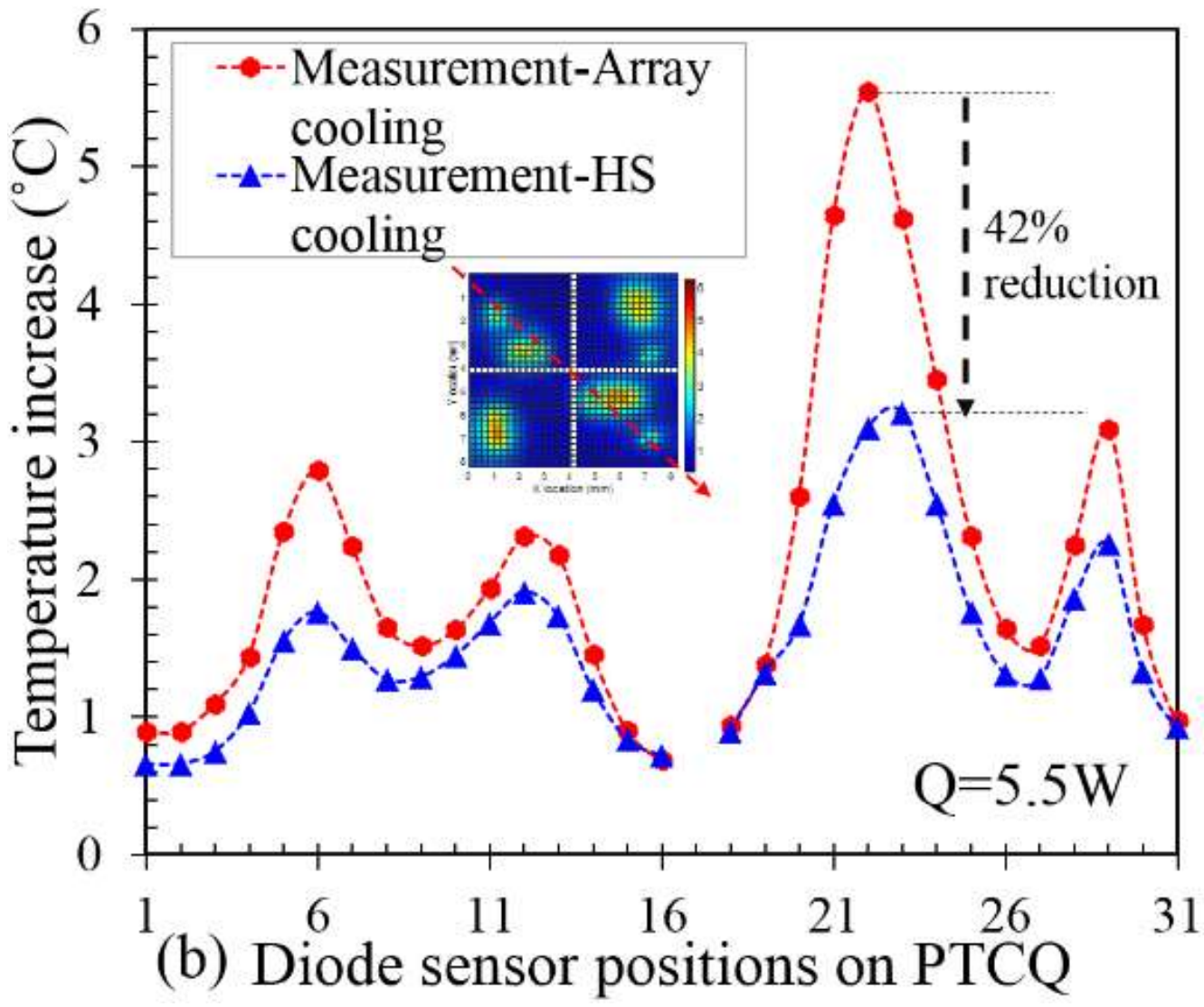

**Figure 7.9:** Temperature profile comparison for (a) test case 1 regular hotspot pattern and (b) test case 2 with various hotspot sizes with the reference cooler. (Notes: the temperature increase is with regard to the inlet fluid)

### 7.3.3. Pressure drop measurements

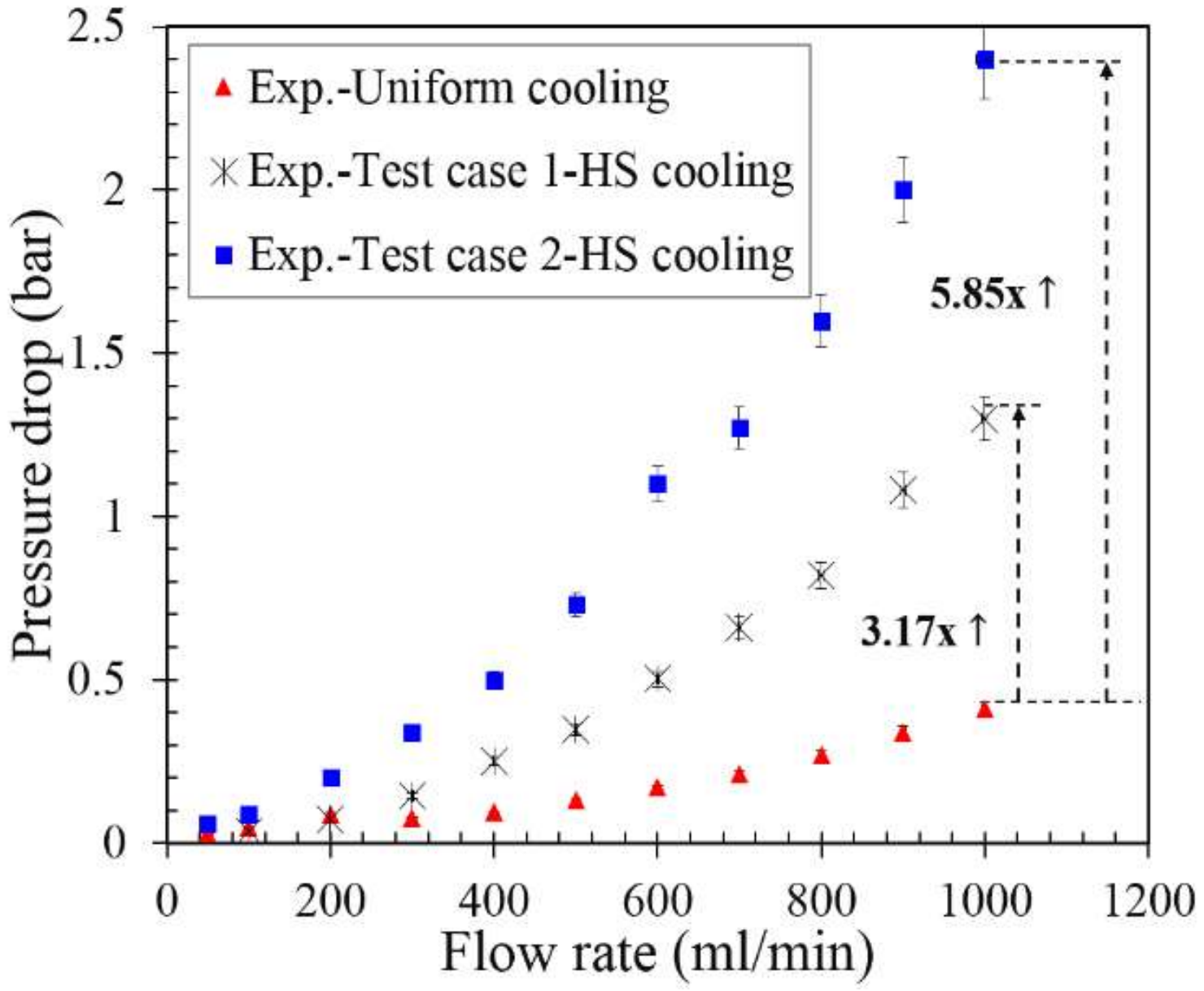

**Figure 7.10:** (a) Schematic diagram of the thermal/flow loop measurement system and (b) the pressure drop measurements for the reference cooler and the two versions of the hotspot target cooler.

The pressure drop measurement for the flow loop system is introduced in chapter 2. The inlet and outlet of the cooler are connected with small tubes for the whole flow loop connection. Therefore, the pressure drop of the inlet/outlet tube and connection is included in the measured pressure drop. The modeling results show that the pressure drop of the cooler is smaller than the tube connection part. Therefore, a de-embedding technique can be used to measure the pressure of the cooler only, without the tube connection. Since the pressure drop over the tube is linearly proportional to the tube length, the pressure drop between the inlet and outlet connection of the cooler can be estimated by measuring the pressure drop for the different tube lengths and extrapolating to zero tube length. The pressure drop over the three coolers has been measured for controlled flow rate values in the range between 50 and 1000 mL/min. As shown in Figure 7.10, the measured pressure drop for the uniform 8×8 array cooler is lower than for the other cooler under the same flow. The pressure drop of hotspot targeted cooler for test case 1 is 3.2 times higher than the uniform array cooling, while the pressure drop for test case 2 is 5.9 times larger under the flow rate of 1000 mL/min. The increase of the pressure drop is due to the reduction of the number of nozzles and the additional hydraulic constriction resistance in the inlet plenum.

In summary, the thermal and hydraulic measurements show that the hotspot targeted cooler can improve the cooling efficiency toward the hotspots, however, at the expense of an increase in pressure drop. In the next section, the hydraulic behavior will be investigated in more detail using validated CFD models.

## 7.4 Modeling methodology and validation

### 7.4.1. Full cooler level CFD model

System-level pumping power is an important factor for the design of the liquid cooler from an energy consumption point of view. Unit cell level models on the level of an individual jet nozzle provide an interesting insight into the thermal and hydraulic behavior of the multi-jet cooler. However, they can only predict the pressure drop between the local level inlet and outlet nozzles on the nozzle plate. In order to assess the cooler hydraulic behavior and to extract the pressure drop between the inlet and outlet, full cooler level CFD models are required. Furthermore, these whole cooler level CFD models can be used to optimize the geometry of the inlet plenum and outlet plenum to reduce the pressure drop in the cooler.

Figure 7.11 shows the full cooler model geometry, extracted from the CAD design file. As illustrated in chapter 2, the model is based on a steady-state conjugate heat transfer CFD model, which takes into account the heat conduction in the solid structures and heat conduction and convection in the liquid domain in the system. The solid domain in the CFD model is the silicon die part, and not the solid part of the plastic manifold. In chapter 5, we show that the thermal conductivity of the cooler material has no impact on the modeling results for temperature distribution in the Si chip, and no difference was found using a thermal insulation boundary condition on the surface of the fluid domain inside the cooler geometry. Therefore, the presented CFD model does not include the plastic part of the cooler. In addition, for the interface between the fluid domain and solid domain (silicon die), there is a boundary layer mesh between the solid part and the fluid part. The transition shear stress transport (SST) model is still chosen as the turbulent model in the CFD simulations, which can cover the laminar and transition flow regimes with good accuracy for jet impingement flows [26]. In order to capture the temperature distribution map with the hotspots, a sufficiently detailed mesh of the heaters is required in the model, as discussed in chapter 5. In Figure 7.11, the full level CFD model of hotspot targeted cooler is shown, and the mesh details are shown in a cross-section of the modeled geometry. The results of the mesh independent analysis for the full cooler CFD model are shown in Figure 7.11(c), performed for the regular hotspots cooler of case 1, at a fixed flow rate of 1000 mL/min. It is observed that the mesh for a number of elements between 4.5M and 5M is mesh-independent. And also, the truncation error estimation from the Richardson extrapolation [32] is around 0.28% and can be used for the modeling study. Since the critical region with the nozzle diameter is the same, therefore, the meshing sensitivity is also applicable for other cases. Based on the meshing sensitivity analysis, the meshing size of the fluid domain is set as 0.12 mm, while the meshing size is 0.04 mm for solid domain. The first layer thickness of the boundary layer is set as 1e-3 mm in Z with 10 layers above the fluid/solid interface, and the layer growth rate is set as 1.2. The total element number is 4.5 M-5 M.

For the boundary conditions of the model, the inlet temperature is set at 10°C. A constant heat flux of 98 W/cm$^2$ is applied to the hotspot areas for test case 1, while the applied heat flux for test case 2 is 75 W/cm$^2$. To match the measurement conditions, a constant velocity is applied on the inlet boundary while the boundary condition for the outlet pressure is set to $P_{out}$=0 Pa. The fluid and solid interface is set as a coupled boundary condition. Since the cooler material is plastic with low thermal conductivity, the boundary walls of the internal cooler channels are assumed to adiabatic walls. This assumption has been validated by full cooler level simulations with different materials, showing no significant impact of the cooler material conductivity or the adiabatic boundary condition. The fluid properties and silicon die properties are not temperature-dependent.

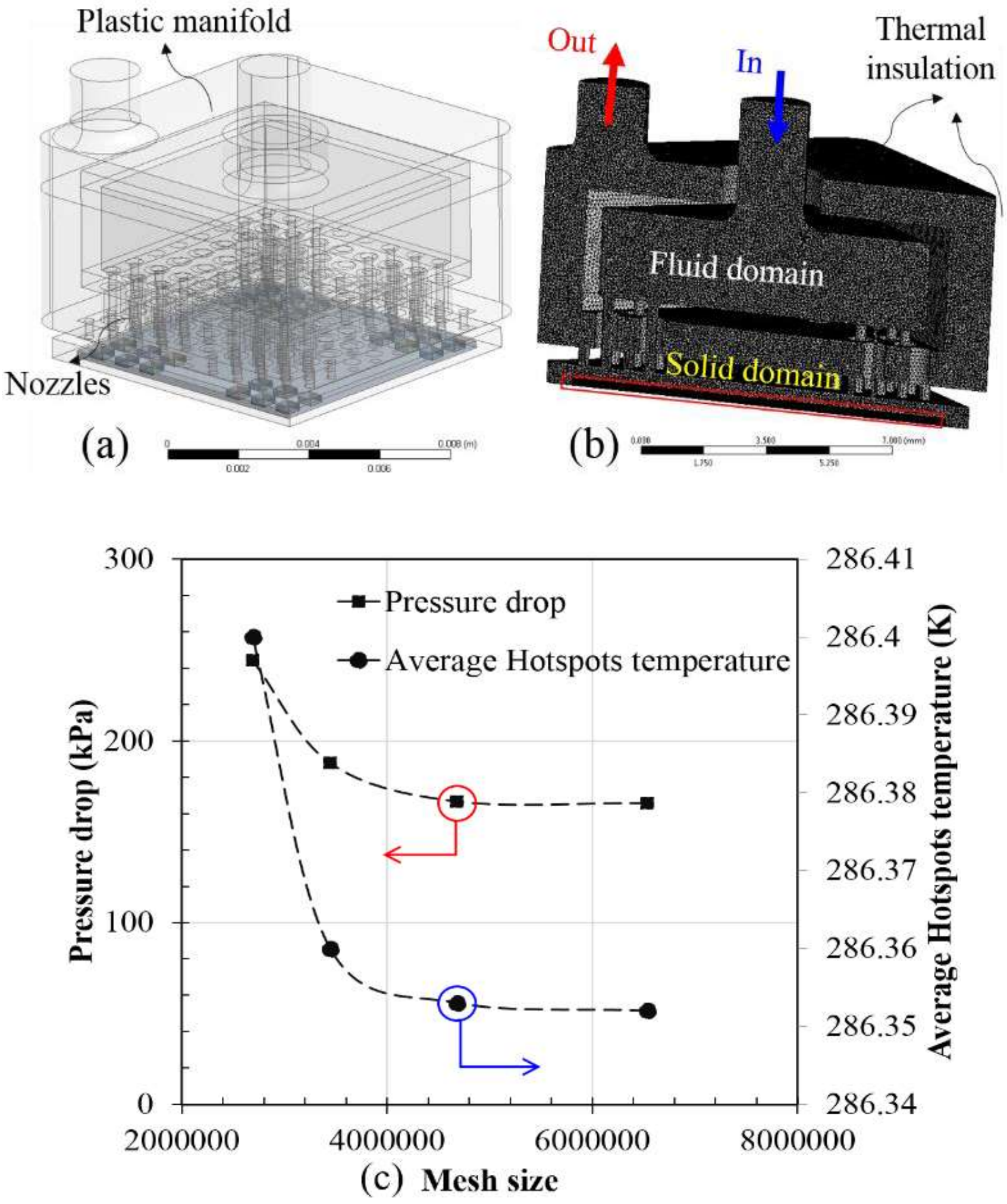

**Figure 7.11:** Full cooler level CFD model: (a) transparent view of the cooler; (b) cross-section of the meshing with test case 1; (c) mesh sensitivity analysis with CFD model of test case 1.

### 7.4.2. Thermal and hydraulic model validation

The temperature distribution map comparisons between the full cooler level CFD modeling and experimental results for the test case 1 and test case 2 are illustrated in Figure 7.12 with a total power dissipation of 4.1W and 5.5 W for a flow rate of 600 mL/min. In general, the comparisons for the temperature map show good qualitative agreement between measurement and simulation results. For the detailed comparison of test case 1, the temperature profile is plotted across the chip diagonal in Figure 7.13(a). It can be seen that the model captures the temperature peaks and the area without power very well, and also shows a good agreement for the temperature profile. Moreover, the average difference between the simulated average chip temperature and the averaged chip temperature based on all 1024 temperature sensors is less than 3% for the uniform nozzle array cooler, while the average difference is 7% for the targeted hotspot cooler. The asymmetrical temperature measurement map shown in Figure 7.13(a) is due to the asymmetrical placement of the outlet tube connector.

For the temperature profile of test case 2, shown in Figure 7.13(b), the comparison also shows a good agreement with the measurements and simulation results. For the averaged chip temperature in the test case 2 calculated from Figure 7.12(b), the comparison between simulation and measurement of the average chip temperature shows 6% and 9% average difference for the uniform array cooler and the targeted hot spot cooler respectively.

As shown in Figure 7.13, the hotspot cooler simulations have a much higher deviation compared to the uniform case. This is because the CFD model is a simplified model where the bottom substrate and solder connections are presented as a boundary condition with an equivalent heat transfer coefficient. In case of the uniform heating and cooling, the heat transfer in the silicon die is primarily one-dimensionally vertical, which is accurately captured by the simplified model. In the case of the hot spots, there is also a significant lateral spreading in the silicon. It could be possible that this lateral spreading is not completely captured.

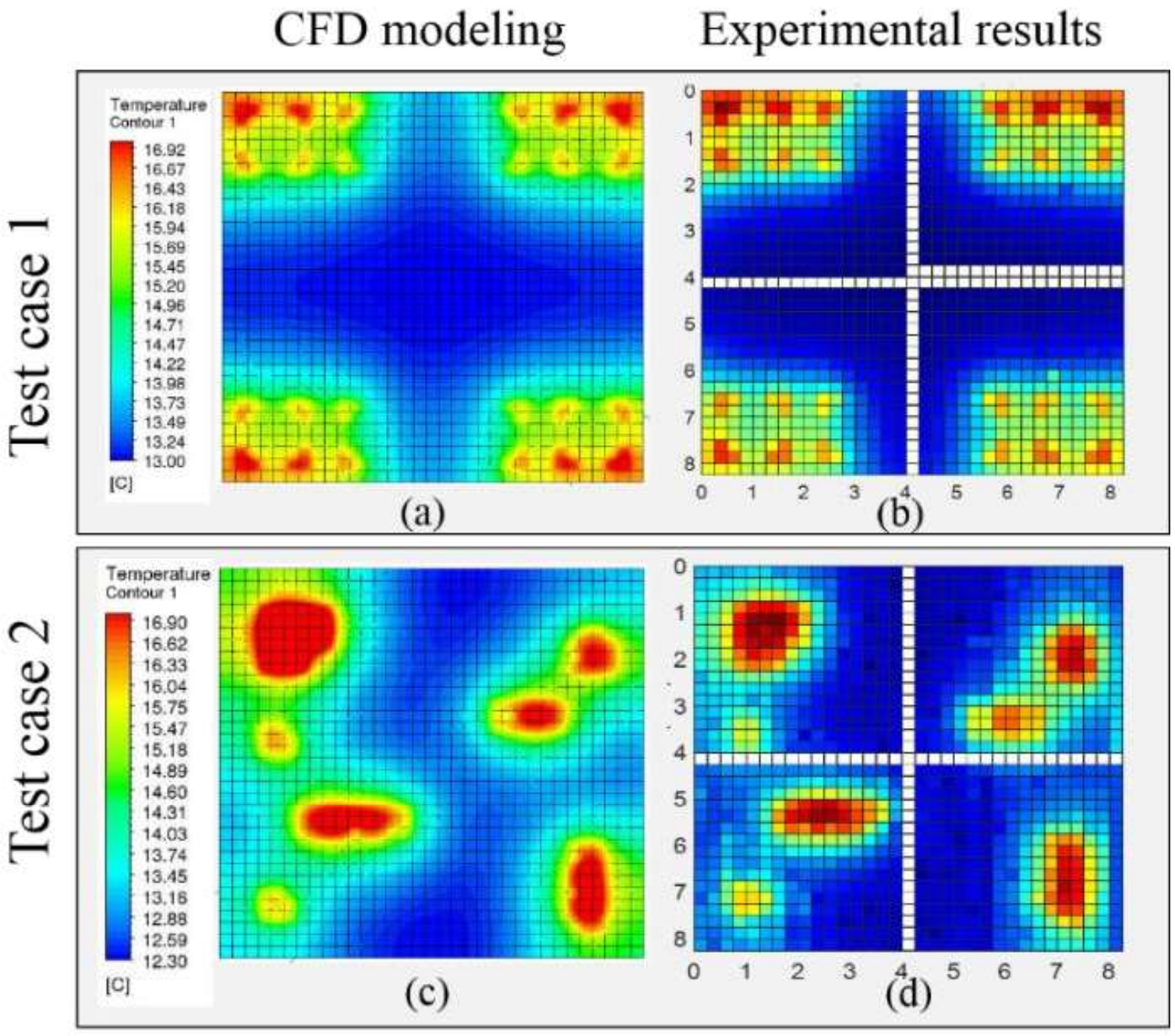

**Figure 7.12:** Temperature map distribution comparison of test case 1 with chip power of Q=4.1W: (a) CFD modeling and (b) experimental results; test case 2 with chip power of 5.5 W; (c) CFD modeling and (d) experimental results (flow rate = 600 mL/min).

Figure 7.14 shows the comparison between the simulated pressure drop between the inlet and outlet connectors and the experimental measurements for the three considered cooler designs. The simulated pressure drop shows a 12.3% average difference from the measured pressure drop at the flow rate of 1000 mL/min for test case 1. In general, the modeling results for uniform array, and HS targeted cooling show good agreement with the experimental results, showing an average difference smaller than 13%.

Based on the acceptable errors of the full CFD cooler model compared with the experimental data, the CFD models with different cooler configurations are successfully validated. The validated CFD models are applied in the next sections to assess the thermal performance gains for design improvements and the trade-off between the thermal performance improvement and the pressure drop penalty in the hotspot targeted cooler.

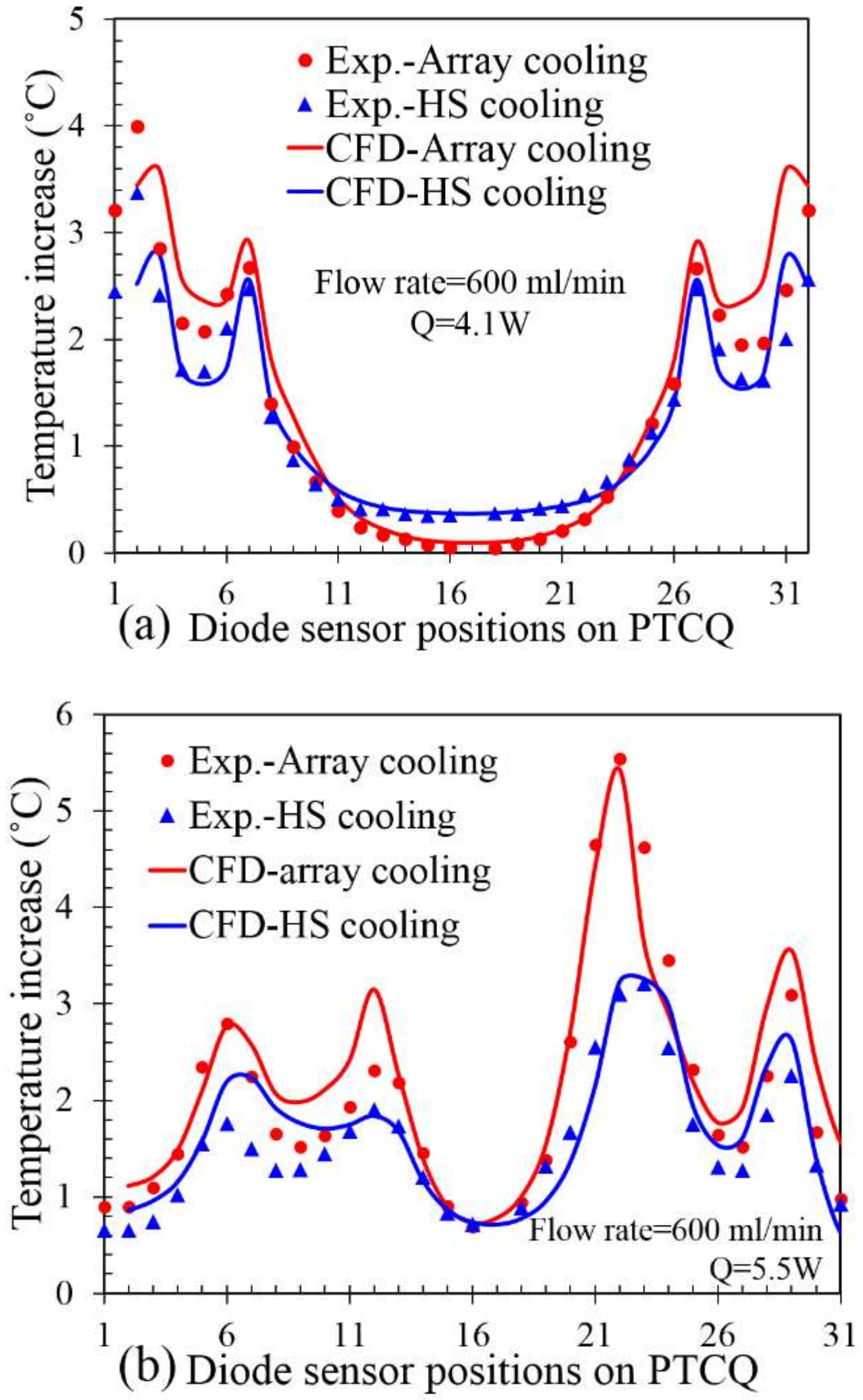

**Figure 7.13:** Temperature measurement results and full cooler CFD model comparison for (a) test case 1 and (b) test case 2.

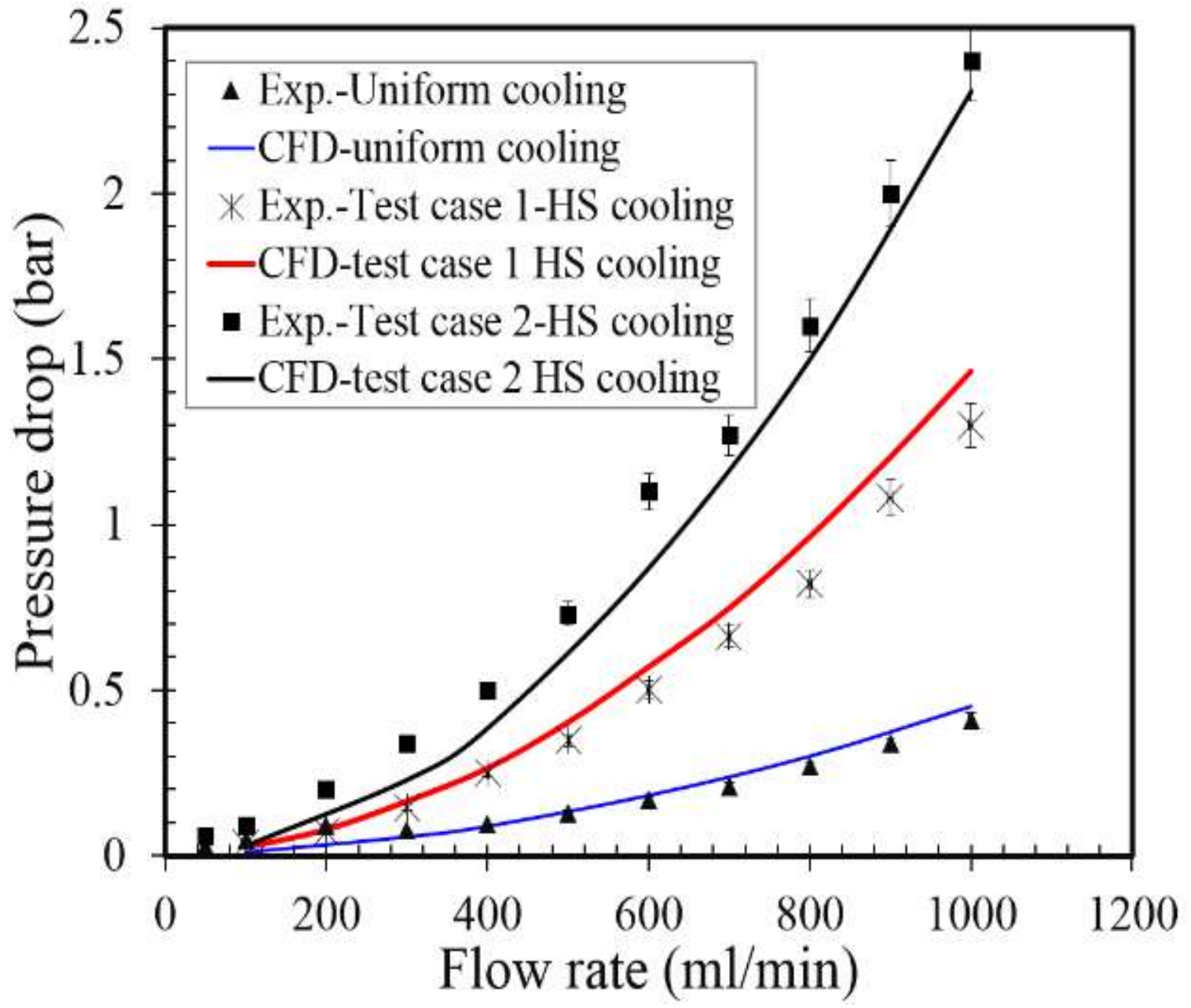

**Figure 7.14:** Experimental and CFD modeling comparison for the pressure drop under different flow rates.

## 7.5 Thermal/hydraulic modeling analysis

### 7.5.1. Local flow rate analysis

For the local flow rate analysis, a unit cell approach is used as a first estimation, to assess the improvement in cooling at the targeted chip areas. Based on the measurement data with $1\times1$ $mm^2$ cooling unit cells shown in Figure 2, the relation between the local heat transfer rate *htc* and local inlet nozzle flow rate $\dot{V}$ is shown with a power-law trend with an exponent of 0.67, derived from equation (1). Therefore, the expected heat transfer coefficient $htc^*$ for the hotspot cooler can be extracted as below:

$$htc^* = m^{0.67}\text{htc} \quad (7.3)$$

$$\dot{V}^* = m\dot{V} \quad (7.4)$$

$$m = \frac{N^2}{M} \quad (7.5)$$

The hot spot area is used for the estimation of the local heat transfer coefficient that is mainly used as a relative comparison with the global heat transfer coefficient.

Similarly, the expected pressure drop for the targeted cooler is shown as below:

$$\Delta\text{p} \sim m^2\dot{V}^2 \quad (7.6)$$

where $\dot{V}$ is the averaged local flow rate per nozzle, $\dot{V}^*$ is the averaged local flow rate for the hotspot targeted cooler. $N^2$ is the total inlet nozzle number with the array cooler. $M$ is the total inlet nozzle number for the hotspot targeted cooler. And $m$ is defined as the ratio between $N^2$ and $M$.

Table 7.1 shows the simplified thermal analysis results for the three test cases. Based on the heat transfer coefficient relation in equation 3, the inlet velocity per nozzle is 9.4 mL/min, 25 mL/min and 40 mL/min for the reference uniform array cooler, test case 1 and test case 2 under a total flow rate of 600 mL/min. The achieved heat transfer coefficient for uniform array cooler is measured at $5.7 \times 10^4$ W/m$^2$ K with the local flow rate per nozzle of 9.4 mL/min. Therefore, for the same measured total flow rate, the achieved heat transfer coefficient for test case 2 is expected to be $15.1 \times 10^4$ W/m$^2$ K with a local flow rate per nozzle of 40 mL/min, at a pressure drop of 1.1 bar.

**Table 7.1:** Simplified thermal analysis using unit cell approach.

| **Case item** | **No.** | **Ratio *m*** | $\dot{V}^*$**(mL/min)** | $htc^*$ **(W/m$^2$ K )** |
|---|---|---|---|---|
| Reference case | 64 | 1 | 9.4 | 5.7 ××$10^4$ |
| Test case 1 | 24 | 2.7 | 25 | $11.1\times10^4$ |
| Test case 2 | 15 | 4.3 | 40 | $15.1\times10^4$ |

Using the full CFD model, the detailed temperature, velocity, and pressure drop information inside the dedicated cooler can be extracted. As for the simulation results of the hotspot targeted cooling, the flow streamlines inside the cooler are shown in Figure 7.15. More flow recirculation is observed inside the hotspot targeted cooler since the flow is concentrated into the reduced number of inlet nozzles. It is also observed that the velocity in the non-heating area is lower since the outlet flow is removed locally through the cooling unit cells near the hotspot areas.

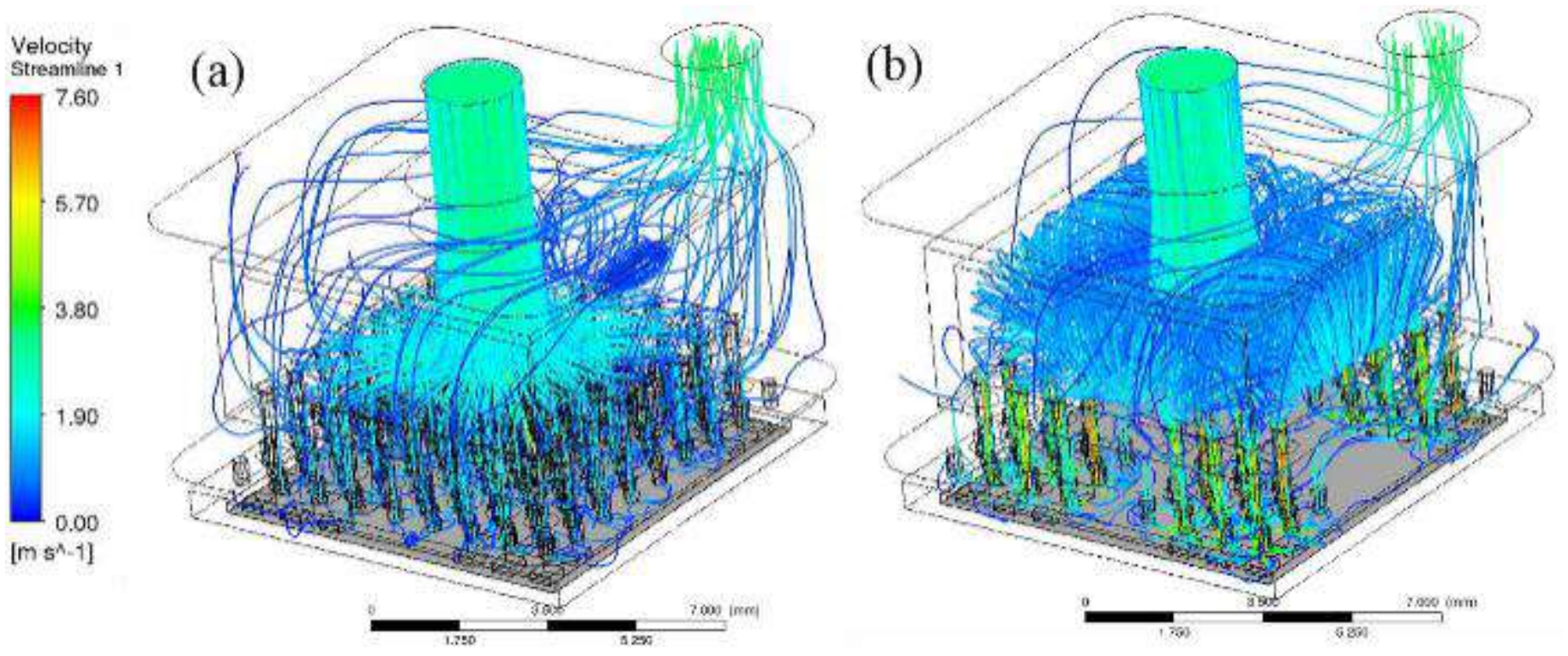

**Figure 7.15:** Flow streamline distribution of hotspots targeted cooler with test case 1: (a) uniform array cooling; (b) hotspots targeted cooling (flow rate = 600 mL/min).

It is expected that the higher local heat transfer coefficient compared to the uniform array cooling case is due to the higher local flow rate with a smaller number of targeted inlet nozzles. Therefore, the local flow rate for the individual inlet nozzles along the chip diagonal is plotted in Figure 7.16.

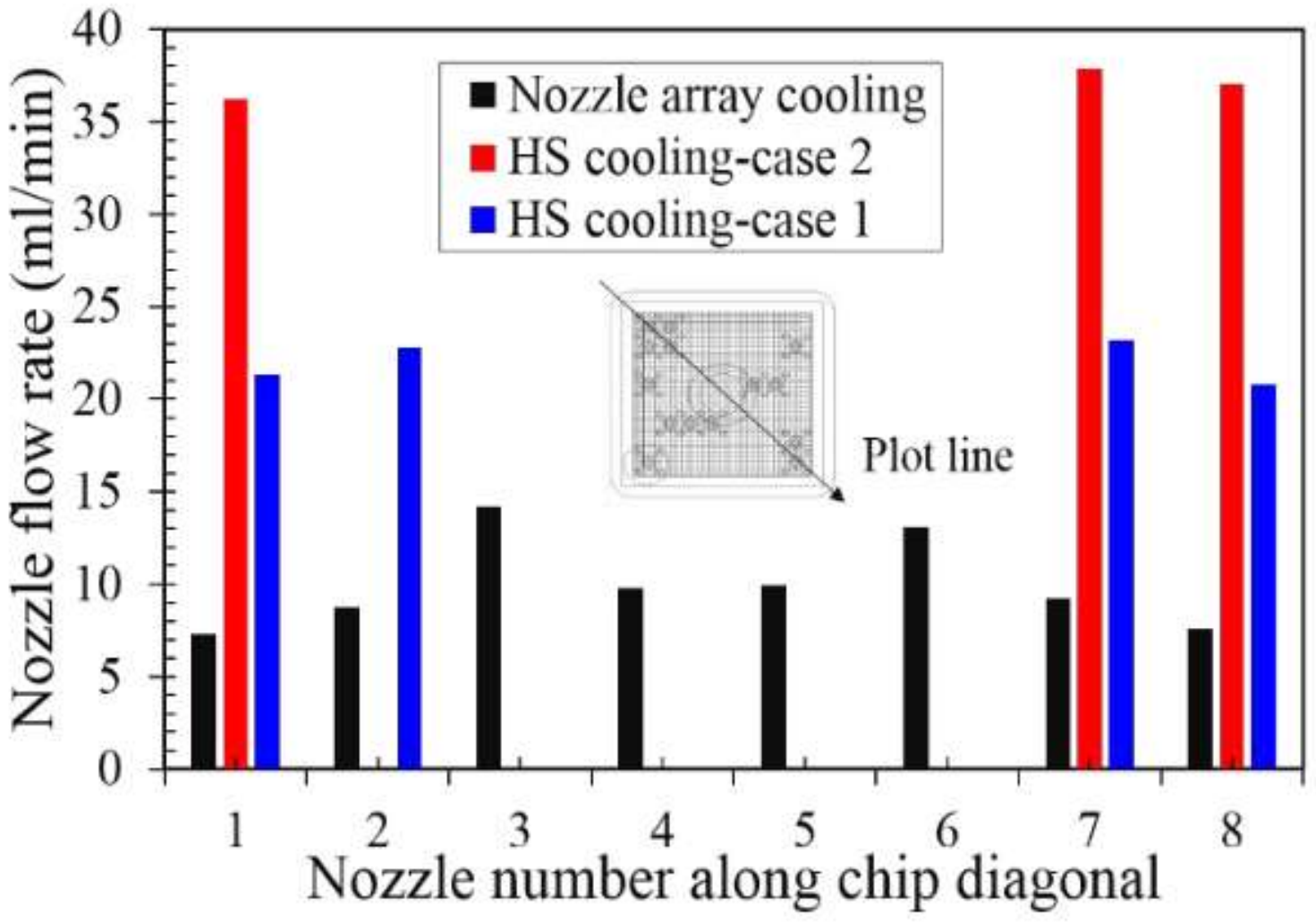

**Figure 7.16:** Nozzle flow rate per nozzle along with the chip diagonal for the three test cases (flow rate =600 mL/min).

### 7.5.2. Temperature uniformity analysis

Given the good agreement between CFD and measurement results, the CFD models are used to assess temperature uniformity for different flow rates. The simulated temperature profiles of the uniform array cooling and hotspot target cooling for test case 2 are shown for different flow rates in Figure 7.17. It can be observed that, for all flow rates the peak temperature of the hotspots with uniform array cooling shown in Figure 7.17(a) is more locally peaked than for the hotspot targeted cooling. Moreover, the peak temperature drops down significantly with hotspots target cooling, as shown in Figure 7.17(b), resulting in better temperature uniformity.

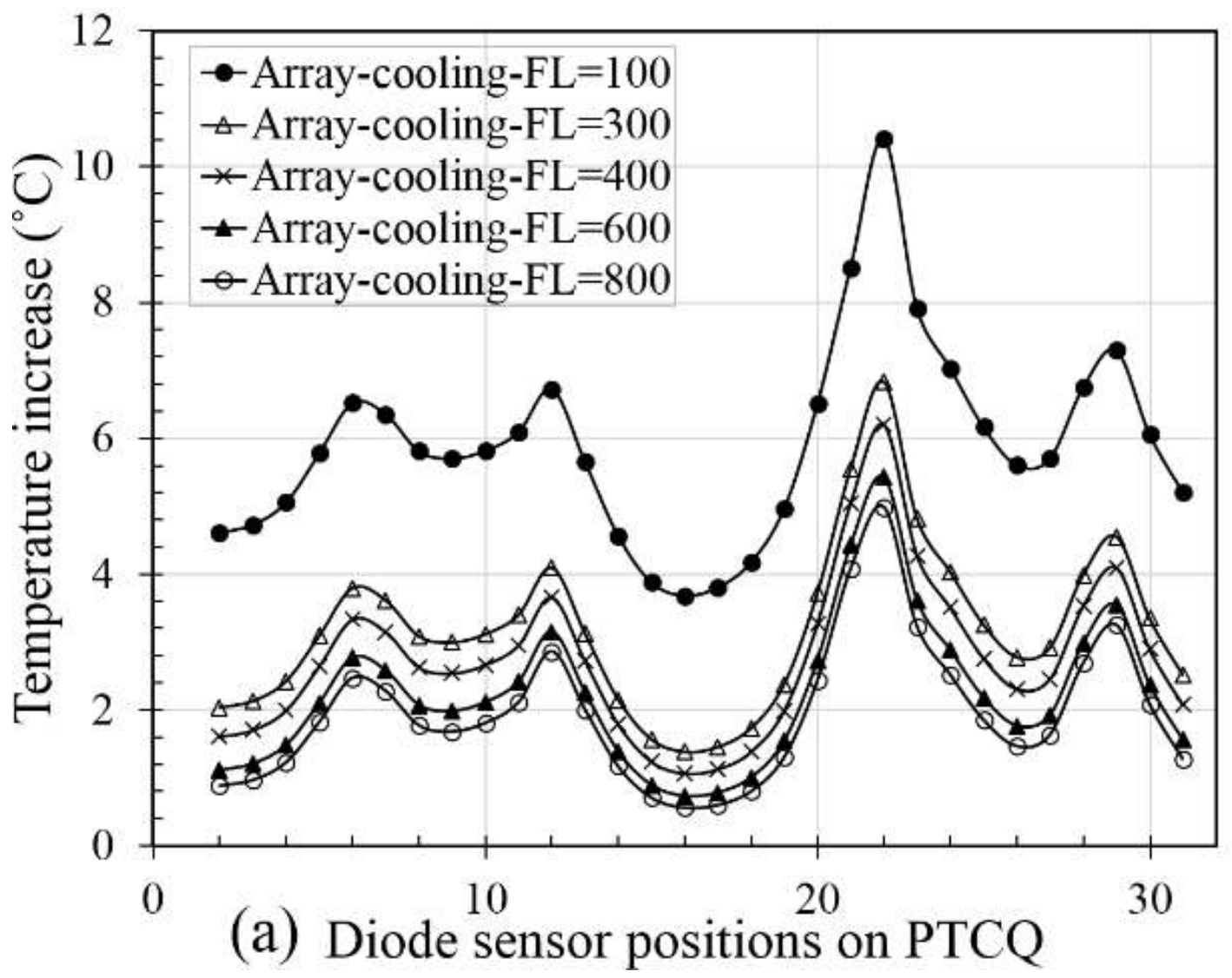

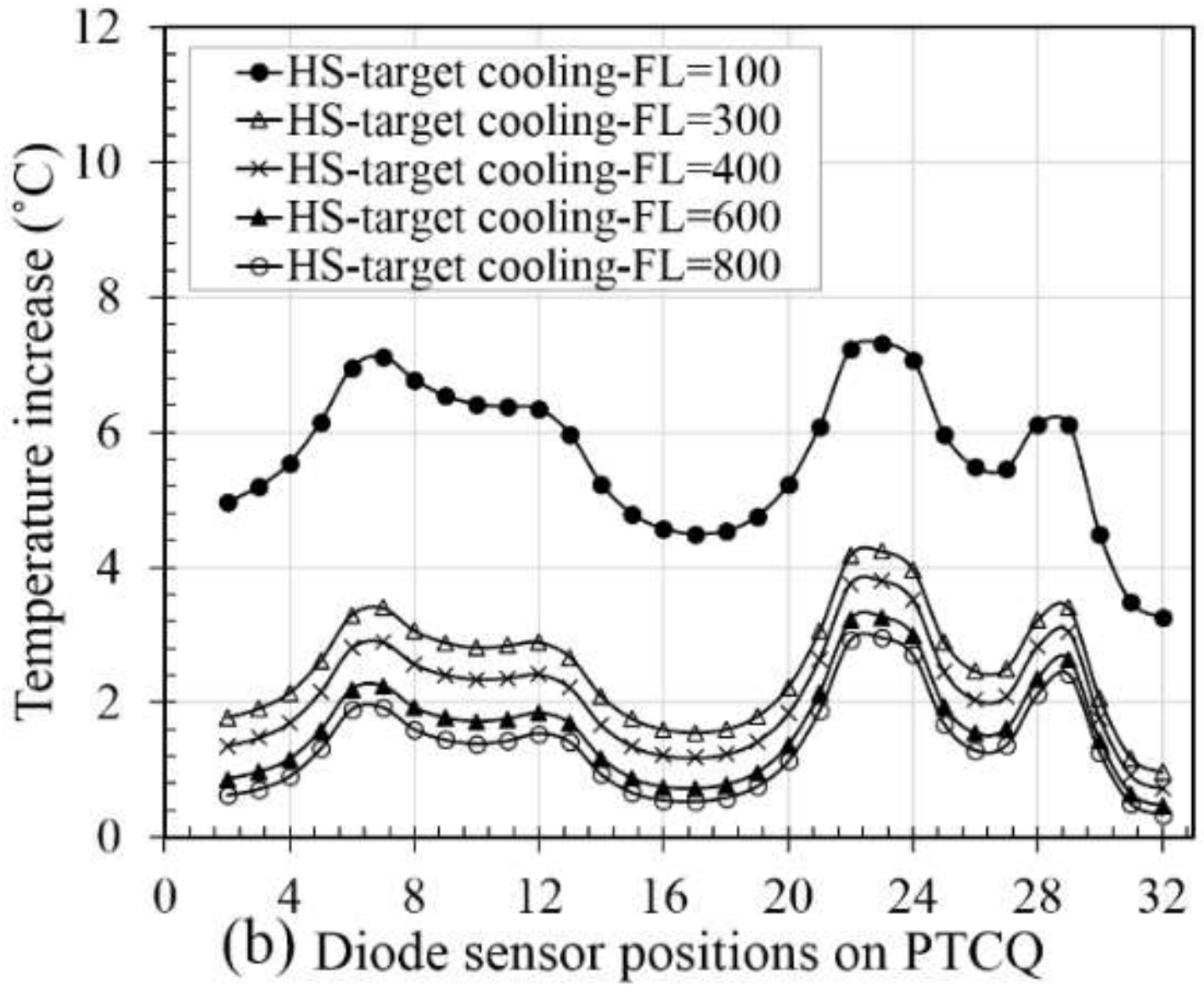

**Figure 7.17:** Temperature profile comparison under different flow rates for test case 2 with (a) CFD modeling results of uniform array cooling and (b) CFD modeling results of hotspots target cooling (Q=5.5 W).

For a more detailed analysis, Figure 18 shows the temperature difference and averaged temperature as a function of the flow rate for the two test cases. The temperature difference is defined as the difference between maximum temperature and minimum temperature. In general, it shows that the required flow rate for the hotspot targeted cooling is smaller compared to the full array cooler in order to achieve the same level of temperature uniformity. As illustrated in Figure 7.18(a), for the same level of temperature uniformity with 3.8 ˚C, the required flow rate for hotspot targeted cooler is only 200 mL/min, which is 3 times lower than for the uniform array cooler. For test case 2 shown in Figure 7.18(b), the required flow rate of the hotspot targeted cooler is about 6 times lower with temperature uniformity of 4 ˚C. Furthermore, it can be observed that the average chip temperature is similar for both cooling solutions.

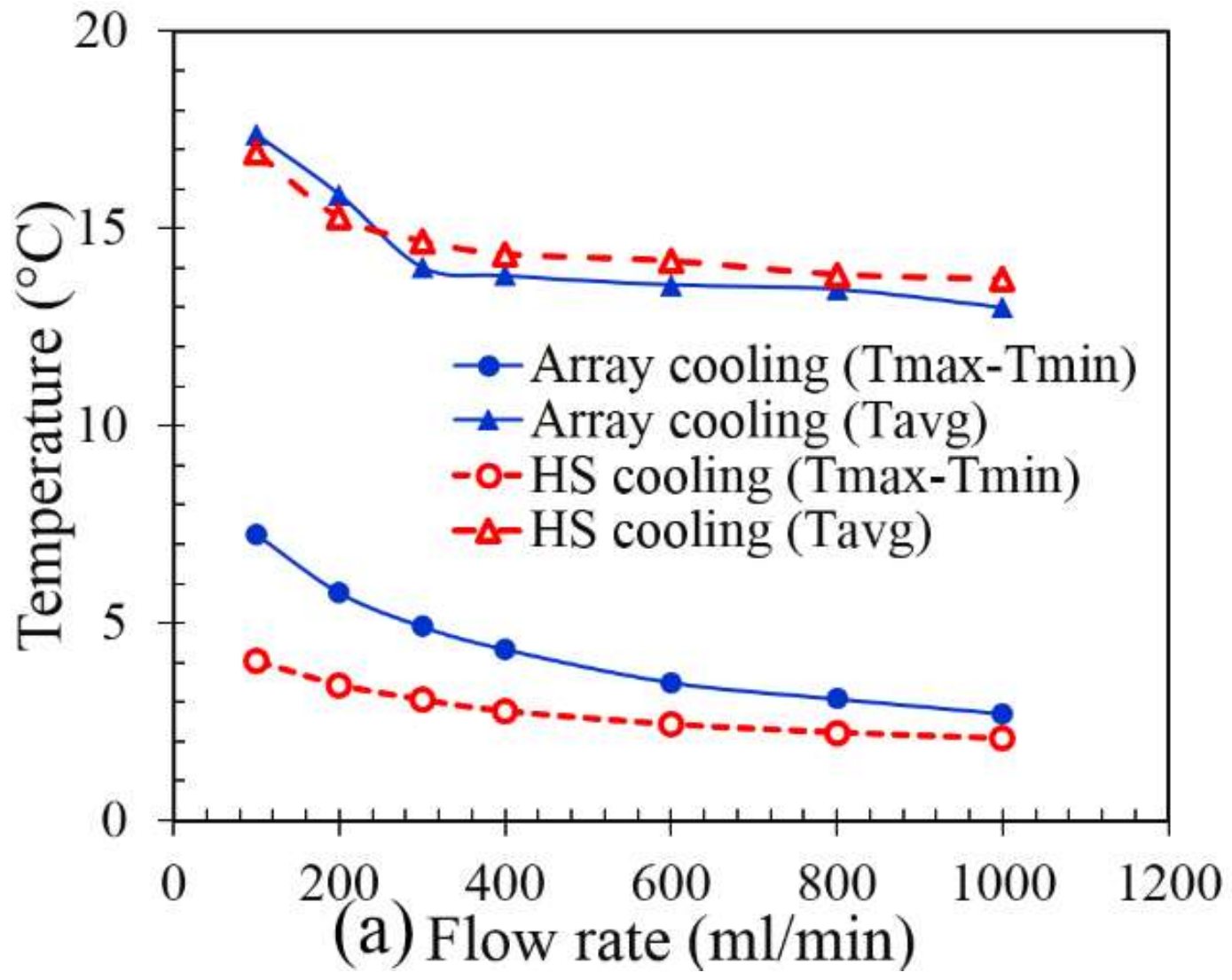

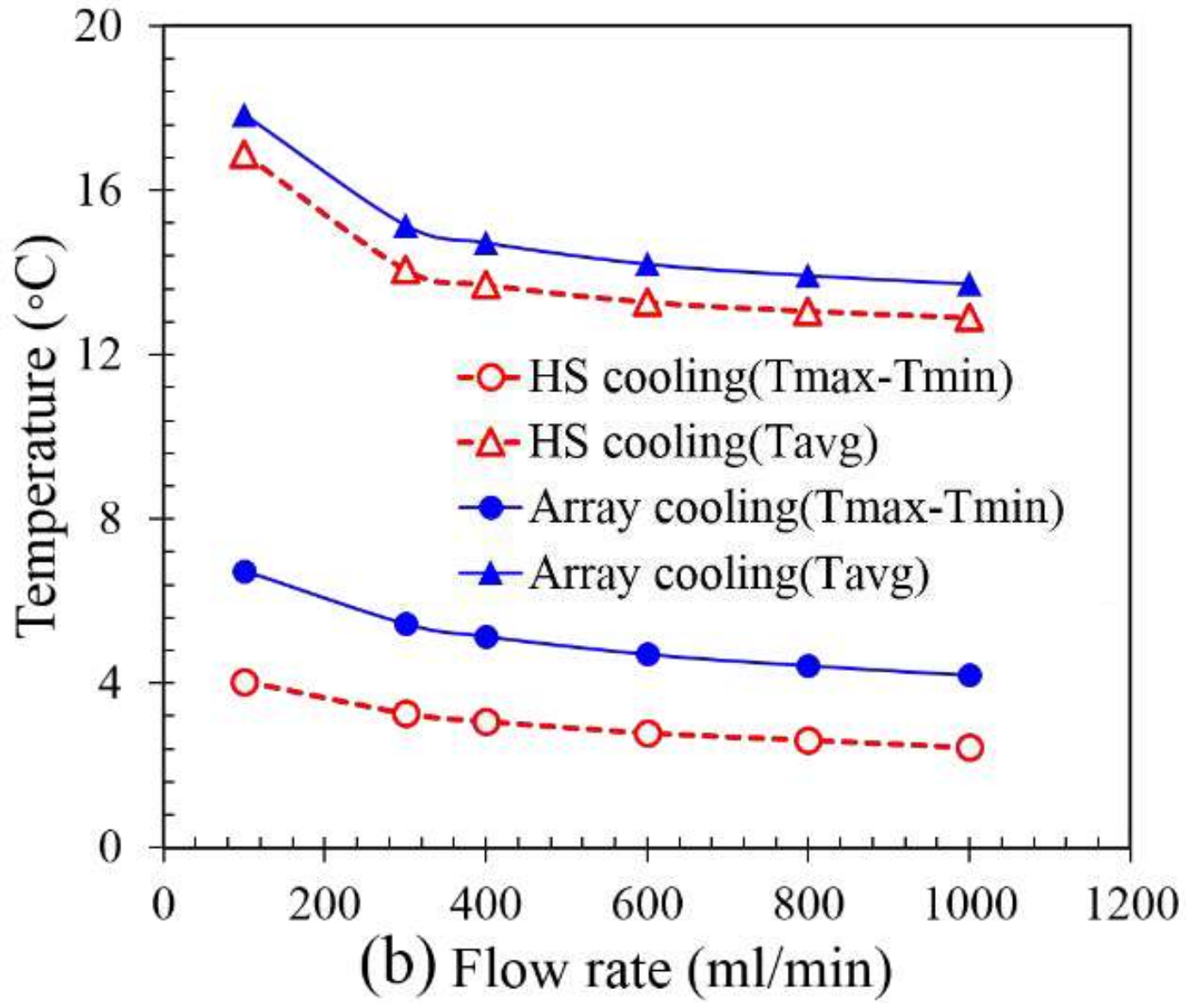

**Figure 7.18:** Temperature uniformity comparison for hotspots cooling and uniform array cooling with (a) test case 1 and (b) test case 2 under different flow rates.

### 7.5.3. Nozzle array distribution configurations

Hotspot targeted cooling with placing the inlet/outlet nozzles only at the hotspot regions shows good cooling performance and temperature uniformity. However, there is the possibility to place the outlet nozzles on the non-heater region to reduce the pressure drop. In Figure 7.19, three different hotspots targeted cooling configurations are compared. Configuration 1 is the uniform nozzle array cooling; configuration 2 is the hotspot targeted cooling outlet present only next to the inlet nozzles, and configuration 3 is the hotspot targeted cooling with the outlet nozzle present across the whole chip surface. The CFD modeling results for the pressure drop for configuration 3 is 0.63 bar at a flow rate of 600 mL/min, which is 1.12x lower than the configuration 2 with closed outlets in the non-heating region.

Moreover, the temperature profile for the three configurations is compared in Figure 7.20. It can be seen that configuration 2 and 3 show a lower peak temperature than 1, and a higher temperature for non-heated region, which results in a lower temperature difference. This is due to the limited heat spreading effects along the non-heater region. For configuration 3, it shows a lower peak temperature than 2, and lower temperature in the non-heated areas, but the smaller difference with lower pressure.

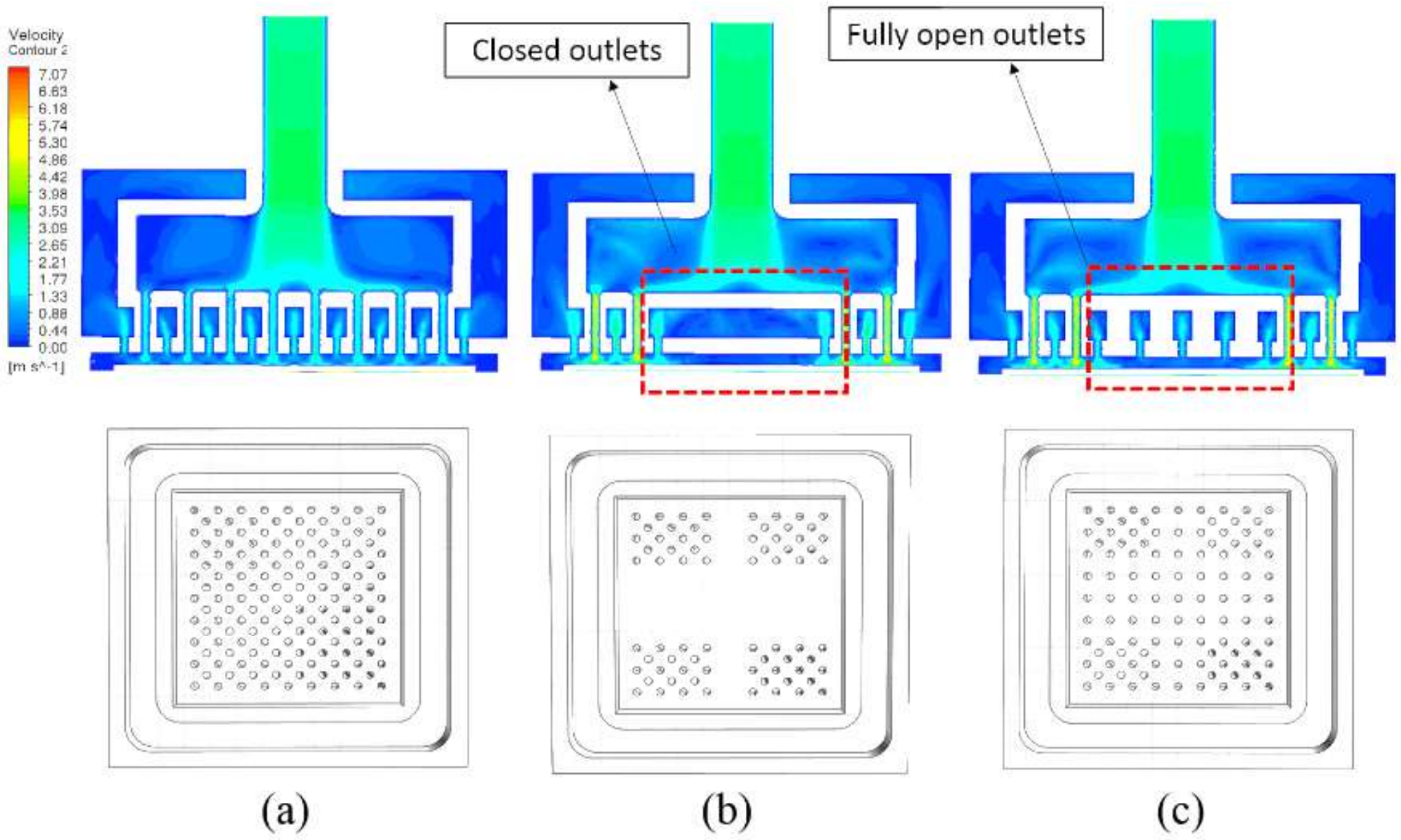

**Figure 7.19:** Different configurations studied: (a) uniform array cooling; (b) hotspot targeted cooling with closed outlets for the non-heating region; (c) hotspot targeted cooling with fully open outlets. (test case 1).

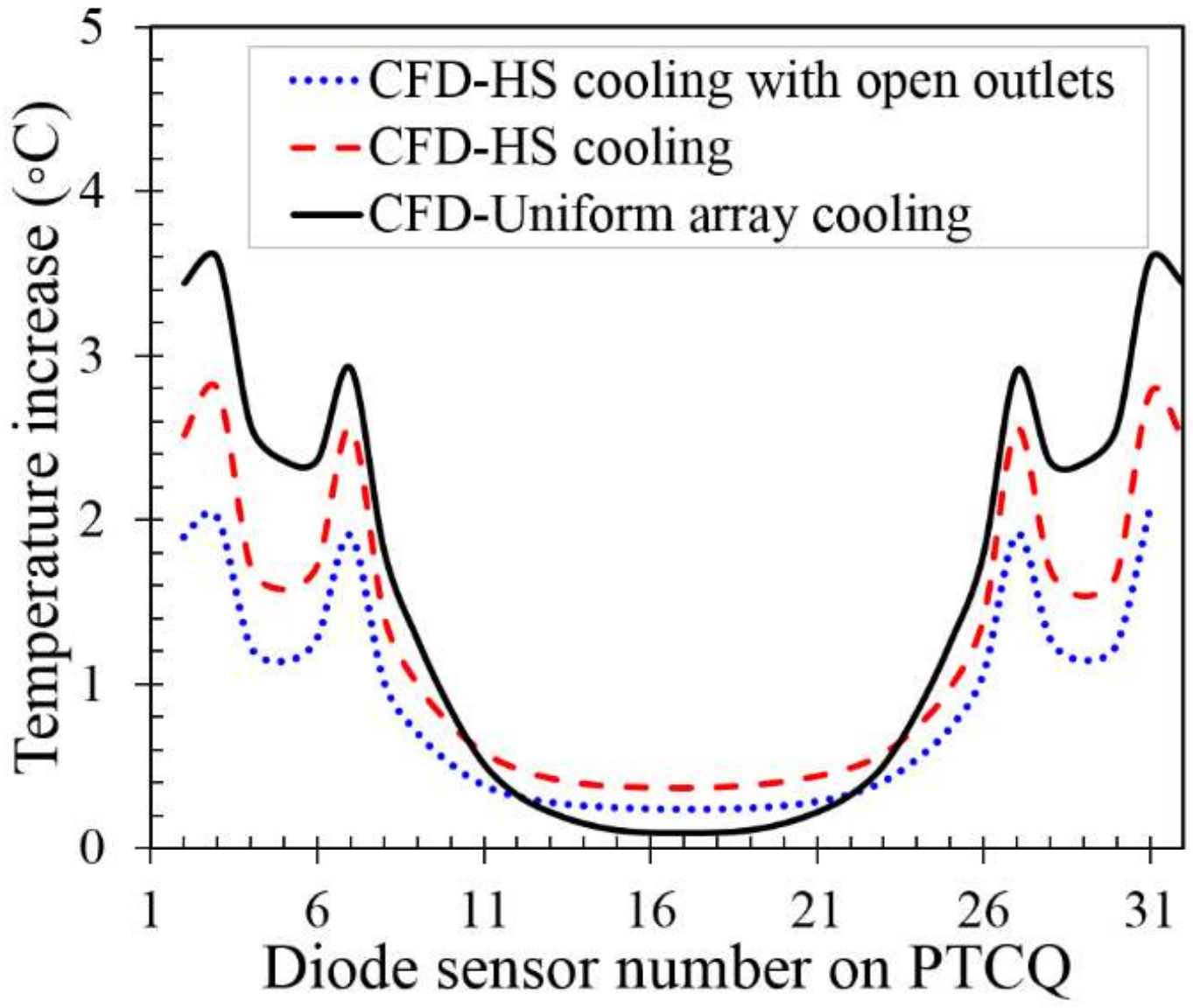

**Figure 7.20:** Temperature profile for the three configurations for test case 1 (Flow rate=600 mL/min, Q=4.1W).

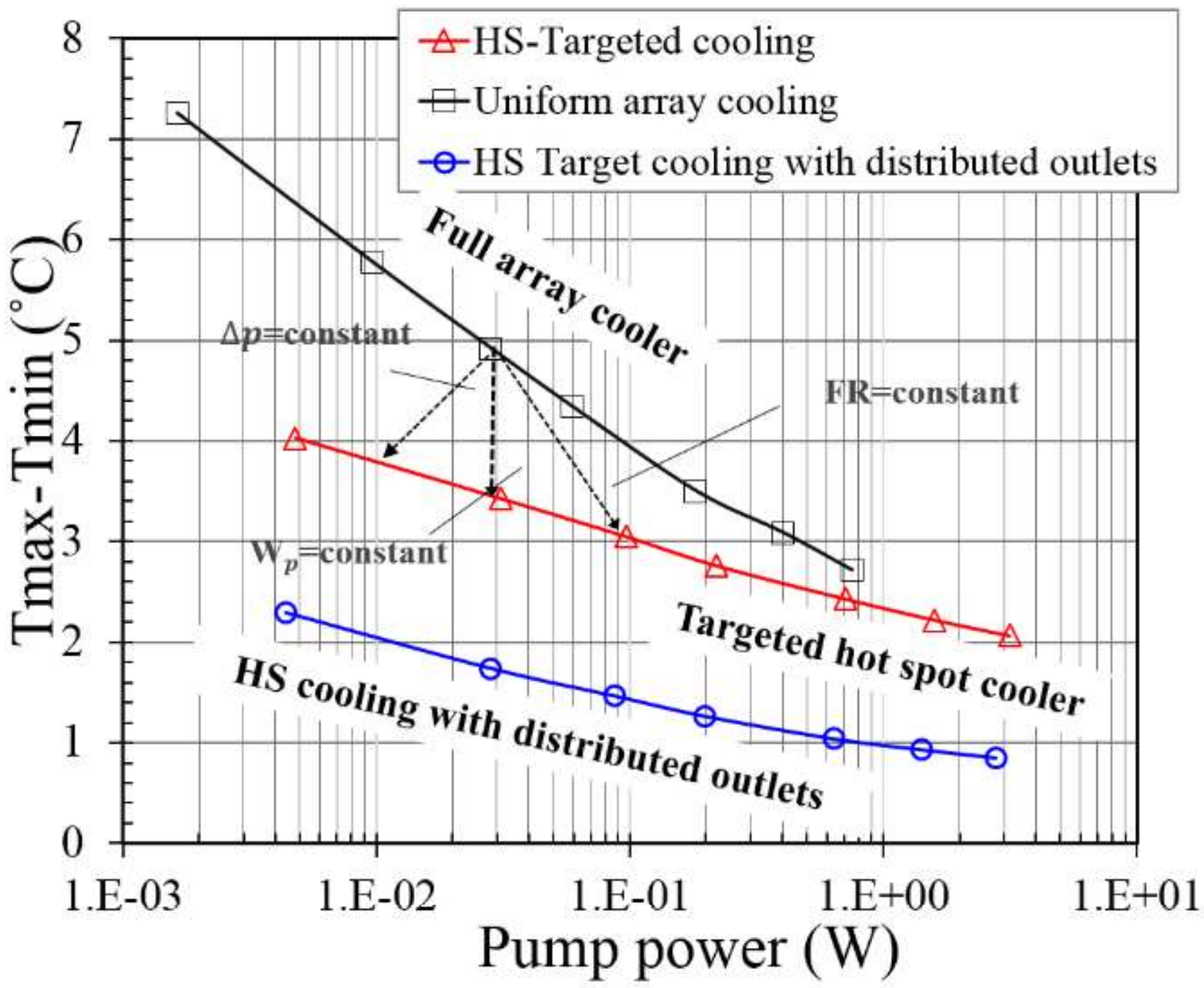

**Figure 7.21:** Thermal/hydraulic trade-off analysis for hotspots targeted cooling. (test case 1)

In Figure 7.21, the comparison between the hotspot targeted cooler with configuration 2 and 3 and the full array cooler is shown. The performance of the three coolers is shown as the curve in terms of the temperature difference as a function of the required pumping power for a range of flow rates. In this benchmarking chart, a better cooler performance is indicated by a lower temperature difference and lower required pump power. The performance of the coolers is now compared for different constraints:

1. Same pressure drop over the cooler
2. Same flow rate
3. Same pumping power

As shown in Table 7.2, for the same pressure drop of 0.18 bar, the chip temperature difference $\Delta T_{uni}$ reduces by a factor of 2.4 for the hotspot targeted cooler of configuration 3 compared to the full array cooler, and it requires 2 times less flow rate and pumping power. For the same flow rate of 600 mL/min, the $\Delta T_{uni}$ reduces by a factor of 1.6, but it requires 4 times larger pressure drop; As for the same pumping power at 0.18 W, the $\Delta T_{uni}$ drops by 63% compared to the full array cooler. In summary, the hotspot targeted cooling with open outlets in the non-heating regions is more energy-efficient compared to the other configurations, despite the higher pressure drop compared to the uniform array cooling. This indicates that the gain in thermal performance due to the targeted cooling by concentrating the liquid coolant at the high heat flux locations outweighs the detrimental impact of the increased required pressure drop and pumping power. Therefore, the hot spot targeted cooler outperforms the uniform array cooler in terms of energy efficiency. This modeling study provides a guideline for the outlet placement during the design of hotspot targeted cooling.

**Table 7.2:** Thermal/hydraulic trade-off modeling results (test case 1).

| | | $\dot{V}$ (mL/min) | ΔP (Bar) | $\dot{W}_p$ (W) | $\Delta T_{uni}$ (˚C) | $T_{avg}$ (˚C) |
|---|---|---|---|---|---|---|
| Reference: full array cooling | | 600 | 0.18 | 0.18 | 3.5 | 13.5 |
| $\dot{V} = C$ | DS2 | 600 | 0.7 | 0.71 | 2.4 | 14.1 |
| | DS3 | 600 | 0.63 | 0.63 | 1.03 | 13.5 |
| $\dot{\Delta p} = C$ | DS2 | 275 | 0.18 | 0.082 | 3.04 | 14.6 |
| | DS3 | 274 | 0.18 | 0.08 | 1.44 | 13.2 |
| $\dot{W}_p = C$ | DS2 | 370 | 0.29 | 0.18 | 2.8 | 14 |
| | DS3 | 380 | 0.28 | 0.18 | 1.3 | 13.1 |

(notes: DS2: hotspot targeted cooling with locally closed outlets; DS3: hotspot targeted cooling with open outlets; C: constant value; $\Delta T_{uni}$ is defined as the difference between the maximum and minimum chip temperature)

In the current study, the nozzle diameter and nozzle pitch have been fixed using the same values as the uniform case, as an illustration. An interesting step for us to further look into is the nozzle configurations with more flexible and optimized designs for the hotspots.

## 7.6 Conclusions

In this chapter, a low-cost energy efficient hotspot targeted cooling concept is introduced. The hotspot targeted cooler demonstrator is fabricated by polymer-based high resolution stereolithography (SLA) with nozzle diameters of 300 µm and a pitch of 1 mm using a water-resistant polymer material. The thermal and pressure drop performance of the demonstrators is characterized by using an advanced programmable thermal test chip and an accurate closed flow loop measurement system. The temperature measurements show a peak temperature reduction of 16% and 42% at a flow rate of 600 mL/min compared to the full array cooler for the targeted hotspot coolers of test case 1 and test case 2 respectively, indicating that concentrating the liquid coolant on the locations where it is needed, can result in a significant reduction of the chip peak temperature due to the locally increased coolant flow rate. On the other hand, the measured pressure drop of the hotspot targeted cooler for test case 1 is 3.2 times higher than the uniform array cooling, while the pressure drop for test case 2 is 5.9 times larger.

Detailed conjugate heat transfer CFD models have been used to assess the local flow distribution and temperature uniformity for the different coolers. The modeling results have been successfully validated, showing a good agreement with the temperature and pressure measurements. The modeling results show that the expected local cooling rate for the hotspot targeted cooling is $m^{0.67}$ times higher than the average cooling rate for the full array cooler, where $m$ is defined as the ratio between the number of inlet nozzles in the full array cooler and in the hotspot targeted cooler. As a result, the hotspot target cooler requires a lower flow rate to achieve the same level of the temperature uniformity compared to the full array cooler. However, the expected pressure drop for the hotspot targeted cooler is $m^2$ times higher than the uniform array cooling. A detailed trade-off between the thermal performance improvements and the higher required pressure drop and pumping power shows that the hotspot targeted cooler outperforms the uniform array cooler in terms of energy efficiency despite the increase in pressure drop. This higher performance is observed for three different bases for comparison: constant flow rate, constant pressure drop, and constant pumping power.

The validated CFD models also show that the hotspot targeted cooler can be further improved by providing outlet nozzles over the full chip area instead of near the inlet nozzles covering the hotspot areas only. The implementation with the additional outlet nozzles achieves a further reduction of the pressure drop across the cooler by 12% and a reduction of the maximum temperature difference by a factor of 2, resulting in an even more energy-efficient design.

**References**

[1] Mudawar, I., "Assessment of high-heat-flux thermal management schemes," in IEEE Transactions on Components and Packaging Technologies, vol. 24, no. 2, pp. 122-141, June 2001.

[2] Arden L. Moore, LiShi, “Emerging challenges and materials for thermal management of electronics”, Materials Today, Volume 17, Issue 4, May 2014, Pages 163-174

[3] Bar-Cohen A, Wang P. Thermal Management of On-Chip Hotspot. ASME. J. Heat Transfer. 2012;134(5):051017

[4] Hamann, H. F., Weger, A., et al., “Hotspot limited microprocessors: Direct temperature and power distribution measurements,” IEEE J. Solid-State Circuits, vol. 42, no. 1, pp. 56–65, Jan. 2007.

[5] P Smakulski, et al., “A review of the capabilities of high heat flux removal by porous materials, microchannels and spray cooling techniques”, Applied Thermal Engineering, Volume 104, 2016, Pages 636-646.

[6] A Radwan, et., “Thermal management of concentrator photovoltaic systems using two-phase flow boiling in double-layer microchannel heat sinks”, Applied Energy, Volume 241, 2019, Pages 404-419.

[7] S. Brunschwiler et al., “Direct liquid jet-impingment cooling with micron-sized nozzle array and distributed return architecture,” in Proc. 10th Intersoc. Conf. Thermal Thermomech. Phenomena Electron. Syst. (ITHERM), San Diego, CA, USA, 2006, pp. 196–203.

[8] S. Krishnamurthy, et al., “Flow Boiling Heat Transfer on Micro Pin Fins Entrenched in a Microchannel”, ASME. J. Heat Transfer.2010;132(4):041007-041007-10.

[9] R. Singh, A. Akbarzadeh, M. Mochizuki, “Sintered porous heat sink for cooling of high-powered microprocessors for server applications”, Int. J. Heat Mass Transfer 52 (2009) 2289–2299.

[10] Chen X, Ye H, et al., “A review of small heat pipes for electronics”, Appl Therm Eng 2016; 96: 1–17.

[11] Y. Han, B. L. Lau, et al., "Thermal Management of Hotspots Using Diamond Heat Spreader on Si Microcooler for GaN Devices," in IEEE Transactions on Components, Packaging and Manufacturing Technology, vol. 5, no. 12, pp. 1740-1746, Dec. 2015.

[12] Z. Gao, Y. Zhang, Y. Fu, et al., "Graphene heat spreader for thermal management of hotspots," 2013 IEEE 63rd Electronic Components and Technology Conference, Las Vegas, NV, 2013, pp. 2075-2078.

[13] Zhao, D., and Tan, G., 2014, "A Review of Thermoelectric Cooling: Materials, Modeling and Applications," Appl. Therm. Eng., 66(1–2), pp. 15–24.

[14] SH Choday, et al., "Prospects of Thin-Film Thermoelectric Devices for Hot-Spot Cooling and On-Chip Energy Harvesting." IEEE Transactions on Components, Packaging and Manufacturing Technology 3.12 (2013): 2059-067.

[15] S Lee, PE Phelan, et al.,"Hotspot Cooling and Harvesting Central Processing Unit Waste Heat Using Thermoelectric Modules." Journal of Electronic Packaging 137.3 (2015): 031010.

[16] Redmond M, Kumar S. "Optimization of Thermoelectric Coolers for Hotspot Cooling in Three-Dimensional Stacked Chips". ASME. J. Electron. Packag. 2014;137(1):011006-011006-6.

[17] SL Li, et al., "Hotspot Cooling in 3DIC Package Utilizing Embedded Thermoelectric Cooler Combined with Silicon Interposer." 2011 6th International Microsystems, Packaging, Assembly and Circuits Technology Conference (IMPACT) (2011): 470-73.

[18] Wang, P., and Bar-Cohen. "On-chip Hotspot Cooling Using Silicon Thermoelectric Microcoolers." Journal of Applied Physics 102.3 (2007): Journal of Applied Physics, 01 August 2007, Vol.102(3).

[19] P. Y. Paik, V. K. Pamula and K. Chakrabarty, "A Digital-Microfluidic Approach to Chip Cooling," in IEEE Design & Test of Computers, vol. 25, no. 4, pp. 372-381, July-Aug. 2008.

[20] H. Oprins, J. Danneels, et al., "Convection heat transfer in electrostatic actuated liquid droplets for electronics cooling", Microelectronics Journal, Volume 39, Issue 7, 2008, Pages 966-974.

[21] H. Oprins, G. Van der Veken, et al., "On-Chip Liquid Cooling With Integrated Pump Technology," in IEEE Transactions on Components and Packaging Technologies, vol. 30, no. 2, pp. 209-217, June 2007.

[22] CS Sharma, et al., "Energy Efficient Hotspot-targeted Embedded Liquid Cooling of Electronics." Applied Energy 138.C (2015): 414-22.

[23] Lee, Y.-J., et al., "Hotspot Mitigating With Obliquely Finned Microchannel Heat Sink-An Experimental Study." IEEE Transactions on Components, Packaging and Manufacturing Technology 3.8 (2013): 1332-341.

[24] T Van Oevelen, M Baelmans, "Design optimization and validation of single-phase rectangular micro channels with axial nonuniform width", Proceedings of the 14th International Heat Transfer Conference, August 2010, Washington DC, pp. 49–58.

[25] Y. Han, et al., "Package-Level Microjet-Based Hotspot Cooling Solution for Microelectronic Devices." IEEE Electron Device Letters 36.5 (2015): 502-04.

[26] Zuckerman, and Lior. "Impingement Heat Transfer: Correlations and Numerical Modeling." Journal Of Heat Transfer-Transactions Of The Asme 127.5 (2005): 544-52.

[27] Tiwei, T., H. Oprins, et al., "High Efficiency Direct Liquid Jet Impingement Cooling of High Power Devices Using a 3D-shaped Polymer Cooler." 2017 IEEE International Electron Devices Meeting (IEDM) (2017): 32.5.1-2.5.4.

[28] Tiwei, T., H. Oprins, et al., "Experimental Characterization of a Chip Level 3D Printed Microjet Liquid Impingement Cooler for High Performance Systems", 2019 IEEE CPMT.

[29] Tiwei, T., H. Oprins, et al., "Experimental Characterization and Model Validation of Liquid Jet Impingement Cooling Using a High Spatial Resolution and Programmable Thermal Test Chip." Applied Thermal Engineering 152 (2019): 308-18.

[30] T. Wei *et al.*, "3D Printed Liquid Jet Impingement Cooler: Demonstration, Opportunities and Challenges," *2018 IEEE 68th Electronic Components and Technology Conference (ECTC)*, San Diego, CA, 2018, pp. 2389-2396.

[31] Source material link: ”Somos watershed xc material datasheet”, https://www.protolabs.co.uk/media/1010047/somos-watershed-en.pdf.

[32] Ferziger, J.H.; Peric, M., 1997, Computational Methods for Fluid Dynamics, Springer Verlag, Berlin, pp.59-61.

# Chapter 8

# 8. Interposer Package Cooling

## 8.1 Introduction

### 8.1.1 Need for advanced cooling solutions for 2.5D Si interposer packages

Three-dimensional through-Si via (TSV) integration has great potential to improve the performance, power consumption, and package footprint by vertically integrating multiple dies [1]. However, the vertical integration with 3D stacked dies will elevate the power density and chip temperature, which requires expensive packaging and cooling solutions [2]. This is due to the thermal bottleneck of the die-die interface materials with low thermal conductivities [3]. Alternatively, 2.5D Si interposer packages with multiple dies integrated side by side, enable more cooling potential for applications combining high power components such as logic, GPU and FPGA, and temperature-sensitive components (DRAM, SerDes). This Si interposer implementation shows potential for high-performance systems with high-bandwidth and high-power applications [4], as demonstrated by the release of Xilinx's FPGA [5] and the AMD Fury X GPU [6].

For typical 2.5D Si interposer packages, a metal lid or heat spreader is attached to the substrate using lid adhesive [7,8]. In literature [7], several thermal solutions are mounted on top of the lidded package to minimize the thermal resistance, such as fin heat sinks and fan-cooled heat sink with or without embedded heat pipes. However, the major thermal bottleneck for conventional cooling solutions is the presence of the thermal interface material (TIM). The thermal resistivity of the most widely used TIMs, such as greases, gels, and phase-change materials (PCMs), can be as low as 10 $mm^2$-K/W [9]. For the state of art nano-TIM, the thermal resistivity can be smaller and even in the range of 1 $mm^2$-K/W with GE's copper nano-spring [10]. However, it is also found that the interfacial thermal resistance (ITR) between TIM and heat sink can vary from 2 $mm^2$-K/W to 20 $mm^2$-K/W due to the mechanical compliance of the TIM [11]. Recently, several embedded cooling techniques without the use of the TIM have been applied on the 2.5D Si interposer packages. In [12], an embedded thermoelectric cooler (TEC) combined with silicon interposer for the electrical path is studied for hot spot cooling, but the power consumption of the TEC driver is a big challenge. In [13,14,15], microfluidic cooling delivery channels are embedded within an interposer package with high aspect ratio TSVs, and microfluidic chip I/Os. However, the I/O density is insufficient for high-bandwidth devices. Moreover, the temperature gradient across and along the channels is hard to avoid, resulting in uneven temperature distribution.

### 8.1.2 Overview of the 3D distributed manifold techniques

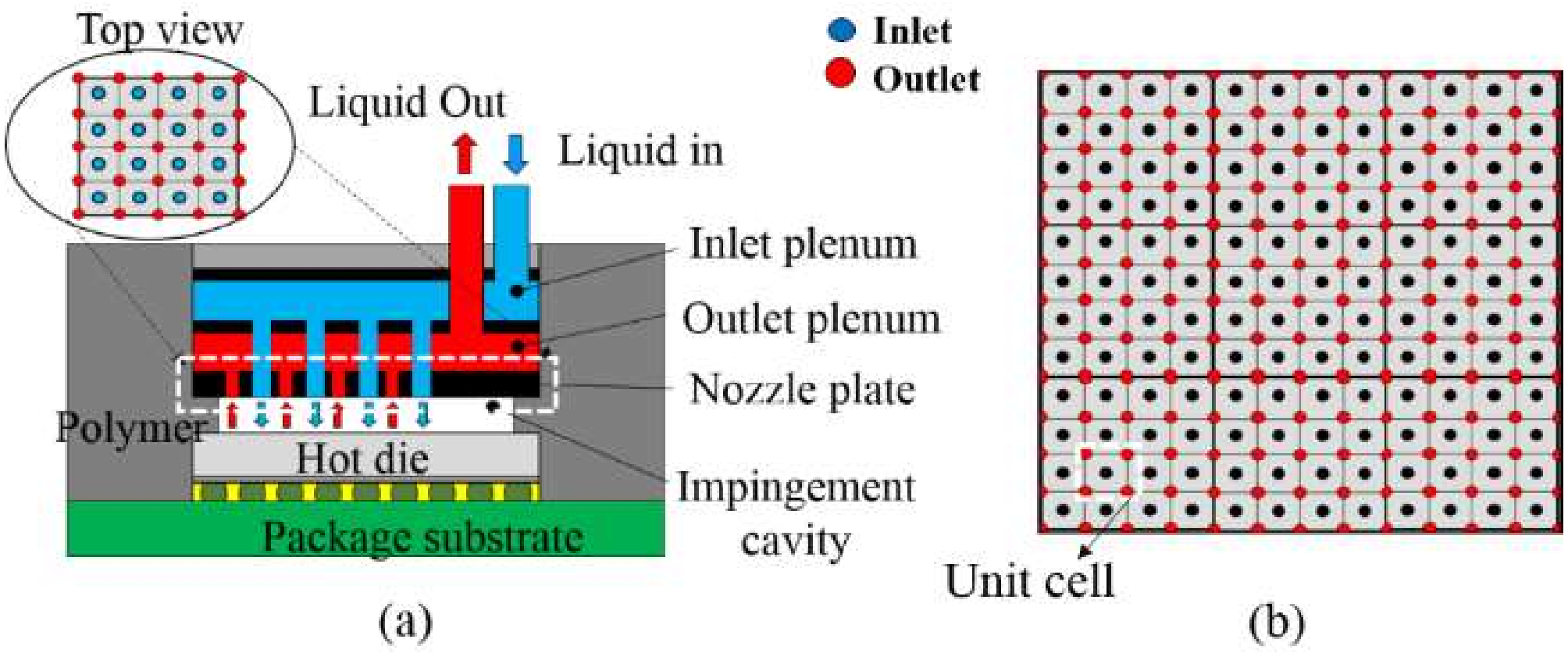

**Figure 8.1**: Scalable impingement jet cooling system with 3D liquid distributed manifold: (a) cross section view of the cooler on a single chip; (b) top view of the nozzle plate with scalable nozzle array.

In the previous chapters, the multi-jet coolers are demonstrated on the 8×8 mm$^2$ thermal test chip, as illustrated in Figure 8.1, showing excellent cooling performance. However, in real high-performance chips, the design of the multi-jet arrays should be scalable to large chip size. This will need an advanced manifold level design for the flow distribution on every inlet nozzle. As discussed in session 8.1, the experiments and numerical modeling results show that the manifold level design of this microfluidic cooler is very important for the overall cooler performance, since it determines the flow uniformity, and system-level pressure and thermal resistance, especially for large area die size applications. Therefore, the design of manifold is very important for the high-performance multi-jet cooler.

With the recent development of the high resolution of additive manufacturing technology, lots of studies move to the 3D manifold with the liquid delivery system, fabricated by 3D printing. Robinson et al. demonstrated a hybrid micro heat sink using impinging micro-jet arrays and microchannels using MICA Freeform process [29], with a predicted effective thermal conductance of 400 kW/m$^2$ K for a flow rate of 0.5 L/min. In [30], we demonstrated for a bare die single chip package that additive manufacturing, or 3D printing can be used to fabricate a highly performant cooler with high density nozzle array and complex internal geometries. Moreover, 3D printing of the cooler shows the potential to integrate the cooling jets directly targeted on each device in the multi-chip module, which can drastically reduce the thermal coupling between different devices. Furthermore, the required pumping power can be significantly reduced due to a streamlined internal channel design that can be fabricated using additive manufacturing, but not with conventional fabrication techniques.

In this chapter, we present the evaluation of the cooling performance of a package level jet impingement cooling solution on dual-chip packages by means of full cooler computational fluid dynamics simulations (CFD) and experiments on test vehicles. For this study, the single chip cooler design introduced in chapter 6 has been optimized for the packages containing two thermal test chips. Furthermore, an improved cooler design of the 3D printed fluid manifold is introduced and benchmarked. Since jet impingement cooling on the bare die is a disruptive cooling technology requiring direct access to the backside of the Si chip, we also consider a less disruptive cooling implementation in which the impingement cooling is applied on the lid. For cooling on the lidded package, the cooling surface area of the lid is larger compared to cooling on the bare die surface. Moreover, the presence of the lid enhances the lateral heat spreading effects inside the package. However, the drawback of this lidded approach is the significant thermal resistance of the TIM between the chip and the lid. The trade-off between the beneficial and detrimental effect of the lid will determine whether the bare die cooling will outperform the cooling on the lidded package.

The chapter is organized as follows: the cooler design considerations are discussed in Section 8.2. The fabricated cooler, the assembly, and the experimental set-up for the thermal and hydraulic characterization are presented in Section 8.3. In section 8.4, the experimental characterization of the package level impingement cooler is analyzed for the bare die and lidded package configurations, including the thermal coupling between the dies in the package. Next, Section 8.5 discusses the experimental and numerical characterization of the novel cooler design with the smaller form factors and the comparison with the reference cooler. Finally, parametric studies of the TIM and lid properties have been performed to assess the trade-off of the beneficial and detrimental impact of the lid for different flow rates.

## 8.2 Design of multi-jet impingement cooler

This section introduces the design considerations for the impingement cooler of the dual-chip package. Firstly, the lidded and bare die thermal test vehicles on 2.5D interposer are introduced. Secondly, the concept of nozzle array scalability with the chip area is introduced in order to estimate the cooler performance based on the extrapolation of previous cooler designs. Next, the different design concepts for the package level impingement cooler are discussed. Finally, the package level CFD modeling approach is presented to study the flow and temperature distribution inside the cooler and the overall cooler performance in detail for the different proposed cooler designs.

### 8.2.1 Lidded and bare die dual-chip test vehicle

In order to compare the performance between the bare die cooling and lidded package cooling, an advanced thermal test vehicle with the lidded package and bare die package versions is introduced. As shown in Figure 8.2, a 35×35 $mm^2$ ball grid array (BGA) package is used, containing a 20×10 $mm^2$ Si interposer with 100 μm thickness and two identically 8×8 $mm^2$ thermal test chips, referred to as PTCQ (Packaging Test Chip version Q), introduced in chapter 2. The interposer stacks are flip-chip soldered on the organic substrate, allowing the cooling solution to be directly applied to the backside of the chips, or on the lid. The schematic of the bare die package is illustrated in Figure 8.2(a). As for the lidded packages shown in Figure 8.2(b), a Cu lid with a thickness of 0.3 mm is attached to the thermal test dies with additional thermal interface material. The thermal interface material is a standard silicone-based material with a specified thermal conductivity of 1.9 W/m-K and a targeted thickness of 80 μm.

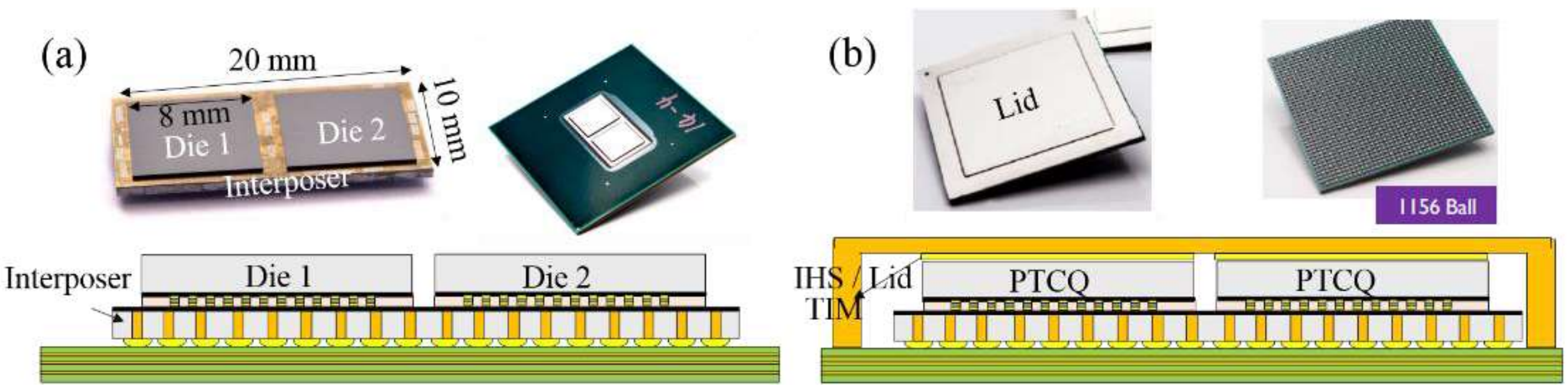

**Figure 8.2**: Bare die package and lidded package: (a) two thermal test chips on Si interposer without a lid; (b) thermal test chips on Si interposer with lid.

### 8.2.2 3D printed dual-chip package cooler design

For the cooler design of the dual-chip package, the objectives are listed below:

- Targeted cooling for both chips;
- High cooling performance and low pressure drop;
- The small form factor of the cooler;

Moreover, the design constraints such as flow loop connections, cooler size limitations, assembly constraints and the manufacturing capabilities should be taken into account. As introduced in the previous chapter, additive manufacturing, or 3D printing enables to use low-cost materials for the cooler fabrication, to print the whole geometry in one piece, to customize the design to match the nozzle array to the chip power map and to fabricate very complex internal structures. This last feature allows the design of complex cooler cavities that cannot be fabricated with conventional fabrication techniques. Moreover, the inlet and outlet divider structures can be printed as hollow cylinders, which can significantly reduce the pressure drop compared to square shape dividers and reduce the number of layers required in the cooler design.

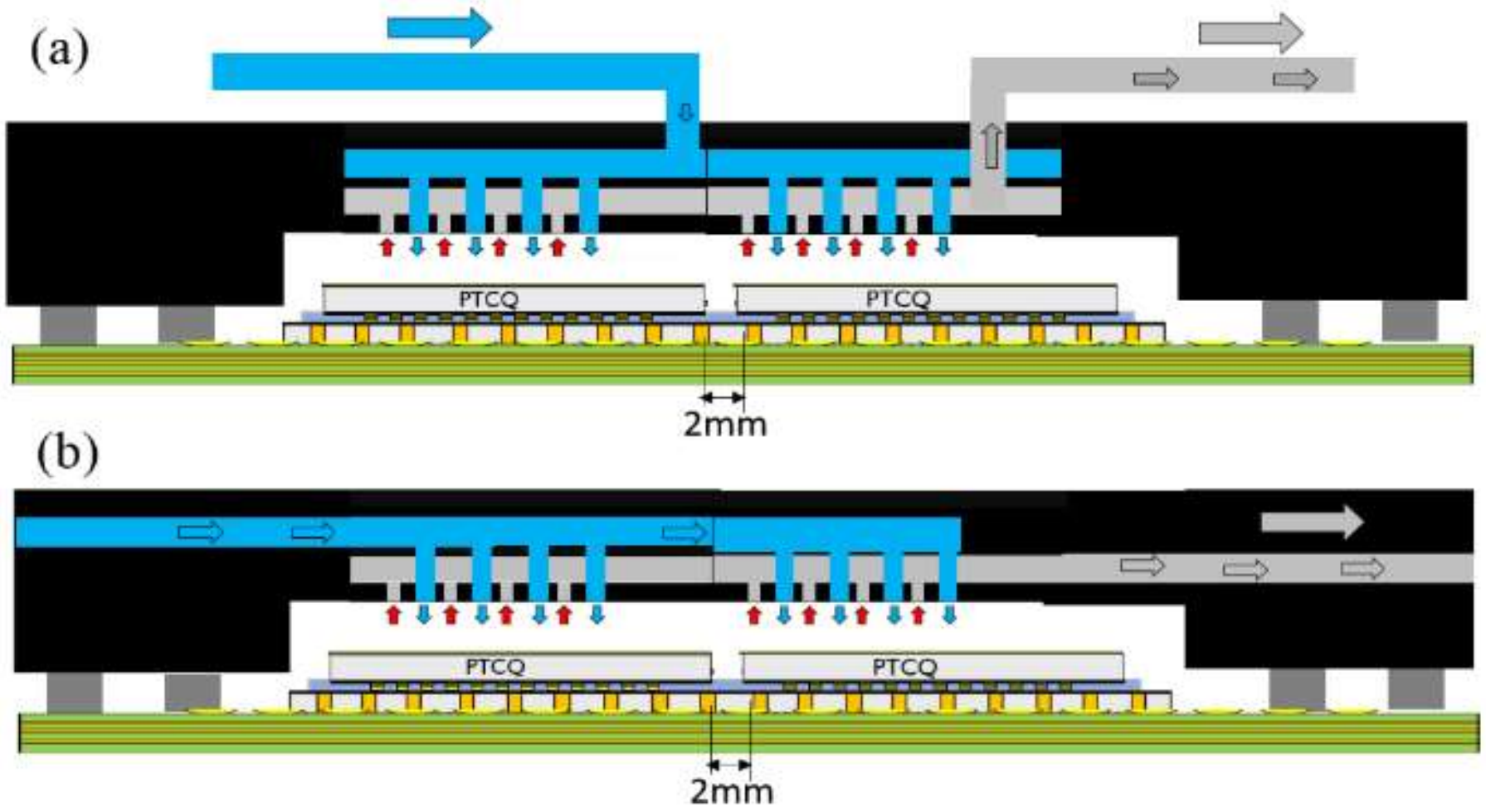

**Figure 8.3**: Cooler schematics for the dual-chip module: (a) vertical feeding design; (b) lateral feeding design.

The schematic concepts of the direct jet impingement cooler for the dual-chip module are shown in Figure 8.3, containing four main parts: inlet plenum, outlet plenum, nozzle plate and impingement cavity. Figure 8.3(a) shows the reference cooler design with vertical coolant supply connectors. The details of the main parts in the cooler structure of vertical feeding design are indicated in Figure 8.4. This design is an extension from the single die cooler demonstrated in chapter 6. Since the optimized geometry parameters for the jet nozzle array of the single die cooler are already investigated in chapter 5 and chapter 6, the design of the package level cooler the for dual-chip module will use the same nozzle array (4×4 inlet nozzle array and 5×5 outlet nozzle array with 600 µm diameter) as for the single chip package cooler, targeted at each of the chips in the module. Figure 8.4 shows the geometry information tabulated in Table 8.1. Table 8.1 lists the geometry comparison between the single chip cooler and the dual-chip module cooler. The cooler is designed to match the package area of 35×35 $mm^2$. The cavity height of the cooler (i.e. the distance between the nozzle plate and the chip surface) is 0.6 mm. The cooler is placed over the stacked chips on the interposer, as shown schematically in Figure 8.5, which also indicates the position of the sealing rings between the cooler and the package substrate. Moreover, the O-ring can also act as a buffer for the mechanical assembly of the cooler, especially for large die packages to compensate for the potential warpage of the assembly.

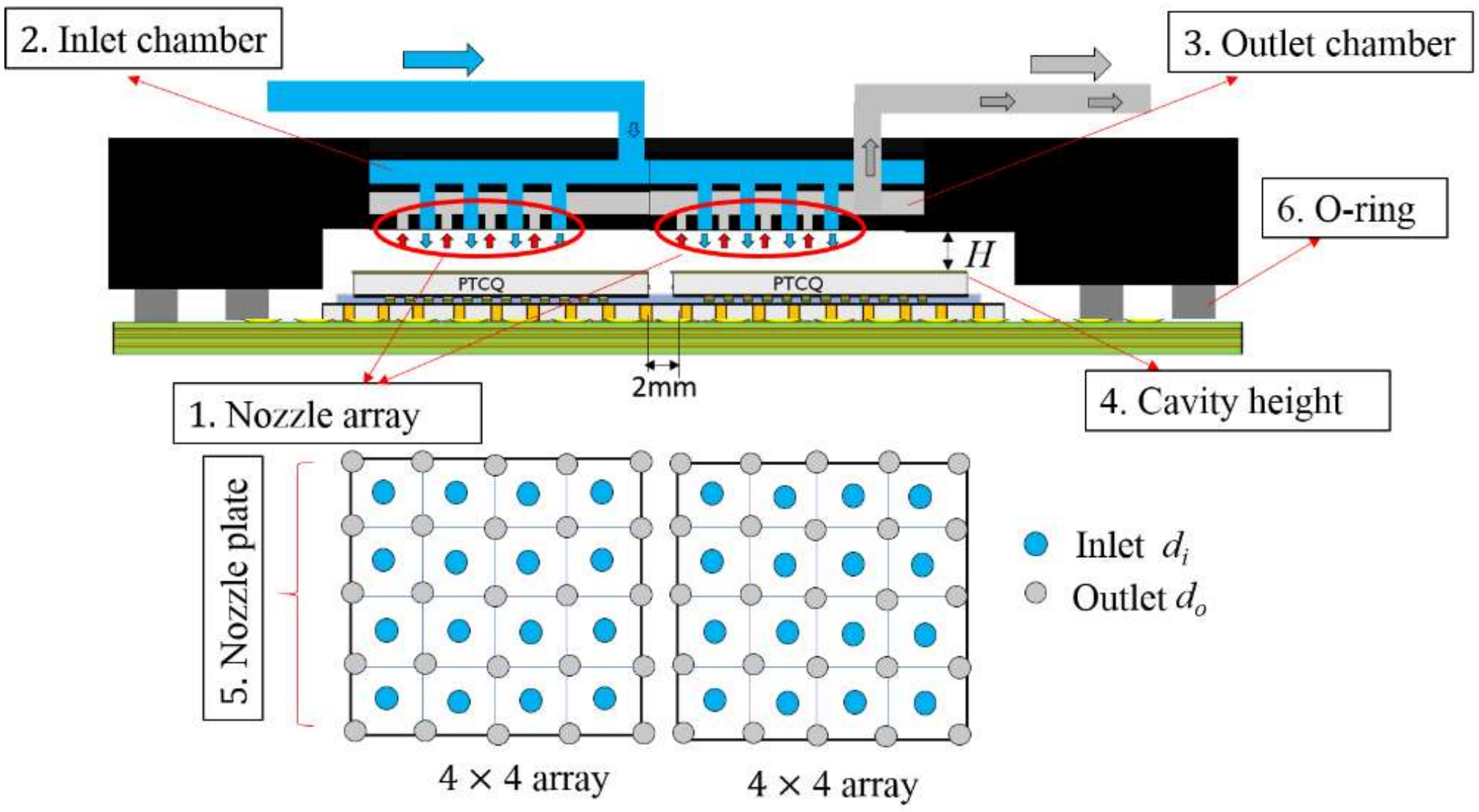

**Figure 8.4**: Cooler geometry parameters for vertical feeding design: 1-nozzle array; 2-inlet chamber; 3-Outlet chamber; 4-Cavity height; 5-Nozzle plate; 6-O-ring.

**Table 8.1.** Geometry comparison between single die cooler and dual-chip module cooler.

| **Geometry** | | **Single chip cooler** | **Dual-chip cooler** |
|---|---|---|---|
| Nozzle array | $N$ | 4×4 | 4×4 per die |
| Inlet chamber height | | 2.5 mm | 2.5 mm |
| Inlet diameter | $d_i$ | 0.6 mm | 0.6 mm |
| Outlet diameter | $d_o$ | 0.6 mm | 0.6 mm |
| Cavity height | $H$ | 0.6 mm | 0.6 mm |
| Nozzle plate thickness | | 0.55 mm | 0.55 mm |
| Cooler size | $x,y,z$ | 14×14×8.7 ($mm^3$) | 35×35×12.6 ($mm^3$) |

As an alternative design, the lateral feeding design is introduced taking full advantage of the additional design options enabled by 3D printing. As shown in Figure 8.3(b), the inlet coolant flow enters the cooler at one side and spreads in the inlet plenum to be distributed over all the nozzles of both chips. This design allows to improve the flow uniformity over the nozzles, to reduce the pressure drop in the cooler through optimized internal design and to reduce the overall cooler thickness significantly.

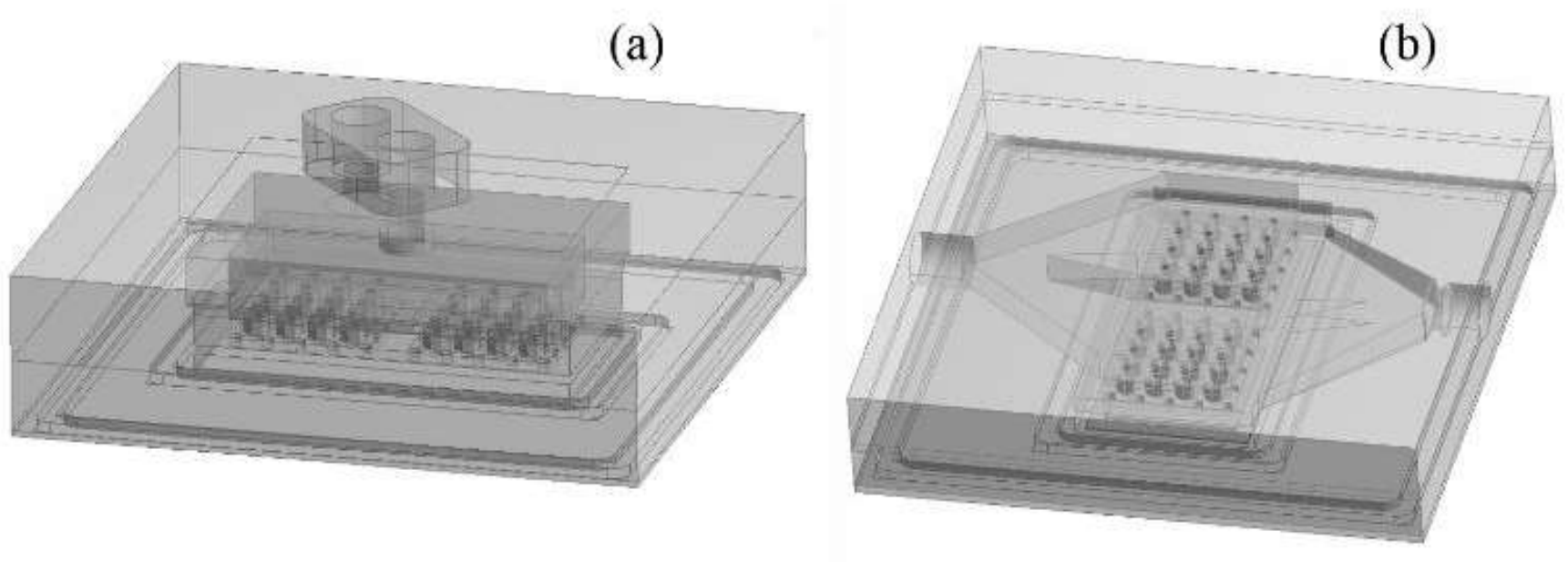

**Figure 8.5**: CAD design of the two different coolers: (a) vertical feeding design; (b) lateral feeding design.

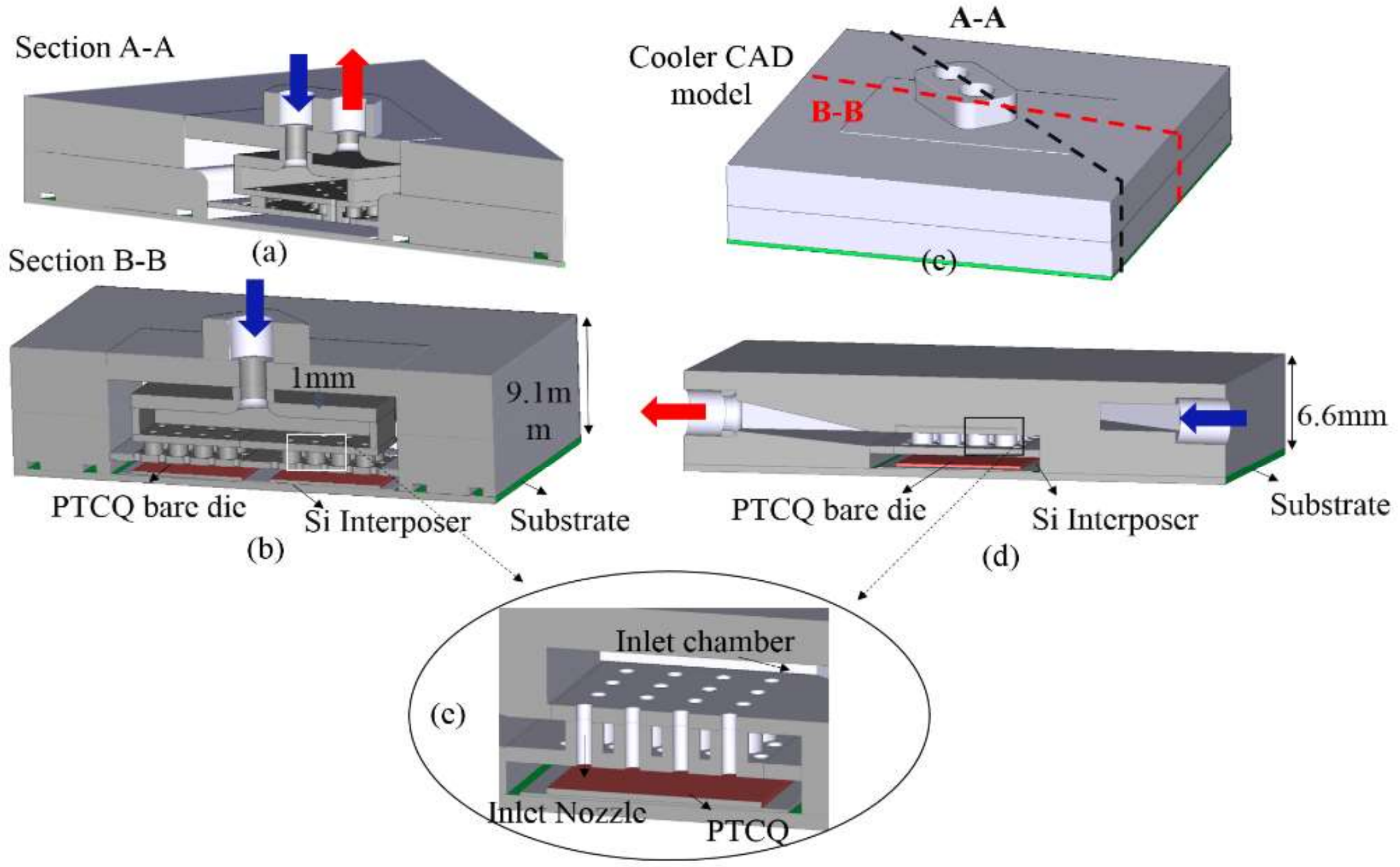

**Figure 8.6**: Cross-section schematics of the two impingement jet cooler configurations: (a) vertical feeding design with cross-section A-A; (b) vertical feeding design with cross-section B-B; (c) full CAD model of vertical feeding design; (d) cross-section of lateral feeding design; (e) enlarged view of the nozzle array and inlet chamber.

Figure 8.6 shows the cross-sections of the inside delivery manifold for the 2 cooler configurations. In the case of the vertical feeding configuration with an inlet and an outlet plenum above each other (Figure 8.6(a) and Figure 8.6(b)), the overall cooler thickness is 12.6 mm, including the tube connection. In the case of the lateral feeding configuration (Figure 8.6(d)), the cooler thickness can be significantly reduced since the two plenums can be integrated on the same level. The enlarged view with the details of nozzle arrays and the inlet chamber is shown in Figure 8.6(e). The overall cooler thickness for this configuration is 6.6 mm, which realizes a reduction of the cooler thickness by a factor of two compared to the standard cooler configuration. The experimental and numerical comparison of the thermal and hydraulic performance of both cooler concepts will be discussed in Section 8.5.

### 8.2.3 Full cooler level model

In order to investigate the thermal and hydraulic performance of the coolers in more detail, conjugate heat transfer and fluid flow simulations are used in this study. The numerical simulations of the full cooler model are performed in this section. The material used for the solid domain is silicon and the water used for the fluid domain. The meshing details of the CFD models for both cooler designs are shown in Figure 8.7(a) and (b), containing typically 5 million elements. The modeling methodology and meshing methods are illustrated in chapter 2 for the full model modeling. The mesh independence of the simulation results has been assessed using the Richardson extrapolation resulting in a truncation error for the chip temperature in the stagnation point of 0.3%.

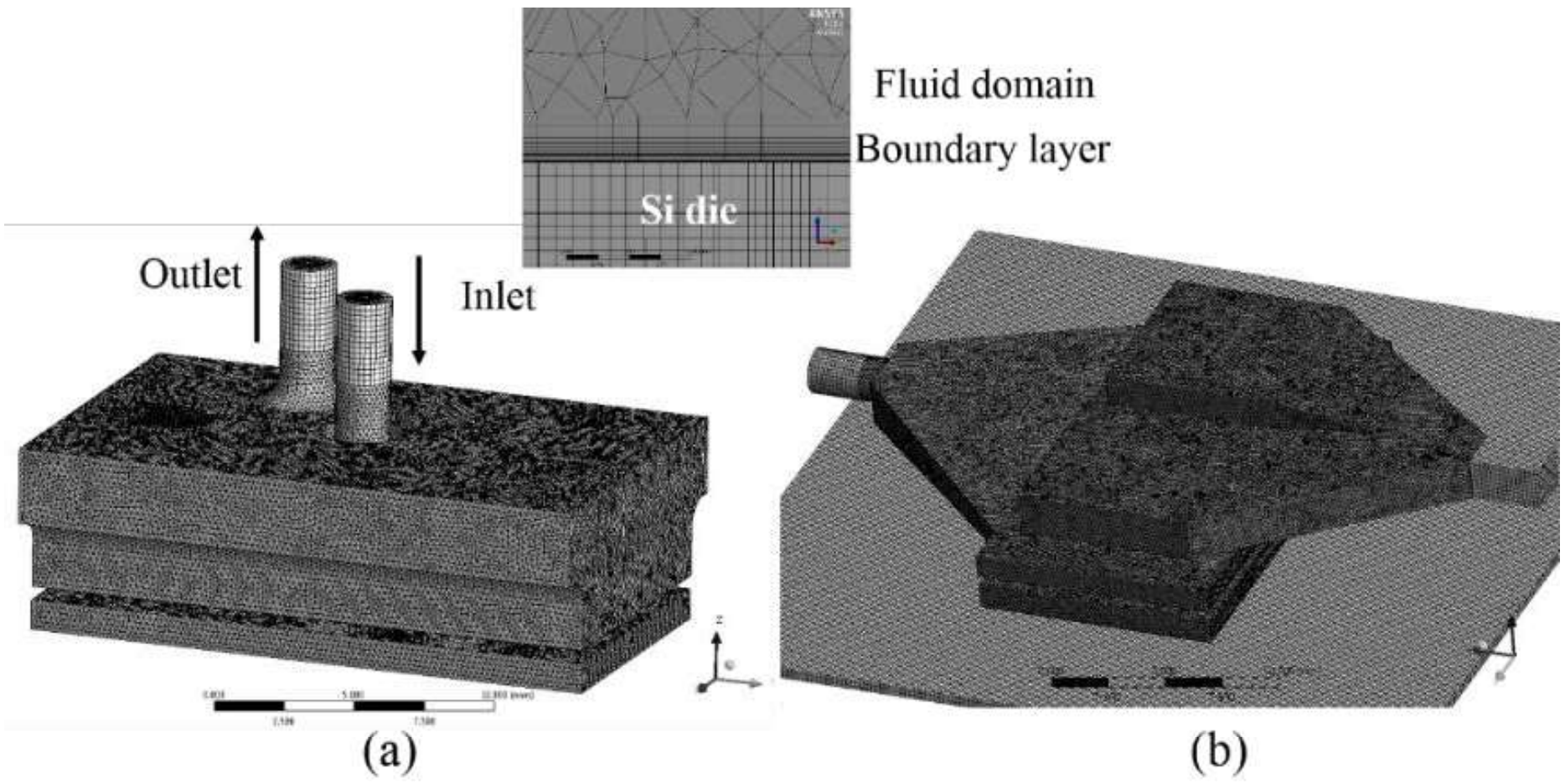

**Figure 8.7**: Details of the CFD model for vertical and lateral cooler design with (a) vertical feeding manifold and (b) lateral feeding manifold.

For the boundary conditions of the CFD model, uniform power dissipation is applied as constant heat flux in the active die while there is no power in the passive die. The bottom part under the active die and bottom die, such as the Cu pillars and underfill material, the package substrate, the solder balls and the PCB are neglected, based on our previous study [32]. The ambient temperature is considered to be at 25ºC. The inlet temperature for the CFD model is set to 10ºC. The flow boundary condition for the inlet is based on the velocity inlet with a specified constant velocity value across the inlet area. The boundary condition for the outlet is set as ‘pressure out’ ($P_{out}$=0). For all the simulations, the net imbalance of overall mass, momentum and energy is kept below 0.02%. The CFD models for the bare die package cooling will be used in Section 8.5 for the thermal and hydraulic performance comparison of the reference cooler design and the improved design.

## 8.3 Cooler assembly and experimental set-up

### 8.3.1. Fabricated cooler assembly

The cooler is printed as a single part by stereolithography (SLA), layer by layer by curing the photosensitive polymer material with exposure of a UV light source, using the same technology as in the previous chapters. The chosen polymer material, Somos WaterShed XC, is a water-resistant material, which shows ABS-like properties and good temperature resistance. The printed coolers are shown in Figure 8.8. The measured average nozzle diameter is 570 μm +/- 20 μm, which only deviates 5% from the nominal design values of 600 μm. The uncured resin in the cavity needs to be removed using a chemical solvent after all the parts are finished. The cooler is finally assembled onto the thermal test board. The final process flow of the dual-chip module cooler with lateral feeding design is summarized in Figure 8.9.

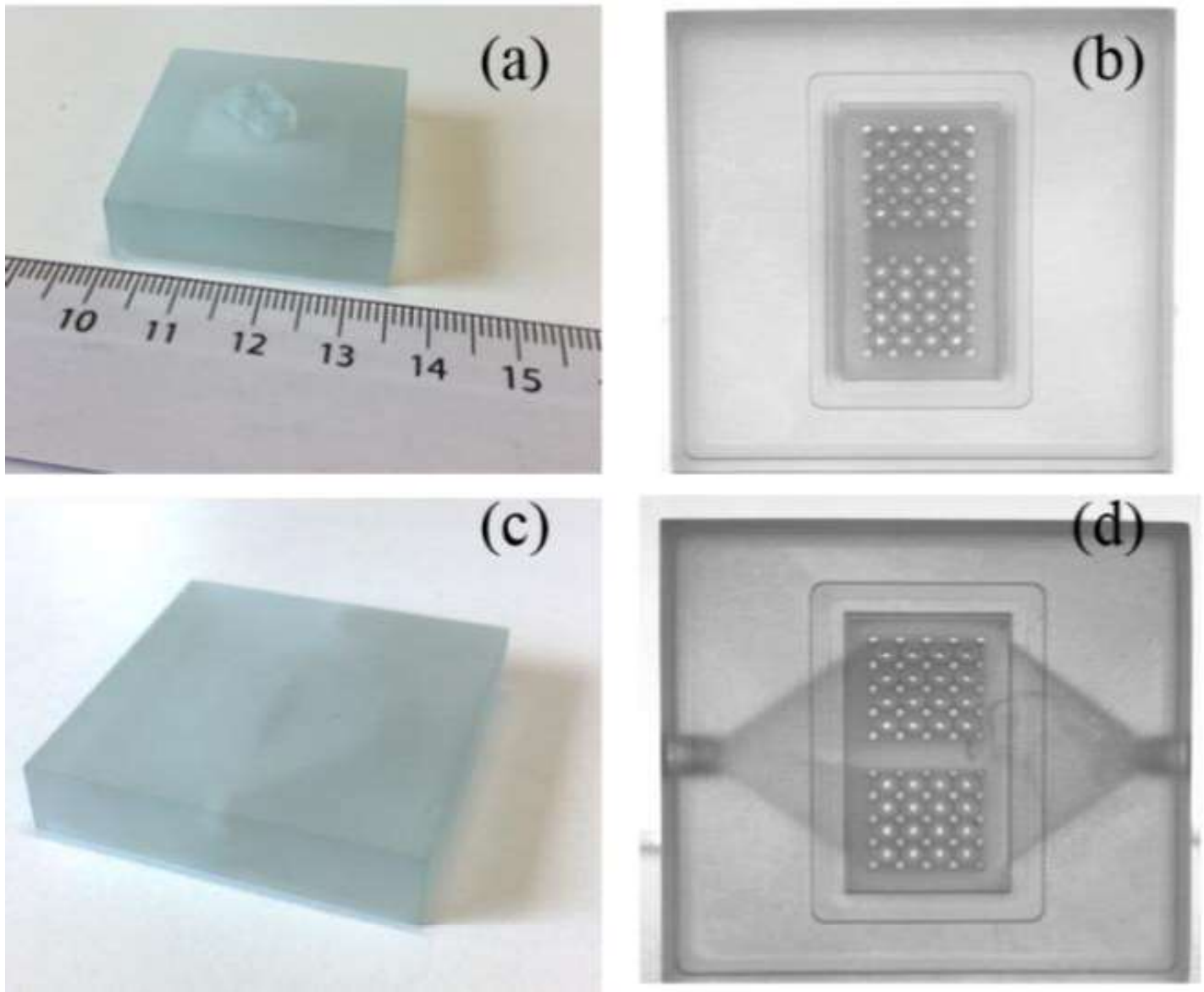

**Figure 8.8**: Dual-chip module cooler demonstrators: side view (a) and bottom view (b) of vertical feeding cooler; side view (c) and bottom view (d) of lateral feeding cooler.

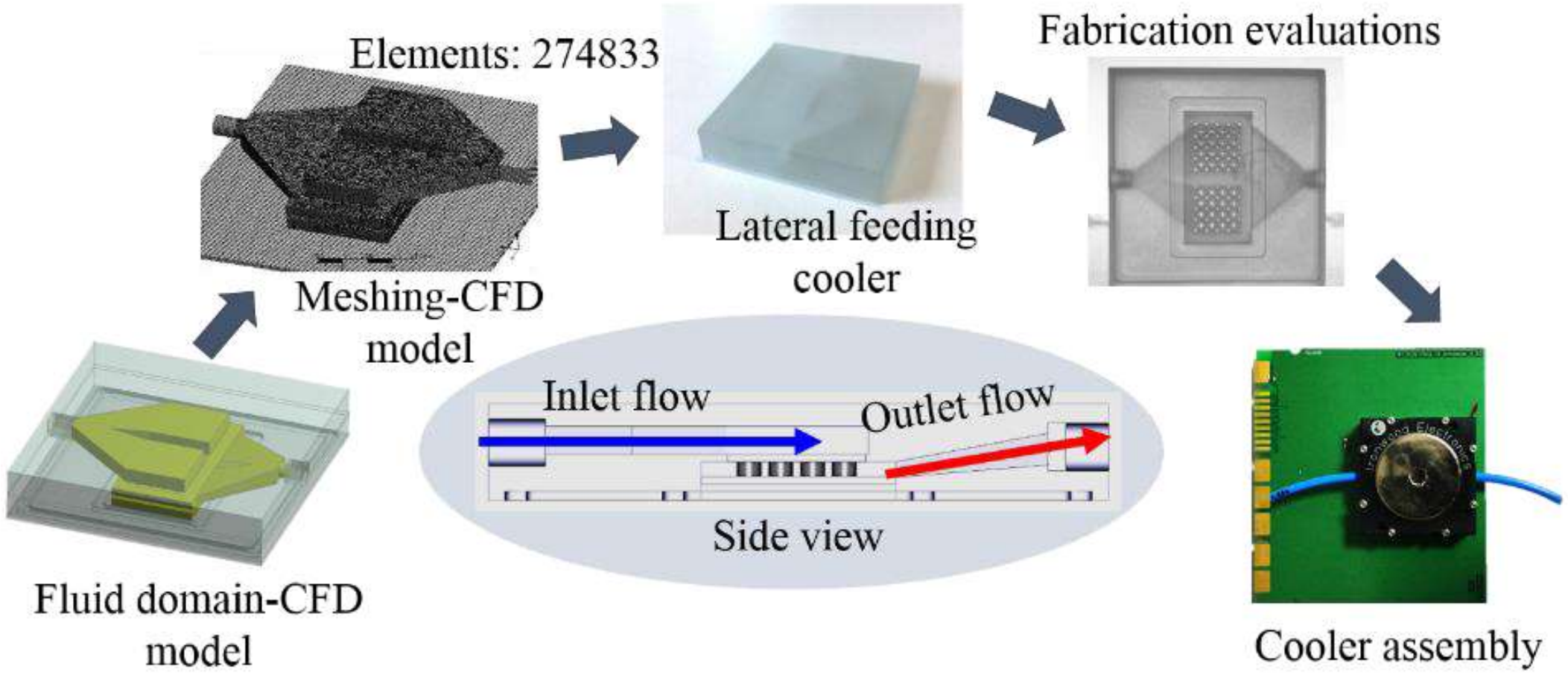

**Figure 8.9**: Design flow for the dual-chip module cooler: lateral feeding design.

### 8.3.2. Experimental set-up

Figure 8.10 shows the experimental set-up for the two different coolers, where the assembled dual-chip module cooler packages are placed in a socket to perform the thermal measurements. A known amount of power is generated in a chosen distribution. In this case, all the heating elements on the chip are activated, while the full chip area temperature distribution is measured in the diodes of the PTCQ test chips. The temperatures are reported as temperature differences with respect to the coolant inlet temperature. The propagated measurement uncertainties are discussed in chapter 2. The thermal performance estimation of the assembled cooling solution also includes the heat losses through the cooler material into the ambient and the heat losses through the bottom side of the assembly, through the test board.

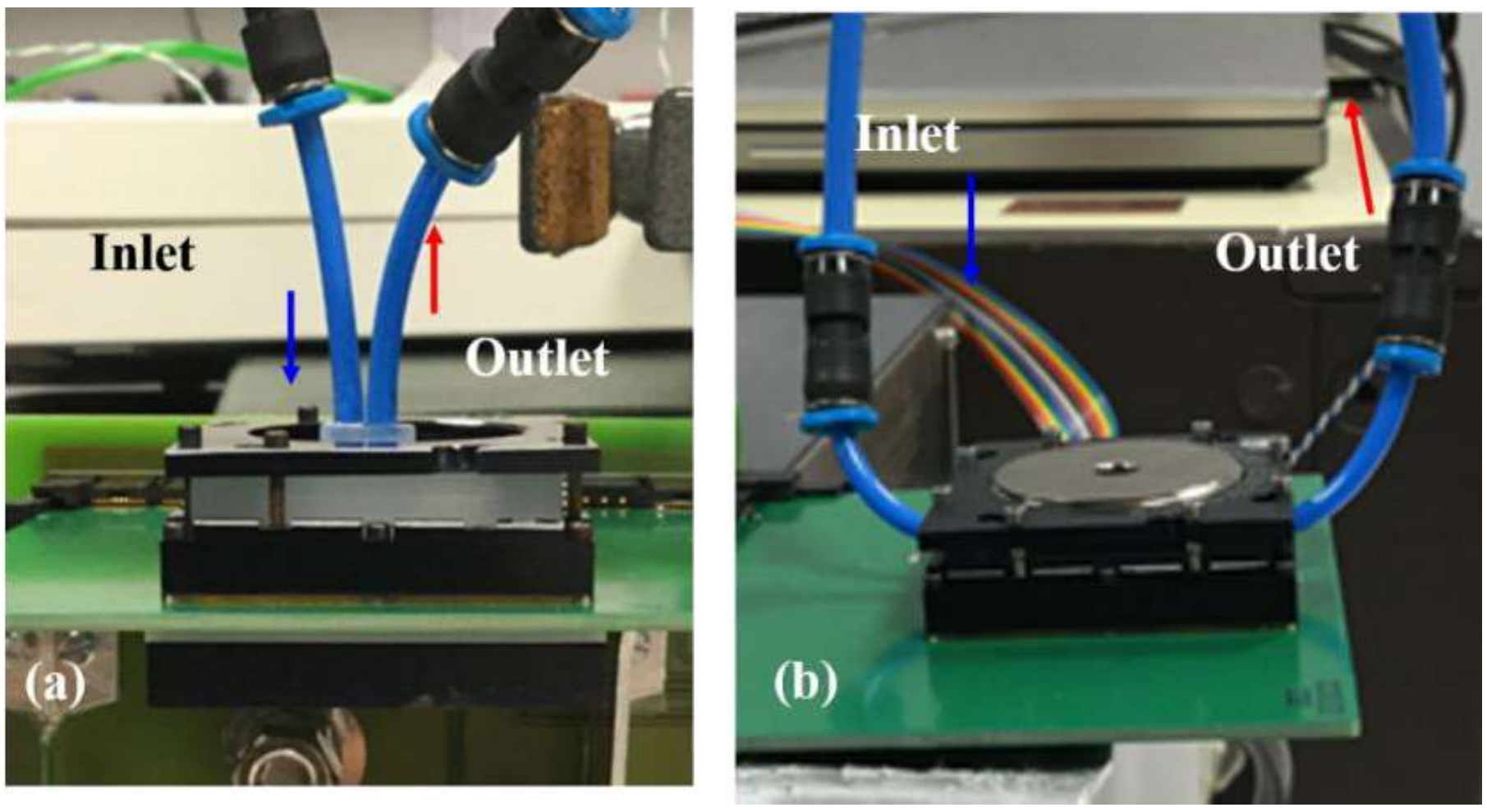

**Figure 8.10**: Experimental set-up for (a) vertical and (b) lateral feeding cooler.

## 8.4 Characterization of standard cooler

### 8.4.1. Reference temperature measurement

One of the objectives of the study is to compare the effect of the impingement cooling solution on the bare die and lidded packages in terms of the thermal resistance of the heated chip and the thermal coupling between the two chips in the package. Since the package type (with lid and without lid) is different, this might have an impact on the heat conduction inside the chip package and consequently, on the thermal resistance and the thermal coupling. Therefore, the packages are first measured without any active liquid cooling applied. There are two main thermal paths towards the ambient for the heat generated in a chip in the package: one is from the top side of the package, through the cooler ($R_{cooler}$); the second parallel thermal path is downward through the package substrate and PCB ($R_{bottom}$). The overall thermal resistance is the parallel connection between the two thermal paths. In the first reference measurements, the top side of the package is insulated, and the heat is removed through the bottom part of the package, enabling the characterization of the bottom thermal part, as shown in Figure 8.11(a). The thermal resistance network is shown in Figure 8.11(b) for the illustration of the heat flow simplified in thermal paths. Figure 8.11(c) shows the measured temperature distribution in the test chip for a power dissipation of 1.7 W. This corresponds to an average thermal resistance of 15.1 K/W, or an average area-normalized thermal resistance of 9.6 $cm^2$-K/W for the bottom thermal path through the package substrate and PCB.

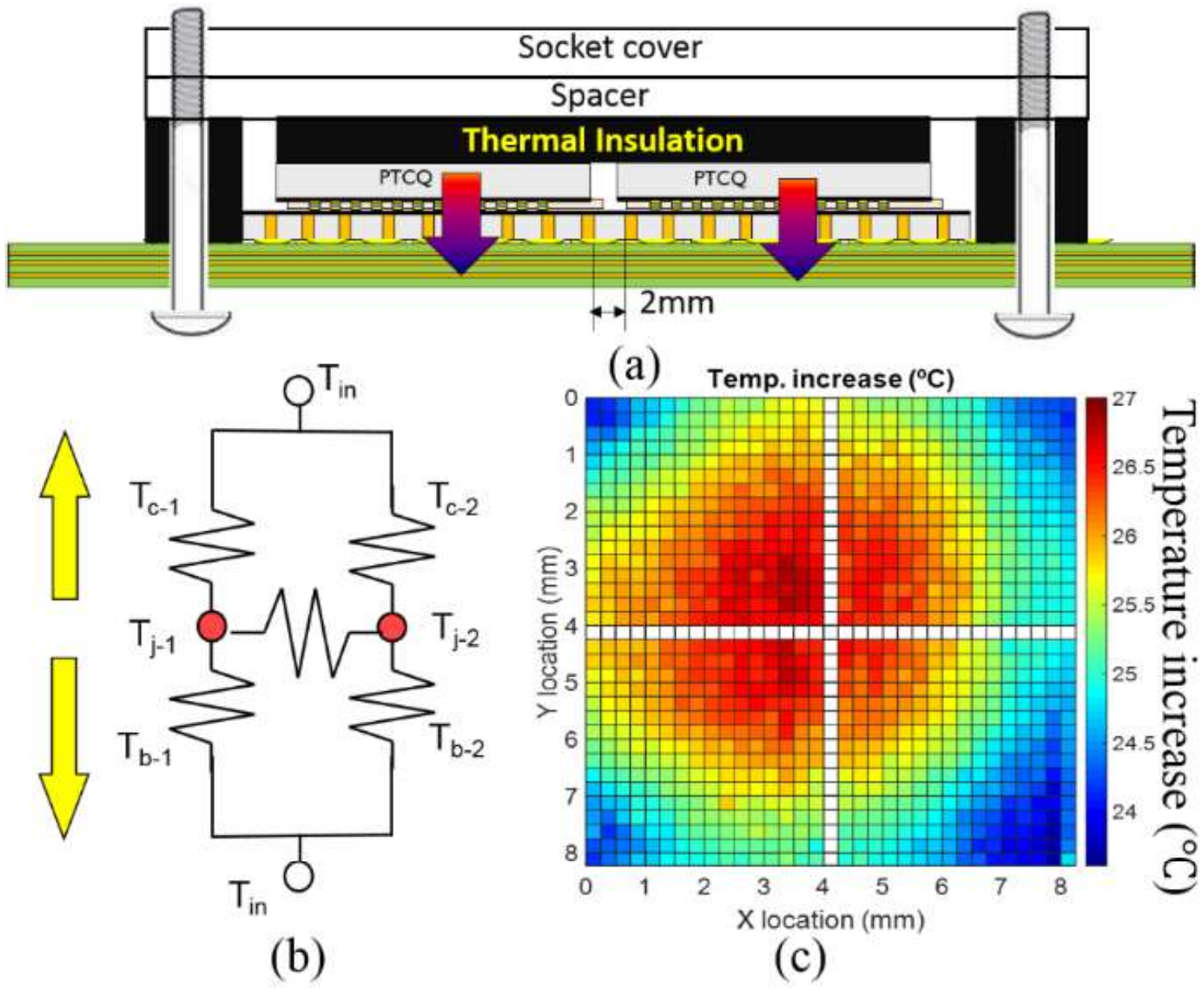

**Figure 8.11**: Reference temperature measurement with thermal insulation: (a) schematic view of the measurement setup with thermal isolation on top of the die surface; (b) thermal resistance network of the thermal path of the interposer package; (c) temperature increase map of the interposer package with thermal insulation.

### 8.4.2. Bare die versus lidded package cooling

For the thermal measurements with the applied cooling, 50 W power is dissipated in each chip of the dual-chip package, while the full temperature map of the chips is measured for a flow rate of 1000 mL/min (for cooler, thus 500 mL/min per chip). The thermal performance of the 3D printed cooler is compared for the lidded and the lidless packages in this section. The measured chip temperature distribution maps for both packages are shown in Figure 8.12. The temperature profile at the center of the chips is shown in Figure 8.13 to allow a more detailed comparison of the thermal behavior. The comparison of the temperature profiles of the two package reveals a significant difference for both the heated chips. The overall thermal resistance of the logic chip is a factor of 2 to 3 higher in the case of the lidded package compared to the lidless package. This large difference is mainly caused by the presence of the thermal interface material.

**Table 8.2:** Thermal comparison between the lidded and lidless packages.

| **Flow rate (mL/min)** | **Average thermal resistance-Total (cm$^2$-K/W)** | | **Average thermal resistance of cooler (cm$^2$-K/W)** | | **Relative heat loss** | |
|---|---|---|---|---|---|---|
| | lidded | lidless | lidded | lidless | lidded | lidless |
| 300 | 0.85 | 0.47 | 0.93 | 0.49 | 9.71% | 5.15% |
| 400 | 0.80 | 0.41 | 0.87 | 0.43 | 9.09% | 4.46% |
| 600 | 0.75 | 0.33 | 0.81 | 0.34 | 8.47% | 3.56% |
| 1000 | 0.68 | 0.26 | 0.73 | 0.27 | 7.62% | 2.78% |

The measurement results for the lidded and lidless packages are summarized in Table 8.2 for different flow rates. The presence of the lid (and mainly the TIM) results in a higher chip temperature, where the relative impact of the lid increases as the flow rate increases since the convective thermal resistance decreases with the flow rate. The additional thermal resistance of the TIM and lid can be estimated as 0.45 cm$^2$-K/W. The thermal conductivity of the TIM is calibrated as 1.9 W/m-K and a targeted thickness of 80 μm. The reported overall thermal resistance values are the result of the combined heat removal through the cooling solution and the heat losses through the package. By combining the results with the reference measurements, the heat losses through the package can be estimated. The relative values for the heat losses are shown in Table 8.2 for the two packages for different coolant flow rate values. These results show that the heat losses are limited to values from 2% for a high flow rate to 5 % for a low coolant flow rate and therefore, the majority of the heat is removed through the cooling solution on top of the package. Table 8.2 also shows the thermal resistance values for the top heat flow part only, after correction for the heat losses. Since the heat losses through the bottom package are small, the type of packaging does not have a significant impact on the thermal resistance values and the thermal coupling in the packages.

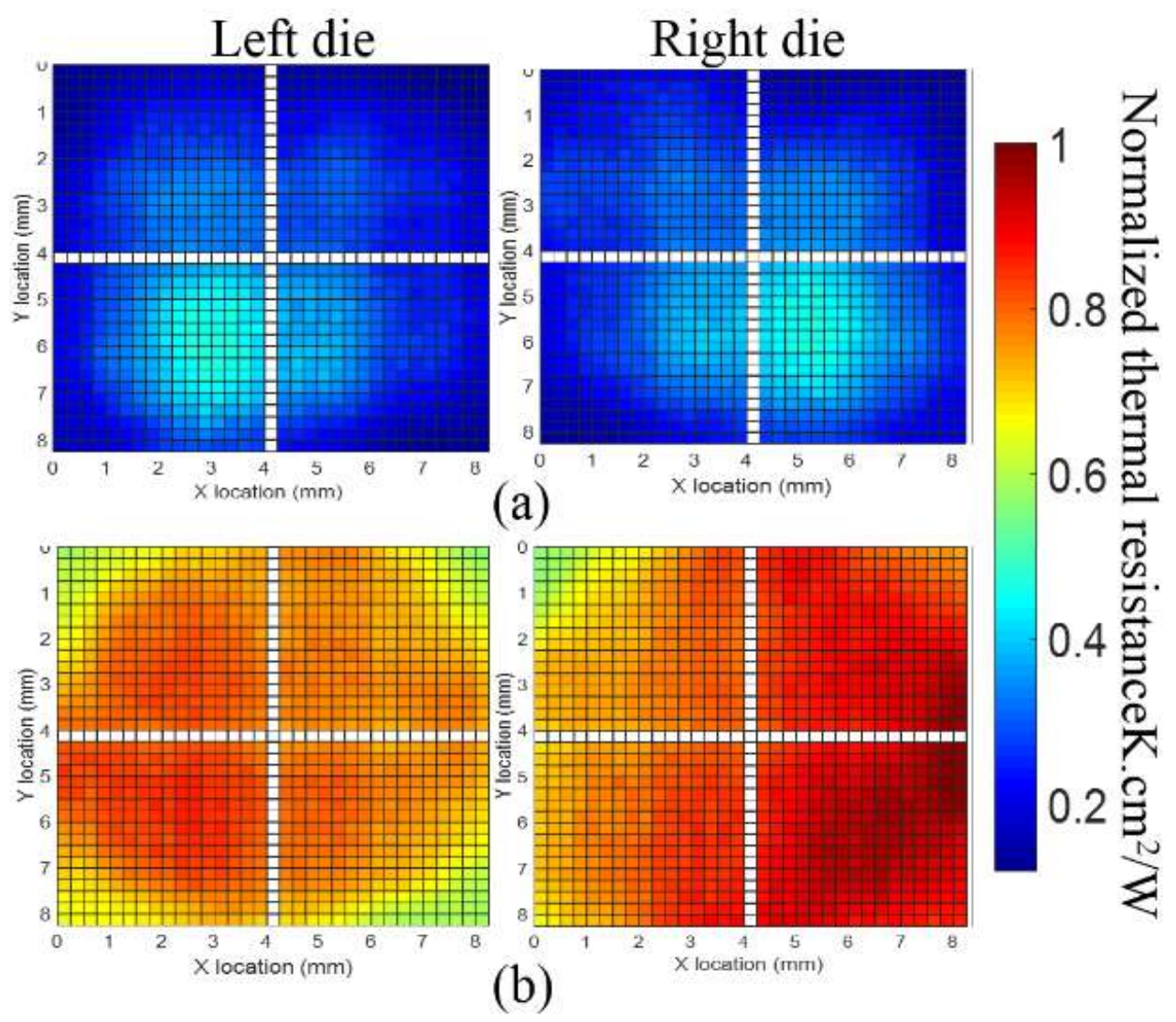

**Figure 8.12**: Temperature distribution on two heated dies with (a) lidless cooling and (b) lidded package cooling (left chip power=50 W, right chip power=50 W, flow rate=1000 mL/min, vertical feeding cooler).

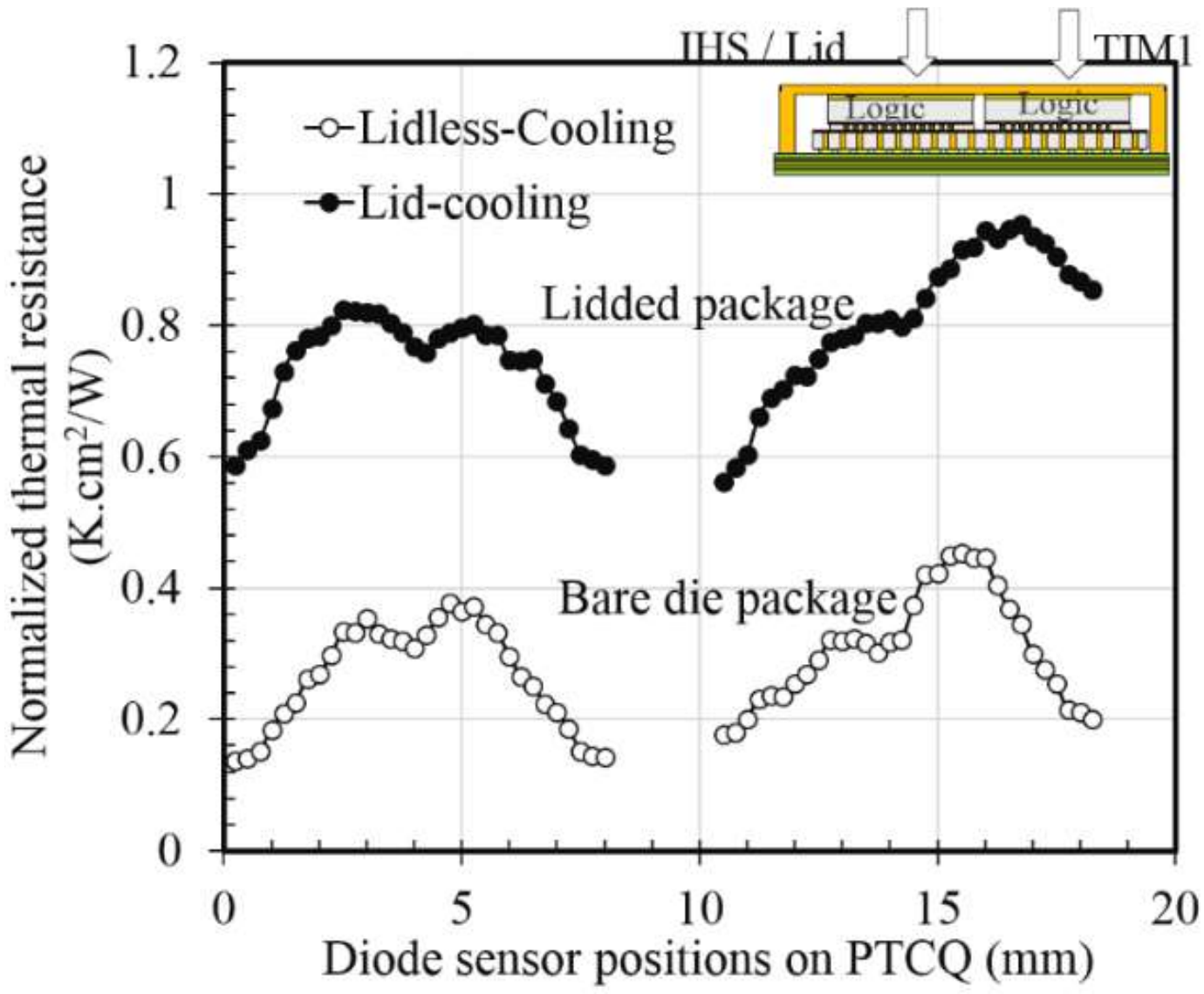

**Figure 8.13**: Thermal measurement comparison between the lidless package cooling and lidded package cooling (single chip power=50W, flow rate=1000 mL/min, vertical feeding cooler).

### 8.4.3. Comparison with single die

The dual-chip module cooler has been characterized for an overall flow rate of 600 mL/min (divided over both chips in the module) and for a power dissipation of 50 W in the left chip and no power dissipation in the right chip. The normalized thermal resistance for vertical feeding of the cooler is shown in Figure 8.14 for a horizontal profile across the two chips. The cooling performance for the single chip cooler [28] with the same nozzle array and the same normalized flow rate, in this case of 300 mL/min per chip, shown in the same figure, results in a similar value of 0.35 $cm^2$-K/W. This comparison shows that the average normalized thermal resistance can be used to extrapolate the thermal performance for different chip sizes or multiple chips, supporting the approach of a scalable cooling solution with a constant intrinsic thermal performance for the unit cooling cell with constant flow rate per unit cell. Therefore, this scalable approach can be used to design specific nozzle arrays for different chip sizes in the module, and this normalization concept has been introduced in chapter 2. It should be noted that there is a discrepancy between the single chip and dual-chip cooling for the actual temperature profile. These local differences are due to the designs of the inlet/outlet chamber in the plenum level resulting in different local flow distribution.

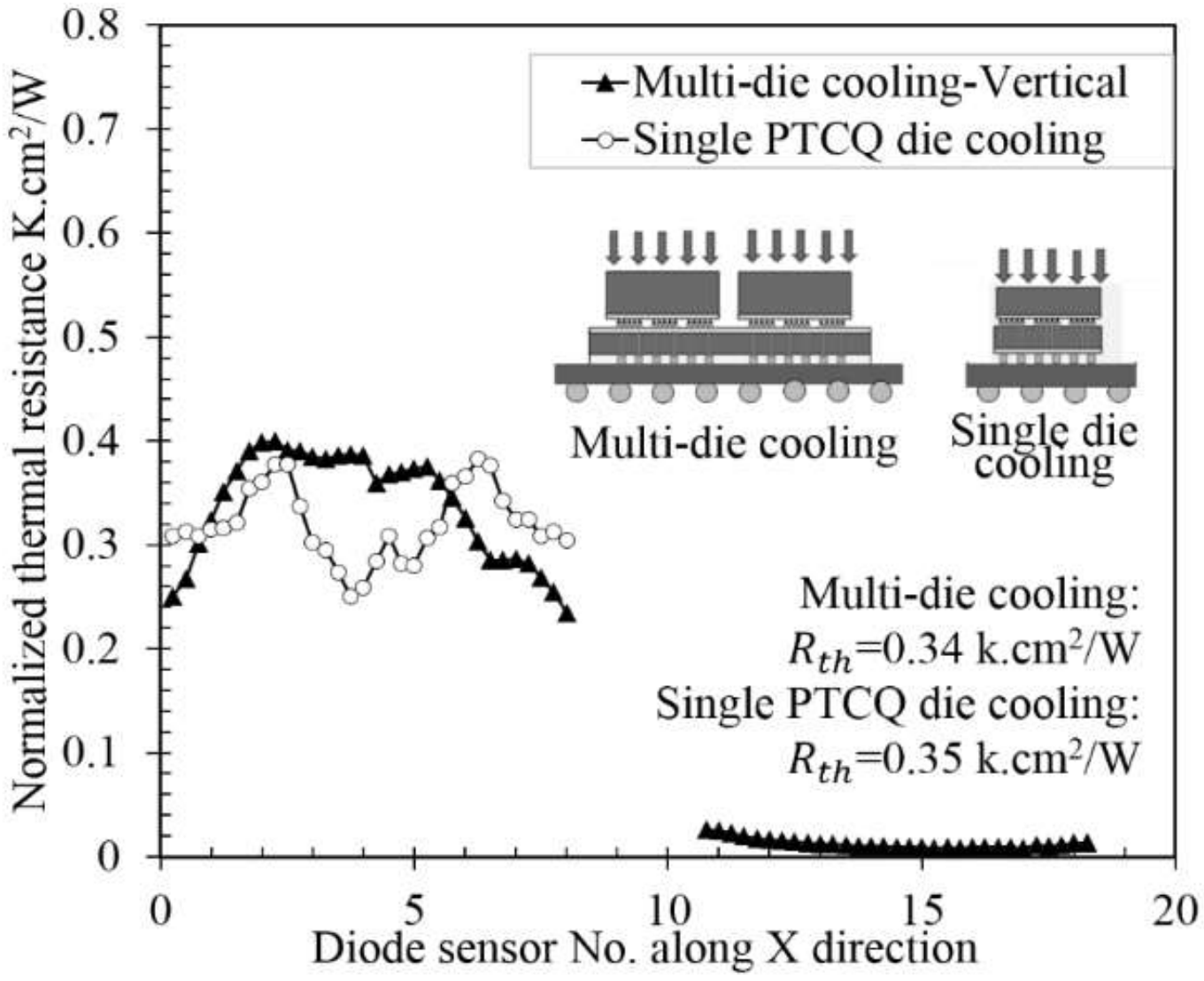

**Figure 8.14**: Experimental comparison with dual-chip cooling and single chip cooling at a flow rate of 600mL/min (Vertical feeing manifold).

### 8.4.4. Thermal coupling effects

In the next step, the thermal coupling between the active die and passive die in the dual-chip module cooled by the package level impingement cooler is investigated. Therefore, the power dissipation in the left chip referred to “logic die” is set as 50W while no power dissipation is applied in the right chip referred as “Memory die”. The temperature distribution map of the bare die cooling with vertical feeding cooler and of the lidded package cooling is shown in Figure 8.15(a) and Figure 8.15(b) respectively for an overall flow rate of 1000 mL/min. The temperature profiles for three different flow rates are plotted in Figure 8.16. It can be observed that the thermal coupling between the active die and the passive die is much higher in the lidded package compared to the lidless package, due to the heat spreading in the Cu lid. In the bare die package, the temperature increase of the passive die is very limited due to the absence of the thermal coupling path of the lid. In the lidded package, the passive die temperature is much higher and also shows a clear temperature gradient from the left side to the right side of the chip.

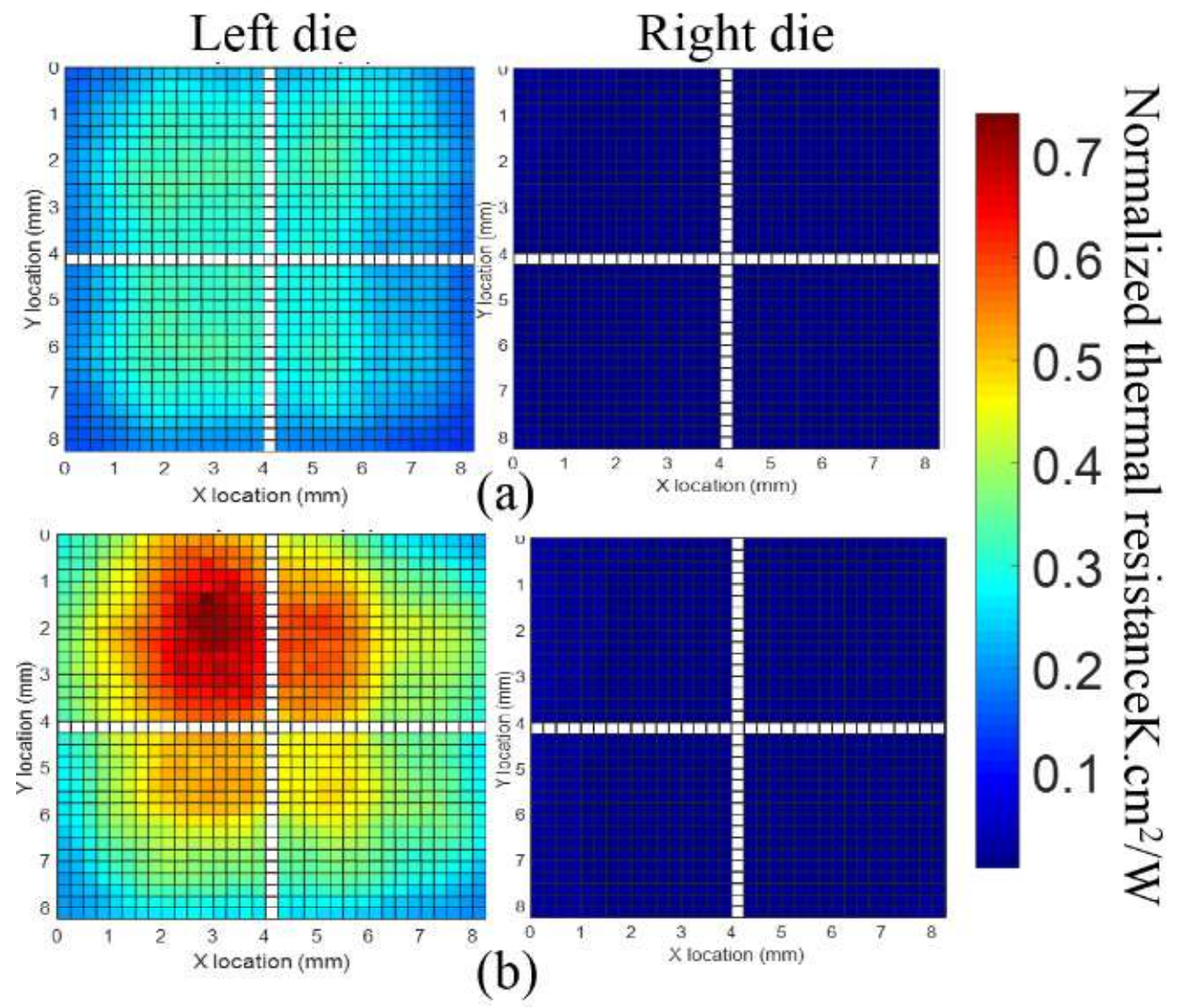

**Figure 8.15**: Normalized thermal resistance measurements (cm2-K/W) for the (a) lidless cooling and (b) lidded package cooling on the interposer package (active die=50 W, passive die=0 W, flow rate =1000 mL/min).

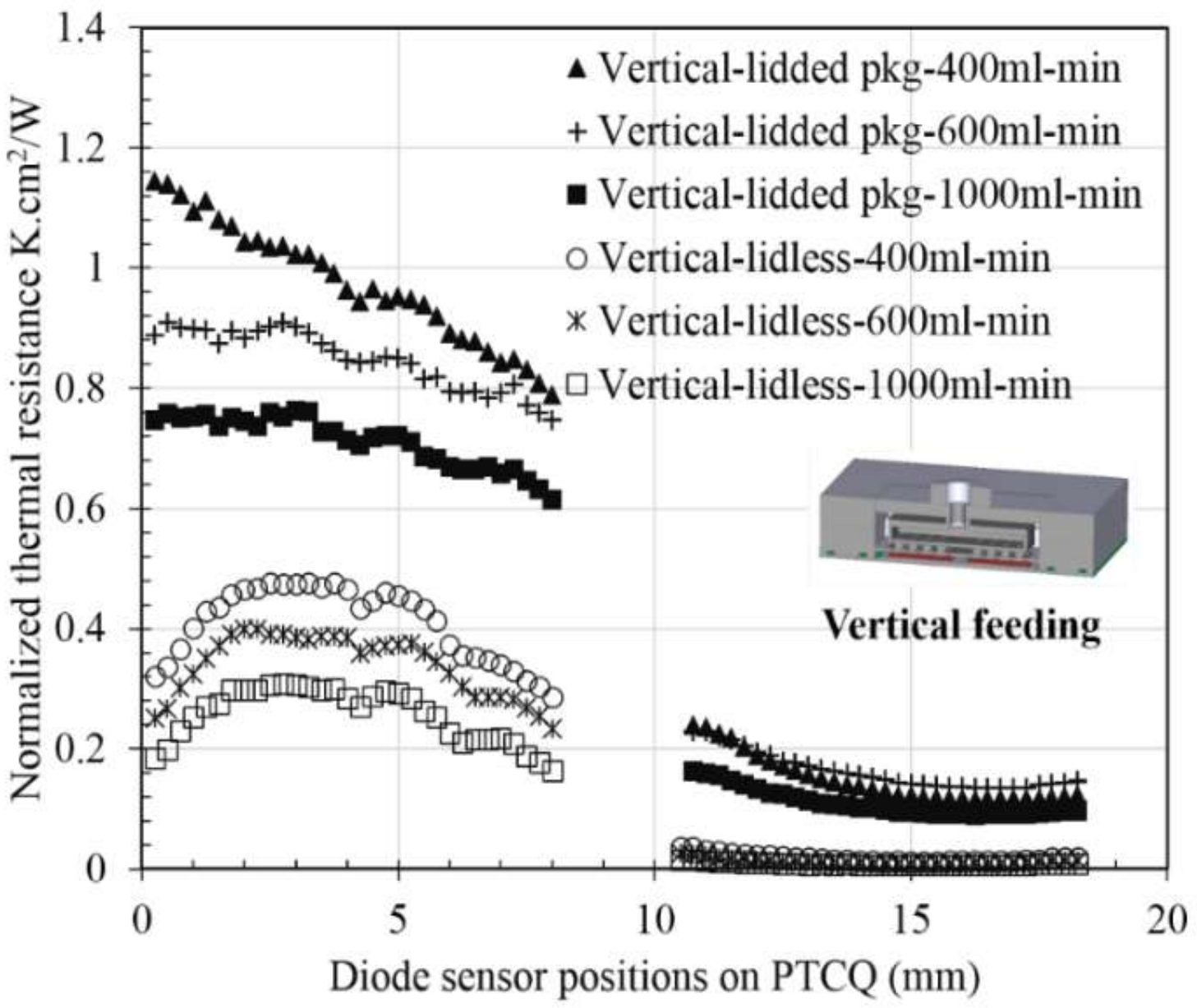

**Figure 8.16**: Normalized thermal resistance (cm$^2$-K/W) measurements profile comparison between lidless cooling and lidded package cooling under different flow rates (active die=50 W, passive die=0 W, flow rate =1000 mL/min).

The thermal coupling in a multi-die module can be expressed as a part of the thermal resistance matrix R, which is used for multi-chip modules describing the thermal interactions between the different heat sources [38-40]:

$$\begin{bmatrix} T_{left} \\ T_{right} \end{bmatrix} \cong \begin{bmatrix} R_{left,left} & R_{Left,Right} \\ R_{Right,Left} & R_{Right,Right} \end{bmatrix} \cdot \begin{bmatrix} P_{left} \\ P_{Right} \end{bmatrix} + \begin{bmatrix} T_{in} \\ T_{in} \end{bmatrix} \tag{9}$$

The resistance $R_{ij}$ in the matrix of (5) is the temperature rise of heat source '$i$' above the ambient temperature (inlet temperature), caused by unit heat dissipation of source "$j$":

$$R_{ij} = \frac{T_i - T_{in}}{P_j} \tag{10}$$

For a dual-die package, 4 thermal resistance terms are required to describe the thermal resistance matrix: the self-heating thermal resistance terms on the diagonal and thermal coupling resistance terms, or mutual heating effects on the cross-diagonal. The thermal resistance (6) should be obtained in case of uniform power dissipation in one of the chips while there is no power dissipation in the other chip(s) in the 3D package. In the case of non-uniform power dissipation, the concept of thermal resistance is not meaningful. Therefore, the thermal coupling between the passive die and active die in this study can be expressed as below based on the average chip temperatures:

$$R_{coupling} = \frac{T_{passive\,-} - T_{in}}{T_{active} - T_{in}} \tag{11}$$

**Table 8.3:** Thermal coupling at different flow rates.

| Flow rate (mL/min) | Lidded pkg | Lidless pkg |
|---|---|---|
| 300 | 15.7% | 7.1% |
| 400 | 15.98% | 5.3% |
| 600 | 17.38% | 4.6% |
| 1000 | 26.1% | 3.4% |

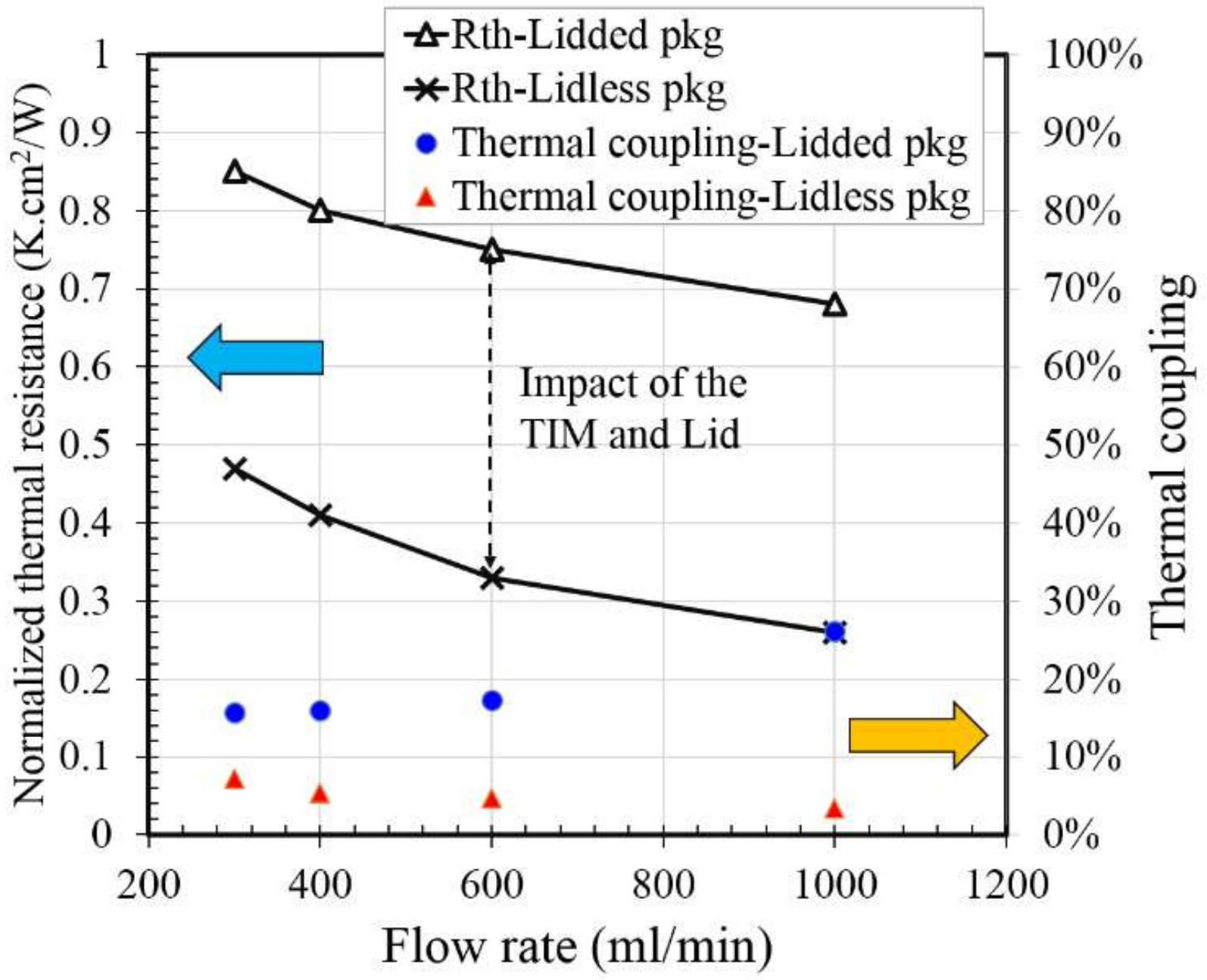

**Figure 8.17**: Normalized thermal resistance and thermal coupling as a function of the total flow rate for the lidded package and lidless package.

Table 8.3 shows the thermal coupling effects for lidless cooling and lidded package cooling at different flow rates. Moreover, the normalized thermal resistance and thermal coupling as a function of the flow rate for the lidded package cooling and lidless cooling are plotted in Figure 8.17. As shown in Figure 8.17 at a flow rate of 1000 mL/min, the thermal coupling is 7.7 times higher in the lidded package, compared to the lidless package. For the lidless package cooling, it can be seen that the thermal coupling reduces for increasing flow rate. This is because the main thermal path is dominated by the high cooling efficiency of microjet cooling without TIM and lid. While for lidded package cooling, the thermal coupling effects become higher with increasing flow rate. The reason for this is that the relative contribution of the conductive thermal resistance from the TIM and lid increases, as the convective resistance decreases. Typical test case values for lidded MCP with heat sink presented in the literature [41], report a thermal coupling value of 31.2%. The presented bare die cooling with a flow rate of 1000 mL/min achieves a 9 times reduction of the thermal coupling.

### 8.4.5. Modeling validation

In this section, the comparison between the CFD modeling results and experimental results is discussed and analyzed. As shown in Figure 8.18, full cooler level CFD modeling results with the temperature profile across the two thermal test die source regions show a similar trend with the experimental data under different flow rates. The lower temperature around the chip edge in the experiments is due to the full submerged of the thermal test die inside the liquid. An additional aspect of the discrepancy between the experimental and modeling results, is the different coolant impact for impingement cooling on the die surface only or also on the chip sides. It is shown in literature [42, 43], that jet impingement hybrid body cooling with a submerged die has better cooling performance than jet impingement surface cooling on the die surface only. This is because the hybrid body cooling can provide extra cooling for the chip by channel cooling to the side surfaces, resulting in a lower temperature at the chip edges. In Figure 8.19, the thermal resistance distribution maps are compared for the experiments and the CFD model results for the lidless interposer package. The nozzle cooling patterns can be clearly distinguished from the modeled temperature distribution that assumes 100% uniform heat dissipation while the actual PTCQ power map is quasi-uniform with 75% heater uniformity. The measured averaged temperature for the “logic” die based on the bare die cooling at a flow rate of 300 mL/min is 0.47 $cm^2$-K/W while the modeling averaged temperature is 0.46 $cm^2$-K/W.

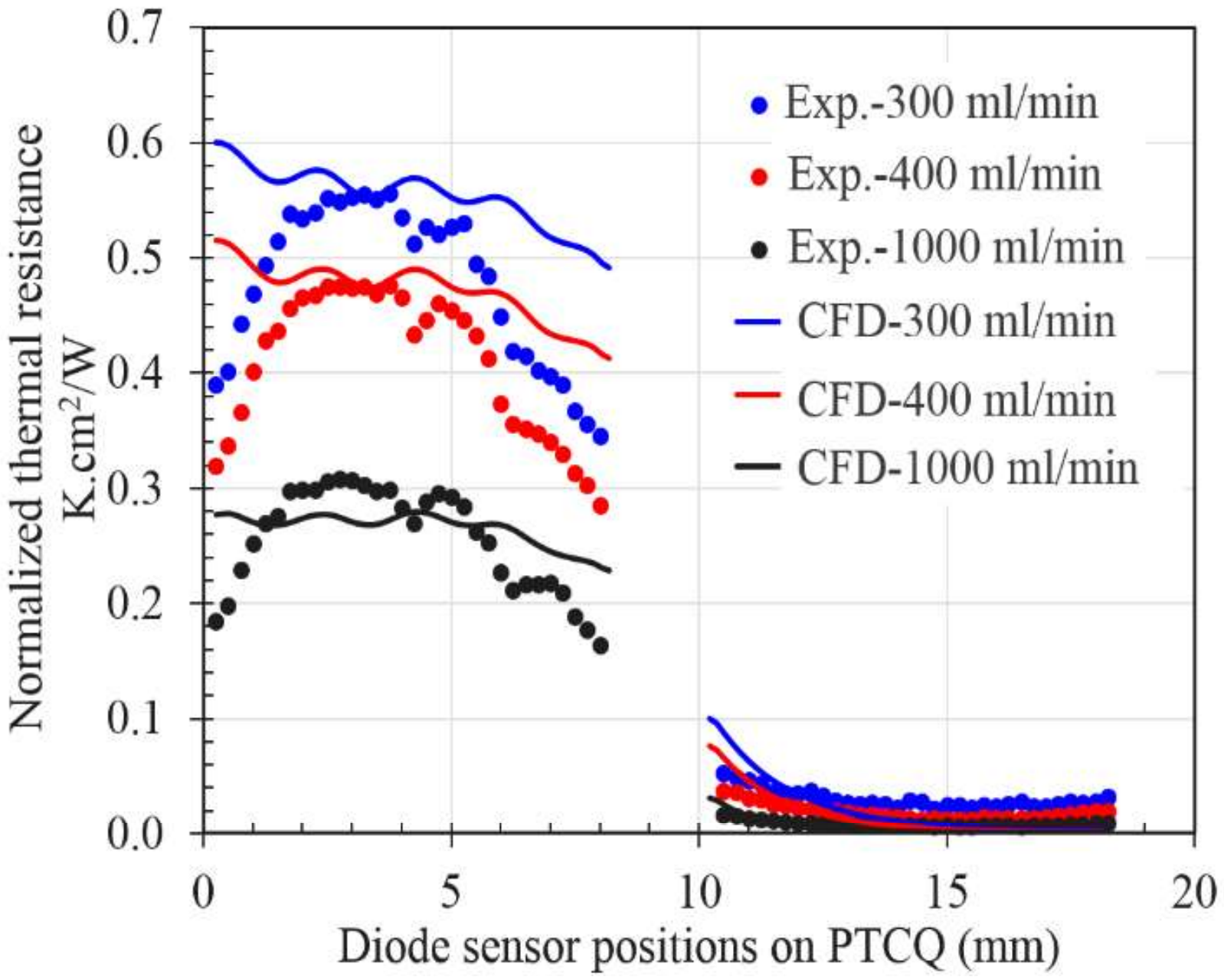

**Figure 8.18**: Experimental validation of CFD modeling for bare die cooling under different flow rates (logic power=50W; memory power=0W).

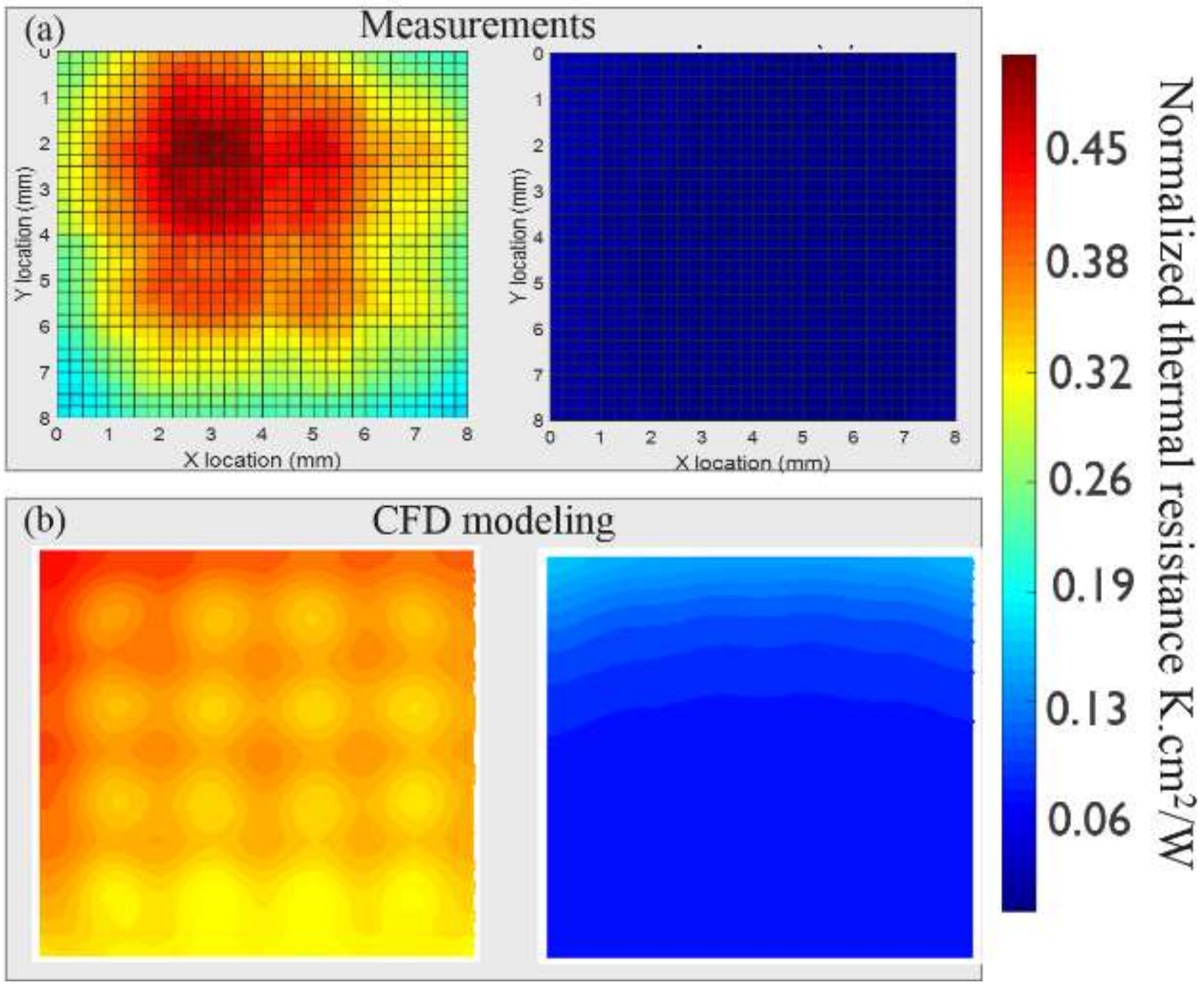

**Figure 8.19**: Normalized thermal resistance ($cm^2$-K/W) distribution comparisons between the measurements and CFD modeling for bare die liquid cooling (logic power=50W; memory power=0W; flow rate = 300 mL/min).

In general, the full cooler level CFD modeling results agree well with the measurement data with respect to the average temperature, however, differences in local temperature distribution become visible at the location of non-heated parts due to the high heat removal rate. For this level of cooling, more details of the chip power map should be included in order to predict the detailed chip temperature map. The lower temperatures around the chip edge in the experiments can be explained due to the absence of the heaters there. The difference between the CFD model and the experimental data for the average chip temperature is 12.6% at a flow rate of 300 mL/min and only 2% at a flow rate of 1000 mL/min. Therefore, we use a uniform heater pattern for the modeling study to save the computation cost.

### 8.5 Lateral feeding cooler performance

In this section, the thermal and hydraulic performance of the lateral feeding cooler design, introduced in Section 8.2, will be evaluated and compared with the standard vertical feeding design.

#### 8.5.1 Experimental comparison

Figure 8.20 shows the comparison of the thermal performance for the vertical feeding design and the lateral feeding design on both the bare die and lidded packages with a power dissipation of 50W in one active chip and a flow rate of 1000 mL/min. Figure 8.20(a) and (b) show the full temperature maps for the lidded packages, while Figure 8.20(c) shows the temperature profiles for the measured cases. It can be seen that the temperature profiles for both coolers are very similar: the difference for the active die temperature is only 4% for both the bare die and lidded cooler. The thermal comparison between the two designs is summarized in Table 8.4. In general, it can be seen that the normalized thermal resistances for the vertical feeding scheme and the lateral feeding scheme are very similar to each other for all flow rates for both the lidded and bare die packages. This comparison proves that the improvement of the plenum shape to minimize the flow resistance does not interfere with the thermal performance of the cooler.

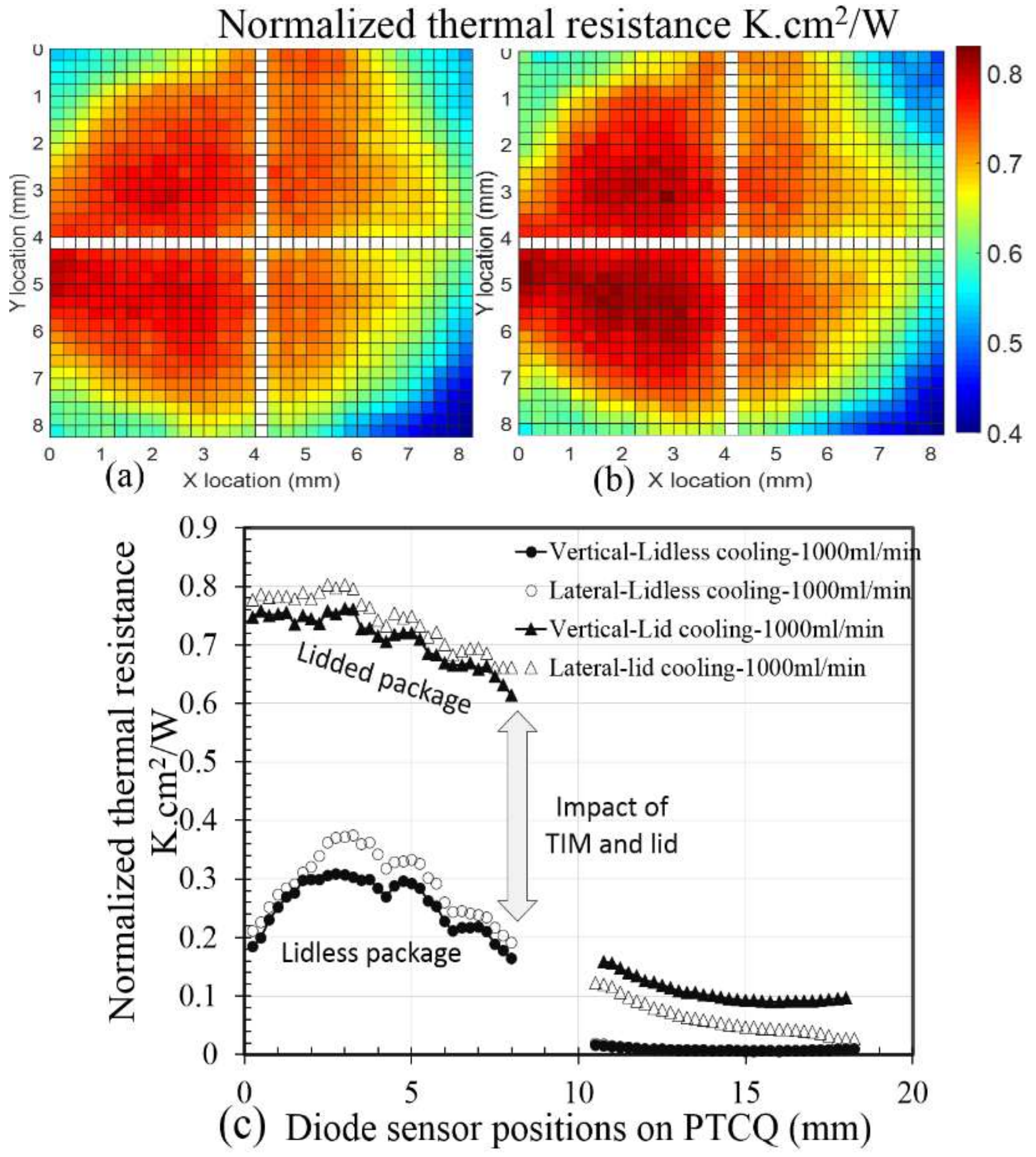

**Figure 8.20**: Thermal measurement comparison in active die between the vertical feeding (a) and lateral feeding (b) design (logic power=50W, memory power=0W, flow rate=1000 mL/min).

**Table 8.4**: Lidless cooling comparison between the vertical feeding and lateral feeding designed cooler at different flow rates.

| **Flow rate (mL/min)** | **Area averaged Thermal resistance ($cm^2$-K/W)** | | **Maximum Thermal resistance ($cm^2$-K/W)** | |
|---|---|---|---|---|
| | Vertical | Lateral | Vertical | Lateral |
| 300 | 0.47 | 0.48 | 0.59 | 0.63 |
| 400 | 0.41 | 0.42 | 0.52 | 0.57 |
| 600 | 0.33 | 0.33 | 0.43 | 0.50 |
| 1000 | 0.26 | 0.27 | 0.34 | 0.37 |

### 8.5.2. Modeling comparison

The thermal and hydraulic performance of the lateral feeding cooler can also be compared to the vertical feeding cooler using the CFD simulations introduced in Section 8.2. Moreover, these simulations can be used to assess the flow distribution in the cooler and the temperature distribution in the chip.

The first part of the comparison is the nozzle inlet velocity uniformity and pressure drop over the cooler between the vertical feeding design and the lateral feeding design. Figure 8.21 shows the comparison of the velocity field inside the vertical and lateral cooler design. The flow streamlines inside the cooler are shown in Figure 8.21(a) and Figure 8.21(b). The cross-section view of the velocity is shown in Figure 8.21(c) and Figure 8.21(d). For the vertical feeding scheme, the coolant is supplied in the center of the cooler. Therefore, the flow velocity will decrease as the flow goes from the central inlet nozzles to the outer inlet nozzles. For the lateral feeding scheme, the entering flow is separated equally into two parts for the distribution of the two dies, resulting in a more uniform distribution over the nozzles. This effect is shown in Figure 8.22(a). The figure compares the distribution of the average inlet nozzle velocity for both cooler designs along a cross-section of the cooler. It can be seen that the velocity distribution of the lateral feeding design is much more uniform than the vertical feeding design. Since the nozzle diameter is kept as the same, therefore, the flow rate uniformity is corresponding to the nozzle velocity distribution. The analysis of the local flow rate for all inlet nozzles shows that the uniformity for the nozzle flow rate is reduced from 25 % to 11 % from the vertical feeding cooler to the lateral feeding cooler. Furthermore, the overall pressure drop over the cooler is much lower for the lateral feeding design, as can be seen from Figure 8.22(b). The improvement of the cooler design results in a reduction of the pressure drop over the cooler of 63% and 53% at flow rates of 100 mL/min and 1000 mL/min, respectively. This pressure drop reduction is caused by the improvement of the internal geometry of the plenum and the elimination of the two 90° bends for the coolant flow in the vertical feeding design. Moreover, the comparison between the CFD modeling and experimental measurement shows a good agreement for the pressure drop across the cooler for both the vertical feeding design and the lateral feeding design.

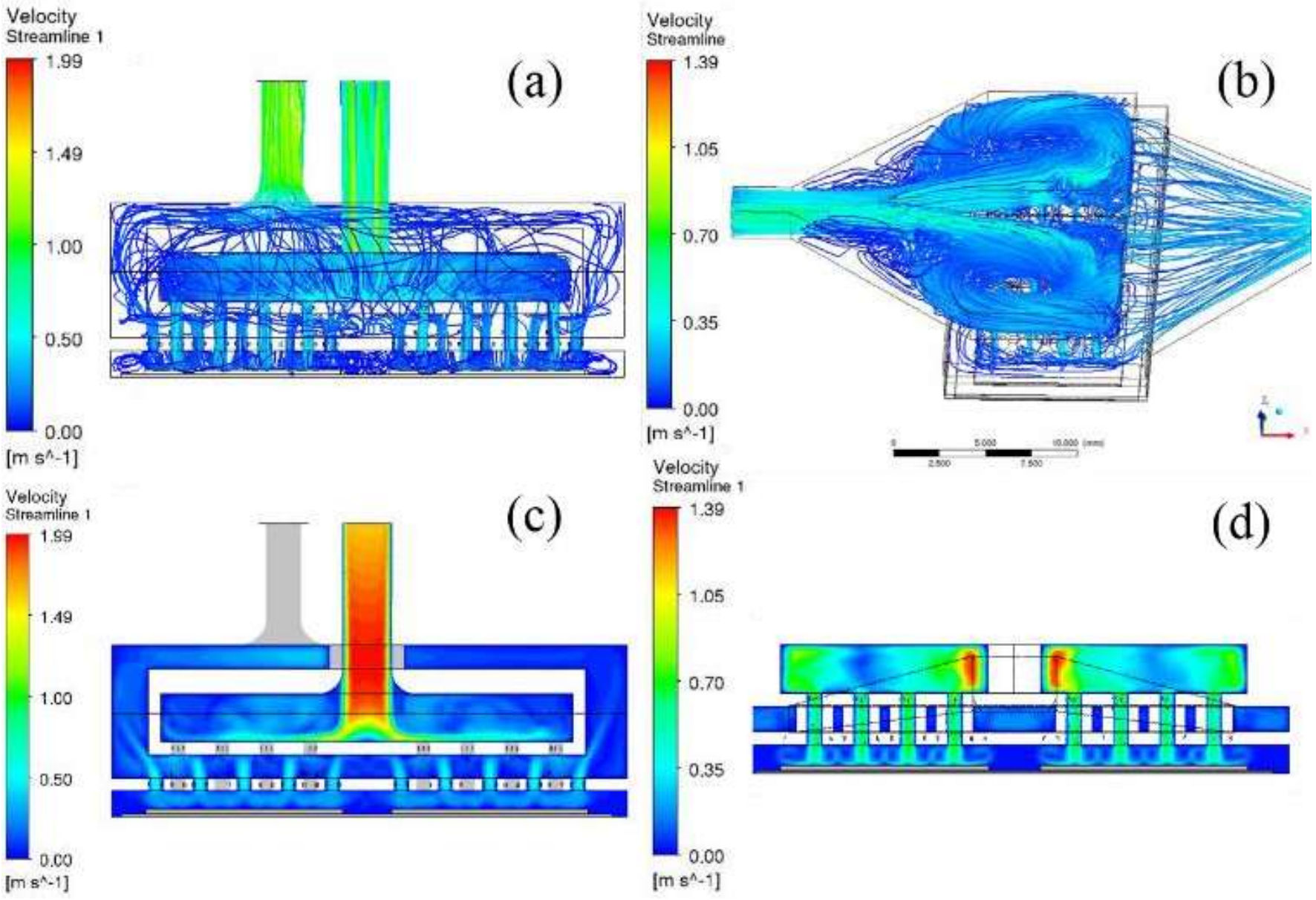

**Figure 8.21**: CFD modeling results comparison between the vertical feeding and lateral feeding design with (a) (b) flow streamline inside the cooler and (c) (d) cross-section of the velocity field.

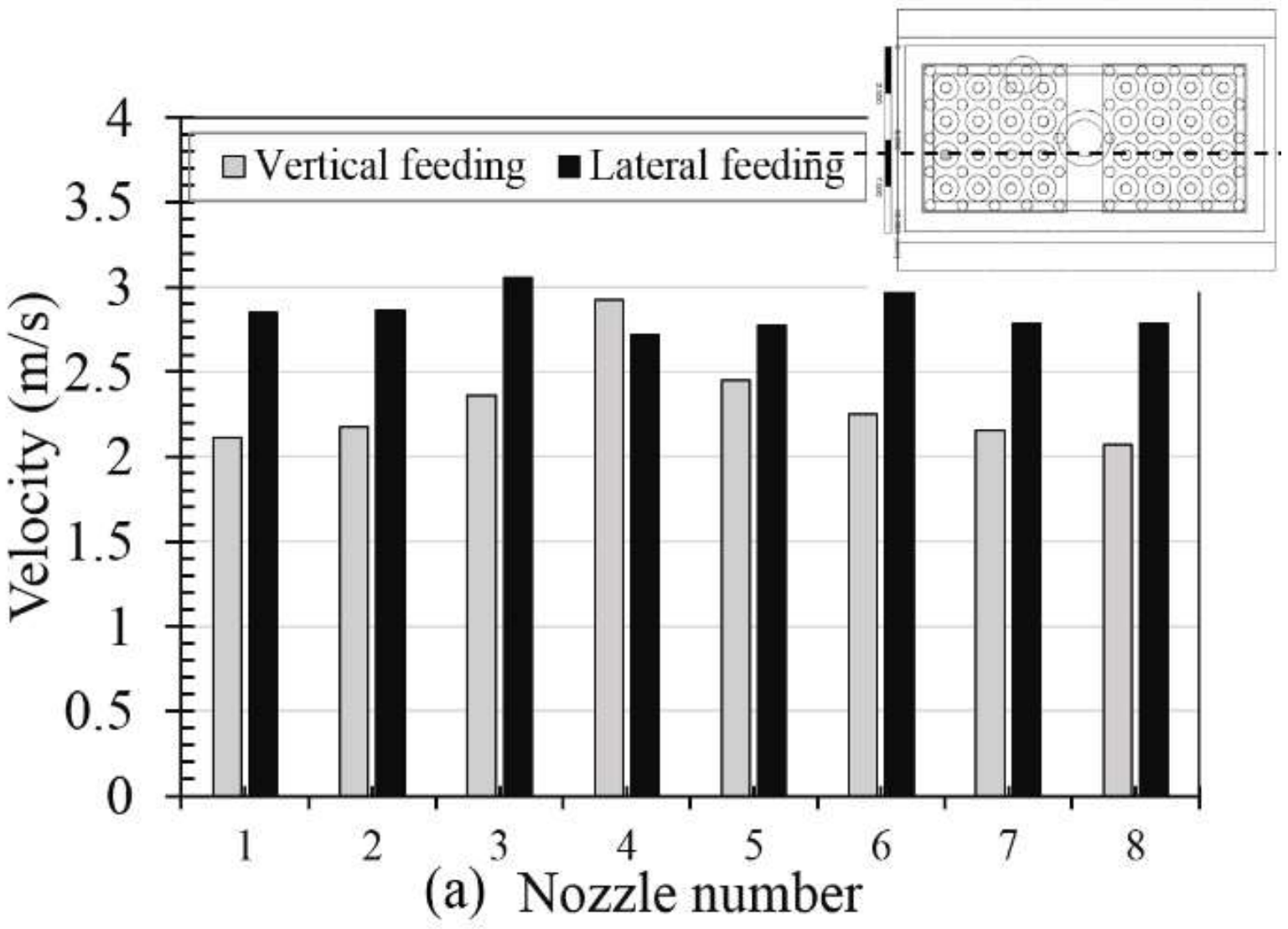

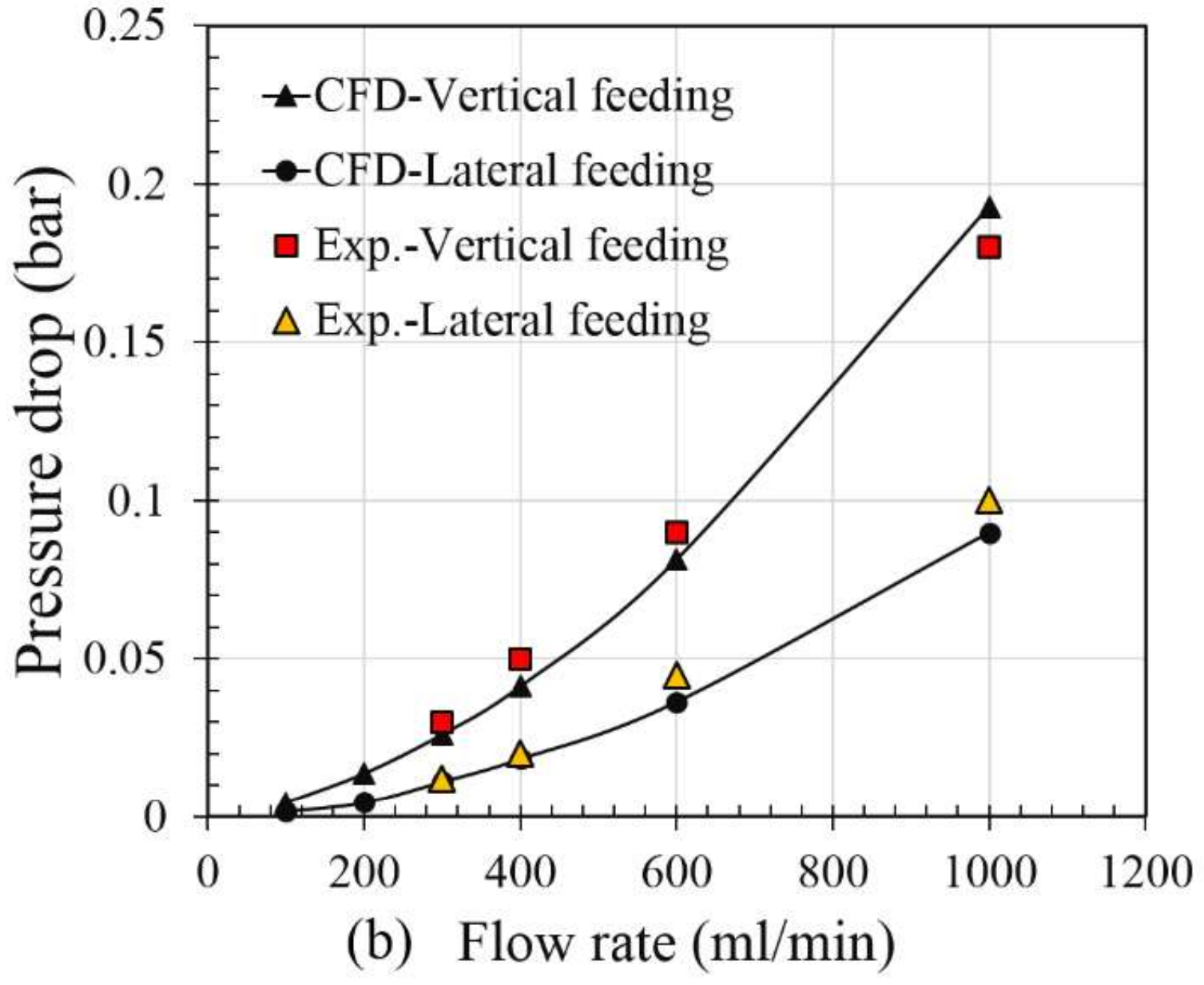

**Figure 8.22**: CFD modeling results with (a) inlet velocity profile and (b) global level pressure drop comparison for the vertical and lateral feeding design.

Figure 8.23 shows the detailed chip temperature distribution map comparison between the vertical and lateral design. For both coolers, the inlet liquid temperature is set as 10 ˚C, and the chip power for the active die is set as 50W. The simulated average thermal resistance is 0.28 $cm^2$-K/W for vertical feeding design, while the lateral feeding shows 0.26 $cm^2$-K/W. It can be seen that the average chip temperature for the two cases is very similar to each other, showing a 7.1% difference, which corresponds well with the experimental results of the previous section. Figure 8.24(a) shows the comparison between the CFD modeling and experiments under different flow rates. It can be seen that the CFD modeling shows good agreement with the experimental data, especially at a higher flow rate 1000mL/min. Moreover, the thermal characteristics of the vertical feeding design and lateral design show similar behaviors. This is due to the same nozzle array design and the same power and velocity boundary condition from the system point of view. The Nusselt number $\overline{\mathrm{Nu}}_\mathrm{j}$ and the $\mathrm{Re_d}$ were calculated based on the jet diameter shown in Figure 8.24(b). The extracted $\overline{\mathrm{Nu}}_\mathrm{j}$- $\mathrm{Re_d}$ correlations are:

$$\text{Vertical feeding design: } \overline{\mathrm{Nu}}_\mathrm{j} = 0.49 Re_d^{0.65} \quad (12)$$

$$\text{Lateral feeding design: } \overline{\mathrm{Nu}}_\mathrm{j} = 0.49 Re_d^{0.64} \quad (13)$$

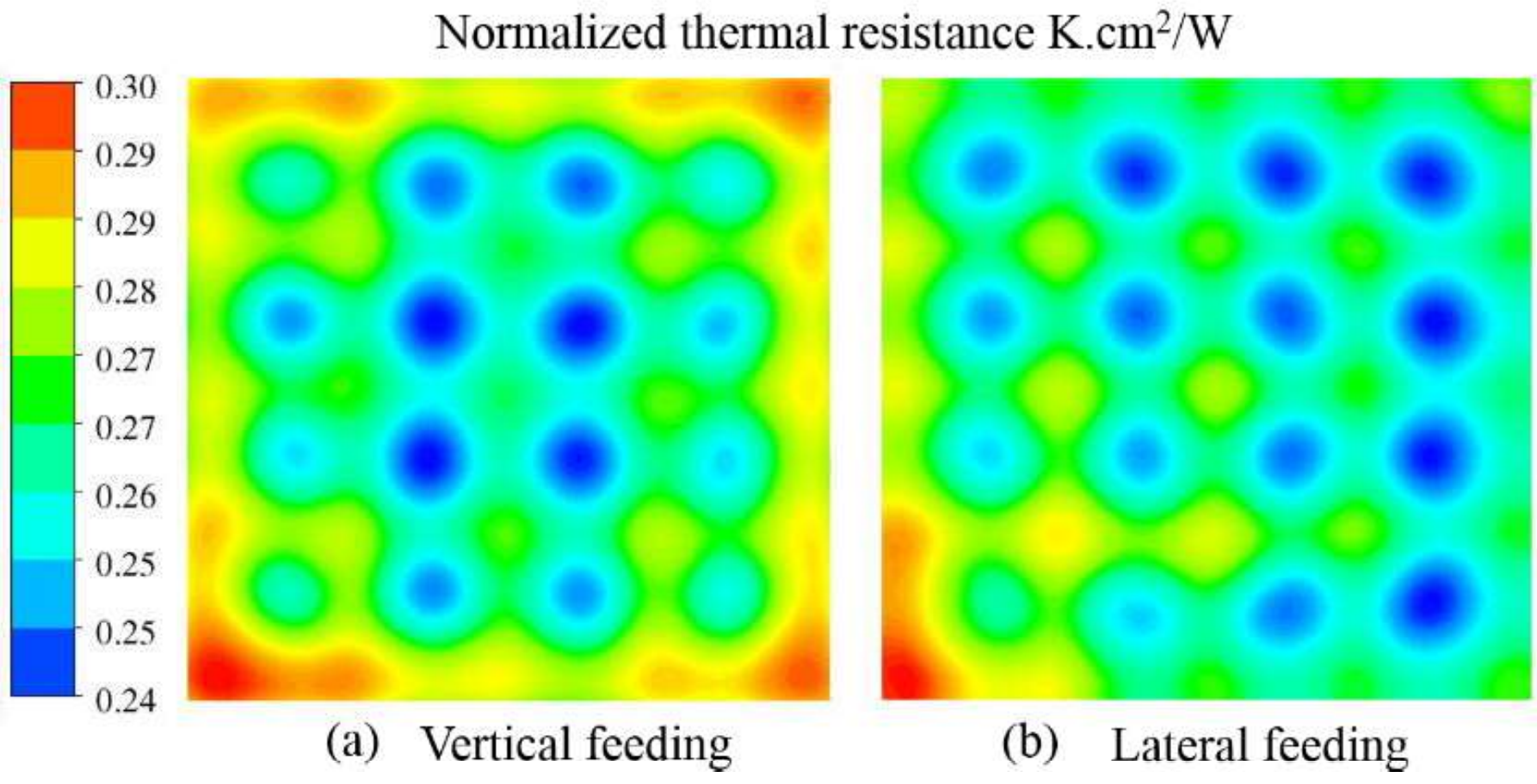

**Figure 8.23**: CFD modeling results with chip temperature distribution comparison between the (a) vertical feeding and (b) lateral feeding design: $R_{vertical}$ = 0.28 cm$^2$-K/W, $R_{lateral}$=0.26 cm$^2$-K/W. (bare die cooling with a flow rate of 1000 mL/min)

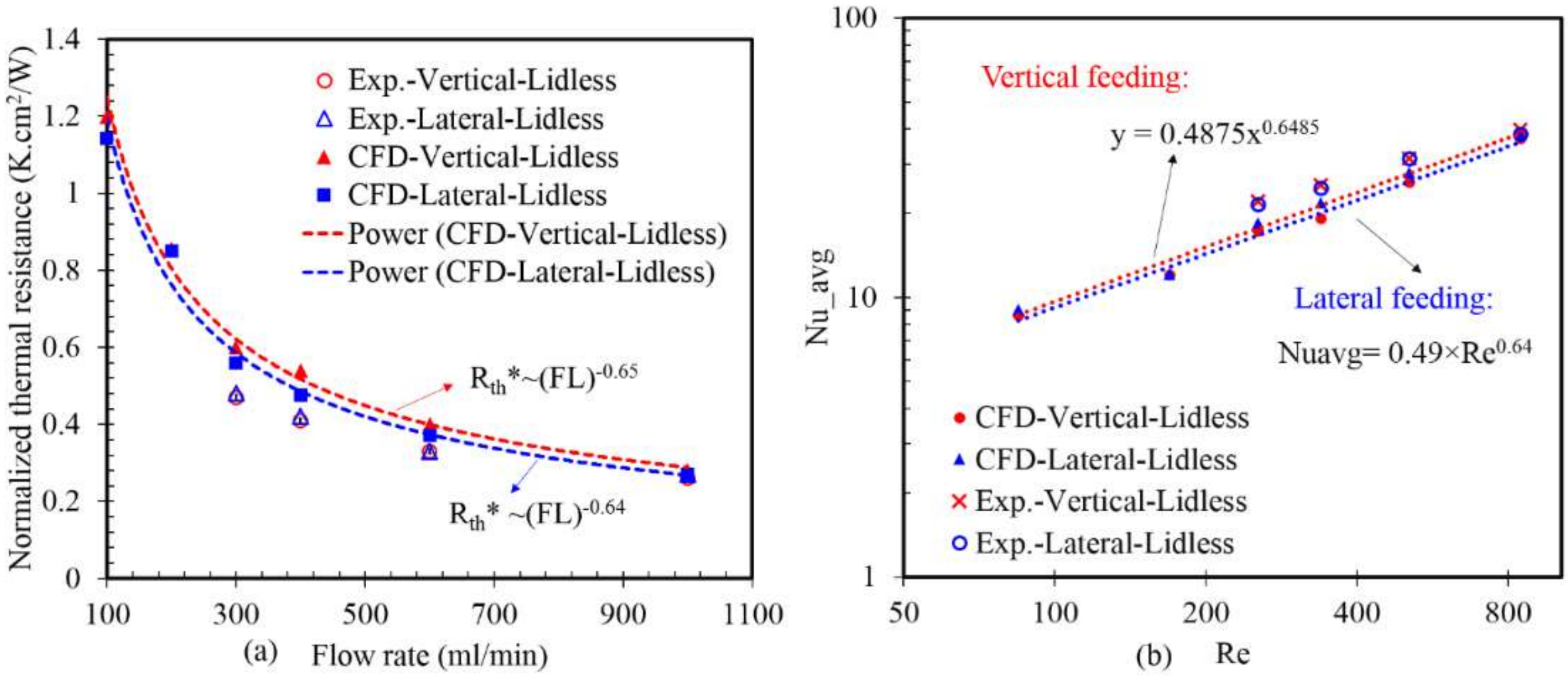

**Figure 8.24**: Comparison of the CFD modeling results and experimental measurements between the vertical feeding and lateral feeding design: (a) Normalized thermal resistance as a function of flow rate; (b) Nusselt number as a function of Reynolds number.

The hydraulic and thermal CFD simulations for the comparison between the cooler designs on the bare die dual-chip package are summarized in Figure 8.25. The chart shows the achieved normalized average thermal resistance as a function of the required pressure drop over the cooler, for flow rates ranging from 100 mL/min to 1000 mL/min. The achieved averaged thermal resistance is similar for both cooler designs at the same flow rate, however, the lateral feeding cooler design requires 50-60% less pressure drop and consequently pump power to realize. At a flow rate of 500 mL/min per chip (1000 mL/min for the cooler), the normalized thermal resistance is 0.26 cm$^2$-K/W. This means that the device temperature increase with respect to the inlet temperature would be 78°C for a heat flux of 300 W/cm$^2$ at a required pressure drop of 0.09 bar. Moreover, the overall thickness of this lateral feeding cooler is 2 times thinner compared to the vertical feeding cooler.

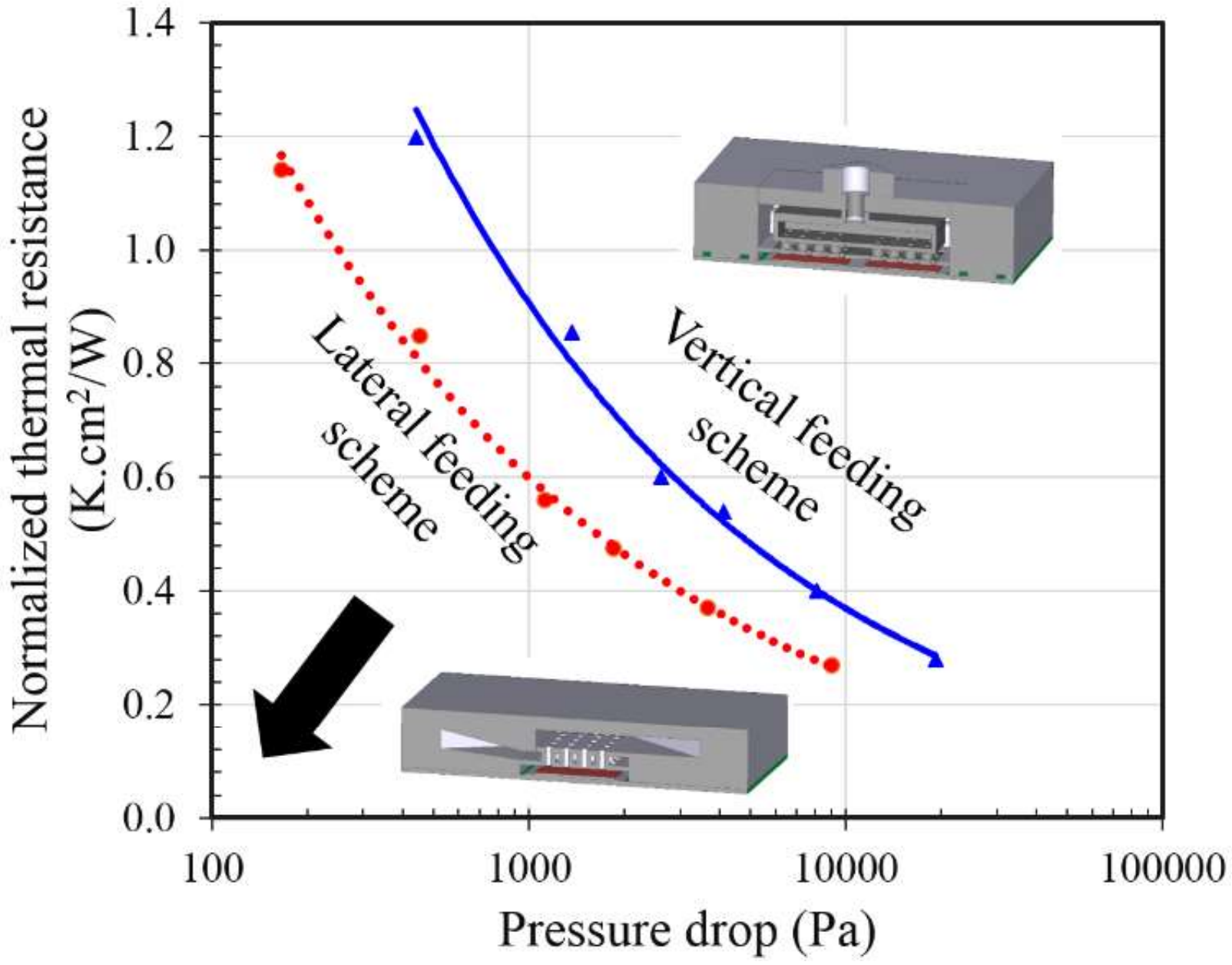

**Figure 8.25**: CFD modeling comparison of the thermal and hydraulic performance for vertical and lateral feeding configurations (lidless cooling).

## 8.6 Thermal interface material considerations

The measurement results above show that the presence of the lid and the TIM have a significant impact on the cooling performance of the 3D printed impingement cooler. The beneficial effect of the lid is the improved thermal spreading, which results in a decrease of the temperature peak and more uniform chip temperature. The detrimental effect of the lid is the additional vertical thermal resistance for the heat conduction through the TIM and the lid. A hybrid finite element modeling simulations FEM/CFD modeling study has been performed to assess this trade-off for the lid for different TIM and lid properties and for different flow rate conditions. For the hybrid CFD/FEM modeling method, a full conduction-convection model is firstly performed using conjugated heat transfer CFD modeling to simulate the heat transfer in the package and the convective heat transfer in the impinging coolant. In order to capture all the heat spreading paths in the structure, not only the lid and TIM, but also the details of the bottom part of the interposer package needs to be included in the model.

In the second step, the heat transfer coefficient distribution on top of the lid is extracted. This distribution is used as a boundary condition input for a conduction model of the complete interposer package using the FEM model in order to perform the DOE for the assessment of the impact of the lid. While changing the properties of the TIM and lid, the assumption is made that the flow distribution and resulted heat transfer coefficient distribution are not affected. This simplification allows us to focus on the conduction heat transfer in the interposer package and lid using the much faster conduction models. Figure 8.26 shows the grid containing 400,000 elements for the finite element modeling study including the lid, the PCB, the solder balls, the package laminate, the interposer, logic and memory chip, the interconnections between the chips and the package, such as BEOL, micro-bump layer, Cu pillars and underfill. The uniform power dissipation is applied as constant heat flux in the "logic" die while there is no power in the memory die. The ambient temperature is considered to be at 25ºC. An equivalent convective heat transfer coefficient of 25 W/m$^2$-K is applied at the bottom of the package to represent the heat transfer from the package towards the PCB.

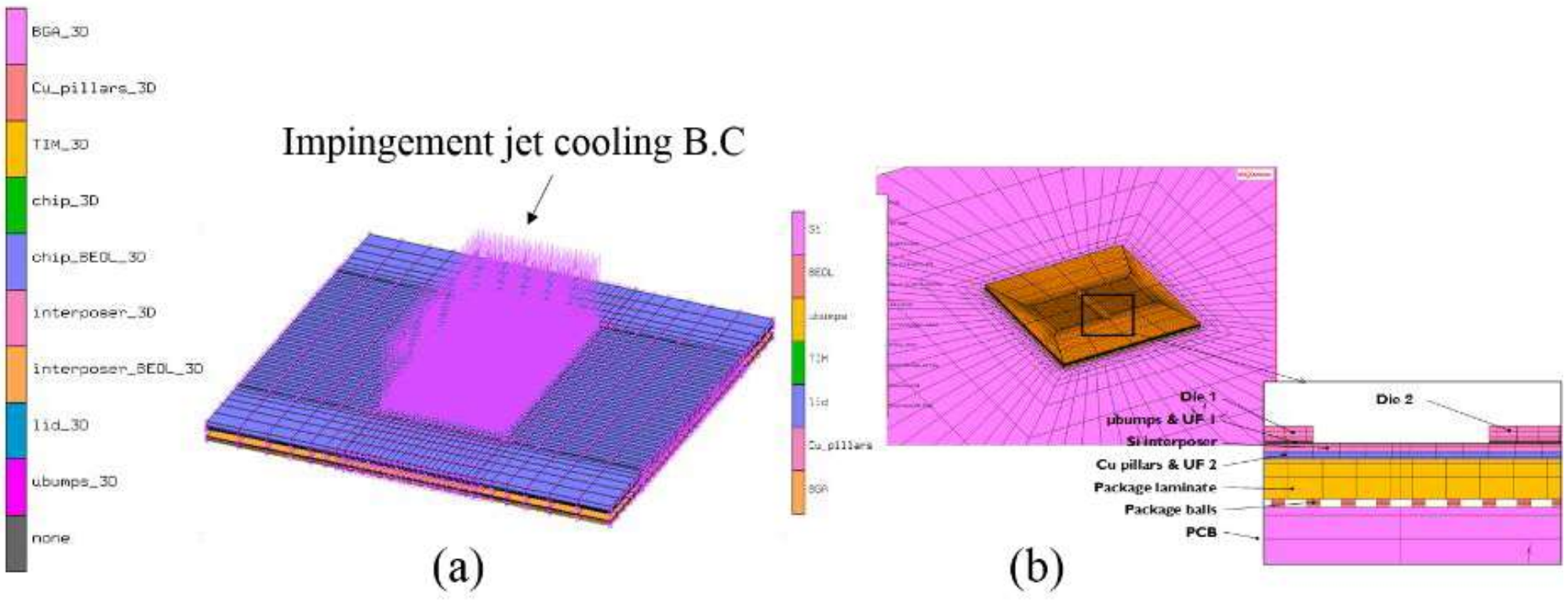

**Figure 8.26**: FEM package model: (a) Extracted heat transfer coefficient map applied on the lid surface in the FEM model; (b) Details of the package elements in the FEM model.

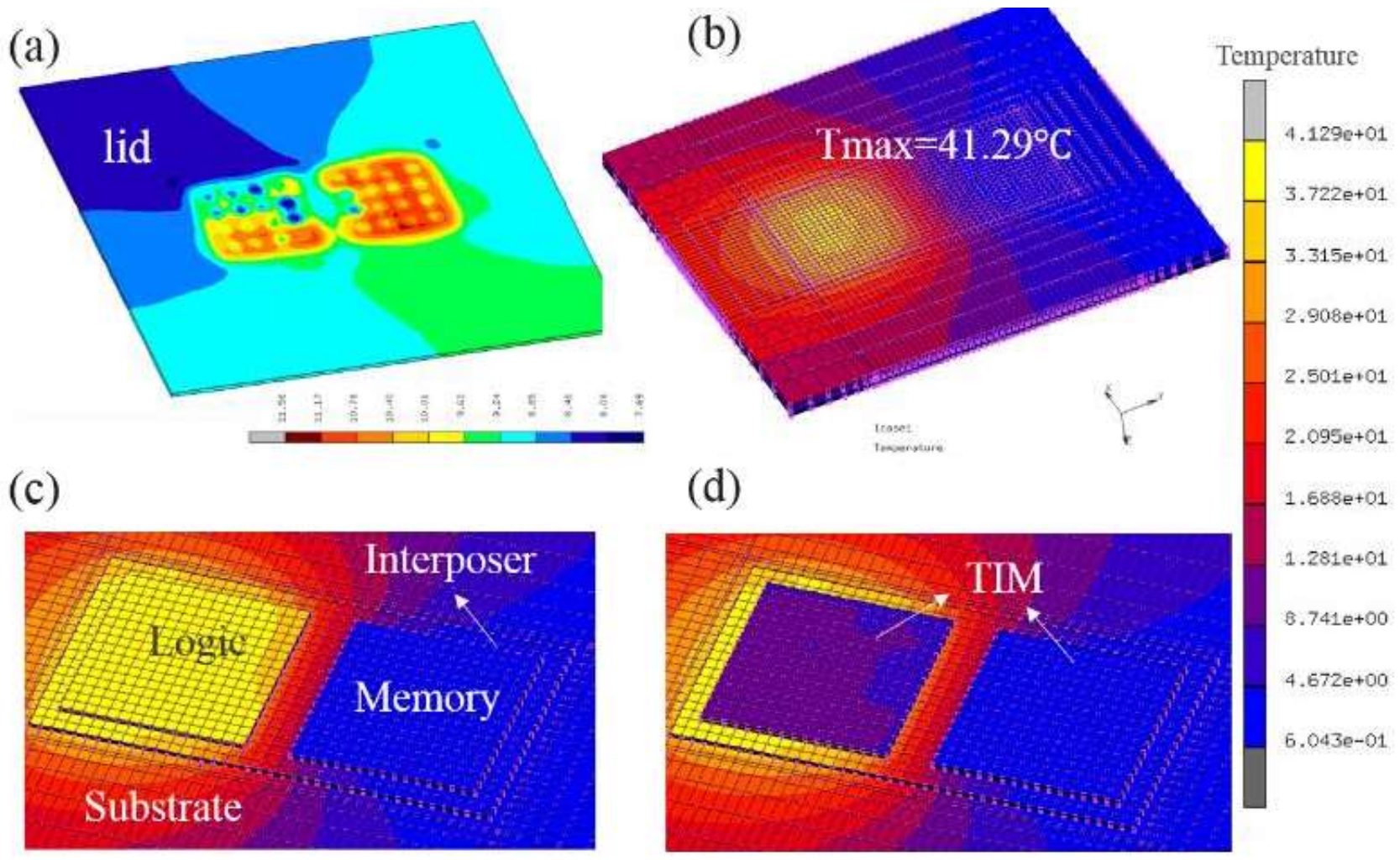

**Figure 8.27**: Modeling results for the hybrid CFD/FEM modeling: (a) Temperature distribution on the lid surface with CFD modeling; (b)(c) and (d) Temperature

distribution on the interposer package with the "logic" and "Memory" die in the FEM model.

To illustrate the hybrid CFD/FEM approach, Figure 8.27(a) shows the heat transfer coefficient extraction results from the full cooler level CFD model. The extracted heat transfer coefficient map is applied on the corresponding lid surface in the FEM model as a convective boundary condition shown in Figure 8.26(a). The temperature map on the lid (left) and die and TIM surface (right) are illustrated in Figure 8.27(c) and (d).

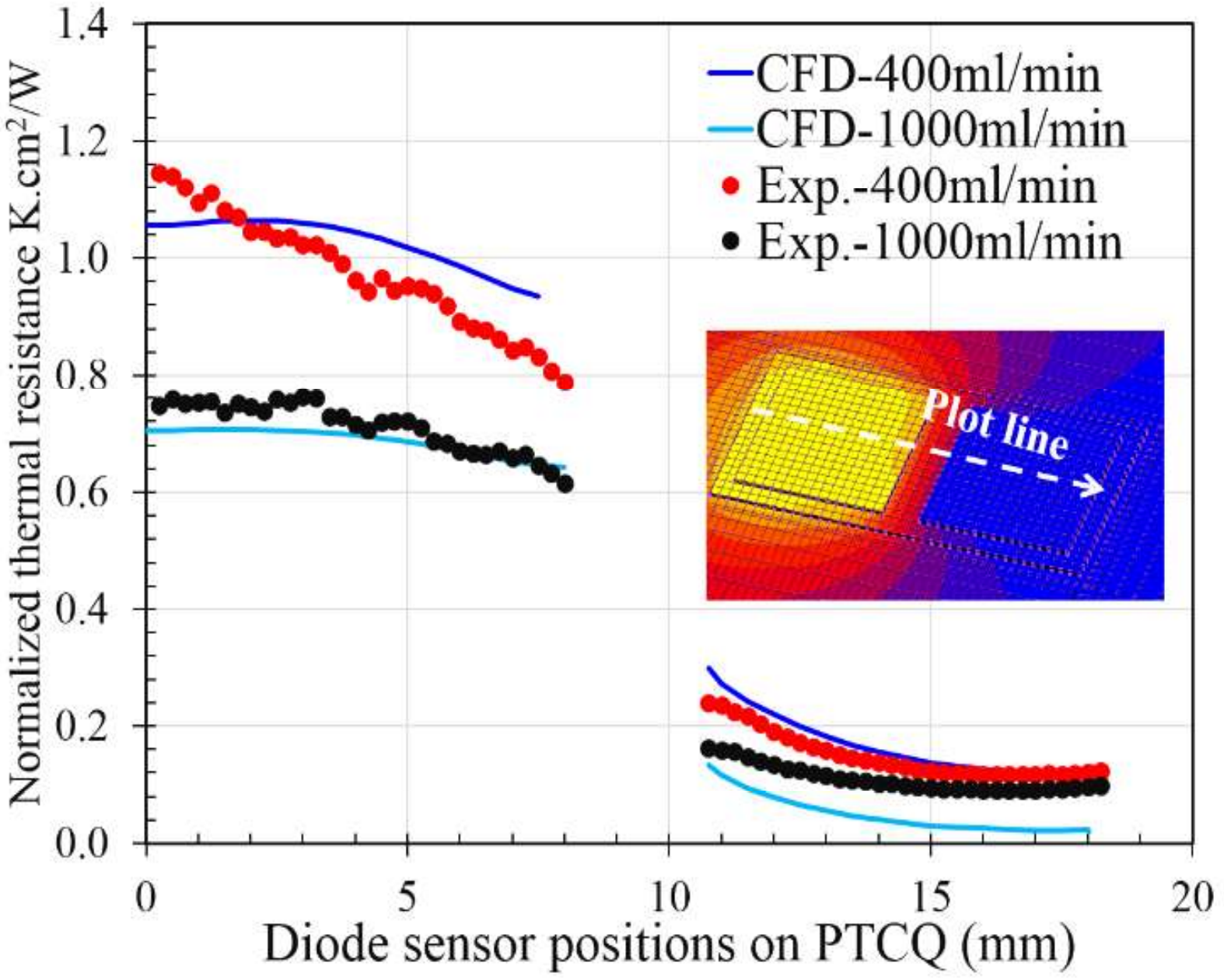

**Figure 8.28**: Modeling validation between the hybrid CFD/FEM modeling approach and the experimental results for lidded packages (logic power=50 W; memory power=0 W).

Figure 8.28 shows the FE modeling results for the lidded package using the extracted heat transfer coefficient from the CFD models at different flow rates as a boundary condition. The comparison with the experimental PTCQ measurements shows a good agreement for both the active heat chip as well as the passive chip. The relative difference of the normalized thermal resistance (defined as the maximum chip temperature difference w.r.t the ambient temperature) between the hybrid model and the experiments are 7.8% and 4.8% for the flow rates of 400 mL/min and 1000 mL/min respectively. Therefore, the CFD model and FE models for the lidded package cooler are successfully validated and can be used for the extrapolation to assess the impact of the lid.

The thermal FE model has been used to assess the impact of the lid and TIM properties for the lidded package and to benchmark the results with lidless package for different flow rates. The TIM used for the demonstrator is a standard silicone-based TIM, while several high-performance TIMs with much lower thermal resistance have been developed [45]. A design of experiments has been performed for the thermal conductivity and thickness of the TIM and lid layer. The parameters ranges used in the DOE are listed in Table 8.5. The total DOE includes 625 simulations for each flow rate.

**Table 8.5**: Simulation DOE properties for the impact of the lid and TIM.

| Parameter | Minimal value | Maximum value |
|---|---|---|
| Lid thickness | 0.05 mm | 1 mm |
| Lid conductivity | 20 W/m-K | 600 W/m-K |
| TIM thickness | 0.02 mm | 400 mm |
| TIM conductivity | 1 W/m-K | 20 W/m-K |

The thermal interface material creates a vertical thermal resistance for heat removal. This thermal resistance scales linearly with the TIM thickness and inverse proportional with the TIM thermal conductivity. The lid, on the other hand, shows a typical thermal spreading behavior: a thicker lid will result in more later spreading, and thus lower temperature values, but at the same time, the vertical thermal conduction resistance increases. Moreover, in the case of the lidded package, the cooling is applied to a larger area compared to the lidless package. This trade-off is now illustrated for a high coolant flow rate of 1000 mL/min.

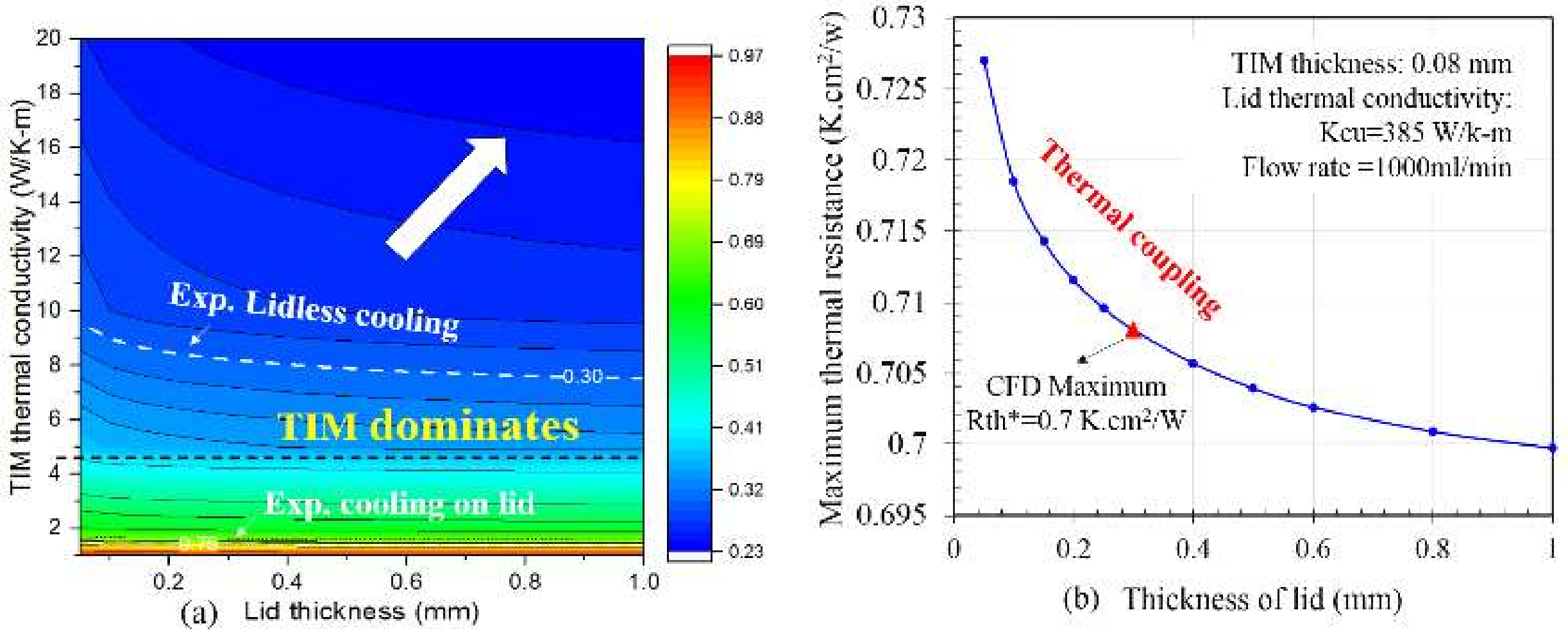

**Figure 8.29**: Tradeoff between the lid thickness and TIM thermal conductivity at a flow rate of 1000 mL/min ($K_{lid}$=385 W/m-K; TIM thickness=80 µm).

Figure 8.29 shows the analysis for the flow rate of 1000 mL/min for a Cu lid and a TIM thickness of 80 µm. Figure 8.29(a) shows the normalized maximum logic temperature as a function of the TIM thermal conductivity and the lid thickness. It can be seen that the impact of the lid thickness is almost negligible for TIM thermal conductivity values smaller than 4 W/m-K. As the thermal conductivity of the TIM increases, the impact of the lid thickness becomes visible. This behavior is illustrated in Figure 8.29(b) for a TIM conductivity of 1.5 W/m-K, where a sharp temperature increase can be observed for lid thickness values below 250 μm. However, due to the high heat removal rate of the impingement cooling on top of the lid, the impact of the lid thickness remains small. The isoline for the value of the lidless cooler with maximum thermal performance (0.30 $cm^2$-K/W) is added in the chart to benchmark the lidless and lidded packages. The measured demonstrator is added as a marker. The comparison shows that a maximum TIM conductivity is 10 W/m-K for an 80 μm thickness (thermal resistance: 8 $mm^2$-K/W) is required for the lidded package cooling to match the performance of the lidless cooler.

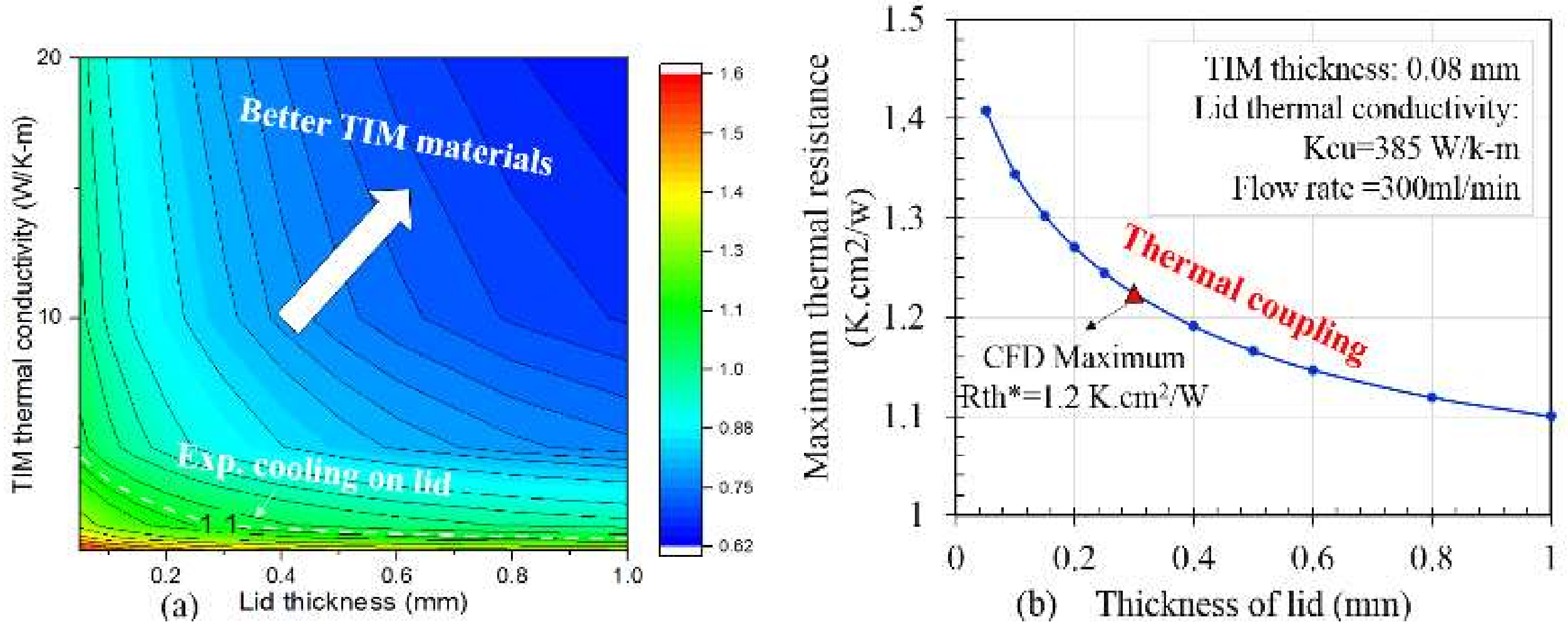

**Figure 8.30**: Tradeoff between the lid thickness and TIM thermal conductivity at a flow rate of 300 mL/min (Lid thickness=300 μm; TIM thickness=50 μm)

In Figure 8.30, the analysis is shown for a flow rate of 300 mL/min, for a Cu lid and a TIM thickness of 80 μm. For this lower flow rate, the spreading effect of the lid is more visible. For TIM conductivity values below 4 W/m-K, the thermal performance remains dominated by the TIM. However, for higher TIM conductivity values, the thermal performance is limited by the reduced thermal spreading in the lid for very thin lid values below 250 μm. Again, the situation of the demonstrator is added as a marker in the chart. For this flow rate, however, the performance of the lidless package (0.56 $cm^2$-K/W) cannot be reached by the lidded package, even for very low TIM thermal resistance values, due to the dominating effect of the thermal spreading in the lid.

The impact of thermal conductivity TIM on the chip temperature profiles in the interposer package is shown in Figure 8.31 for a TIM thickness of 80 μm, a lid thickness of 300 μm and a flow rate of 1000 mL/min. The measured profiles for the lidless cooler are added as a reference. This figure shows the temperature profiles for each data point in the chart of Figure 8.29(a). For higher TIM thermal conductivity values, lower logic temperatures are observed. However, increased relative thermal coupling is observed for higher TIM thermal conductivity values. This chart shows that, for a TIM with sufficiently high thermal conductivity, the lidded package cooling can achieve the same cooling performance as the lidless package cooling at this high flow rate.

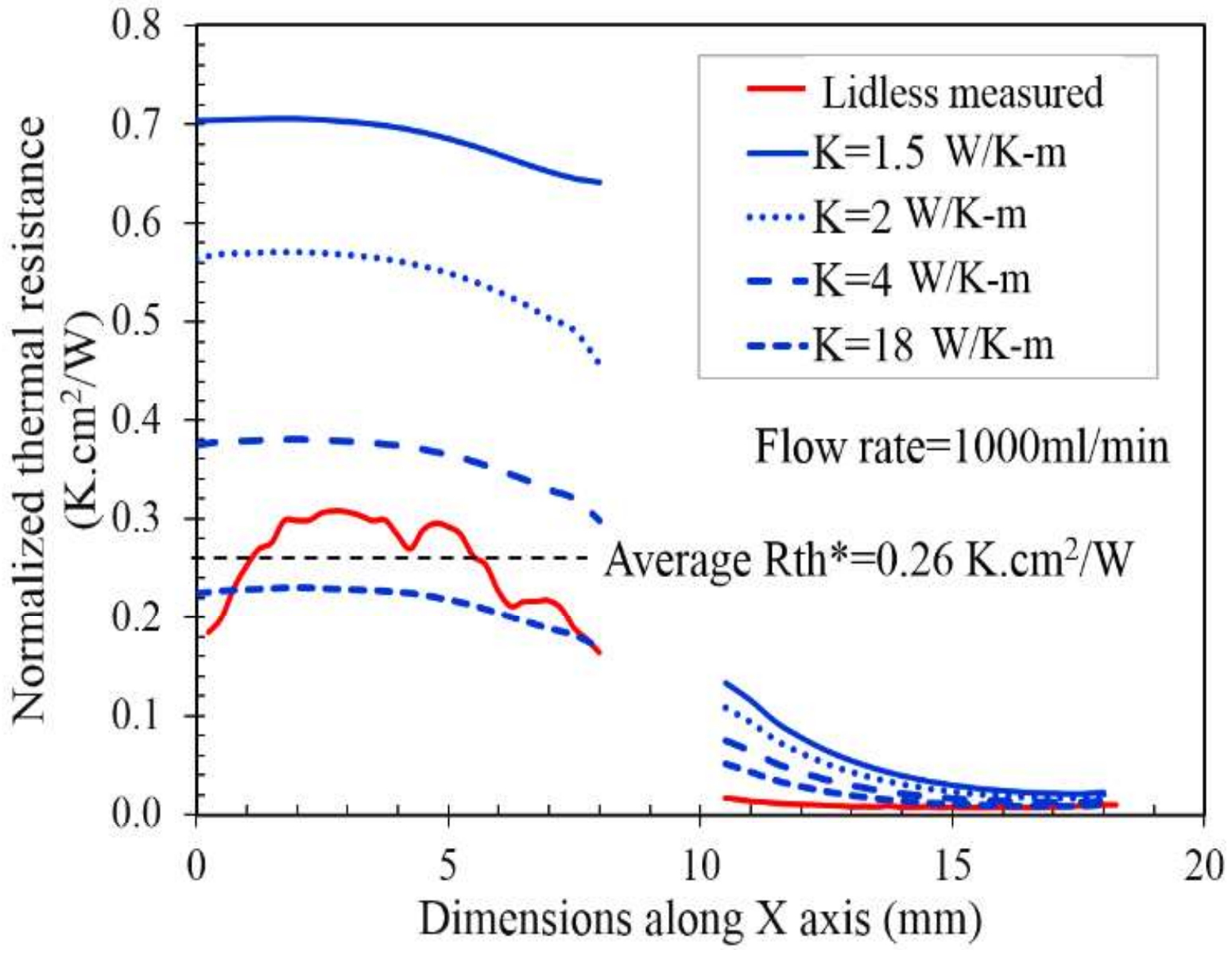

**Figure 8.31**: Impact of TIM thermal conductivity on the thermal resistance of the impingement cooler on the lidded package for a flow rate of 1000 mL/min and the benchmarking with the lidless cooling (red curve).

## 8.7 Conclusion

In this chapter, we demonstrate for the first time the design, modeling, fabrication and experimental thermal, and hydraulic characterization of package-level 3D printed direct liquid micro-jet array impingement cooling applied to the dual-chip module used in power electronics. A scalable design methodology for the chip area is proposed and experimentally validated by comparison of the thermal performance of the dual-chip package cooler with earlier single chip cooler data. The cooler has been designed for a dual-chip package that contains two advanced thermal test chips, taking advantage of the capabilities of additive manufacturing to create complex internal structures and to fabricate the cooler as a single part. The coolers, fabricated using high-resolution stereolithography with the water-resistant, have been assembled on the bare die and lidded versions of the test vehicle.

For the bare die package, a very low thermal resistance of 0.26 $cm^2$-K/W is measured at a cooler flow rate of 1000 mL/min for two heated chips. The presence of the lid (and mainly the TIM) results in a higher chip temperature, where the relative impact of the lid increases as the flow rate increases. Furthermore, it is demonstrated that the bare die jet impingement cooling on the dual-chip package can realize a very low thermal coupling between the chip of only 4%, which is 9 times lower than typically reported values for multi-chip modules.

Furthermore, in this chapter, we introduced a novel lateral coolant feeding design for the dual-chip package cooler, which enables a reduction of the cooler height by a factor of two. The experimental comparison shows that the lateral feeding cooler achieves a very similar thermal performance as the standard vertical feeding cooler. The modeling comparison shows that the lateral feeding design can achieve a more uniform flow distribution over the nozzles in the cooler. Furthermore, this lateral feeding design can realize 50-60% reduction of the pressure and required pumping power for the same thermal performance and at the same time, achieve a reduction of the cooler thickness by a factor of 2 compared to the reference vertical feeding design. An optimized 3D printed fluid delivery manifold design with lateral feeding structure has a thermal resistance from junction to coolant inlet temperature of 0.26 $cm^2$-K/W, which can cool down the heat flux up to 300 $W/cm^2$ for a 78 °C temperature increase for a flow rate of 500 mL/min per chip and a pressure drop of 0.09 bar.

Moreover, an extensive DOE has been performed to assess the trade-off of the lid for different TIMs and flow rate conditions. The parameter sensitivity studies show that with a sufficiently low thermal resistance of the TIM (below 10 $mm^2$-K/W), the lidded package cooling can achieve the same cooling performance as the lidless package cooling at this high flow rate.

The next step is to optimize the inlet distributor to have better flow uniformity. One best option is using the topology optimization [44] design to tailor the flow for every chip module. Moreover, we can also design the intermediate layer to split the flow more uniformly. Other important aspects for future work are to address the potential reliability concerns.

The results of this chapter are partially published in the following publication:

**Tiwei Wei**, Herman Oprins, et al., "Experimental and numerical investigation of direct liquid jet impinging cooling using 3D printed manifolds on lidded and lidless packages for 2.5D integrated systems [J]", Applied Thermal Engineering, Volume 164, 5 January 2020, 11453. (IF= 4.026)

**Tiwei Wei,** Herman Oprins, et al., "Thermal analysis of polymer 3D printed jet impingement coolers for high performance 2.5D Si interposer packages [C]", IEEE-ITherm 2019.

## References

[1] G. Van der Plas et al., "Design Issues and Considerations for Low-Cost 3-D TSV IC Technology", in IEEE Journal of Solid-State Circuits, vol. 46, no. 1, Jan. 2011, pp. 293-307.

[2] D. Stow, I. Akgun, R. Barnes, P. Gu and Y. Xie, "Cost and Thermal Analysis of High-Performance 2.5D and 3D Integrated Circuit Design Space", 2016 IEEE Computer Society Annual Symposium on VLSI (ISVLSI), Pittsburgh, PA, 2016, pp. 637-642.

[3] Villamor, A. , et al., "The Power Electronics Industry Is Showing Steady Growth and High Dynamism", Yole Développement, Lyon, France.

[4] A. S. Bahman, K. Ma and F. Blaabjerg, "A Lumped Thermal Model Including Thermal Coupling and Thermal Boundary Conditions for High-Power IGBT Modules", in IEEE Transactions on Power Electronics, vol. 33, no. 3, March 2018, pp. 2518-2530.

[5] K. Saban, Xilinx, "Stacked Silicon Interconnect Technology Delivers Breakthrough FPGA Capacity, Bandwidth, and Power Efficiency". Xilinx white paper, 2012.

[6] B. Black, "Die-stacking is happening: AMD Fury X GPU," in 3D ASIP, Dec 2015.

[7] H. Hung-Hsien, C. Tsai, J. Tsai, M. Shih, D. Tang and C. Hung, "From Package to System Thermal Characterization and Design of High Power 2.5-D IC", 2019 International Conference on Electronics Packaging (ICEP), Niigata, Japan, 2019, pp. 36-41.

[8] S. Shao et al., "Comprehensive Study on 2.5D Package Design for Board-Level Reliability in Thermal Cycling and Power Cycling", 2018 IEEE 68th Electronic Components and Technology Conference (ECTC), San Diego, CA, 2018, pp. 1668-1675.

[9] S. Narumanchi, M. Mihalic, K. Kelly, and G. Eesley, "Thermal interface materials for power electronics applications", 2008 11th IEEE Intersoc. Conf. Therm. Thermomechanical Phenom. Electron. Syst. I-THERM, 2008, pp. 395–404.

[10] A. Bar-Cohen, K. Matin, and S. Narumanchi, "Nanothermal Interface Materials: Technology Review and Recent Results", J. Electron. Packag., 2015, vol. 137, no. 4, pp. 040803.

[11] Gowda A, Esler D, Tonapi S, et al., “Micron and Submicron- Scale Characterization of Interfaces in Thermal Interface Material Systems”, ASME. J. Electron. Packag. 2006; 128(2): pp.130-136.

[12] S. Li, C. Hsu, C. Liu, M. Dai, et al., “Hot spot cooling in 3DIC package utilizing embedded thermoelectric cooler combined with silicon interposer”, 2011 6th International Microsystems, Packaging, Assembly and Circuits Technology Conference (IMPACT), Taipei, 2011, pp. 470-473.

[13] C. R. King, D. Sekar, M. S. Bakir, B. Dang, J. Pikarsky and J. D. Meindl, “3D stacking of chips with electrical and microfluidic I/O interconnects”, 2008 58th Electronic Components and Technology Conference, FL, 2008, pp. 1-7.

[14] T. Brunschwiler, B. Michel, H. Rothuizen, U. Kloter, B. Wunderle, H. Oppermann and H. Reichl, “Interlayer Cooling Potential in Vertically Integrated Packages”, Microsystem Technol. p.57, Vol.15(1), 2009.

[15] T. E. Sarvey et al., “Embedded cooling technologies for densely integrated electronic systems”, 2015 IEEE Custom Integrated Circuits Conference (CICC), San Jose, CA, 2015, pp. 1-8.

[16] N.Zuckerman, N.Lior, “Jet Impingement Heat Transfer: Physics, Correlations, and Numerical Modeling”, Advances in Heat Transfer, Volume 39, 2006, pp. 565-631.

[17] C.-Y. Li and S. V. Garimella, “Prandtl-Number Effects and Generalized Correlations for Confined and Submerged Jet Impingement”, International Journal of Heat and Mass Transfer, Vol. 44, No. 18, 2001, pp. 3471-3480.

[18] D. C. Wadsworth, I. Mudawar, “Cooling of a Multichip Electronic Module by means of Confined Two-Dimensional Jets of Dielectric Liquid”, Journal of Heat Transfer, NOVEMBER 1990, Vol. 112, pp. 891-898.

[19] Fabbri M, Jiang S, Dhir VK. “A Comparative Study of Cooling of High Power Density Electronics Using Sprays and Microjets”, ASME. J. Heat Transfer. 2005;127(1), pp. 38-48.

[20] G. M. Harpole and J. E. Eninger, “Micro-channel heat exchanger optimization”, 1991 Proceedings, Seventh IEEE Semiconductor Thermal Measurement and Management Symposium, Phoenix, AZ, USA, 1991, pp. 59-63.

[21] YH Kim, et al., “Forced Air Cooling by Using Manifold Microchannel Heat Sinks”, KSME International Journal, July 1998, Volume 12, Issue 4, pp. 709–718.

[22] J.H. Ryu, et al., “Three-dimensional numerical optimization of a manifold microchannel heat sink”, Int. J. Heat Mass Transf. 46 (9) (2003), pp. 1553–1562.

[23] L. Boteler, N. Jankowski, P. McCluskey, B. Morgan, “Numerical investigation and sensitivity analysis of manifold microchannel coolers”, Int. J. Heat Mass Transf. 55 (25–26) (2012), pp. 7698–7708.

[24] K. Olesen, R. Bredtmann, *et al.*, “‘ShowerPower’ New Cooling Concept for Automotive Applications”, in Proc. Automot. Power Electron., no. June 2006, pp. 1–9.

[25] J. Jorg, S. et al., “Direct single impinging jet cooling of amosfet power electronic module”, IEEE T POWER ELECTR, vol. 33 (5), 2018, pp. 4224–4237.

[26] T. Brunschwiler et al., “Direct liquid jet-impingement cooling with micronsized nozzle array and distributed return architecture”, in Proc. Therm. Thermomech. Phenom. Electron. Syst., 2006, pp. 196–203.

[27] G. Natarajan and R. J. Bezama, “Microjet cooler with distributed returns”, Heat Transf. Eng., vol. 28, no. 8–9, July 2010, pp. 779–787.

[28] T.-W. Wei, H. Oprins, et al., “High efficiency polymer based direct multi-jet impingement cooling solution for high power devices”, IEEE Transactions on Power Electronics. doi: 10.1109/TPEL.2018.2872904.

[29] Robinson, A.J. and Kempers, R. and Colenbrander, J. and Bushnell, N. and Chen, R., “A single phase hybrid micro heat sink using impinging micro-jet arrays and microchannels”, Applied Thermal Engineering, 136, 2018, pp.408-418.

[30] T.-W. Wei, H. Oprins, *et al.*, “Experimental Characterization of a Chip Level 3D Printed Microjet Liquid Impingement Cooler for High Performance Systems”, in IEEE Transactions on Components, Packaging and Manufacturing Technology. doi: 10.1109/TCPMT.2019.2905610

[31] H. Oprins, V Cherman, et al., “Experimental Characterization of the Vertical and Lateral Heat Transfer in Three-Dimensional Stacked Die Packages”, ASME. J. Electron. Packag. 2016;138(1):010902-010902-10.

[32] Tiwei Wei, Herman Oprins, et al., “Experimental characterization and model validation of liquid jet impingement cooling using a high spatial resolution and programmable thermal test chip”, Applied thermal engineering, Volume 152, April 2019, pp. 308-318.

[33] Sauciuc, Ioan, et al.,"Thermal Performance and Key Challenges for Future CPU Cooling Technologies." IPACK2005 , Advances in Electronic Packaging, San Francisco, California, USA. July 17–22, 2005. pp. 353-364.

[34] Ralph Remsburg, Joe Hager, "Direct Integration of IGBT Power Modules to Liquid Cooling Arrays", Electric Vehicle Symposium, 2007.

[35] Principles Elements of Power Electronics, barry W. Williams, "cooling of power switching semiconductor devices", 2006, pp 199-200.

[36] Wei T, Oprins H, Cherman V, Beyne E, Baelmans M. Conjugate Heat Transfer and Fluid Flow Modeling for Liquid Microjet Impingement Cooling with Alternating Feeding and Draining Channels. Fluids. 2019; 4(3):145.

[37] WaterShed® XC 11122 – Protolabs, https://www.protolabs.co.uk/media/1010047/somos-watershed-en.pdf.

[38] Mital, M., Pang, Y., Scott, E.P., "Evaluation of Thermal Resistance Matrix Method for an Embedded Power Electronic Module", IEEE Transactions on Components and Packaging Technologies, vol.31, no.2, June 2008, pp.382,387.

[39] Lall, B.S.; Guenin, B.N.; Molnar, R.J., "Methodology for thermal evaluation of multichip modules", IEEE Transactions on Components, Packaging, and Manufacturing Technology, Part A, , vol.18, no.4, Dec 1995, pp.758-764.

[40] Jain, A., Jones, R.E., Chatterjee, et al., "Analytical and Numerical Modeling of the Thermal Performance of Three-Dimensional Integrated Circuits", IEEE Transactions on Components and Packaging Technologies, vol.33, no.1 (2010), pp. 56-63.

[41] Je-Young Chang, et al., "Estimating the Thermal Interaction between Multiple Side-by-Side Chips on a Multi-Chip Package", Electronics Cooling, Volume 20, 2014.

[42] Ruikang Wu, et al., "Thermal modeling and comparative analysis of jet impingement liquid cooling for high power electronics", International Journal of Heat and Mass Transfer, Vol.137:42-51,2019.

[43] Songkran Wiriyasart, et al., "Liquid impingement cooling of cold plate heat sink with different fin configurations: High heat flux applications", International Journal of Heat and Mass Transfer, Vol.140:281-292,2019.

[44] Ercan M. Dede, Jaewook Lee, et al., "Multiphysics Simulation: Electromechanical System Applications and Optimization", 2014, DOI: 10.1007/978-1-4471-5640-6.

[45] A. Bar-Cohen, K. Matin, and S. Narumanchi, "Nanothermal Interface Materials: Technology Review and Recent Results," J. Electron. Packag., 2015, vol. 137, no. 4, p. 040803.

# Chapter 9

# 9. Large Die Cooling

## 9.1 Introduction

With the increasing demand on the functionality and higher computation performance for high performance chips, the die size is increasing very fast. The die size has increased from 12 mm$^2$ in 1971 to 688 mm$^2$ in 2019 for Intel microprocessors, and from 270 mm$^2$ in 1998 to 696 mm$^2$ in 2019 for IBM microprocessors [1]. Table 9.1 lists the die sizes for typical applications, such as CPU, GPU, and FPGA. For traditional microchannel cooling with coolant flow parallel to the chip surface, it is very challenging to maintain a small temperature gradient over the chip area for large die size applications. Previously, the 3D printed cooler is demonstrated for the single PTCQ test chip of 8 mm × 8 mm. In this chapter, the liquid jet impingement cooling with scalable nozzle array concept is applied to large die size applications, as indicated in Figure 9.1 for a die size large than 500 mm$^2$ and power dissipation higher than 250 W. Based on the normalization concept, which is validated in chapter 8, the normalized thermal resistance of the nozzle array cooling is area independent for a constant nozzle flow rate. Therefore, the cooling performance of the large die size cooler can be extrapolated from the characterized results of 3D printed cooler discussed in chapter 5.

**Table 9.1:** Typical die size for the high-performance applications.

| **Company** | **Products** | **Die size (mm$^2$)** | **Node** | **TDP (Watts)** |
|---|---|---|---|---|
| Nvidia [2] | Volta GPU (GV100) | 815 | 12 nm | 250 |
| Xilinx [3] | Virtex VU19P FPGA | 900 | 16 nm | -- |
| Intel [4] | Nervana Spring Crest NNP-T | 688 | 16 nm | 150-250W |
| AMD [5] | EPYC 7601 | 213 | 14 nm | 180W |
| Cerebras Systems [6] | AI | 46,225 | 16 nm | 20 kW |
| IBM [7] | z15 | 696 | 14 nm | -- |

This chapter will demonstrate the application of the multi-jet cooling concept for a realistic die size and chip power. Specifically, this chapter will present the design, fabrication, experimental characterization and reliability evaluation of a package level multi-jet cooler for large die sizes, fabricated using 3D printing. In general, there are two versions for the large die cooler design. The first cooler version, referred to as the reference cooler, is the scaled-up design of the 8×8 mm$^2$ chip cooler from Chapter 5 to much larger die size. The thermal and hydraulic performance of the reference large die cooler with and without lid is characterized and analyzed in section 9.3. The second version of the cooler, referred to as the improved cooler design, has an additional distribution layer to improve the coolant flow uniformity. In section 9.4, the thermo-hydraulic performance of the improved large die cooler is characterized and compared with the reference cooler. In the last section, a longer-term thermal measurement of 1000 hours for the reference large die cooler is performed and evaluated.

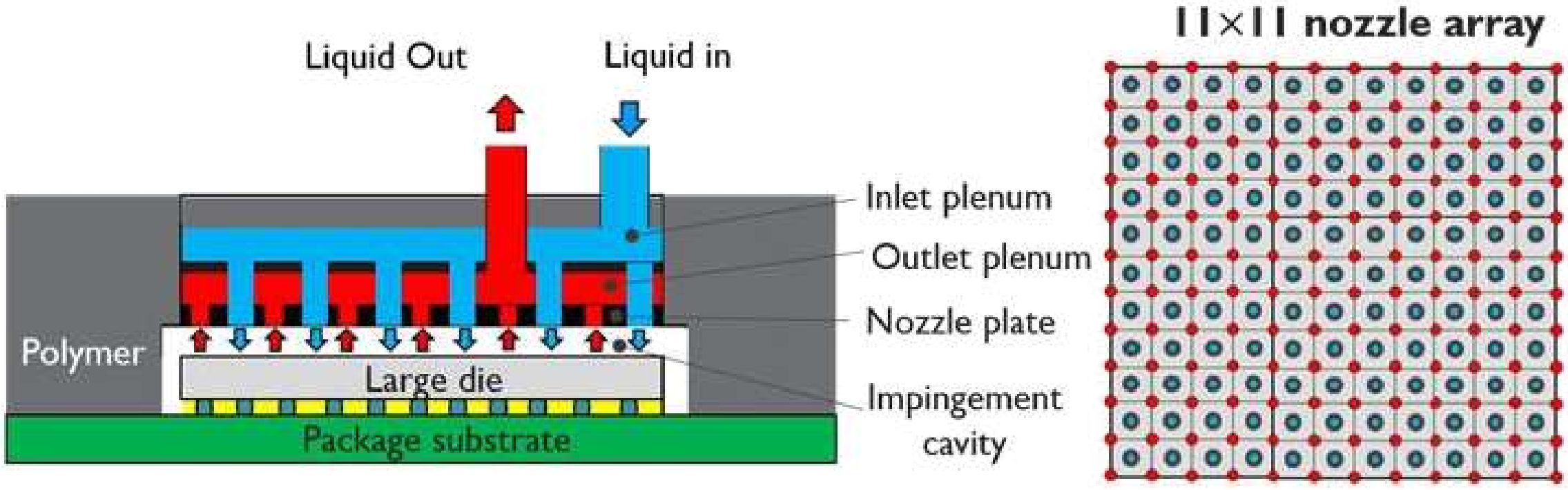

**Figure 9.1:** Large die cooler design: (a) schematic of the large die cooler and (b) top view of the nozzle plate with large nozzle array.

## 9.2 Large die cooler design and demonstration

### 9.2.1 Large die size thermal test vehicle

In order to characterize the thermal performance of the 3D printed large die cooler, a thermal test chip from Global Foundries with a size of 23×23 mm$^2$ is used that contains 16 metal meander resistor heaters and 25 temperature sensor resistors with a calibrated temperature coefficient of resistance (TCR) of 3553 ± 2 ppm/°C at a reference temperature of 25°C. Figure 9.2 shows the image of the thermal test die with the heater zones layout and the locations of the temperature sensors in the large die. The heater ID numbers are aligned with the temperature sensor ID numbers in Figure 9.2(c). The heaters generate a total power of 250-275W (depending temperature of the cooled chip since the heater resistance is temperature dependent) for an applied voltage of 50V. The test chip is flip-chip mounted on a package laminate substrate of 55×55 mm$^2$.

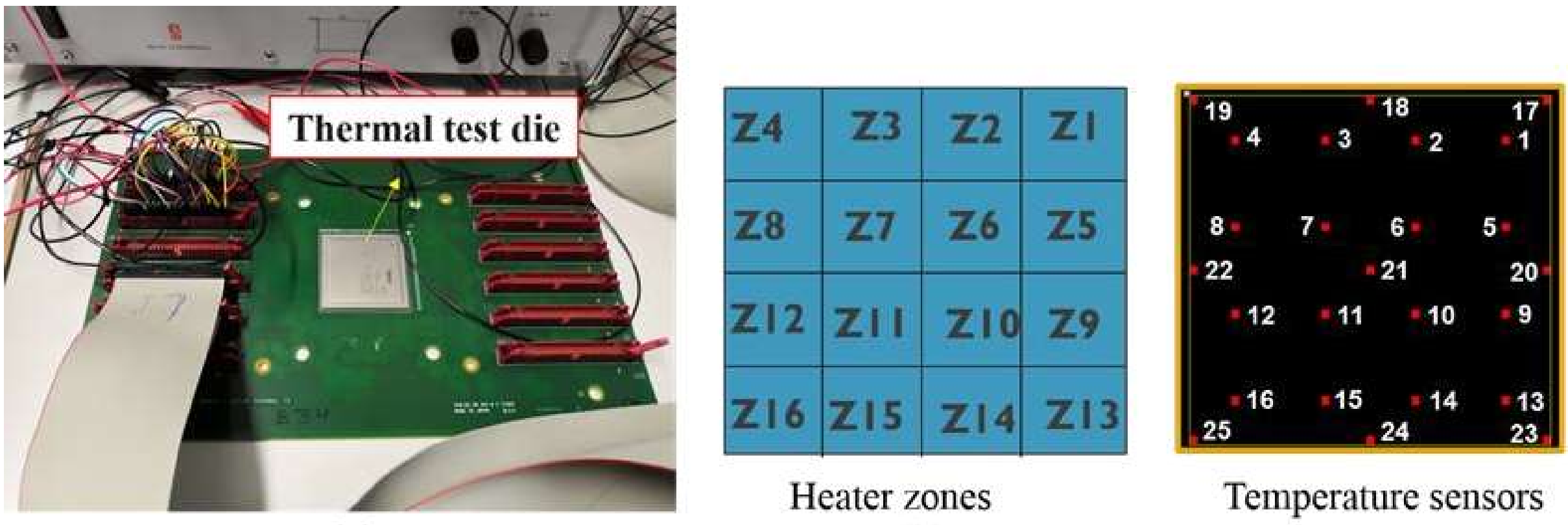

Heater zones    Temperature sensors

(a) (b) (c)

**Figure 9.2:** 23 mm×23 mm thermal test chip with 16 heater zones and 25 temperature sensors: (a) Lidded package with 55×55 mm$^2$ package substrate, is assembled to the test board; (b) Heater zones layout; (c) Temperature sensors layout.

The thermal performance of the large die cooler will be evaluated both on a lidded package (1 mm thick Cu lid) as well as on a bare die package with an exposed chip backside accessible for direct liquid cooling. Figure 9.3 shows the images of the lidded package and lidless package. For the lidded package in Figure 9.3(a), the thermal interface material (TIM) is used as the interface between the lid and lidless. The thickness of the TIM is around 20 μm (14 μm in the die center, thicker at the edges), with a thermal conductivity of 2.3 W/m-K. Figure 9.3(b) shows the bare die package after removing the TIM and lid. The cooling performance difference between bare die package and lidded package is expected to be less than for the interposer package (Section 8.2.1), due to the lower TIM thermal resistance (thinner TIM and higher thermal conductivity). The thermal performance comparison between the lidded package and lidless package will be discussed in section 9.3.3.

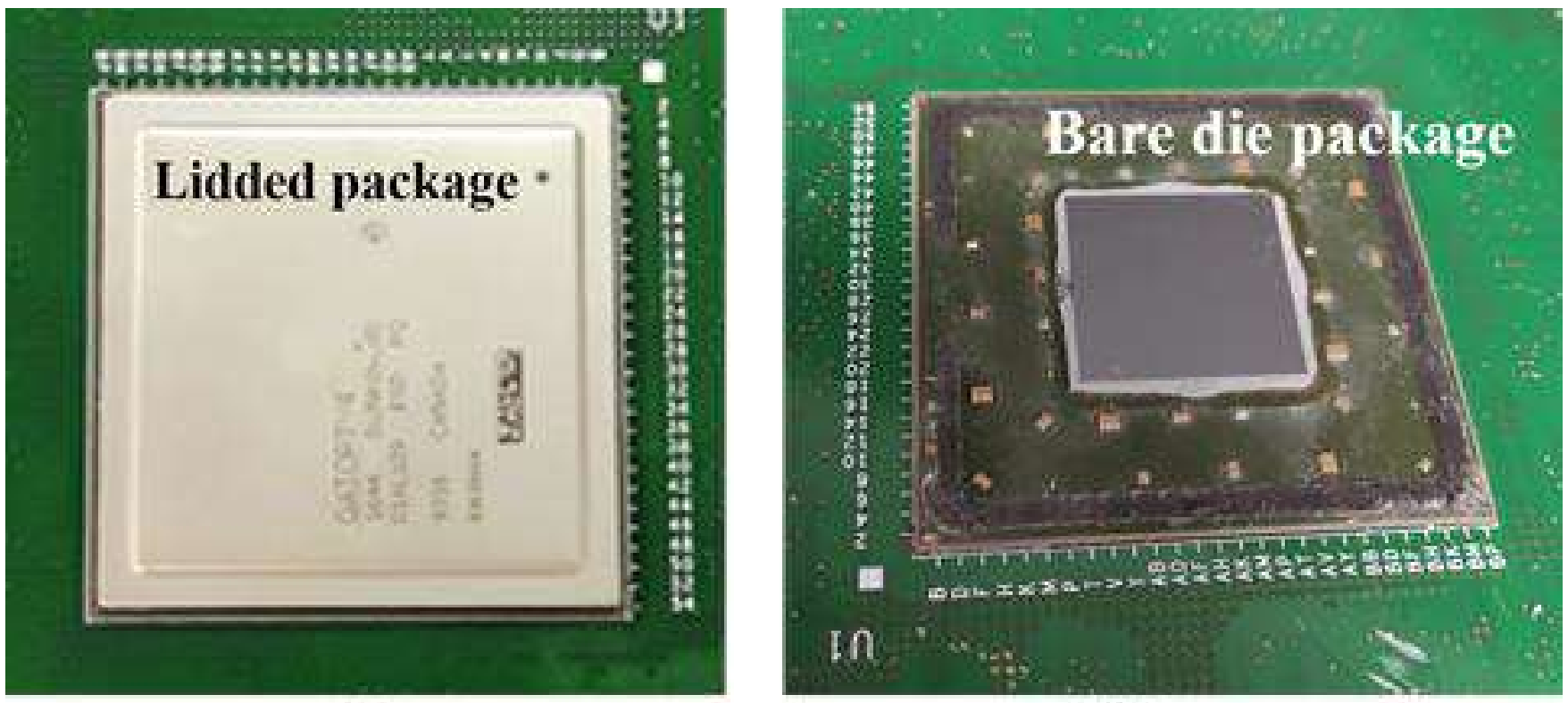

(a) (b)

**Figure 9.3:** Large die test vehicle: (a) Lidded package; (b) Large die package without the lid.

### 9.2.2 Large die cooler design and challenges

The main design considerations for the large die cooler are the inlet coolant flow uniformity and the possible die and package warpage of the large die assembly. Two large die cooler configurations matching the dimensions of the chip package are shown in Figure 9.4: design 1 is the reference cooler with a nozzle array below the coolant entrance connection (Figure 9.4(a)), and design 2 is an improved design with an additional flow distribution layer for improved flow uniformity (Figure 9.4(b)).

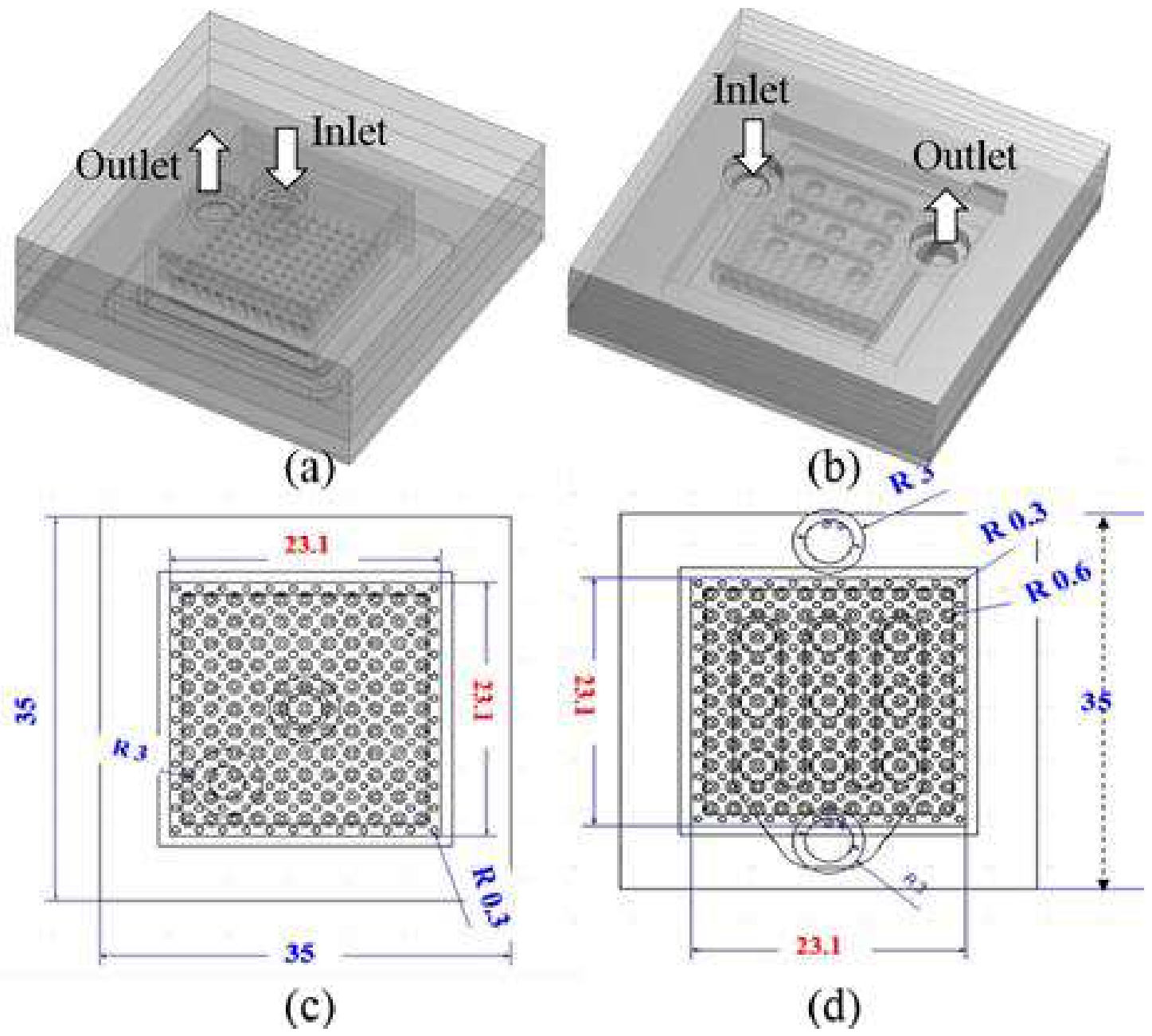

**Figure 9.4:** CAD structure for the large die cooler: (a) normal design with vertical feeding; (b) design 2 with additional distribution layer; (c) and (d) bottom view of two designed large die coolers.

For both the designs, the nozzle array is a scaled version of the 4×4 nozzle array cooler for the PTCQ test chip with a nozzle pitch of 2 mm and a nozzle diameter of 600 µm. The scaled nozzle plate contains an 11×11 inlet array and a 12×12 array of outlets distributed in between the inlets. The bottom view of the nozzle plate is shown in Figure 9.4. To limit the fabrication risks of this demonstrator, the more conservative design of the 4×4 cooler design has been chosen, rather than the 8×8 which is more challenging to fabricate. Figure 9.4(c) indicates the location of the coolant entrance connection from the top view, while the coolant exit connection is located at the left-bottom corner. For the top view of design 2 with additional distribution layer, the coolant entrance and exit connections are designed at two opposite side of the cooler, as shown in Figure 9.4(d). Moreover, there is an additional distribution layer with a 3×3 array of vertical feeding tubes, which are distributed uniformly on the top of the 11 ×11 microjet nozzle arrays. In order to visualize the internal flow delivery channels, the cross-section view for the two designed coolers are illustrated in Figure 9.5. The additional distribution layer indicated in Figure 9.5(b) is designed as a second flow feeding system to improve the flow uniformity, and therefore the temperature uniformity on the large die. The presence of this additional layer in the cooler geometry might result in an additional pressure drop over the cooler. This will be evaluated numerically and experimentally.

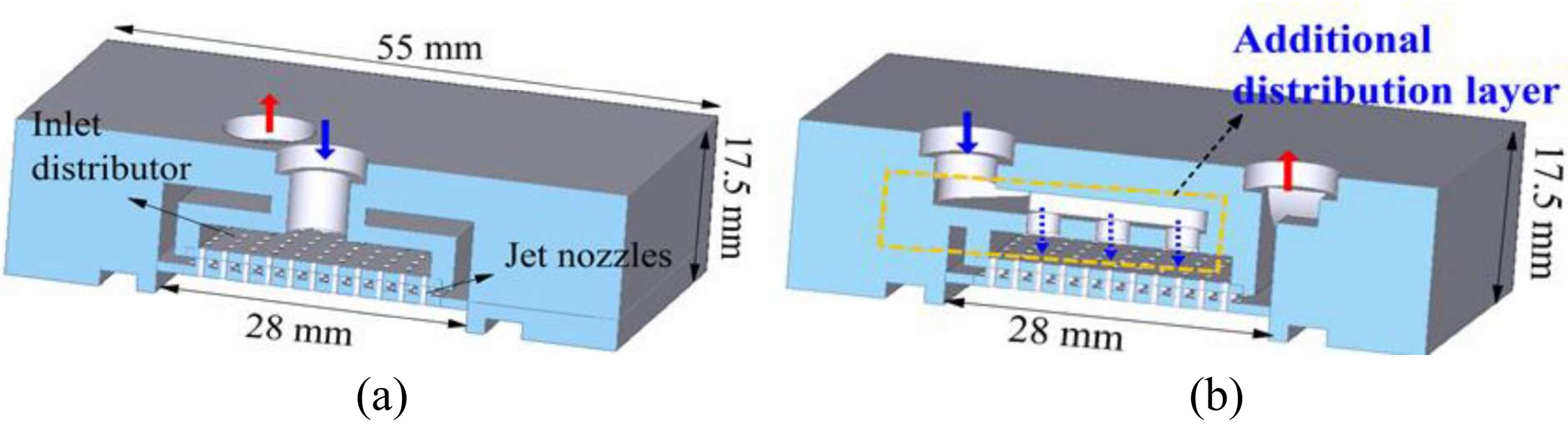

**Figure 9.5:** Cross-section view of two microfluidic cooler configurations: (a) reference design 1 with nozzle array below coolant entrance connection; (b) design 2 with additional distribution layer to improve flow uniformity over chip surface.

The cavity height, defined as the nozzle-to-chip surface distance is 600 µm. For both designs, the total cooler size is 55×55×17.5 mm$^3$. To accommodate the package warpage, O-rings are used, for which a dedicated groove is foreseen in the cooler design, as shown in Figure 9.6. The O-ring can be used as a buffer layer to compensate for the warpage during assembly. The geometry comparison between the single jet cooler (chapter 6), the interposer cooler (chapter 8) and the large die cooler is listed in Table 9.2. For all the three demonstrators, the cooling unit cell is all based on the same design: 2 ×2 mm$^2$ with inlet and outlet nozzle diameters of 0.6 mm. As discussed in chapter 8, the normalized $\boldsymbol{R_{th}}$ for the interposer cooler and single PTCQ die cooler are consistent with each other, which means that the intrinsic behavior of the interposer cooler and single die cooler is the same. Based on the validated normalization concept, the small die cooler design can be scaled to the large die cooler design.

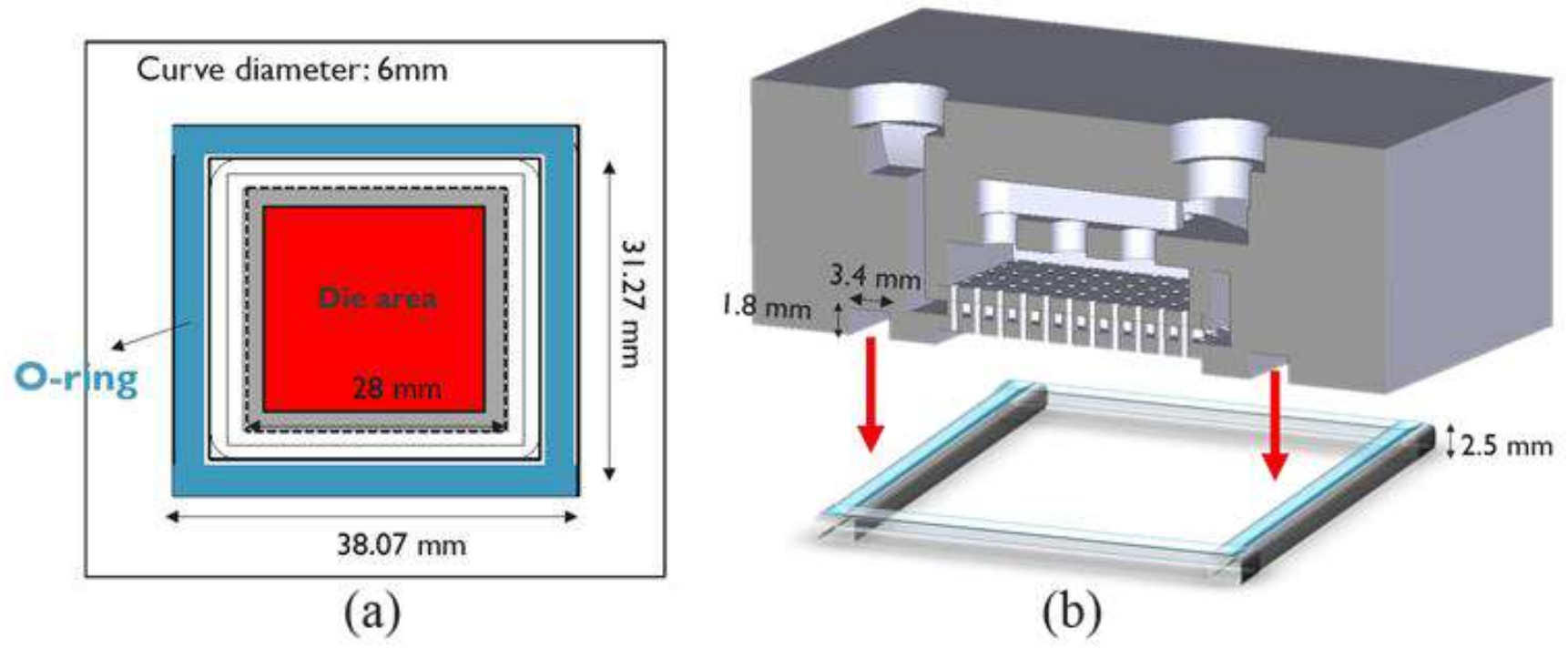

**Figure 9.6:** O-ring arrangement and groove design for the large die warpage reduction during cooler assembly.

**Table 9.2:** Scale the single die cooler design to large die cooler.

| Geometry | | Single chip cooler | Interposer cooler | Large die cooler |
|---|---|---|---|---|
| Nozzle array | $N$ | 4×4 | 4×4 per die | 11×11 |
| Inlet chamber height | | 2.5 mm | 2.5 mm | 3 mm |
| Inlet diameter | $d_i$ | 0.6 mm | 0.6 mm | 0.6 mm |
| Outlet diameter | $d_o$ | 0.6 mm | 0.6 mm | 0.6 mm |
| Cavity height | $H$ | 0.6 mm | 0.6 mm | 0.6 mm |
| Nozzle plate thickness | $t$ | 0.55 mm | 0.55 mm | 0.55 mm |
| Cooler size | $x,y,z$ | 14×14×8.7 ($mm^3$) | 35×35×9.1 ($mm^3$) | 55×55×17.5 ($mm^3$) |

### 9.2.3 Cooler performance modeling

Since the thermal and flow uniformities are very important for the large die cooler design, the full cooler level CFD model for the large die cooler is used in this study. Figure 9.7 (a) and (b) show the extracted fluid domain from the CAD structure in Figure 9.4, where the solid plastic structure is made invisible. The additional distribution layer can be seen clearly in Figure 9.7(b). The meshing methodology introduced in chapter 2 is used in the large die cooler models.

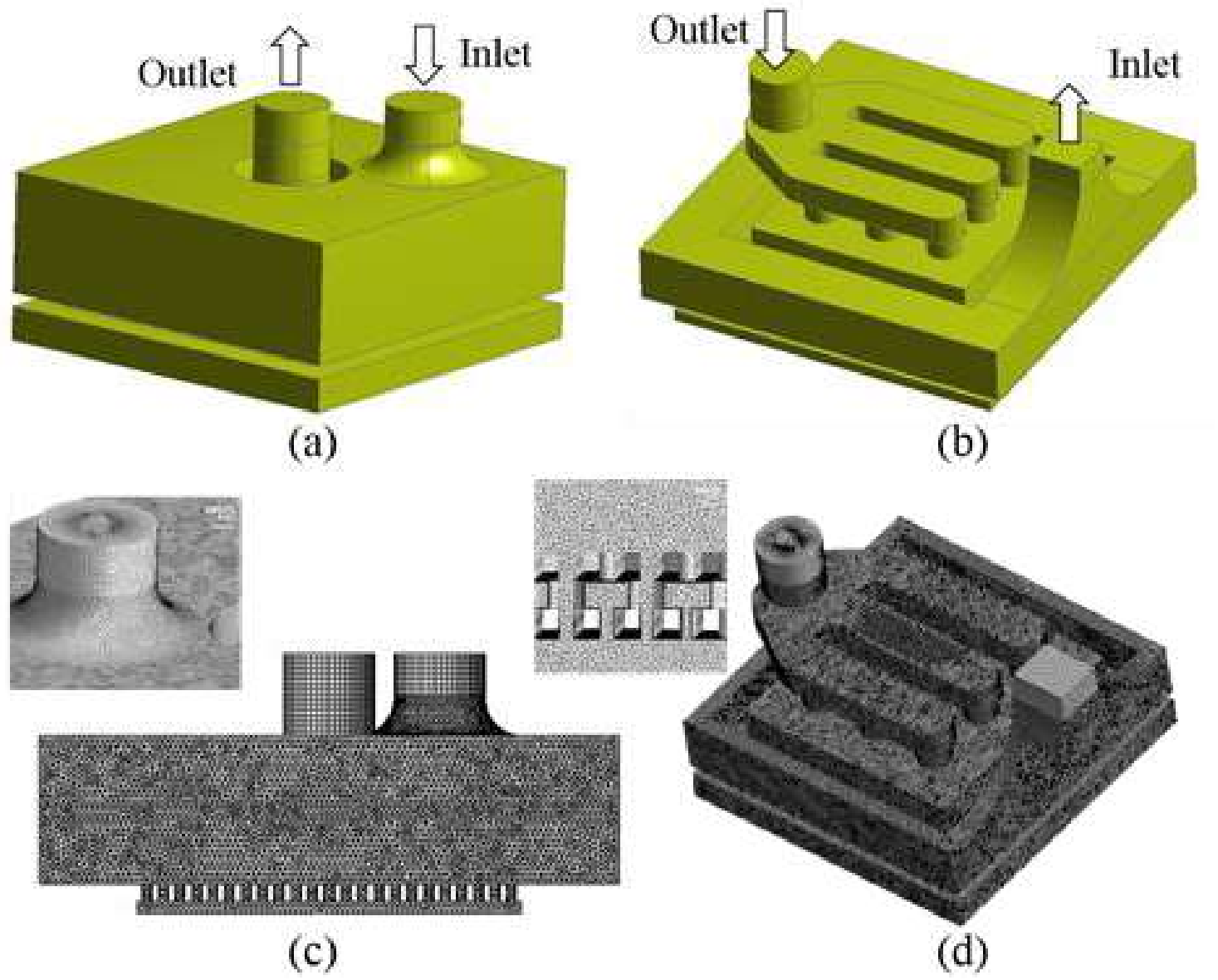

**Figure 9.7:** Modeling of the large die coolers: (a) and (b) the extracted fluid domain without the plastic cooler structure; (c) and (d) meshing details of the CFD models for the designed two cooler configurations.

Figure 9.8 shows the flow streamline distributions inside the designed coolers with vertical feeding and design 2 with additional distribution layer. A low thermal resistance of 0.06 K/W or 0.32 K.cm$^2$/W for a flow rate of 3.25 LPM is predicted for the reference cooler. The chip temperature distributions in Figure 9.9 show that the introduction of the distribution layer in design 2 results in a similar average temperature compared to design 1, while Figure 9.10 shows that it results in a better flow rate uniformity over the inlet nozzle array. The nozzle flow rate standard deviation reduces from 16% in design 1 to 4% in design 2. As a result, the temperature non-uniformity, defined as the difference between the maximum and minimum temperature is reduced from 6.5°C to 3.7°C, demonstrating that this additional layer achieves a better temperature uniformity, however, at the expense of a higher pressure drop, and more structural complexity that excludes conventional fabrication techniques.

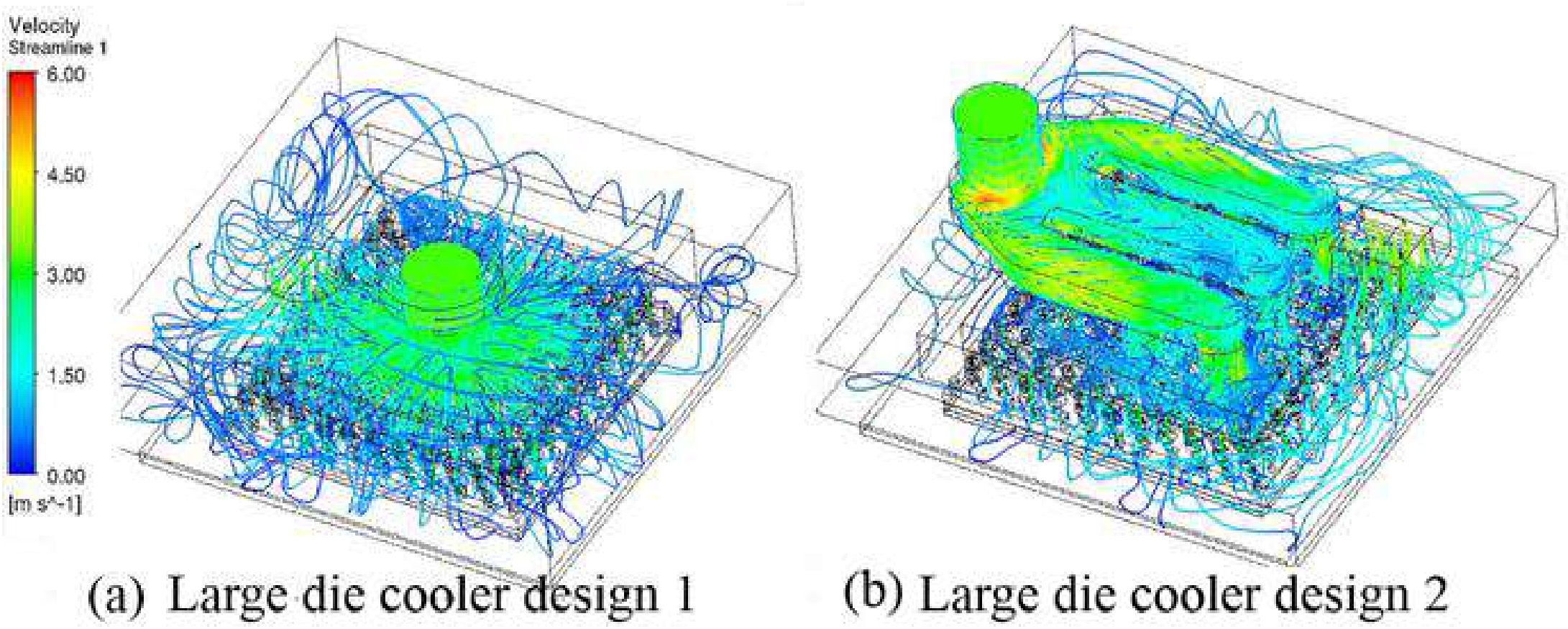

**Figure 9.8:** CFD modeling results with the flow streamline distributions inside the designed cooler at a flow rate of 3.25 LPM: (a) large die cooler design 1; (b) large die cooler design 2.

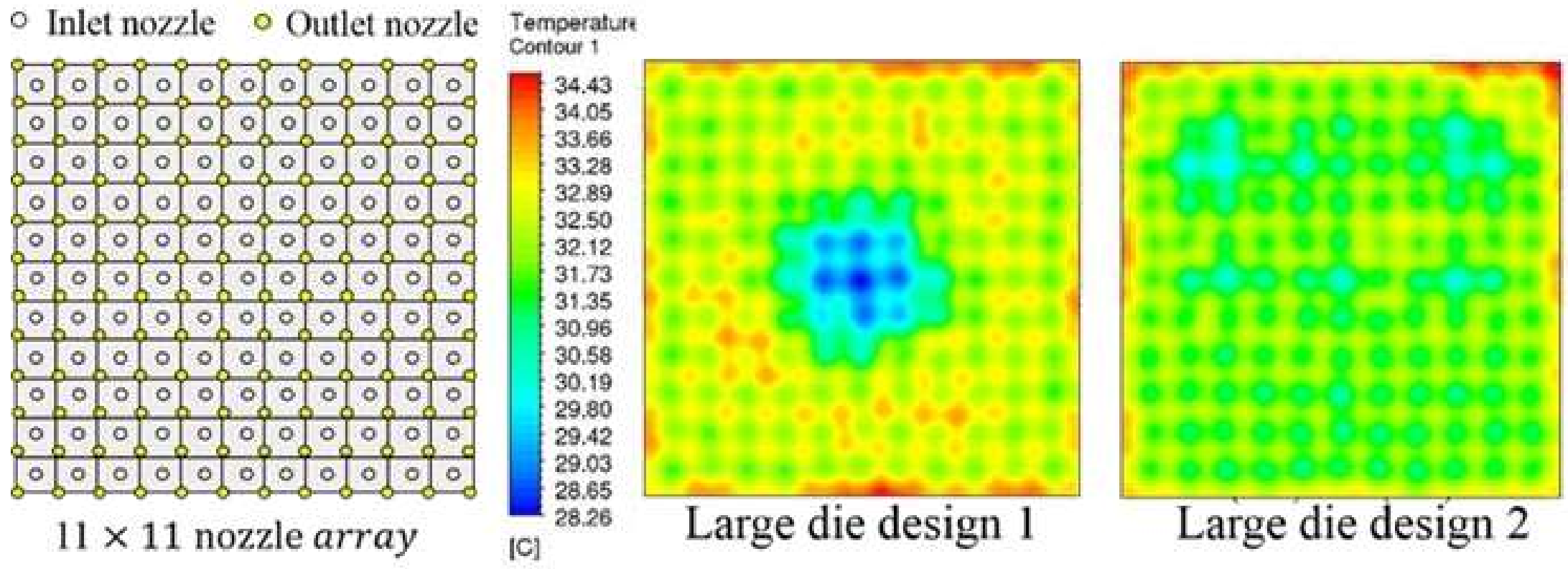

**Figure 9.9:** Temperature distribution comparison with CFD modeling results: design 2 with additional distribution layer shows better temperature uniformity (FL=3.25 LPM).

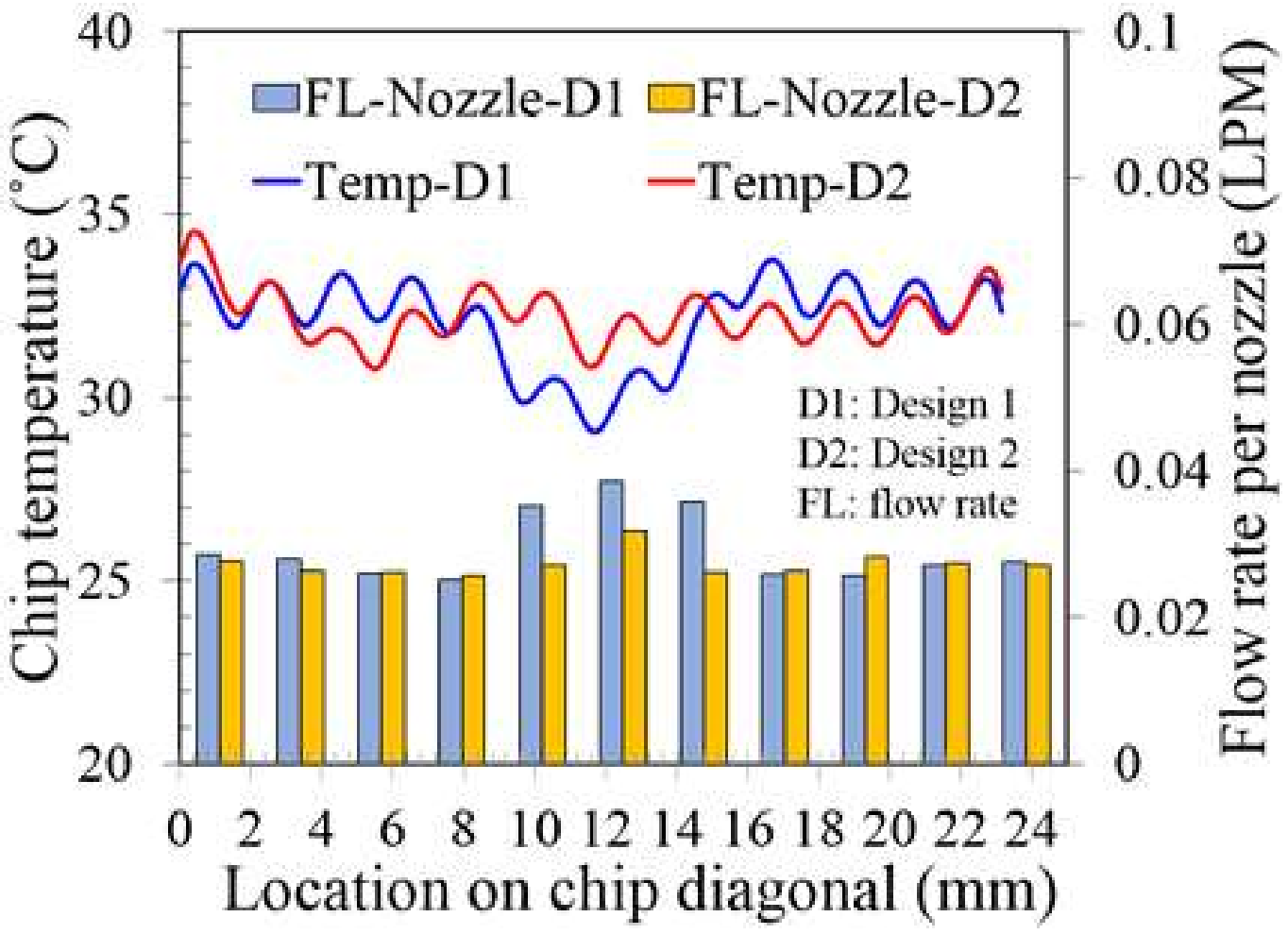

**Figure 9.10:** CFD simulation results: temperature and flow rate per nozzle distribution for cooler design 1 and 2 along the diagonal line at 3.25 LPM. (diagonal from top left to bottom right)

### 9.2.4 Demonstration of 3D printed large die coolers

The 3D printed large die coolers are fabricated using high resolution Stereolithography (SLA), using the water-resistant material Sonos WaterShed XC 11122 [8], as introduced in chapter 6. Figure 9.11 shows the side view of the two 3D printed large die coolers, matched with the package size for the large die. Using microscopy, the bottom view with the full 11×11 inlet nozzle array and 12×12 inlet nozzle array can be evaluated, shown in Figure 9.12(a). The fabrication tolerance of the fabricated nozzles is also assessed using 2D and reconstructed 3D microscope images, as shown in Figure 9.12(b). The measured average nozzle diameter is 630 µm, which deviates only 5% from the nominal design value of 600 µm. The cross-section view of the transparent coolers in Figure 9.13 reveals the successfully fabricated internal structures of the two cooler designs, that could not be fabricated with conventional fabrication techniques. This demonstrates that additive manufacturing can be used for the fabrication of complex large die cooler geometries with micro-scale features.

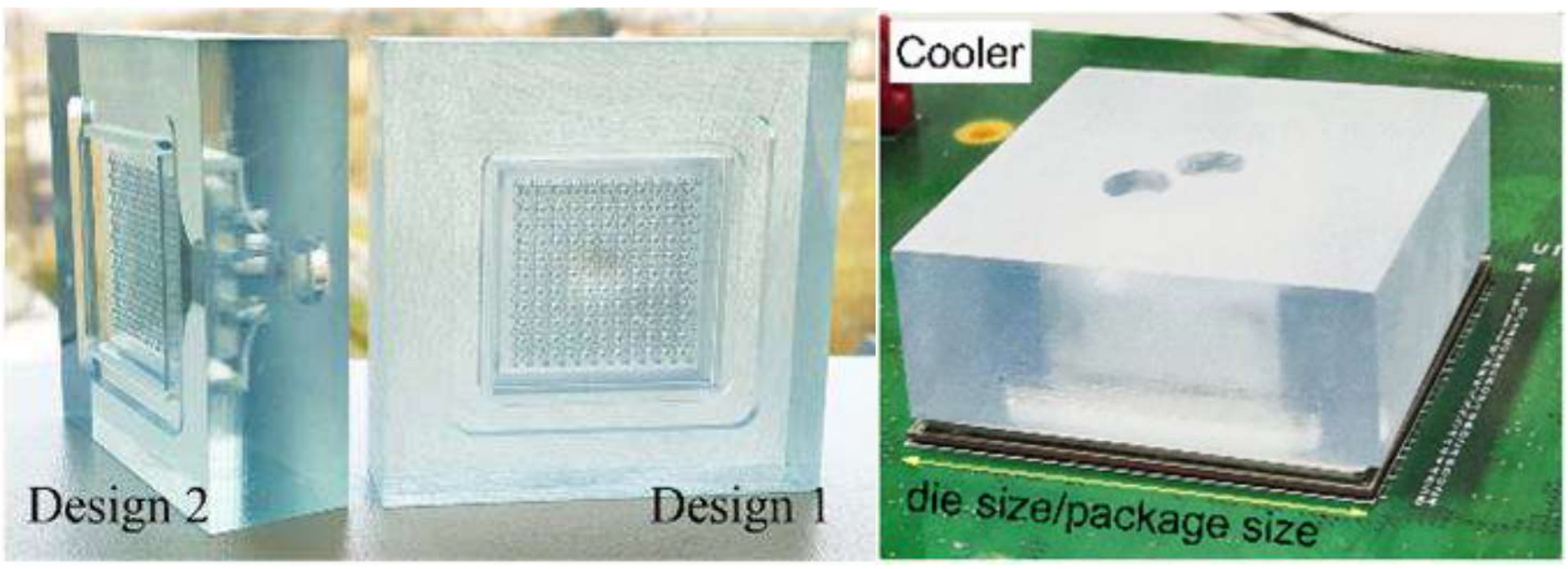

**Figure 9.11:** Images of the demonstrated coolers with design 1 and design 2: the cooler size is matched with large die/package size.

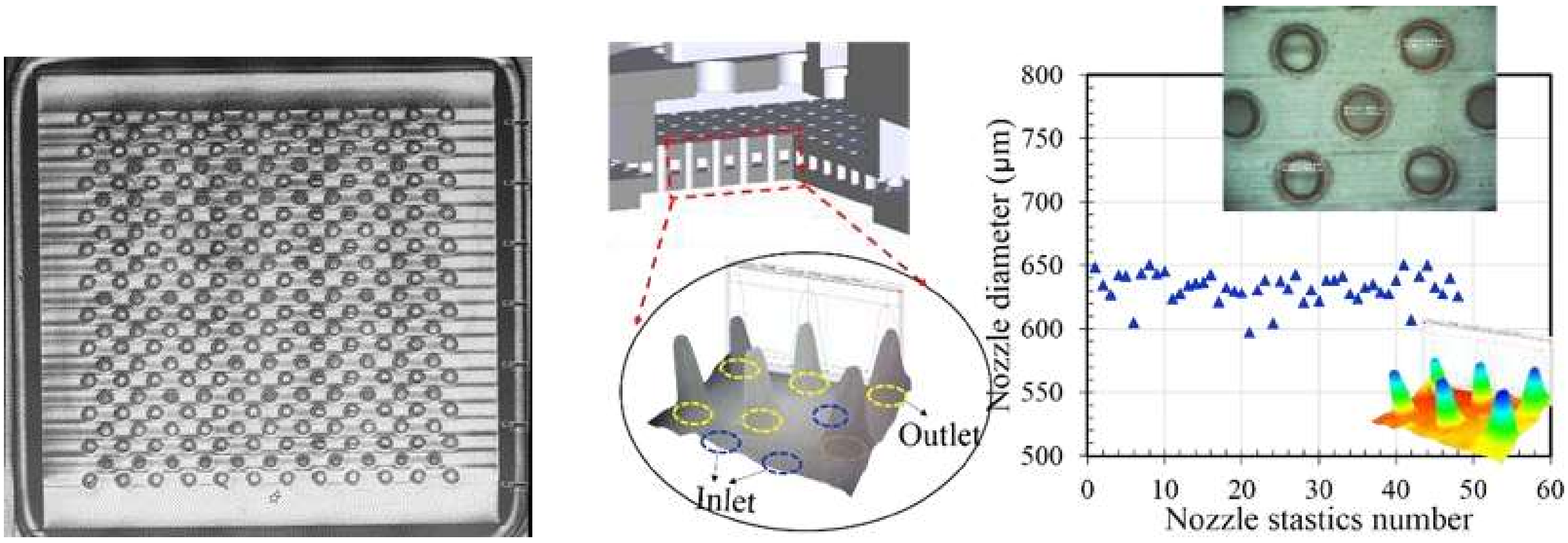

(a) (b)

**Figure 9.12:** (a) Bottom view of the full 11×11 inlet nozzle arrays and 12×12 outlet nozzle arrays; (b) Nozzle geometry and nozzle diameter evaluation with 2D and reconstructed 3D microscope image: 5% deviation from the nominal design value of 600 μm.

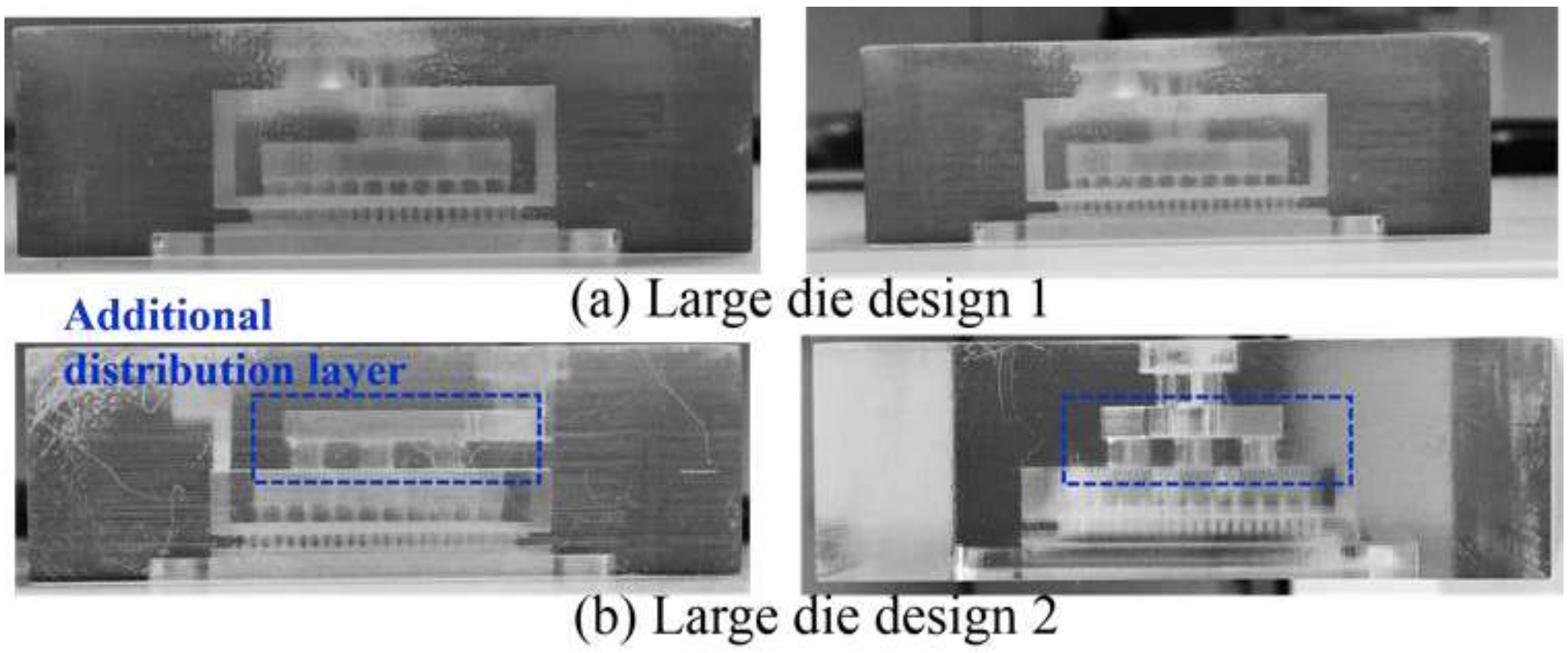

**Figure 9.13:** Side view of the transparent fabricated microfluidic heatsinks with internal liquid delivery microchannels: (a) Design 1 with vertical feeding; (b) The additional distribution layer in design 2 is successfully fabricated.

### 9.3 Experimental characterization

#### 9.3.1 Large die cooler setup and calibration

For the thermal and hydraulic characterization of the large die coolers, the 3D printed coolers are connected to the closed-loop fluidic circuit, which is introduced in chapter 2. The details of the fluidic and electrical connections to the large die test chip are shown in Figure 9.14(a). Moreover, the schematic of the electrical control for the heaters in the large die is shown in Figure 9.14(b). In general, all the heaters are connected to a single power supply, with 4 parallel chains of 4 heaters in series to dissipate uniform power in the chip. Off-chip resistors are designed to measure the current in each of the four branches in order to measure the actual dissipated power in each resistor heater.

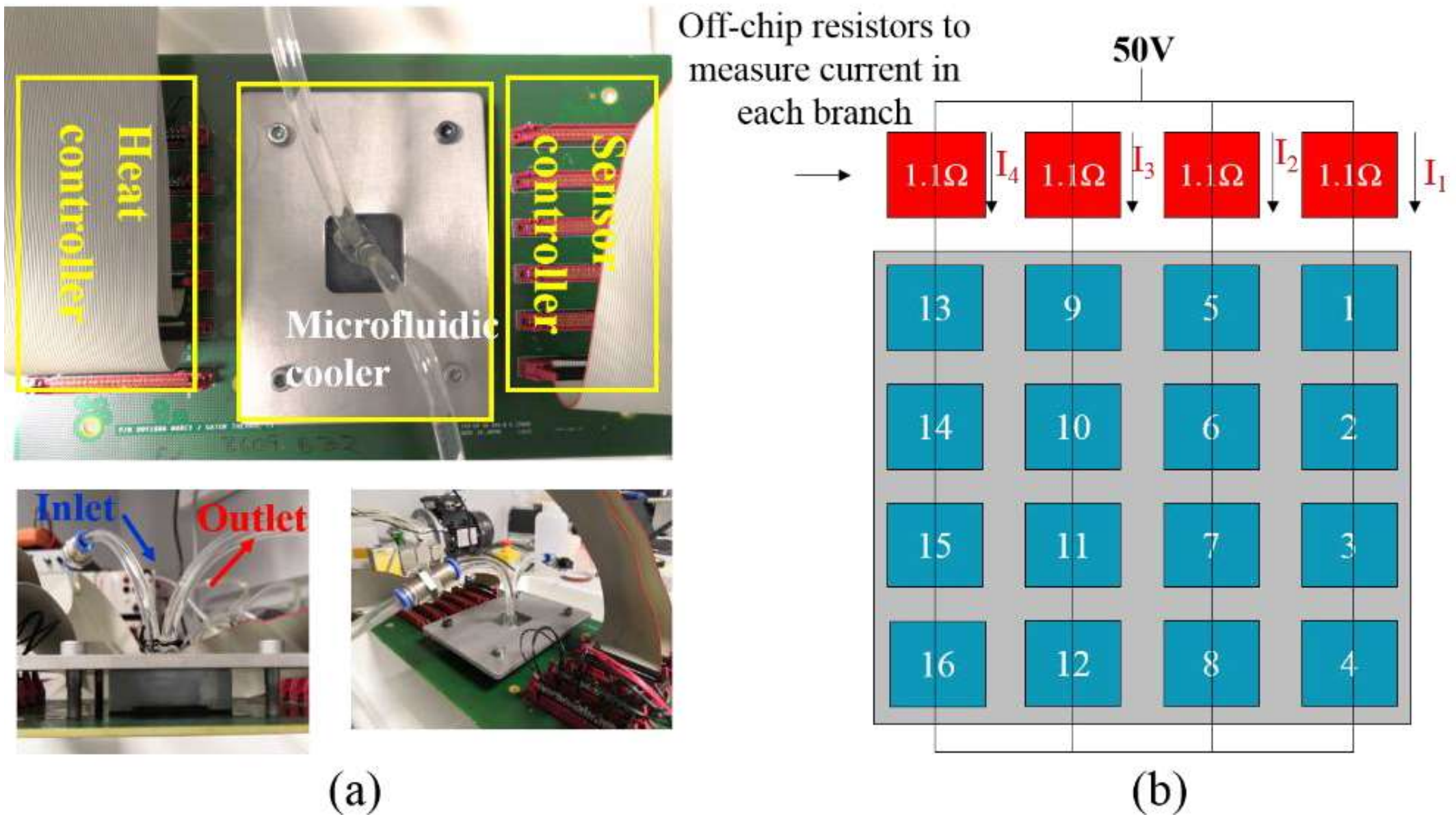

**Figure 9.14:** Measurements set-up: (a) Details of the fluidic and electrical connections to the test vehicle in the flow loop; (b) schematic of the electrical control for the 16 heaters in the large die.

In the first step, the total heater power for different applied voltages is measured based on the electrical connections shown in Figure 9.14(b). Figure 9.15 illustrates the steps for calculating the total heater power for a total applied voltage of 50V, at flow rate of 1 L/min. Since the resistance is temperature dependent, the resistance and power depend on the applied voltage and flow rate. First, the voltage across each heater resistor is measured for a fixed total voltage, shown in Figure 9.15(a). For this analysis, the small voltage drop over the off-chip resistors are taken into account; Secondly, the current across each branch is calculated through the measured voltage of the off-chip resistors; Next, the resistance for each heater is extracted in Figure 9.15(b). Finally, the power distribution for all the heaters is extracted in Figure 9.15(c). The total heater power can be calculated by adding the 16 resistor values, resulting in 266.8 W for a total applied voltage of 50V, at flow rate of 1L/min. Table 9.3 lists the measured total heater powers with regard to different applied voltages, for a constant flow rate 1L/min. In addition, the measured total heater power values at 50V heater voltage for different flow rate values are also listed in Table 9.4. The variation of the heater power is due to the heater resistance change with the heater temperature, influenced by the flow rate.

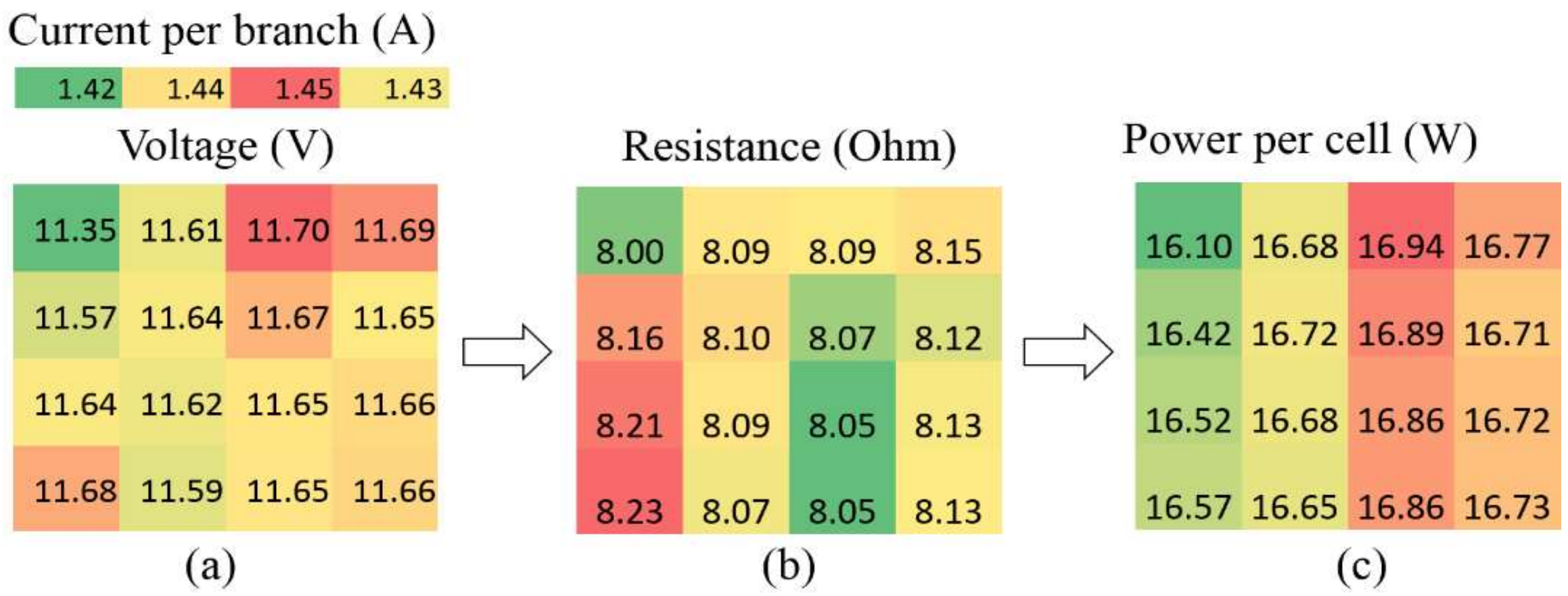

**Figure 9.15:** Heater power distribution evaluations on 4×4 heater array: (a) Measure voltage (V) across each heater resistor; (b) Extracted resistance for each heater based on the current per branch; (c) Calculate power in each resistor (Flow rate=1 L/min, total applied voltage = 50V).

**Table 9.3:** Measured total heater power for different applied voltages (FL=1LPM).

| Heater voltage (V) | Lidded package Power (W) | Lidless package Power (W) |
|---|---|---|
| 30 | 96.50 | 96.50 |
| 35 | 130.63 | 130.63 |
| 40 | 169.59 | 169.59 |
| 45 | 213.70 | 213.70 |
| 50 | 266.81 | 261.50 |

**Table 9.4:** heater power at 50 V heater voltage with lidded package and lidless cooling for different flow rates.

| Flow rate (LPM) | Lidded package Power (W) | Lidless package Power (W) |
|---|---|---|
| 0.5 | 252.04 | 248.39 |
| 1 | 261.50 | 266.81 |
| 1.5 | 265.67 | 270.71 |
| 2 | 267.76 | 273.29 |
| 2.5 | 268.97 | 274.85 |
| 3 | 269.75 | 275.37 |
| 3.25 | 269.94 | 275.73 |

In addition, the deviation from the average power for each heater power is shown in Figure 9.16. The deviation of the power distribution for the heaters can be used to analyze the temperature uniformity in the large die. It can be seen that there is a consistent distribution for different power values from 4% to +2%.

(a) 30 V

| | | | |
|---|---|---|---|
| -3.5% | 0.1% | 1.5% | 1.2% |
| -1.7% | 0.2% | 1.2% | 1.0% |
| -1.2% | -0.1% | 0.9% | 0.9% |
| -1.2% | -0.5% | 0.7% | 0.7% |

(b) 40 V

| | | | |
|---|---|---|---|
| -3.8% | 0.0% | 1.6% | 1.2% |
| -1.9% | 0.2% | 1.3% | 0.9% |
| -1.4% | -0.1% | 1.1% | 0.9% |
| -1.2% | -0.4% | 0.9% | 0.8% |

(c) 45V

| | | | |
|---|---|---|---|
| -3.9% | 0.0% | 1.6% | 1.1% |
| -2.0% | 0.2% | 1.3% | 0.8% |
| -1.5% | 0.0% | 1.1% | 0.9% |
| -1.3% | -0.3% | 1.1% | 0.8% |

(d) 50V

| | | | |
|---|---|---|---|
| -4.0% | 0.0% | 1.7% | 1.1% |
| -2.2% | 0.2% | 1.4% | 0.8% |
| -1.6% | 0.0% | 1.2% | 0.9% |
| -1.3% | -0.2% | 1.2% | 0.9% |

**Figure 9.16:** Local heater power deviation from average power value for different applied voltage: (a) 30V; (b) 40V; (c) 45V; and (d) 50V (variation respect to average power).

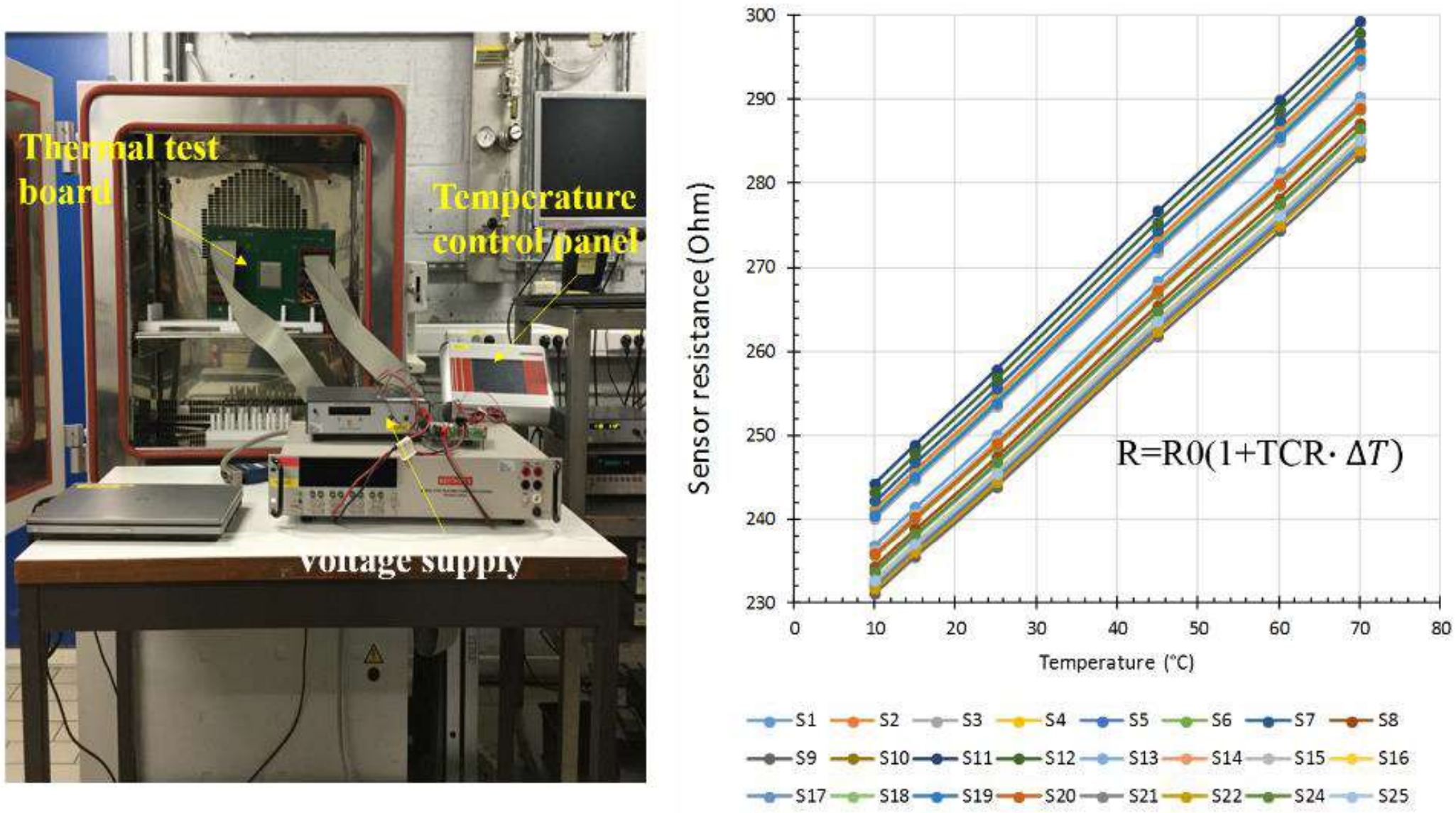

**Figure 9.17:** Calibration of the TSRs in the large die thermal test vehicle: (a) TSR calibration setup; (b) Sensor resistance as function of temperature for 25 sensors.

In the second step, the 25 temperature-sensitive resistors (TSR) are calibrated as a function of temperature. The experimental setup and the sensor resistance as function of the temperature are illustrated in Figure 9.17. The range of the calibrated temperature is from 10°C to 75°C. The calibrated TCR is 3553 ± 2 ppm/°C at the reference temperature of 25°C, and 3750 ± 2 ppm/°C at the reference temperature of 10°C. Therefore, the temperature increase of the sensor ΔT can be determined by the following equation:

$$\Delta T = \frac{R - R_0}{R_0 \cdot TCR} \qquad (9.1)$$

where $R_0$ is the resistance at the reference temperature. The TCR is the temperature coefficient of the resistance.

### 9.3.2 Characterization of the reference heatsink

The distribution of the measured chip temperature increases with respect to the coolant inlet temperature in the bare die package for a chip power of 275W and a coolant (DI water) flow rate of 3.25 LPM is shown in Figure 9.18(a). At this power, an average chip temperature increase as low as 17.5°C is achieved with a low temperature non-uniformity of 6% distribution for a low pressure drop of 0.7 bar, demonstrating the efficiency of the microfluidic cooler. Figure 9.18(b) shows the measured chip temperature profile along the chip diagonal for three different flow rates for a total applied voltage of 50V. The average temperature increase in this chapter is defined as the average chip temperature with regard to the inlet temperature.

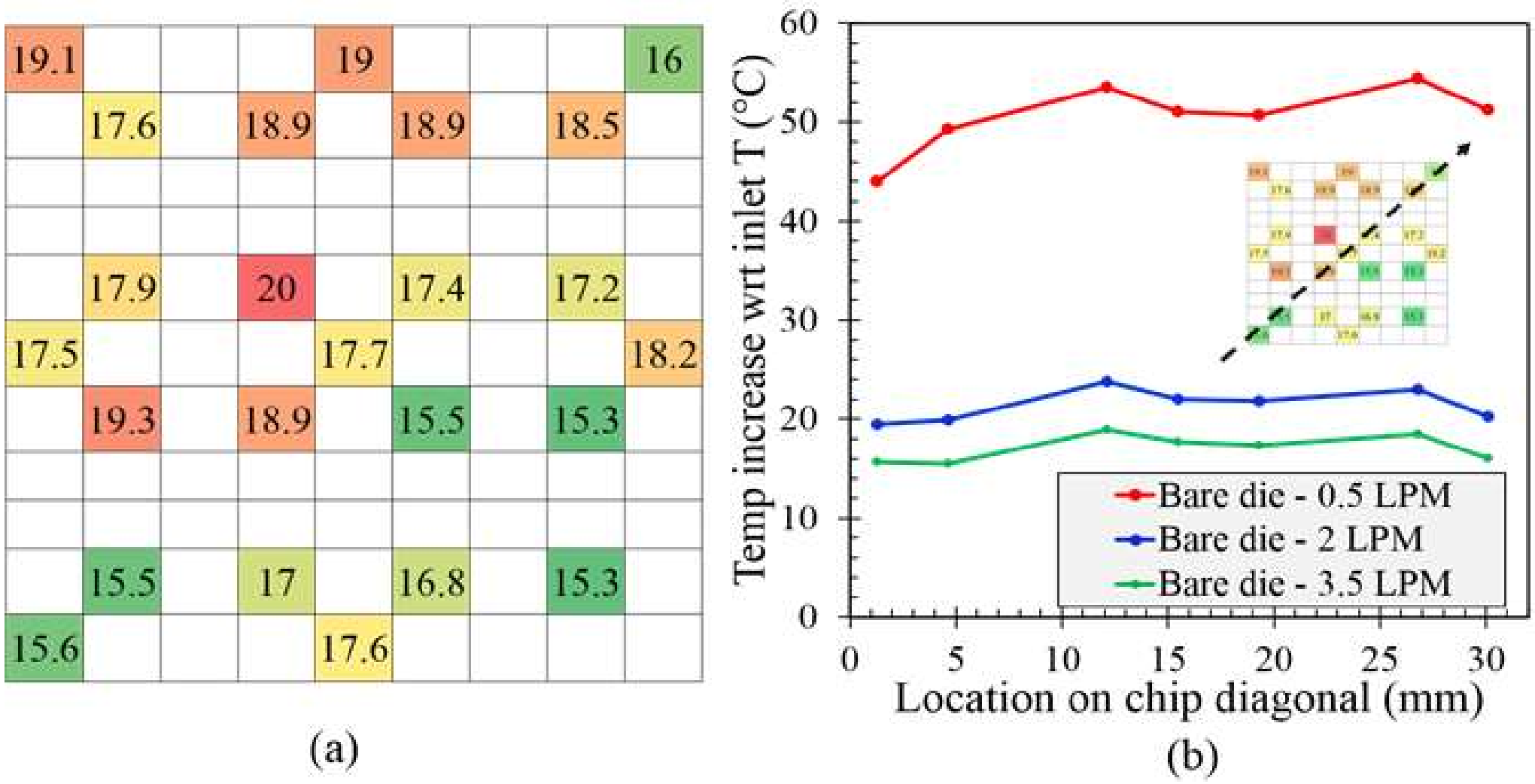

**Figure 9.18:** Temperature measurement on the test chip for a total applied voltage of 50V using the reference large die cooler with design 1: (a) Temperature distribution map for a constant flow rate of 3.25 LPM; (b) Temperature profile along the large die diagonal with total applied voltage of 50V.

The evolution of the average temperature increase as a function of the chip power for constant flow rate of 1 L/min is plotted in Figure 9.19(a). The thermal resistance can be extracted from the correlations between the measured chip temperature increase and chip power. Moreover, the evolution of the average temperature increase as a function of the flow rate for a total applied voltage of 50V is plotted in Figure 9.19(b). The temperature increases as a function of flow rate exhibits a power law relation, with an exponent of -0.55. This trend is consistent with the results with an exponent of -0.54 shown in Figure 6.25 of chapter 4, for 3D printed 4×4 nozzle array cooling. The normalized thermal resistance comparison in Figure 9.20 shows that the scaled large die cooler has the same normalized thermal performance as the 4×4 array cooler on the 8×8 mm$^2$ PTCQ. Therefore, the implementation of the multi-jet cooling on the large die size with 11×11 nozzle array further validates the normalization concept introduced in chapter 2. This thermal performance can be extrapolated to large die size with 530 mm$^2$: for an assumed maximum allowable chip temperature of 80 ℃, a total chip power of 1363 W can be cooled with a flow rate of 26 mL/min per nozzle.

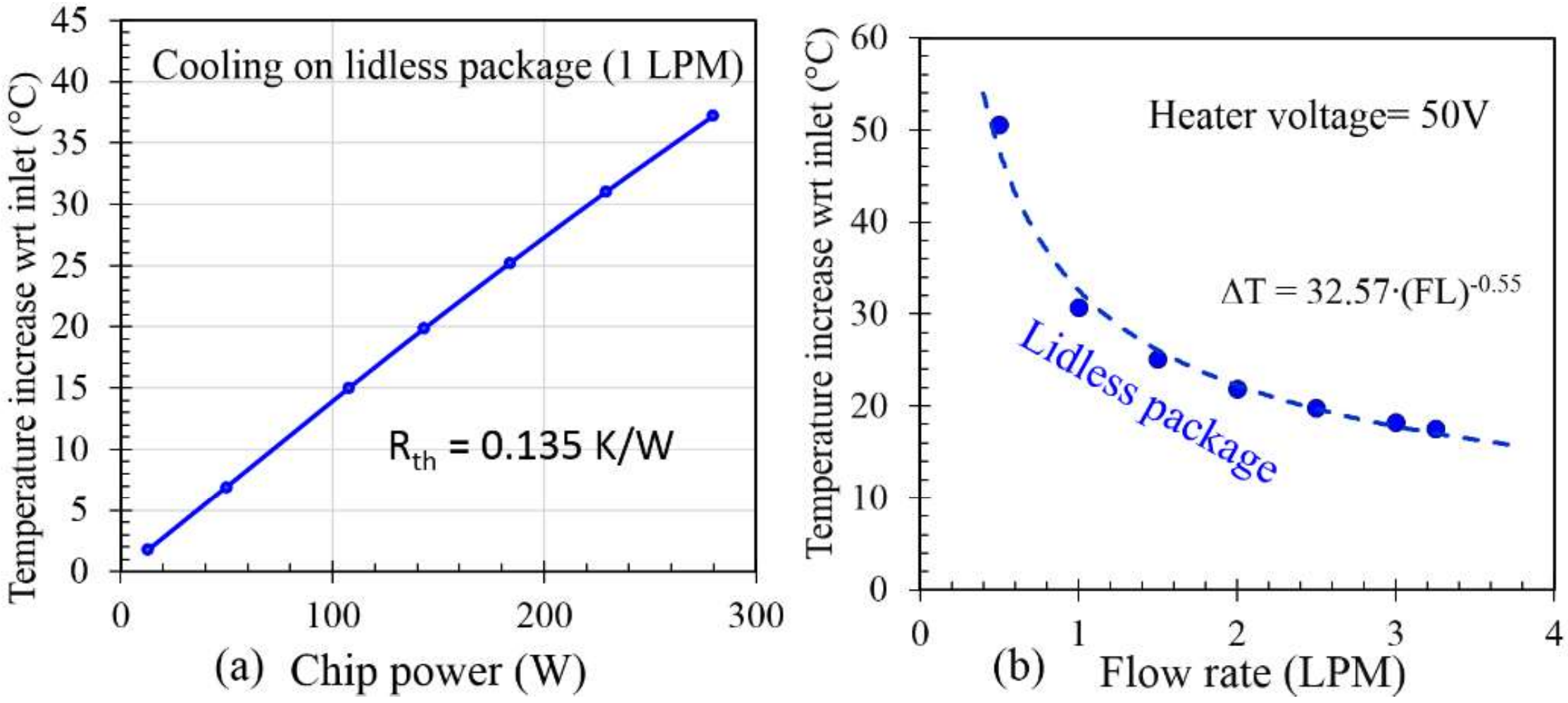

**Figure 9.19:** Temperature measurements comparison with bare die package at (a) different chip power and (b) different flow rates at total heater voltage of 50V.

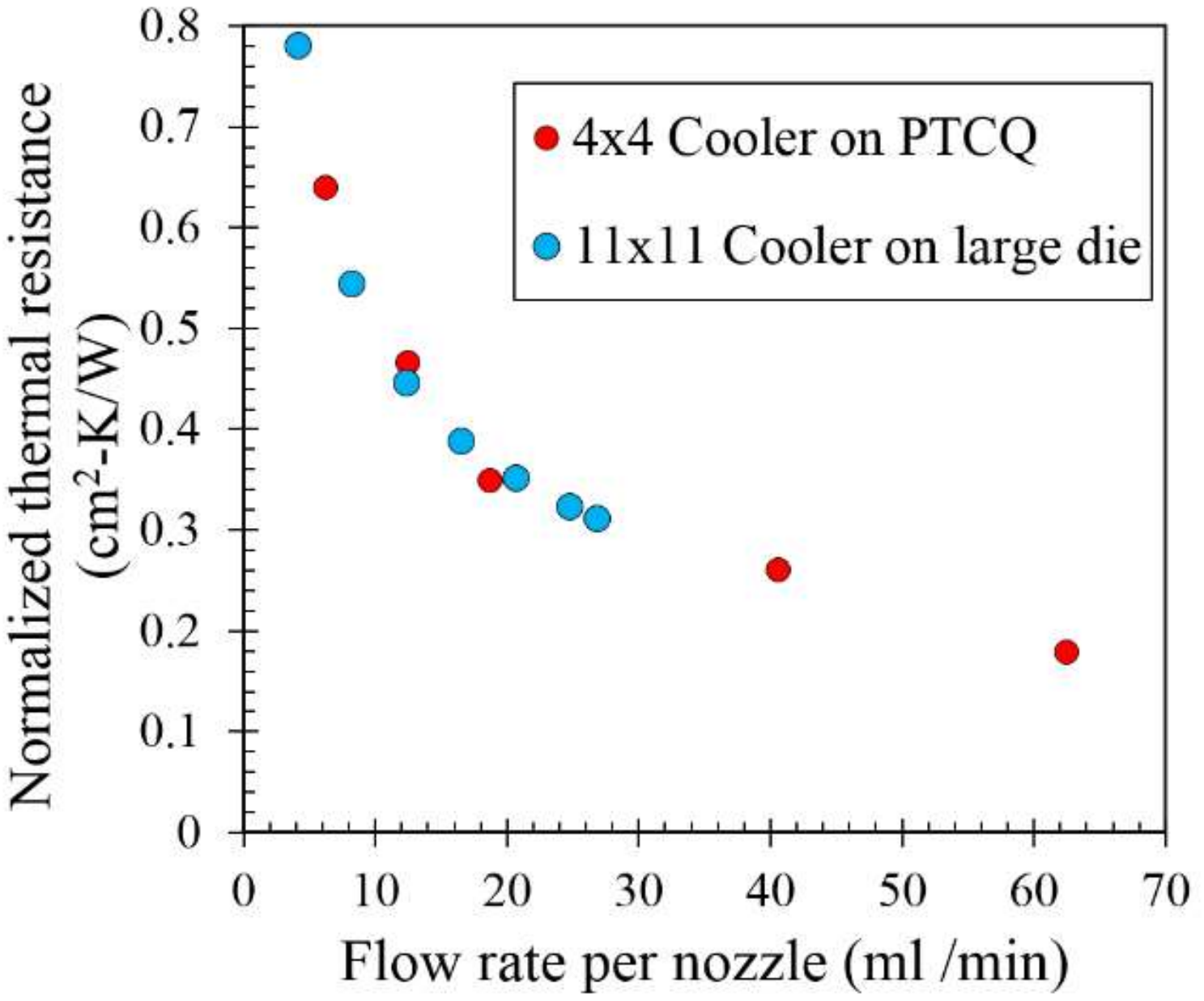

**Figure 9.20:** Scaled large die cooler shows the same thermal performance as the 4x4 array cooler on the 8×8 mm$^2$ PTCQ with the same nozzle geometry.

### 9.3.3 Comparison lidded package vs bare die

In chapter 8, the thermal impact of the TIM on the cooling of the 2.5D interposer packages is discussed. For the 2.5D interposer package, the thermal conductivity of the TIM is about 1.9 W/m-K for a thickness of 90 µm. This results in a large difference between the bare die and lidded package of a factor of 2.5 to 3. As introduced in section 9.2.1, the thermal conductivity of the TIM applied in the large die package is 2.3 W/m-K with about 20 µm thickness. Therefore, the difference between the bare die package and lidded package is expected to smaller. In this section, the temperature difference will be characterized and quantified experimentally. The measured temperature distribution map for the lidded package and lidless package at different flow rates for a total applied voltage of 50V are shown in Figure 9.21 and Figure 9.22.

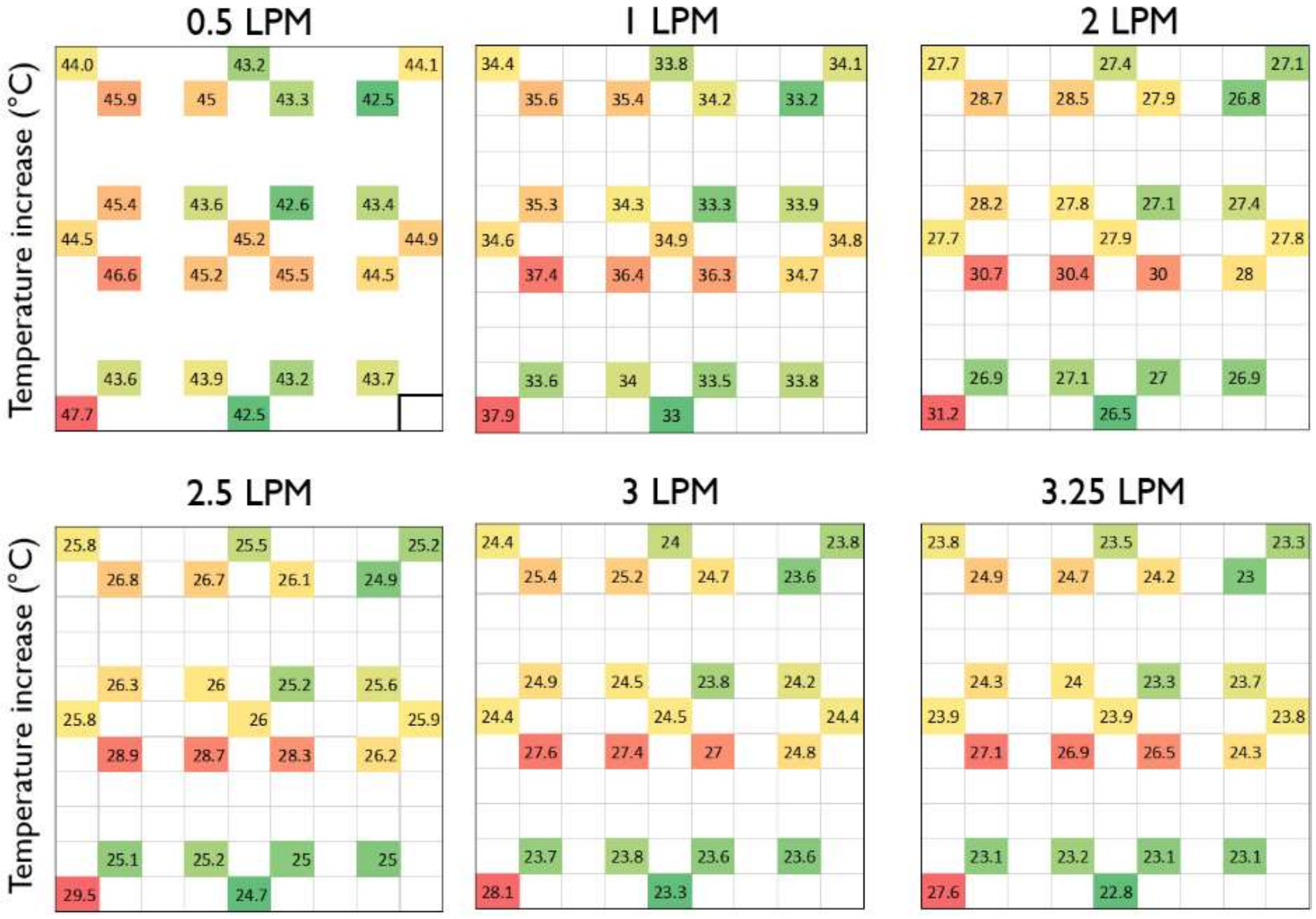

**Figure 9.21:** Thermal measurements cooler on lidded package at different flow rates for a total applied voltage of 50V.

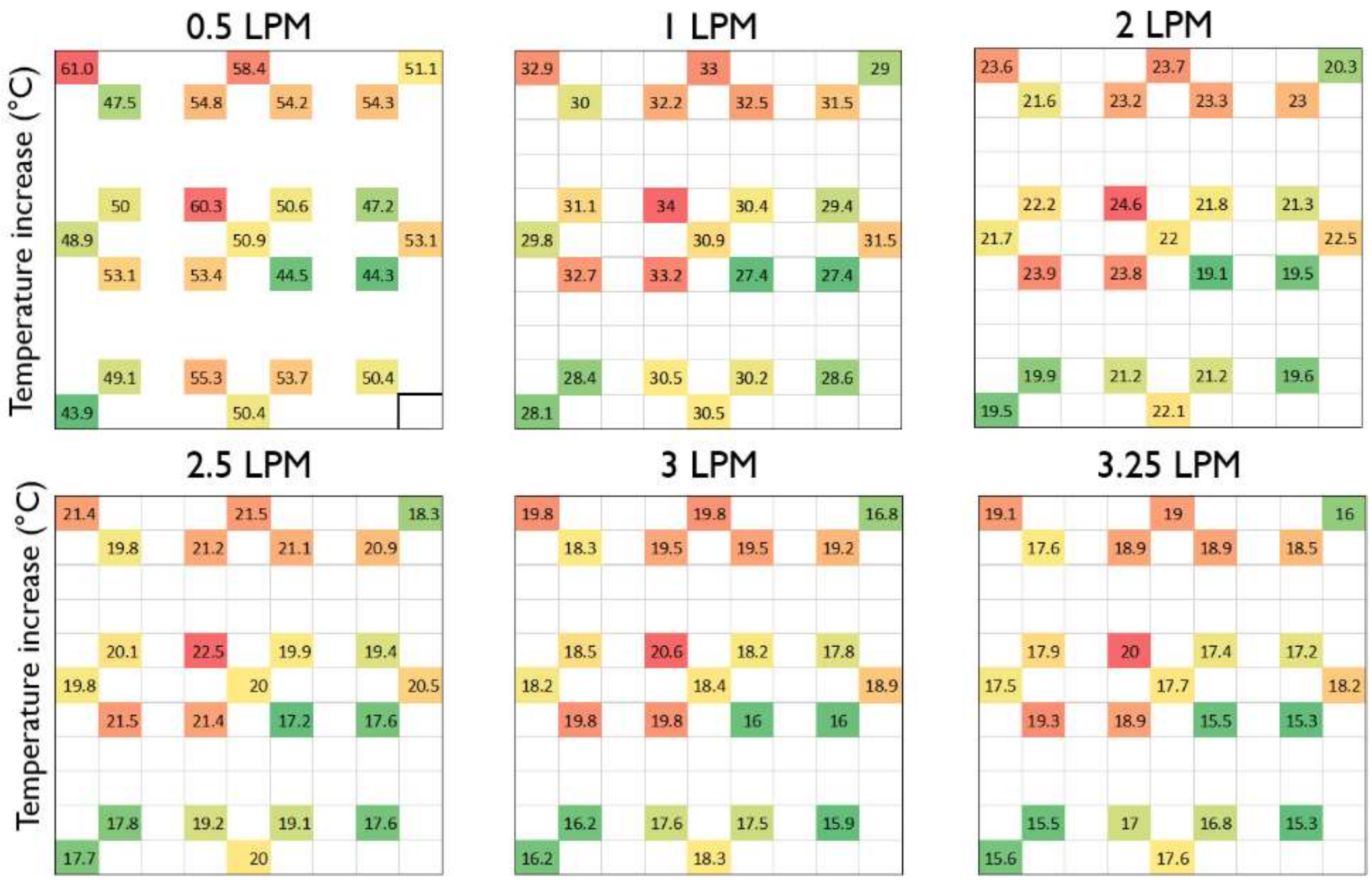

**Figure 9.22:** Thermal measurements cooler on lidless package at different flow rates for a total applied voltage of 50V.

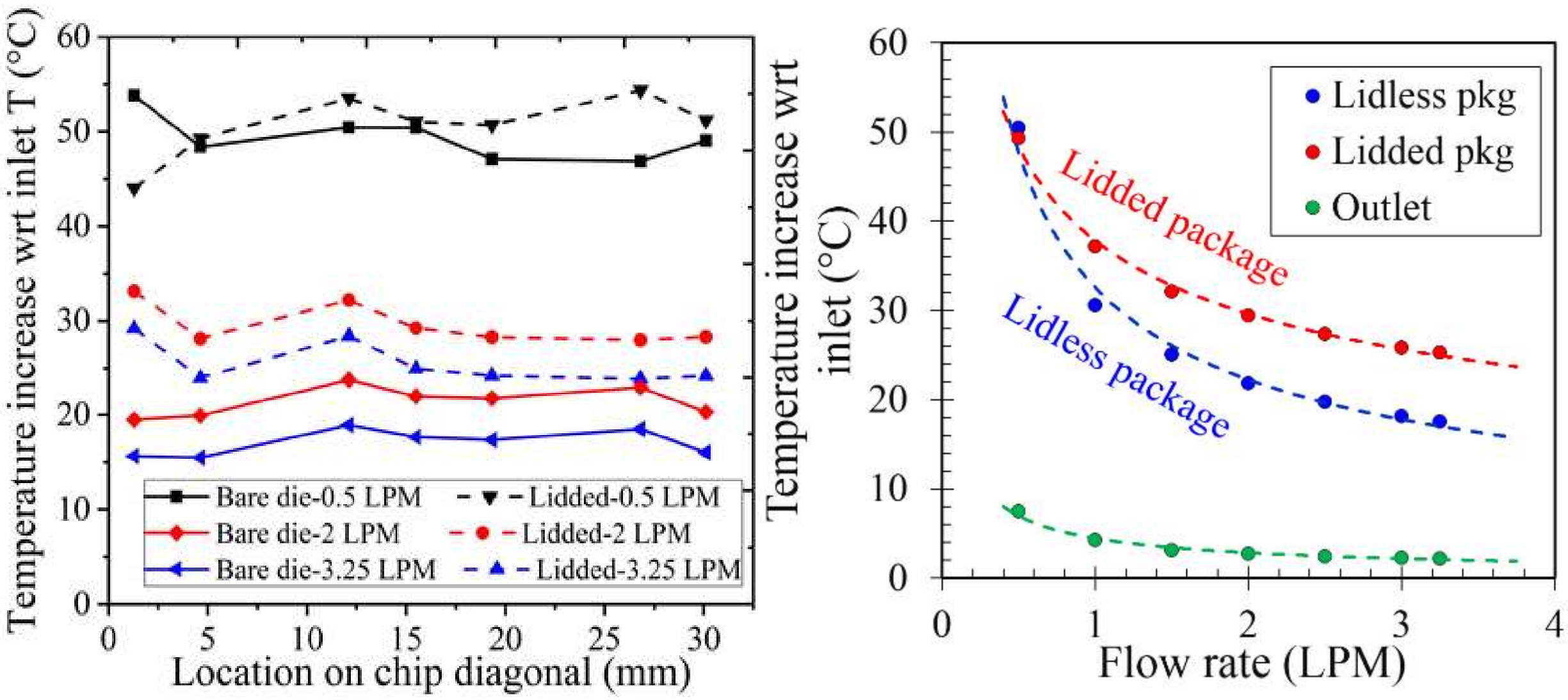

(a) (b)

**Figure 9.23:** Temperature increase with regard to the inlet temperature comparison with the lidded and bare die package for a total applied voltage of 50V: (a) Temperature profile comparison on the chip diagonal; (b) average temperature comparison as function of the flow rate.

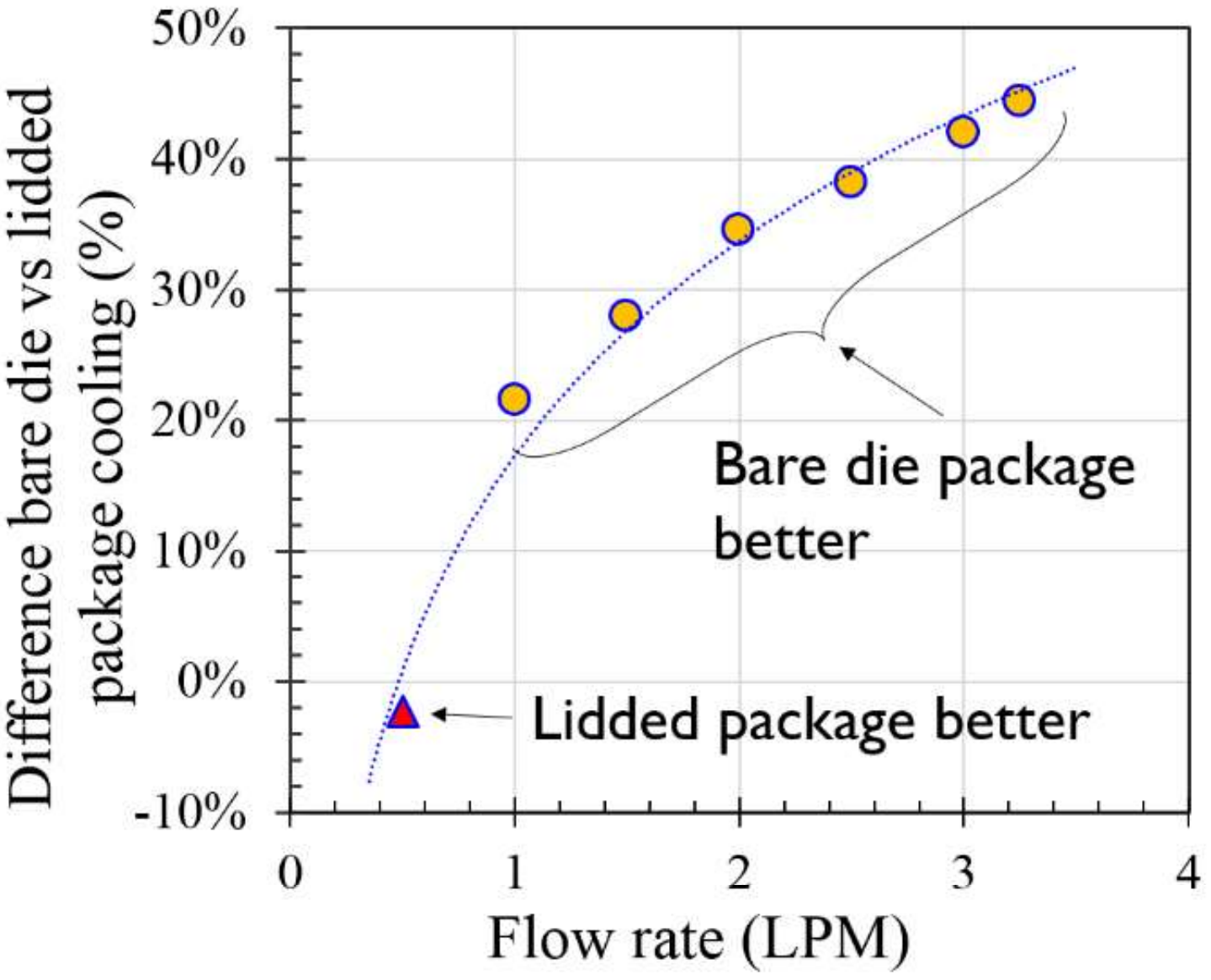

**Figure 9.24:** Temperature difference between the bare die cooling compared with the lidded package cooling.

Figure 9.23 shows the comparison of the temperature distribution map and temperature profile along the chip diagonal for a total applied voltage of 50V. The measurements show that the cooling on the lidded package has a better thermal performance at very low flow rates, due to the lateral thermal spreading in the metal lid, while the bare die cooling outperforms the cooling on the lid for flow rate values above 0.7 LPM. This is because the effect of the high cooling rate on the lid, reduces the heat spreading in the lid for lidded package cooling. At a flow rate of 2 LPM, the temperature increase of the cooler on the bare die package is 35% lower compared to the lidded package. For a higher flow rate of 3.25 LPM, the temperature difference is increased to 44%. This is caused by the additional thermal resistance of the lid and mainly the thermal interface material in case of the lidded package. The temperature difference between the bare die cooling and lidded package cooling is summarized in Figure 9.24. It can be seen that the temperature difference becomes larger with the increase of the flow rate.

For the comparison of the bare die package and lidded package cooling, the temperature uniformity is also investigated, as illustrated in Figure 9.25. The bare die cooling shows about 8.1% variation while the lidded package shows about 6.4% temperature variation. Moreover, the standard deviation of the chip temperature for the lidded package and bare die package is also studied, as shown in Figure 9.26 to evaluate the temperature uniformity. The heat spreading effect is represented with the standard temperature deviation across the chip surface. The standard deviation for lidded package at a low rate of 0.5 LPM is about 2.6 times higher than the bare die cooling. This is due to the lateral heat spreading effect in the lid that is dominated at the low flow rate. The difference of the standard deviation for both cases becomes smaller with the increasing of the flow rate, which is because the high heat flow rate on the top of lid results in more vertical heat transfer through the lid, thus, relatively less lateral heat spreading.

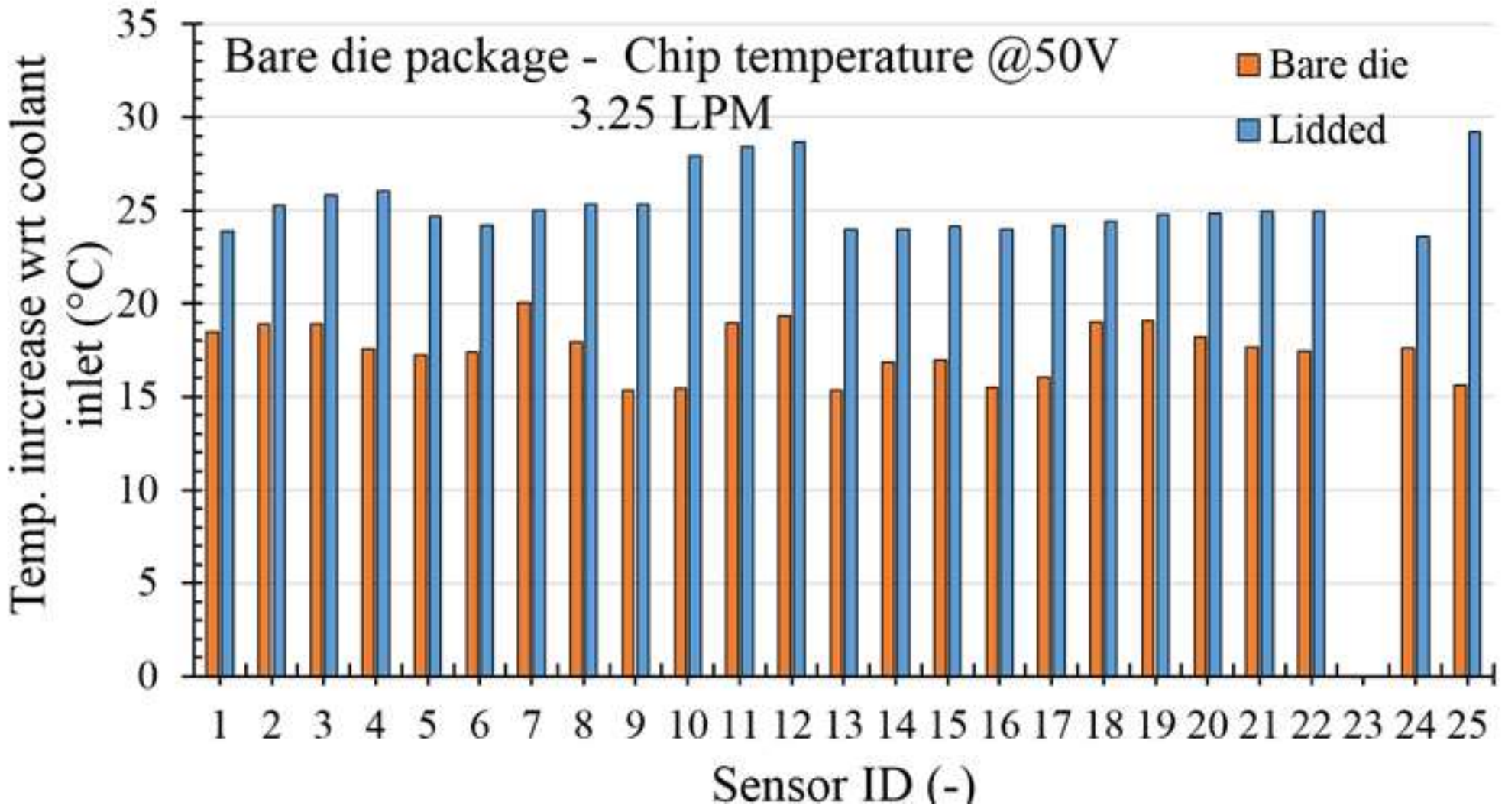

**Figure 9.25:** Temperature increase variations for different temperature sensors at a flow rate of 3.25 LPM for a total heater voltage of 50V.

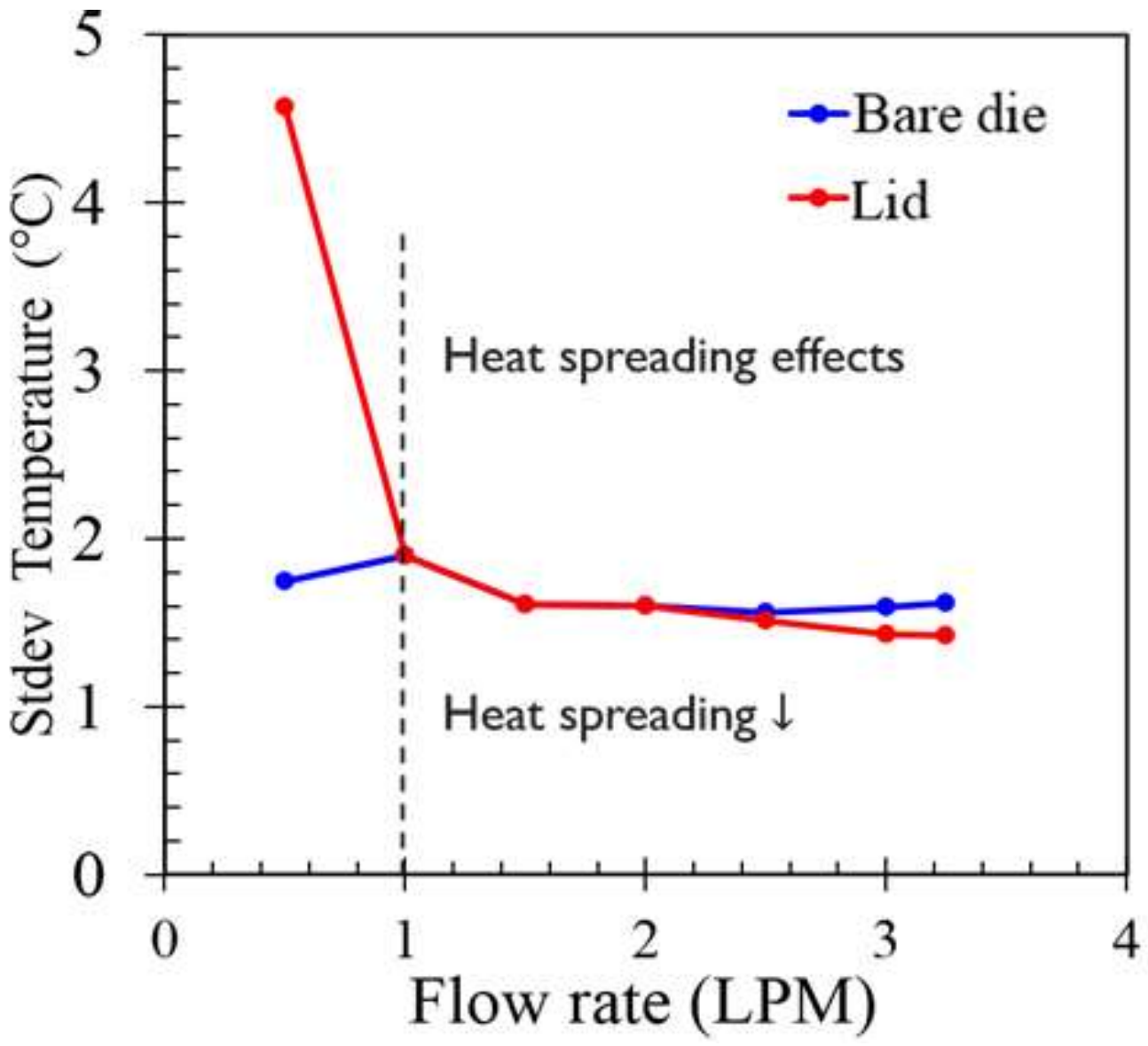

**Figure 9.26:** Standard deviation of the chip temperature as function of the flow rate for bare die cooling and lidded package cooling.

### 9.3.4 Temperature non-uniformity analysis

In Figure 9.25, the temperature non-uniformity for the lidded package cooling and bare die cooling are analyzed for different flow rate. In this section, the source of the temperature non-uniformity will be investigated. Basically, there are two factors that can influence the temperature uniformity: the power non-uniformity of the heaters on the test chip and the coolant flow non-uniformity in the nozzle array of the large die cooler. First, the influence of the power non-uniformity on the temperature non-uniformity is discussed. In this study, the local cooling effect is defined as the ratio between the sensor temperature increase and corresponding heater power. Figure 9.27 shows the cooling effects analysis for an applied voltage of 50V at flow rate of 0.5 LPM. In order to extract the cooling effect with the ratio between the TSR sensor temperature and heater power, the following steps are performed:

(1) The temperature at the heater locations is extracted;

(2) The temperature deviation from the average chip temperature is calculated;

(3) The normalized power deviation distribution for the heaters are also extracted;

(4) The ratio between the temperature deviation and power deviation is calculated;

(5) The average ratio per quarter is used for the comparison of the cooler orientation;

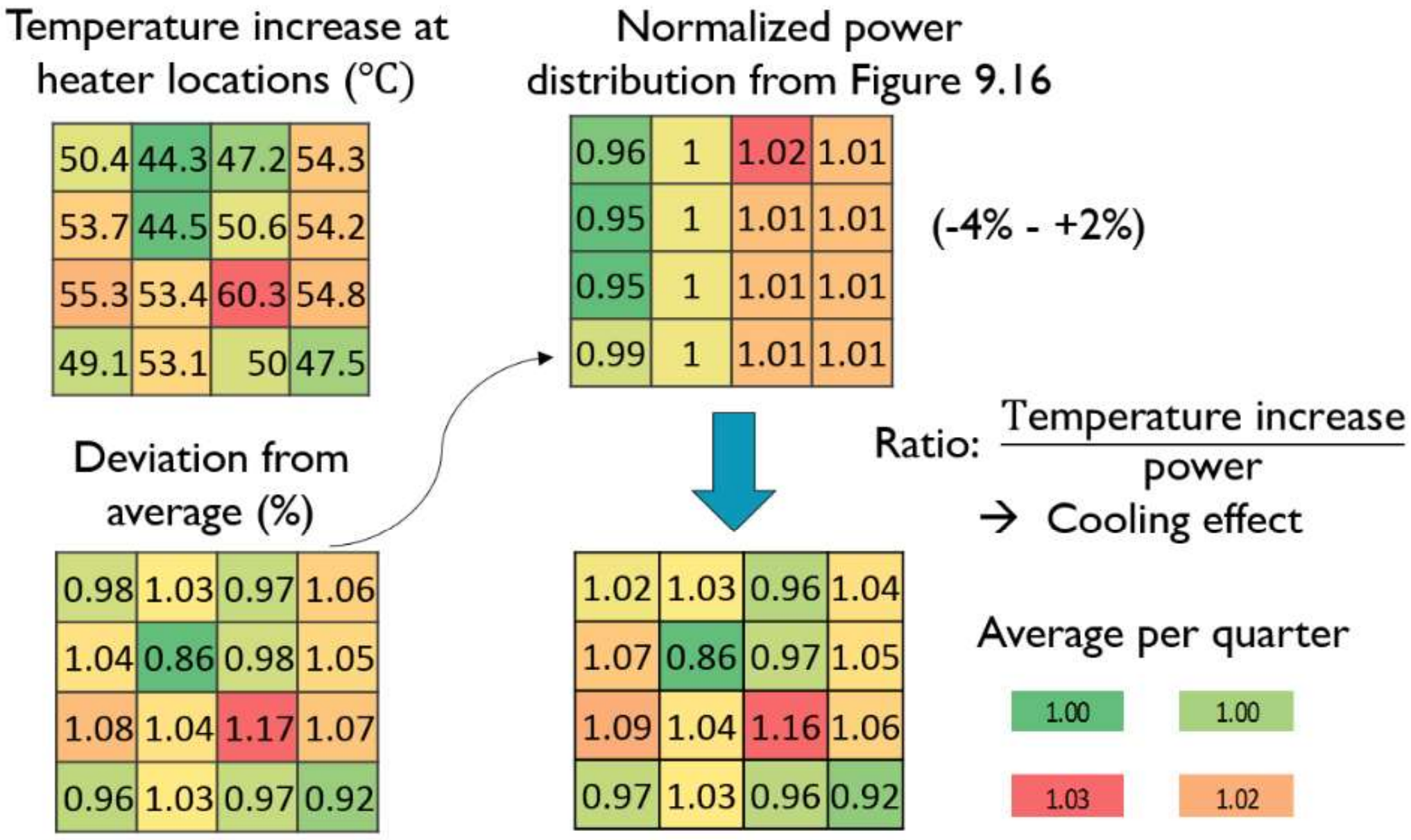

**Figure 9.27:** Temperature analysis with the cooling effect with ratio of temperature / power for the reference cooler (50V, 0.5 LPM).

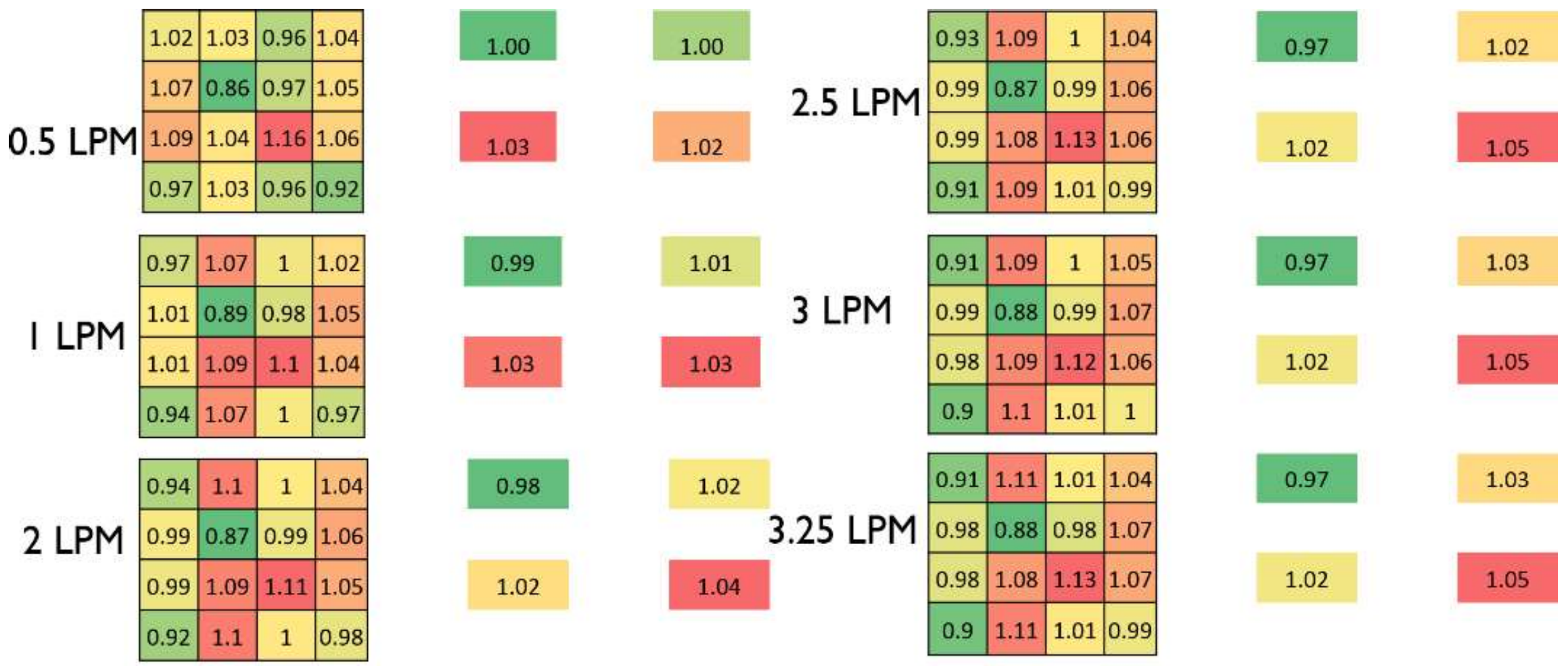

**Figure 9.28:** Consistent cooling effects distribution at different flow rates for an applied voltage of 50V.

Figure 9.28 shows the cooling effects at different flow rates for an applied voltage of 50V, which shows a consistent distribution for different flow rates. The ratio between the temperature variation and power variation suggests that this non-uniformity is not caused by the power non-uniformity.

The other contribution to the chip temperature non-uniformity is the distribution of the coolant flow in the cooler on the test chip. The cooler is rotated 90° to differentiate between the impact of flow and power non-uniformity, as illustrated in Figure 9.29. Figure 9.30 shows the T/P ratio comparison between the 90° counterclockwise rotated and the reference orientation. In the comparison, a rotation is observed in the temperature measurements which indicates that the temperature non-uniformity is caused by the flow non-uniformity. The alignment between the temperature variation and cooler inlet/outlet configuration in Figure 9.31 shows that the location of the worst value of the T/P ratio is opposite to the position of the exit connection. This is because the outlet flow should be collected through the outlet manifold and flow out through the exit outlet and as a result a lower flow rate is expected at that position, which is confirmed by the CFD model. In conclusion, the temperature non-uniformity is dominated by the coolant flow non-uniformity with the position of the exit port. This can be addressed by placing multiple exit ports at each corner of the cooler, resulting in more uniform flow.

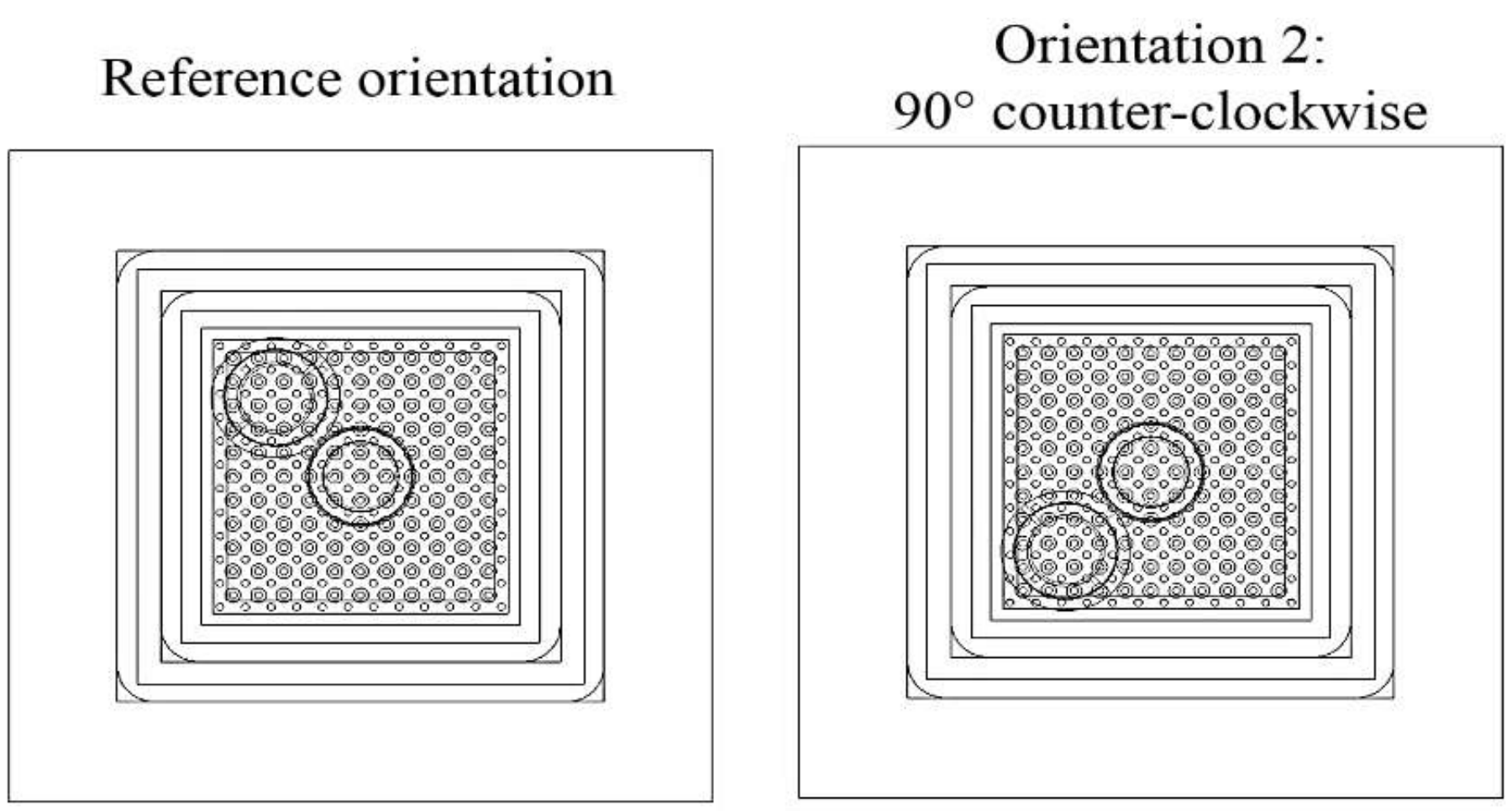

**Figure 9.29:** Cooler orientation comparison: rotate cooler to differentiate between flow and power non-uniformity.

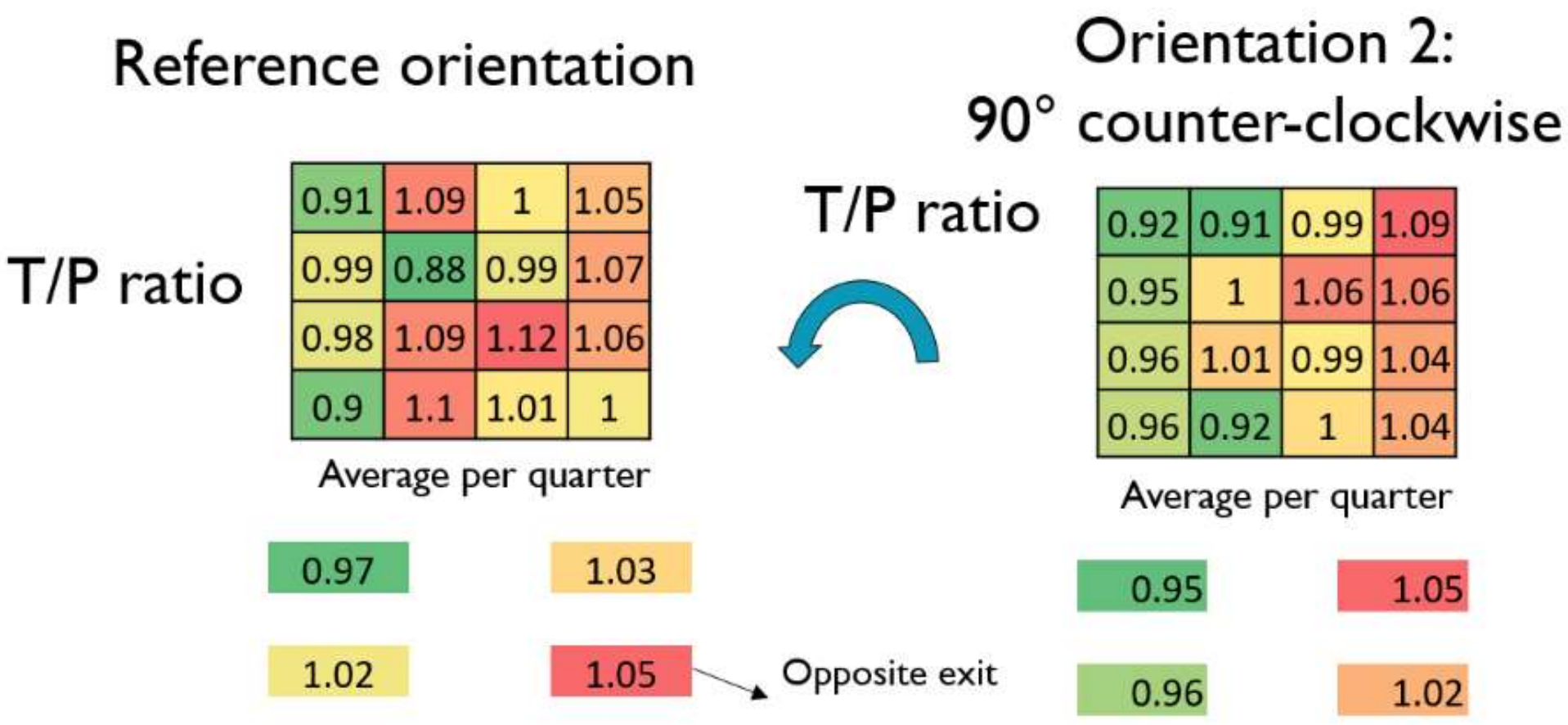

**Figure 9.30:** Cooling effects (T/P ratio) comparison between the reference orientation and $90^0$ counterclockwise orientation.

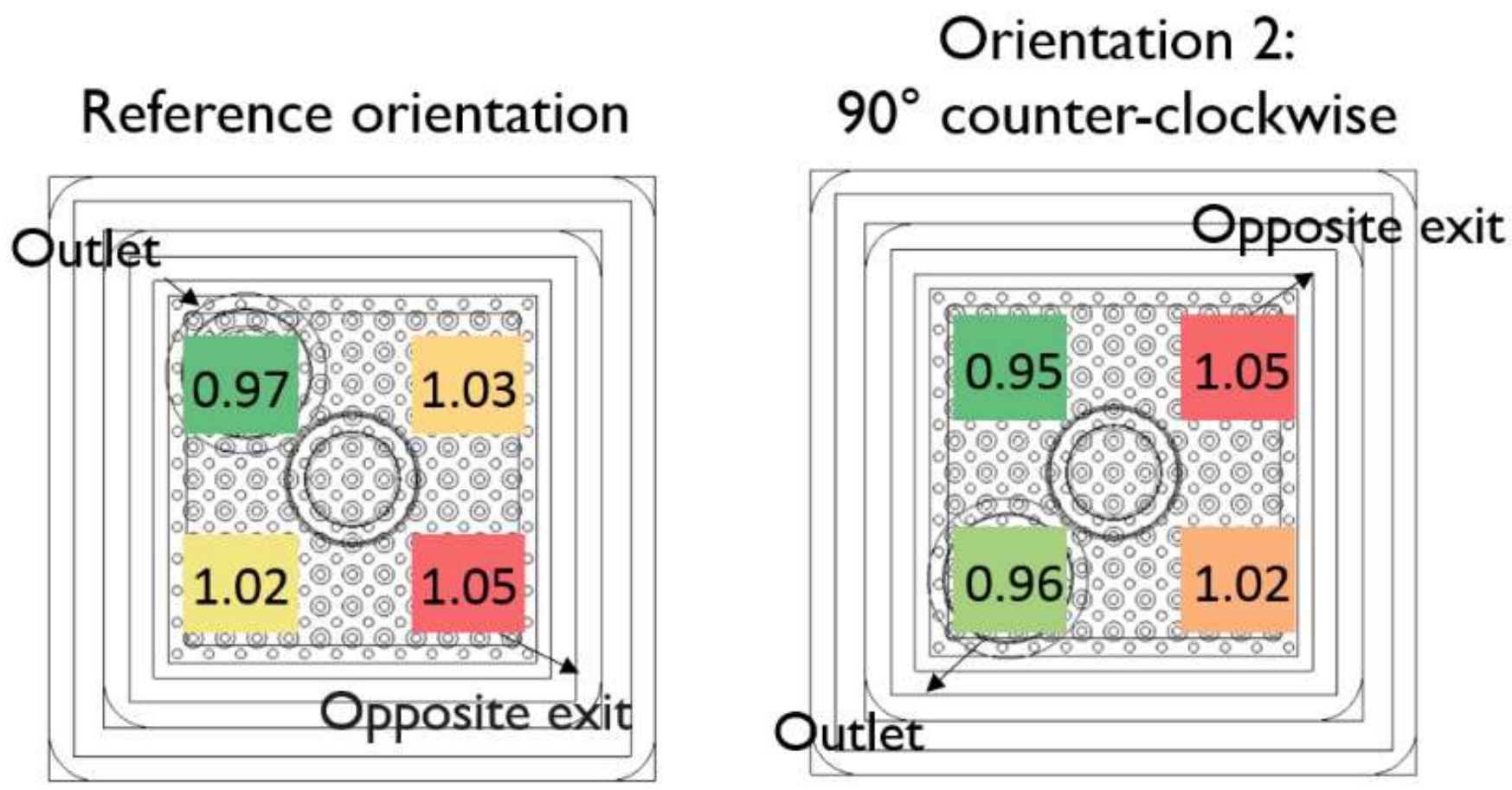

(a) (b)

**Figure 9.31:** Alignment with the temperature variation map with the location of the exit port: (a) reference orientation; (b) $90^0$ counterclockwise orientation.

## 9.4 Improved large die cooler design

This section will discuss the temperature and hydraulic performance of the additional distribution layer (design 2), applied on the lidless package. The temperature distribution for the improved design is measured for different flow rates, from 0.5 LPM to 3.25 LPM. As shown in Figure 9.32, the temperature distribution measured with 24 sensors (one sensor failed) shows better temperature uniformity than the first design in Figure 9.22.

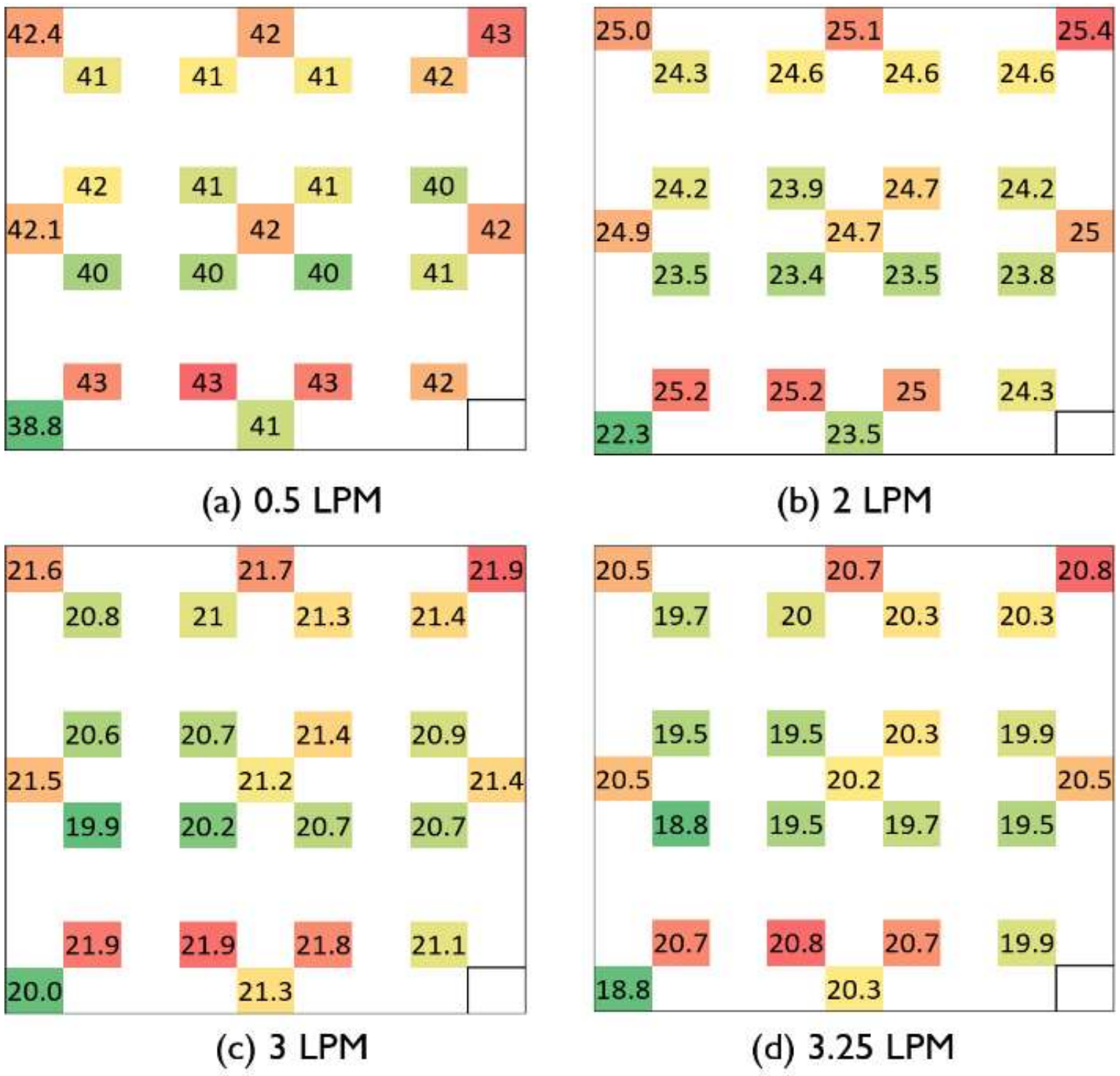

**Figure 9.32:** Measured temperature distribution map (°C) for different flow rates: (a) 0.5 LPM; (b) 2 LPM; (c) 3 LPM; (d) 3.25 LPM.

Moreover, the temperature profiles across the chip diagonal with different locations are plotted in Figure 9.33. Figure 9.33(a) shows the temperature profile from the top left to bottom right across the chip diagonal. The temperature profile for design 1 shows larger temperature gradient especially at low flow rate of 0.5 LPM, while the temperature distribution for design 2 is more uniform. This is due to the uniform inlet flow distribution by using the additional manifold layer. In addition, it can be seen that the improved design shows lower temperature than design 1 under the flow rate of 0.5 LPM. This is because that the heat conduction dominates under the low flow rate.. A similar trend is observed in Figure 9.33(b) across the other chip diagonal.

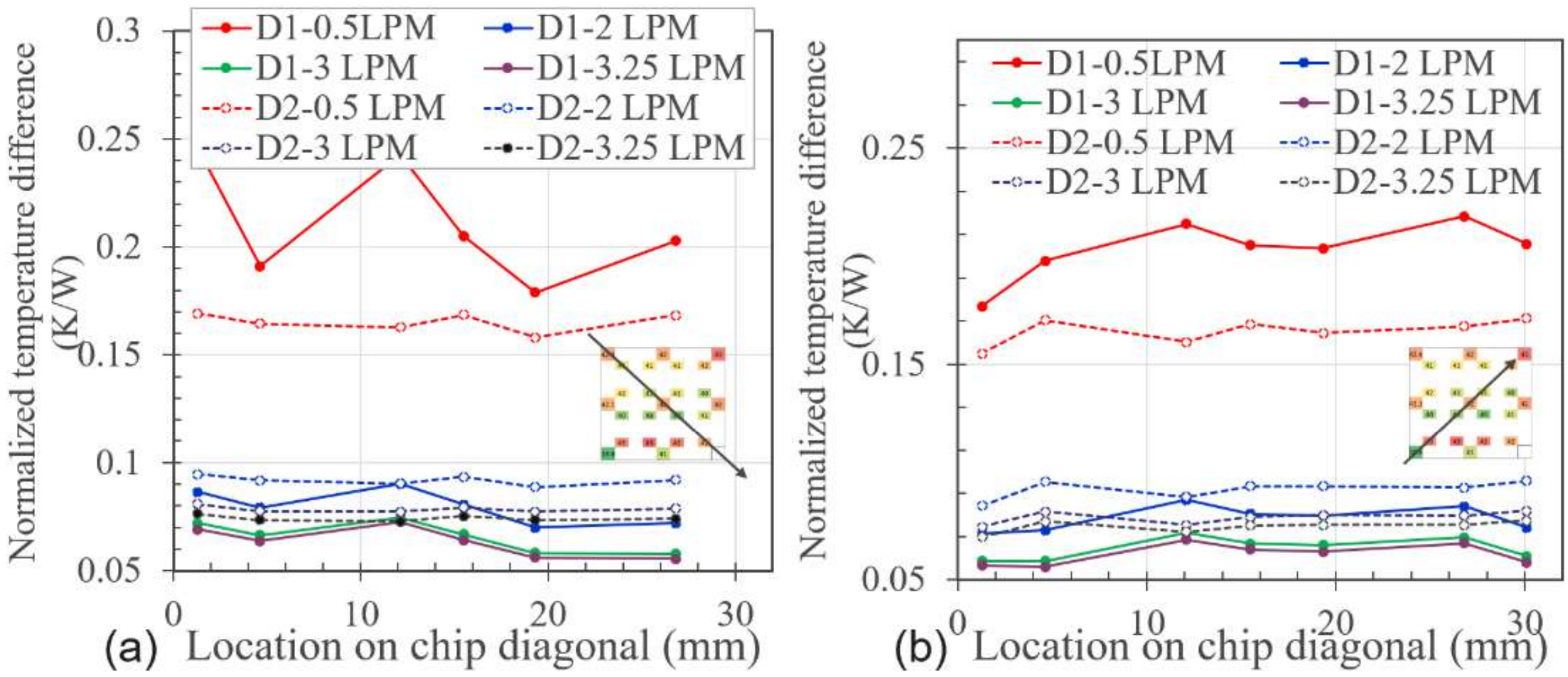

**Figure 9.33:** Temperature profile comparison between design 1 and design 2: (a) from top left to bottom right along the chip diagonal; (b) from bottom left to top right along the chip diagonal position.

Furthermore, the comparison of the thermal performance of the reference cooler (design 1) and the cooler with the additional distribution layer (design 2) is shown in Figure 9.34, in terms of the average chip temperature and the temperature uniformity. In general, the comparison shows that the average temperature difference is about 17% for a flow rate of 3 LPM, between the design 1 and design 2. It can be also observed that design 2 achieves a better chip temperature uniformity compared to the reference design, showing a factor of 4 times improvement for a flow rate of 0.5 LPM and 2.3 times improvement for a flow rate of 3 LPM.

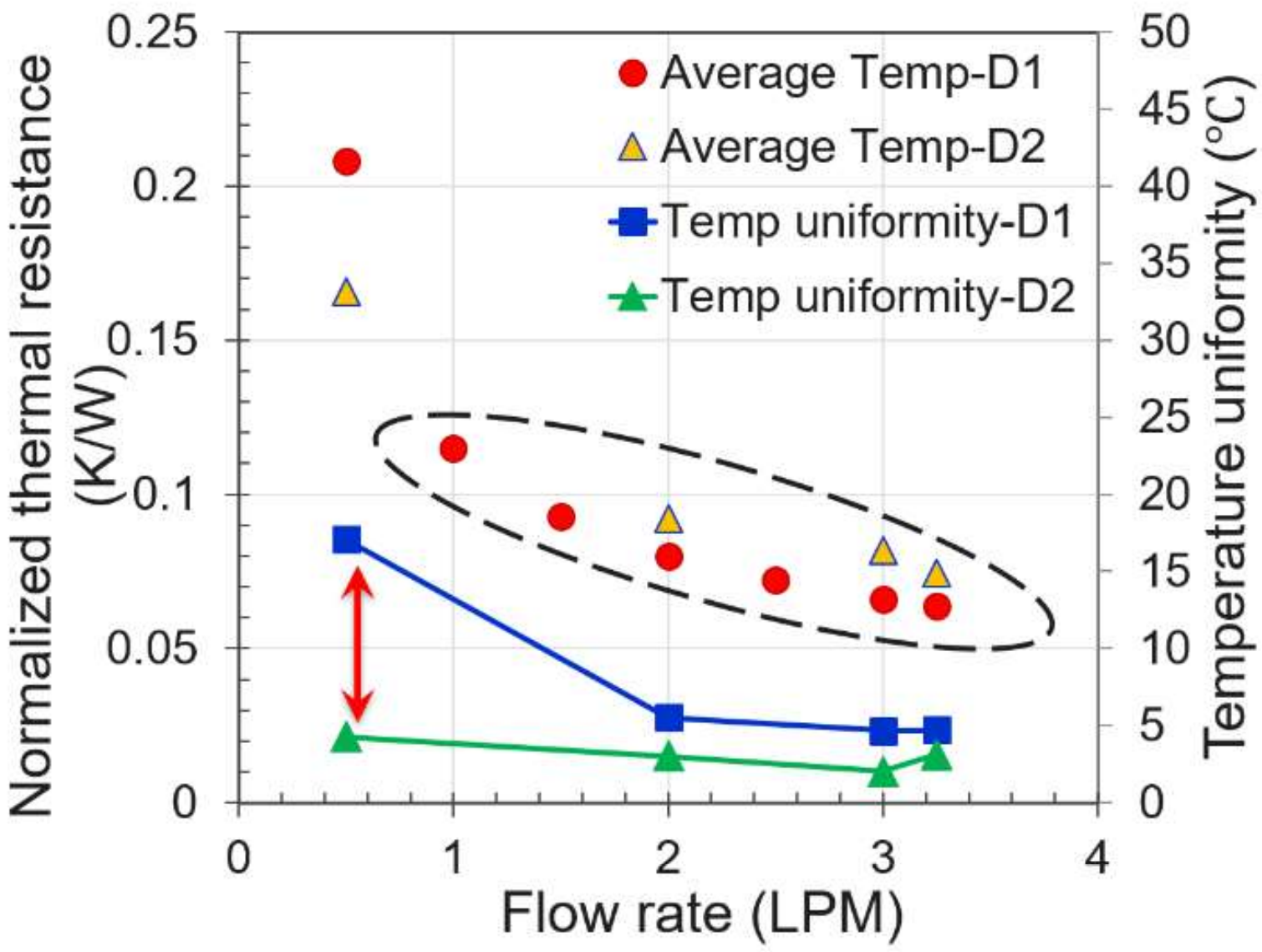

**Figure 9.34:** Averaged chip temperature (normalized thermal resistance) and temperature uniformity comparison between the design 1 and design 2.

Figure 9.35 shows the pressure drop experimental comparison for different flow rates. The pressure drop is measured between the inlet and outlet tube connectors. The pressure measurements show that despite the presence of the additional distribution layer in the cooler, the impact on the measured overall pressure drop is insignificant. These measurement results prove that the unique fabrication capabilities of additive manufacturing enable the design and fabrication of better large die coolers resulting in more uniform coolant flow distribution and temperature distribution, while limiting the pressure drop penalty caused by the additional required layers.

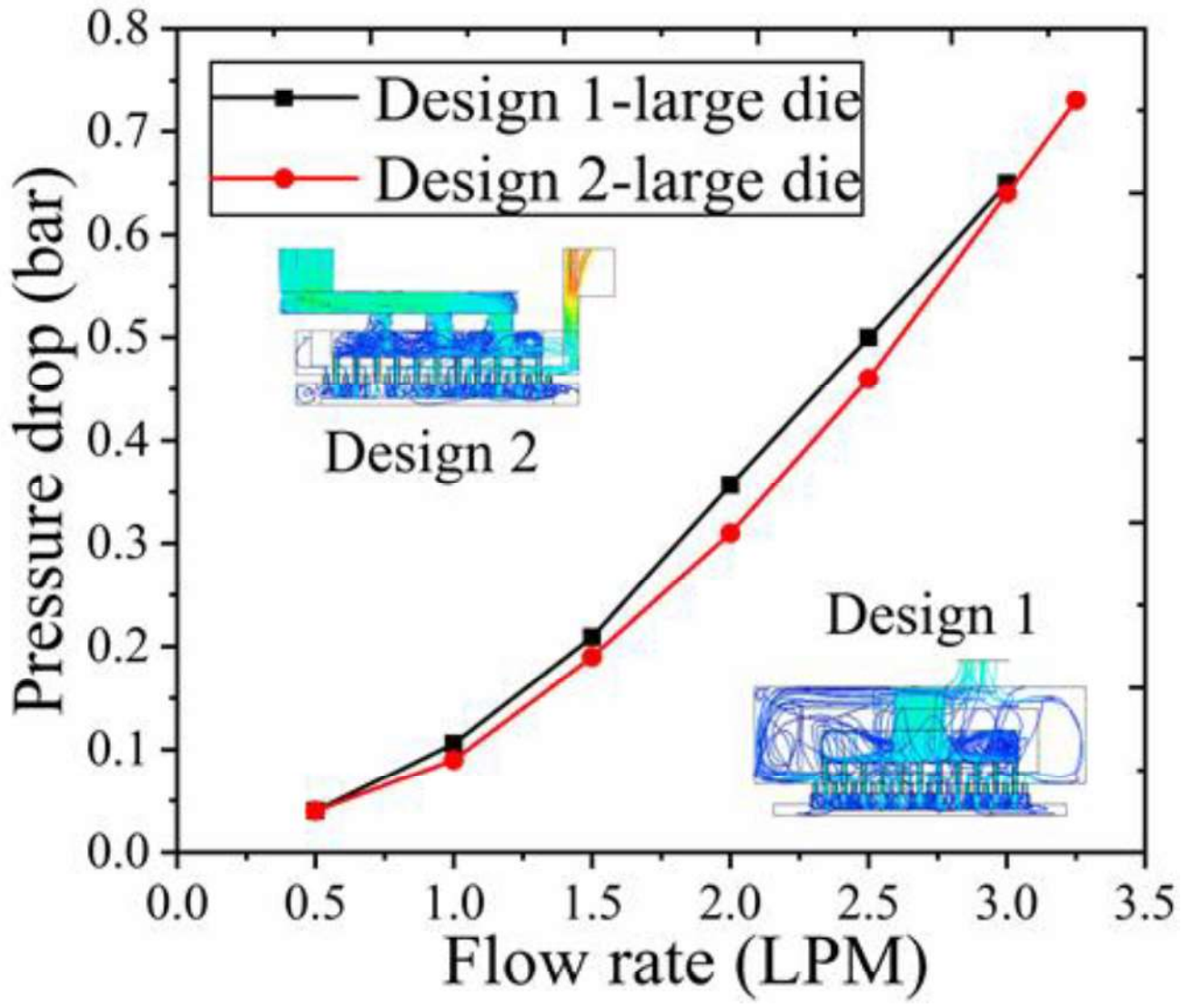

**Figure 9.35:** Measured pressure drop as function of flow rate for the two designed 3D printed large die coolers.

### 9.5 Model validation

Figure 9.36 illustrates the thermal resistance comparison between the measurement data (markers) and the modeling results (solid lines). In general, the experimental data for the chip temperature is lower than the CFD model, especially for low flow rates below 2 LPM. The small difference is attributed to the simplified CFD model, where the heaters on the large die model are assumed as uniform heating. Secondly, the bottom package including the substrate and PCB of the large die model is neglected in order to simply the model. Thus, the heat conduction path through the bottom of the heated die can also help to reduce the chip temperature. However, the difference between experiments and CFD modeling becomes smaller for higher flow rate, such as 3 LPM. This is because the heat conduction through the bottom package can be neglected compared with the heat convection on the chip surface by impingement jet cooling.

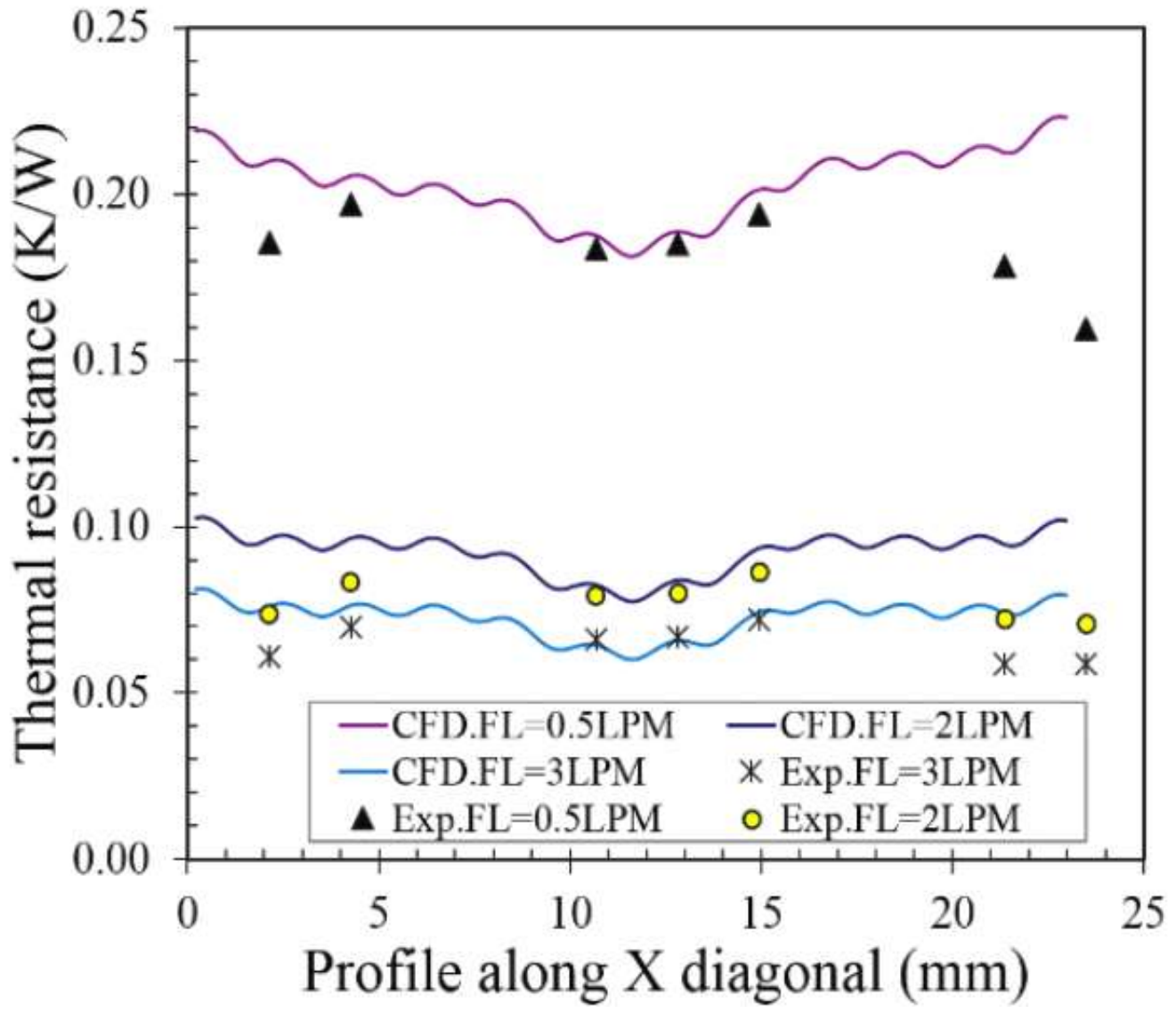

**Figure 9.36:** Experimental and CFD modeling results for the temperature profile comparison with lidless cooling for design 1.

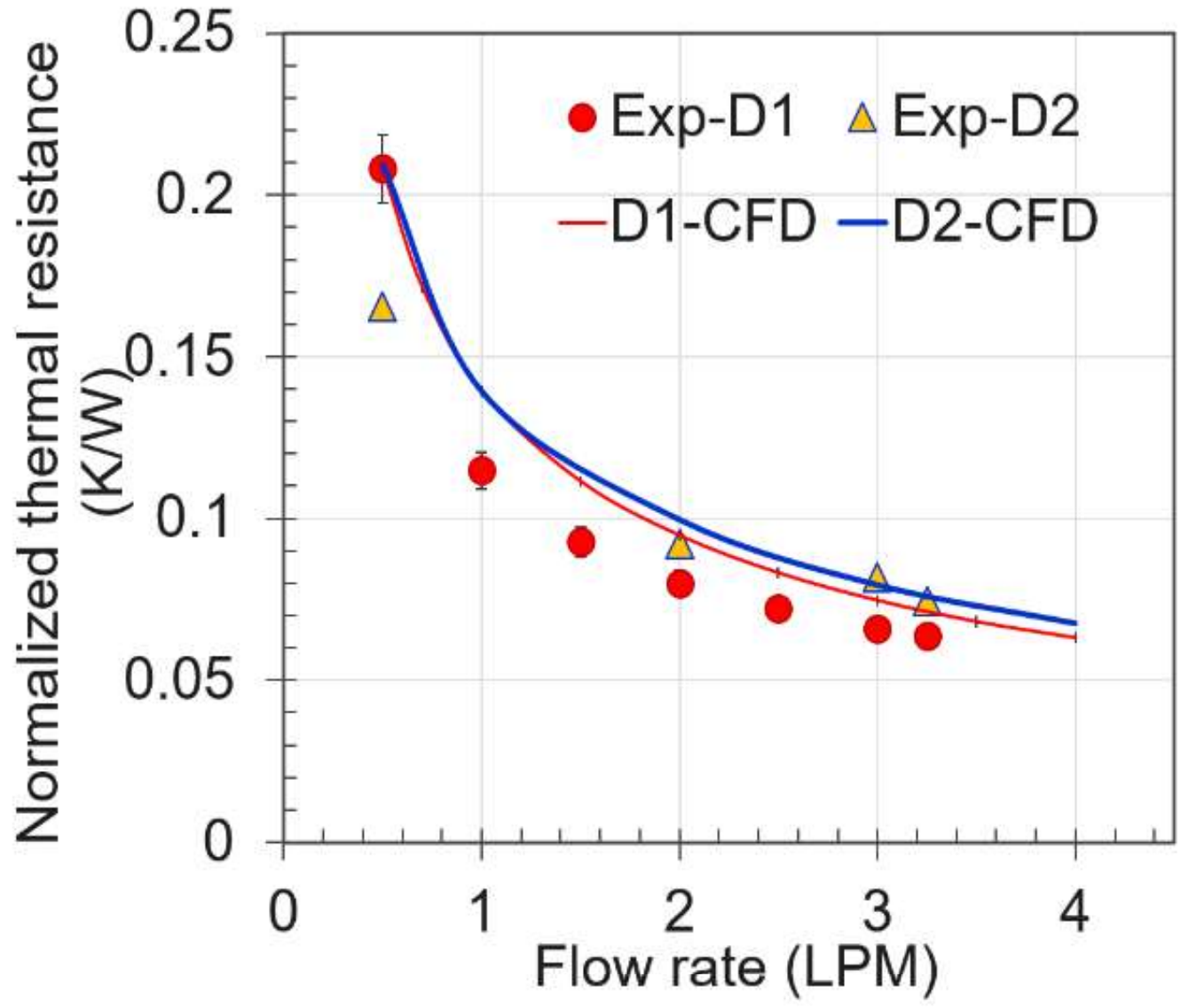

**Figure 9.37:** Experimental and CFD modeling results for the normalized thermal resistance as function of the flow rate.

## 9.6 Reliability investigations

### 9.6.1 Long-term cooling measurements

For direct on-chip liquid cooling, there are several concerns such as the cooling performance over time and potential reliability issues of the devices. In order to evaluate these aspects of the 3D printed microfluidic cooler, a longer-term measurement of the cooler is being conducted during 50 days in a closed loop system with DI water as coolant, where the chip temperature and ambient temperature are monitored. The cooler geometry and nozzle diameter are inspected before and after the long-term measurements. In order to perform the long-term measurement for the large die liquid cooling, a simplified set-up with an integrated pump and heat exchanger is developed for the thermal and flow measurement, shown in Figure 9.38. Temperature measurements are performed in all 25 sensors of the test chip during the long-term measurement. The test board is placed in the plastic tray to check for potential leakages that might occur during the test.

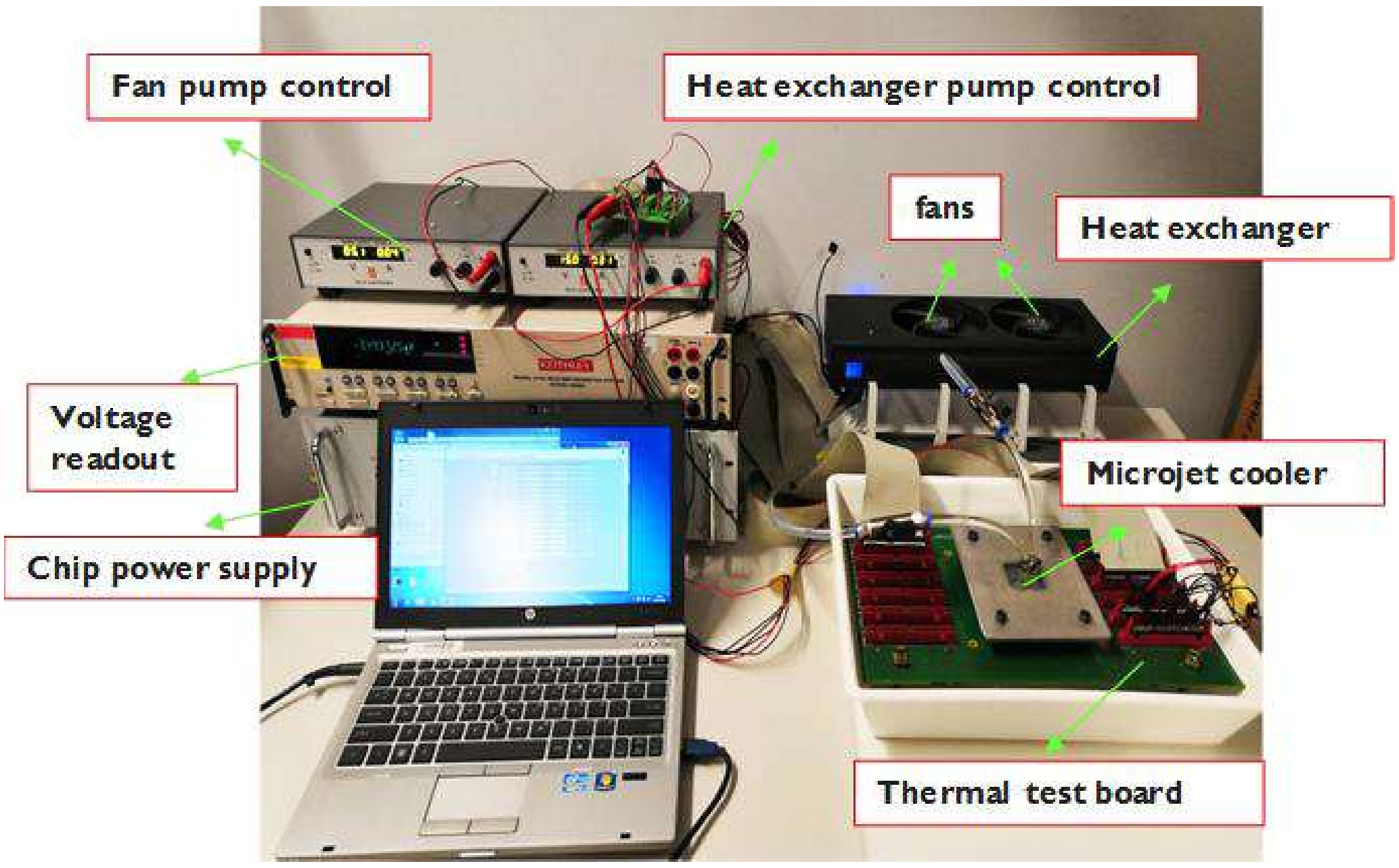

**Figure 9.38:** Simplified set-up developed for thermal and flow measurement.

For the test conditions, the measured actual power in the heaters of the test chip is 90 W for an applied voltage of 30V. The pump voltage is 12V and the heat exchanger voltage is 10 V. Since the flow rate is controlled by the pump voltage, the flow rate in this experiment is estimated by the thermal performance, which is estimated as 1.5 L/min based on the performance reported in Figure 9.23(b).

During the long-term measurement, the ambient temperature is also monitored as shown in Figure 9.39. It can be seen that the trend of the absolute temperature for all the sensors is consistent with the trend of the ambient temperature. Therefore, the reported temperature increase is defined as the average chip temperature with respect to the ambient temperature. Figure 9.40 shows that the thermal performance of the 3D printed large die cooler remains constant over the measurement period of 1000 hours. During this period, no reliability issues have been observed. Moreover, the temperature profile along the chip diagonal during the long-term measurement for every 400 hours are also plotted in Figure 9.41, showing stable thermal performance.

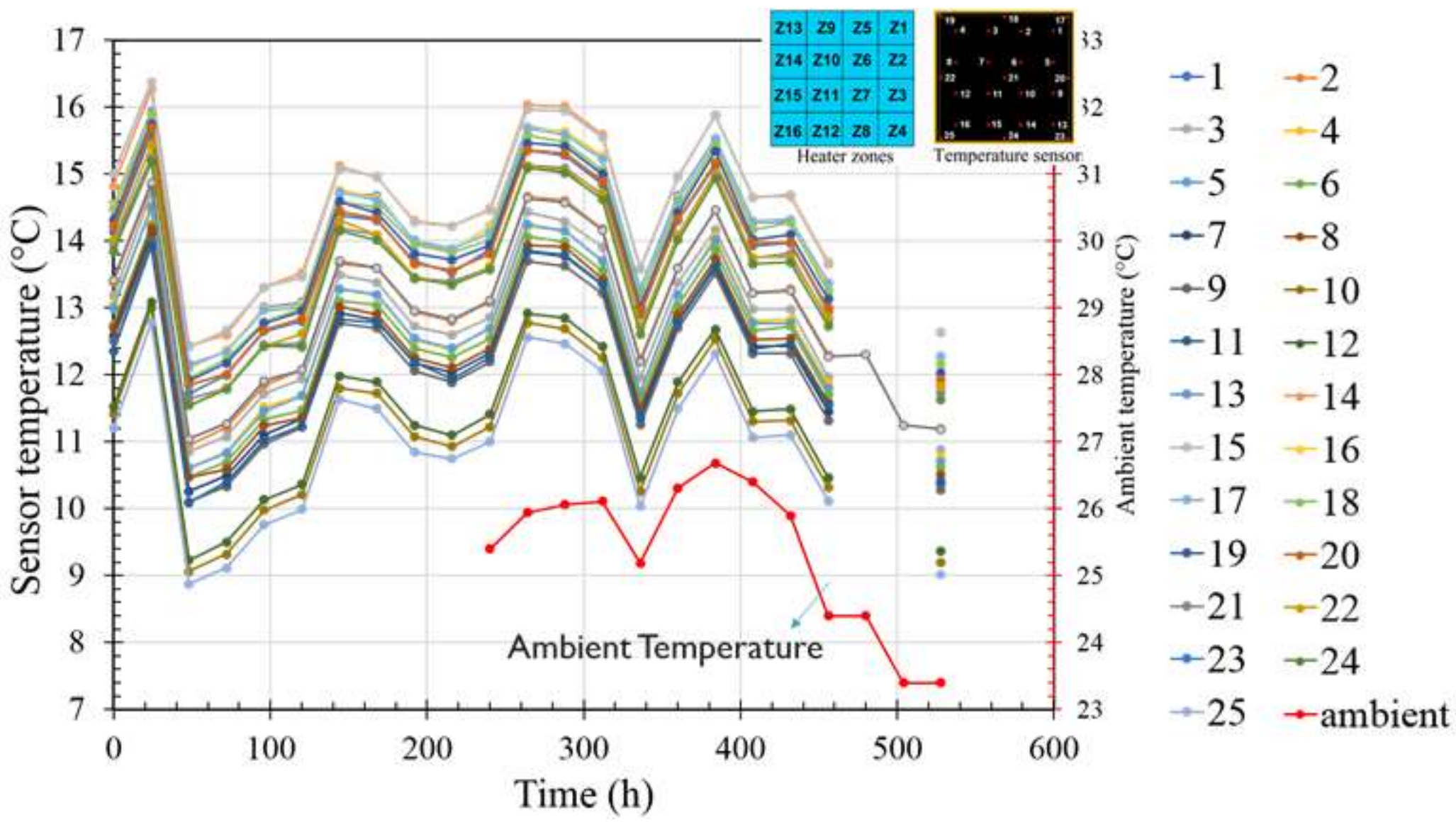

**Figure 9.39:** Trend for all temperature sensors and the ambient temperature during the long-term measurements, compared with the ambient temperature.

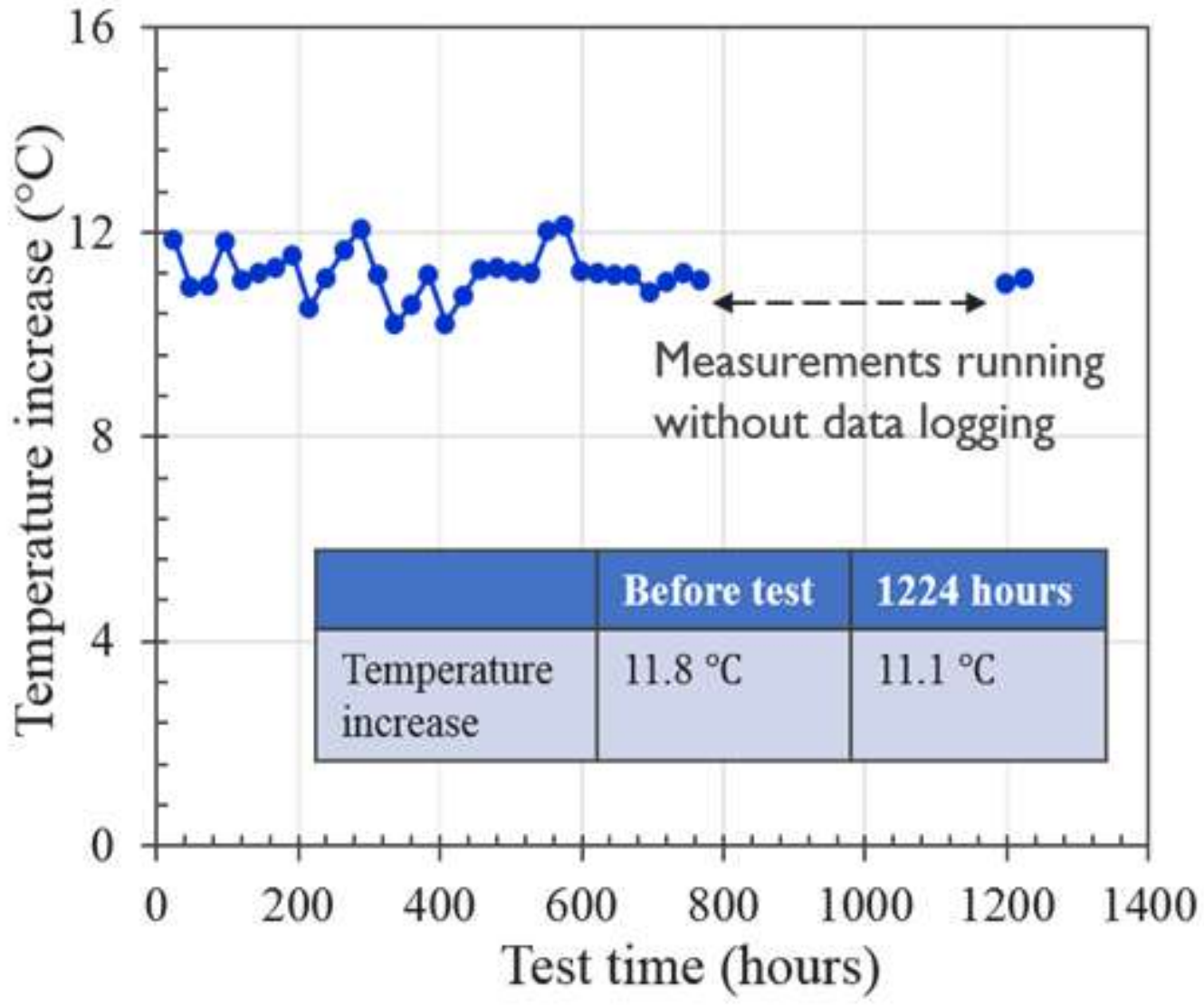

**Figure 9.40:** long term temperature measurement with regard to the coolant temperature of the impingement jet cooler over 1000 hours.

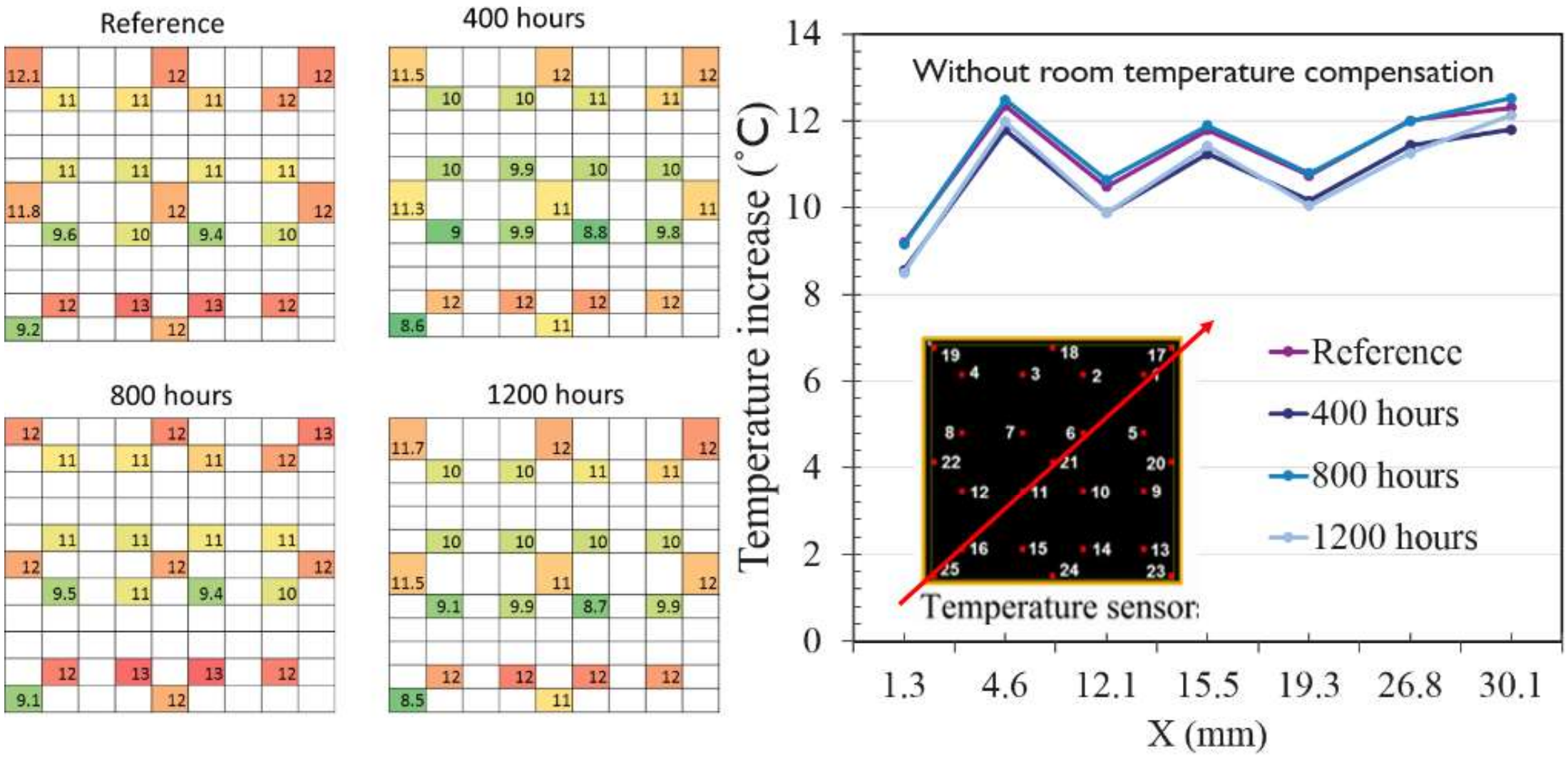

**Figure 9.41:** Temperature profiles comparison during the long-term measurements for every 400 hours.

### 9.6.2 Cooler geometry impact

In order to evaluate the nozzle diameter variation before and after the long-term measurement, the nozzle diameter is measured. From the cross-section analysis in Figure 9.42, no clogging of the nozzles or internal channels is observed despite the lack of filters in the simple test setup. Also, no erosion of the nozzles is observed. In addition, there is no significant difference for the nozzle diameters before and after the measurements, as show in Figure 9.43.

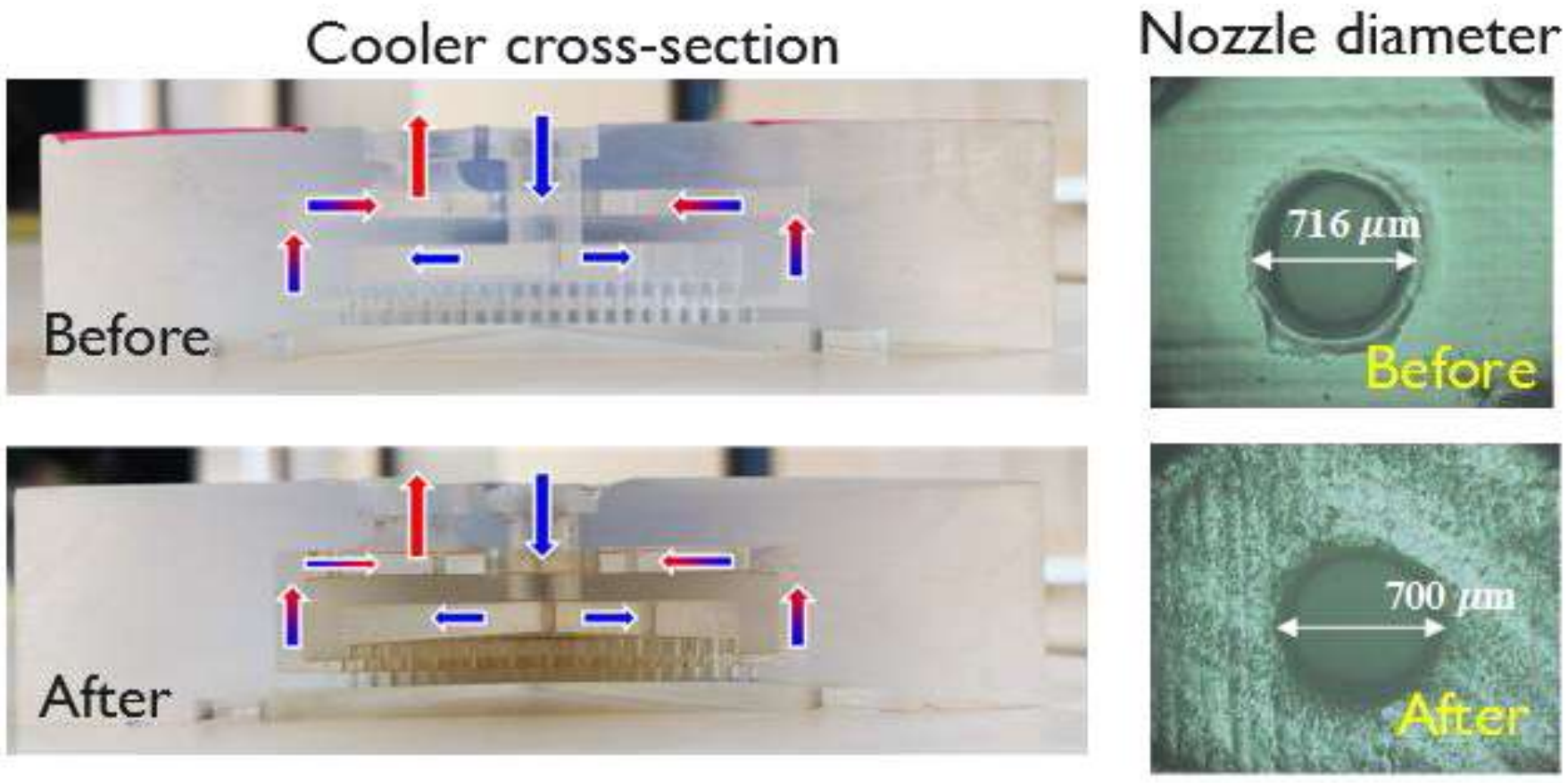

**Figure 9.42:** Cross section analysis after long term measurement with DI water for the 3D printed plastic cooler.

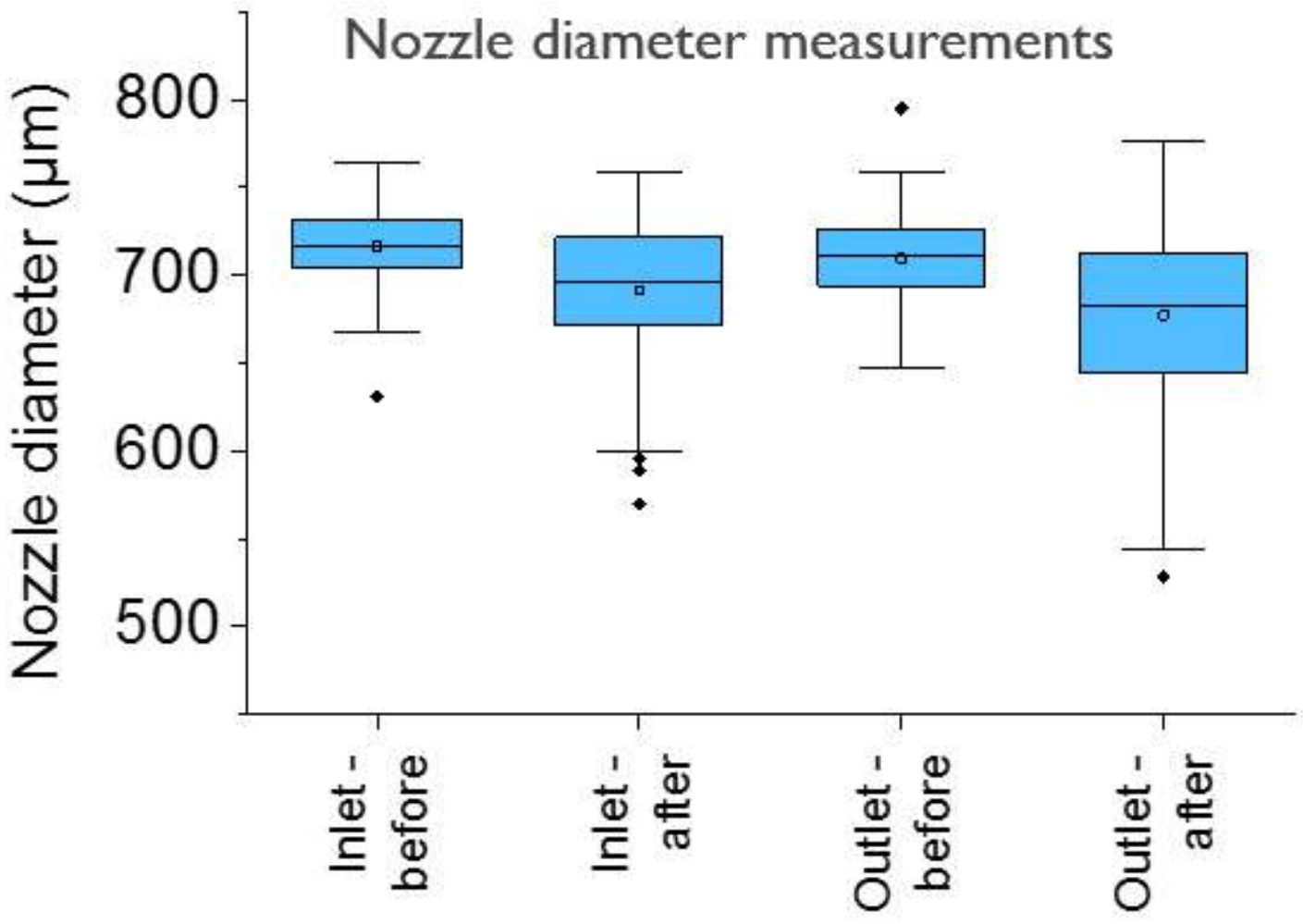

**Figure 9.43:** No significant difference for the nozzle diameters before and after the measurements.

## 9.7 Conclusions

To deal with the increasing challenges for the large die cooling with high power, this chapter experimentally and numerically investigates the design methodology, fabrication limitations, and the cooling performance of the large die cooler. Thanks to the large die test vehicle with 23×23mm² large die with 275W power dissipation, provided by GlobalFoundries, two versions of large die cooler are designed and fabricated. For both the designs, the nozzle array is a scaled version of the 4×4 nozzle array cooler for the PTCQ test chip with a nozzle pitch of 2 mm and a nozzle diameter of 600 µm. The scaled nozzle plate contains an 11×11 inlet array and a 12×12 array of outlets distributed in between the inlets.

Firstly, the experimental characterization based on the reference large die cooler is conducted. The measurements show that the average chip temperature increase is 17.5°C with a pressure drop of 0.7 bar, for a coolant flow rate of 3.25 LPM,. At that flow rate, the cooling on the bare die outperforms the cooling on the lidded package by 35%. The temperature non-uniformity is investigated in detail, which shows that there is a significant impact from the coolant flow distribution. After that, the improved large die cooler with additional distribution layer is experimentally characterized and compared with the reference large die cooler. The comparison shows that improved design with additional layer achieves a better chip temperature uniformity compared to the reference design, showing a factor of 4 times improvement for a flow rate of 0.5 LPM and 2.3 times improvement for a flow rate of 3 LPM.

Lastly, a longer-term measurement of 1000 hours of the cooler has been conducted in a closed-loop liquid coolant system with DI water, where the chip temperature and ambient temperature were monitored. The cooler geometry and nozzle diameters are inspected before and after the long-term measurements. The measurements show that the thermal performance of the microfluidic cooler remains constant over the measurement period of 1000 hours. During this period, no reliability issues have been observed.

In summary, this chapter demonstrates that additive manufacturing enables the accurate fabrication of complex internal structures in multiple layers inside the 3D printed large die cooler as one single piece, allowing the creation of additional structures to improve the flow and temperature uniformity without significant increase of the pressure drop over the cooler, with constant cooling performance over the measurement period of 1000 hours.

**References**

[1] Haissam El-Aawar, Increasing the Transistor Count by Constructing A Two-Layer Crystal Square on A Single Chip,

[2] Joel Hruska, "Intel Details Its Nervana Inference and Training AI Cards", https://www.extremetech.com/computing/296990-intel-nervana-nnp-i-nnp-t-a-training-inference, August 21, 2019

[3] Reaching new heights with the world's largest FPGA, https://www.xilinx.com/products/silicon-devices/fpga/virtex-ultrascale-plus-vu19p.html#advantage, 2019.

[4] Nervana Neural Network Processor - Intel AI, https://www.intel.ai/nervana-nnp/, 2019.

[5] AMD EPYC™ 760, https://www.amd.com/en/products/cpu/amd-epyc-7601, 2019.

[6] CS-1 is powered by the Cerebras Wafer Scale Engine - the largest chip ever built, https://www.cerebras.net/, 2019.

[7] IBM z15 (8561) Technical Guide, https://www.redbooks.ibm.com/redpieces/pdfs/sg248851.pdf, 2019.

[8] DSM Somos WaterShed® 11122, https://www.protolabs.co.uk/media/1010047/somos-watershed-en.pdf.

# Chapter 10

# 10. Advanced Manifold Level Design Methodology

## 10.1 Problem statement

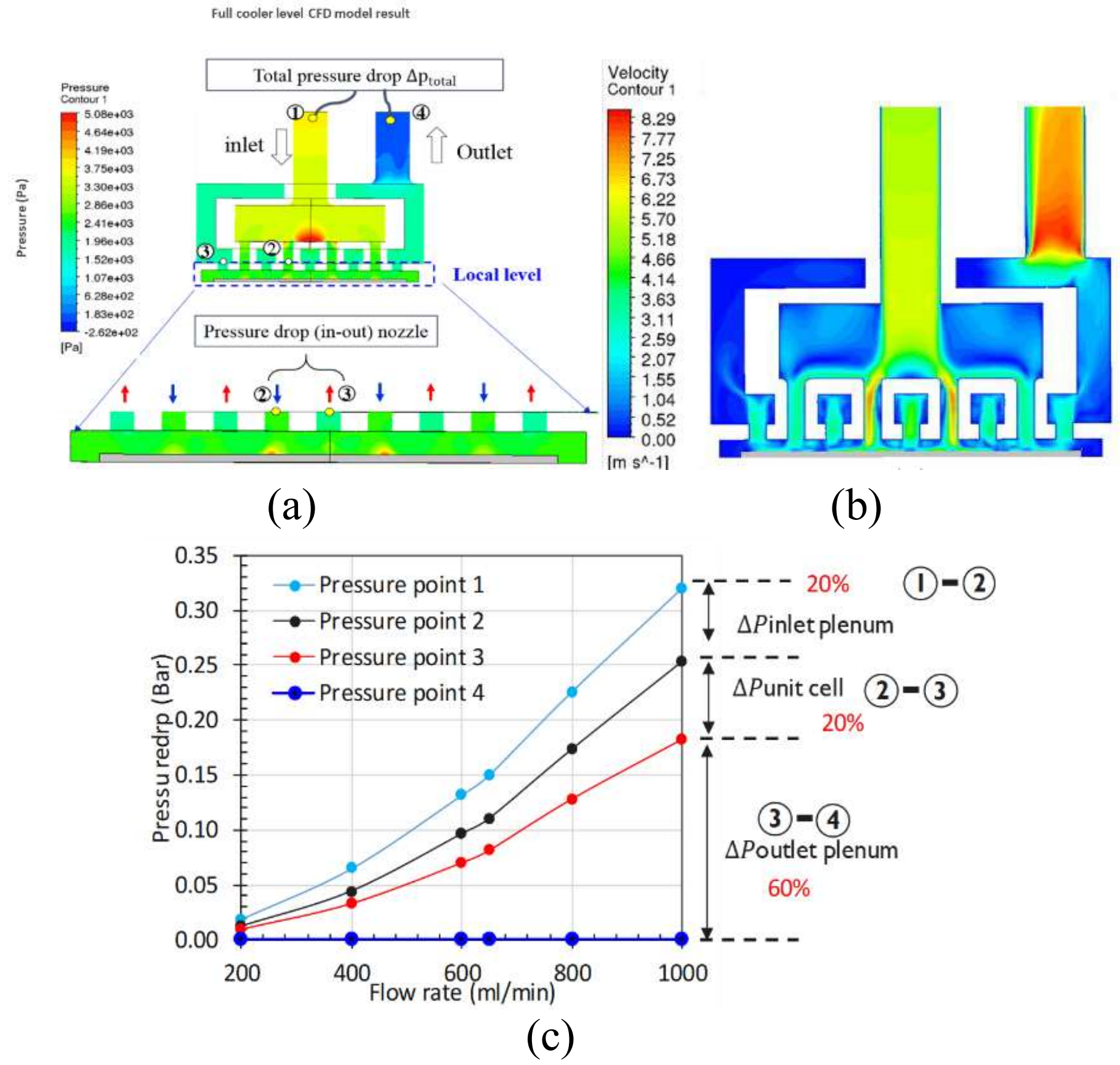

**Figure 10.1:** Hydraulic Impact of the manifold level: (a) pressure distribution. (b) velocity distribution and (c) analysis of the contributions of the different parts in the coolant flow (from chapter 2: Figure 2.12).

In the section of 2.1.3.4 B of chapter 2, the importance of the manifold level optimization is discussed. It is shown that the manifold level design of this microfluidic cooler is very important for the overall cooler performance, since it determines the flow uniformity, and system level pressure, especially for large area die size applications. As illustrated in Figure 10.1, the pressure drop analysis of the 3D printed full cooler level shows that the manifold level is responsible for majority (80%) of cooler pressure drop, including the inlet and outlet manifold. Furthermore, the inlet manifold defines the coolant flow distribution over the chip. The flow uniformity can further determine the temperature gradient across the chip surface, which is important to improve the design. This chapter will focus on the design improvements of the inlet manifold of the cooler geometry in order to improve the flow and temperature gradient, and also the pressure drop of the cooler.

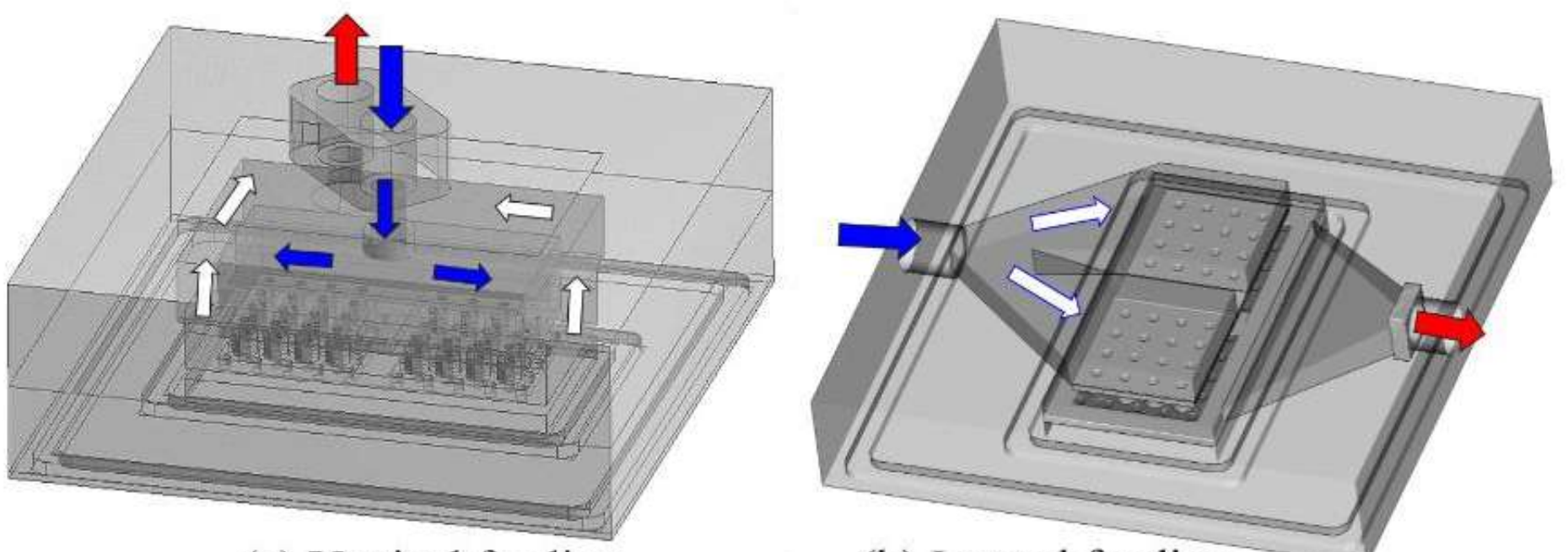

**Figure 10.2:** CAD design and extracted 3D models for the 2.5D interposer cooler design: (a) vertical feeding; (b) lateral feeding.

For the test case of the manifold design, there are mainly two design schemes used in the previous study. In chapter 8, the vertical feeding and lateral feeding scheme are introduced and demonstrated for the cooling of 2.5D interposer package. The two designed demonstrators and the extracted 3D models are summarized in Figure 10.2. As for the vertical feeding scheme, the inlet flow is vertically feed from the top part of the inlet manifold and distributed over the bottom nozzles. For the lateral feeding, the inlet feeding flow is coming from the inlet manifold left/right. In chapter 8, the experimental and numerical comparison show that the cooler with the lateral feeding manifold requires 60% less pumping power with respect to the vertical feeding manifold for the same high thermal performance while reducing the overall thickness of the cooling solution by a factor of 2. This study shows that the improvement of the inlet manifold is very crucial for the pressure reduction and flow uniformity. In this chapter, two design methodologies are introduced: conceptual design innovation and topology optimization. Conceptual design innovation is based on the innovative design, by changing the cooler manifold shape, while topology optimization is an automated design method with defined objectives.

## 10.2 Conceptual design innovation

As for the conceptual design innovation [1], there were already lots of examples implemented for the microchannel cooling, such as pin fins [2, 3] and hybrid microchannel/jet cooling [4], and branched flow networks [5]. Since 3D printing can enable the freeform fabrication with many design options. In this section, three innovative designs are proposed and compared with the initial standard design. The thermal and hydraulic performance are analyzed based on the CFD modeling results. In section 10.2.1, the mushroom manifold design is proposed to reduce the pressure drop and improve the flow/temperature gradient. The pressure drop reduction is due to the increase of the outlet manifold volume; In section 10.2.2, the isolated jet nozzles are studied by reducing the nozzle to chip surface distance; In section 10.2.3, the finger-shape manifold design is proposed to reduce the cooler thickness. All the conceptual design will be compared with the standard design shown in chapter 6.

### 10.2.1 Mushroom manifold

As illustrated in the flow distributions of the multi-jet impingement cooling in Figure 10.1(c), the inlet flow goes into the inlet chamber, showing a mushroom shape. However, the flow at the top corner of the manifold consumes lots of pressure. Therefore, a mushroom shape inlet manifold is proposed to reduce the pressure drop. The CAD design structure is shown in Figure 10.3, with the internal visualization and the cross-section view of the cooler.

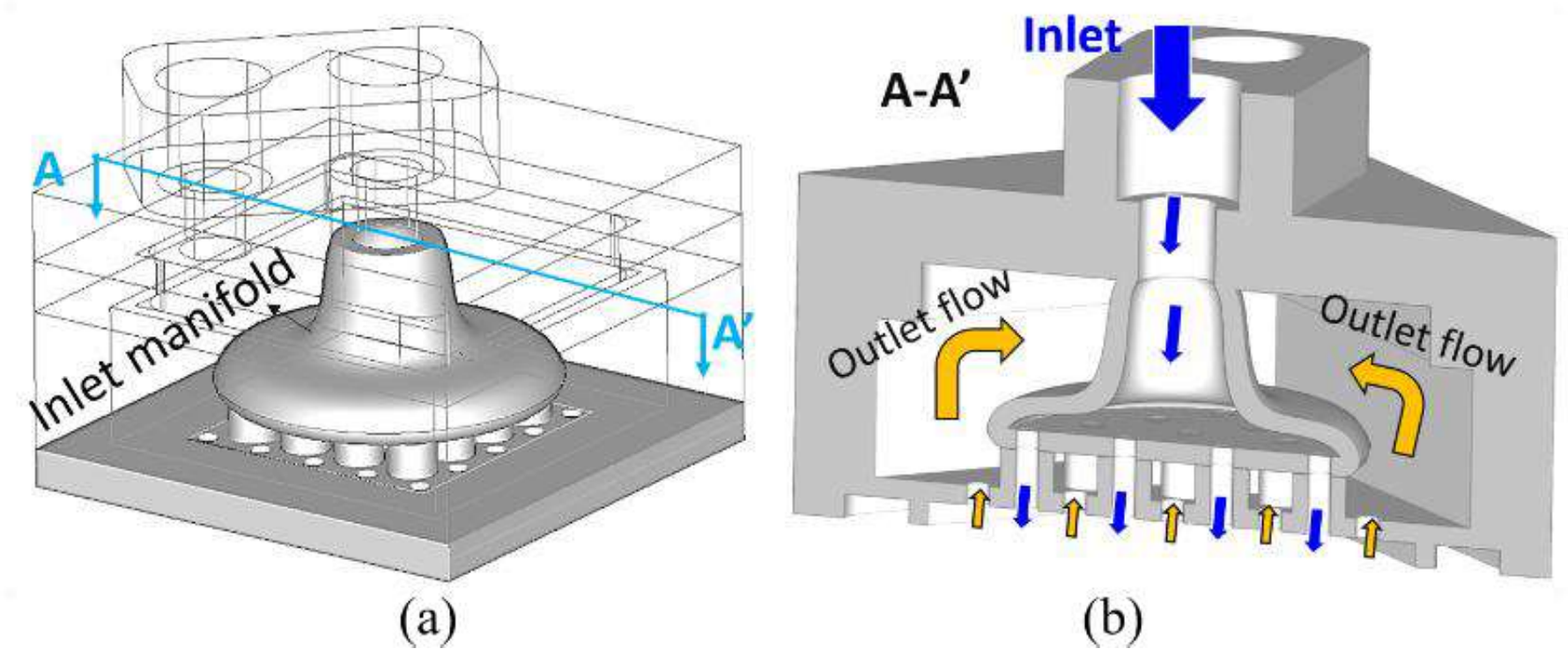

**Figure 10.3:** Mushroom manifold design: (a) internal visualization of the mushroom inlet manifold design; (b) cross section view of the new design with indication of the flow directions.

For the comparison of the mushroom design and standard design, the CFD modeling is conducted to evaluate the chip temperature and pressure drop. The flow rate used in this comparison is 1 L/min. The chip power applied is 50 W, on an 8×8 $mm^2$ chip. The flow distribution for the two designs are shown in Figure 10.4. It can be seen that much more design space is transferred to the outlet manifold. Moreover, the velocity distribution across the inlet nozzles shows better flow uniformity. In Table 10.1, the averaged chip temperature, temperature gradient and pressure drop for both designs are compared. It shows that the mushroom design can reduce the pressure drop by a factor of 1.4, 1.1 x reduction for the average chip temperature and also 1 x for the temperature gradient.

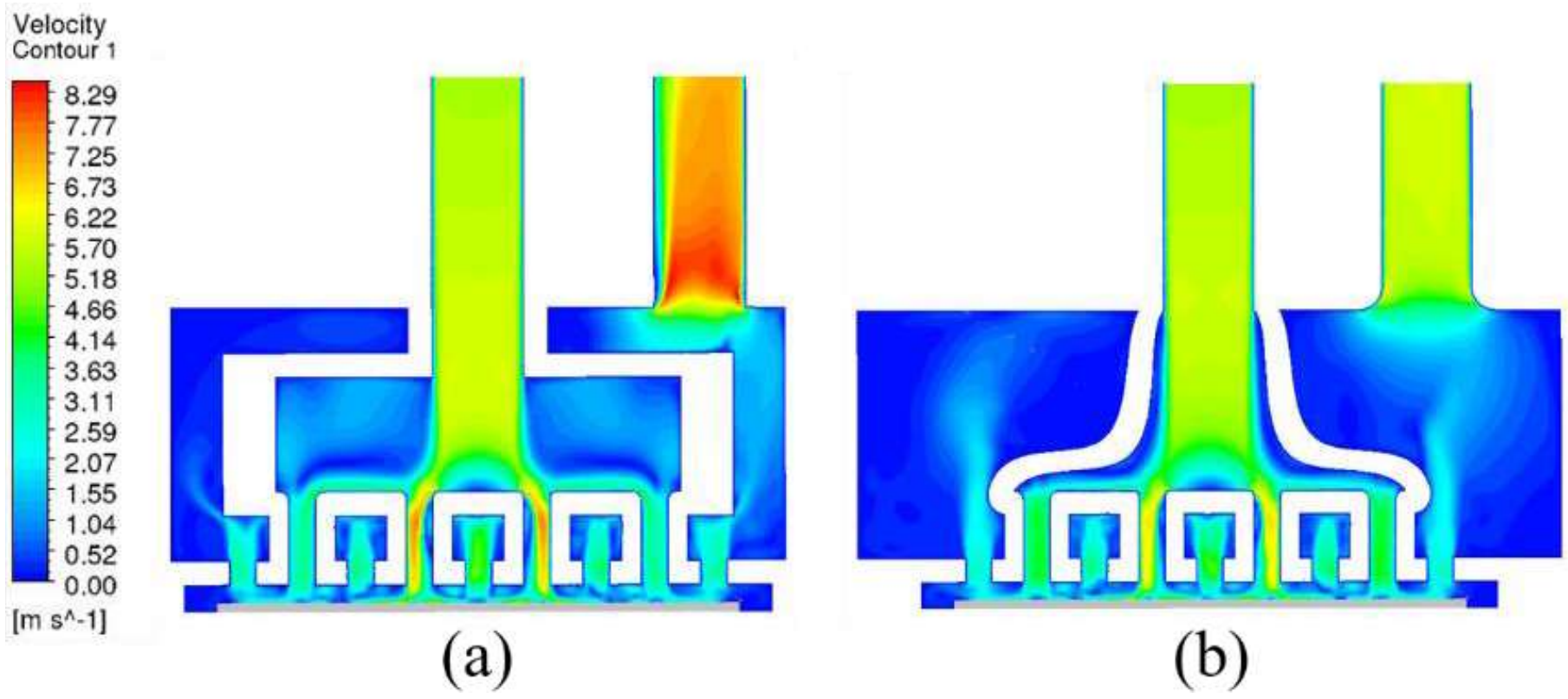

**Figure 10.4:** Velocity distribution comparison for the (a) standard cooler design and (b) mushroom design (chip power=50W, flow rate=1 L/min).

**Table 10.1:** Performance comparison between the standard design and mushroom design.

| Design | Pressure drop (Pa) | Averaged chip temperature (℃) | Temperature gradient (℃) |
|---|---|---|---|
| Standard design | 45128.7 | 24.92 | 3.4 |
| Mushroom design | 30467.6 | 21.85 | 3.12 |

### 10.2.2 Isolated nozzles

The standard design shown in Figure 10.5(a) has locally distributed outlets, which is intended to remove the outlet liquid fast. However, the outlet flow needs to go through the outlet nozzle, resulting in additional pressure drop inside the outlet nozzles. In order to avoid this extra pressure drop, isolated inlet nozzles are used for this study. As shown in Figure 10.5(b), the outlet nozzles are replaced by an open area inside the manifold.

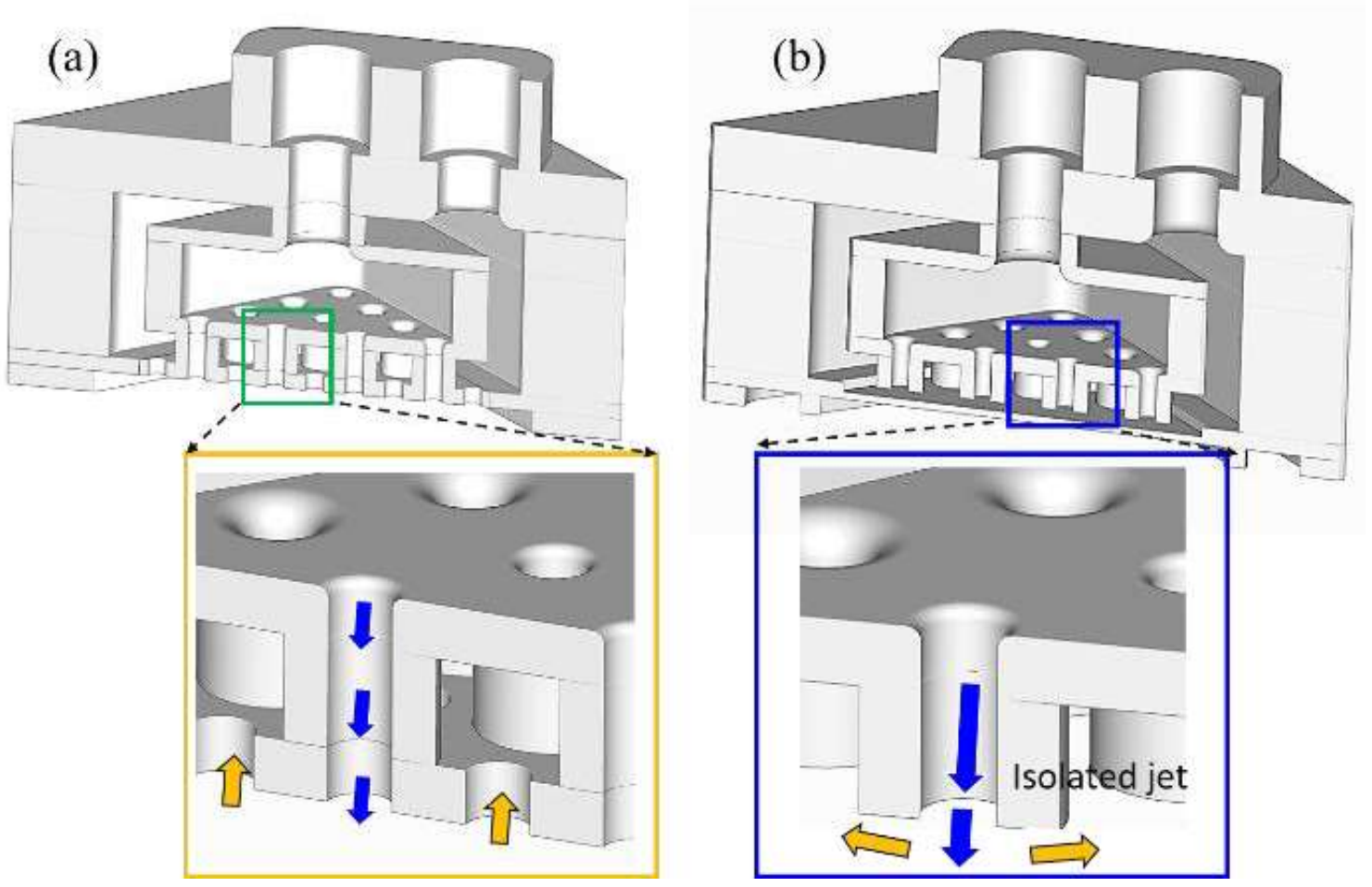

**Figure 10.5:** (a) Initial standard design with locally distributed outlets; (b) Isolated jet for outlet manifold level design.

The temperature and pressure drop comparison between the initial vertical feeding design and isolated jet design is summarized in Table 10.2. In general, the isolated jet cooling shows worse thermal performance than the initial design with locally distributed outlets. This can be explained in Figure 10.6, where the confined outlet nozzle plate can confine the wall jet region on the cooling surface, resulting a lower temperature. The open area outlet flow can result in cross flow effects inside the cavity, which can influence the temperature gradient of the chip.

**Table 10.2:** Performance comparison between the initial design and mushroom design.

| Design | Pressure drop (Pa) | Averaged chip temperature (℃) | Temperature gradient (℃) |
|---|---|---|---|
| Standard design | 45128.7 | 24.92 | 3.4 |
| Isolated jet | 42591.7 | 26.5 | 9.9 |

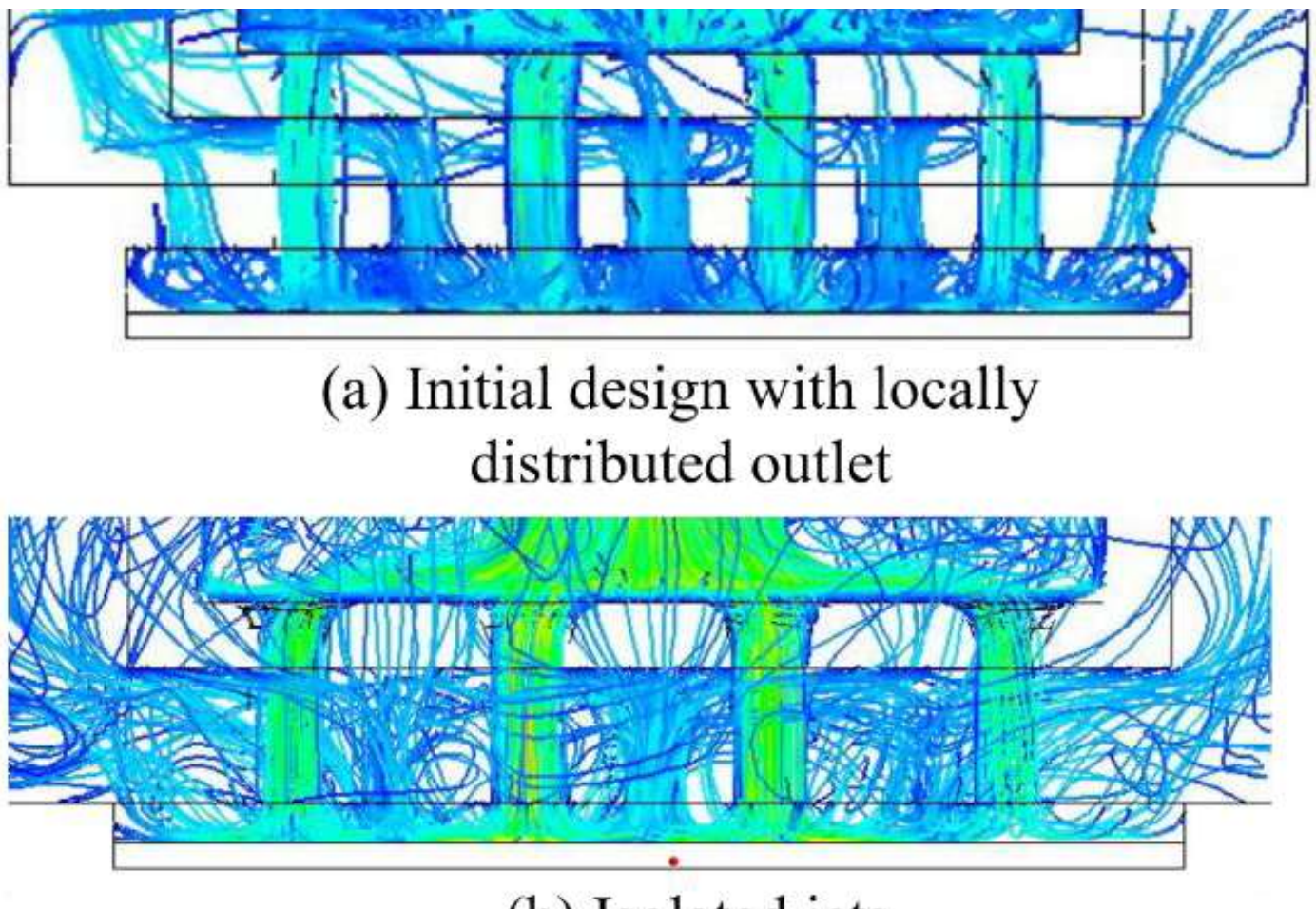
(a) Initial design with locally distributed outlet

(b) Isolated jets

**Figure 10.6:** (a) Initial standard design with locally distributed outlets; (b) Isolated jet for outlet manifold level design.

### 10.2.3 Finger-shape manifold

In order to compatible with the chip package, the cooler size should be match with the chip size, and also, the cooler thickness should also be as thinner as possible. One possible solution is to reduce the inlet manifold thickness and cavity height, as indicated in Figure 10.7(b). The inlet chamber thickness is only 0.8 mm while the inlet chamber thickness is 2.5 mm for standard design.

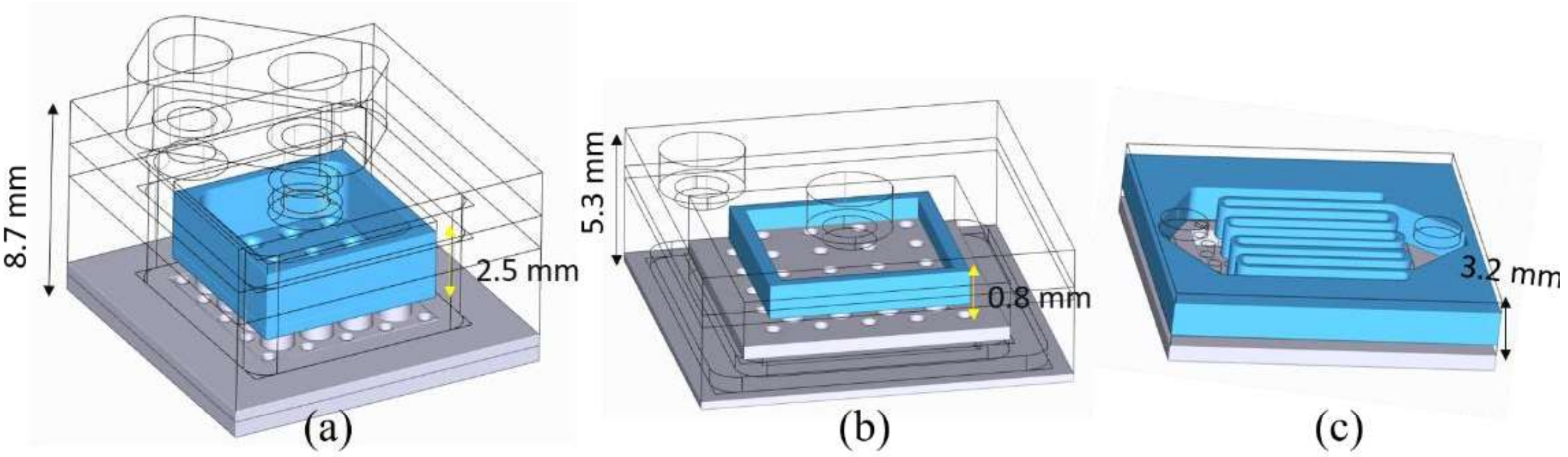

**Figure 10.7:** (a) Standard design with high cavity; (b) thin manifold design; (c) snake shape design.

The other possible solution is to locate the inlet/outlet flow pact horizontally, such as lateral feeding design introduced in Chapter 9. The lateral feeding design shows lots of advantages comparing with vertical feeding design, as discussed in section 10.1. However, this design still needs two layers: one for inlet manifold and the other for outlet manifold. In order to compatible with the chip packaging design, thinner cooler thickness is needed. Therefore, a finger shape design is proposed in Figure 10.7(c), which combing the inlet manifold and outlet manifold into one manifold layer. The cooler thickness can be further reduced comparing with lateral feeding design, with a factor of 2.8.

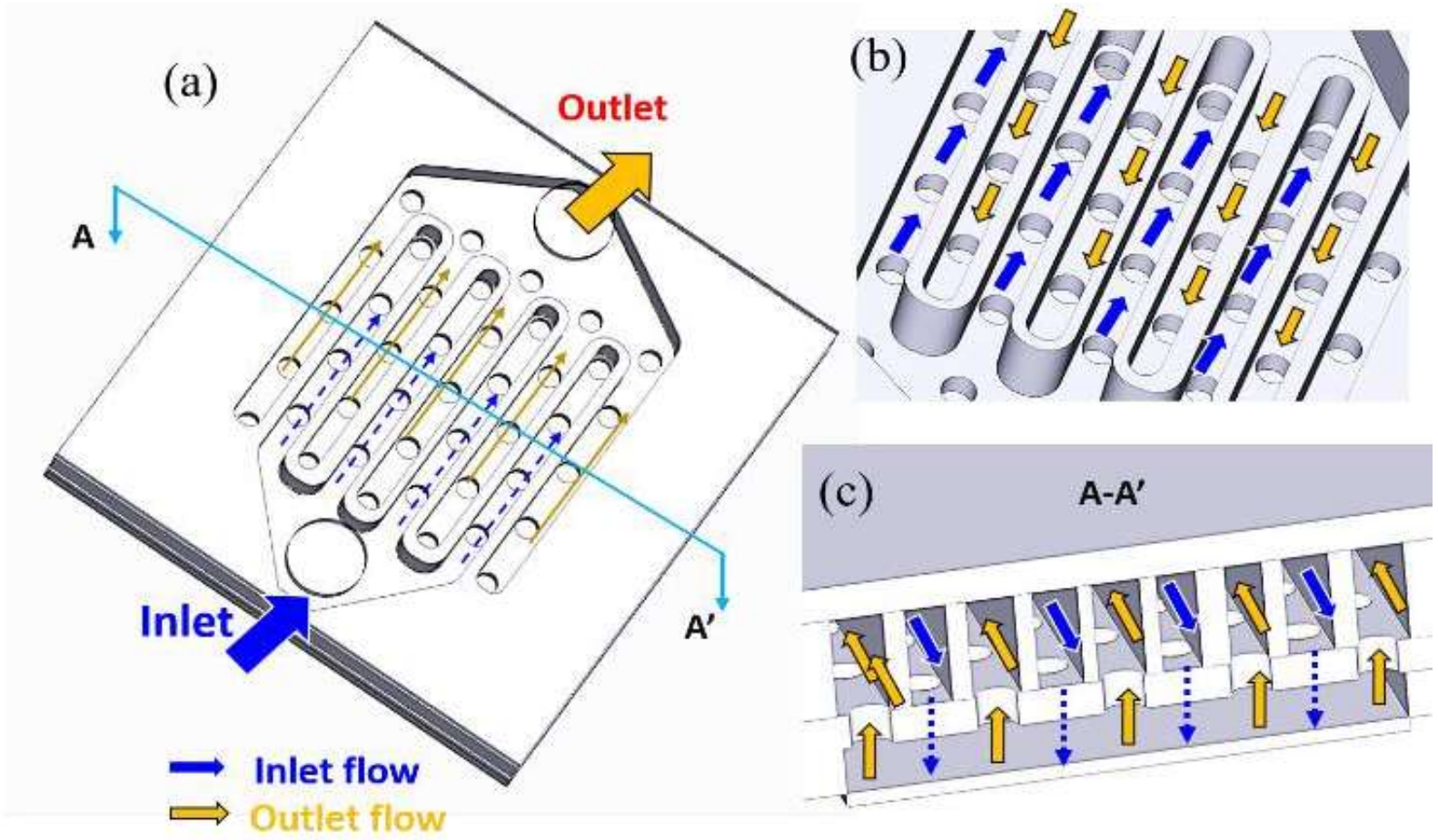

**Figure 10.8:** Schematic of the finger-shape manifold design:(a) entire CAD design structure; (b) snake shape design with inlet and outlet manifold; (c) snake manifold combined with the microjet nozzles.

In order to get better understanding the finger-shape designed concept, the 3D CAD structures with different views are shown in Figure 10.8. The inlet manifold and outlet manifold are separated by the solid wall. The full scale CFD model is also performed for the finger-shape design. As illustrated in Figure 10.9, the CFD model and meshed model with the fluid domain are extracted from the CAD structure. The flow rate is 1 L/min, under chip power of 50 W.

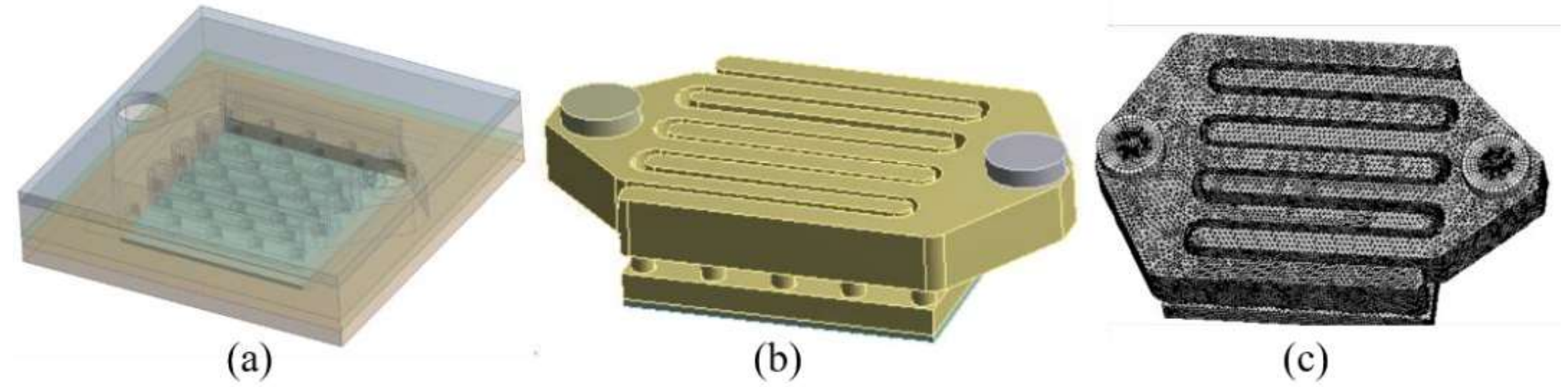

**Figure 10.9:** CFD model of the finger-shape design:(a) solid cooler structure; (b) fluid domain inside the cooler; (c) meshing of the cooler.

The flow distribution is visualized from the flow streamline shown in Figure 10.10.

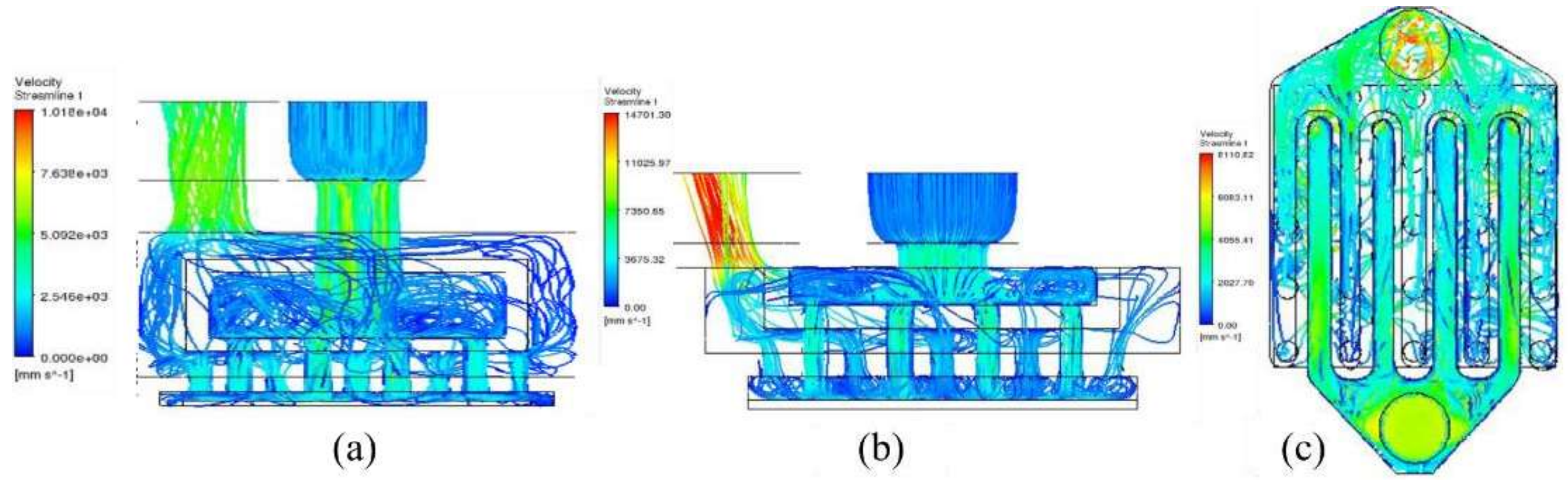

**Figure 10.10:** Flow distribution comparison for the (a) standard design, (b) thin manifold design and (c) finger-shape design: chip power=50W, flow rate=1 L/min (unit not clear)

The temperature comparison in Figure 10.11 shows that the thin manifold design with vertical feeding results in a lower temperature in the chip center while the hottest temperature is around the chip corner. For the finger-shape design, the lowest temperature is at the end of the inlet manifold, where we expected a recirculation at those locations. The highest chip temperature is located at the end of the outlet manifold, showing less flow rate at those locations.

In Table 10.3, the thermal and hydraulic performance are compared, including the averaged chip temperature, pressure drop and temperature gradient. The finger-shape design shows a 1.7 times cooler thickness reduction from 5 mm to 3 mm. The pressure drop can be reduced by a factor of 2.5. Moreover, the temperature gradient can be improved by a factor of 1.4. In general, the finger-shape manifold design shows great advantage of the cooler thickness and the pressure drop reduction.

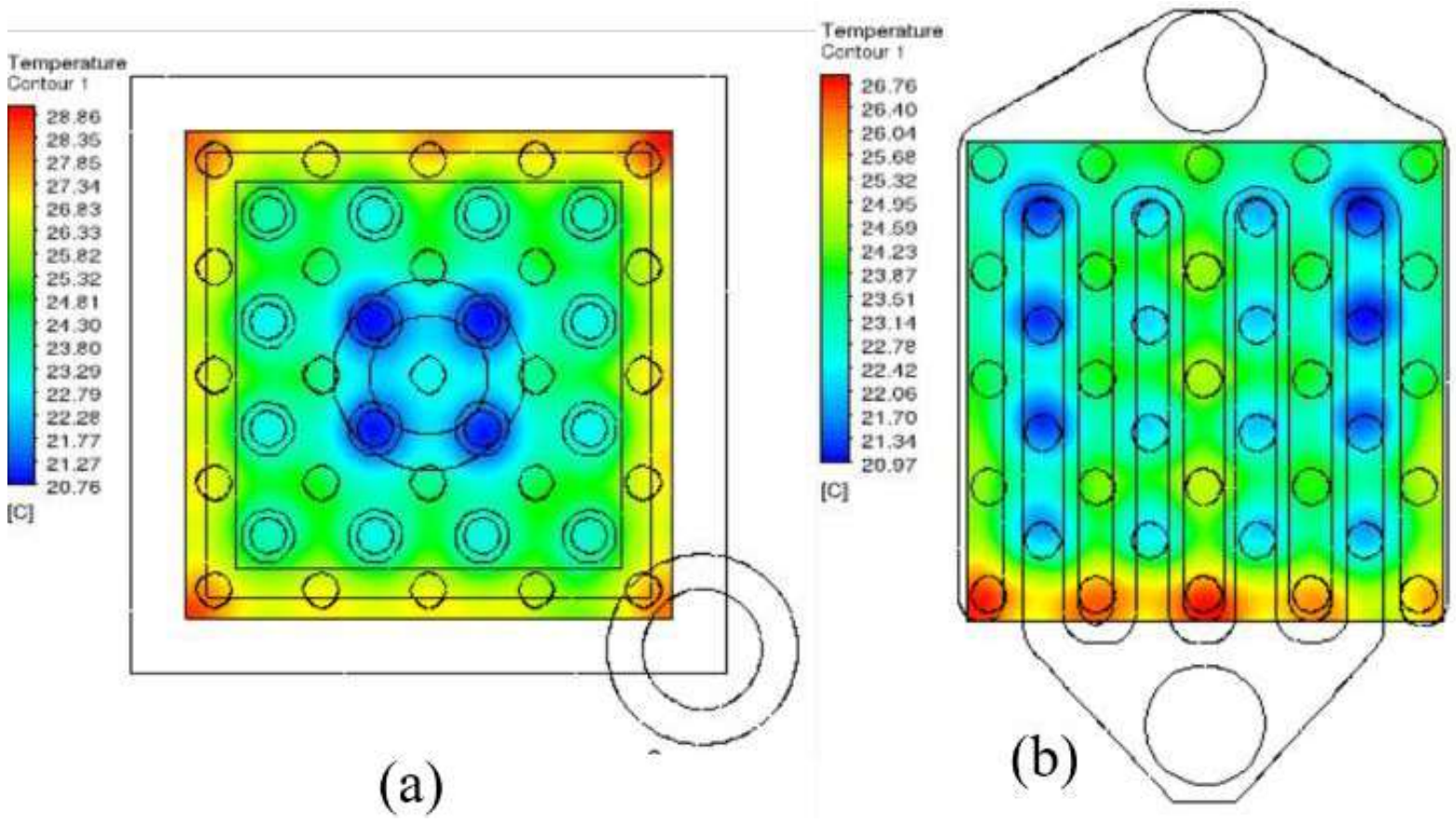

**Figure 10.11:** Temperature distribution comparison between (a) thin cavity design and (b) finger-shape design: chip power=50W, flow rate=1 L/min.

**Table 10.3:** Performance comparison between the initial design and mushroom design: use the previous design.

| Design | Pressure drop (Pa) | Averaged chip temperature (℃) | Temperature gradient (℃) |
|---|---|---|---|
| Standard design | 45128.7 | 24.92 | 3.4 |
| Thin manifold design | 162674 | 24.74 | 8.1 |
| Snake shape design | 64970.3 | 23.42 | 5.79 |

## 10.3 Topology optimization methodology

The first section with conceptual design is based on the innovation. In this section, we will use the mathematic way to improve the design.

### 10.3.1 Literature overview

For a typical microchannel heat sink design with different parameters (fin width, fin length or depth), parameter optimization is a widely used design method for improving the heatsink performance. The parameter optimization [6] is referred to the fine-tuning of the size of the heatsink with single or multi-objectives. The shape optimization deals with optimizing the shape of existing boundaries, the goal of topology optimization is to create and delete boundaries [7]. This research field initiated with the work of Bendsøe and Kikuchi [8].

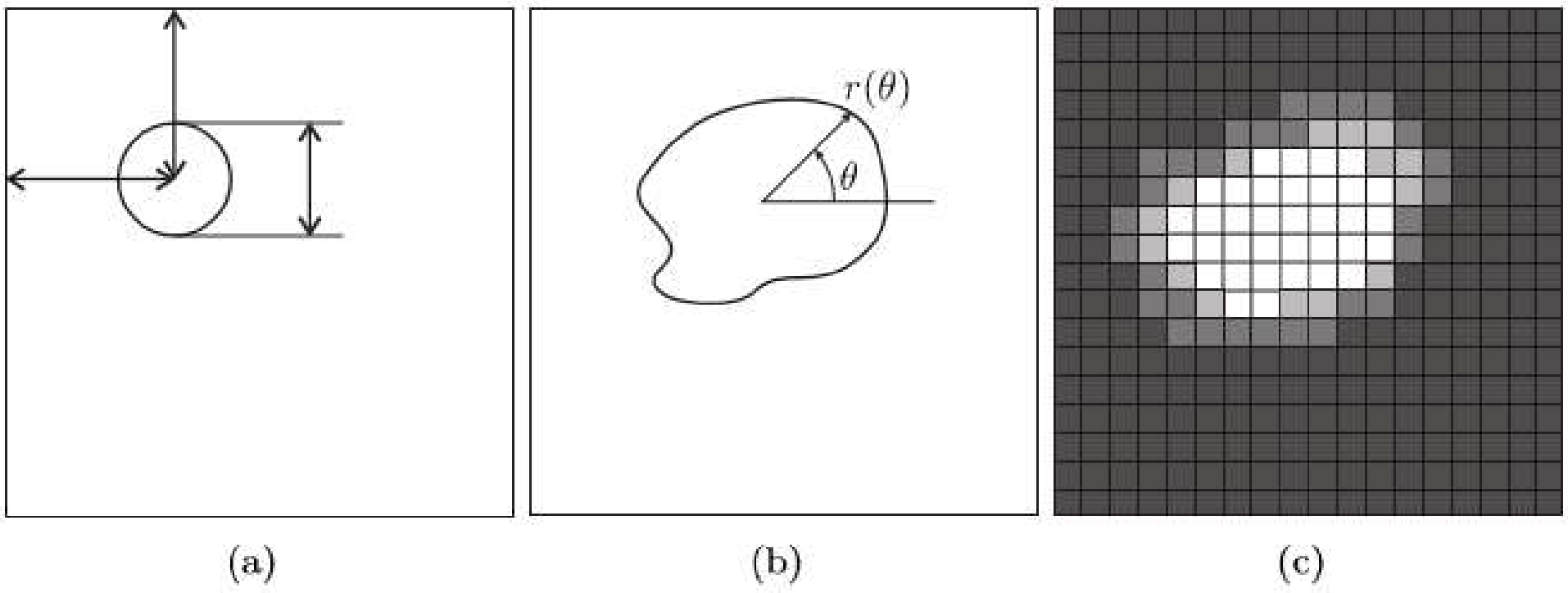

**Figure 10.12:** Conceptual figure of the parameterization of a hole in a domain, based on: (a) sizing, (b) shape and (c) topology [32].

Different from the shape optimization and size optimization, Topology optimization allows the topological changes of the full available volume. Moreover, high resolution 3D printing can realize the complex structure generated from the topology design, while classical manufacturing methods are not suitable. The topology optimization method was first introduced to solve structural problems since the 1980s [9-11].

In recent years, topology optimization design has been applied in optimal heat transfer and fluid flow systems, including the purely heat conduction problems, fluid flow problems and conjugate heat transfer problems [12]. For the topology optimization of the hydraulic problems, lots of work have been implemented to optimize the collectors and distributors [12,33]. The earlier research was conducted by Borrvall and Peterson [28]. After that, lots of literature studies focus on advances in hydraulic optimization and their application to hydraulic optimization of thermal components, such as air-cooled heat sink design for natural convection cooling. Dede et al. [13] implemented the topology optimization method in 3D to demonstrate the improved design of a heatsink for air cooling, and the experimental results indicate that the optimized heat sink design has a higher coefficient of performance (COP) compared with benchmark plate and pin-fin heat sink geometries. Alexandersen et al. [14] presents a density-based topology optimization method for the design of 3D heat sinks cooled by natural convection, with 20–330 million degrees of freedom. Moreover, topology optimization has been also applied to microchannel cooling design with fluid and thermal conjugate simulations [15-19]. Shi zeng et al., [15] applied the topology optimization method for the optimization of a liquid-cooled microchannel heat sink with fin structures. It is shown that the design based on the results of the topology optimization outperforms size-optimized straight channel heat sinks. Van Oevelen [17] implemented the topology optimization method for maximizing the heat transfer of a microchannel heatsink with a constant temperature heat source with a thermo-hydraulic model.

As mentioned in section 10.1, the optimization of the inlet manifold is very important for the pressure drop reduction of the cooler. Also, the inlet manifold defines the flow uniformity across the nozzle arrays, which impacts the temperature gradient over the chip surface. In this chapter, we will focus on the hydraulic design of the inlet by using topology optimization. The main objective of the inlet manifold optimization is to improve flow uniformity over the nozzles and reduce the pressure drop between inlet and outlet. For the literature studies of the manifold level topology optimization, several investigations have been conducted for the flow distribution in the fluidic channels design with flow rate equality constraints [21] or with user-specified outlet velocity [22]. In some cases, minimizing the viscous dissipation in the flow is also combined with constraints on the target flow rate. Seiji Kubo, et al., [20] implemented the topology optimization method on the Z-type and U-type manifolds that ensure sufficient flow uniformity among a five-microchannel array while minimizing pressure drop in a microfluidic device. However, this topology optimization concept has not been done yet for jet impingement cooling.

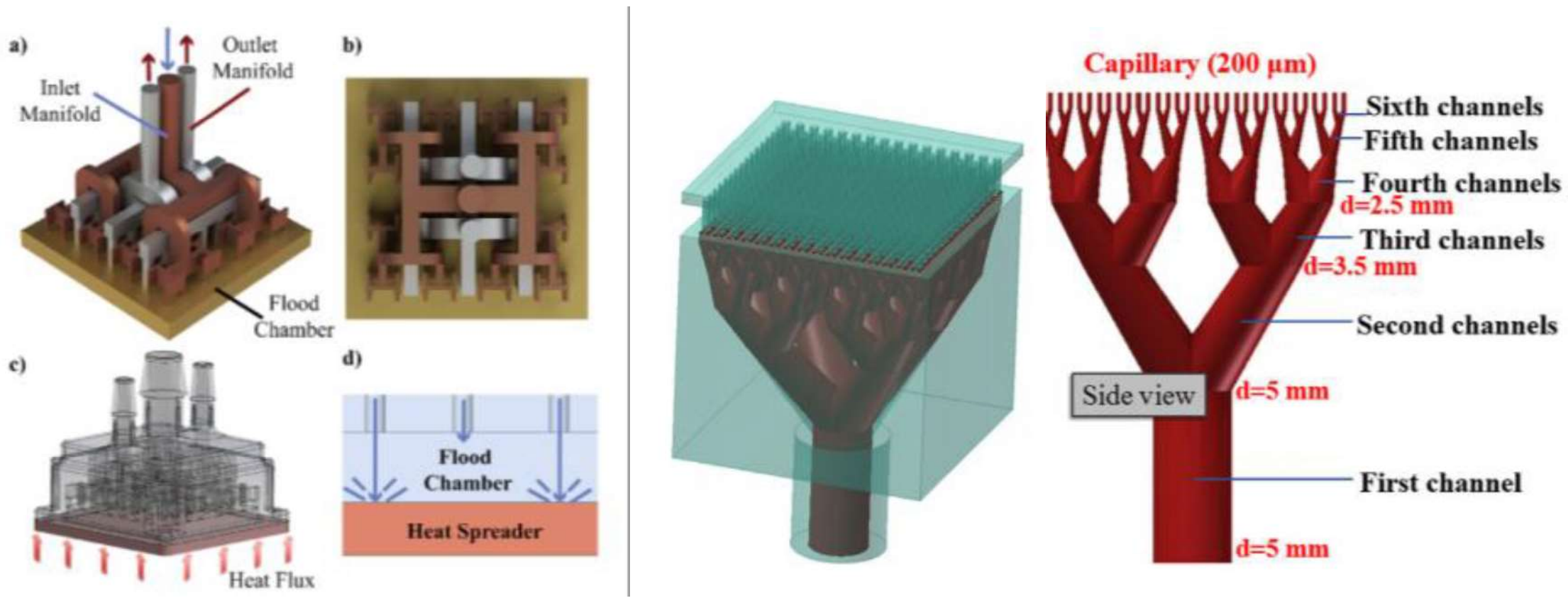

**Figure 10.13:** (a) 3D hybrid micro-jet heat sinks formed by a pair of fractal channel manifolds, used as liquid inlet and outlet conduit [26]; (b) Tree-like channel [31].

Some literatures on alternative designs are also used to improve the flow distribution, such as bifurcation H-design with equal distance [23,24,25]. In literature, a hybrid micro-jet heat sinks formed by a pair of fractal channel manifolds, used as liquid inlet and outlet conduits is 3D printed, shown in Figure 10.13(a). The experiments and numerical modeling results show that the flow uniformity can be improved with the penalty of the pressure drop [26,27]. A porous module with a tree-like micro-channel shown in Figure 10.13(b) was manufactured using a metal additive manufacturing method [31]. The liquid water was uniformly distributed from the central coolant inlet to the whole heated surface by the treelike channel. However, branched design with one branch will add extra level layer that will increase the cooler size, and also the pressure drop will be higher.

In this chapter, the "Topology optimization" methodology (2D) has been applied for the inlet manifold design in the impingement cooler. The current code is adapted from a topology optimization example from FEniCS' Pyadjoint package [23,24]. We base ourselves on their methodology to design our jet cooling inlet manifold design.

### 10.3.2 Topology optimization model

The 2D model for the vertical feeding and lateral feeding design with the boundary conditions is illustrated in Figure 10.14. The dimension of the computational domain of the inlet manifold is 10 mm × 3 mm for the vertical feeding design, while the lateral feeding design is 10 mm × 2 mm. The nozzle diameter of the inlet and outlet are both set as 0.6 mm as the test case. The 2D flow simulation and optimization is performed in the open source software FEniCS/Pyadjoint for 2D steady state conditions.

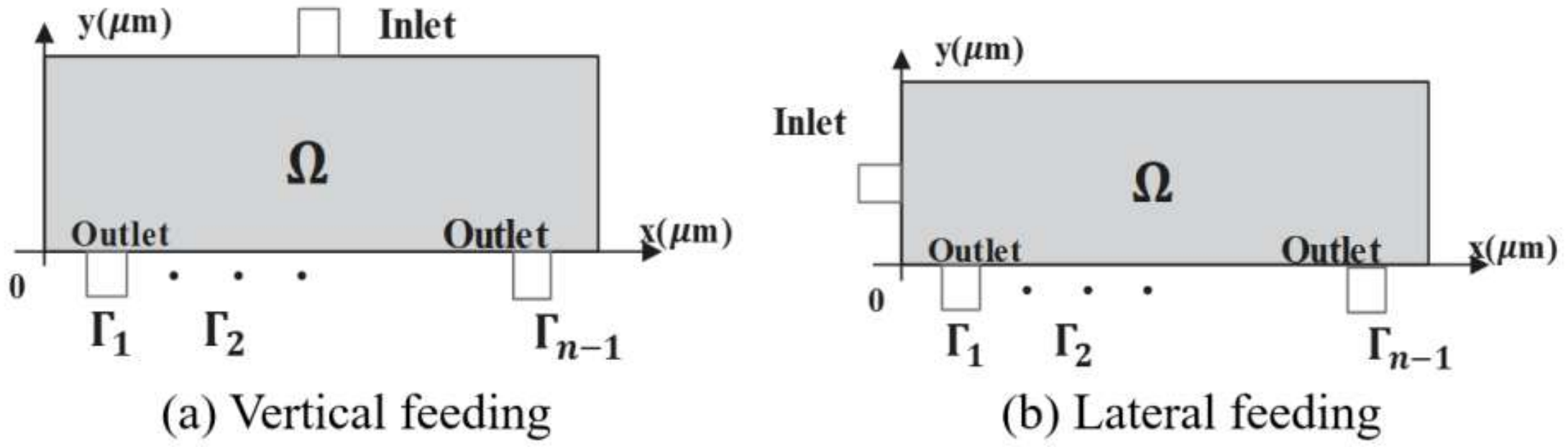

**Figure 10.14:** 2D model for the inlet manifold design with inlet and outlet locations: (a) vertical feeding; (b) lateral feeding configurations.

The viscous dissipation of the flow is governed by the Navier–Stokes equations [25], including momentum and continuity equations, are presented as below:

$$\rho_m \left(\frac{\partial u}{\partial t} + u \cdot \nabla u\right) = -\nabla p + \mu \nabla^2 u + f \tag{10.1}$$

$$\frac{\partial \rho_m}{\partial t} + \nabla \cdot (\rho_m u) = 0, \tag{10.2}$$

where, $\rho_m$ is the fluid density, $u$ and $p$ are the velocity and pressure field, $\mu$ is the fluid dynamic viscosity, and f is the fluid body force.

For the initial study, the model equations are reduced to Stokes equations, which govern steady-state flow of an incompressible Newtonian fluid at low Reynolds numbers, in which inertial effects are negligible. The Stokes equations are given by

$$-\mu \nabla^2 u + \nabla p = f \tag{10.3}$$

$$\nabla \cdot u = 0. \tag{10.4}$$

In order to describe both the solid and fluid domain with a unique equation, a mixed model called "Darcy's law" is used that describes the fluid flow through a porous medium. The porous media model can be used to control the material permeability to act as solid in the limit of a very impermeable porous medium. Therefore, the Stokes Flow equations with velocity Dirichlet conditions can be expressed as below [23,24]:

$$\alpha(\varepsilon)\, u \text{-} \mu \nabla^2 u + \nabla p = f \quad in\ \Omega \tag{10.5a}$$

$$\nabla \cdot u = 0 \qquad in\ \Omega \tag{10.5b}$$

$$u = b \qquad at\ \delta\Omega\ , \tag{10.5c}$$

with ε representing the design variables, controlling the material phase ($\varepsilon(x)$=1 means fluid present, $\varepsilon(x)$=0 means solid).

Moreover, control constraints are used to restrict the available fluid volume as shown below:

$$0 \leq \varepsilon(x) \leq 1 \quad \forall x \in \Omega \tag{10.6}$$

$$\int_{\Omega} \varepsilon \leq V \tag{10.7}$$

$$V = \frac{V_{fluid}}{V_{total}} \tag{10.8}$$

with V the volume bound on the control and $\alpha(\varepsilon)$ the inverse permeability as a function of the control variables. The volume constraint thus controls the portion of volume of fluid $V$ that is desired at the end of the optimization in relation to the total volume of the initial domain design $V_{total}$.

The inverse permeability of the porous media $\alpha(\varepsilon)$ is expressed as

$$\alpha(\varepsilon) = \bar{\alpha} + (\alpha - \bar{\alpha})\varepsilon\frac{1+q}{\varepsilon+q}, \text{ with} \tag{10.9}$$

$$\alpha = \frac{2.5\mu}{0.01^2};\ \overline{\alpha} = \frac{2.5\mu}{100^2}; \tag{10.10}$$

The controlling parameter q is a pseudo-density, varying from 0 to 1, that controls the material permeability in between solid and fluid [20,22]. The inverse permeability can be used to control the fluid ''resistance'' of the porous material to make porous material more or less interesting than fully fluid or solid material for the optimizer.

***Cost function for minimizing the rate of viscous dissipation:***

The first objective function considered is the minimization of the viscous dissipation in the fluid flow, where the viscous power dissipation is equivalent to the pumping power that needs to be applied over the channel. This objective function is shown as below [23,24]:

$$\underline{\text{Objective function 1}}: J_1 = \frac{1}{2}\int_{\Omega} \alpha(\varepsilon)\, u \cdot u d_{\Omega} + \frac{\mu}{2}\int_{\Omega} \nabla u \cdot \nabla u d_{\Omega} - \int_{\Omega} f \cdot u d_{\Omega} \tag{10.11}$$

where $u$ is the velocity field, and $f$ is the fluid body force. As previously described, $\mu$ is the fluid dynamic viscosity, and $\alpha(\varepsilon)$ is the inverse permeability of individual fluid flow cells as a function of the pseudo-density $q$.

***Cost function for the flow rate uniformity:***

The second objective is to minimize the difference of the flow rate at every outlet with the mean outlet flow rate, in order to improve the flow uniformity.

$$\underline{\text{Objective function 2}}: J_2 = \frac{1}{2}\sum_{i=1}^{m} (q_i - \bar{q})^2 \tag{10.12a}$$

$$q_i = \frac{1}{2}\int_{\delta\Omega i} \vec{u} \cdot \vec{n} d_{\Gamma}; \bar{q} = \frac{1}{n}\sum_{1}^{n} q_i \tag{10.12b}$$

where $q_i$ is the individual outlet flow rate and $\bar{q}$ is the averaged flow rate based on the outlet nozzles.

Therefore, the final objective with the pressure drop and flow uniformity constraint can be expressed as below:

$$\text{Min}\, J(\varepsilon, X(\varepsilon)) = (1 - \beta)\, \lambda_1 \cdot \frac{1}{2}\sum_{i=1}^{m} (q_i - \bar{q})^2 + \beta\ \lambda_2 \cdot \Phi \tag{10.13a}$$

$$\lambda_1 = \frac{1}{{q_{in}}^2};\ \lambda_2 = \cdot\, \mu \cdot \frac{{u_{ref}}^2}{L^2}; \tag{10.13b}$$

where the $\lambda_1$ and $\lambda_2$ are factors that makes the objective functions dimensionless and of order unity, and beta is the weighting factor that balances the two objectives.

Thus, the topology optimization problem we face is in fact a multi-objective optimization problem. We will deal with it by minimizing a weighted objective function J, subjected to the flow equations (10.5), defined as:

**Minimize:**

$\min_{\epsilon, q} : J(\varepsilon, q)$ $\varepsilon$ is the design variable

**Subject to:**

$c_1(\varepsilon, q) = 0$ Flow equations

**Constraints:**

$0 \leq \varepsilon \leq 1$ Box constraints

$V = \frac{V_{fluid}(\varepsilon)}{V_{total}}$ V is the impost volume fraction

$q = (\vec{u}, p)$ State variables

$\alpha(\varepsilon) = \overline{\alpha} + (\alpha - \overline{\alpha})\varepsilon\frac{1+q}{\varepsilon+q}$ Penalization parameterization

For the topology optimization, the ***dolfin*** and ***dolfin_adjoint modules*** are imported into the FEniCS Finite Element simulation library. ***Taylor-Hood finite element*** are used to discretize the Stokes equations. The minimal mesh size for the model is 0.03 mm. We choose an initial guess of $q = 0.01$ for the control and use it to solve *the* model equations.

For the topology optimization, two boundary condition constraints are investigated. The first is to apply a constant uniform velocity profile at the outlets. This concept can be used for the hotspot- targeted cooling, where different sizes or different power of hotspots need different velocity. The risk with using such a boundary condition for the optimization is that different pressures arise at the outlet, while physically the pressure should equal to the pressure at the chambers with the jets. In other words, those circumstances might only be realized using small pumps at the different outlets. For the second boundary condition, a fixed outlet pressure is combined with a penalty on the flow rate equality.

### 10.3.3 Numerical implementation

Figure 10.15 lists the design flow for the inlet manifold topology optimization. By using topology optimization, the fluid flow is better guided towards the outlet nozzle locations. In Figure 10.15, the final porous density distribution map is shown, where the black represents the solid structure and the white represents the fluid structure.

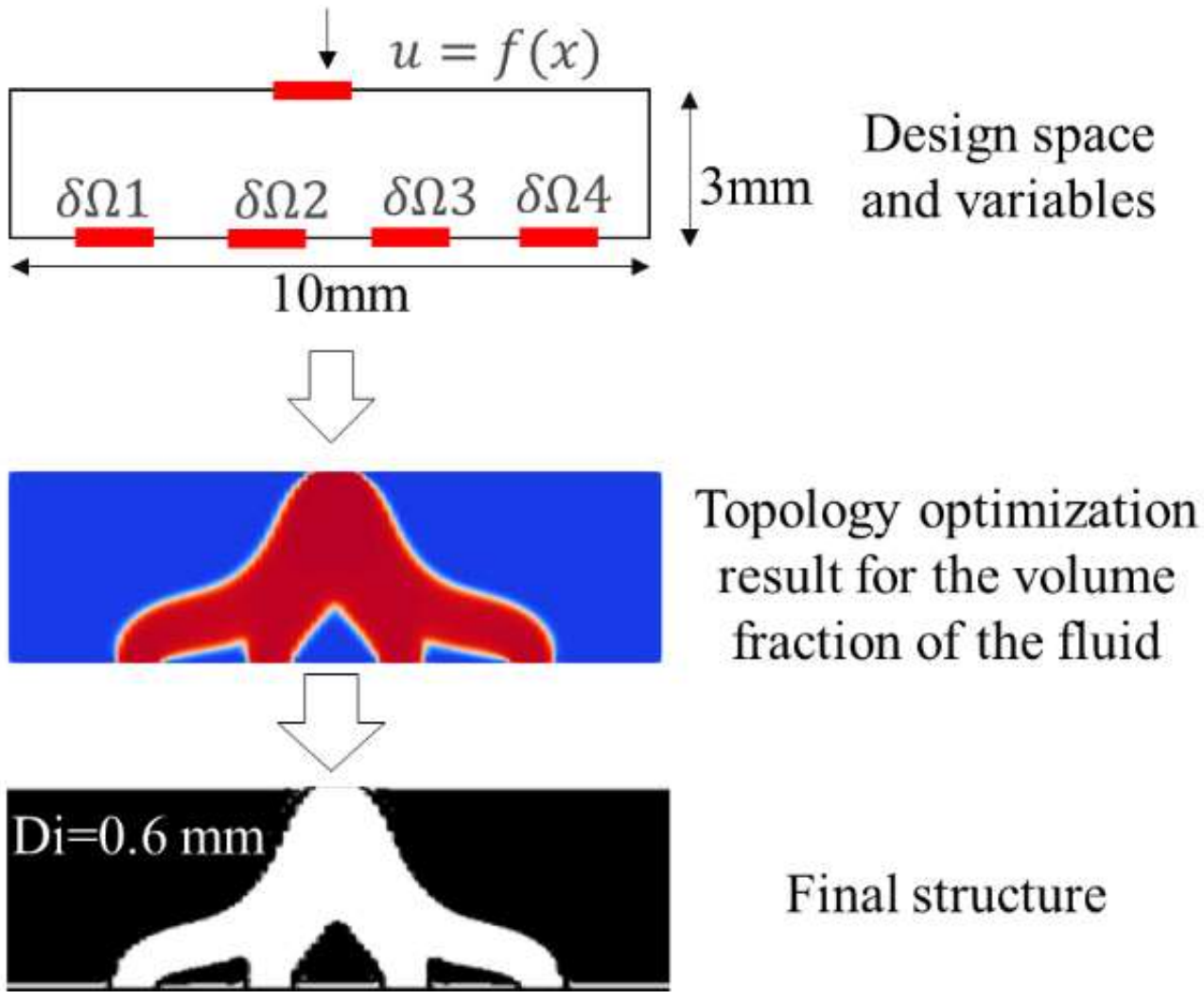

**Figure 10.15:** Inlet manifold topology optimization design flow.

The final obtained black/white structure is compared with the initial design domain. The comparison presented in Figure 10.16 shows that the improved design leaves more design space for the outlet manifold, which can reduce the pressure drop in outlet part.

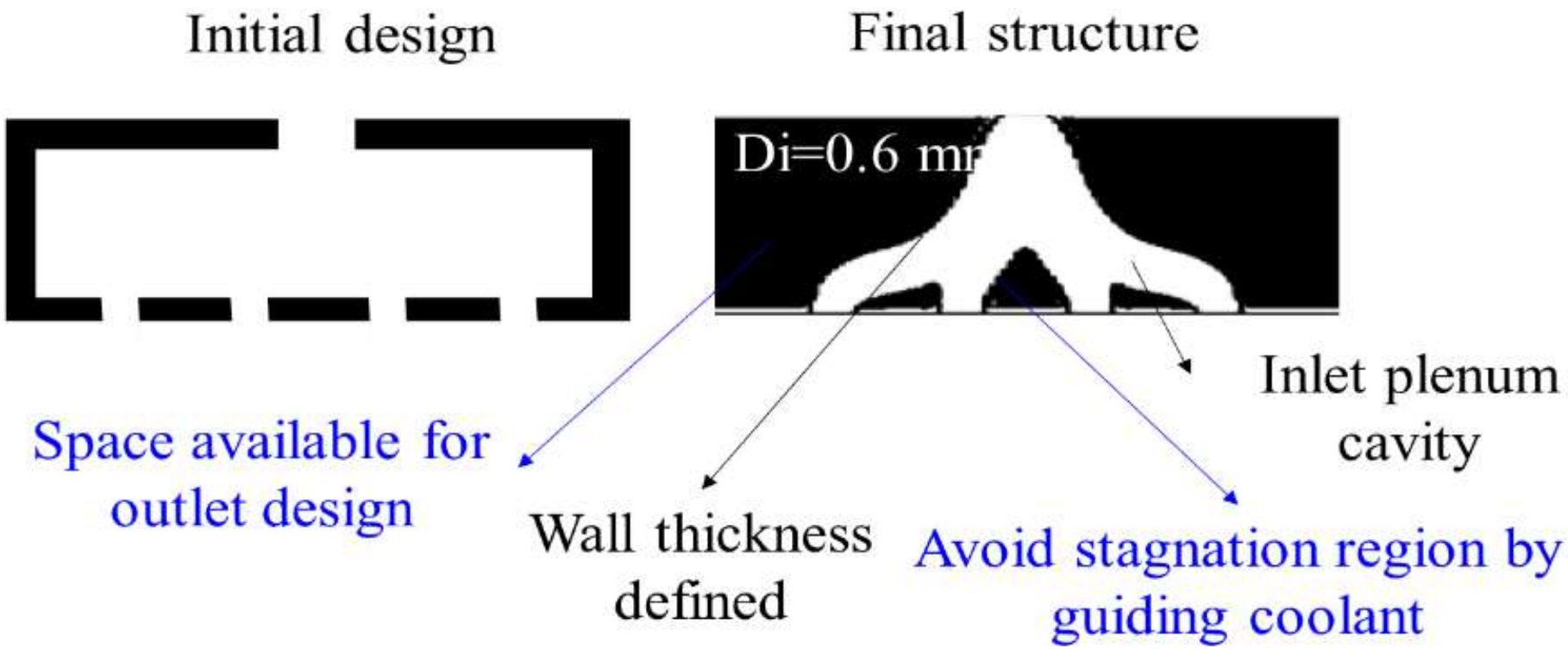

**Figure 10.16:** Comparison with the initial design and improved cooler design (white represents fluid and black represents solid).

### 10.3.4 Discussion and benchmarking

10.3.4.1 Reference case: 2D analysis from Initial Design

In order to benchmark with the improved design using the 2D topology optimization, the 2D models for the initial design are analyzed using ANSYS Fluent. The flow distribution and pressure distribution are all illustrated in Figure 10.17. The test case is chosen for a 4×4 array with a constant uniform inlet velocity of 20 mm/s. As shown in Figure 10.17(a), the flow rate distribution over the 2D outlet nozzles of the array, showing a higher flow rate in the central part of the array with 30%. Moreover, the pressure is built-up at the stagnation regions at the bottom of the cavity and at the sudden expansion at the entrance. For the lateral feeding design, the extracted flow rate distribution percentage across the four outlets is 32%, 28%, 24% and 16%, resulting in a significant flow non-uniformity across all the nozzles. It can also be seen that, the pressure drop of vertical feeding case is higher than lateral feeding case, which was already shown for the 3D design in chapter 8.

Most importantly, part of the inlet manifold volume remains unused, which limited the design space for the outlet chamber. In this work, the design improvement with the vertical feeding and lateral feeding will be both investigated by using topology optimization.

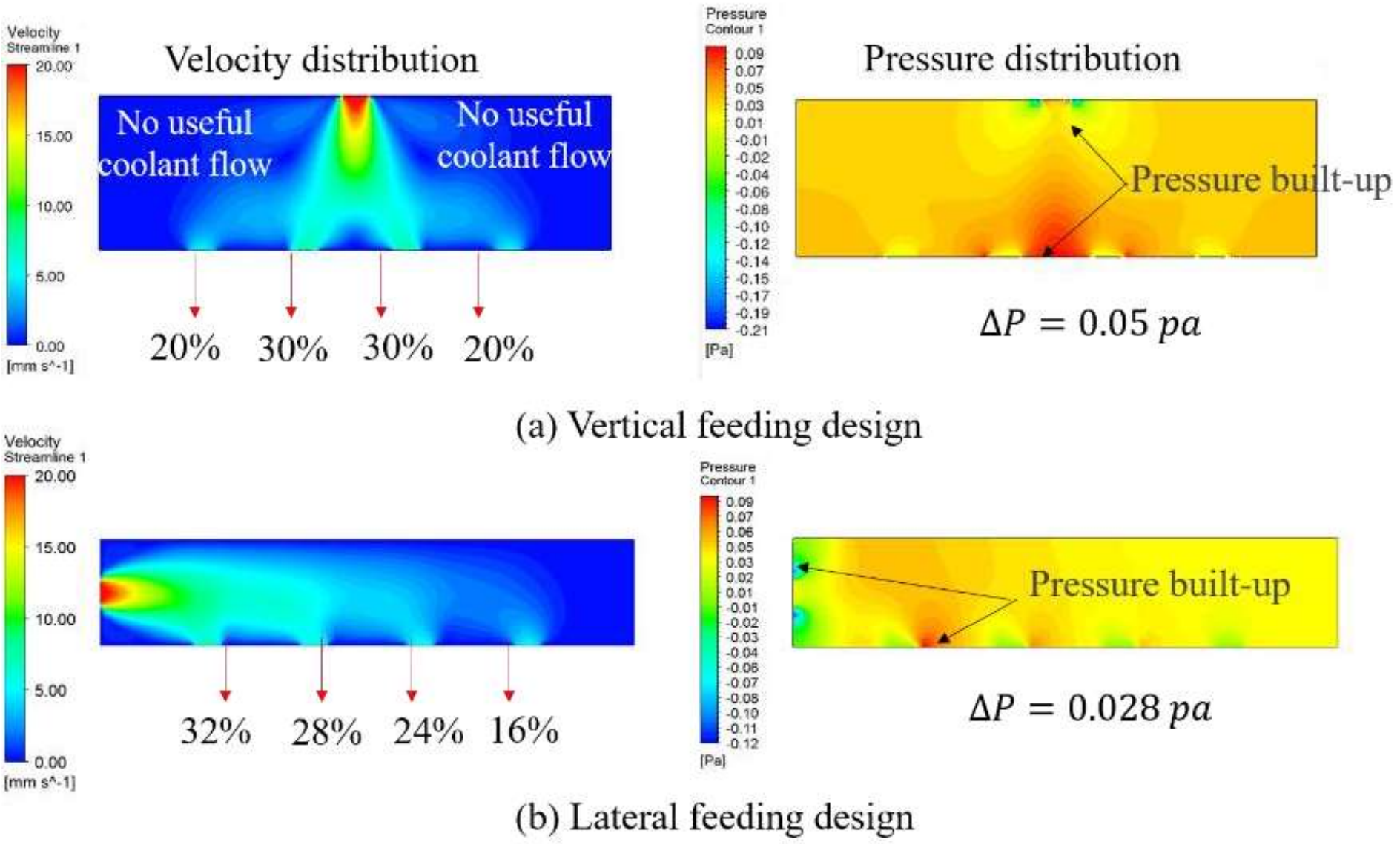

**Figure 10.17:** 2D CFD analysis of initial cooler design of 4x4 array with (a) vertical feeding; (b) lateral feeding (Vin=20 mm/s).

10.3.4.2 Specific velocity for outlet nozzles

In this section, the applied boundary condition is imposing the velocity at the inlet and outlet nozzles.

| | | |
|---|---|---|
| Objective: | Min $J(\varepsilon, q(\varepsilon))= \Phi$ | |
| B.C: | $u = b$ | $at\ \Gamma_{in}$; |
| | $V_1=V_2=V_3=V_4=0$ | $at\ \Gamma_{out}$; |

*(a) Vertical feeding manifold for different nozzle numbers*

Figure 10.18 shows the comparison between the topologically optimized results and the initial design for different numbers of nozzles: 3, 4, and 6. The flow streamlines and pressure distribution of the initial designed structure with full fluid volume are illustrated. It can be seen that the topology optimization can be used to design the channels with the equal ratio of flow rate at the outlets, with different nozzle number. The improved design also shows better uniformity than the initial design.

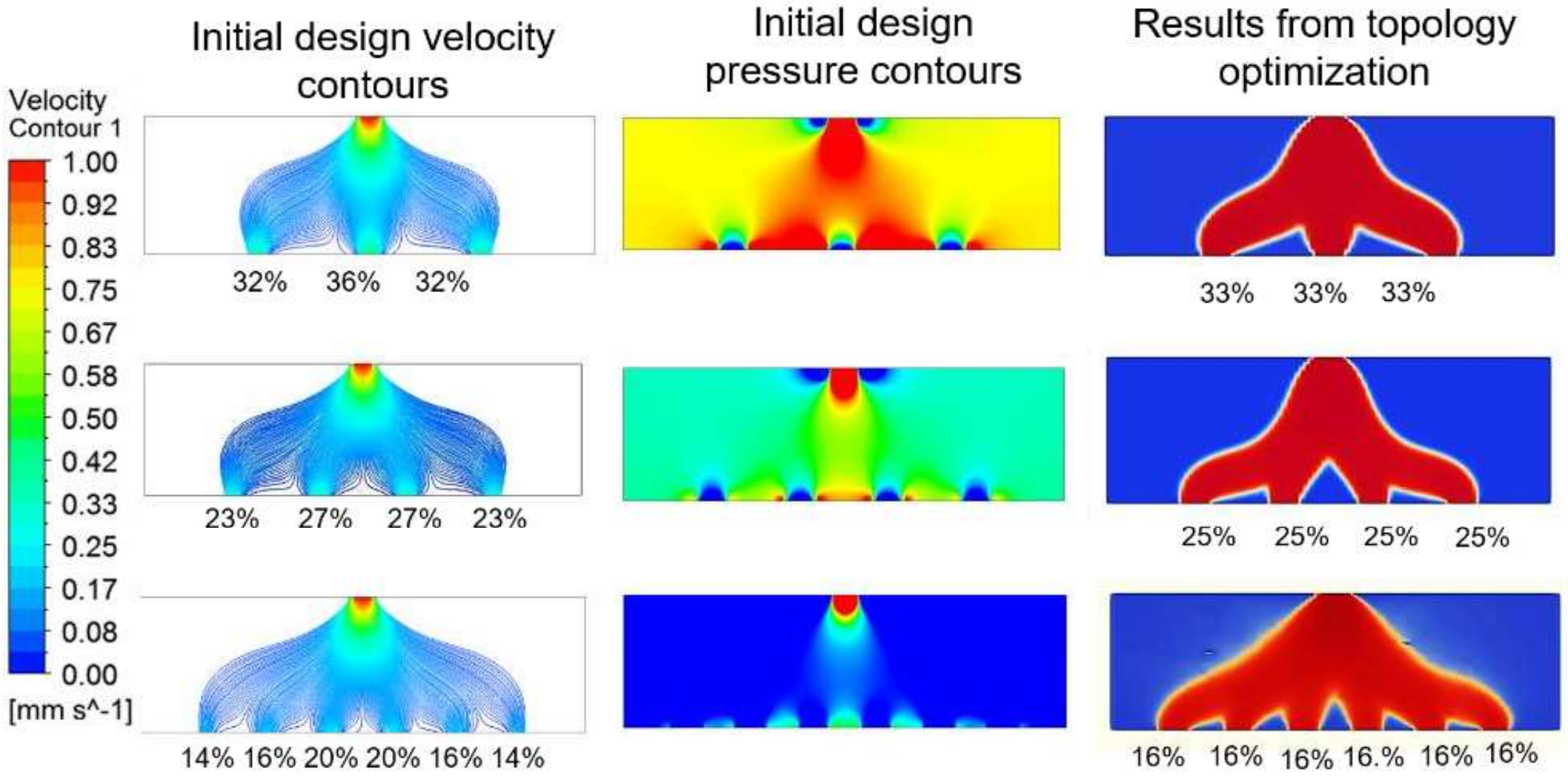

**Figure 10.18:** Velocity and pressure distribution of the inlet manifold with 1 mm/s.

*(b) Lateral feeding manifold for different nozzle array*

Inlet manifold with lateral feeding has thinner thickness, which is much easier for package level cooler integration. However, feeding from one side might result in higher flow non-uniformity over the nozzle array. By using the topology optimization, the flow uniformity can be improved shown in Figure 10.19.

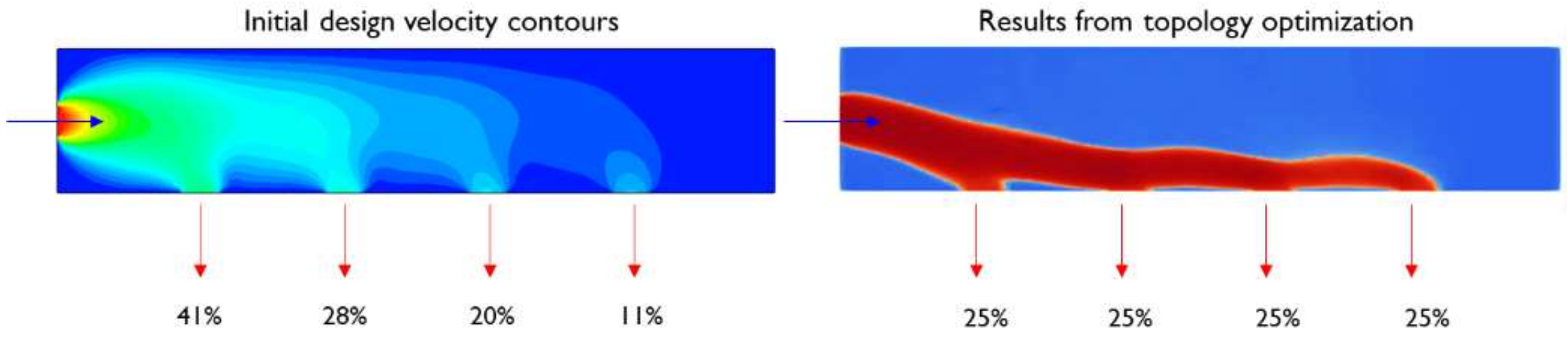

**Figure 10.19:** Example for 2D topology optimization for lateral feeding manifold for 4×4 array: (a) ANSYS Fluent modeling results with initial design; (b) topology optimization results in FEniCS (Vin=1mm/s)

*(c) Manifold designs for specified nozzle flow rates*

Moreover, the topology optimization can be also used for specified nozzle flow rates, as illustrated in Figure 10.20. The flow feeding channel can be tailed based on the requirement of the flow rate with widen channel for the large flow rate and narrow channel for the small flow rate.

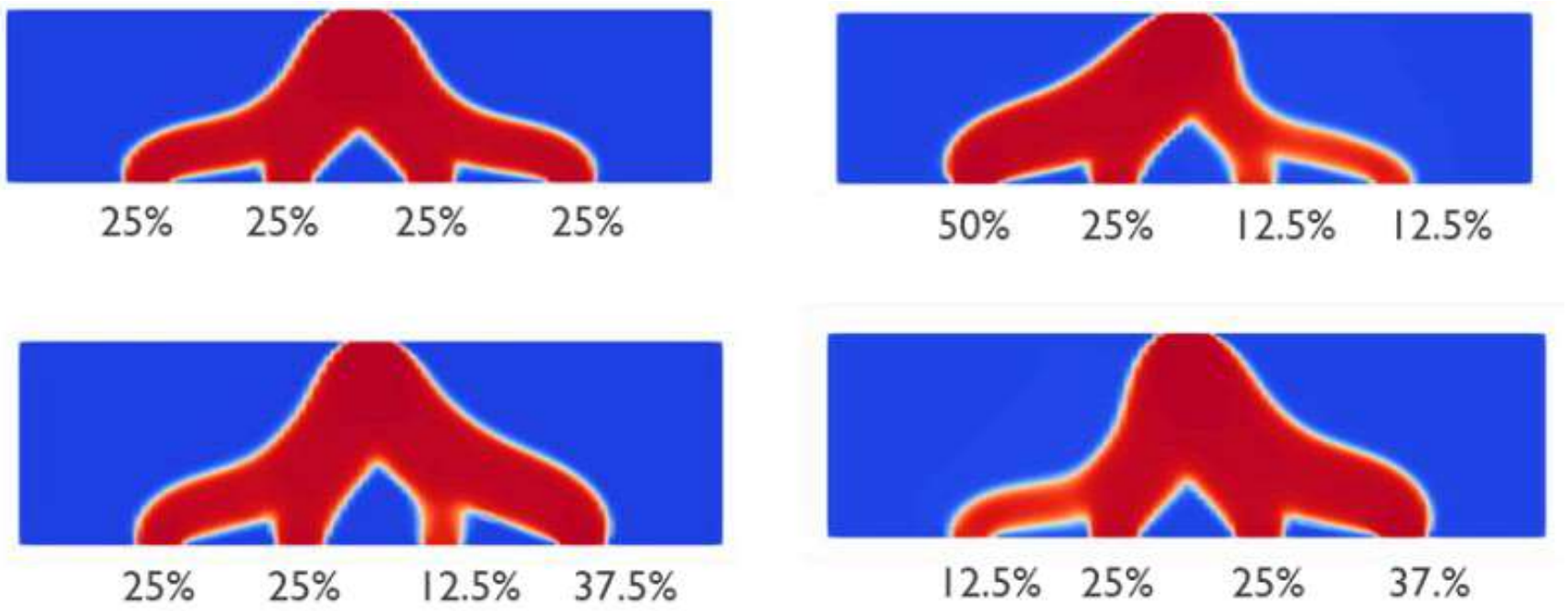

**Figure 10.20:** Topology optimized structures with specified outlet nozzle flow rates in FEniCS.

10.3.4.3 Pressure out with equal flow rate constraint

As illustrated in Figure 10.1, all the liquid flow coming from the inlet manifold will be collected by the outlet manifold. The pressure drop for all the outlet nozzle should be the same, referred to pressure outlet boundary condition for all the outlet nozzles. In the following part, the pressure out boundary condition with outlet flow uniformity constraint is studied.

Objective: Min $J(\varepsilon, q(\varepsilon))=\lambda_1\ (1-\beta)\cdot\ \frac{1}{2}\sum_{i=1}^{m}(q_i-\bar{q})^2+\lambda_2\ \beta\ \Phi$

B.C: $u = b \quad at\ \Gamma_{in};$

$p_1=p_2=p_3=p_4=0 \quad at\ \Gamma_{out};$

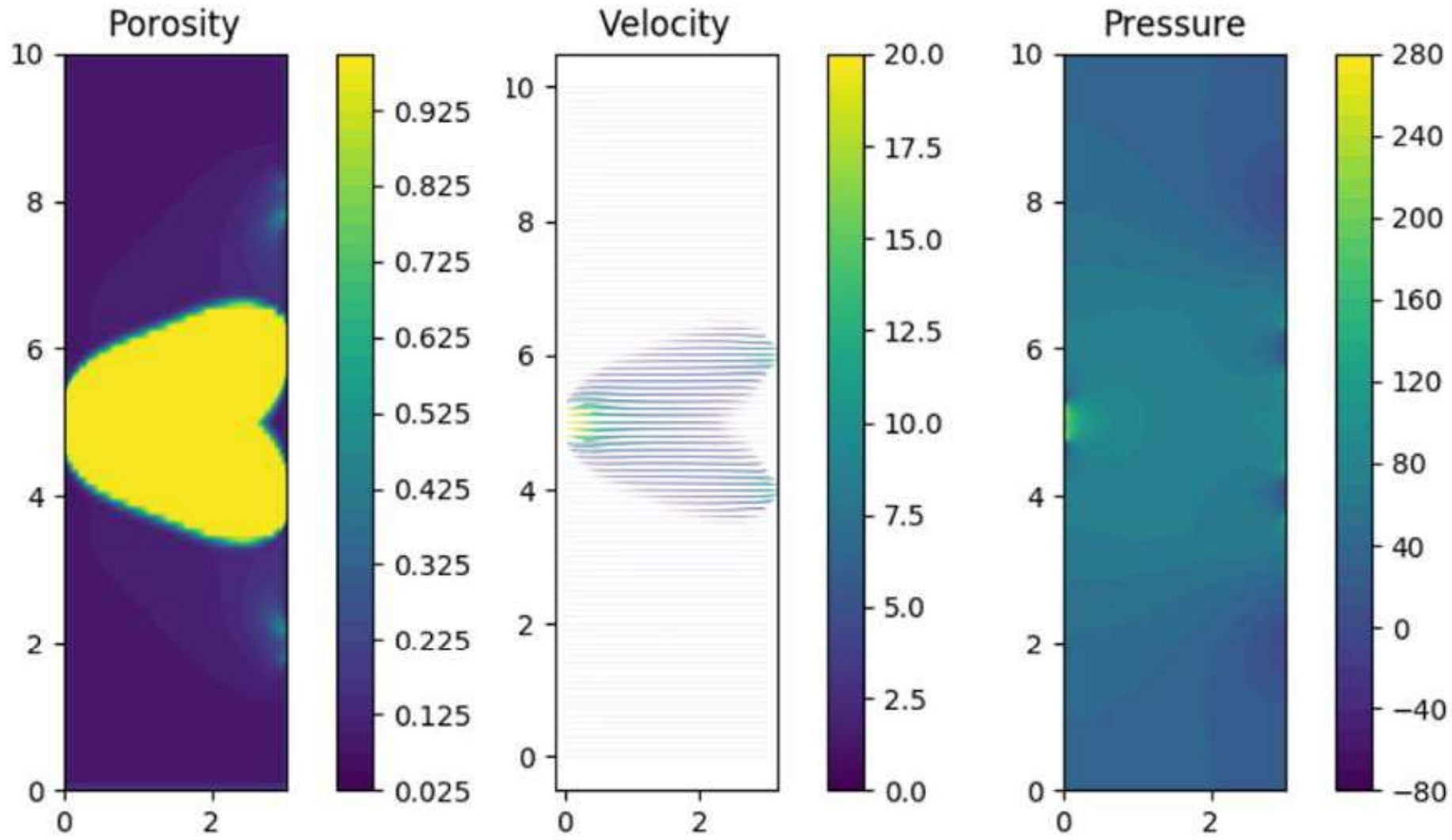

**Figure 10.21:** Adjoint optimization results for the porosity, pressure and velocity distribution under pressure out boundary condition without outlet flow uniformity constraint (Vin=20mm/s; $\lambda_1$=0; $\lambda_2$=1).

In Figure 10.21, the first objective function with minimizing the viscous dissipation is applied by setting $\beta = 0$. It can be seen that most of the flow goes toward to the center outlet nozzles, which has the lowest dissipation power with pressure drop of 0.28 Pa. However, the flow uniformity is worse with 96% of the flow going through the outlet 2 and outlet 3.

In the second step, the value of beta is adjusted. It can be seen that the flow rate percentage for the four outlets are 24%, 26%, 26% and 24%, with the penalty of the high pressure drop up to 0.4 Pa. In Figure 10.23, the final topological structure with the improved design are benchmarked with the initial design structure with whole volume chamber. The flow rate percentage comparison between the initial design and improved design are plotted in Figure 10.24, referring to flow uniformity to Figure 10.17(a).

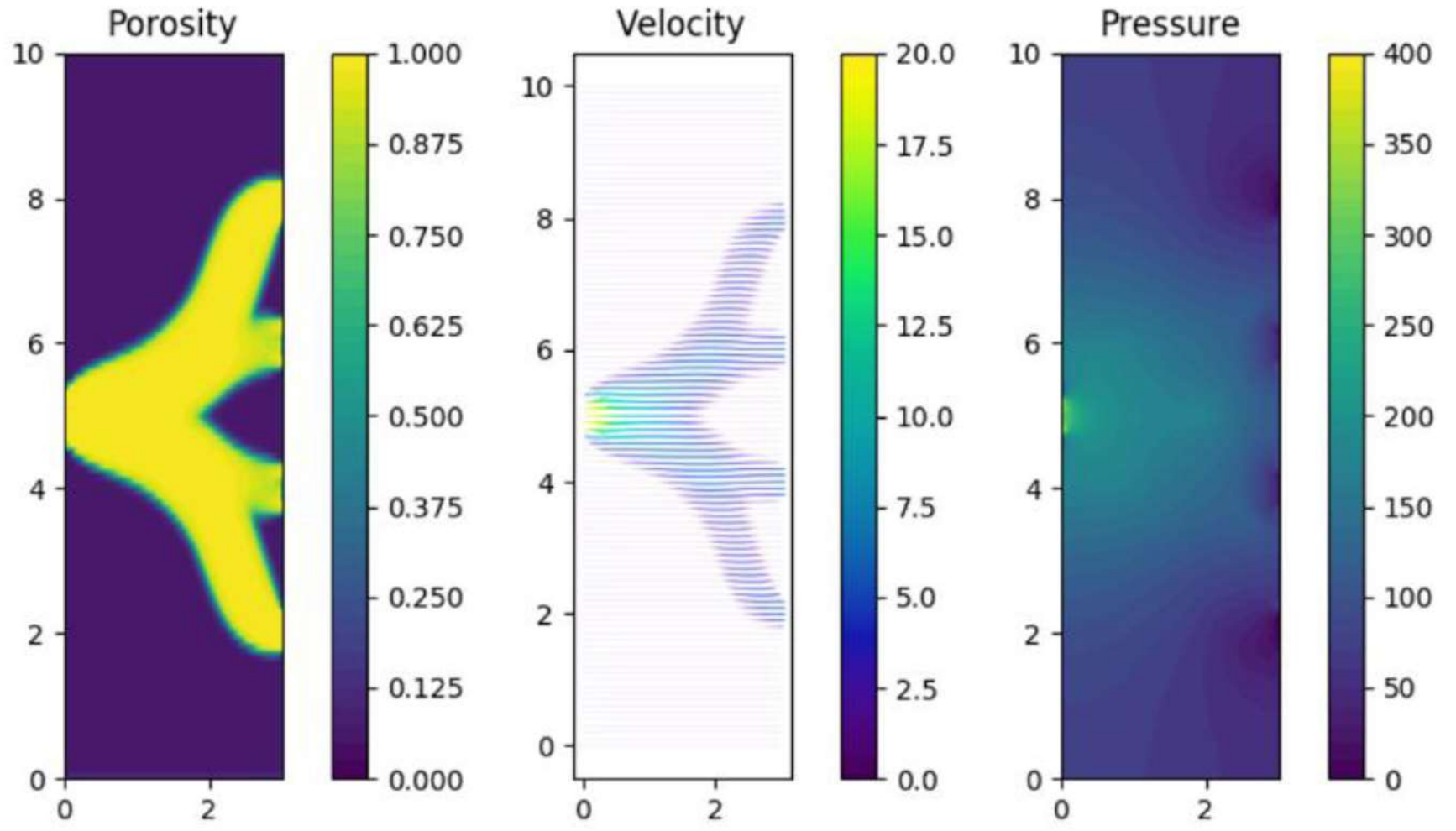

**Figure 10.22:** Adjoint optimization results for the porosity, pressure and velocity distribution under pressure out boundary condition with outlet flow uniformity constraint (Vin=20mm/s; $\lambda_1$=100; $\lambda_2$=1).

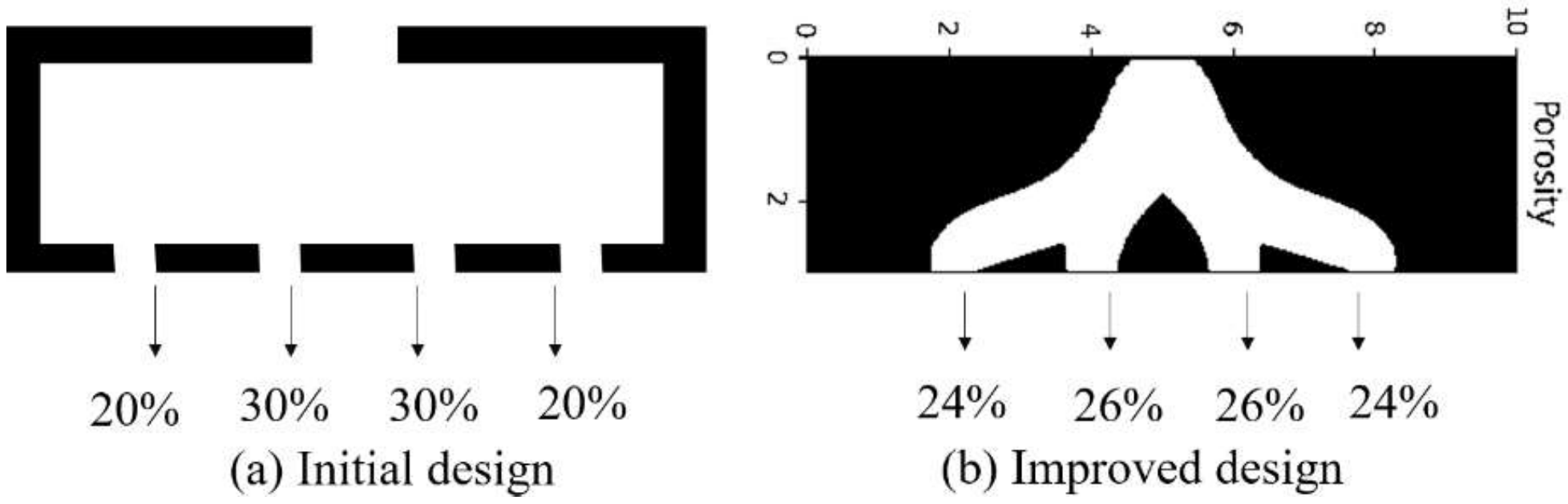

**Figure 10.23:** Topological structure comparison with flow rate distribution percentage for the initial design and improved design (Vin=20mm/s; $\lambda_1$=100; $\lambda_2$=1).

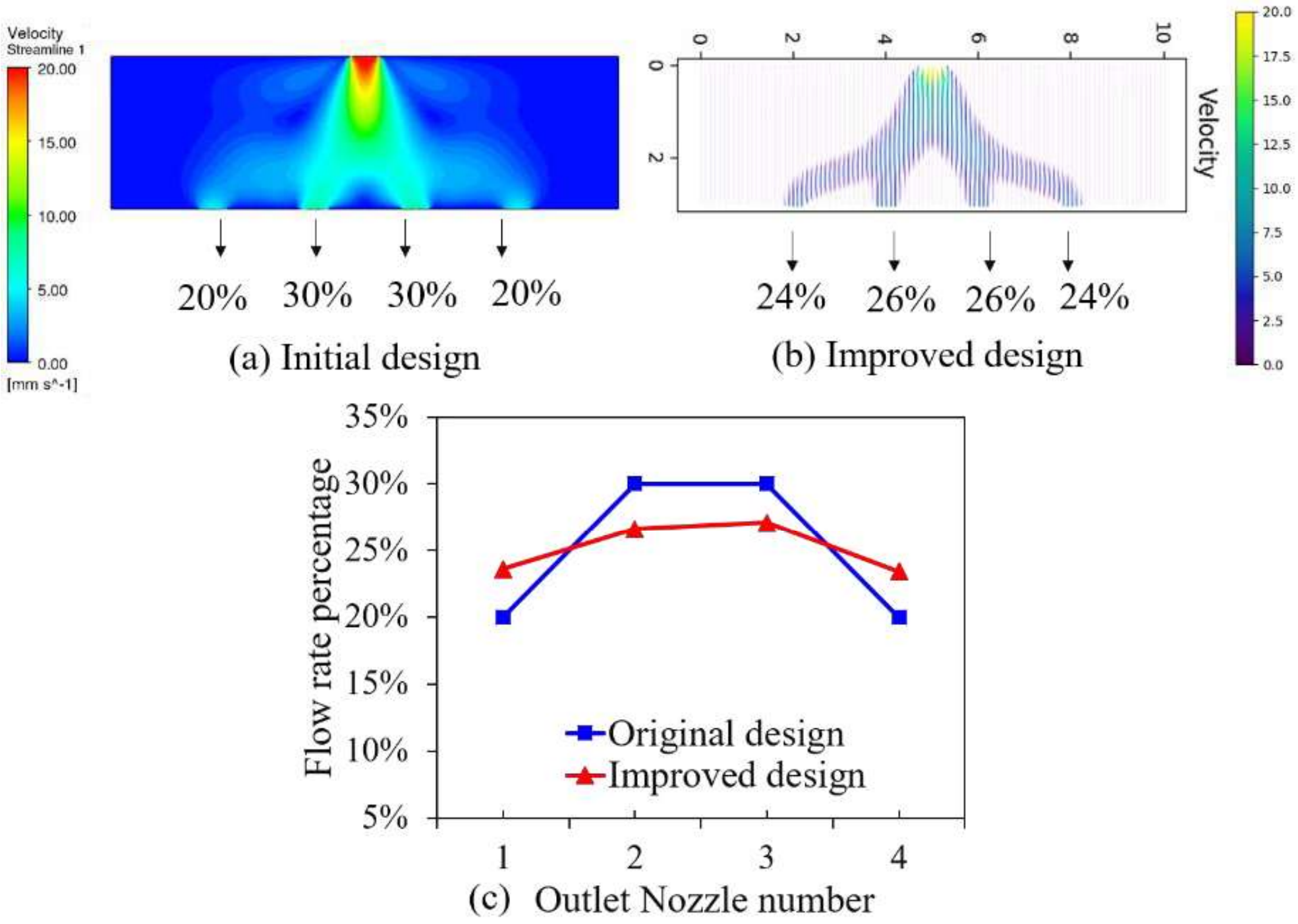

**Figure 10.24:** Comparison between the initial design (ANSYS modeling) and the improved design with topology optimization (FEniCS optimization): (a) initial design; (b) improved design; (c) flow rate percentage comparison (Vin=20mm/s; $\lambda_1$=100; $\lambda_2$=1).

As we known that the selection of the weighting factor $\beta$ influences the optimized results, based on the following objective function, shown in 10.25(a). When the $\lambda_1 = 0$, it means that only the first objective function with minimizing dissipation power is applied. For $\lambda_1 = \lambda_2 = 50$, the modeling result shown in Figure 10.25(b) is similar with Figure 10.25(a). Therefore, the flow is mostly concentrated into the outlet 2 and outlet 3. As for the $\lambda_1 = 1$ and $\lambda_2 = 100$, the objective focuses more on the flow rate uniformity, shown in Figure 10.25(c).

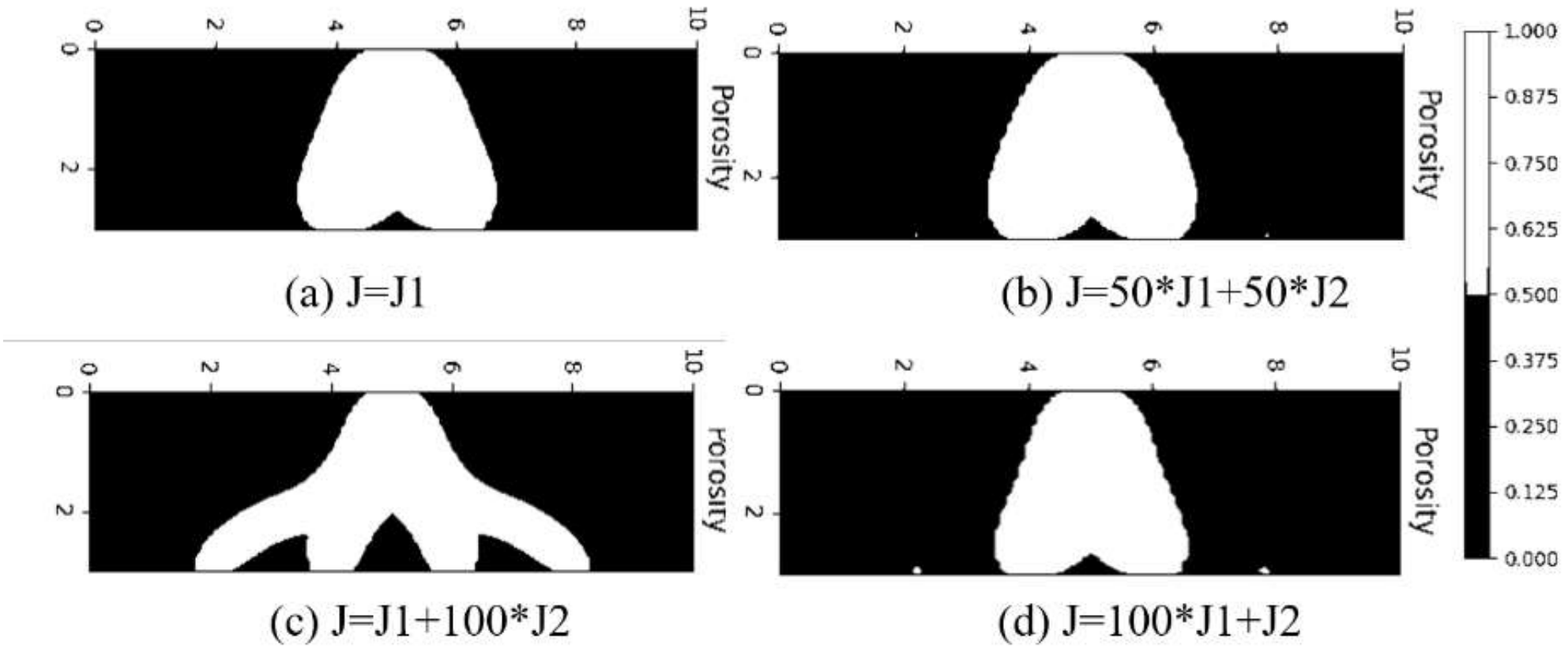

**Figure 10.25:** Topology optimization with minimize dissipation power and outlet flow uniformity, under different penalization correlations.

In addition, the changing of the volume constraint defined as the fraction between the fluid and solid part, can also influence the topology results. As shown in Figure 10.26, three different volume constraints are applied in the constraint equation, resulting in three different topological shapes. The flow uniformity and pressure drop for the three constraints are evaluated in Figure 10.27, where volume constraint $V$=1/3 has better flow uniformity, with relatively higher pressure drop than others.

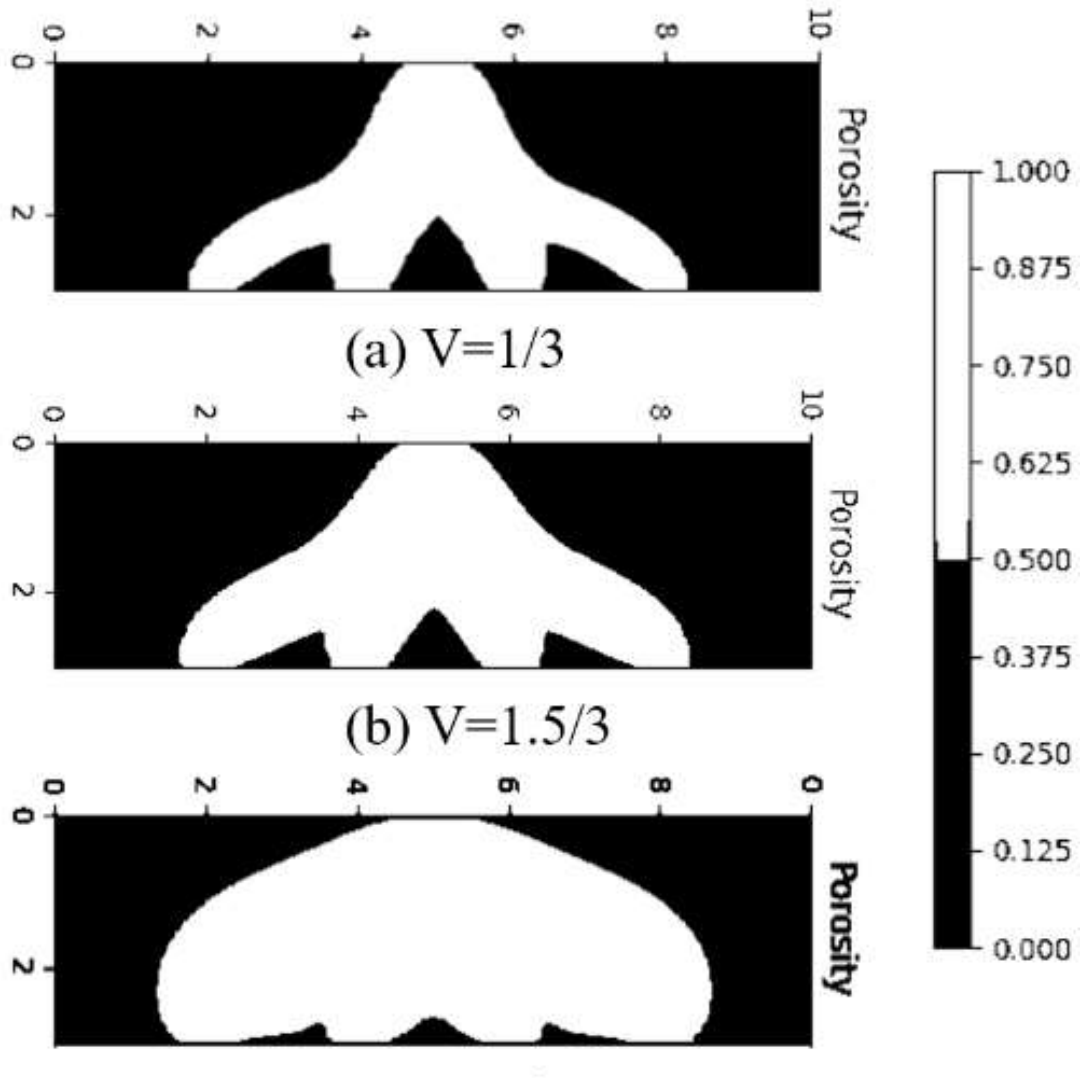

**Figure 10.26:** Topology optimization with minimize dissipation power and outlet flow uniformity, under different volume fraction V.

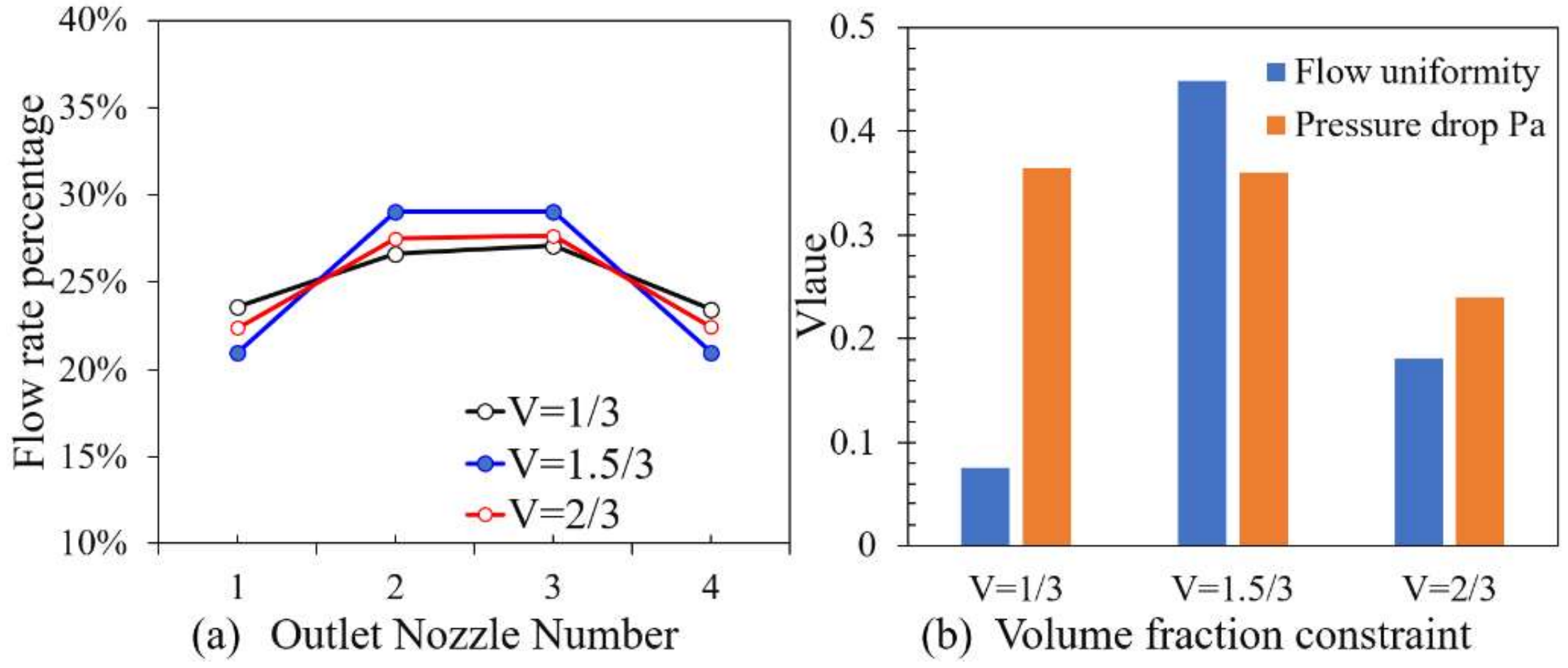

**Figure 10.27:** Comparison of flow uniformity and pressured drop for different volume fraction value.

## 10.4 Conclusion

Manifold level design is very crucial for the thermal/hydraulic performance of the cooler. In this chapter, the conceptual design and improved design based on topology optimization are introduced. For the conceptual design, three different designs are proposed and compared, based on the average temperature, temperature gradient and pressure drop. Moreover, topology optimization with 2D is introduced as a design tool to improve the inlet manifold geometry with regard to the coolant flow distribution and pressure drop. Two different boundary condition with specific velocity and pressure out with equal flow rate are applied on the outlets. The porosity, pressure drop and velocity distribution are characterized and compared under different constraint, such as the penalization factor and volume fraction.

The investigations prove that topology optimization can be used to improve the flow uniformity. The future work will be implementing the current 2D model to 3D model, targeting at the fabrication of the topological heat sink.

**References**

[1] Nadjahi, Chayan, Hasna Louahlia, and Stéphane Lemasson. "A review of thermal management and innovative cooling strategies for data center." Sustainable Computing: Informatics and Systems 19 (2018): 14-28.

[2] Peles, Yoav, et al. "Forced convective heat transfer across a pin fin micro heat sink." International Journal of Heat and Mass Transfer 48.17 (2005): 3615-3627.

[3] Adewumi, O. O., Tunde Bello-Ochende, and Josua P. Meyer. "Constructal design of combined microchannel and micro pin fins for electronic cooling." *International Journal of Heat and Mass Transfer* 66 (2013): 315-323.

[4] Sung, Myung Ki, and Issam Mudawar. "Experimental and numerical investigation of single-phase heat transfer using a hybrid jet-impingement/micro-channel cooling scheme." *International journal of heat and mass transfer* 49.3-4 (2006): 682-694.

[5] Calame, J. P., et al. "Investigation of hierarchically branched-microchannel coolers fabricated by deep reactive ion etching for electronics cooling applications." *Journal of heat transfer* 131.5 (2009).

[6] Copeland, David. "Optimization of parallel plate heatsinks for forced convection." *Sixteenth Annual IEEE Semiconductor Thermal Measurement and Management Symposium (Cat. No. 00CH37068)*. IEEE, 2000.

[7] Van Oevelen, Tijs. "Optimal heat sink design for liquid cooling of electronics." (2014).

[8] M.P.BendsøeandN.Kikuchi. Generatingoptimaltopologiesinstructural design using a homogenization method. Computer Methods in Applied Mechanics and Engineering, 71:197–224, 1988.

[9] Lavrov N, Lurie K, Cherkaev A (1980) Non-uniform rod of extremal torsional stiffness. Mech Solids 15(5):74–80.

[10] Glowinski R (1984) Numerical simulation for some applied problems originating from continuum mechanics. Trends in applications of pure mathematics to mechanics, Symp., Palaiseau/France 1983. Lect Notes Phys 195:96–145.

[11] Goodman J, Kohn R, Reyna L (1986) Numerical study of a relaxed variational problem from optimal design. Comp Meth Appl Mech Eng 57:107–127.

[12] Bendsøe MP, Kikuchi N (1988) Generating optimal topologies in structural design using a homogenization method. Comput Methods Appl Mech Eng 71(2):197–224.

[13] T.Dbouk, "A review about the engineering design of optimal heat transfer systems using topology optimization", Applied Thermal Engineering,Volume 112, 5 February 2017, Pages 841-854.

[6] Dede E, Joshi S, Zhou F (2015) Topology optimization, additive layer manufacturing, and experimental testing of an air-cooled heat sink. J Mech Des 137(11):111403(1–9).

[14] Alexandersen J, Sigmund O, Aage N (2016) Large-scale three-dimensional topology optimization of heat sinks cooled by natural convection. Int J Heat Mass Transfer 100:876–891.

[15] Shi Zeng, Poh Seng Lee, "Topology optimization of liquid-cooled microchannel heat sinks: An experimental and numerical study", International Journal of Heat and Mass Transfer. Volume 142, October 2019, 118401

[16] Buckinx (2017). Liquid cold plate: channel geometry for constant temperature distribution. Reference: KULeuven

[17] Van Oevelen, Tijs, and Martine Baelmans. "Numerical topology optimization of heat sinks." International Heat Transfer Conference Digital Library. Begel House Inc., 2014.

[18] Lacko, Paul; Buckinx, Geert; Baelmans, Martine; 2018. A Macro-scale Topology Optimization Method for Flows Through Solid Structure Arrays.EngOpt 2018 Proceedings of the 6th International Conference on Engineering Optimization; 2018; pp. 153 - 163 Publisher: Springer, Cham.

[19] Sicheng Sun, et al., Itherm 2019: Micro heat sink optimization.

[20] Kubo, S., Yaji, K., Yamada, T. et al. “A level set-based topology optimization method for optimal manifold designs with flow uniformity in plate-type microchannel reactors”, Struct Multidisc Optim (2017) 55: 1311.

[21] Zhenyu Liu, et al., “Topology optimization of fluid channels with flow rate equality constraints”, Struct Multidisc Optim (2011) 44:31–37, DOI 10.1007/s00158-010-0591.

[22] Teng Zhou, et al., “Design of microfluidic channel networks with specified output flow rates using the CFD-based optimization method”, Microfluidics and Nanofluidics, 21(1), 11, January 2017.

[23] Siddiqui, Osman K., and Syed M. Zubair. "Efficient energy utilization through proper design of microchannel heat exchanger manifolds: A comprehensive review." Renewable and Sustainable Energy Reviews 74 (2017): 969-1002.

[24] Liu, Hong, Peiwen Li, and Jon Van Lew. "CFD study on flow distribution uniformity in fuel distributors having multiple structural bifurcations of flow channels." international journal of hydrogen energy 35.17 (2010): 9186-9198.

[25] Gonzalez-Valle, C. Ulises, Saurabh Samir, and Bladimir Ramos-Alvarado. "Experimental investigation of the cooling performance of 3-D printed hybrid water-cooled heat sinks." Applied Thermal Engineering 168 (2020): 114823.

[26] L.E. Paniagua-Guerra, S. Sehgal, C.U. Gonzalez-Valle, B. Ramos-Alvarado, Fractal channel manifolds for microjet liquid-cooled heat sinks, Int. J. Heat Mass Transf. 138 (2019) 257–266.

[27] L.E. Paniagua-Guerra, B. Ramos-Alvarado, Efficient hybrid microjet liquid cooled heat sinks made of photopolymer resin: thermo-fluid characteristics and entropy generation analysis, Int. J. Heat Mass Transf. 146 (2020) 118844.

[28] T. Borrvall and J. Petersson. Topology optimization of fluids in Stokes flow. International Journal for Numerical Methods in Fluids, 41(1):77–107, 2003. doi:10.1002/fld.426.

[29] C. Taylor and P. Hood. A numerical solution of the Navier-Stokes equations using the finite element technique. Computers & Fluids, 1(1):73–100, 1973. doi:10.1016/0045-7930(73)90027-3.

[30] Zuckerman, N., and N. Lior. "Jet impingement heat transfer: physics, correlations, and numerical modeling." *Advances in heat transfer* 39 (2006): 565-631.[31] Huang, Gan, et al. "Biomimetic self-pumping transpiration cooling for additive manufactured porous module with tree-like micro-channel." International Journal of Heat and Mass Transfer 131 (2019): 403-410.

[31] Huang, Gan, et al. "Biomimetic self-pumping transpiration cooling for additive manufactured porous module with tree-like micro-channel." International Journal of Heat and Mass Transfer 131 (2019): 403-410.

[32] Van Oevelen, Tijs. "Optimal heat sink design for liquid cooling of electronics." (2014).

[33] Zeng, Shi and Poh Seng Lee. “A Header Design Method for Target Flow Distribution among Parallel Channels Based on Topology Optimization.” 2018 17th IEEE Intersociety Conference on Thermal and Thermomechanical Phenomena in Electronic Systems (ITherm) (2018): 156-163.

# Chapter 11

# 11. General Conclusions and Recommendations

In this Chapter, an overview of the major findings and conclusions of this thesis are presented. Recommendations for further research are presented in Section 11.2.

### 11.1 General conclusions

To cope with the increasing cooling demands for future high-performance devices and 3D systems, conventional liquid cooling solutions such as (microchannel) cold plates are no longer sufficient. Drawbacks of these conventional cold plates, with a coolant flow parallel to the chip surface, are the presence of the thermal interface material (TIM), which represents a major thermal bottleneck, and the temperature gradient across the chip surface. Alternative advanced liquid cooling solutions have been proposed such as inter-tier and intra-tier cooling for 3D systems. These solutions are however not compatible with the fine pitch requirements for high bandwidth communication between different tiers of a 3D system.

Liquid jet impingement cooling is an efficient cooling technique where the liquid coolant is directly ejected from nozzles on the chip backside resulting in a high cooling efficiency due to the absence of the TIM and the lateral temperature gradient. In literature, several Si-fabrication based impingement coolers with nozzle diameters of a few tens of µm have been presented for common returns, distributed returns or combination of micro-channels and impingement nozzles. The drawback of this Si processing of the cooler is the high fabrication cost. Other fabrication methods for nozzle diameters of a few hundred µm have been presented for ceramic and metal. Low cost fabrication methods, including injection molding and 3D printing have been introduced for much larger nozzle diameters (mm range) with larger cooler dimensions. These dimensions and processes are however not compatible with the chip packaging process flow. This PhD focuses on the modeling, design, fabrication and characterization of a micro-scale liquid impingement cooler using advanced, yet cost-efficient, fabrication techniques. In the framework of my Ph.D. work, the main conclusions are summarized in the following parts.

As the first achievement of this thesis, an extensive literature review about multi-jet impingement coolers has been summarized systematically in **Section 1.2**, including the cooler material, nozzle array geometry, and the achieved thermal performance. The graphical representations of the geometrical, thermal and hydraulic specifications of the cooler described in the literature (figure 1.x and figure 1.xx)illustrate the trend of the nozzle density on the chip area as a function of the nozzle diameter, and the trend of the normalized required pumping power in the cooler as a function of the dissipated heat flux in the chip. Multi-jet impingement coolers can achieve very high heat transfer coefficients. However, in the case of the small nozzle diameters, high pressure and consequently high pumping power is required.

***Modeling study***

A multi-level modeling methodology based on nozzle level unit cell models and full cooler level models has been introduced **Section 3.1** for the thermal and hydraulic assessment of impingement jet cooling. Different comparison metrics including the coefficient of performance (COP) and the trade-off char of the thermal resistance as function of the required pumping power, are introduced to investigate the combined impact of the jet array design parameters on the thermo-hydraulic cooler performance. These parameters mainly include the nozzle density, the cavity height, and the nozzle diameter. The modeling results show that it is not necessary to scale up the number of unit cells and to shrink the nozzle diameter accordingly to improve the thermal performance for a fixed cavity height: a saturation of the thermal performance improvement is observed beyond a specific nozzle density, making the required diameters compatible with polymer fabrication methods.

Besides, dimensionless analysis is performed to describe the performance of the cooler in terms of normalized parameters in **Section 3.2**. The impact of the dimensionless variables including the $d_i$/L, $d_o$/L, H/L, t/L and $t_c$/L are studied fundamentally. Moreover, the individual trend for every single variable is analyzed and the relations are extracted. Then, the *Nu-Re* and *k-Re* correlations have been fitted based on a large DOE of unit cell CFD simulations. The two correlations are validated by our own experimental results, and as well as experimental data from the literature. Finally, fast prediction models for the thermal and hydraulic performance of the cooler, based on the dimensionless analysis.

The full cooler level conjugate heat transfer and fluid flow CFD models indicate that the thermal conductivity of the cooler material has no impact on the thermal performance of the impingement cooler and that the heat transfer is dominated by the convection in the coolant, enabling the use of plastic materials with low thermal conductivity for the cooler, as shown in **Section 5.1**. It is demonstrated that polymer is a valuable alternative material for the fabrication of the impingement cooler instead of expensive Si-based fabrication methods. The full model modeling results also show that it is necessary to include a high-level of detail for the heat sources to capture the chip temperature profile accurately for high cooling rates, as discussed in **Section 5.3**.

***Impingement cooler demonstrations***

For the cooler demonstration, a single jet demonstrator is firstly designed and fabricated as a proof of concept in **Chapter 4**, for the fundamental understanding of the impingement jet cooling, from numerical modeling and experimental characterization point of view. The stagnation region, the recirculation regions and the wall jet regions for typical impingement jet cooling are idetified. Moreover, the symmetry or periodic behavioralong the symmetry boundaries has been investigated and compared. The symmetry boundary condition is chosen in the modeling study after the comparison Most importantly, the correlation between Nusselt number and Reynolds number for the single jet cooling has been extracted both from CFD modeling results and experimental characterization, which shows a good agreement.

Next, the concept of the multi-jet cooling is demonstrated in **Chapter 5** as a proof of concept to prove the improved energy efficiency of the multi-jet cooling compared to single jet cooling. A polymer-based 4×4 array jet impingement cooler with 600 μm diameter nozzles has been design to match the dimensions of the 8x8 $mm^2$ advanced thermal test chip and has been fabricated using mechanical machining in PVC. The experimental characterization shows a very low thermal resistance of 0.25 K/W (0.16 $cm^2$.K/W) and good temperature uniformity across the chip surface. The benchmarking study with literature data for impingement coolers with a large range of inlet diameters shows a very good thermal performance of the fabricated polymer cooler for a low required pumping power. The benchmarking study confirms furthermore that multi-jet cooling is more efficient than single jet cooling and that direct cooling on the backside of the semiconductor device is more efficient than cooling the substrate or base plate. The modeling analysis shows that our proposed impingement jet cooler with distributed outlets achieve better cooling performance than the coolers with common outlets since the cross-flow effects can be reduced.

In order to further improve the thermal/hydraulic performance of the multi-jet impingement cooler, cost-effective 3D printing technologies including Stereolithography (SLA) and digital light processing (DLP), are investigated for the demonstration of the chip level 3D printed microjet cooler with sub-mm nozzle dimensions, discussed in **Chapter 6**. The conclusion from the first experiment shows that polymer-based 3D printing can create the complex internal geometries for package-level impingement coolers, but that the material aspects (defect-free fabrication and water resistance) are very important. With the high-resolution SLA, the coolers with 3×3, 4×4 and 8×8 inlet nozzle arrays are successfully printed using one single process without assembly of the individual parts. The polymer material Somos WaterShed with excellent temperature resistance is selected as the printing material. The experimental studies show that a very good thermal performance for the 8×8 cooler with 1×1mm$^2$ cooling cells can be achieved as low as 0.13 cm$^2$-K/W for a flow rate of 1000 ml/min. The observed trend with increasing performance is $R_{3\times3} < R_{4\times4} < R_{8\times8}$. The experimental results based on the three 3D printed coolers are used to successfully validate the predictive model developed in **Chapter 3**.

***Cooling application test cases***

Exploiting the flexible fabrication of 3D printed cooler, the hotspot targeted jet impingement cooling concept is introduced and successfully demonstrated in **Chapter 7**, with a chip-level jet impingement cooler with a 1 mm nozzle pitch and 300 μm nozzle diameter fabricated using high-resolution SLA. A detailed trade-off between the thermal performance improvements and the higher required pressure drop and pumping power shows that the hotspot targeted cooler outperforms the uniform array cooler in terms of energy efficiency despite the increase in pressure drop. The validated CFD models also show that the hotspot targeted cooler can be further improved by providing outlet nozzles over the full chip area instead of near the inlet nozzles covering the hotspot areas only.

As a second test case, the package-level 3D printed direct liquid micro-jet array impingement cooling concept is applied to a 2.5D interposer packages with and without metal lid which is discussed in **Chapter 8**. The experimental results show that the presence of the lid (and mainly the TIM) results in a higher chip temperature, where the relative impact of the lid increases as the flow rate increases. Furthermore, it is demonstrated that the bare die jet impingement cooling on the dual-chip package can realize a very low thermal coupling between the chip of only 4%, which is 9 times lower than typically reported values for multi-chip modules. Moreover, an improved 3D printed fluid delivery manifold design with a lateral feeding structure can realize 50-60% reduction of the pressure and required pumping power for the same thermal performance and at the same time, achieve a reduction of the cooler thickness by a factor of 2 compared to the reference vertical feeding design. Finally, the parameter sensitivity studies show a low thermal resistance of the TIM (below 3 mm$^2$-K/W) is required in order for the lidded package cooling to achieve the same cooling performance as the lidless package cooling at a flow rate of 1 LPM.

For the third test case, the package-level 3D printed multi-jet cooling concept with sub-mm microjets, is applied to a 23×23 mm² large die with high power dissipation in **Chapter 9**. The experimental results show that 3D printed large die cooler achieves a chip temperature increase of 17.5°C at a chip power of 285 W for a flow rate of 3.25 LPM, Moreover, a long-term measurement has been conducted, showing that the thermal performance of the 3D printed large die cooler remains constant over the measurement period of 1000 hours. During this period, no reliability issues have been observed. Moreover, an innovative design with an additional distribution layer for the large die cooling has been proposed and demonstrated, to improve the flow non-uniformity issues for the large die application. It is experimentally shown that the improved design achieves a better chip temperature uniformity compared to the reference design.

The conceptual design and improved design based on topology optimization for the inlet manifold are introduced in **Chapter 10**. For the conceptual design, three different designs are proposed and compared, based on the average temperature, temperature gradient and pressure drop. Moreover, topology optimization with 2D is introduced as a design tool to improve the inlet manifold geometry with regard to the coolant flow distribution and pressure drop. The investigations prove that topology optimization can be used to improve the flow uniformity, as well as save design space for the manifold level design.

## 11.2 Recommendations for further Research

In this thesis, package level jet impingement has been demonstrated and applied to different configurations, showing high cooling efficiency. However, there are still several aspects of the cooling solutions needed to be developed further. The most important aspect is the material compatibility between cooler material, coolant, package materials and the reliability requirements of the application. Other aspects include the further continuation of the cooling design optimization, the experimental characterization and the cooling applications.

### 11.2.1 Cooler material aspects

As for the cooler material properties, low CTE, high Heat Deflection Temperature (HDT) and manufacturability are needed for reliable cooler. For the current used 3D printed polymer cooler, the CTE of the polymer is very high, which is not compatible with Si and Package laminate. For the cooler assembly options: the first option is to assemble the package to the board, and then mount the cooler on the package by using glue or clamping. For this type of assembly, there is no harsh temperature requirement for cooler material. For the second assembly option, the cooler is first mounted and sealed on the package, and then assemble the package with cooler on the PCB. However, this cooler assembly option needs to survive the reflow temperature (250°C). Regarding with the cooler materials, two alterative solutions are proposed in the next step: glass cooler and polymer with lower CTE.

***Printed cooler on substrate***

A preliminary study with the CTE modified cooler is in the collaboration with PMA, KU Leuven [2]. The Silica and ceramic fillers are added into the polymers to lower the CTE of the polymer material. On the other hand, the filler concentration should be limited to keep the material printable. Therefore, a systematic DOE is necessary to optimize the 3D printing process. For the 3D printing, the Digital Image Correlation (DIC) technique was used to measure the CTE of 3 composite materials for 3D printing of impingement cooler, as shown in Figure 11.1.

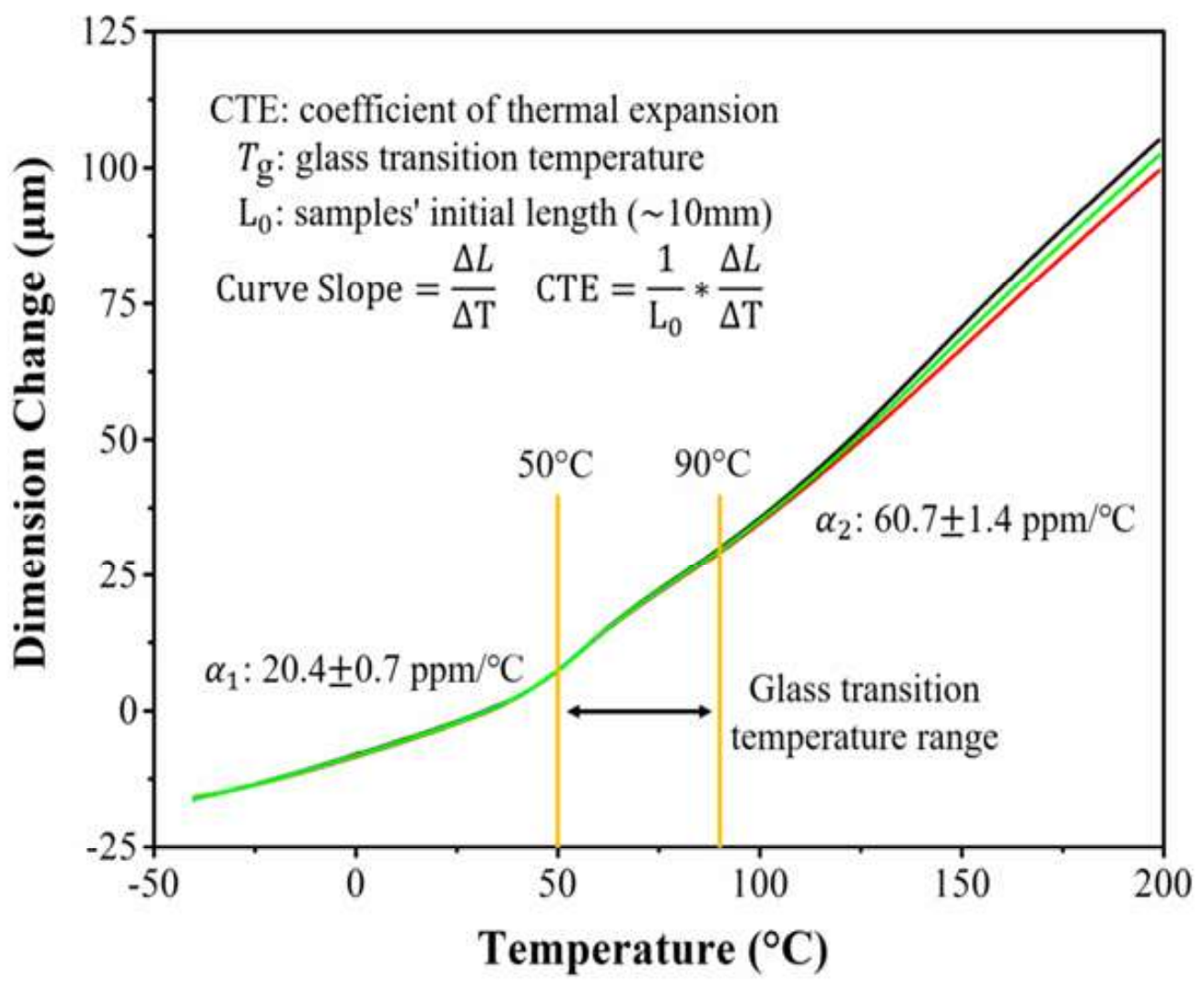

**Figure 11.1:** Dimensional change with temperature of 3 composite samples (Manufactured under the same conditions with the SEM samples) [2].

As for the material modification process: A mixture of resin with silica was used as printing feedstock, to reduce mismatch of thermal expansion coefficient (CTE) between the part and PCB. The silica loading of 60 vol% is appropriate to achieve a compromise between viscosity of the suspension and CTE. Adhesion forces between printed parts and PCBs were measured, showing a significant correlation with the PCB surface roughness. Thermal cycling test indicated that the tailored materials owned excellent CTE compatibility with PCB. Ten samples were tested in the temperature shock chamber. After 100 cycles, only one sample failed during the test. It means 90% of the samples can survive after 100 cycles from -40°C to 80°C. This result indicates the prepared materials owned good CTE compatibility with the PCB substrate.

In general, the current developed 3D printed cooler has still high CTE value of 60.7 ppm/ºC, but it shows a significant improvement (2.5x) compared to the reference material. The next step is to further reduce the CTE of the 3D printed material with high mechanical and thermal reliability.

***Low CTE Glass cooler***

Glass has interesting material properties: lower CTE than printed polymers and more compatible with Si and package. Moreover, Glass also has the high temperature resistance. For the fabrication of the glass cooler, subtractive manufacturing technique is used to create the complex internal structures. The laser beam is focused locally to modify the density inside the glass. Therefore, the inside cavity thickness should be limited to make sure the laser beam pass through the glass. Moreover, the cavities created can be removed by additional chemical etching. For the demonstration of the glass cooler, 4×4 nozzle array is designed, with nozzle diameter of 0.6 mm. Different from the 3D printed cooler, the inlet chamber thickness is reduced to 1 mm for laser beam modification. The CAD design structure is shown in Figure 11.2. The fabricated glass cooler is shown in Figure 11.3, showing front view, bottom view and the side view. It is shown that the nozzle diameter of the glass has very small variations comparing with the nominal design. The SEM images with the nozzle and microchannel of the glass cooler are shown in Figure 11.4.

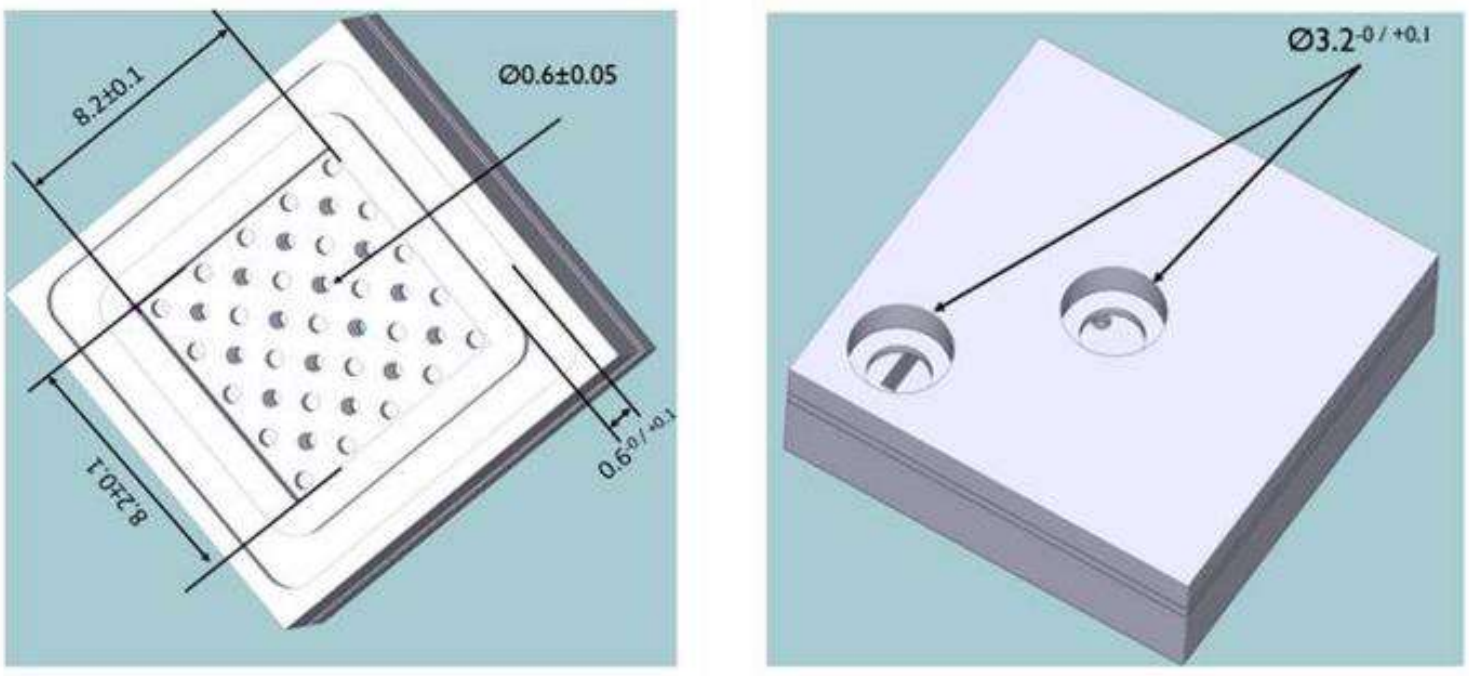

**Figure 11.2:** Demonstrated for cooler geometry:  4x4 nozzle array design.

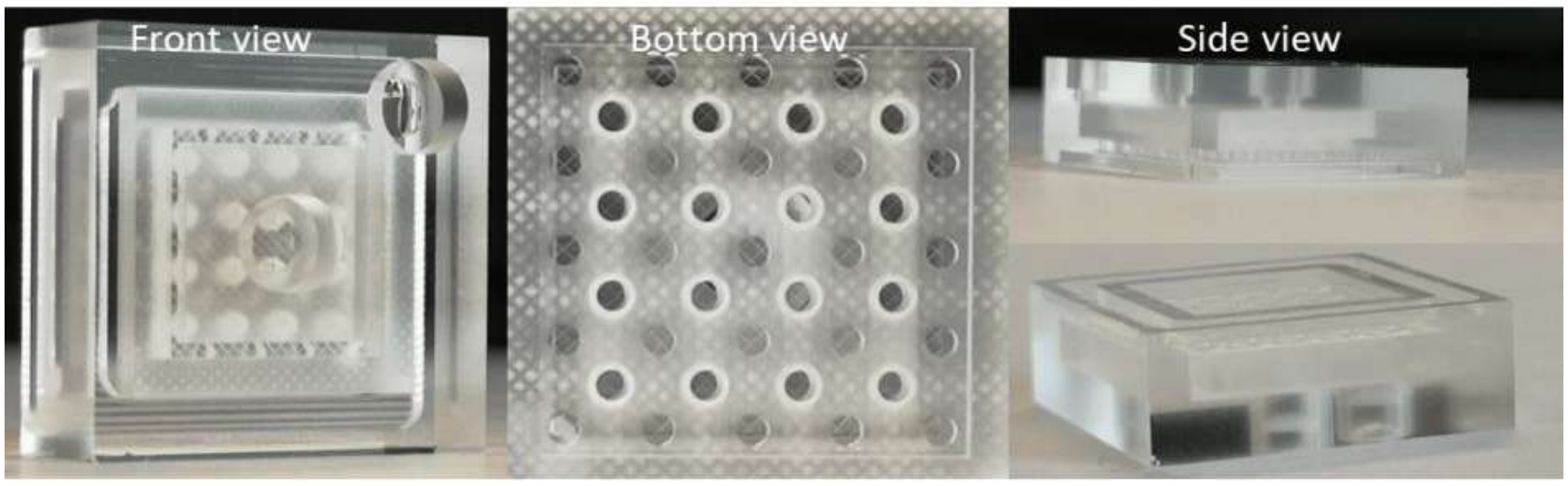

**Figure 11.3:** Demonstrated for cooler geometry with 4×4 nozzle array design.

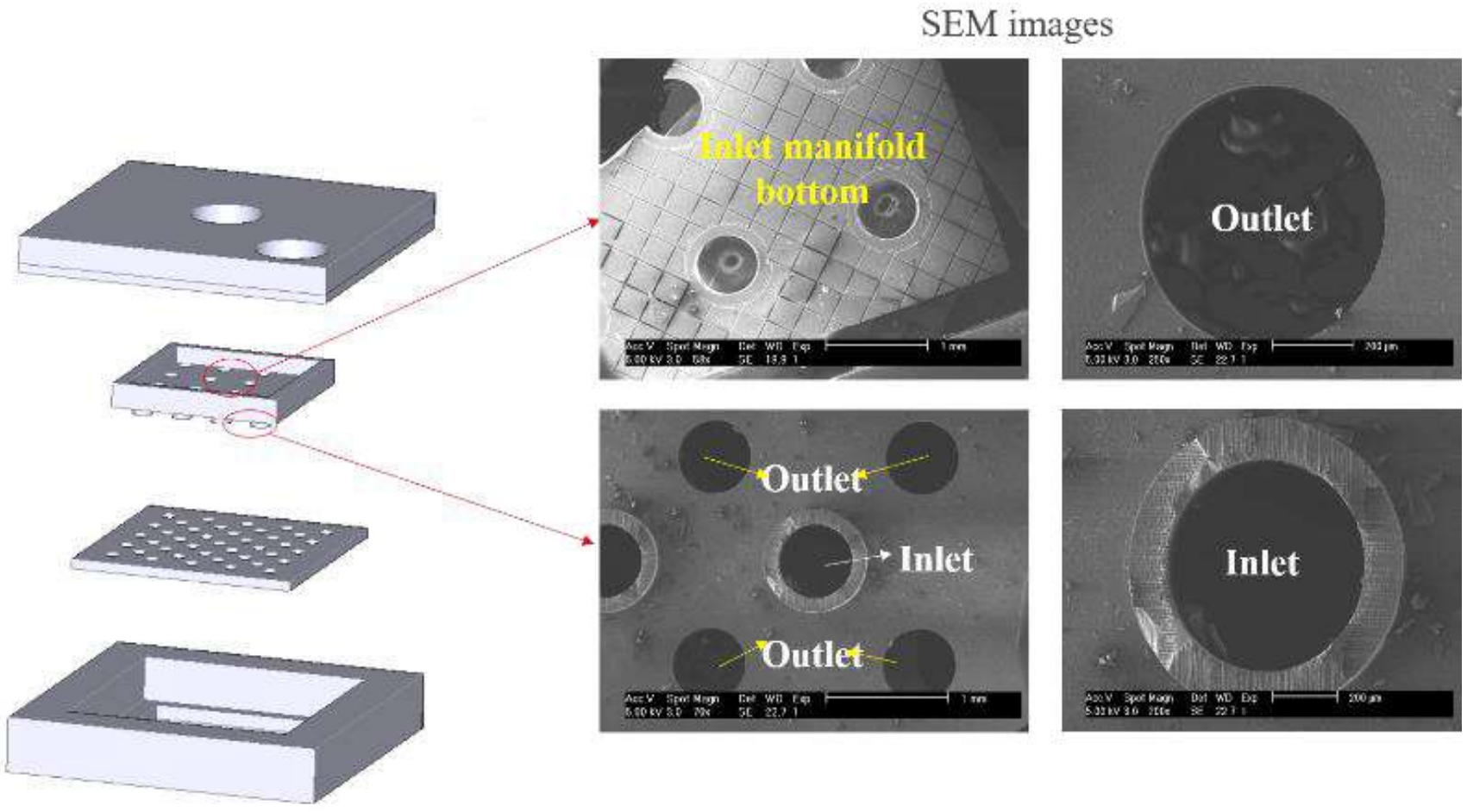

**Figure 11.4:** Glass cooler demonstrator with SEM images.

For the assembly of the glass cooler, the PTCQ+ thermal test vehicle is first soldered to PCB, and then the glass cooler is glued to the package. Also, the In-/Outlet tubes are glued to the glass cooler. The final assembled glass cooler and 3D printed cooler are compared in Figure 11.5. For the thermal measurements of the glass cooler, the set water flow rate is 1 L/min, with water inlet temperature of 10°C. The full power is applied to PTCQ+ chip with 55W. The temperature profile comparison between the glass cooler and 3D printed cooler is shown in Figure 11.6. Higher temperature gradients across the chip area are observed in the case of the glass cooler. The difference in volumes of internal cavities could be the cause of different cooling performance.

The next step for the study would be the investigation of the reliability of the glass-on-substrate and 3D printed cooler-on-substrate, under thermal cycling test.

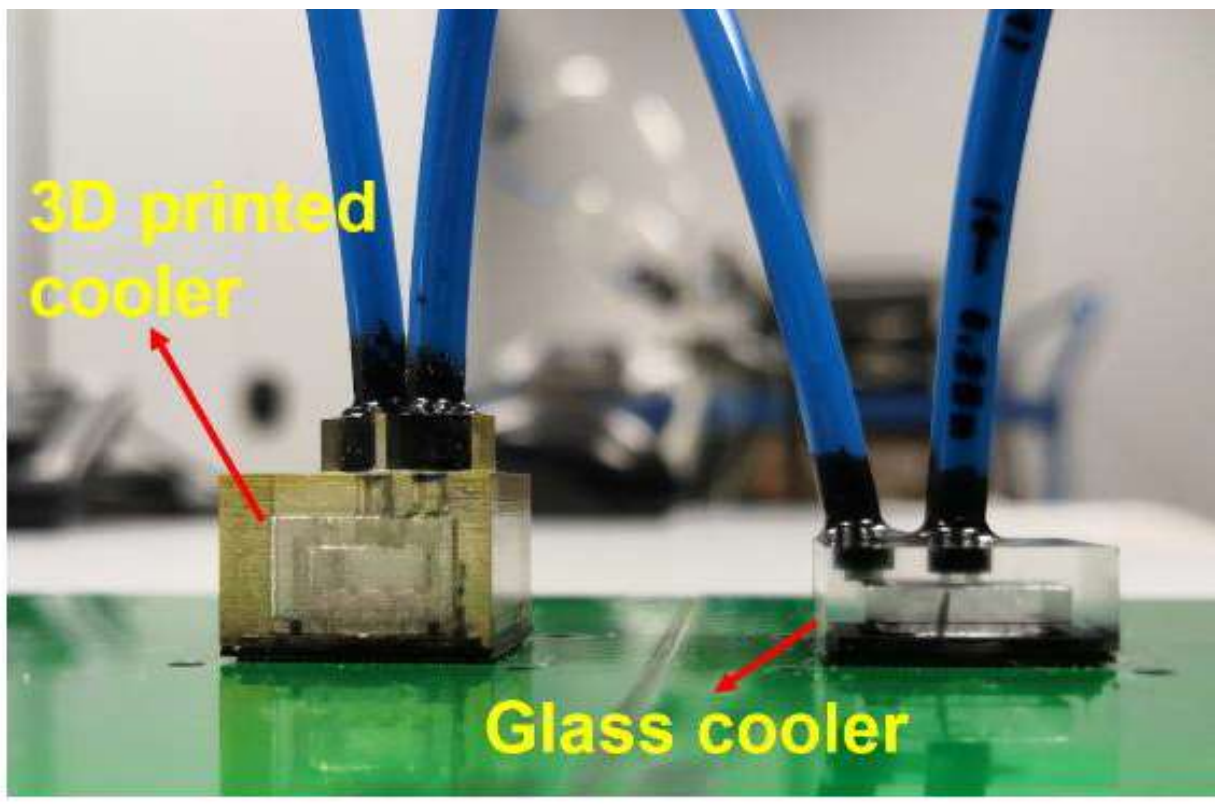

**Figure 11.5:** Glass cooler demonstrator with SEM images.

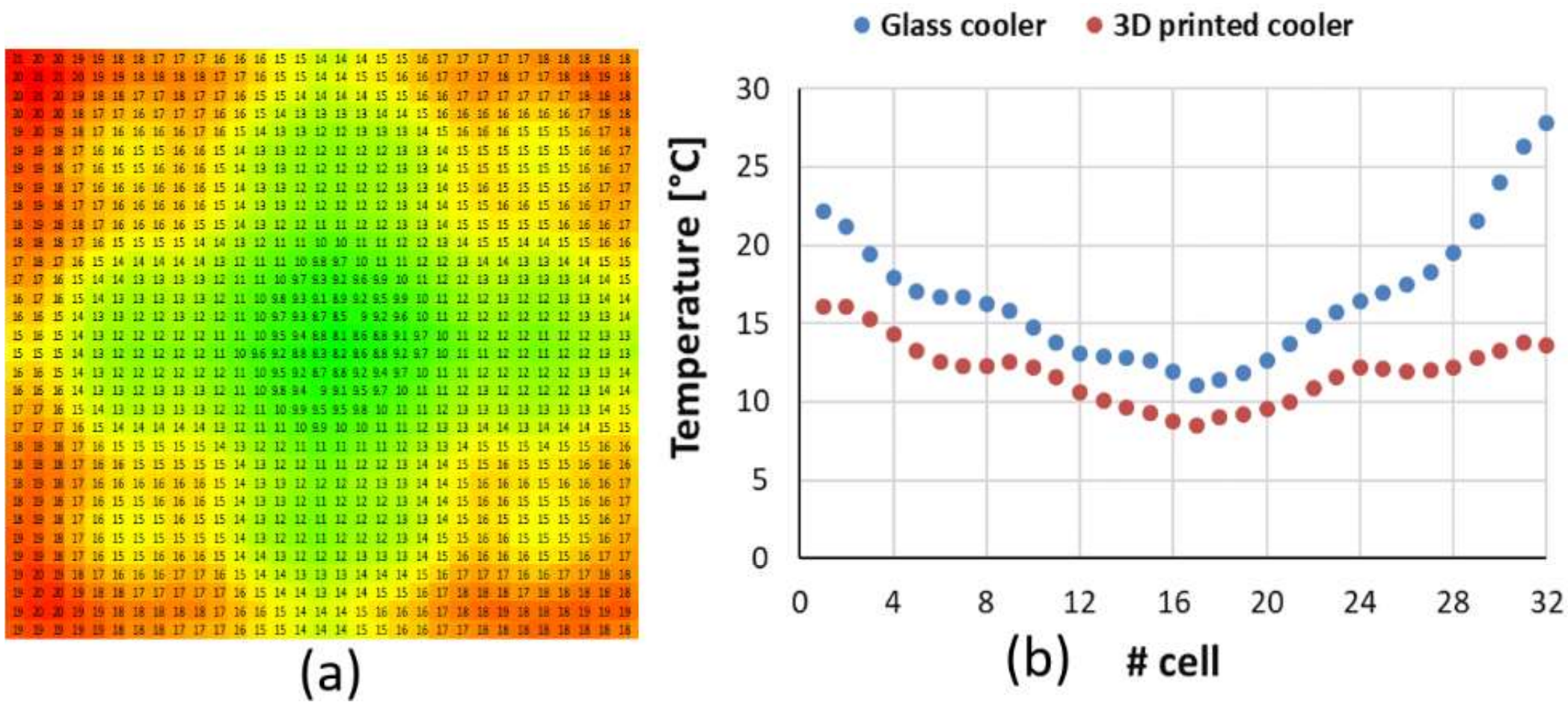

**Figure 11.6:** Experimental measurements with the glass cooler and 3D printed cooler glued on substrate.

### 11.2.2 Design optimization: 3D TO

Application of topology optimization techniques for the complex internal geometry of the chip package level impingement cooler. The objective is to fabricate the improved cooler geometry using high-resolution 3D printing and to characterize the thermal performance of the cooler using imec's high resolution thermal test chip. The current topology design is based on the 2D model. In the next step, the 3D model with topology optimization will be investigated, demonstrated and experimentally characterized. The thermal and hydraulic performance will be benchmarked with the standard 3D printed cooler design. Figure 11.7 shows an example of the 3D printed heatsink using 3D topology optimization method.

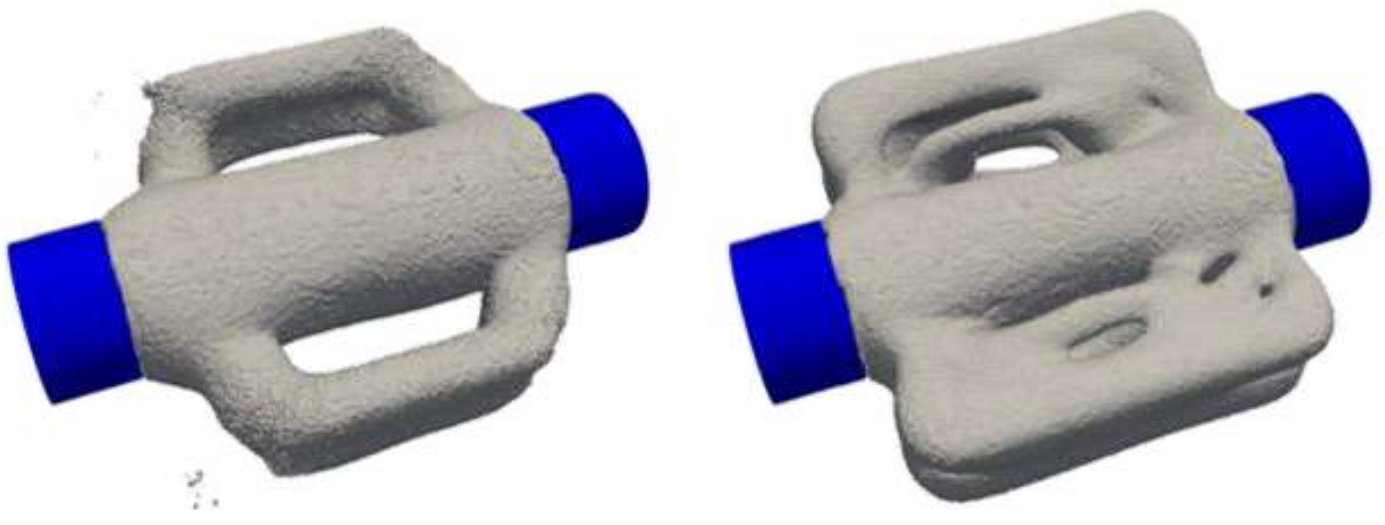

(a) $w_1/w_2 = 5:1$ (b) $w_1/w_2 = 1:1$

**Figure 11.7:** Examples of the 3D printed heatsink using topology optimization [3]

### 11.2.3 Experimental characterization

The previous thermal test vehicle PTCQ has limitation to dissipate different power value for different heater cells. An advanced thermal test vehicle PTCQ+ is developed in imec, where the joule heat dissipation levels can be programmed for every cell. Figure 11.8(a) shows the 16×16 heater array and temperature sensor arrangement across the die size of 4 ×4 mm$^2$. Figure 11.8(b) shows the details of the heater meander structure inside the 240×240 $\mu$m$^2$ cell, where the diode is located in the center of the cell. The heater is fabricated through BEOL (back end of line) process, and located in metal 5 layer, as shown in Figure 11.8(c). The new PTCQ+ test vehicle provides more options for the cooling performance investigation.

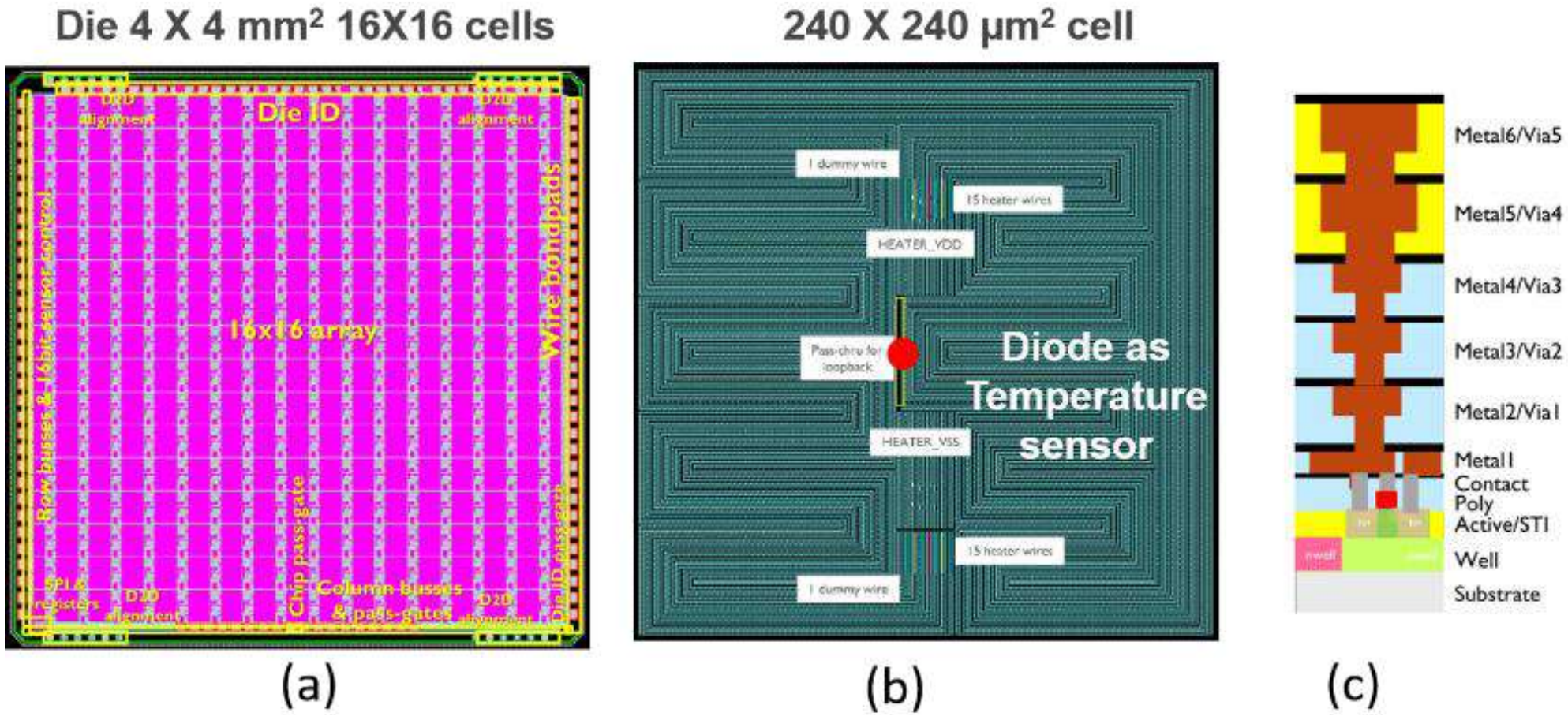

**Figure 11.8:** Details of the new thermal text vehicle PTCQ+: (a) heater map over the whole chip area; (b) heater structure inside the 240×240 μm2 cell; (c) Cross-section of the PTCQ+ with indication of the heater location in metal 5 [1].

### 11.2.4 Cooler applications

***Hotspots targeted cooling for arbitrary power distribution***

The hotspots targeted cooling concept has been introduced in chapter 7. A systematical study based on the defined hotspots pattern is conducted both numerically and experimentally. However, chapter 7 only focuses on the same power density at the hotspots and zero power density for the background region. In real application, there are different power density levels in the microprocessor.

Based on the PTCQ+ test vehicle, the hot spots can be programmed with different power value, as shown in Figure 11.9(a). For the design methodology, the local heat transfer coefficient below a jet depends on diameter and nozzle flow rate, where the relations for heat transfer and pressure drop can be extracted from unit cell model. Below is the example for 1 mm pitch nozzles:

$$htc = 8440 \cdot d^{-0.4157} \cdot \dot{m_{nz}}^{(0.7843-0.6624 \cdot d)} \quad 11.1$$

$$\Delta p = 0.655 \cdot d^{-4} \cdot \dot{m_{nz}}^{1.76} \quad 11.2$$

where *htc* is the heat transfer coefficient for the single cooling cell. $m_{nz}$ is the flow rate per nozzle, d is the nozzle diameter to be optimized.

Based on the fitting, the nozzle diameter can be determined by the generic design optimization methodology for the arbitrary power distribution map. Below is an example for 25 × 25 mm$^2$ chip with 25 × 25 nozzle array, and the random power map is generated by programming each cell between 50 and 350 W/cm$^2$ power density. The final optimized nozzle diameter distribution map is shown in Figure 11.9(b).

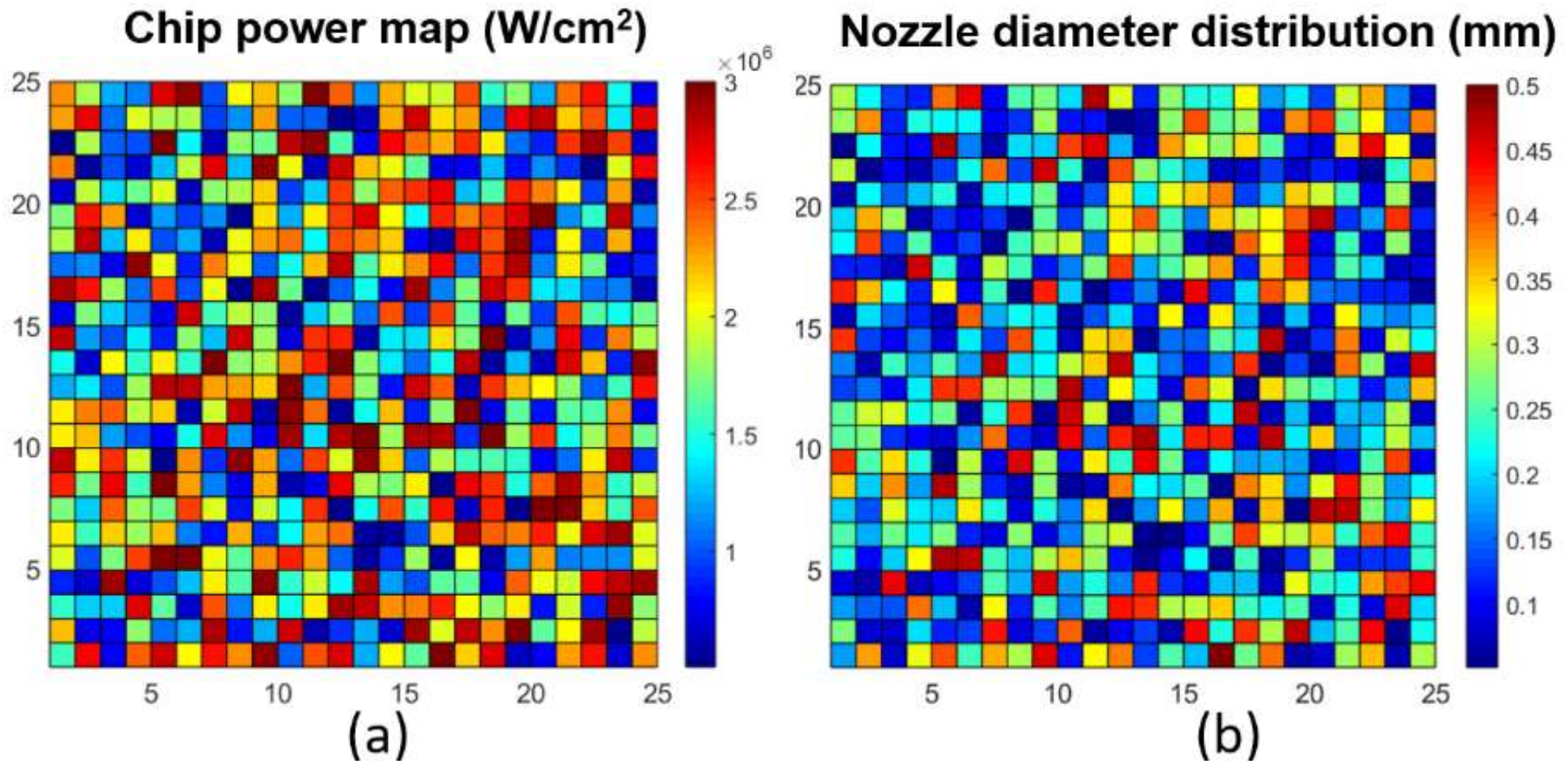

**Figure 11.9:** Nozzle array design methodology presented to define nozzle diameters to obtain uniform temperature distribution for arbitrary power map.

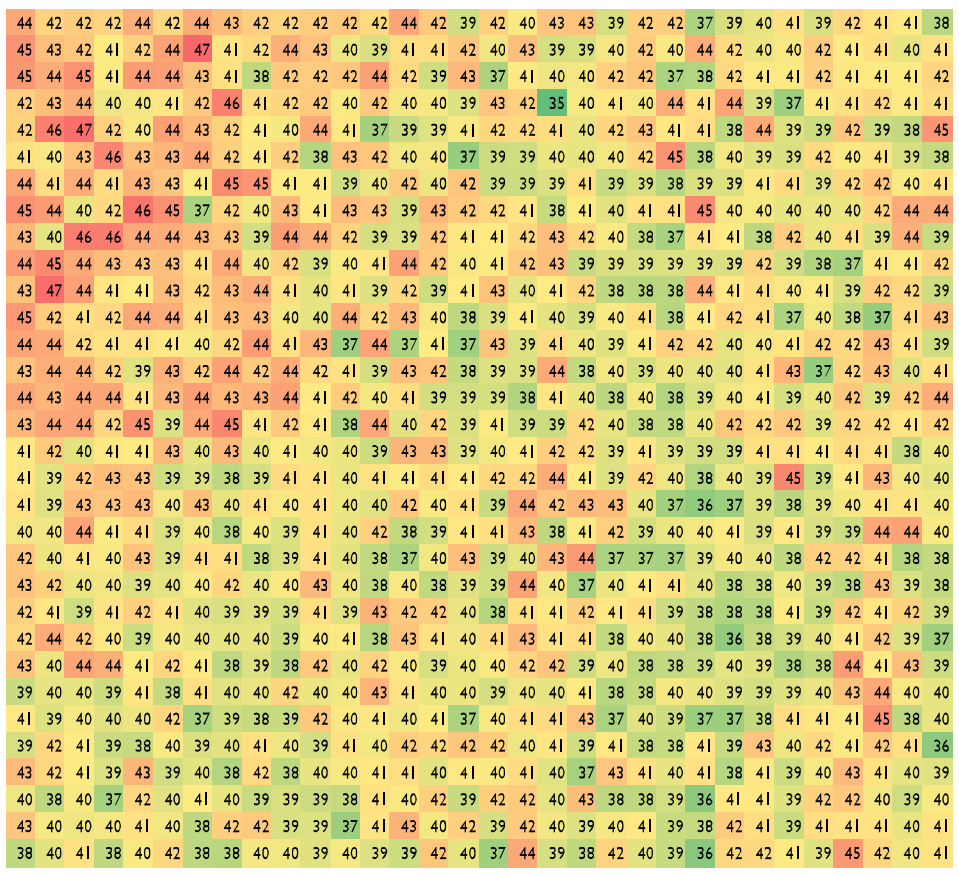

**Figure 11.10:** Distribution of heat dissipated in every cell [mW] with fully uniform heating in PTCQ+.

Figure 11.10 shows the distribution of heat dissipated in every cell, with fully uniform heating in PTCQ + thermal test chip. The power delivered to the chip (measured by source) is 55W. Therefore, the power dissipated in every cell is 40mW ± 15% ($3\sigma$), where the observed variation is due to measurement accuracy. For the future work, the new demonstrator based on the defined arbitrary power density map will be fabricated by 3D printing or other advanced fabrication techniques. The performance will be benchmarked with the standard uniform array cooling.

***Active flow control in nozzles for dynamic power profiles***

The cooling solution, fabricated using low-cost plastic fabrication techniques, demonstrates a high thermal performance, good temperature uniformity and a reduction in cooler size while it only requires a low pumping power for the coolant flow circulation. To increase the number of application options for this promising cooling method, the next step is to introduce the active control of the flow rate in the individual liquid jets to match the temporal and spatial coolant flow rate distribution with the heat load of the chip and to improve the cooler design in order to reduce the cooler drop and improve the flow and temperature distribution.

Development of an active flow control actuation method and control strategy to control the flow rate in the jets depending on the local cooling need in order to maintain a constant chip temperature and to improve the energy efficiency of the cooler and the closed loop liquid cooling system. The design of the actuation mechanism should be compact to be integrated in the package level cooler. Demonstration of flow control on an advanced thermal test chip is used for the model validation and the experimental characterization of the cooler.

**Reference**

[1] Vladimir Cherman, et al., "Thermal, Mechanical and Reliability Assessment of Hybrid Bonded Wafers, Bonded at 2.5µm Pitch", ECTC 2020.

[2] Fei, Guanghai, T-W Wei, et al. "Preliminary Study on Hybrid Manufacturing of the Electronic-Mechanical Integrated Systems (EMIS) via the LCD Stereolithography Technology."

[3] Sicheng Sun, et al., "Micro heat sink optimization.", ITherm 2019.

# Appendix A: Turbulence Model Investigation

## A.1 Unsteady RANS modeling

In the numerical modeling analysis, we first check this physical problem with an unsteady solver, which is unsteady RANS (URANS) simulation to analyze the unsteady state behavior. The time step, $10^{-7}$ s in this case, is calculated based on the cell size and inlet velocity. Two velocity points with the stagnation point and recirculation point are monitored during the URANS modeling. As shown in Figure A.1, after 600 time-steps, the flow is fully developed, and from then on, there is no velocity fluctuation observed, which reveals the steady phenomenon. Therefore, this flow problem is steady at Re = 2048. So, for all the following simulations, we choose the RANS solver instead of URANS solver.

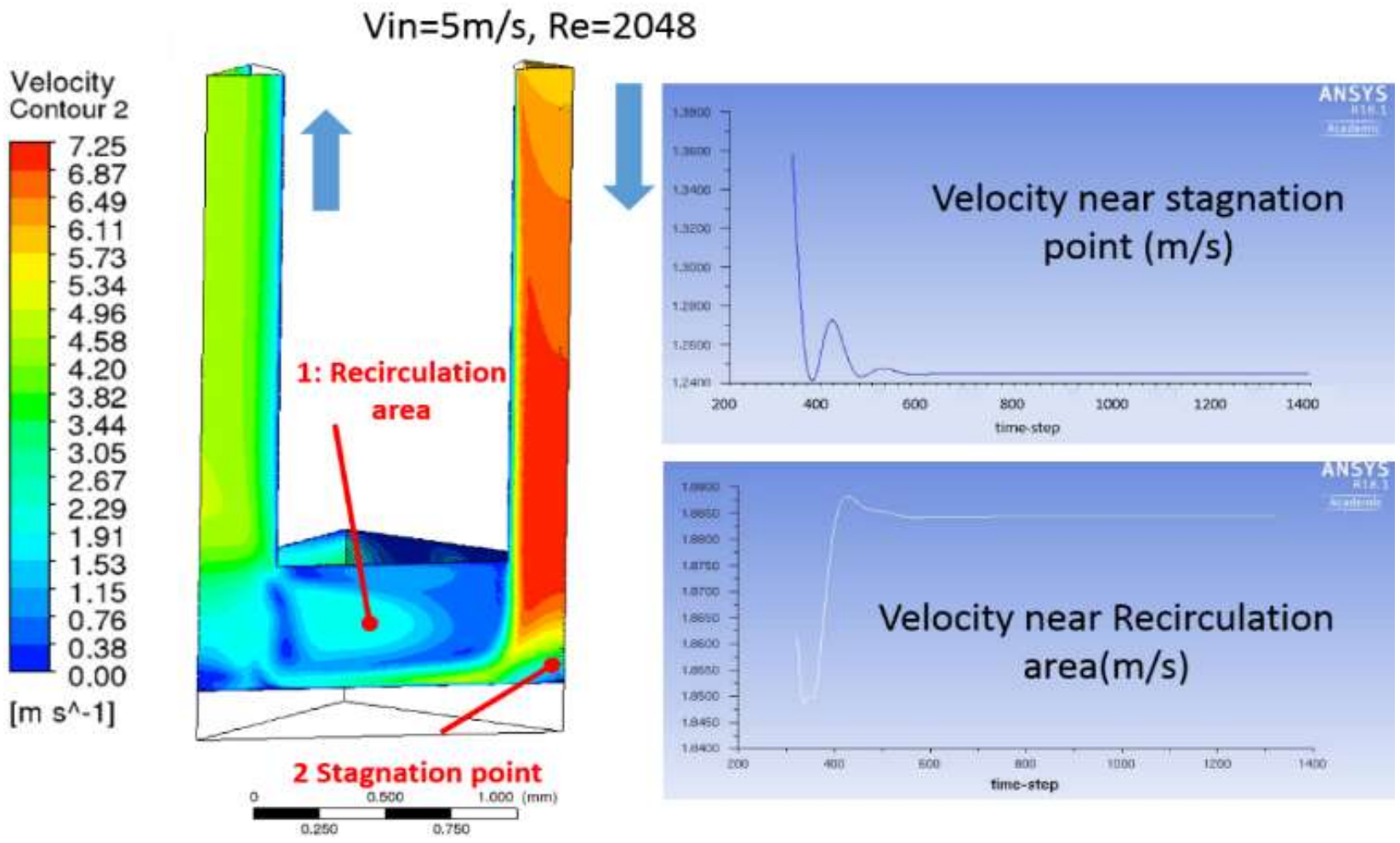

**Figure A.1:** Unsteady flow simulation—unsteady Reynolds-averaged Navier–Stokes (URANS) for $V_{in}$ = 5 m/s and $Re_d$ = 2048. The velocity at two points as function of flow time (time step) is plotted.

## A.2 3D-RANS modeling

Three-dimensional RANS models with different numerical modeling techniques are simulated and compared, including $k$ - $\omega$ SST model, laminar model, $k$ -ε model, Transition SST and Spalart–Allmaras (SA) One-Equation Model. The Reynolds number $Re_d$ is ranging from 30 to 4000, which covers the practical range for this electronic cooling application. The simulation results for $Re_d$ = 1024 are shown in Figure A.2. It can be seen that the temperature distributions for different turbulence models are different even though the flow streamlines show similar behavior. The temperature distributions of the SA model and the $k$-ε model show lower temperatures than the other models. In addition, the temperature patterns for $k$-$\omega$ SST, transition SST and the laminar model show similar temperature distributions at $\mathrm{Re}_d$ = 1024. The strong temperature differences using one turbulence model or another will be explained in details later based on the $Nu_{\mathrm{avg}}$-$\mathrm{Re}_d$ and $f$-$\mathrm{Re}_d$ correlation curve.

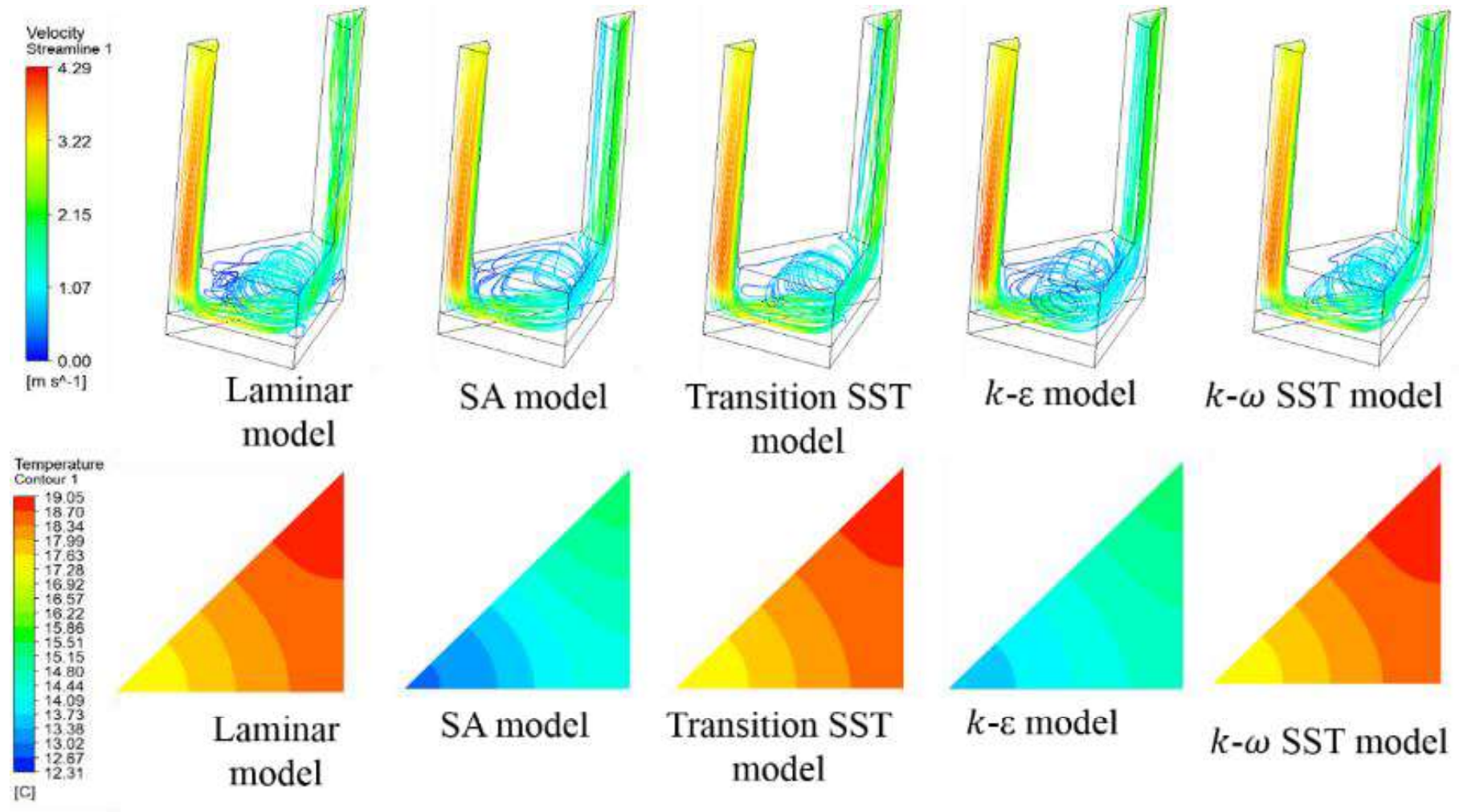

**Figure A.2:** Conjugate flow and thermal unit cell modeling for $\mathrm{Re}_d$ = 1024: top row—velocity streamlines in the unit cell model, and bottom row—temperature distribution in the active region of the Si chip for different turbulent models.

### A.3 LES modeling

As mentioned in the chapter 2, LES is regarded as one of the most promising approaches for the simulation of turbulent flows with given length scale $\delta$, which is often connected to the mesh size. In [1], a well-resolved three-dimensional LES impinging jet model shows a very good heat transfer coefficient prediction. Therefore, we extend the impingement CFD RANS simulations to an unsteady LES simulation as a benchmark. The time step is set to $10^{-3}$ s. As shown in Figure A.3, the flow is fully developed after around 10 iterations. In order to compare with the RANS model, the mean values of the variable are calculated by time-averaging of instantaneous results from 0.1 s to 0.5 s. The velocity and temperature simulation results with the LES model are shown in Figure 7. By using the LES model, the smaller scale flow behavior can also be captured. The simulated $Nu_{\mathrm{avg}}$ and $Nu_0$ for different RANS models are compared with the LES model at $\mathrm{Re}_d$ = 1024, as listed in Table 3. It can be seen that, on the one hand, the laminar model, $k$-$\omega$ SST and Transition SST model can produce better results than any of the high-Re models, matching $Nu_{\mathrm{avg}}$ and $Nu_0$ within 1% compared to the LES model, however by reducing the calculation time by a factor of 6. On the other hand, $k$- ε model and SA model show large $Nu_{\text{avg}}$ prediction errors up to 80%, and the $Nu_0$ prediction errors are above 100%.

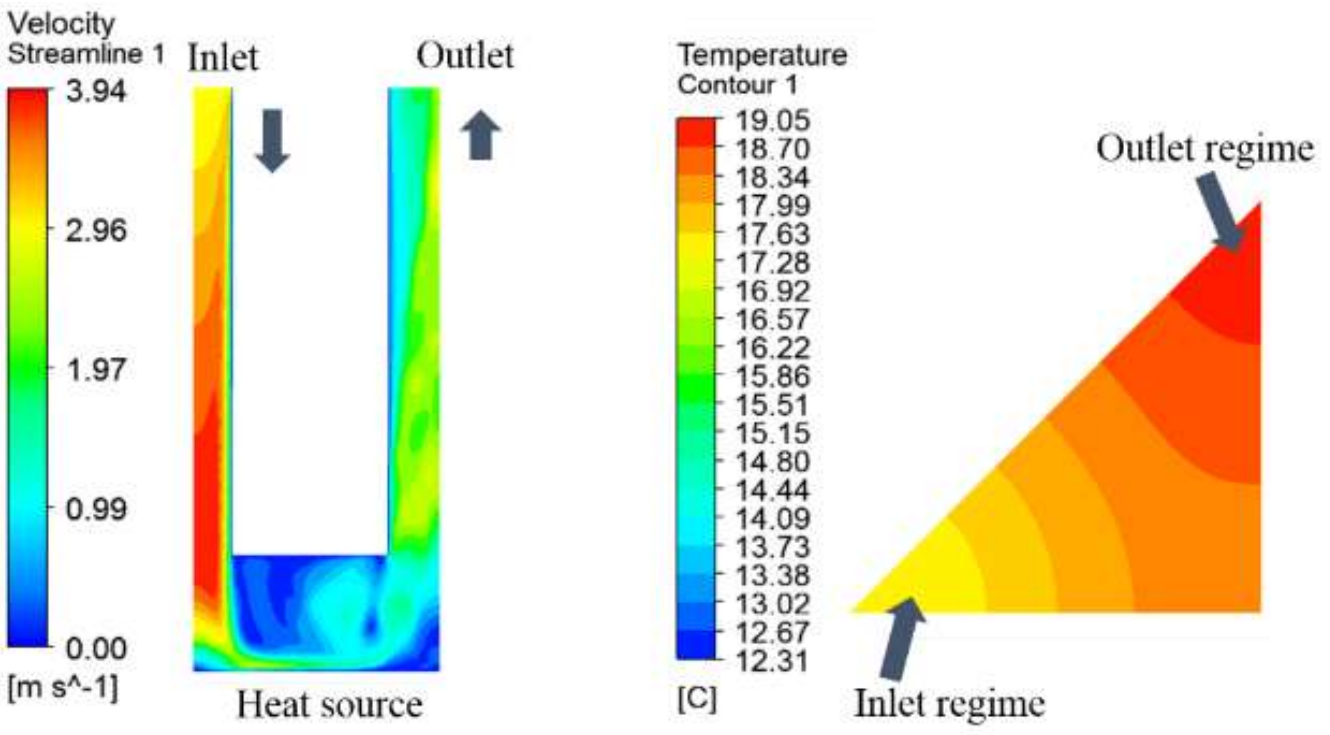

(a)

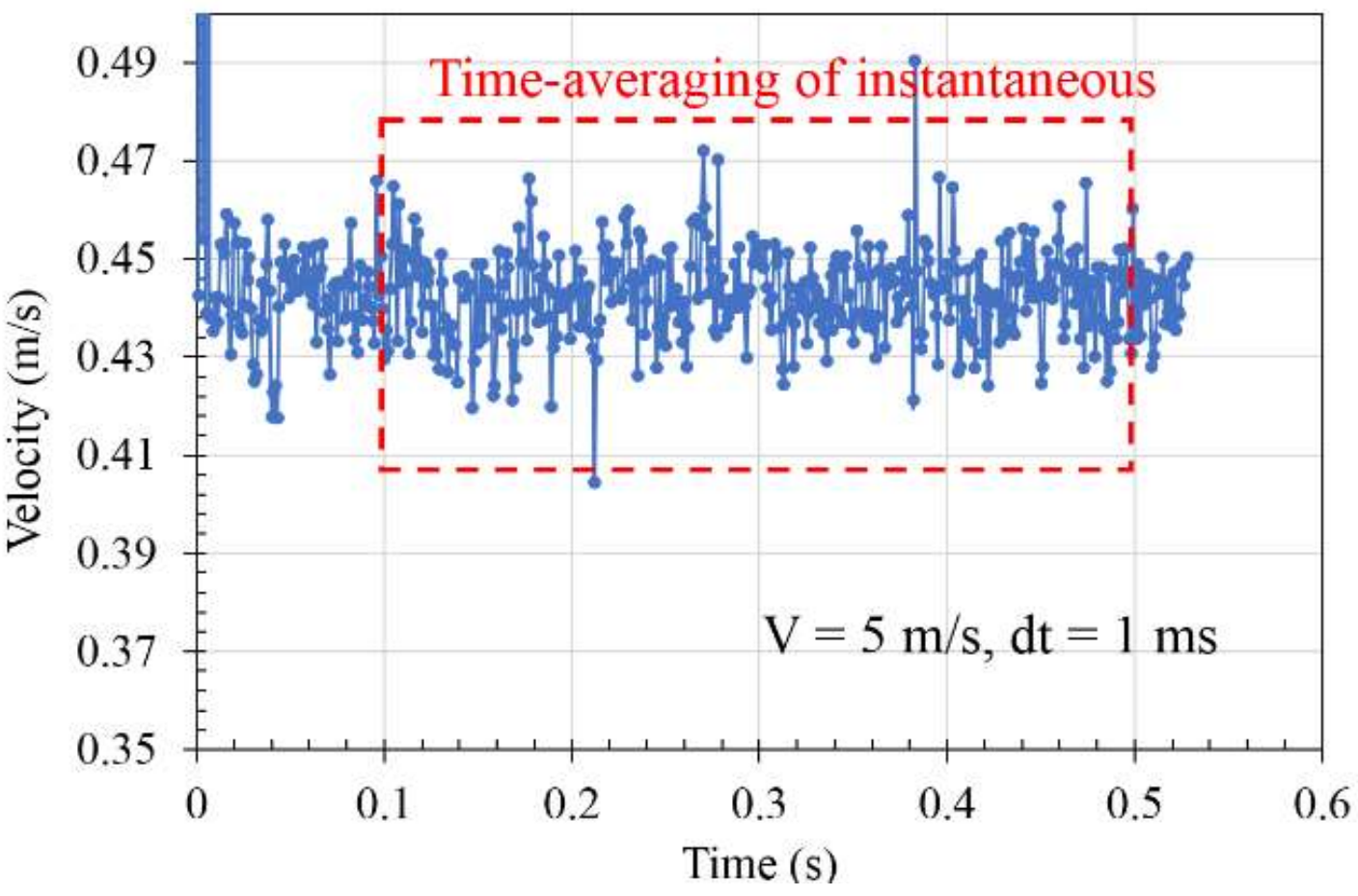

(b)

**Figure A.3:** Large Eddy Simulations (LES) modeling results: (a) velocity and temperature distribution at $Re_d$ = 1024; (b) flow velocity development as function of the time. (4 × 4 nozzle array, di = 0.6 mm)

**Table A.1:** Conjugate flow and thermal modeling under $Re_d$ = 1024.

| **Turbulent Model** | **Re** | $Nu_{\text{avg}}$ | $Nu_{\text{avg}}$ **Difference** | $Nu_0$ | $Nu_0$ **Difference** |
|---|---|---|---|---|---|
| LES model | 1024 | 35.14 | 0 | 39.42 | 0 |
| Laminar model | 1024 | 34.92 | 0.6% | 39.42 | 0.1% |
| $k$-$\omega$ SST | 1024 | 34.84 | 0.9% | 39.37 | 0.1% |
| Transition SST | 1024 | 35.05 | 0.3% | 39.53 | 0.3% |
| $k$-ε model | 1024 | 64.94 | 84.8% | 81.95 | 107.9% |
| SA model | 1024 | 69.27 | 97.1% | 98.28 | 149.3% |

## A.4 Turbulence Modeling comparisons

Detailed comparisons between the $Nu_{\text{avg}}-\text{Re}_d$ correlation, $Nu_0-\text{Re}_d$ and $f-\text{Re}_d$ correlation are shown in Figures A.4–A.5. It can be seen that there is no transition observed for the $Nu$–$\text{Re}_d$ correlation while there is a clear transition point visible for the pressure coefficient correlation $f-\text{Re}_d$. For the $Nu$–Re correlations shown in Figures A.4 and A.5, it can be seen that the laminar model, $k$-$\omega$ SST model and transition SST model show maximum prediction errors of 5% compared with the LES model. For the SA model or $k$-ε model, the difference increases from 5% to 20% as the Reynolds number increases. The reason why the $k$-$\omega$ SST and $k$-ε models are so different is because the $k$-ε model is based on the wall function model, which is good for high Reynolds number case. For the $k$-$\omega$ SST and transition SST models, both are based on the low Reynolds near wall model, the calculation starts from the near wall cells. As for the $f$–$\text{Re}_d$ correlations in Figure A.6, the laminar flow model shows a large difference compared with the LES model around the transition point $\text{Re}_d$ = 650. On the other hand, the $k$-$\omega$ SST model and the transition SST model match very well with the LES model after the transition point. The reason is that the transition SST model in ANSYS Fluent extends the traditional SST $k$-$\omega$ transport equations by tracking two additional variables for intermittency and transition onset using empirical correlations developed by Menter et al. [2]. Various authors have shown that the $k$-$\omega$ SST model shows unsatisfactory performance for jets, both free jets [3] and impinging jets [4]. This arises due to the eddy-viscosity hypothesis used in two-equation turbulence models, that over-predict the mixing rate in the CFD simulation [5]. However, for integral quantities of interest like the heat transfer, the interaction between the liquid fronts on the surface engendered by the jets is a critical criterion. This integral quantity of interest is still well predicted by the $k-\omega$ SST model.

In summary, as for the unit cell model, the transition SST model and $k$-$\omega$ SST model both can predict the average chip temperature, the stagnation temperature on the chip, and also the pressure drop with less than 5% difference, compared with the reference LES model, when the $\text{Re}_d$ is in the range between 30 to 4000.

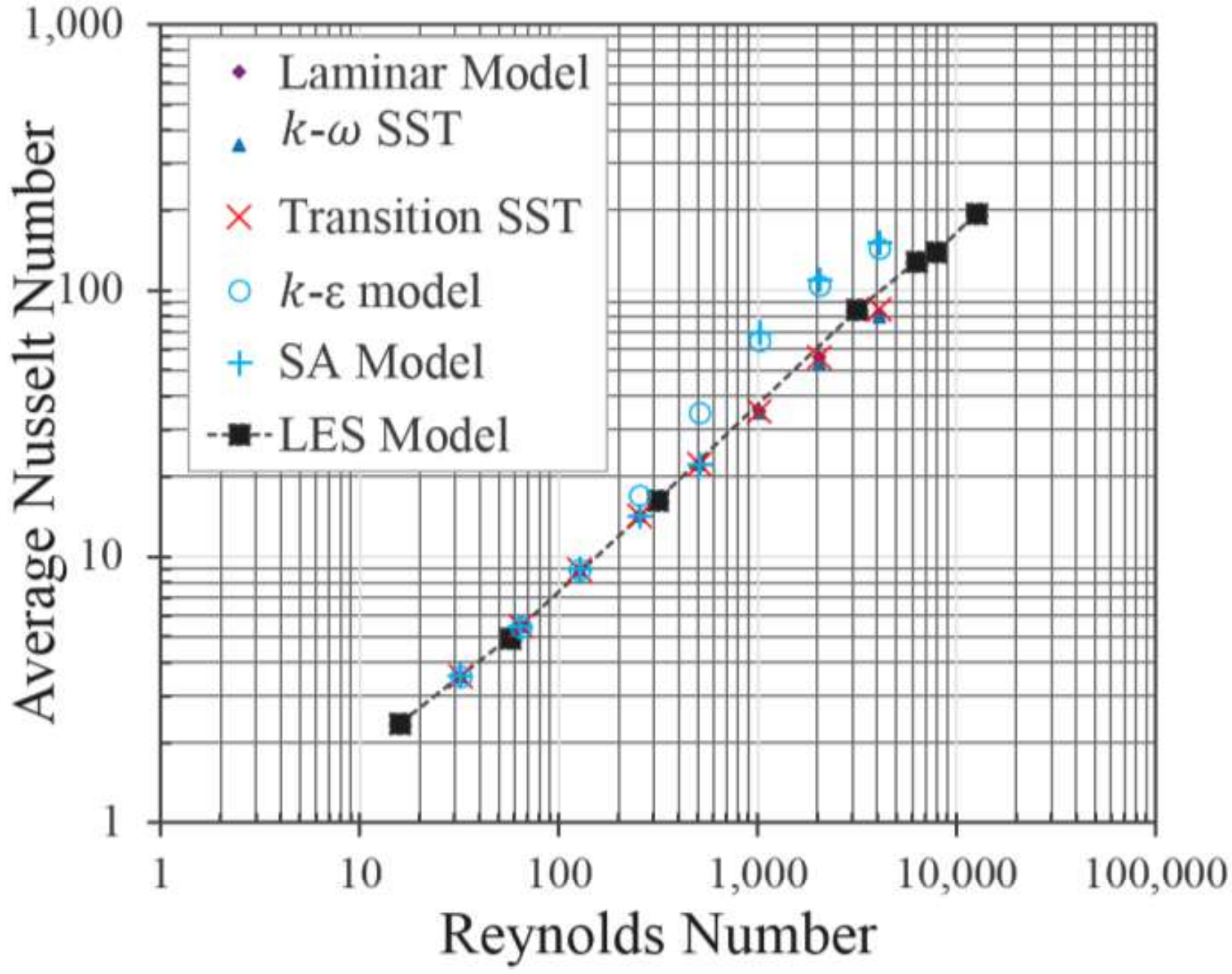

**Figure A.4:** $Nu_{\text{avg}}$–$\text{Re}_d$ correlations: Turbulence model comparison with different RANS models and benchmarked with LES model ($30 \leq \text{Re}_d \leq 4000$).

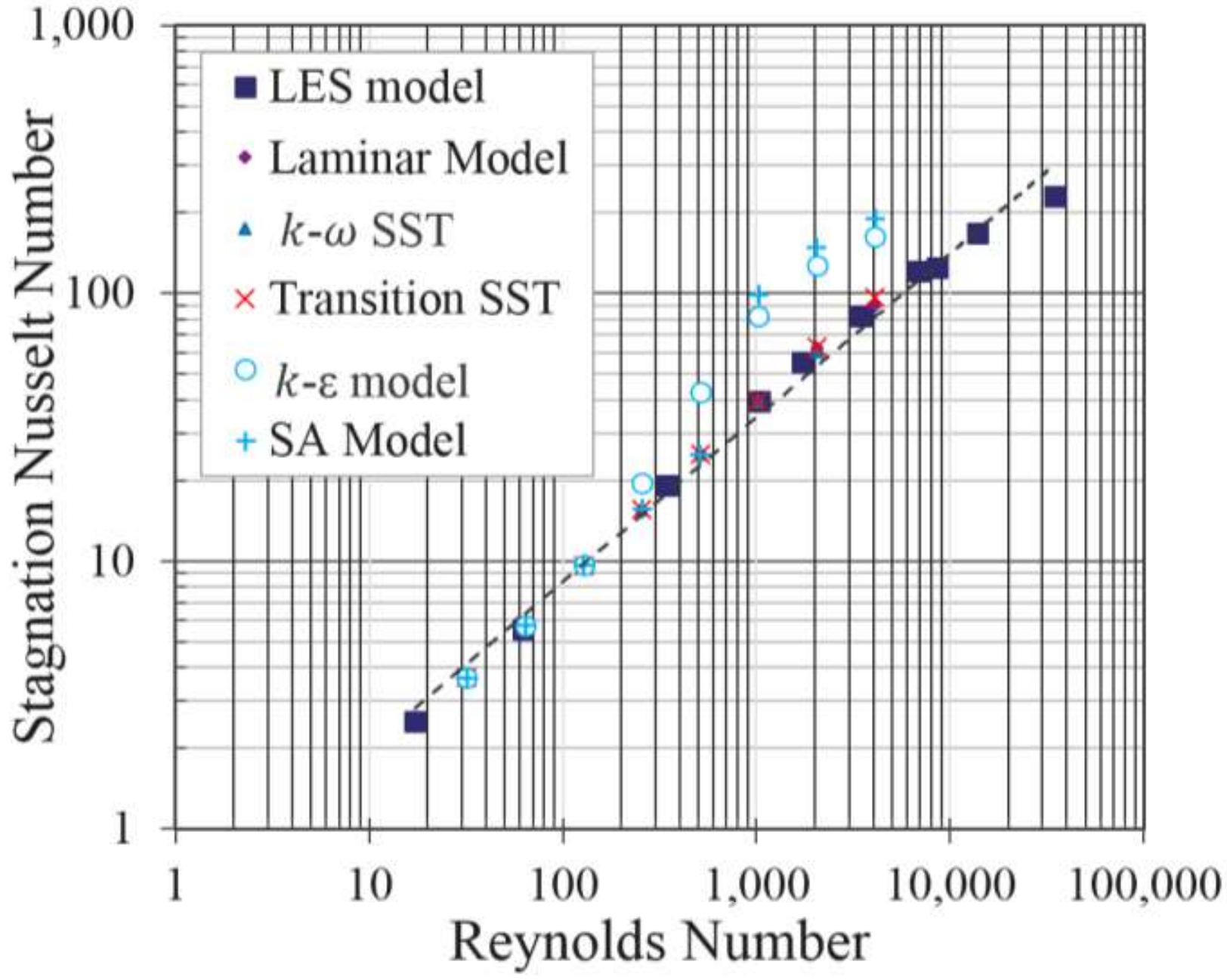

**Figure A.5:** $Nu_0$–$\text{Re}_d$ correlations: Turbulence model comparison with different RANS models and benchmarked with LES model ($30 \leq \text{Re}_d \leq 4000$).

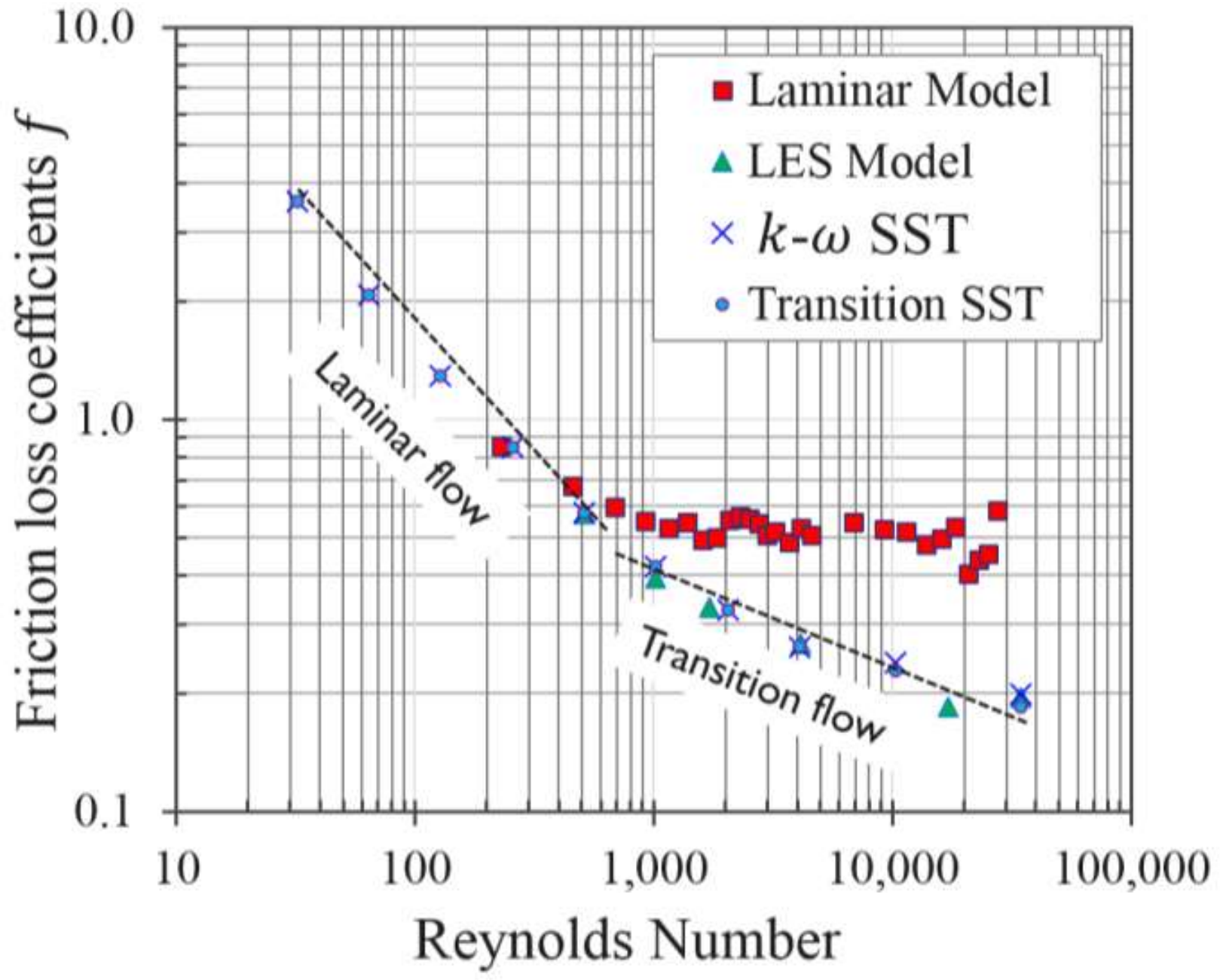

**Figure A.6:** $f-\mathrm{Re}_d$ correlations: Turbulence model comparison with different RANS models and benchmarked with LES model ($30 \leq \mathrm{Re}_d \leq 4000$).

In Figure A.7, the measured average chip temperature values are compared with the full CFD model and unit cell model results. The measured flow rate is ranging from 100 mL/min to 1000 mL/min, resulting in a $\mathrm{Re}_d$ number from 130 to 1400. The heat flux applied on the thermal test chip is 80 W/cm$^2$. Similar with the unit cell model analysis, different turbulence models are used for the full cooler level model, including laminar model, $k$-ε model, $k$-$\omega$ model, Transition SST model and SA model. It can be seen that the full CFD model with SA model overestimates the Nusselt number by a factor of 4 comparing with the experimental result. Moreover, the full model with $k$-ε model also shows very high prediction errors compared with the experiments. As expected, the LES model shows good agreement with the measurements. In general, the comparison shows that the laminar model, the $k$-$\omega$ model and the transition SST model show good agreement with the measured chip temperature, for the $\mathrm{Re}_d$ number from 130 to 1400.

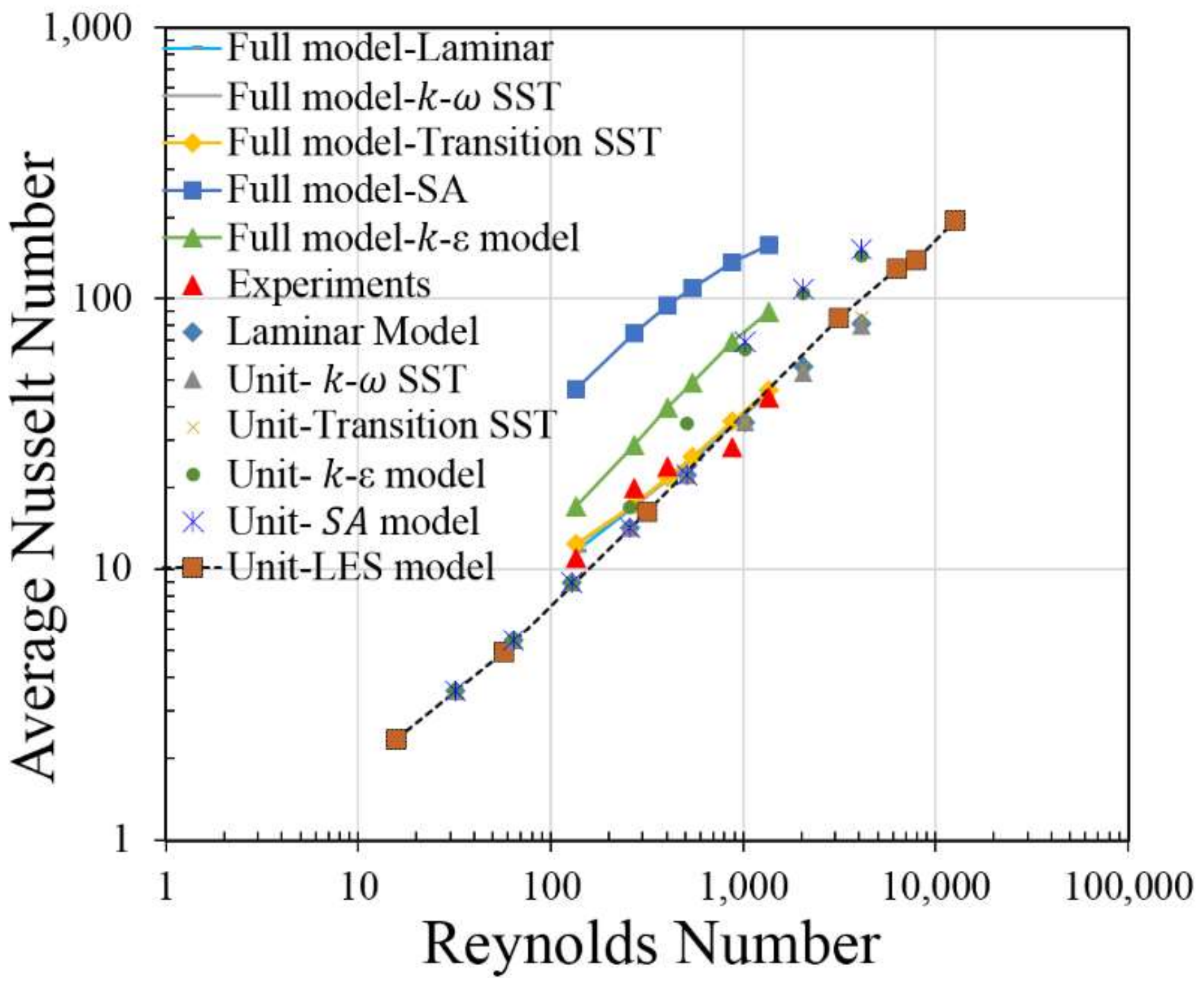

**Figure A.7:** Model comparison with full CFD model, unit cell model and experimental results data on $Nu_{\mathrm{avg}}$-$\mathrm{Re}_d$ number.

### A.5 Conclusions

This appendix A presents the conjugate flow and heat transfer modeling for microjet cooling with locally distributed outlets. We analyzed the turbulence models for their applicability in unit cell level models and full cooler level models for these liquid microjet impingement cooling applications. The results of the different turbulence models based on steady state CFD RANS models for a microjet unit cell are benchmarked with LES results. It is concluded that the transition SST model and $k$-$\omega$ SST model both can accurately predict the average chip temperature, the stagnation temperature on the chip, and the pressure drop with less than 5% difference, compared with the reference LES model. Moreover, the unit cell model is validated with the full cooler level model for different flow rate conditions. However, the usability of the unit cell model changes with the flow rate. A test case with a microjet cooler has been demonstrated by using 3D printing technology in order to validate the numerical simulations of the turbulence models. The experimental results are compared with the unit cell model and full cooler model with different numerical modeling methods.

In summary, the transition SST model and $k$-$\omega$ SST model both show excellent ability to predict the local or average Nu, as well as the local level pressure coefficient $f$ with less than 5% difference in the range of $30 < \mathrm{Re_d} < 4000$, compared with the reference

LES model. For the comparison with experimental measurements, the LES model, transition SST model and $k$-$\omega$ SST model all show less than 25% prediction error as the $\mathrm{Re}_d$ number ranging from 130 to 1400.

**References**

1. Zuckerman, N. Jet Impingement Heat Transfer: Physics, Correlations, and Numerical Modeling. *Adv. Heat Transf.* 2006, *39*, 565–631.
2. Menter, F.R.; Langtry, R.B.; Likki, S.R.; Suzen, Y.B.; Huang, P.G.; Völker, S. A correlation-based transition model using local variable—part I: Model formulation. *J. Turbomach.* 2006, *128*, 413–422.
3. Mishra, A.A.; Iaccarino, G. Uncertainty estimation for reynolds-averaged navier–stokes predictions of high-speed aircraft nozzle jets. *AIAA J.* 2017, *55*, 3999–4004.
4. Granados-Ortiz, F.J.; Arroyo, C.P.; Puigt, G.; Lai, C.H.; Airiau, C. On the influence of uncertainty in computational simulations of a high-speed jet flow from an aircraft exhaust. *Comput. Fluids* 2019, *180*, 139–158.
5. Mishra, A.A.; Mukhopadhaya, J.; Iaccarino, G.; Alonso, J. 2018. Uncertainty Estimation Module for Turbulence Model Predictions in SU2. *AIAA J.* 2018, *57*, 1066–1077.

# Appendix B: Grid Sensitivity Analysis

## B.1 Meshing independent criteria

In order to investigate the hydraulic and thermal phenomena in the cooler numerically, conjugated heat transfer computational fluid dynamic (CFD) models have been created in Ansys-Fluent for both the unit cell model and full cooler level model.

For the meshing of the CFD models, hybrid meshing is chosen in the simulation. The fluid domain mesh is chosen as tetrahedron mesh cells. The first layer thickness for the boundary layer along the wall is 1 μm to make sure the grid is fine enough to get $y^+$ in the viscous sublayer. The grid convergence index (GCI) is used for the meshing sensitivity analysis.

Grid convergence order:

$$p = \frac{\ln\left(\frac{f_3 - f_2}{f_2 - f_1}\right)}{\ln r} \quad (1)$$

where $p$ is the order of computational method. These solutions ($f_3$; $f_2$; $f_1$) are computed over three different grid levels ($\bar{h}_3$; $\bar{h}_2$; $\bar{h}_1$), which are subsequently refined according to a constant grid refinement ratio $r$, shown as $\bar{h}_1 = \frac{\bar{h}_2}{r} = \frac{\bar{h}_3}{r^2}$.

Once the order of convergence $p$ is known, the Grid convergence index (GCI) can be calculated by using two subsequent results. In particular, if $f_3$ and $f_2$ are used and the final reported result is $f_3$, the one on the coarsest grid is defined as below:

$$GCI = \frac{F_s r^p}{r^p - 1} \left| \frac{f_3 - f_2}{f_2} \right| \quad (2)$$

where the $F_s$ is a safety factor. Fs = 1:25 in case three grid levels. It is also important to be sure that the selected grid levels are in the asymptotic range of convergence for the computed solution. The check for asymptotic range is evaluated using the equation as below:

$$\frac{GCI_{23}}{r^p GCI_{12}} \approx 1 \quad (3)$$

where $GCI_{23}$ and $GCI_{12}$ are the values of GCI computed by considering, respectively, $f_2$; $f_3$ and $f_1$; $f_2$.

### B.2 GCI analysis for different test cases

#### B.2.1 Unit cell meshing sensitivity

For the unit cell model, extensive design of experiments are performed for the dimensionless analysis and parametric analysis. Therefore, meshing sensitivity is needed to make sure the modeling results are mesh independent. Table B.1 lists the GCI analysis for different nozzle array for a fixed cavity height. The final chosen mesh size is 0.02 mm for the nozzle array from 4×4 to 32×32. This is due to the meshing of the cavity height is more dominated.

**Table B.1:** GCI analysis for the unit cell model with different nozzle array ($d_i$/L=0.3, H=0.6 mm, Vin=1.47m/s).

| Nozzle array | Mesh size | $GCI_{12}$ | Asymptotic range of convergence |
|---|---|---|---|
| $4 \times 4$ | 0.02 | 0.0000 | 1.0000 |
| $6 \times 6$ | 0.02 | -0.0002 | 0.9990 |
| $8 \times 8$ | 0.02 | 0.0002 | 0.9999 |
| $10 \times 10$ | 0.02 | -0.0013 | 0.9994 |
| $12 \times 12$ | 0.02 | 0.0001 | 1.0000 |
| $20 \times 20$ | 0.02 | 0.0008 | 0.9995 |
| $32 \times 32$ | 0.02 | 0.0008 | 0.9991 |

#### B.2.2 Full cooler level model meshing sensitivity

For the full cooler level model, the meshing sensitivity analysis is also performed for the different demonstrators, including the single jet cooler model, micromachined cooler model, 3D printed cooler model, hotspots targeted cooler model, 2.5D interposer model and the large die cooler model. For the fluid domain of the full cooler model, the minimal mesh size is determined by the nozzle diameter. Since the design for micromachined cooler, 3D printed cooler, interposer cooler and large die cooler are all based on the cooling unit cell with 2×2 $mm^2$ for a nozzle diameter of 0.6 mm. For the meshing of the solid domain (silicon die) with 0.2 mm thickness, the minimal mesh size is set as 0.02 mm.

#### B.2.2.1 Single jet cooler model

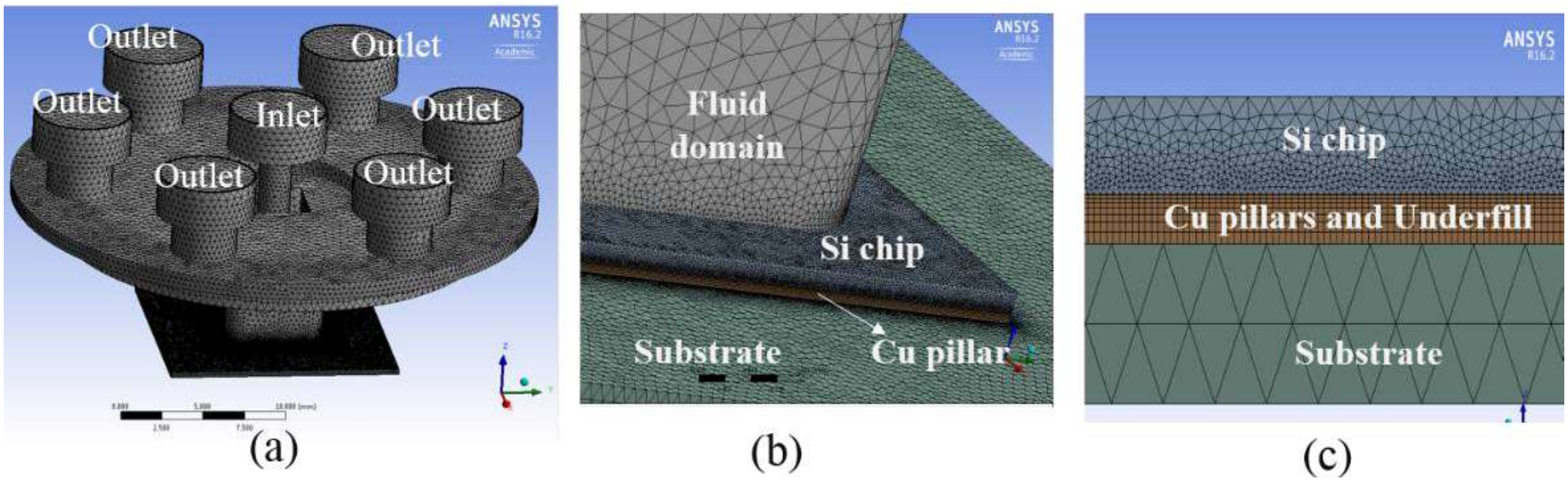

**Figure B.1:** Meshing details of the full single jet cooler model: (a) fluid domain meshing with one inlet nozzle and 6 outlet nozzles; (b) and (c) details of the bottom package mesh including the Cu pillar and large size substrate.

In Table B.3, the GCI analysis for both the stagnation temperature and the average temperature is investigated. For the GCI analysis of the single jet model, the final chosen minimal mesh size is 0.15 mm. The total element number for single jet model is 2.5 million.

**Table B.2:** Grid convergence index analysis for the model of the single jet cooler.

| Grid size | Nozzle exit average velocity -Line 1 | Stagnation temperature-Line 2 | Element number |
|---|---|---|---|
| 1H (75um) | 3.24m/s | 19.78°C | 7M |
| 2H (150um) | 3.24m/s | 19.81°C | 2.5M |
| 4H (300um) | 3.23m/s | 20.95°C | 1.3M |

**Table B.3:** Grid convergence index analysis for the model of the single jet cooler.

| Temperature | $GCI_{12}$ | Asymptotic range of convergence |
|---|---|---|
| Stagnation Temp | 0.0019 | 0.9984 |
| Averaged Temp | 0.0043 | 1.0012 |

### B.2.2.2 3D printed cooler model

For the meshing of the 3D printed cooler, the detailed heater cells are also included in the full model. Since the nozzle array is 4×4 jet array, the cooling unit cell is 2×2 $mm^2$ for a nozzle diameter of 0.6 mm. The meshing sensitivity can be used for the other full model with the same critical inlet nozzle diameter and cooling unit cell.

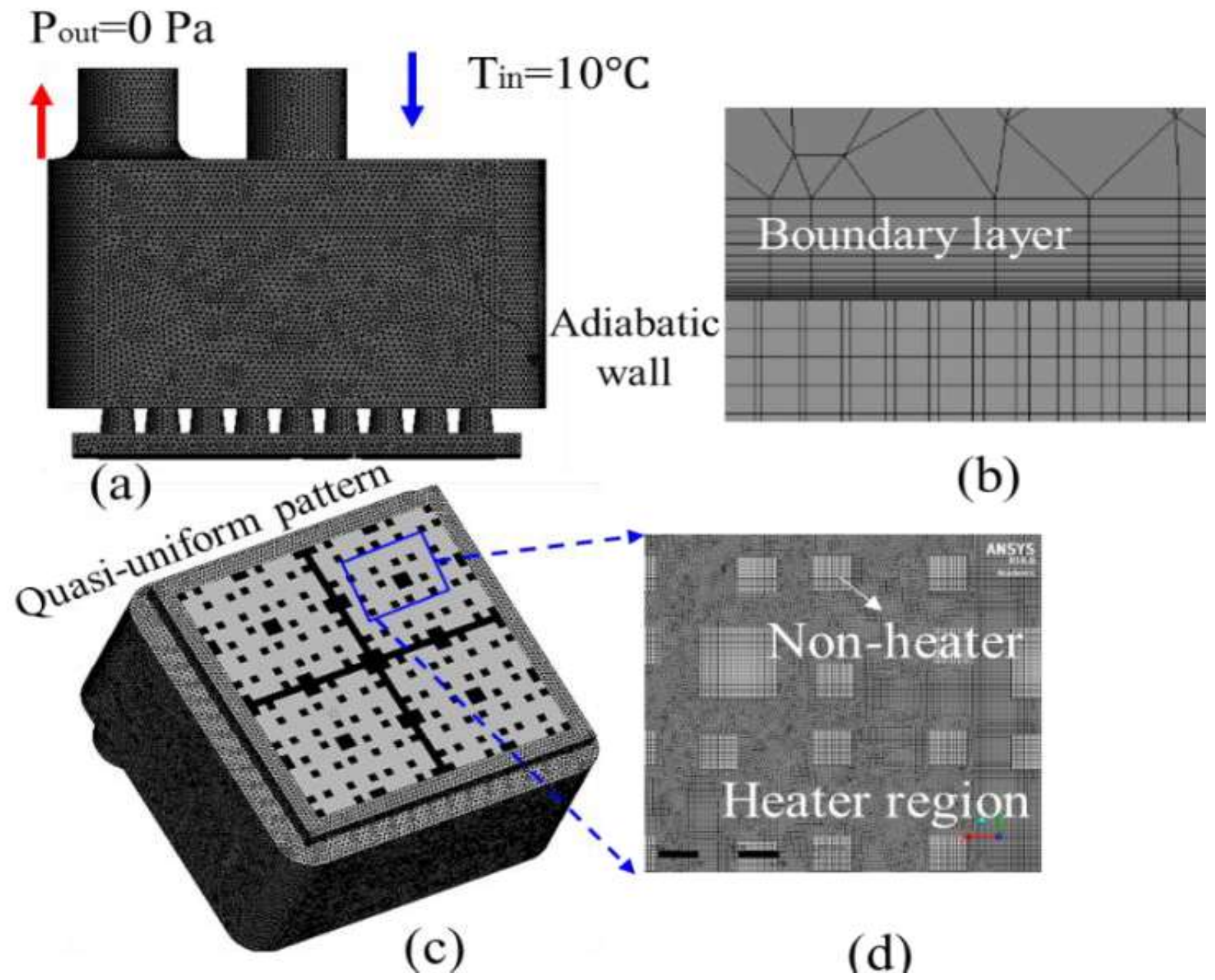

**Figure B.2:** CFD models: a) transparent view and b) meshing of full 4×4 nozzle array models; c) meshing of a single jet cooler with one inlet nozzle and 6 outlet nozzles; d), e) and f) details of the boundary layer and heater cell meshing.

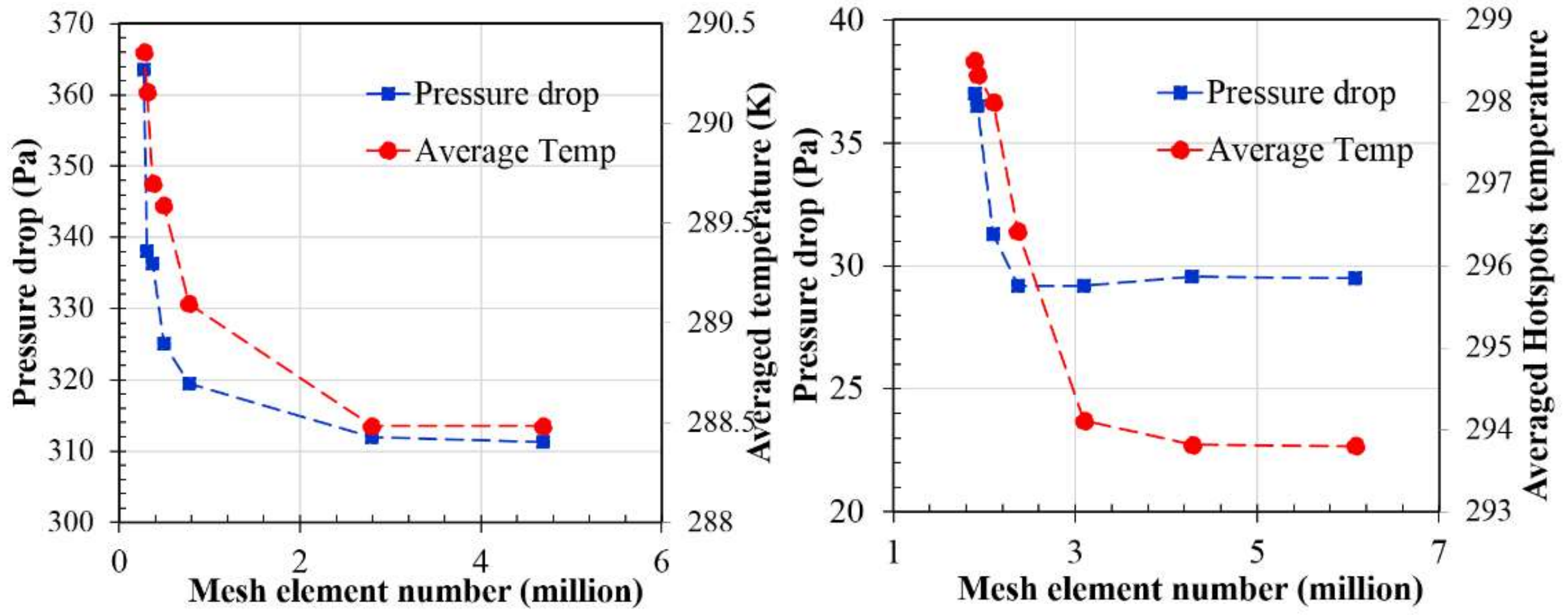

(a) 3×3 nozzle array; (b) 4×4 nozzle array

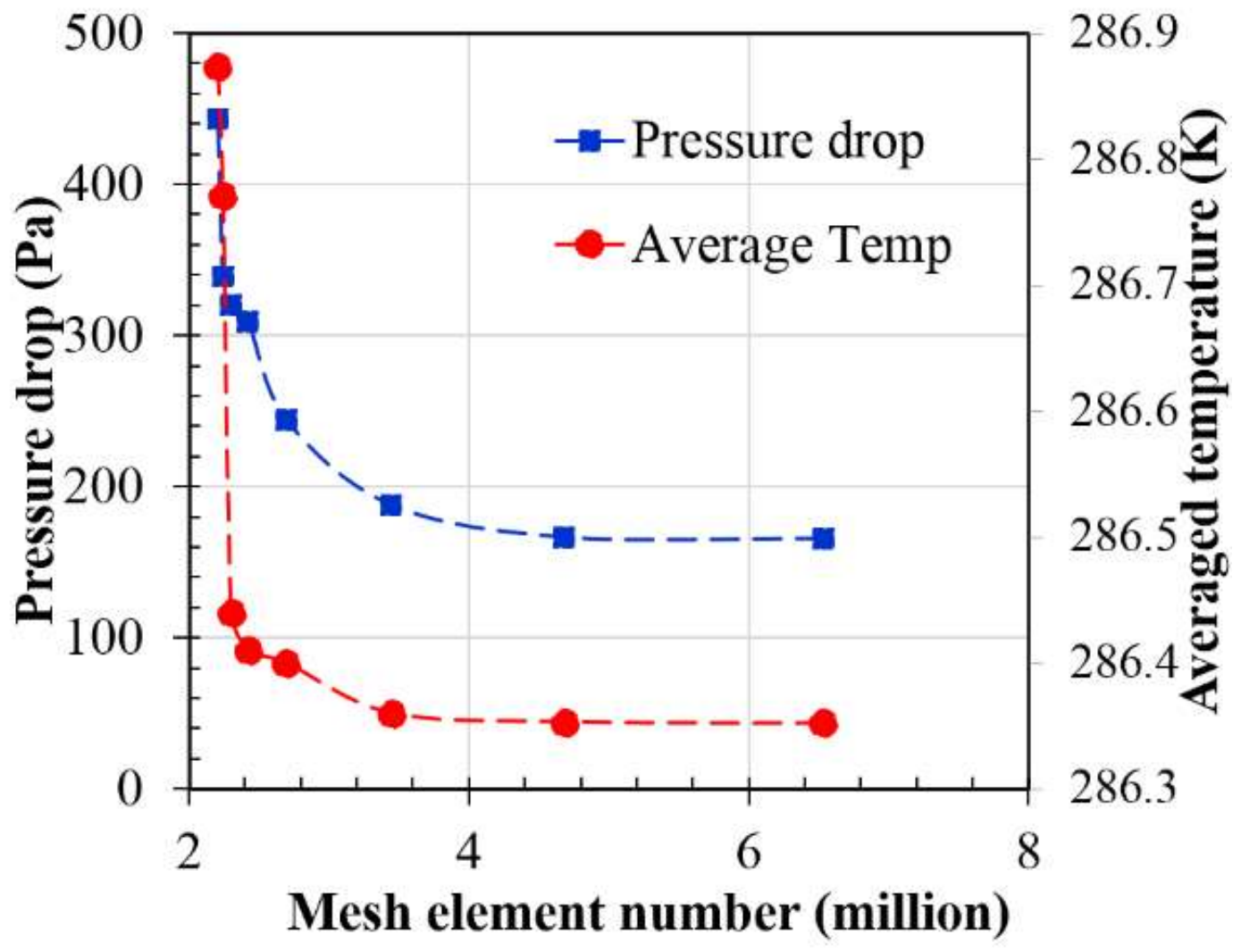

(c) 8×8 nozzle array

**Figure B.3:** Meshing sensitivity of 4×4 full cooler model: 4.2 million for 4x4 3D printed jet cooler

Based on the meshing sensitivity analysis, the minimal mesh size for the 4×4 array cooler is chosen as 0.12 mm. The minimal mesh size for the 3×3 array cooler is 0.12 mm. And also, the minimal mesh size for the 8×8 array cooler is 0.12 mm. Therefore, the final element number is 4.2 million element size for 4×4 nozzle array, 2.8 million element size for 3×3 nozzle array, and 4.7 million element size for 8×8 nozzle array.

**Table B.4:** 3×3 nozzle array.

| Element number | Mesh size | Averaged temp (K) | Pressure drop (kPa) |
|---|---|---|---|
| 4688151 | 0.1 | 288.483 | 311.254 |
| 2791158 | 0.12 | 288.4825 | 311.922 |
| 770137 | 0.2 | 289.094 | 319.456 |
| 491047 | 0.25 | 289.588 | 324.999 |
| 371075 | 0.3 | 289.698 | 336.329 |
| 309965 | 0.35 | 290.158 | 338 |
| 273964 | 0.4 | 290.357 | 363.549 |

**Table B.5:** 4×4 nozzle array

| Element number | Mesh size | Averaged temp (K) | Pressure drop (kPa) |
|---|---|---|---|
| 6077080 | 0.1 | 293.8 | 29.5 |
| 4287043 | 0.12 | 293.817 | 29.5644 |
| 3097673 | 0.15 | 294.11 | 29.2076 |
| 2367794 | 0.2 | 296.421 | 29.1959 |
| 2101682 | 0.25 | 297.994 | 31.2765 |
| 1925688 | 0.35 | 298.335 | 36.5023 |
| 1891717 | 0.4 | 298.5 | 37 |

**Table B.6:** 8×8 nozzle array.

| Element number | Mesh size | Averaged temp (K) | Pressure drop (kPa) |
|---|---|---|---|
| 6533921 | 0.1 | 286.352 | 165.73 |
| 4681619 | 0.12 | 286.353 | 166.603 |
| 3447331 | 0.15 | 286.36 | 187.79 |
| 2693583 | 0.2 | 286.4 | 244.38 |
| 2421858 | 0.25 | 286.41 | 308.937 |
| 2302413 | 0.3 | 286.44 | 320 |
| 224238 | 0.35 | 286.771 | 339.331 |

**Table B.7: G**rid convergence index (GCI) based on the averaged chip temp.

| Nozzle array | $GCI_{12}$ | $GCI_{23}$ | Asymptotic range of convergence |
|---|---|---|---|
| 3×3 array | 0.0051 | 0.0106 | 0.9979 |
| 4×4 array | -0.0428 | -0.0336 | 0.9912 |
| 8×8 array | 0.0002 | 0.0023 | 0.9998 |

### B.2.2.3 Micromachined cooler model

Based on the mesh sensitivity analysis of the 3D printed cooler model with cooling unit cell of 2×2 $mm^2$, the minimal mesh size should be the same for the micromachined cooler. The results are determined by the minimal size of the full cooler model, which is the critical parameter of inlet nozzle diameter 0.6 mm. Therefore, the minimal mesh size should be 0.12 mm for the cooling unit cell of 2×2 $mm^2$ with nozzle diameter of 0.6 mm, for the design of 4×4 nozzle array cooler. The final element number is 4.9 million for 4×4 micromachined cooler, while the solid part with detailed heaters are 0.02 mm.

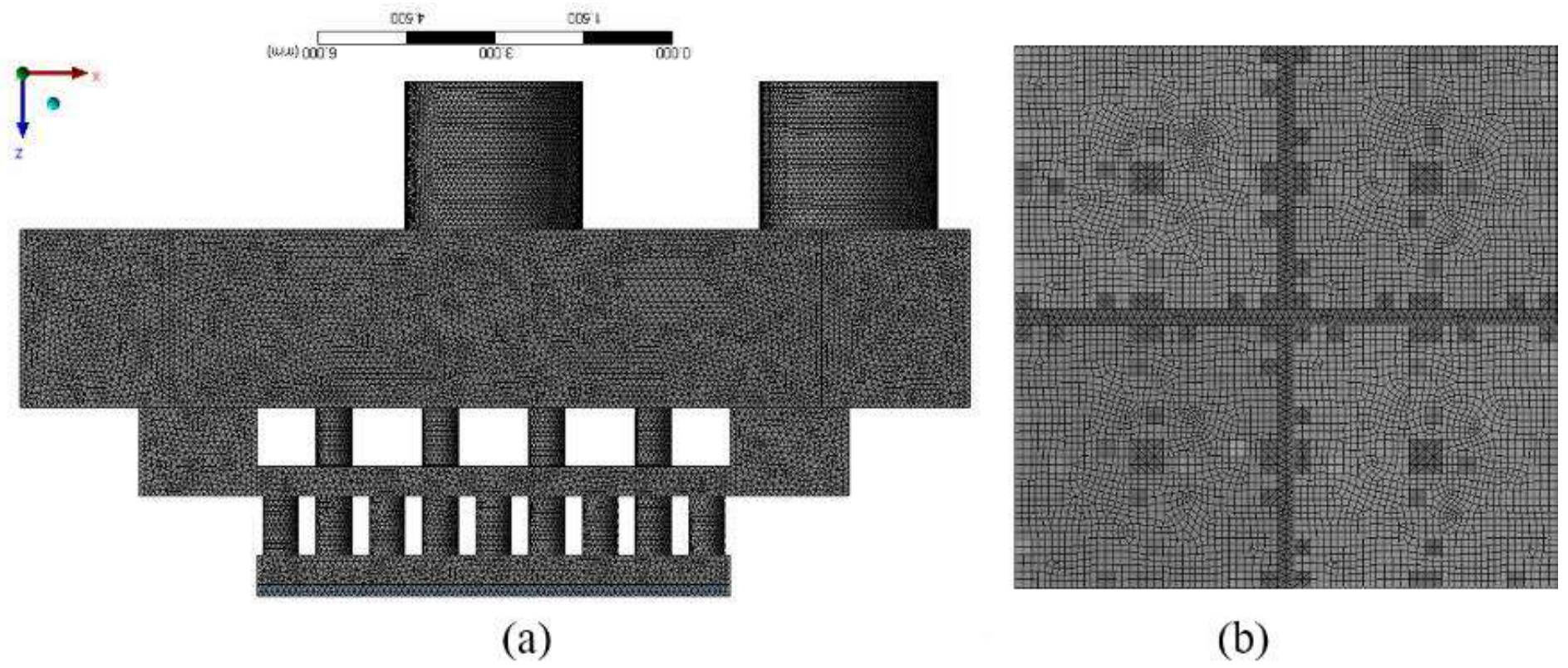

**Figure B.4:** Meshing details for the micromachined cooler and the top view of the thermal test die.

### B.2.2.4 Hotspots targeted cooler model

**Table B.9:** Meshing comparison between the design 1 and design 2.

| | Reference case | Test case 1 | Test case 2 |
|---|---|---|---|
| Fluid domain | 0.12 mm | 0.12 mm | 0.12 mm |
| Solid domain | 0.02 mm | 0.02 mm | 0.02 mm |
| First layer thickness | 1e-3 mm | 1e-3 mm | 1e-3 mm |
| Layer number | 10 | 10 | 10 |
| Growth ratio | 1.5 | 1.5 | 1.5 |
| Total element size | 4.7 M | 4.3 M | 4.5 M |

For the hotspots targeted cooler, the nozzle array is 8×8 nozzle with cooling unit cell of 1×1 $mm^2$, for a nozzle diameter of 0.3 mm. It is observed that the mesh for a number of elements between 4.5M and 5 M is mesh-independent. In section 2.2.2, the meshing sensitivity of the 8×8 nozzle array is analyzed, showing minimal mesh size of 0.12 mm. Since the critical region with the nozzle diameter is the same, therefore, the meshing sensitivity is also applicable for other cases. The minimal mesh size for the hotspots targeted cooler is chosen as 0.12 mm, with total meshing element number of 4.7 M.

### B.2.2.5 2.5D interposer cooler model

The meshing details for the interposer cooler with two designs are illustrated in Figure B.5. As discussed in Section 2.2.2, the 2.5D interposer cooler includes two 3D printed cooler with 4×4 nozzle array, where the minimal mesh size is 0.12 mm for single die 3D printed cooler model. Thus, the minimal size for the interposer cooler is also selected as 0.12 mm.

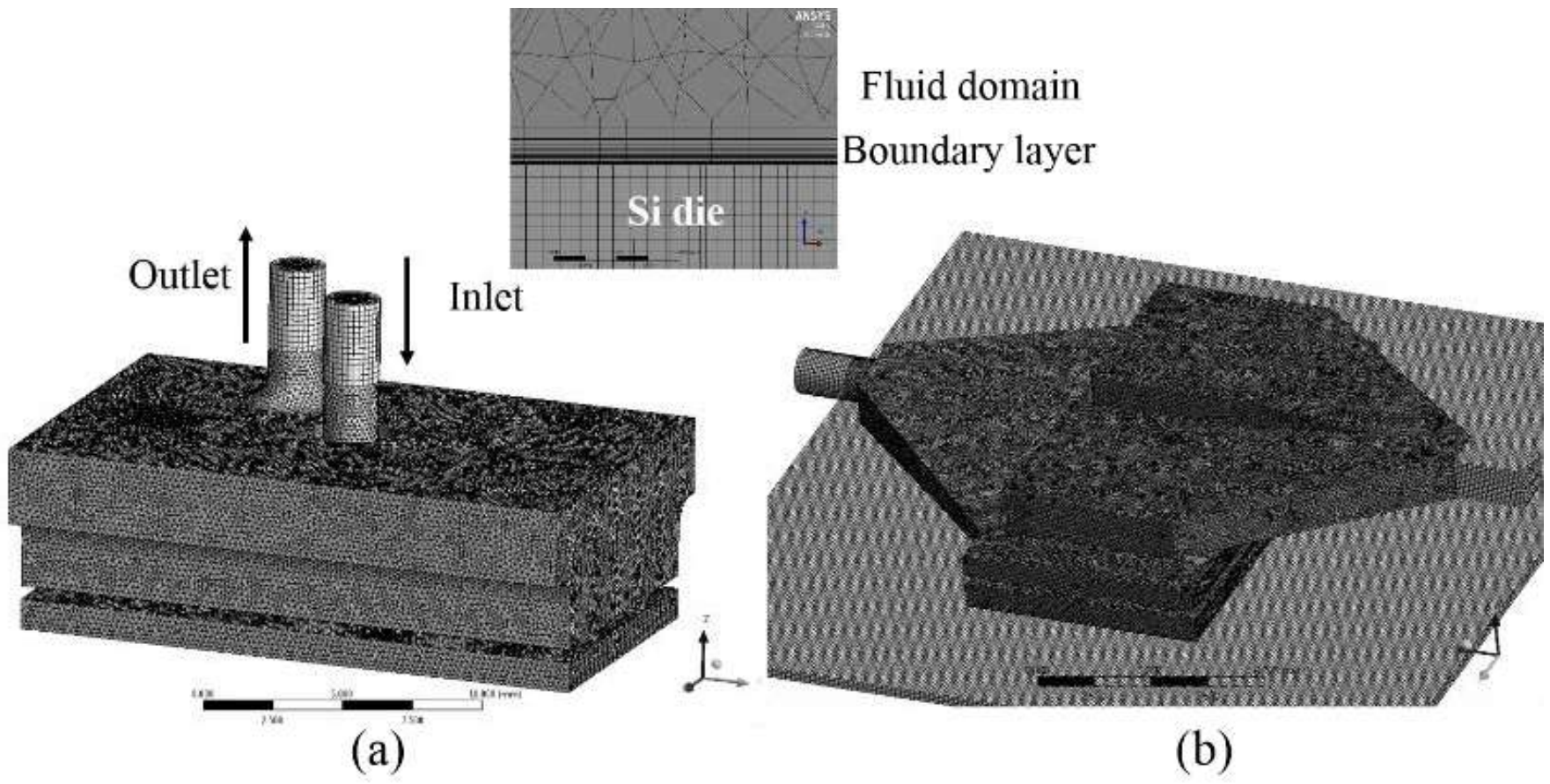

**Figure B.5:** Details of the CFD model for vertical and lateral cooler design with a) vertical feeding manifold and b) lateral feeding manifold.

**Table B.10:** Meshing comparison between the design 1 and design 2.

| | Vertical feeding | Lateral feeding |
|---|---|---|
| Fluid domain | 0.12 mm | 0.12 mm |
| First layer thickness | 1e-3 mm | 1e-3 mm |
| Layer number | 10 | 10 |
| Growth ratio | 1.5 | 1.5 |
| Total element size | 35537266 | 8662192 |

### B.2.2.6 Large die cooler model

**Table B.11:** Meshing comparison between the design 1 and design 2.

| | Design 1 | Design 2 |
|---|---|---|
| Fluid domain | 0.15 mm | 0.15 mm |
| First layer thickness | 1e-3 mm | 1e-3 mm |
| Layer number | 10 | 10 |
| Growth ratio | 1.5 | 1.5 |
| Total element size | 18,279,453 | 15,201,961 |

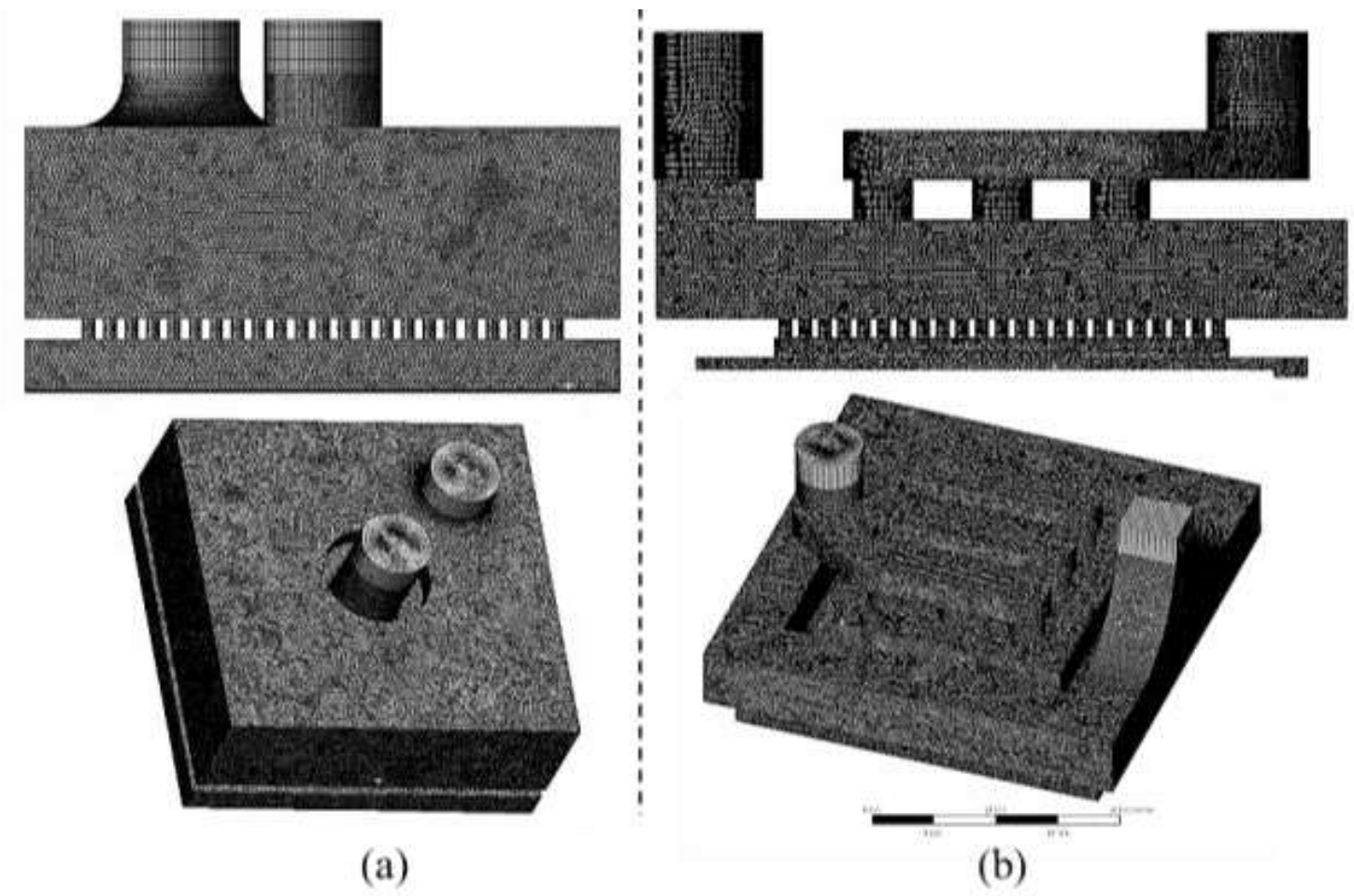

**Figure B.6:** The mesh is extracted from the 3D printed: 4×4 jet array model:

# List of Publications

## Journal Publications

1. Tiwei Wei, Herman Oprins, et al., "High-Efficiency Polymer-Based Direct Multi-Jet Impingement Cooling Solution for High-Power Devices," in IEEE Transactions on Power Electronics, vol. 34, no. 7, pp. 6601-6612, July 2019.
2. Tiwei Wei, Herman Oprins, et al., "Experimental characterization and model validation of liquid jet impingement cooling using a high spatial resolution and programmable thermal test chip [J]", Applied Thermal Engineering, vol. 152, pp. 308-318, April 2019.
3. Tiwei Wei, Herman Oprins, et al., "Experimental Characterization of a Chip Level 3D printed Microjet Liquid Impingement Cooler for High Performance Systems [J]", in IEEE Transactions on Components, Packaging and Manufacturing Technology, vol. 9, no. 9, pp. 1815-1824, Sept. 2019.
4. Tiwei Wei, Herman Oprins, et al., "Experimental and numerical investigation of direct liquid jet impinging cooling using 3D printed manifolds on lidded and lidless packages for 2.5D integrated systems [J]", Applied Thermal Engineering, Volume 164, 5 January 2020, 11453.
5. Tiwei Wei, Herman Oprins, et al., "Low-cost Energy Efficient On-chip Hotspot Targeted Microjet Cooling for High Power Electronics [J]", IEEE Transactions on Components, Packaging and Manufacturing Technology, Digital Object Identifier: 10.1109/TCPMT.2019.2948522.
6. Tiwei Wei, Herman Oprins, et al., "Conjugate Heat Transfer and Fluid Flow Modeling for Liquid Microjet Impingement Cooling with Alternating Feeding and Draining Channels". Fluids 2019, 4, 145.
7. Herman Oprins, Tiwei Wei, et al., "A cold shower for chips [J]", Chip Scale Review, Volume 22, Number 6, November - December, 2018: Page.26.

## Conference Proceedings

1. Tiwei Wei, Herman Oprins, et al., "Demonstration of Package Level 3D-printed Direct Jet Impingement Cooling applied to High power, Large Die Applications", ECTC-2020. (accepted)
2. Tiwei Wei, Herman Oprins, et al., "Thermal analysis of polymer 3D printed jet impingement coolers for high performance 2.5D Si interposer packages [C]", *IEEE-ITherm* 2019.
3. Tiwei Wei, Herman Oprins, et al., "First Demonstration of a Low

Cost/Customizable Chip Level 3D Printed Microjet Hotspot-Targeted Cooler for High Power Applications [C]", *2019 Electronic Components and Technology Conference ECTC.*

4. Tiwei Wei, Herman Oprins, et al., "High efficiency direct liquid jet impingement cooling of high power devices using a 3D-shaped polymer cooler [C]", *2017 IEEE International Electron Devices Meeting (IEDM).*
5. Tiwei Wei, Herman Oprins, et al., "3D Printed Liquid Jet Impingement Cooler: Demonstration, Opportunities and Challenges [C]", *2018 Electronic Components and Technology Conference (ECTC).*
8. Tiwei Wei, Herman Oprins, et al., "Nozzle array scaling analysis of the thermal performance of liquid jet impingement coolers for high performance electronic applications [c]", *The International Heat Transfer Conferences (IHTC)*, 2018.